\documentclass [11pt, twoside,tightenlines] {uwthesis}[2012/06/19]
\usepackage{amssymb,latexsym,mathbbol}
\usepackage{amsmath,amsbsy}
\usepackage{epsfig,bm}
\usepackage{graphicx,comment}
\usepackage[figuresright]{rotating}
\usepackage{subfigure}
\usepackage{caption}
\usepackage{hyperref}
\usepackage{color}
\usepackage{units}
\usepackage{lipsum}
\usepackage{afterpage}
\usepackage{amsthm}
\usepackage{multirow}
\usepackage{cite}
\usepackage{bibentry}

\newcommand{\gsim}{\raisebox{-0.7ex}{$\stackrel{\textstyle >}{\sim}$ }}
\newcommand{\lsim}{\raisebox{-0.7ex}{$\stackrel{\textstyle <}{\sim}$ }}

\def\si{^1 \hskip -0.03in S _0}
\def\siii{^3 \hskip -0.025in S _1}
\def\diii{^3 \hskip -0.03in D _1}

\def\pislash{ {\pi\hskip-0.6em /} }
\def\nopi{ {\rm EFT}(\pislash) }

\usepackage{alltt}  %

\begin{document}
 
%
%

\prelimpages
 
%
%
\Title{Formal Developments for Lattice QCD
\\
with Applications to Hadronic Systems}
\Author{Zohreh Davoudi}
\Year{2014}
\Program{Department of Physics}

\Chair{Martin J. Savage}{Professor}{Department of Physics}
\Signature{David B. Kaplan}
\Signature{Stephen R. Sharpe}




{\Degreetext{A dissertation%
  \\
  submitted in partial fulfillment of the\\ requirements for the degree of}
 \def\thefootnote{\fnsymbol{footnote}}
 \let\footnoterule\relax
 \titlepage
 }
\setcounter{footnote}{0}

 
%
%

%
%

\setcounter{page}{-1}
\abstract{%
In order to make reliable predictions with controlled uncertainties for a wide range of nuclear phenomena, a theoretical bottom-up approach, by which hadrons emerge from the underlying theory of strong interactions, quantum chromodynamics (QCD), is desired. The strongly interacting quarks and gluons at low energies are responsible for all the dynamics of nucleons and their clusters, the nuclei. The theoretical framework and the combination of analytical and numerical tools used to carry out a rigorous non-perturbative study of these systems from QCD is called lattice QCD. The result of a lattice QCD calculation corresponds to that of nature only in the limit when the volume of the spacetime is taken to infinity and the spacing between discretized points on the lattice is taken to zero. A better understanding of these discretization and volume effects, not only provides the connection to the infinite-volume continuum observables, but also leads to optimized calculations that can be performed with available computational resources. This thesis includes various formal developments in this direction, along with proposals for novel improvements, to be used in the upcoming LQCD studies of nuclear and hadronic systems.

As the space(time) is discretized on a (hyper)cubic lattice in (most of) lattice QCD calculations, the lattice correlation functions are not fully rotationally invariant. This is known to lead to mixing between operators (those interpolating the states or inserting external currents) of higher dimensions with those of lower dimensions where the coefficients of latter diverge with powers of inverse lattice spacing, $a$, as the continuum limit is approached. This issue has long posed computational challenges in lattice spectroscopy of higher spin states, as well as in the lattice extractions of higher moments of hadron structure functions. We have shown, through analytical perturbative investigations of field theories, including QCD, on the lattice that a novel choice of operators, smeared over several lattice sites and deduced from a continuum angular momentum, has a smooth continuum limit. The scaling of the lower dimensional operators is proven to be no worse than $a^2$, explaining the success of recent numerical studies of excited state spectroscopy of hadrons with similar choices of operators. These results are presented in chapter \ref{chap:operators} of this thesis.

Due to Euclidean nature of lattice correlation function, the physical scattering parameters must be obtained via an analytical continuation to Minkowski spacetime. However, this continuation is practically impossible in the infinite-volume limit of lattice correlation function except at the kinematic thresholds. A formalism due to Lu\"scher overcomes this issue by making the connection between the finite-volume spectrum of two interacting particles and their infinite-volume scattering phase shifts. We have extended the L\"uscher methodology, using an effective field theory approach, to the two-nucleon systems with arbitrary spin, parity and total momentum (in the limit of exact isospin symmetry) and have studied its application to the deuteron system, the lightest bound states of the nucleons, by careful analysis of the finite-volume symmetries. A proposal is presented that enables future precision lattice QCD extraction of the small D/S ratio of the deuteron that is known to be due to the action of non-central forces. By investigating another scenario, we show how significant volume improvement can be achieved in the masses of nucleons and in the binding energy of the deuteron with certain choices of boundary conditions in a lattice QCD calculation of these quantities. These results are discussed in chapters \ref{chap:NN}, \ref{chap:deuteron} and \ref{chap:TBC}.

In order to account for electromagnetic effects in hadronic systems, lattice QCD calculations have started to include quantum electrodynamic (QED). These effects are particularly interesting in studies of mass splittings between charged and neutral members of isospin multiplets, e.g. neutral and charged pions. Due to the infinite range of QED interactions large volume effects plaque these studies. Using a non-relativistic effective theory for electromagnetic interactions of hadrons, we analytically calculate, and numerically estimate, the first few finite-volume corrections (up to $1/L^4$ where $L$ is the spatial extent of the volume) to the masses of hadrons and nuclei at leading order in the QED coupling constant, but to all orders in the short-distance strong interaction effects. These results are presented in chapter \ref{chap:EM}.}

%
%
\tableofcontents
\listoffigures
 
%
 
%
%
\acknowledgments{
  
I would like to thank my advisor, Prof. Martin J. Savage, for four years of incredible effort to teach, advise and support me. I entered the graduate school with a true passionate for physics but a limited vision of what the important problems are to invest on; the problems that would immediately impact our understanding of nature and its laws. Today, I am leaving the graduate school with a magnified passionate for physics and with a far better understanding of the physical world, and what my role is going to be in scientifically exploring it as a physicist. This non-trivial transition could not have happened in such a short period of time if I did not have the opportunity to have him as an incredible teacher when it came to Physics and as a great advisor when it came to every aspect of my academic life. I have enjoyed every moment of my time as his student, even in the toughest times he put me through, since I knew how much he cared to build my strength, expand my knowledge and turn me to a better physicist.

I am indebted to my parents for what I have achieved today. Their role in every achievement in my life, and in signifying my desire to become a scientist at the first place, can not be overestimated. They always set high standards for me, and although always appreciated, better than anybody else, what I have achieved, it never stopped them from encouraging me to work better and fight harder. 

I am grateful to my beloved husband for his influential role in making this journey happier and easier for me. His encouragement and support have been always more than effective in making me develop trust in myself and in my abilities. He always knew when to give me some distance to let me focus on my work and when to interrupt me to make sure I maintain my sanity by offering me a break; after all it is not the easiest thing in this world to be a physicists' life partner. His understanding and cooperation can not be overstated.

I would like to thank all my professors, those whom I had learned from and advised by in Iran in particular Prof. Neda Sadooghi, and those whom I have had the opportunity to talk to and learn great physics from in the US, including Profs. David Kaplan, Stephen Sharpe, Silas Beane, Andrea Karch, Dam Son, Eric Adelberger and Joseph Kapusta and many others. I would like to thank all my colleagues and collaborators whom I have talked to and enjoyed discussing physics with, in particular Thomas Luu, Maxwell Hansen, Emmanuel Chang and Michael Wagman. I am specially grateful for a few enjoyable years of working with Ra\'ul Brice\~no with whom I experienced the joy of trying to figure things out together. I will never forget our intense arguments, our coordinated progress in understanding different aspects of the projects and accomplishing things as a result of all these.

I am grateful to my wonderful friends, Behzad and Kimia, for providing mental and intellectual support through the past few years, with whom I felt part of a family once again and Seattle felt like a home. I deeply appreciate the encouragement and emotional support of my dear siblings, Farideh, Monireh, Omid, Saeedeh and Saeed, throughout years and am thankful to them for being always caring for my life and my career.
}

%
%
\dedication{

\
\

This dissertation is dedicated to all the amazing teachers I have encountered throughout my life, the first of whom being my parents, Hassan Davoudi and Fariba Lotfi, and the most recent of whom being my PhD advisor, Martin J. Savage. Teachers from whom I learned that thoughts and ideas are central to my life and my identity as a human, and that what can not be substituted by anything else in this world is a thinking imaginative mind.
}

%

%
%

\textpages

\chapter {INTRODUCTION}
\label{chap:intro}

Since the discovery of the nucleus in 1911 by Ernest Rutherford, which established  the field of nuclear physics, our understanding of the properties of nuclei, their structure and and their underlying interactions has constantly evolved. Yet, it remains an  exciting frontier of research and discovery in the modern theoretical and experimental physics. While the development of quantum mechanics revolutionized this field in its early stages, the establishment of the standard model (SM) of particles and interactions led to the next prominent revolution that although started as early as in early 1970s, it has not yet completed forty years later. The realization of the strong interactions among quarks and gluons -- the building blocks of nucleons -- as the underlying mechanism for the structure of nucleons and the forces among them, has opened up new frontiers in nuclear physics -- from studies of the spectra of exotic strongly interacting particles to the explorations of  new phases of matter, e.g., in heavy ion collisions experiments. Moreover, the theory of strong interactions, quantum chromodynamics (QCD), is starting to base traditional studies of nuclear structure and reactions on a rigorous ground such that uncontrolled approximations due to the lack of the knowledge of the physics at short distances will be eliminated in modern-day nuclear calculations.

It is known that strong interactions are described by a local, non-Abelian, SU(3) gauge theory, within which all hadronic phenomena can, in principle, be predicted once a few input parameters are set to their physical values. These parameters include the masses of those quarks that are kinematically allowed in a given process, and the strength of the QCD bare coupling constant, or in turn the QCD scale, $\Lambda_{\rm QCD}$. Including electromagnetism, which plays an important role in nuclear physics, only introduces one more parameter, the strength of the quantum electrodynamics (QED) coupling. 
As will be discussed in Sec. \ref{QCD}, QCD is an asymptotically free theory meaning that its coupling becomes weak at high energies -- at energies above the QCD scale.  Moreover, at low energies, the theory is confining. In this limit, quarks and gluons form clusters of hadrons; mesons and baryons - entities that are neutral with respect to the color charges of the underlying gauge symmetry group. This remarkable feature, along with the running of the QCD coupling towards larger values, prohibits the use of standard perturbative methods in studying (most of) nuclear physics phenomena from first-principle QCD calculations. This is in sharp contrast with electroweak interactions whose contributions can be accounted for perturbatively.

Historically, the first attempts to study nuclear and hadronic phenomena in a model-independent way were the invention and development of effective field theories (EFTs). Such theories, exhibit the symmetry and symmetry breaking patterns of QCD and are formulated in terms of low-energy degrees of freedom of QCD, namely the low-lying hadrons. As long as the detail of the short distance interactions are not of interest, their effects can be \emph{effectively} included via several unknown parameters at low energies. The required precision of the calculations dictates the number of unknown parameters to be determined via fits to experimental data. In several important cases, such as in three (multi)-hadron systems, there are not enough data available to constrain these parameters well. Despite this limitation, such theories have greatly changed the perspective of nuclear physics with regard to few(many)-nucleon calculations. Examples of these theories will be presented in Sec. \ref{LE-QCD}. What is prominent about these EFTs is their interplay with the first-principle lattice QCD (LQCD) calculations as they (will) play a role in filling the gap between lattice calculations of few-hadron systems and conventional nuclear calculations of many-body systems.  

To date, the only fully predictive, non-perturbative method for studying QCD at low energies is lattice QCD. This method is based on a numerical evaluation of the QCD path integral using Monte Carlo techniques as will be introduced in Sec. \ref{LQCD}. In lattice QCD calculations fields are evaluated on a discrete set of spacetime points (sites) and are interacting via a discrete number of link variables that are connecting the adjacent sites, see Sec. \ref{LQCD}. The spacing between the two adjacent sites, $a$, must be sufficiently small to resolve sub-hadronic scales. Its value is not a direct input of the calculations as it is not an independent parameter - it is closely related to the input value of the bare coupling constant through the renormalization group. Moreover, as only a finite number of spacetime points can be computed on any finite computing machines, the volume of spacetime is truncated to finite extents. 
 Physical observables are obtained upon taking the continuum limit as well as the infinite-volume limit. Such extrapolations can be done by performing calculations at multiple lattice spacings and in multiple lattice volumes. However, multiple calculations are expensive due significant computational costs of these calculations. Here the role of EFTs become prominent as they provide the knowledge of the analytic dependence of quantities on lattice spacing, volume and the masses of light quarks, in several cases. This latter dependence matters as most LQCD calculations, in particular those of multi-hadron systems, have not been performed at the physical values of light-quark masses due to limited computational resources. This thesis contains formal topics with regard to such continuum limit and infinite-volume extrapolations. The results presented in this thesis have been/will be utilized to build/improve our understanding of the quantities calculated with LQCD in connection to physical observables.

The most commonly used lattice geometries are hyper-cubes (see Sec. \ref{LQCD}). However, the conceptual and practical problems arising from the explicit breaking of the spacetime symmetries of the continuum theory, down to those of a hyper-cubic lattice theory, remain partly a challenge in the continuum extrapolation of classes of observables calculated using LQCD. One of these challenges is the enhancement of the lower-dimension operators by powers of inverse lattice spacing which obscures the extraction of e.g., the excited spectrum of hadrons with well-defined angular momentum in continuum, as well as the determination of the matrix element of higher \emph{twist} operators in studies of hadronic structure functions. One knows, however, that as the lattice becomes finer, the full spacetime symmetries of the continuum are in fact approximately recovered for observables involving wavelengths that are large compared with the scale of discretization. In chapter \ref{chap:operators}, we explore in detail a novel idea for the construction of lattice operators that guarantees a smooth extrapolation of the correlation functions to the continuum limit. The results presented in this chapter are based on the following publication
\begin{itemize}
\item  Z.~Davoudi and M.~J.~Savage,
  \emph{Restoration of Rotational Symmetry in the Continuum Limit of Lattice Field Theories},
  Phys.\ Rev.\ D {\bf 86}, 054505
  arXiv:1204.4146 [hep-lat] (2012).
  \end{itemize}

Besides the challenges arising from the breakdown of the rotational symmetry due to the cubic boundaries of the volume, the infinite-volume limit of lattice quantities is even less trivial to perform in many cases due to the following reason. To be able to evaluate expectation values in the background of QCD vacuum (the path integral approach) using a Monte Carlo sampling method, it is essential to transform from the Minkowski to the Euclidean spacetime. Consequently lattice correlation functions do not immediately correspond to physical correlation functions. The connection between these two quantities must be set in a non-trivial manner for some physically interesting cases such as scattering processes. This is the subject of the finite-volume formalism for LQCD which will be briefly reviewed in Sec. \ref{IV-intro} of this introduction. What can be immediately extracted from the lattice correlation functions are the FV spectra. The idea is that the calculated discrete energy eigenvalues of, e.g., the interacting two-particle states in a finite volume can be utilized to extract the scattering amplitudes, as long as the multi-particle inelastic channels are not kinematically accessible -- as formulated by \emph{Martin Lu\"scher} and extended by several others to more general cases. Examples of the application of this method in modern-day LQCD calculations of two-hadron systems will be presented in Sec. \ref{IV-intro}. A derivation of a general form of the L\"uscher relation, applicable to coupled-channel systems in the moving frame and with arbitrary partial waves, is presented in this section. This derivation closely follows that of presented in the this publication

\begin{itemize}
\item R.~A.~Brice\~no and Z.~Davoudi,
  \emph{Moving Multi-Channel Systems in a Finite Volume with Application to Proton-Proton Fusion},
  Phys.\ Rev.\ D {\bf 88}, 094507,
    arXiv:1204.1110 [hep-lat] (2012). 
\end{itemize}

Chapters \ref{chap:NN} and \ref{chap:deuteron} of this thesis is devoted to the development of a formal framework that enables the extraction of phase shifts and scattering parameters of general two-nucleon systems with arbitrary spin, parity and center of mass (CM) momenta. This formalism directly impacts our ability to extract the properties of the lightest bound state of nucleons, and one of the most unusual ones, the deuteron, from first-principle lattice QCD calculations. In particular, it provides guidance for the the upcoming LQCD calculations to optimize the extraction of the S-D mixing parameter in the deuteron channel. These chapters are based on the following publications
\begin{itemize}
\item R.~A.~Brice\~no, Z.~Davoudi and T.~C.~Luu,
  \emph{Two-nucleon systems in a finite volume: (I) Quantization conditions},
  Phys.\ Rev.\ D {\bf 88}, 034502,
  arXiv:1305.4903 [hep-lat] (2013).
\item
  R.~A.~Brice\~no, Z.~Davoudi, T.~Luu and M.~J.~Savage,
  \emph{Two-nucleon systems in a finite volume: (II) 3S1-3D1 coupled channels and the deuteron}, Phys.\ Rev.\ D {\bf 88}, 114507, arXiv:1309.3556 [hep-lat] (2013).  
\end{itemize}

In a closely related approach in chapter \ref{chap:TBC}, by imposing a particular type of boundary conditions (BCs), namely twisted BCs (TBCs), on the fields in a finite cubic volume, it is shown that not only can one optimize the extraction of the scattering parameters of any general two-hadron states from the FV spectrum, but also can improve the volume dependence of the extracted masses and two-body binding energies by tuning the BCs. The results presented in this chapter are based on the following publication
\begin{itemize}
\item
  R.~A.~Brice\~no, Z.~Davoudi, T.~Luu and M.~J.~Savage,
  \emph{Two-Baryons Systems with Twisted Boundary Conditions}, 
  Phys.\ Rev.\ D {\bf 89}, 074509,
   arXiv:1311.7686 [hep-lat] (2013).
\end{itemize}

As lattice QCD calculations have begun to include QED interactions in studies of hadronic systems, it is important to understand how the volume dependence of quantities are affected by QED effects. In particular, as electromagnetism -- due to the zero mass of the photon -- is an infinite-range interaction, such volume effects are expected to be large. This is in contrast with the strong interactions where the long range of the interaction is due to the light but massive pions (and other exchanged mesons). As a result, as will be shown in Sec. \ref{IV-intro}, although the volume corrections to the masses of hadrons due to QCD induced interactions are exponentially suppressed, the volume corrections due to QED interactions are  power law. By utilizing an effective description of the QED interactions of the hadrons (and light nuclei) at low energies, such infrared (IR) corrections to the masses of these particles can be systematically calculated in relation to their electric charge, charge radius, magnetic moment and polarizabilities. These results are presented in chapter \ref{chap:EM} and are based on the following publication 
\begin{itemize}
\item {Z.~Davoudi and M.~J.~Savage,
  \emph{Finite-Volume Electromagnetic Corrections to the Masses of Mesons, Baryons and Nuclei},
  arXiv: 1402.6741 [hep-lat] (2011).}
\end{itemize}

The following sections of this chapter provide the required background, and give a brief status report of the field prior to these publications which help to put the presented results in their own context.

\section{Quantum Chromodynamics and its ``Effective'' Description
\label {QCD}}
Although the strong interactions were long believed to be responsible for interactions among constituents of nucleons, the weakly interacting feature of the theory, postulated based on the results of deep-inelastic experiments in 1960s, had posed a mystery. The reason was that no gauge theory at the time was known that exhibits a strongly interacting feature at low energies while becomes almost free at high momentum transfers. It was only in early 1970 that 't Hooft, Politzer, Gross and Wilczek found out that the \emph{non-Abelian} gauge theories \cite{PhysRev.96.191}, in four dimensions, possess the desired property \cite{thooft, Politzer:1974fr, Gross:1973id, Gross:1973ju, Gross:1974cs}. The beautiful \emph{quark} model of Gell-Mann and Zweig that was developed in 1960s to describe the rich spectrum of strongly interacting particles, was then integrated into the underlying non-Abelian gauge theory. Interestingly, the extra charge of the quarks, namely their \emph{color}, that was proposed to ensure that the Fermi statistics of spin 1/2 particles is obeyed, had a natural interpretation in this gauge theory as quarks would now belong to a multiplet of the group (in its fundamental representation) and carry a new index. Both the quark model and the experimental evidence, e.g. $e^++e^- \to {\rm hadrons}$ cross section, suggested that there exist three distinct colors, constraining the dimensionality of the gauge group to be three. The guiding principle in constructing the Lagrange density describing quarks and gauge fields, has been the \emph{principle of local gauge invariance} -- a principle that had already played an important role in the description of the simpler gauge theory of QED. 

According to the principle of local gauge invariance, in order for the free Lagrangian density of a quark multiplet $q^a$ with $a=1,2,3$ (all of the same mass $m$),\footnote{The summation over repeated indices is to be understood throughout.}
\begin{eqnarray}
\mathcal{L}_{free}=\bar{q}_a(i\gamma^{\mu}\partial_{\mu}-m)q^a,
\label{L-q-free}
\end{eqnarray}
to be invariant under a local rotation in the internal space of the quark multiplet by a  unimodular unitary transformation $U$ -- namely a $SU(3)$ transformation,
\begin{eqnarray}
q^a \to q^{a'}={U^{a'}}_b q^b \equiv {(e^{-igT^i\alpha^i(x)})^{a'}}_b q^b,
\label{q-trans}
\end{eqnarray}
there must necessarily exist a vector field $A_{\mu}\equiv A_{\mu}^iT^i$ with $i=1,2,\dots,8$ which  minimally couples to the quark fields through the covariant derivative
\begin{eqnarray}
D_{\mu}q^a\equiv \partial_{\mu}q^a+igA_{\mu}^i {(T^i)^a}_b q^b,
\end{eqnarray}
and whose gauge transformation takes the following form
\begin{eqnarray}
A_{\mu} \to A'_{\mu}=U A_{\mu}U^{\dagger}+\frac{i}{g}\partial_{\mu}U U^{\dagger}.
\label{A-trans}
\end{eqnarray}
$T^i$s are the generators of $SU(3)$ Lie algebra, $T^i=\frac{1}{2}\lambda^i$ with $\lambda^i$ being the usual Gell-Mann matrices, and which are normalized as ${\rm Tr}(T^iT^j)=\frac{1}{2}\delta^{ij}$. These generators satisfy the commutation relations $[T^i,T^j]=if^{ijk}T^k$ where $f^{ijk}$ are the structure constants of $SU(3)$. $\alpha_i$ in Eq. (\ref{q-trans}) is the continuous parameter of transformation and $g$ characterizes the strength of the coupling between quarks and the gauge fields.

To maintain the acquired gauge invariance, the Lagrange density corresponding to the gauge fields themselves must be constructed gauge invariantly. This, first of all, means that the eight $A_{\mu}^i$ fields must be massless, the quanta of which are the familiar gluons. Secondly, in analogy with the electromagnetic (EM) interactions, one can form a field strength tensor $G_{\mu \nu}=\frac{1}{ig}[D_{\mu},D_{\nu}]$ whose transformation properties can be easily deduced using Eq. (\ref{A-trans}),
\begin{eqnarray}
G_{\mu \nu} \to G_{\mu \nu}'=U G_{\mu \nu}U^{\dagger}.
\end{eqnarray}

There are only two dimension four gauge-invariant operators that can be built out of this tensor. One of which is even under the CP transformation,
\begin{eqnarray}
\mathcal{L}_{gauge}^{(CP)}=-\frac{1}{2} {\rm Tr}(G_{\mu \nu} G^{\mu \nu}),
\label{CP-even}
\end{eqnarray}
and its normalization is chosen such that, upon replacing the $SU(3)$ transformations with an Abelian $U(1)$ transformation, the QED Lagrangian is recovered.\footnote{This also justifies the factor of $\frac{1}{ig}$ in the definition of $G_{\mu \nu}$ as it would result in the usual normalization of the kinetic term of gluons.} The $CP$ odd term,
\begin{eqnarray}
\mathcal{L}_{gauge}^{({CP\hskip-0.8em /}~)}=\bar{\theta}\frac{g^2 N_f}{32 \pi^2} \epsilon_{\mu \nu \alpha \beta} {\rm Tr}(G^{\mu \nu} G^{\alpha \beta}),
\label{CP-odd}
\end{eqnarray}
is irrelevant for most of QCD phenomenology as the experimental value of its corresponding strength, characterized by the parameter $\bar{\theta}$, is unexpectedly close to zero, $\bar{\theta}\lesssim 10^{-9}$.\footnote{The convention used for the normalization of this term ensures that, in the absence of massive quarks, the contribution from such term vanishes upon setting $\bar\theta=2\alpha$, where $\alpha$ is the parameter of the $U(1)_A$ transformation, $q \to e^{i\alpha \gamma_5}$, whose current, $J_5^{\mu} \equiv \bar{q} \gamma^{\mu} \gamma_5 q$, is anomalous.} $N_f$ denotes the number of quark flavors (up, down, strange, etc.), and $\epsilon_{\mu \nu \alpha \beta}$ is the fully anti-symmetric Levi-Civita tensor.

The Lagrange density of QCD, neglecting the CP-odd contribution and taking into account different quark flavor sectors, can be written in the explicit form,
\begin{eqnarray}
\mathcal{L}_{QCD}&=&\sum_{f=1}^{N_f}\left[\bar{q}_f(i\gamma^{\mu}\partial_{\mu}-m_f)q_f-gA_{\mu}^i \bar{q}_f \gamma^{\mu} T^i q_f\right]
\nonumber\\
&& -\frac{1}{4}F^i_{\mu \nu}F^{i\mu \nu}+\frac{g}{2}f_{ijk}F_{\mu \nu}^i A^{i \mu}A^{j \nu}
-\frac{g^2}{4}f_{ijk}f_{klm}A_{\mu}^jA_{\nu}^kA^{l \mu}A^{m \nu},
\label{L-QCD}
\end{eqnarray}
where $F^i_{\mu \nu} \equiv \partial_{\mu}A^i_{\nu}-\partial_{\nu}A^i_{\mu}$. The striking feature of this Lagrange density is the self interactions among gluons which makes the vacuum of the theory nontrivial compared to QED. This is not a surprise as in any non-Abelian gauge theory, the gauge field $A_{\mu}^i$ carries a characteristic charge (color in the case of QCD) corresponding to the internal space of the gauge group, and must be able to interact with other charged members of the gauge multiplet. The other feature of the QCD Lagrange density is that the coupling of gauge fields to the quark fields cannot be arbitrary and is constrained by the Lie algebra of the group to be the same among quarks with different colors and from different families, and should match that of self-gluon couplings. This is again in contrast with QED where, although the interaction Lagrangian has a universal form, different matter fields can couple to the EM field with different strengths, characterized by their distinct electric charges. 

The two important properties of QCD, asymptotic freedom and color confinement, can be deduced from an analytical approach based on perturbation theory. The former, as is a standard textbook calculation, is obtained by looking at the running of the QCD coupling constant with energy from a weak-coupling expansion of the QCD $\beta$-function (see Sec. \ref{HE-QCD}) using the Feynman diagram technology. The latter property can be studied using a strong-coupling expansion of the potential between two static quarks. In the following, we discuss several features of QCD at high and low energies in more details.

\subsection{QCD at high energies
\label{HE-QCD}}
Due to (ultra-violet) UV divergences in any perturbative calculation of QCD when the quantum corrections are included, a reference energy scale must be introduced to renormalize the theory. In a sense, the value of any quantity is measured compared with a reference energy scale and so the divergent contributions cancel out when quantities are calculated at two energy scales relative to each other. This however means that one must know the relation that governs the evolution of the quantity of interest at the reference energy scale down to the energy scale relevant to a given physical process. Such relations are the familiar Callan-Symanzik or renormalization group equations \cite{Callan:1970yg, Symanzik:1970rt}. Here we are only interested in the evolution of the QCD coupling constant with the energy scale $\mu$, characterized by the so-called $\beta$-function,
\begin{eqnarray}
\beta(\alpha_s) \equiv \mu^2 \frac{\partial}{\partial \mu^2}\alpha_s(\mu).
\end{eqnarray}

The fields and parameters of the Lagrange density that one starts with are \emph{bare} quantities, meaning that they suffer from UV divergences. These can be replaced with the renormalized quantities whose divergences are removed by fixing their values at the reference scale $\mu$, using some chosen renormalization conditions. These finite quantities which now carry a $\mu$-dependence can then be used in perturbation theory in a well-defined expansion. The beauty of perturbative QCD, as well as other renormalizable theories, is that a finite number of such conditions suffices to remove all the UV divergences that occur to all order in perturbation theory.

This well-defined procedure can be carried out for the \emph{effective} coupling constant felt at energy scale $\mu$ where not only e.g. the three-gluon vertex must be replaced by its renormalized value but also the external gluonic legs must be corrected by the corresponding wavefunction renormalization factors. Then a two-loop calculation shows that
\begin{eqnarray}
\beta(\alpha_s)=-(b_0 \alpha_s^2+b_1 \alpha_s^3)+\mathcal{O}(\alpha_s^4),
\label{QCD-beta}
\end{eqnarray}
with $b_0=\frac{1}{12\pi}(33-2N_f)$ and $b_1=\frac{1}{24\pi^2}(153-19N_f)$ \cite{Beringer:1900zz}. For the current discussion let us ignore the NLO correction to the $\beta$-function and solve Eq. (\ref{QCD-beta}). Explicitly, we want to know given the coupling constant at scale $\mu$, what the value of the coupling would be at scale $\mu'$. It easily follows that
\begin{eqnarray}
\alpha_s(\mu')=\frac{\alpha_s(\mu)}{1+b_0 \alpha_s(\mu) \log \frac{\mu'^2}{\mu^2}}.
\label{alpha-s-I}
\end{eqnarray}
\begin{figure}[h!]
\begin{centering}
\includegraphics[scale=1.05]{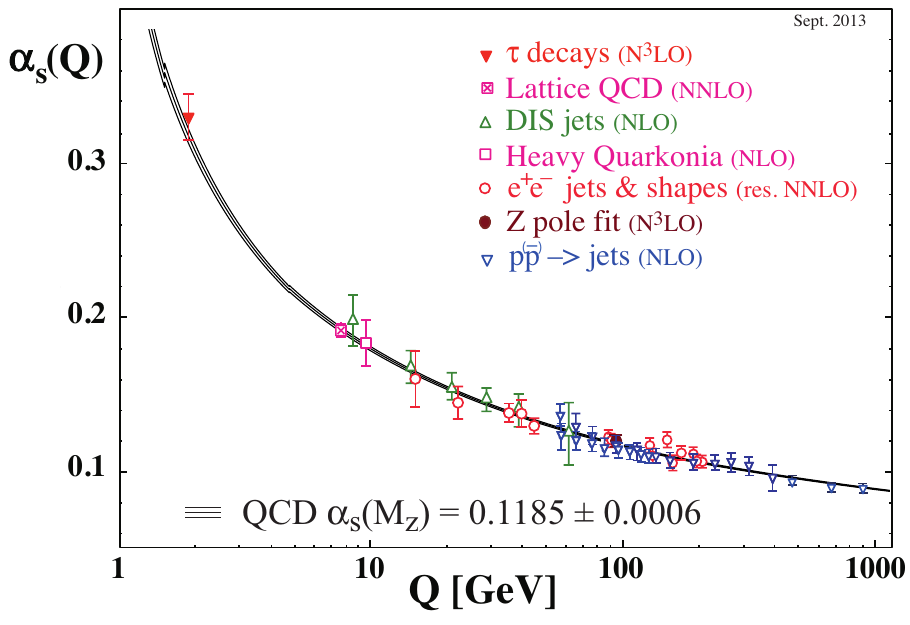}
\par\end{centering}
\caption{{\small The coupling of QCD as a function of a characteristic energy scale $\mu=Q$, obtained from matching the QCD perturbative calculation to a given order (as given in brackets) to the experimental measurements of several quantities. There is also one point which is obtained by matching to a lattice QCD calculation  \cite{Beringer:1900zz}. Figure is reproduced with the permission of Michael Barnett on behalf of the Particle Data Group.}}
\label{fig:alpha-s}
\end{figure}
Given the positive sign of $b_1$ for QCD with $N_f=6$, it is evident that $\alpha_s(\mu')$ decreases as $\mu'$ increases, indicating the theory tends to become free at asymptotically high energies. Experimental determinations of $\alpha_s$ for a range of energies have resulted in values that lie on the predicted scale-dependence curve to an extremely well precision, as is shown in Fig. \ref{fig:alpha-s}. To parametrize the characteristic scale at which the theory becomes strong, we can define the scale $\Lambda_{QCD}$ such that $b_0 \alpha_s(\mu) \log \frac{\mu^2}{\Lambda_{QCD}^2}=1$, then one can rewrite Eq. (\ref{alpha-s-I}) as following
\begin{eqnarray}
\alpha_s(\mu')=\frac{1}{2b_0 \log \frac{\mu'}{\Lambda_{QCD}}}.
\label{alpha-s-II}
\end{eqnarray}
As can be seen, perturbation theory is only valid if $\mu' \gg \Lambda_{QCD}$. Experimentally $\Lambda_{QCD} \approx 200~{\rm MeV}$ which is of the order of the inverse size of the light hadrons. This is consistent with our realization of hadrons being composed of strongly interacting constituents when low-energy probes are used. In fact at low energies, these hadrons are the effective degrees of freedom of QCD, and the details of their properties and interactions, although sensitive to the short distance theory of QCD, can be studied in a systematic low-energy expansion. This requires understanding QCD symmetries and the mechanism for the breaking of some of these symmetries. We discuss this topic in the next section, Sec. \ref{LE-QCD}.

\subsection{QCD at low energies
\label{LE-QCD}}
\begin{figure}[t!]
\begin{centering}
\includegraphics[scale=0.365]{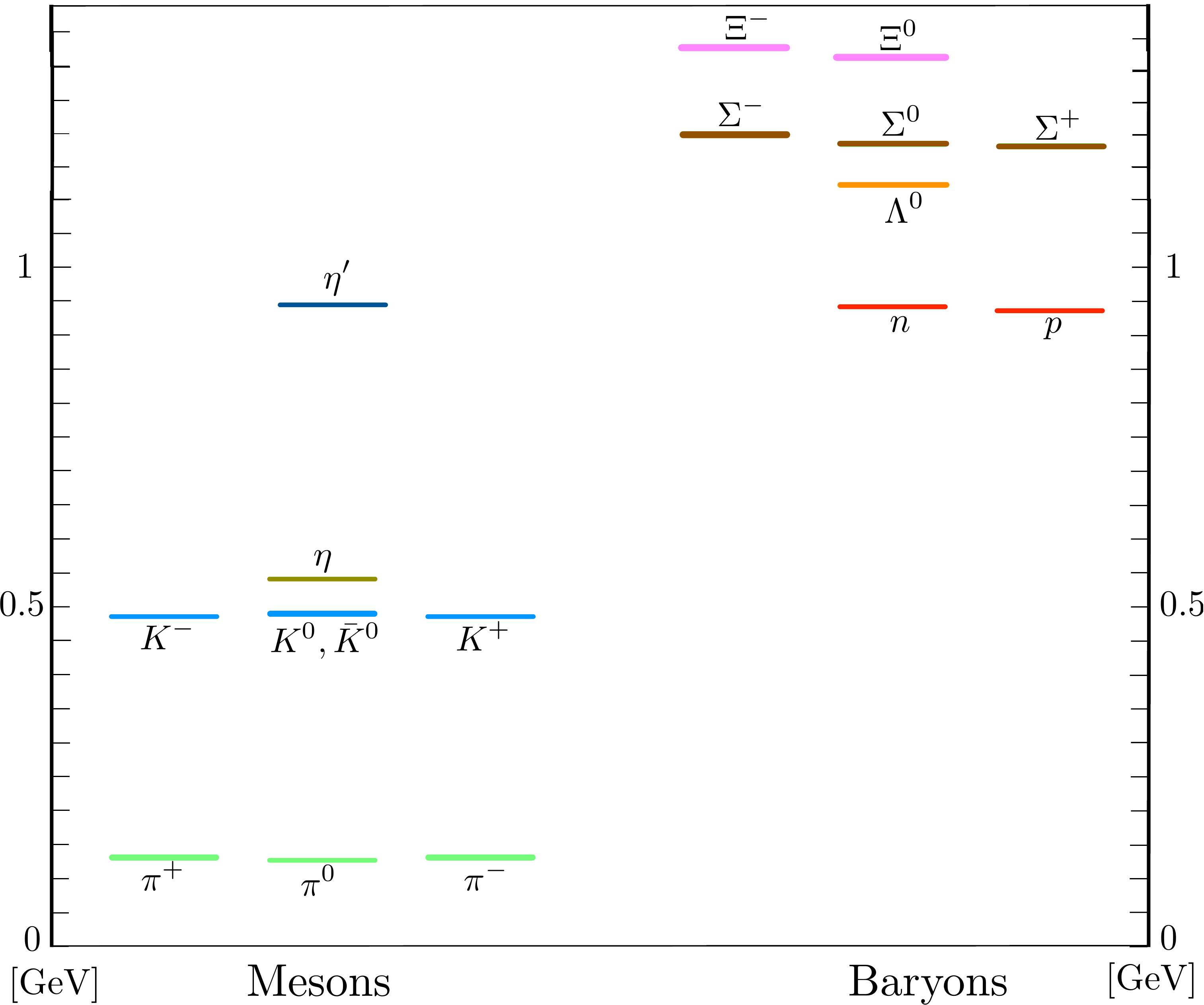}
\par\end{centering}
\caption{{\small The masses of the the few lightest mesons with $J^{P}=0^{-}$ and baryons with $J^{P}=\frac{1}{2}^{+}$ in GeV. The error bars associated with the experimental measurements of masses are not included and each line only represents the central values as reported in Ref. \cite{Beringer:1900zz}.}}
\label{fig:had-spec}
\end{figure}

Although quarks and gluons do not show up as explicit degrees of freedom in the spectrum at energies of the order of $\Lambda_{QCD}$, the imprint of their interactions can be found in the spectrum of hadrons. For example, the low-lying spectrum of (negative parity) mesons and (positive parity) baryons, as illustrated in Fig. \ref{fig:had-spec}, exhibits several interesting patterns whose origin can be understood via the fundamental theory of QCD. As is seen, pions are noticeably lighter than the rest of hadrons and come in an almost degenerate triplet. The next multiplet of mesons, while remain low in mass compared to baryons, are not as light as pions. On the other hand, the $\eta'$ meson that has the same quark content as that of $\eta$ in the quark model is surprisingly heavier than $\eta$. Baryons have masses at the order of $\gsim1~{\rm GeV}$ and like mesons come in various nearly degenerate multiplets. Moreover, the parity partners of mesons and baryons have been observed to have different masses, e.g., the difference in the mass of the nucleons $\sim 940~{\rm MeV}$ and their negative parity counterpart $N(1535)$ is as large as $600~{\rm MeV}$.

To understand these features all together, it suffices to study the underlying symmetries of the QCD Lagrangian. In the limit of zero quark masses (chiral limit), the left-handed and right-handed quarks of each flavor do not mix with each other through QCD  interactions, 
\begin{eqnarray}
\mathcal{L}_{QCD}^{(q;1)}&=&\sum_{f=1}^{N_f}\left[\bar{q}_{L,f}(i\gamma^{\mu}D_{\mu})q_{L,f}+\bar{q}_{R,f}(i\gamma^{\mu}D_{\mu})q_{R,f}\right],
\label{L-kin}
\end{eqnarray}
where each quark is decomposed to components that have specific handedness, $q=q_L+q_R$ with $q_L=\frac{1-\gamma_5}{2}q$ and $q_R=\frac{1+\gamma_5}{2}q$. Due to the heavy mass of the charm quark, $m_c \approx 1.3~{\rm GeV}$, only up, down and strange quarks play a significant role in the dynamic of strongly interacting systems at low to medium energies. With three flavors of quarks the Lagrangian in Eq. (\ref{L-kin}) is seen to be invariant under $U(3)_L \times U(3)_R$ symmetry, which however breaks down to $U(1)_V \times SU(3)_L \times SU(3)_R$ symmetry. Let us discuss these symmetries and their reduction in more details.

\begin{itemize}
\item The $U(1)_A$ symmetry is broken due to the chiral anomaly. The chiral anomaly refers to the non-conservation of the number of massless left-handed fermions compared with the right-handed fermions due to the non-invariance of the quantum expectation values (as opposed to the classical Lagrangian) under an axial $U(1)$ transformation $q \to e^{i\alpha \gamma_5}q$, where the corresponding isosinglet axial-vector current $J^{\mu5}=\bar{q}\gamma^{\mu}\gamma^{5} q$ is not conserved \cite{Adler:1969gk, Adler:1969er}. This already gives a hint to why the mass of the isosinglet pseudo-scalar meson $\eta'$ is noticeably different than that of isovector pseudo-scalar mesons. However in order to understand the small mass of these latter mesons, further investigation of symmetries is required.

\item The invariance under $U(1)_V$ is realized by the transformation $q \to e^{i\alpha}q$ with the corresponding conserved isosinglet vector current $J^{\mu}=\bar{q}\gamma^{\mu}q$, and is manifested by the conservation of the net baryon number.

\item Finally, independent $SU(3)$ transformations of left-handed and right-handed quarks, represented by $q \to L~q$ and $q \to R~q$, leave the Lagrangian in Eq. (\ref{L-kin}) invariant, where $L$ and $R$ are $SU(3)$ matrices, $L \in SU(3)_L$ and $R \in SU(3)_R$, and $q$ denotes a quark triplet in the 3 representation of $SU(3)$. This is called the chiral symmetry of QCD which plays an important role in constructing an effective low-energy theory of hadrons, namely chiral perturbation theory ($\chi$PT), at energies of the order of $\Lambda_{QCD}$.

\end{itemize}

There are two features of QCD that deprive nature from the exact chiral symmetry. The first one is the presence of a non-vanishing quark condensate,
\begin{eqnarray}
\left\langle 0 \left| \bar{q}_{R,j} q_{L,i} \right| 0 \right\rangle = - \Lambda^3 \delta_{ij},~~~~~~~~~i,j=1,2,3.
\label{condensate}
\end{eqnarray}
resulting in a \emph{spontaneous} breaking of the chiral symmetry \cite{Nambu:1961tp, Nambu:1961fr}. This means that although the action of theory is chirally invariant, the vacuum state\footnote{Note that the $\bar{q}q$ pair in Eq. (\ref{condensate}) has the same quantum numbers as vacuum.} does not respect the chiral symmetry. This is manifested in the change of condensate as a chiral transformation is performed on the quark fields,
\begin{eqnarray}
\left\langle 0 \left| \bar{q}_{R,j} q_{L,i} \right| 0 \right\rangle \to L_{ii'}R^{\dagger}_{j'j} \left\langle 0 \left| \bar{q}_{R,j'} q_{L,i'} \right| 0 \right\rangle = - \Lambda^3(LR^{\dagger})_{ij}.
\label{cond-trans}
\end{eqnarray}
Only if $L=R$ does the condensate remain invariant, reducing the symmetry group to its $SU(3)_V$ subgroup with the corresponding conserved isovector vector current $J^{\mu,a}=\bar{q} \gamma^{\mu} T^a q$. For each $8$ generators of the broken subset of symmetries, deduced from the isovector axial vector current $J^{\mu 5,a}=\bar{q} \gamma^{\mu} \gamma^5 T^a q$, there exists a corresponding massless Goldstone boson which must be found in the spectrum of mesons with quantum numbers of the generators broken symmetry. Such massless excitations can be parametrized by a field, $\Sigma$, that lives in the $(3,\bar{3})$ representation of $SU(3)$. From Eq. (\ref{cond-trans}), it is clear that $LR^{\dagger}$ produces a different vacuum than that of Eq. (\ref{condensate}) for $L \neq R$ and therefore it can be readily identified as $\Sigma$. The $\Sigma$ field can be explicitly parametrized by,
\begin{eqnarray}
\Sigma \equiv e^{2i \bm{\pi}(x) / f},
\label{Sigma}
\end{eqnarray}
where $\bm{\pi}(x)$ can be related to the pseudo-scalar meson octets,
\begin{eqnarray}
\bm{\pi}\equiv\pi^aT^a=
\left(\begin{array}{ccc}
\frac{\pi^0}{\sqrt{2}}+\frac{\eta}{\sqrt{6}} & \pi^+ & K^+\\
\pi^- & -\frac{\pi^0}{\sqrt{2}}+\frac{\eta}{\sqrt{6}} & K^0\\
K^- & \bar{K}^0 & -\frac{2\eta}{\sqrt{6}}
\end{array}\right).
\label{Sigma}
\end{eqnarray}
$T^{a}$ with $a=1,\dots,8$ are the $8$ generators of $SU(3)$ and $f$ is a constant with dimension mass whose value is matched to the pion weak decay constant $f=f_{\pi}=130.41 \pm 20~{\rm MeV}$ \cite{Beringer:1900zz}. The effective interactions of these Goldstone bosons at low energies can then be studied by forming the most general Lagrangian that is invariant under the chiral symmetry. The significance of each term in this Lagrangian is determined through a systematic expansion with respect to the ratio of the typical momentum in a process to the scale of chiral symmetry breaking, $\Lambda_{\chi}\sim1~{\rm GeV}$. We will come back to this topic in Sec. \ref{subsec:ChiPT}.

The second feature of QCD which \emph{explicitly} breaks the chiral symmetry is  non-vanishing masses of quarks. It is evident from the QCD Lagrangian that the mass term mixes quarks of different chiralities,
\begin{eqnarray}
\mathcal{L}_{QCD}^{(q;2)}&=&\sum_{f=1}^{N_f}\left[\bar{q}_{L,f}m_{q_f}q_{R,f}+\bar{q}_{R,f}m_{q_f}q_{L,f}\right],
\label{L-mq}
\end{eqnarray}
and manifestly spoils the chiral symmetry. However, the masses of light quarks,\footnote{These masses are the Particle Data Group average of several lattice QCD determinations that are converted to a renormalized mass in the $\overline{MS}$ scheme at scale $\mu=2~{\rm GeV}$ \cite{Beringer:1900zz}.} 
\begin{eqnarray}
m_u=2.15 \pm 0.15~{\rm MeV},~m_d=4.70 \pm 0.20~{\rm MeV},~m_s=93.5 \pm 2.5~{\rm MeV},
\label{mq}
\end{eqnarray}
in particular those of up and down quarks, are much smaller than the scale of the spontaneous chiral symmetry breaking. As a result, chiral symmetry remains an approximate symmetry of QCD before the spontaneous chiral symmetry breaking occurs. The spontaneous symmetry breaking (SSB) mechanism then generates 8 nearly massless bosons or namely 8 \emph{pseudo}-Goldstone bosons (pGBs). In Sec. \ref{subsec:ChiPT} we will show how the quark mass contributions can be included in the chirally invariant Lagrangian of pseudo-scalar bosons.

The first immediate evidence of pGBs is in the spectrum of hadrons. As discussed, the pseudo-scalar octets are unusually light compared with the rest of hadrons and whose parity quantum number is consistent with that of expected for the generators of the broken symmetry. This also means that the hadrons will no longer be degenerate with their parity partners when the chiral symmetry is broken.\footnote{Since the Hamiltonian of QCD is invariant under parity (ignoring nearly vanishing $CP$ violating interactions in Eq. (\ref{CP-odd})), the vacuum state and its parity partner are both the eigenstate of the Hamiltonian with the same eigenvalues. However, it can be shown that these two degenerate states are eigenstates of the axial charge $\hat{Q}^{5,a} \equiv \int d^3 x j^{05,a}(\mathbf{x},t)$ with eigenvalues that differ in sign. After the spontaneous symmetry breaking, only one of these vacua is picked, resulting in breaking the degeneracy between the parity partners.} We note in particular that the $\eta'$ does not correspond to a SSB mechanism and so its mass is not protected to be small.\footnote{Due to the heavier mass of the strange quark, the explicit chiral symmetry breaking is severe for the case of $SU(3)$ symmetry compared with its $SU(2)$ subgroup. As a result, the pGB features of pions are more prominent than that of strange mesons, see Fig. \ref{fig:had-spec}.}  The other evidence for the existence of pGBs of a spontaneously broken symmetry had been observed experimentally through pion-pion and pion-nucleon scattering experiments even in pre-QCD era. The $\pi\pi$ scattering cross sections had been observed to vanish at low energies. On the other hand, the most naive effective interaction among pions and nucleons at low energies consistent with the parity of pions and nucleons failed to describe pion-nucleon cross sections. Both of these cross sections could be reproduced if pions would only derivatively couple to other hadrons. This is of course only consistent with the identification of pions as the pGBs of a broken symmetry with an explicit \emph{shift} symmetry as is evident from Eq. (\ref{Sigma}). We will present these interactions in the following subsection.

\subsubsection{Chiral perturbation theory for mesons and baryons
\label{subsec:ChiPT}}
\noindent \emph{Lagrangian for pseudo-Goldstone bosons:} The Lagrangian describing the dynamics of pGBs can be constructed from field $\Sigma$ in Eq. (\ref{Sigma}) order by order in powers of $\frac{\partial}{\Lambda_{\chi}} \sim \frac{p}{\Lambda_{\chi}}$ and $\frac{m_q}{\Lambda_{\chi}}$, where $p$ is the typical momentum of the process and $m_{pGB}$ denote the mass of the pGBs.\footnote{The mass of the next meson that is not a pGBs can be taken as the scale $\Lambda_{\chi}$ for which this effective approach breaks down. This is the $\rho$ meson with $m_{\rho}=775.26\pm0.25~{\rm MeV}$. It gives rise to an expansion parameter that is not typically small, $\frac{m_K}{m_{\rho}}\sim0.6$, consistent with the expectation that the $SU(3)$ symmetry breaking is fairly severe given the mass of the strange quark. For processes that only involve pions and nucleons, one can restrict the effective interactions to only respect the $SU(2)$ chiral symmetry for which the expansion parameter can only be as large as $\frac{m_{\pi}}{\Lambda_{\chi}}=\frac{m_{\pi}}{m_K}\sim0.3$ for low-energy processes.} For pseudo-scalar mesons to be Goldstone boson, they must only interact derivatively. However as they are only pGBs due to non-vanishing mass of quarks, they can also couple non-derivatively through insertions of the quark mass matrix defined as
\begin{eqnarray}
M \equiv \left(\begin{array}{ccc}
m_u & 0 & 0\\
0 & m_d & 0\\
0 & 0 & m_s
\end{array}\right).
\label{Mq-matrix}
\end{eqnarray}
By promoting $M$ to a dynamical field, namely a \emph{spurion} field, which transforms under $SU(3)$ chiral symmetry as $M \to LMR^{\dagger}$, its non-zero  value can be interpreted as causing a SSB similar to the field $\Sigma$. This provides the necessary ingredients to write down the leading order (LO), $\mathcal{O}\left(\frac{p^2}{\Lambda_{\chi}^2},\frac{m_{pGB}^2}{\Lambda_{\chi}^2}\right)$, chiral Lagrangian of pseudo-scalar mesons as following \cite{Gasser:1983yg, Gasser:1984gg},
\begin{eqnarray}
\mathcal{L}_{pGB}^{(2)}=\frac{f_{\pi}^2}{8}~{\rm Tr}\left[\partial_{\mu} \Sigma \partial^{\mu} \Sigma^{\dagger}+2B (M\Sigma^{\dagger}+M^{\dagger} \Sigma)\right].
\label{L-pGB}
\end{eqnarray}
This Lagrangian is invariant under the Lorentz and $SU(3)$ chiral symmetry and its normalization is chosen in such a way to reproduce the canonical normalization of the kinetic term for pseudo-scalar mesons. It only contains one more parameter, or low-energy coefficient (LEC), beside $f_{\pi}$ which, upon a straightforward expansion in the pGB fields in EQ. (\ref{L-pGB}), can be related to the mass of mesons, e.g., $B=\frac{m_{\pi}^2}{m_u+m_d}$. This indicates that each insertion of the quark mass matrix counts as $\sim m_{pGB}^2$ in this treatment.
The value of parameter $B$ can be directly matched to the value of the quark condensate using the Feynman-Hellman theorem and is readily found to be
\begin{eqnarray}
B=-\frac{2}{f_{\pi}^2}\left\langle 0 \left| \bar{u}u \right| 0 \right\rangle.
\label{B}
\end{eqnarray}

At next to LO, $\mathcal{O}\left(\frac{p^4}{\Lambda_{\chi}^4},\frac{p^2m_{pGB}^2}{\Lambda_{\chi}^4},\frac{m_{pGB}^4}{\Lambda_{\chi}^4}\right)$, there are 8  distinct chirally invariant operators with up to 4 derivatives and up to two insertions of quark mass matrix whose corresponding LECs, the Gasser-Leutwyler coefficients \cite{Gasser:1984gg}, must be matched to experimental data on meson-meson scattering. In doing such matching, the loop effects with insertions of the leading operators in Eq. (\ref{L-pGB}) must be taken into account. This is because these loop contributions are enhanced compared with the tree-level contributions of the next order by factors of $\log(\mu/m_{pGB})$. $\mu$ is the renormalization scale in a mass-independent normalization scheme such as $\overline{MS}$ that is used to renormalize the amplitudes when encountering the UV divergences in loops \cite{Kaplan:1995uv}, see for example Fig. \ref{fig:pipi}. In particular, it is notable that the scale-dependence of LECs at any order in a systematic EFT is canceled by that of introduced by the chiral loops of previous order so that the amplitudes calculated at that order is rendered scale independent.

The EFT procedure just described is a powerful method for the following reasons:
\begin{itemize}
\item
Firstly, the dynamics of pGB is highly constrained by the chiral symmetry such that the interactions of all pseudo-scalar mesons can be put in a universal form, e.g., Eq. (\ref{L-pGB}), eliminating the need to introduce several LECs at each order for different members of the multiplet. This feature remains true in constructing the interactions of pGBs with baryons as is discussed below.

\item
Once LECs that occur at a given order in EFT are matched to one or several observables, the EFT interactions can be applied in studying a wide range of phenomena where such operators contribute, giving the EFT a predictive power. For example as we just observed, the value of parameter $f$ that was determined by matching to the weak decay rate of the pion, can now be used to fully predict the $\pi\pi$ scattering cross section at LO using Eq. (\ref{L-pGB}). A straightforward calculation shows that \cite{Colangelo:2001df}
\begin{eqnarray}
t^{(0)}_0(s)=\frac{2s-m_{\pi}^2}{16\pi f_{\pi}^2},
~t^{(1)}_1(s)=\frac{s-4m_{\pi}^2}{48\pi f_{\pi}^2},
~t^{(2)}_0(s)=-\frac{s-2m_{\pi}^2}{16\pi f_{\pi}^2},
\label{pipi-I012}
\end{eqnarray}
at LO in chiral expansion, where $t^{(I)}_l(s)$ denotes the $\pi\pi$ scattering partial-wave amplitude in isospin channel $I$ and partial-wave channel $l$, and $s$ denotes the total invariant mass of the $\pi\pi$ system.

\item
Despite phenomenological models with an arbitrary number of parameters -- that are fit to experimental data -- with which no well-defined systematic uncertainty can be associated, the EFT approach enables the quantification of errors in calculated quantities in a systematic way. These errors result from neglecting higher order terms in the low-momentum expansion.

\item
\begin{figure}[t!]
\begin{centering}
\includegraphics[scale=0.4235]{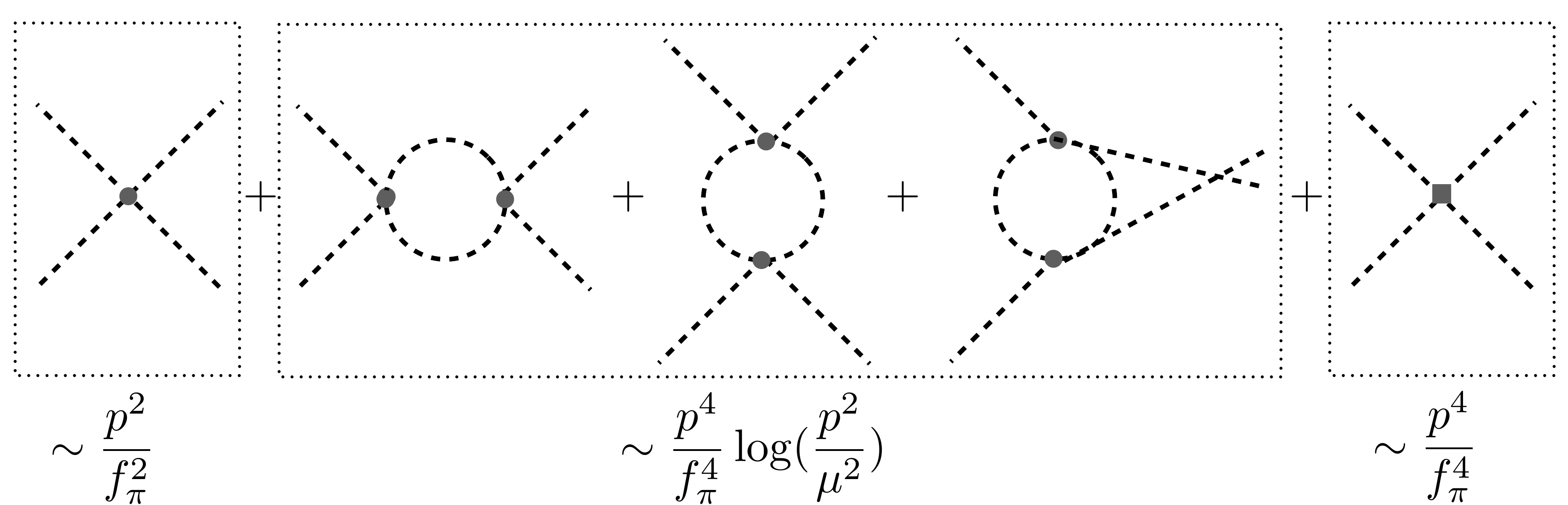}
\par\end{centering}
\caption{{\small Diagrams contributing to pion-pion scattering in $\chi$PT} up to NLO. Dashed lines represent pions, the grey dot denotes the LO tree level vertex obtained from expanding Eq. (\ref{L-pGB}) in pion fields, while the grey square denotes the NLO vertex and depends on the Gasser-Leutwyler coefficients. The power-counting of each diagram is given in the figure.}
\label{fig:pipi}
\end{figure}
\begin{figure}[t!]
\begin{centering}
\includegraphics[scale=0.45]{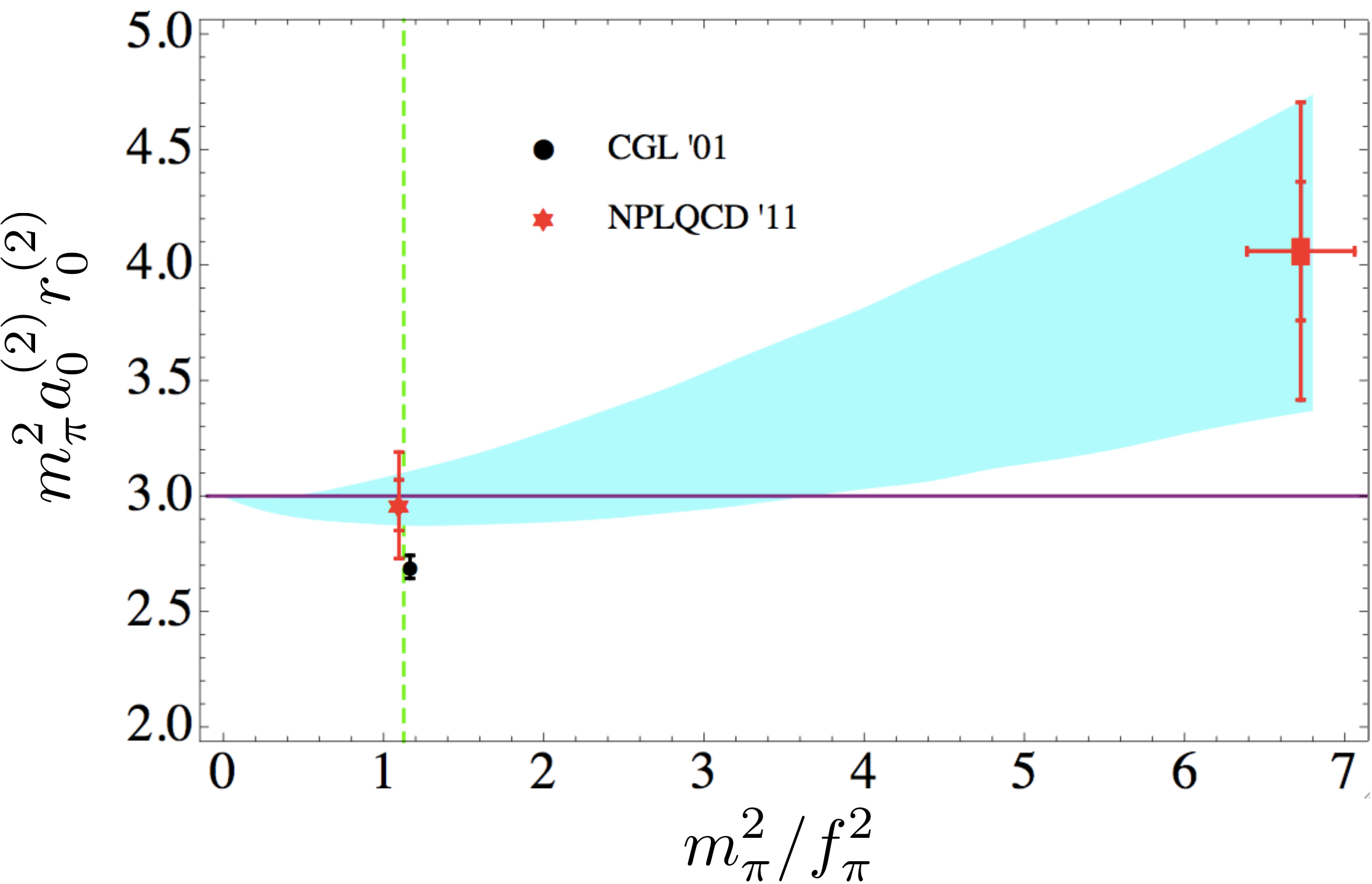}
\par\end{centering}
\caption{{\small The LQCD determination of $m_{\pi}^2a^{(2)}_0r^{(2)}_0$ for the $I=2$ $\pi \pi$ scattering at the physical point (the red star on the physical line denoted by a dashed green line). The band represents the $68\%$ confidence interval interpolation of the LQCD result (the red rectangle) at $m_{\pi}=390~{\rm MeV}$ \cite{Beane:2011sc}. The horizontal purple line denotes the LO $\chi$PT prediction in the chiral limit. The Roy equation prediction \cite{Colangelo:2001df} is shown by the black circle on the physical line. Figure is reproduced with the permission of the NPLQCD collaboration.}}
\label{fig:NPLQCD-pipi}
\end{figure}
Such EFT technique determines the light-quark mass dependence of observables order by order in the EFT expansion. This is particularly important as it enables making predications for physical observable from lattice QCD calculations that are performed at heavier quark masses. As long as the quark masses used in those calculations produce pGB with masses within the range of validity of chiral perturbation theory, the $\chi$PT expressions might be used to interpolate to the physical values of quantities. A nice example of which is the determinations of the $I=2$ S-wave $\pi\pi$ scattering length, $a^{(2)}_0$, and effective range, $r^{(2)}_0$,  at the physical point (physical values of light-quark masses) using the LQCD input at $m_{\pi}=390~{\rm MeV}$ as performed by the NPLQCD collaboration \cite{Beane:2011sc}. The S-wave scattering length and effective range are defined via the effective range expansion (ERE) at low energies,
\begin{align}
& k^* \cot \delta_0=-\frac{1}{a_0}+\frac{1}{2}r_0 k^{*2}+\dots~,
\label{a-r-def}
\end{align}
where $k^*$ denotes the momentum of each pion in the CM frame. These LQCD results have been used in the chiral expansions of these quantities at NLO in two-flavor $\chi$PT,\footnote{We are using the nuclear physics convention for the sign of scattering length where a positive scattering length corresponds to an attractive interaction.}
\begin{align}
& m_{\pi}a^{(2)}_0 = \frac{m_{\pi}^2}{f_{\pi}^2} \left[\frac{1} {8\pi}+\frac{3}{128 \pi^2}\frac{m_{\pi}^2}{f_{\pi}^2} \log\left(\frac{m_{\pi}^2}{f_{\pi}^2}\right)+C_1\frac{m_{\pi}^2}{f_{\pi}^2}\right],
\nonumber\\
& m_{\pi}r^{(2)}_0 = \frac{24 \pi f_{\pi}^2}{m_{\pi}^2}+C_2+\frac{17}{6 \pi} \log\left(\frac{f_{\pi}^2}{m_{\pi}^2}\right),
\label{ar-pipi-I}
\end{align}
where $C_1$ and $C_2$ are two combinations of Gasser-Leutwyler coefficients, renormalized at scale $\mu=f_{\pi}$, and are fit to LQCD data at this pion mass.  This results in impressively precise determinations of scattering length and effective range at the physical point,\footnote{The numbers in parentheses denote the statistical and systematic uncertainties of various sources as explained in Ref. \cite{Beane:2011sc}.}
\begin{eqnarray}
m_{\pi}a^{(2)}_0=0.0417(07)(02)(16),~ m_{\pi}r^{(2)}_0=72.0(5.3)(5.3)(2.7),
\label{ar-pipi-II}
\end{eqnarray}
in $1\sigma$ agreement with the determination of these parameters from the Roy (dispersion relation) analysis \cite{Roy:1971tc} of experimental data with the $\chi$PT input \cite{Ananthanarayan:2000ht, Colangelo:2001df}. This example demonstrates the role of EFT in empowering the LQCD calculations at yet unphysical pion masses  to make predictions for the physical point. Besides the low-energy scattering parameters \cite{Li:2007ey, Aoki:2007rd, Beane:2010hg, Beane:2011xf, Beane:2011sc, Beane:2011iw, Beane:2012ey,Yamazaki:2012hi, Lang:2012sv, Beane:2013br,  Pelissier:2011ib,  Lang:2011mn, Pelissier:2012pi, Ozaki:2012ce, Buchoff:2012ja, Dudek:2012xn, Dudek:2012gj, Lang:2014tia}, the masses of hadrons and their decay constants are among quantities that are being extensively studied through a combination of LQCD and EFTs (For reviews on these calculations see Refs. \cite{Aoki:2013ldr, Laiho:2009eu, Lin, Prelovsek:2013cta}). We will present an example of the use of EFTs in deducing the FV corrections to the mass of the nucleons in Sec. \ref{IV-intro}.
\end{itemize}

\noindent \emph{Lagrangian for Baryon octets:} Let us first focus on the the case of $SU(2)_L \times SU(2)_R$ in constructing the Lagrangian. We present the general result for the case of $SU(3)$ chiral symmetry later. First note that the transformation property of nucleons doublet, $N=\left(\begin{array}{c} p\\ n
\end{array}\right)$, under $SU(2)_L \times SU(2)_R$ is not constrained - in contrary to the pGB field $\Sigma$, so we can take the freedom to choose it. The simplest transformation, $N_L \to L  N_L$ and $N_R \to R N_R$, where left-handed and right-handed nucleons transform separately, turns out to not be the most convenient one. We can require the same transformation for the left-handed and right-handed components,
\begin{eqnarray}
\widetilde{N}_L \to U  \widetilde{N}_L,~ \widetilde{N}_R \to U \widetilde{N}_R,
\label{N-tilde}
\end{eqnarray}
where $U$ is an element of $SU(2)$. This can be achieved if we redefine (dress) the nucleon field as following
\begin{eqnarray}
\widetilde{N}_L\equiv \xi N_R,~ \widetilde{N}_R \equiv \xi^{\dagger} N_L,
\label{N-dressed}
\end{eqnarray}
where $\xi=\sqrt{\Sigma}$ can be seen to transform under $SU(2)_L \times SU(2)_R$ as
\begin{eqnarray}
\xi \to L \xi U^{\dagger}=U \xi R^{\dagger}.
\label{xi-trans}
\end{eqnarray}

In order for the familiar free nucleon Lagrangian $\overline{N} i \gamma^{\mu} \partial_{\mu} N$ to remain invariant under (local) transformation (\ref{N-tilde}), the minimal coupling to a vector field $\mathcal{V}_{\mu}$ must be introduced to assure the covariant derivative $D_{\mu}=\partial_{\mu}+\mathcal{V}_{\mu}$ transforms properly under the chiral transformation, $D_{\mu}N \to U(D_{\mu}N)$.\footnote{We reassign the notation $N$ to nucleon fields that transform as in Eq. (\ref{N-tilde}).} This can be seen to be satisfied if $\mathcal{V}_{\mu}$ is chosen to be
\begin{eqnarray}
\mathcal{V}_{\mu}=\frac{1}{2}(\xi^{\dagger} \partial_{\mu} \xi+\xi \partial_{\mu} \xi^{\dagger}),
\label{Vector}
\end{eqnarray}
given its transformation property $\mathcal{V}_{\mu} \to U \mathcal{V}_{\mu} U^{\dagger}+U\partial_{\mu}U^{\dagger}$. Another chiral invariant term in the nucleon Lagrangian is possible by forming the following combination
\begin{eqnarray}
\mathcal{A}_{\mu}=\frac{i}{2}(\xi^{\dagger} \partial_{\mu} \xi-\xi \partial_{\mu} \xi^{\dagger}).
\label{Axial}
\end{eqnarray}
Since this combination transforms as $\mathcal{A}_{\mu} \to U \mathcal{A}_{\mu} U^{\dagger}$, it can directly couple to nucleons at LO in a chirally invariant way although its coefficient is not protected by the minimal coupling mechanism as for the vector fields. Then the leading chiral Lagrangian describing nucleons and their interactions with the pGBs (through field $\xi$) can be written as \cite{Gasser:1987rb}
\begin{eqnarray}
\mathcal{L}^{(1)}_{N,pGB}=\overline{N}(i\gamma^{\mu}D_{\mu}-M_{N}+g_{A}\gamma^{\mu}\gamma^{5} \mathcal{A}_{\mu})N,
\label{L-NpGB}
\end{eqnarray}
where the only new LEC is $g_{A}$ whose value can be matched to the neutron semi-leptonic weak decay, $g_{A}=1.2701(25)$ \cite{Beringer:1900zz}.

Extending the formalism to the case of $SU(3)$ chiral symmetry is now straightforward by noting that the baryon octet fields,
\begin{eqnarray}
B=
\left(\begin{array}{ccc}
\frac{\Sigma^0}{\sqrt{2}}+\frac{\Lambda}{\sqrt{6}} & \Sigma^+ & p\\
\Sigma^- & -\frac{\Sigma^0}{\sqrt{2}}+\frac{\Lambda}{\sqrt{6}} & n\\
\Xi^- & \Xi^0 & -\frac{2\Lambda}{\sqrt{6}}
\end{array}\right),
\label{B-octet}
\end{eqnarray}
can be made to transform as $B \to U B U^{\dagger}$, and as a result the Lagrangian in Eq. (\ref{L-NpGB}) can be generalized to \cite{Krause:1990xc}
\begin{eqnarray}
\mathcal{L}^{(1)}_{B,pGB}={\rm Tr}\left[\overline{B}(i\gamma^{\mu}D_{\mu}-M_{B})B\right]-D{\rm Tr}\left[\bar{B}\gamma^{\mu}\gamma^{5}\{\mathcal{A}_{\mu},B\}\right]-F{\rm Tr}\left[\bar{B}\gamma^{\mu}\gamma^{5}[\mathcal{A}_{\mu},B]\right],
\label{L-BpGB}
\end{eqnarray}
where $D$ and $B$ are two new LECs that can be determined by matching to semi-leptonic weak decay decays of baryons octets, $D \approx 0.8$ and $F \approx 0.5$ \cite{Borasoy:1998pe}. At NLO, the insertions of the quark mass matrix must be taken into account  . Given the transformation properties of $B$, $\xi$ and $M$ as discussed above, the most general Lagrangian at this order can be readily formed,
\begin{align}
&\mathcal{L}^{(2)}_{B,pGB}=a_1{\rm Tr}\left[\overline{B}(\xi^{\dagger}M\xi^{\dagger}+{\rm h.c.})B\right]+a_2{\rm Tr}\left[\bar{B}B(\xi^{\dagger}M\xi^{\dagger}+{\rm h.c.})\right]+a_3{\rm Tr}\left[\bar{B}B \right] {\rm Tr} \left[M \Sigma+{\rm h.c.} \right],
\nonumber\\
\label{L-BpGB-Mass}
\end{align}
in which three new LECs are introduced. 

An apparent problem in developing a power-counting scheme for EFTs with baryons is that the mass of the baryons is of the order of the chiral symmetry breaking scale, and as a result an expansion in $M_{B}/\Lambda_{\chi}$ is meaningless. To resolve this issue \cite{Jenkins:1990jv}, one should notice that in the heavy field limit, the momentum transfer between baryons and pGBs remains small. So by performing a field redefinition, the large contribution to the baryon momentum, $P_{\mu}=mv_{\mu}+l_{\mu}$, due to its mass can be canceled, leaving a small residual momentum $l_{\mu}$, where $v_{\mu}$ is the baryon four velocity. Let us focus on the case of nucleons and rewrite the $N$ field as
\begin{align}
& N=e^{-iM_Nv.x} (N_{l}+N_{h}),
\label{N-redefinition}
\end{align}
where 
\begin{align}
& N_{l}=e^{iM_Nv.x} \mathcal{P}^+_v N,~N_h=e^{iM_Nv.x} \mathcal{P}^-_v N,
\label{N-redefinition}
\end{align}
with projection operators $\mathcal{P}^{\pm}_v=\frac{1\pm\gamma^{\mu}v_{\mu}}{2}$. Then it is straightforward to see that in the heavy field limit, when $v=(1,0,0,0)$, $\mathcal{P}^+_v$ projects out the upper components of the nucleon spinor with energy $E-M_N$ while $\mathcal{P}^+_v$ project the lower components of the nucleon spinor with energy $E-M_N$. With this decomposition, the only dynamical field that survives as $M_N \to \infty$ is $N_l$ whose corresponding Lagrangian can be written as \cite{Jenkins:1990jv}
\begin{eqnarray}
\hat{\mathcal{L}}^{(1)}_{N,pGB}=\overline{N}_l(iD_{0}-g_{A}\bm{\sigma} \cdot \bm{\mathcal{A}})N_l,
\label{L-NpGB}
\end{eqnarray}
at LO in $1/M_N$ expansion where $\bm{\sigma}$ are Pauli matrices of $SU(2)$ in the spin space. Note that the mass term in Eq. (\ref{L-NpGB}) is now canceled via such non-relativistic (NR) reduction. This formalism, that is known in literature as heavy-baryon $\chi$PT (HB$\chi$PT), makes the EFT calculations involving baryons considerably easy specially at higher orders. For future use, let us make explicit the interactions among nucleons and pions in this Lagrangian by expanding the $\xi$ field in Eq. (\ref{L-NpGB}) in powers of pion fields. After neglecting terms with more than two pion fields, one arrives at
\begin{eqnarray}
\hat{\mathcal{L}}^{(1)}_{\pi N}=\overline{N}_l\left[i\partial_{0}-\frac{1}{4f_{\pi}^2} \bm{\tau} \cdot (\bm{\pi} \times \partial_{0} \bm{\pi})-\frac{g_{A}}{2f_{\pi}} \bm{\tau} \cdot (\bm{\sigma} \cdot \bm{\partial}) \bm{\pi}\right]N_l,
\label{L-piN}
\end{eqnarray}
where $\bm{\tau}$ are the Pauli matrices of $SU(2)$ in the isospin space. Several interesting processes can be studied with this Lagrangian including the pion-nucleon scattering and the quark-mass dependence of nucleon mass. We will use this Lagrangian in the next section to evaluate the FV corrections to the mass of nucleons, and later in chapter \ref{chap:TBC} to improve such volume corrections by modifying the quark-field boundary conditions in a finite volume.

The interactions of pGBs and baryons with external fields such as EM field can be also included in the EFT. For the case of electromagnetism, for example, a minimal coupling of hadrons to the photon field $A_{\mu}$ will account for such interactions at LO. It is notable that the quark electric charge matrix $Q$,
\begin{eqnarray}
Q=\left(\begin{array}{ccc}
\frac{2}{3} & 0 & 0\\
0 & -\frac{1}{3} & 0\\
0 & 0 & -\frac{1}{3}
\end{array}\right),
\label{L-piN}
\end{eqnarray}
breaks chiral symmetry explicitly just as the quark mass matrix and its inclusion in the chiral Lagrangian follows in a similar fashion. We will not discuss this extension of EFT Lagrangian here and refer the reader to various comprehensive reviews on $\chi$PT and its applications as can be found in Refs. \cite{Ecker:1994gg, Pich:1995bw, Scherer:2002tk, Kaplan:2005es, Machleidt:2011zz}. In studying EM FV corrections to the mass of hadrons in chapter \ref{chap:EM}, we introduce a simple NR EFT that captures the features of the EFTs coupled to EM fields.

\subsubsection{Effective field theories for nucleons
\label{subsec:NEFT}}
\indent 

\emph{EFT potentials and Weinberg power counting}: In early 1990s, Weinberg proposed that the phenomenological potentials of nuclear physics \cite{Jackson:1975be, Partovi:1969wd, Partovi:1972bj, Lacombe:1980dr, Machleidt:1989tm} can be replaced with potentials that are systematically constructed from chiral EFT interactions \cite{Weinberg:1990rz, Weinberg:1991um, Weinberg:1992yk}. The uncertainties of the nuclear few- and many-body calculations due to neglecting higher order terms in the EFT forces can then, in principle, be systematically estimated. This procedure goes as follows:
\begin{enumerate}
\item 
 Write down, order by order in $\chi$PT, the potential among two nucleons. At LO, there is no contribution from three (and more) nucleon forces. The one-pion exchange (OPE) potential, which was also included in the phenomenological NN potentials to account for the long-range force among nucleons, contributes at LO. In the static limit
 \begin{eqnarray}
V_{1\pi}^{(LO)}(\bm{p},\bm{p}')=-\frac{g_A^2}{2 f_{\pi}^2} \bm{\tau}_1 \cdot \bm{\tau}_2
\frac{\bm{\sigma}_1 \cdot \bm{q}~\bm{\sigma}_2 \cdot \bm{q}}{\bm{q}^2+m_{\pi}^2},
\label{V-OPE}
\end{eqnarray}
where $\bm{p}$ and $\bm{p}'$ are the three momenta of the two interacting nucleons and $\bm q$ is the three momentum of the exchanged pion, see Fig. \ref{fig:V-NN-LO}. It consists of both central and tensor force and therefore can account for $L=0$ and $L=2$ angular-momentum mixing in the deuteron (total spin $S=1$ and total isospin $I=0$) wavefunction, see chapter \ref{chap:NN}.
\begin{figure}[h!]
\begin{centering}
\includegraphics[scale=0.465]{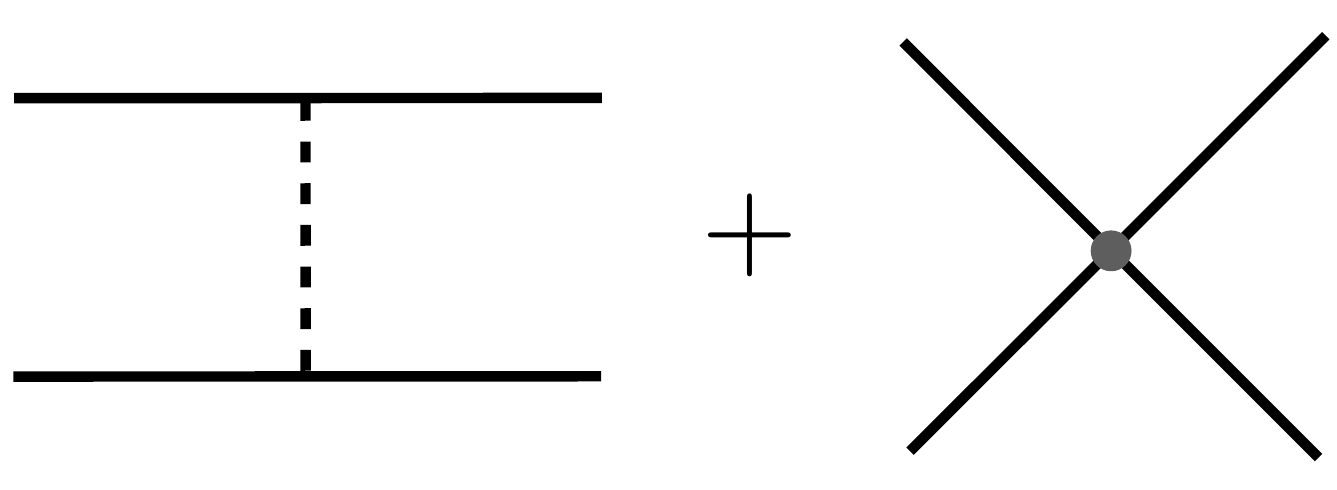}
\par\end{centering}
\caption{{\small The LO contributions to the NN potential in the Weinberg power counting. Solid (dashed) line represents the nucleon (pion). The black dot denotes the four-nucleon contact, $C_S$ or $C_T$.}}
\label{fig:V-NN-LO}
\end{figure}

In order to describe the short-range nuclear force and to renormalize away the $\delta$-function singularity of the OPE potential (in position space), two four-nucleon contact operators, with coefficients $C_S$ and $C_T$ must be introduced at the same order, giving rise to the potential 
 \begin{eqnarray}
V_{CT}^{(LO)}(\bm{p},\bm{p}')=C_S + C_T \bm{\sigma}_1 \cdot \bm{\sigma}_2.
\label{V-CT}
\end{eqnarray}

One keeps going to higher orders in the $p/\Lambda_{\chi}$ expansion, by including multi-pion exchange potentials with leading as well as higher order pion-nucleon vertices, and by including as many contact interactions needed to renormalize the UV singularities at any given order. 

\item
Given the potential, calculate the NR scattering amplitude, $\mathcal{M}$, by solving the NR Lippmann-Schwinger equation,
 \begin{eqnarray}
i \mathcal{M}({\bm{p}',\bm{p}}) = V({\bm{p}',\bm{p}}) + \int d^3 p'' V({\bm{p}',\bm{p}''}) \frac{M_N}{\bm{p}^2-\bm{p}''^2+i \epsilon} i \mathcal{M}({{\bm{p}'',\bm{p}}}).
\label{LS}
\end{eqnarray}

\item
By fitting to the well-known scattering phase shifts in various NN channels, constrain the LECs of the EFT potentials, including those of the contact terms.

\item
Solve the many-body problem by inputting these constrained EFT potentials to make predictions for the properties of few and many-body nuclear system, see e.g. Refs. \cite{Gezerlis:2013ipa, Kruger:2013kua, Tews:2013wma, Gezerlis:2014zia, Lynn:2014zia, Roggero:2014lga, Lee:2004si, Borasoy:2005yc, Borasoy:2006qn}.
\end{enumerate}

\begin{figure}[t!]
\begin{centering}
\includegraphics[scale=0.435]{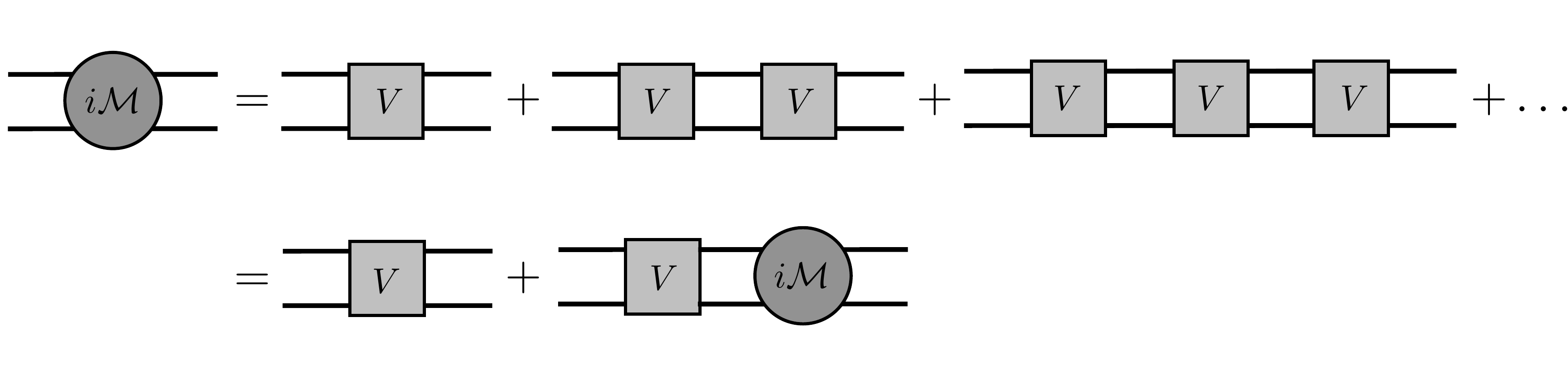}
\par\end{centering}
\caption{{\small The diagrammatic representation of the Lippmann-Schwinger equation. Solid lines represent the nucleons.}}
\label{fig:LS}
\end{figure}

Unfortunately, Weinberg procedure, despite producing potentials in a systematic way, does not give rise to a consistent power counting in all the two-nucleon channels, and the phase shifts obtained with this method typically diverge as the cutoff used to regularize the divergences is taken to infinity. This undesired scale dependence of physical quantities in the Weinberg power counting can be understood from Eq. (\ref{LS}) where, for example inputting the LO potential in the integral equation results in a summation of the LO interactions to all orders, see Fig. (\ref{fig:LS}). Therefore the amplitude obtained at this order is not a true LO amplitude as it contains higher order loops. Unfortunately these higher order terms, e.g. two-pion exchange, etc., suffer from singularities that cannot be renormalized given the absence of the contact interactions at this order. These interactions only appear in the expansion of the potential at higher orders which are not included in the LO potential that is used in the Lippmann-Schwinger equation. This means that the calculated amplitude is divergent as the cutoff is taken to infinity, and in this sense this procedure cannot be regarded as a genuine EFT approach. In practice, the uncertainty associated with the determination of a given quantity with this method is estimated by varying the cutoff scale in the calculation. For a nice review of chiral nuclear forces, see Ref. \cite{Machleidt:2011zz}.

\
\

\begin{figure}[h!]
\begin{centering}
\includegraphics[scale=0.425]{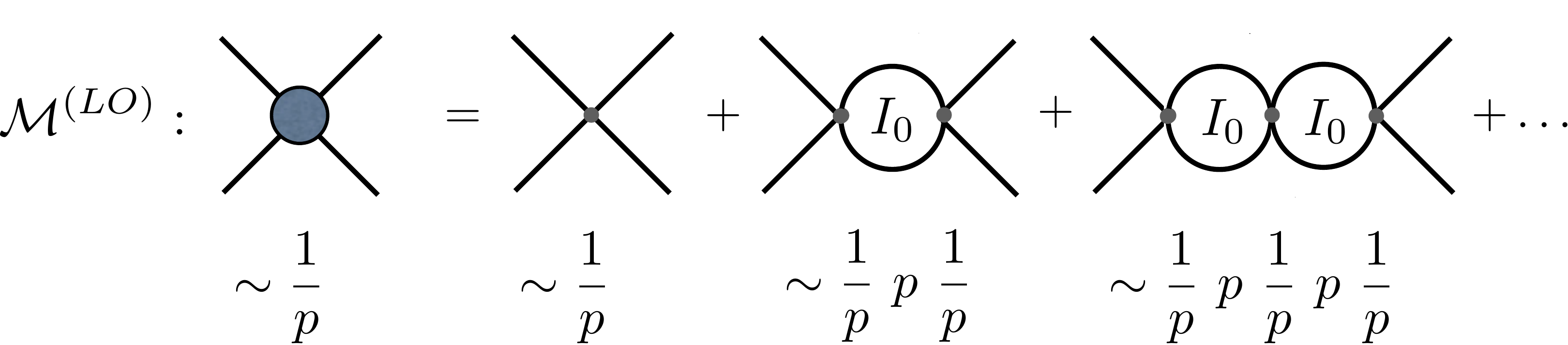}
\par\end{centering}
\caption{{\small The contribution from the leading contact interaction $C_0$ is summed up to all orders, giving rise to the the leading scattering amplitude consistent with unnaturally large scattering amplitude in the S-wave NN channels. Note that due to the PDS scheme, each term in the expansion comes at the same order, $\mathcal{O}(\frac{1}{p})$, in contrast with the $\overline{MS}$ or momentum subtraction schemes, see Refs. \cite{Kaplan:1998tg, Kaplan:1998we}.}}
\label{fig:C0-KSW}
\end{figure}

\emph{NN interactions with Kaplan-Savage-Wise power counting}: Instead of working with potentials, one can directly relate scattering amplitudes to the interaction Lagrangian order by order in a low-energy expansion scheme. In the two-body elastic scattering at low energies,\footnote{Below the t-channel cut, $E^*=m_{\pi}/2$.} the relevant intrinsic scales are the scattering length and effective range -- and shape parameters as defined in Eq. (\ref{NN-ERE}). One might think that given the effective range expansion, the scattering amplitude can be straightforwardly written as expansions in $pa$ and $pr$ where $p$ is the typical momentum of the process, i.e. the energy of each particle in the CM frame $p \sim k^*$ or the mass of the pions $p \sim m_{\pi}$. In fact this turns out to be the case in many of the scattering channels. However, the S-wave NN scattering represents unnatural features arising from seemingly fine-tuned interactions. This is manifested in the large scattering length of the system, for example in the $^1 S_0$ channel where $a^{(^1 S_0)}=-23.714 \pm 0.013~{\rm fm} \gg \frac{1}{m_{\pi}}$. The same feature is seen in the ${^3S_1}-{^3D_1}$ coupled channel where $a^{({^3S_1}-{^3D_1})}=5.423 \pm 0.005~{\rm fm} \gg \frac{1}{m_{\pi}}$, giving rise to a near threshold bound state, the deuteron, whose binding energy, $\sim 2~{\rm MeV}$, is much smaller than that set by the typical QCD scale, $\Lambda_{QCD}$. The unnaturalness in these channels indicates that the LO scattering amplitude does not count as $\sim p^0$, but is instead of $\sim p^{-1}$,
 \begin{eqnarray}
\mathcal{M} = \frac{4 \pi}{M_N} \frac{1}{k^* \cot \delta - ik^*}
= - \frac{4 \pi}{M_N} \frac{1}{1/a+ik^*}\left[ 1 + \frac{r/2}{(1/a+ik^*)} k^{*2} + \dots \right],
\label{M-EFE}
\end{eqnarray}
where $k*=\sqrt{M_N E^*}$ with $E^*$ being the CM energy of two-nucleon system. A sensible power counting at low energies must be able to reproduce this effective range expansion of the amplitude. Clearly, the OPE interaction of nucleons comes at $\sim p^0$ and cannot be counted as a LO interaction. The momentum-independent contact interaction, with coefficient $C_0$ then must be responsible for the LO amplitude, provided that it scales as $\sim p^{-1}$. This requires the chain of bubble diagrams with insertions of this leading operator to all scale at most as $\sim p^{-1}$ or otherwise one loses control over these contributions. A suitable regularization scheme to ensure this scaling is the dimensional regularization with the power-divergence subtraction (PDS) scheme, as proposed by Kaplan, Savage and Wise (KSW) in late 1990s \cite{Kaplan:1998tg, Kaplan:1998we}. Explicitly the LO amplitude, according to Fig. \ref{fig:C0-KSW}, can be written as
 \begin{eqnarray}
\mathcal{M}^{(LO)} = \frac{-C_0}{1-I_0^{PDS}C_0}= \frac{-C_0}{1+\frac{M_N}{4\pi}(\mu+ik^*)C_0},
\label{M-LO-KSW}
\end{eqnarray}
where 
 \begin{eqnarray}
I_0 &=& -i (\frac{\mu}{2})^{4-d} \int \frac{d^d q}{(2\pi)^d} \frac{i}{q^0-\frac{\bm{q}^2}{2M_N}+i\epsilon}\frac{i}{E^*-q^0-\frac{\bm{q}^2}{2M_N}+i\epsilon}
\nonumber\\
&=&-M_N (-M_NE^*-i \epsilon)^{(d-3)/2} \Gamma \left(\frac{3-d}{2}\right) \frac{(\mu/2)^{4-d}}{(4\pi)^{(d-1)/2}},
\label{I0-KSW}
\end{eqnarray}
with $d$ being the dimensionality of spacetime and $\mu$ being the renormalization scale. As is seen, although $I_0$ does not have any singularity in $d=1+3$ dimensions, it is singular in a lower dimension $d=1+2$ (corresponding to the power divergence of $I_0$ in $4$ dimensions that is absent in the dimensional regularization). PDS scheme prescribes that this pole must be subtracted from $I_0$ and therefore making the result $\mu$ dependent,
 \begin{eqnarray}
I_0^{(PDS)}=I_0|_{d=4}-I_0^{({\rm div.})}|_{d=3} = - \frac{M_N}{4\pi}(\mu+ik^*),
\label{I0-KSW-PDS}
\end{eqnarray}
giving rise to Eq. (\ref{M-LO-KSW}). Now by comparing Eq. (\ref{M-LO-KSW}) and Eq. (\ref{M-EFE}) one will find that $C_0$ indeed scales as $p^{-1}$
 \begin{eqnarray}
C_0(\mu)=\frac{4\pi}{M_N}\left( \frac{1}{-\mu+1/a} \right),
\label{C0-KSW}
\end{eqnarray}
given that $\mu \sim p \gg 1/a$.

At NLO, not only the OPE contributes,\footnote{For scattering processes above the t-channel cut, $E^*>m_{\pi}/2$.} but also the insertions of both $C_2 \nabla^2$ and $D_2 m_{\pi}^2$ operators must be taken into account. Given that each loop scales as $\sim p$, these coefficients must scale as $\sim \frac{1}{p^2}$. In addition, the initial and final nucleon legs must be dressed by the chain of $C_0$ bubbles as they give rise to $\mathcal{O}(1)=(\frac{1}{p})(p)$ contributions. We will not discuss these contributions in details here, however we can already see that this expansion, with the devised power counting, systematically treats pion exchanges perturbatively, and the scale dependence of the amplitudes at each order is completely removed by the introduction of corresponding \emph{counter terms}, i.e. coefficients of the contact interactions that are representative of the short-distant physics of the problem. It is also notable that the pion-mass dependence of the NN interactions systematically arises from these EFT interactions.\footnote{At energies well below the pion mass, the pions can be integrated out from the EFT, giving rise to the \emph{pionless EFT}. In chapter \ref{chap:NN}, we will work with this EFT, along with the use of a dimer field, to reproduce the ERE in NN systems.} For a nice review of EFTs for nucleons, see Ref. \cite{Kaplan:2005es}.

Although the KSW EFT for nucleons has been shown to be a powerful method in studies of the electroweak transitions in the few-body systems, as well as in developing an EFT for three-nucleon systems, it suffers from a slow convergence in the $^1 S_0$ channel and is not converging in the ${^3S_1}-{^3D_1}$ channel \cite{Fleming:1999ee, Fleming:1999bs}. In the coupled ${^3S_1}-{^3D_1}$ channels, the piece in the OPE potential that survives in the chiral limit is large enough to ruin the convergence of an EFT with perturbative pions. This problem is however alleviated in the Weinberg power counting where pions are treated nonperturbatively. This has led the authors of Ref. \cite{Beane:2001bc} to propose a better power-counting scheme which requires an expansion around the chiral limit. The community remains in need for a better EFT for nuclear interactions which does not suffer from the drawbacks of the approaches mentioned here. Nonetheless, these EFTs have widely been used in studying a variety of nuclear systems \cite{KalantarNayestanaki:2011wz, Barrett:2013nh, Roth:2011ar, Epelbaum:2009pd, Epelbaum:2011md, Epelbaum:2012iu, Otsuka:2009cs, Holt:2010yb, Holt:2013vqa, Hagen:2012sh, Hagen:2012fb, Roth:2011vt, Hergert:2012nb, Soma:2012zd, Wienholtz:2013nya, Kaiser:2001jx, Epelbaum:2008vj, Hebeler:2009iv, Hebeler:2010xb, Hebeler:2010jx, Tews:2012fj, Kruger:2013kua, Holt:2012yv, Gezerlis:2013ipa}. The uncertainties on three- and multi-nucleon force parameters remain a significant source of  uncertainty in some of these calculations. Due to limited experimental data, the help of LQCD to constrain these parameters will be crucial in the upcoming years.

\section{Lattice Quantum Chromodynamics
\label{LQCD}}
\begin{figure}[t!]
\begin{centering}
\includegraphics[scale=0.465]{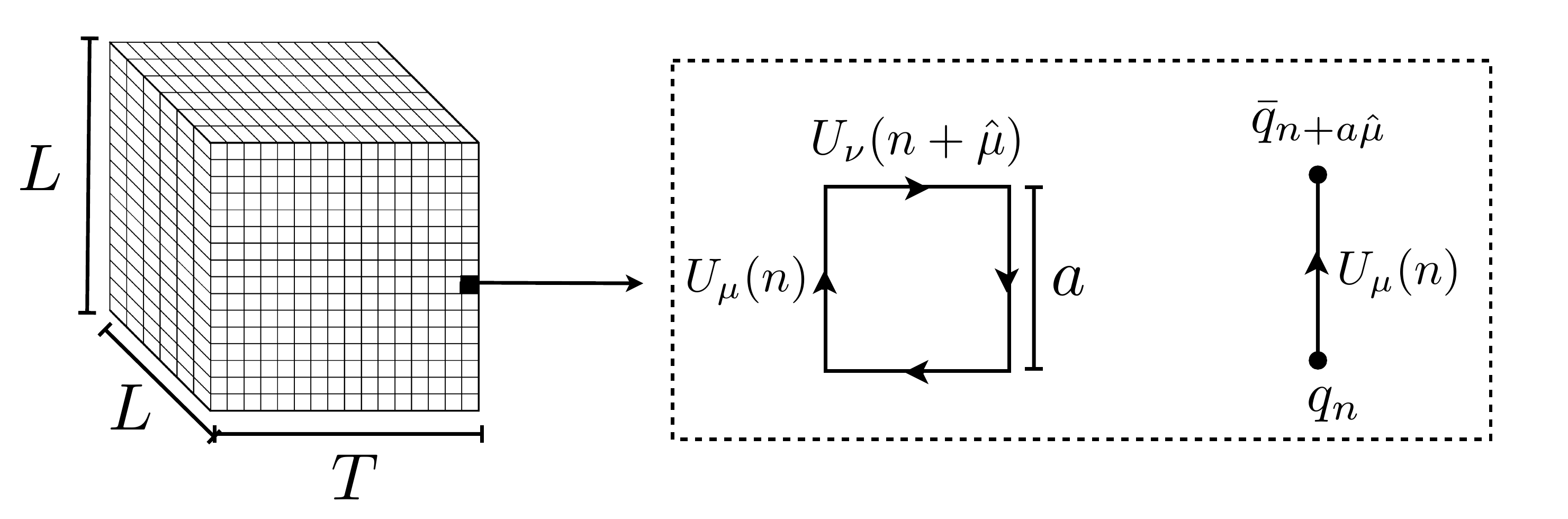}
\par\end{centering}
\caption{{\small A $2+1$ dimensional cubic lattice is shown in the left panel. The (trace of) plaquette and the product of quark, the link variable and the antiquark (right panel) are two examples of gauge-invariant constituents of the lattice gauge theories in their compact formalism.}}
\label{fig:Lattice}
\end{figure}

A non-perturbative approach in solving QCD, without making any assumption about the strength of the coupling or the energy scale, is via the path integral formalism. In this formalism, physical quantities are evaluated by taking  expectation values of the corresponding operators in the background of the QCD vacuum,
\begin{eqnarray}
\langle \hat{\mathcal{O}} \rangle = \frac{1}{\mathcal{Z}} \int \mathcal{D}A_{\mu} \mathcal{D}q \mathcal{D}\bar{q} ~ e^{iS_{QCD}} ~ \hat{\mathcal{O}},
\label{path-integral}
\end{eqnarray}
where $\mathcal{Z} = \int \mathcal{D}A_{\mu} \mathcal{D}q \mathcal{D}\bar{q} ~ e^{iS_{QCD}}$ denotes the QCD partition function, $S_{QCD}=\int d^4x ~ \mathcal{L}_{QCD}$ is the action and $\mathcal{L}_{QCD}$ is given in Eq. (\ref{L-QCD}). Evaluating this path integral in practice requires several steps to be followed:

\emph{1) A discrete action}: The path integral in Eq. (\ref{path-integral}) is only defined rigorously if the degrees of freedom of the theory are discrete. Numerical evaluations become plausible in practice, firstly, with a measure that is nonoscillatory. This can be achieved by a Wick rotation of the coordinates to Euclidean spacetime, $t \to i\tau$ so that $iS_{QCD} \to -S_{QCD}^{(E)}$ where $S_{(QCD)}^{(E)}$ is purely real. Secondly, the number of degrees of freedom of the integration must be finite, requiring the spacetime to be truncated to a finite region in both spatial and temporal directions and to be discretized.  Lattices with geometry of a hypercube are the most convenient choices in LQCD calculations, see Fig. \ref{fig:Lattice}, although the anisotropic cubic lattices with lattice spacing in the temporal direction being finer than that of the spatial direction are being also used. The spacing between two adjacent lattice sites, $a$, must be small compared with the hadronic scale, $a \ll \Lambda_{QCD}^{-1}$, while the spatial extent of the volume, $L$, must be large compared with the Compton wavelength of the pions which sets the range of hadronic interactions, $L \gg m_{\pi}^{-1}$, see Sec. \ref{IV-intro}.

Quark fields are placed on the lattice sites, and a choice for defining the gauge fields, as plotted in Fig. \ref{fig:Lattice}, is through the Wilson link variables,
\begin{eqnarray}
U_{\mu}(n) \equiv e^{i g A_{\mu}(n)}.
\label{Links}
\end{eqnarray}
These are the elements of the $SU(3)$ Lie group and transform under a local gauge transformation as
\begin{eqnarray}
U_{\mu}(n) \to U_{\mu}(n)'=V(n) U_{\mu}(n) V^{\dagger}(n+\hat{\mu}),
\label{Links-trans}
\end{eqnarray}
where $V$ is an element of the Lie group. The use of link variables, which is called the \emph{compact} formulation of lattice gauge theories, is a convenient choice as it makes the implementation of gauge invariance on the lattice straightforward. In fact, the  only gauge invariant quantities are the gauge links starting and ending at the quark fields, and the trace of any closed loop formed by the gauge links, Fig. \ref{fig:Lattice}.
 With these gauge invariant blocks, we can write down a Lagrangian for QCD interactions on the lattice that recovers the Lagrangian in Eq. (\ref{L-QCD}) once the continuum limit is taken. A common choice of action is the Wilson action \cite{PhysRevD.10.2445} which uses the elementary plaquette, defined as $P_{\mu \nu;n} \equiv U_{\mu}(n) U_{\nu}(n+\hat{\mu}) U^{\dagger}_{\mu}(n+\hat{\nu}) U^{\dagger}_{\nu}(n)$, see Fig. \ref{fig:Lattice}, for gluons and the Wilson fermions formulation for the quarks, 
\begin{eqnarray}
S^{(E)}_{{\rm Wilson}}&=&\frac{\beta}{N_c} \sum_{n} \sum_{\mu < \nu} \Re {\rm Tr} [\mathbb{1}-P_{\mu \nu;n}]
\nonumber\\
&-& \sum_{n} \bar{q}_n [\overline{m}^{(0)}+4] q_n + 
\sum_{n} \sum_{\mu} \left[ \bar{q}_n \frac{r-\gamma_{\mu}}{2} U_{\mu}(n) q_{n+\hat{\mu}} + \bar{q}_n \frac{r+\gamma_{\mu}}{2} U_{\mu}^{\dagger}(n-\hat{\mu}) q_{n-\hat{\mu}} \right],
\nonumber\\
\label{L-Wilson}
\end{eqnarray}
where $n$ runs over all the $N_s^3 \times N_t$ lattice points and $\beta \equiv \frac{2 N_c}{g^2}$ is the lattice coupling constant with $N_c=3$ for QCD. Note that the action is written in terms of dimensionless fields and parameters. Explicitly, the continuum field $q$ at point $na$ is replaced by $a^{-3/2} q_n$ and the continuum bare mass of the quarks $m^{(0)}$ is replaced by $a^{-1}\overline{m}^{(0)}$. $r$ is the Wilson parameter whose value is commonly set to $1$ in the calculations. The sum over quark flavors is left implicit.

The gluonic part of the action clearly recovers the continuum action in Eq. (\ref{CP-even}) up to corrections that scale as $a^2$, and leads to the following lattice propagator in momentum space
\begin{eqnarray}
a^{-2}\mathcal{D}^{(G)}_{{\rm Wilson}}(\overline{k})=\frac{i}{4\sum_{\mu}\sin^{2}\left(\overline{k}_{\mu}/2\right)},
\label{Gluon-prop}
\end{eqnarray}
where the Feynman gauge is used to fix the gauge and $\overline{k}_{\mu} = k_{\mu} a$.\footnote{Due to the use of manifestly gauge-invariant path integral, the compact formulation of lattice gauge theories does not require gauge fixing. This is in contrast with a non-compact formulation where the gauge fields remain the explicit degrees of freedom and the continuum action is discretized directly. This is the popular formulation used in pure lattice QED calculations and so requires fixing the gauge, see chapter \ref{chap:EM}.} The fermionic part of the action is nothing but what is  expected from a \emph{naive} discretization of the Dirac operator in Eq. (\ref{L-q-free}) -- with the inclusion of link variables to render the discrete derivative gauge invariant --  plus an additional contribution proportional to $r$. This latter contribution is introduced by Wilson to circumvent the so-called fermion doubling problem, due to which the continuum limit of the naive Dirac fermions leads to $2^4$ degenerate fermions. One way to see this problem is by studying the Wilson quark propagator that we use extensively in chapter \ref{chap:operators} to study the lattice operators perturbatively, and can be derived readily from the action in Eq. (\ref{L-Wilson}),
\begin{eqnarray}
a^{-1}\mathcal{D}^{(F)}_{{\rm Wilson}}=\frac{-i\sum_{\mu}\gamma_{\mu}\sin\left(\overline{k}_{\mu}\right)+2r\sum\limits_{\mu}\sin^{2}\left(\overline{k}_{\mu}/2\right)+\overline{m}^{(0)}}
{\sum_{\mu}\sin^{2}\left(\overline{k}_{\mu}\right)+(2r\sum\limits_{\mu}\sin^{2}\left(\overline{k}_{\mu}/2\right)+\overline{m}^{(0)})^2}.
\label{Quark-prop}
\end{eqnarray}
When $r = 0$, the poles of the propagator for $\overline{m}^{(0)} = 0$ occur at 16 distinct momenta, $\bar{k}$,
\begin{eqnarray}
(0,0,0,0),(\pm \pi,0,0,0),(0,\pm \pi,0,0), \dots ,(\pm \pi,\pm \pi,\pm \pi,\pm \pi),
\label{doublers}
\end{eqnarray}
however, only the $\bar{k}=0$ pole is the desired continuum pole. By adding the Wilson term, the doublers that correspond to $|\bar{k}| \sim \pi$ acquire a mass that, in the continuum limit, scale as $m(r) \sim r/a$ and will therefore decouple from theory due to their heavy mass.

The downside of the Wilson action is that it breaks the chiral symmetry explicitly. As is evident from action (\ref{L-Wilson}), the terms proportional to $r$ behave similar to the quark mass term and mix the left-handed and right-handed quarks. As it turns out, these are universal problems with most of the discretized fermionic actions that can be nicely summarized via the Nielsen-Ninomiya theorem \cite{Nielsen198120}. The theorem states that a lattice Dirac operator cannot simultaneously 1) be a periodic function of momentum and analytic except at $\mathbf{p} \neq \mathbf{0}$, 2) be proportional to $\gamma_{\mu} p_{\mu}$ in the continuum limit, and 3) anticommute with $\gamma_5$. It is clear that the naive Dirac operator satisfies both 2 and 3 but fails to meet the first condition given the presence of doublers. The Wilson Dirac operator on the other hand satisfies 1 and 2 but it does not anticommute with $\gamma_5$ signifying its chiral symmetry breaking feature. Solutions to the lattice fermions' puzzle include the \emph{domain-wall} fermions \cite{Kaplan:1992bt} and overlap fermions \cite{Narayanan:1993ss, Narayanan:1994gw} that both belong to the category of the Ginsberg-Wilson fermions.\footnote{Domain-wall fermions only satisfy the Ginsberg-Wilson relation in a particular limit, i.e. when the domain-walls separation is infinite.} Ginsberg and Wilson relation \cite{Ginsparg:1981bj} redefines the chiral symmetry on the lattice,
\begin{eqnarray}
\{D,\gamma_5\}=a D \gamma_5 D,
\label{GW}
\end{eqnarray}
with $D$ being a dimensionful Dirac operator, and therefore breaks the last condition in the Nielsen-Ninomiya theorem. It however ensures that the chiral features of the continuum fermions, including the chiral anomaly, are exactly reproduced as long as the operator $D$ satisfies this relation. Unfortunately, numerical simulations of both domain-wall and overlap fermions comes with additional cost compared with Wilson fermions.\footnote{Simulating  domain-wall fermions includes adding an extra dimension to the calculation of the quark propagators while simulating overlap fermions requires inversion of an extra operator beside the overlap operator, see Refs. \cite{Kaplan:2009yg, Kennedy:2006ax, Jansen:1992tw} for more details.} Nonetheless, the use of the chiral lattice fermions in LQCD calculations has become more common as the computational resources improve. A nice review of fermions and chiral symmetry on the lattice can be found in Ref. \cite{Kaplan:2009yg}.

\emph{2) Generate gauge-filed configurations}: Now that we have a discrete action with the desired continuum limit, let us go back to the path intergarl we aim to evaluate,
\begin{eqnarray}
\langle \hat{\mathcal{O}} \rangle = \frac{1}{\mathcal{Z}} \int \mathcal{D}U_{\mu} \mathcal{D}q \mathcal{D}\bar{q} ~ e^{-S^{(G)}_{\rm lattice}[U]-S^{(F)}_{\rm lattice}[U,q,\bar{q}]} ~ \hat{\mathcal{O}}[U,q,\bar{q}],
\label{path-integral-lattice-I}
\end{eqnarray}
where we have split the action to the purely gauge part and the fermionic part, and have left the superscripts $E$ for the Euclidean action implicit. This expectation value can be written as
\begin{eqnarray}
\langle \hat{\mathcal{O}} \rangle = \frac{1}{\mathcal{Z}} \int \mathcal{D}U_{\mu} ~ e^{-S^{(G)}_{\rm lattice}[U]} \mathcal{Z}_F[U]~ \langle \hat{\mathcal{O}} \rangle_F
\label{path-integral-lattice-II},
\end{eqnarray}
where the path integral over gauge links $U$ are separated from that of the fermionic path integrals with
\begin{eqnarray}
\langle \hat{\mathcal{O}} \rangle_F= \frac{1}{\mathcal{Z}_F} \int \mathcal{D}q \mathcal{D}\bar{q} ~ e^{-S^{(F)}_{\rm lattice}[U,q,\bar{q}]}\mathcal{O}[q,\bar{q},U],
\label{O-F}
\end{eqnarray}
and $\mathcal{Z}_F$ is the partition function of the fermions which will still depend on the value of the gauge link. By expressing the fermionic action as $S^{(F)}_{\rm lattice}=\sum_{n,m} \bar{q}_n D_{n,m} q_{m}$, where $D_{n,m}$ is the matrix element of one of the chosen lattice operators discussed above in position space, the fermionic partition function can be written as
\begin{eqnarray}
\mathcal{Z}_F= \int \mathcal{D}q \mathcal{D}\bar{q} ~ e^{-S^{(F)}_{\rm lattice}[U,q,\bar{q}]}= \prod_f \det D_f,
\label{Z-F}
\end{eqnarray}
where the product of the determinant of Dirac operator matrix, $D_f$, corresponding to each dynamical flavor is explicit. Now from Eq. (\ref{path-integral-lattice-II}) it is clear that once the fermionic expectation value $\langle \hat{\mathcal{O}} \rangle_F$ is computed, the full expectation value can be computed using a Monte Carlo sampling integration with the probability measure $ \frac{1}{\mathcal{Z}}e^{-S^{(G)}_{\rm lattice}[U]} \prod_f \det D_f$. An important property of the lattice Dirac operators, the $\gamma_5$-hermiticity $D^{\dagger}=\gamma_5 D \gamma_5$, ensures that the determinant of the Dirac operator is real, providing a well-defined sampling weight in the numerical evaluation of the expectation values. LQCD calculations with dynamical fermions require computing the gauge-field configuration with a distribution that depends on the fermion determinant -- the determinant of the Dirac operator which is a large matrix with dimensionality $(12 N_s^3 \times N_t )^2$ (on each spacetime point on the lattice there are 3 color and 4 spinor degrees of freedom for each flavor of quarks). After each configuration generation both the gauge part and the determinant part must be updated simultaneously to generate the next configuration.\footnote{As a result, early LQCD calculations were limited to the \emph{quenched} approximation where the fermion determinant is set to one to reduce the computational cost of the gauge-field configurations. Unfortunately quenching is an uncontrolled approximation and only describes QCD if the quarks were infinitely heavy. Nowadays, the growth in the computational resources available to LQCD calculations has enabled abandoning this approximation and has made the use of dynamical configurations viable in most calculations.}

When a large number of almost statistically uncorrelated gauge field configurations, $N$, are generated, the statistical average
\begin{eqnarray}
\langle \hat{\mathcal{O}} \rangle = \frac{1}{N} \sum^{N}_{i} \langle \hat{\mathcal{O}} \rangle_F [U^{(i)}],
\label{expec-value}
\end{eqnarray}
is an estimator of the the expectation value in Eq. (\ref{path-integral-lattice-II}), where $U^{(i)}$ is the $i^{th}$ generated configuration.

\begin{figure}[t!]
\begin{center}
\includegraphics[scale=0.365]{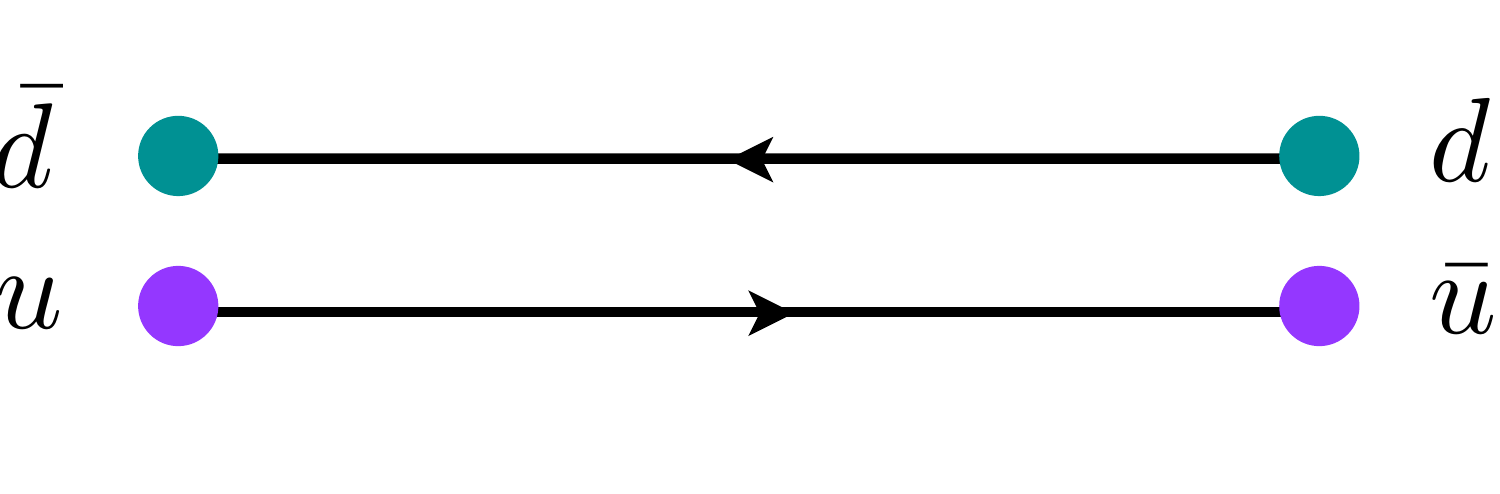}
\caption
{{\small The Wick contractions in the evaluation of the $\pi^+$ two-point correlation functions.}
}
\label{fig:Pion-contraction}
\end{center}
\end{figure}

\emph{3) Form the correlation functions}: The next step of the calculation is observable dependent and requires both analytical and numerical evaluation to determine $\langle \hat{\mathcal{O}} \rangle_F$. Here we are interested in the n-point correlation functions of (multi) hadrons from which one can extract masses and the low-lying energies. Let $\hat{O}^{\dagger}$ denote the interpolating operator that creates a (multi-)hadron states from the vacuum of QCD and $\hat{O}$ be an interpolator that annihilates the state. With the notation used in Eq. (\ref{path-integral-lattice-I}), $\hat{\mathcal{O}} \equiv \hat{O}\hat{O}^{\dagger}$. In order for an interpolating operator to have overlap with a desired state, it must share the same quantum numbers, e.g. the particle number, flavor, spin, parity, charge conjugation, etc., as that of the state. For example the $\pi^+$ state can be created by a bilinear quark operator $O^{{\pi^+}\dagger}=\overline{u} \gamma_5 d$. In order to calculate the correlation function, we need to perform the fermionic path integral that appears in the expectation value $\langle \hat{\mathcal{O}} \rangle_F$ which is a usual Grassmann integration. This part is called the quark \emph{Wick} contractions and for the case of $\pi^+$ two-point correlation function can be performed as following
\begin{eqnarray}
\langle \hat{O}^{\pi^+}(n)\hat{O}^{\pi^+ \dagger}(0) \rangle_F&=&
\langle \overline{d}_{a,\alpha}(n) \gamma^5_{\alpha \beta} u_{\beta}^a(n) ~~
\overline{u}_{b,\alpha'}(0) \gamma^5_{\alpha' \beta'} d_{\beta'}^b(0) \rangle_F
\nonumber\\
&=&
- \gamma^5_{\alpha \beta} \gamma^5_{\alpha' \beta'} ~
 \langle d_{\beta'}^b(0) \overline{d}_{a,\alpha}(n) \rangle_d ~
\langle u_{\beta}^a(n) \overline{u}_{b,\alpha'}(0)   \rangle_u
\nonumber\\
&=&
- \gamma^5_{\alpha \beta} \gamma^5_{\alpha' \beta'}~
(D^{-1}_d)^b_{a,\beta' \alpha}(0,n) (D^{-1}_u)^a_{b,\beta \alpha'}(n,0)
\nonumber\\
&=&
-{\rm Tr} \left[ \gamma^5 D_u^{-1}(n,0) \gamma^5 D_d^{-1}(0,n) \right]
\nonumber\\
&=& -{\rm Tr} \left[D_u^{-1}(n,0) D_d^{-1}(n,0) \right],
\label{pi-contraction}
\end{eqnarray}
where we have chosen to create the pion at the origin and annihilate it at coordinate $n$. The trace is taken over spin and color degrees of freedom and the negative sign has been resulted from anti-commutation of the Dirac fields in the second line. In the last line the $\gamma^5$-hermiticity of the Dirac operator has been used. The resulting correlation function has been pictorially shown in Fig. \ref{fig:Pion-contraction}. 
The value of the inverse Dirac operator depends on the value of the link variable, therefore for each gauge-field configuration generated in the previous step, the inverse of the Dirac operator must be evaluated.\footnote{When the value of the light-quark masses that are used are close to their physical values, the small eigenvalues of the Dirac operator causes difficulties in numerical evaluations of the inverse matrix given the limited statistics. This is among the reasons for the numerical limitations faced by the LQCD community in approaching the physical point.}

\begin{figure}[t!]
\begin{center}
\includegraphics[scale=0.335]{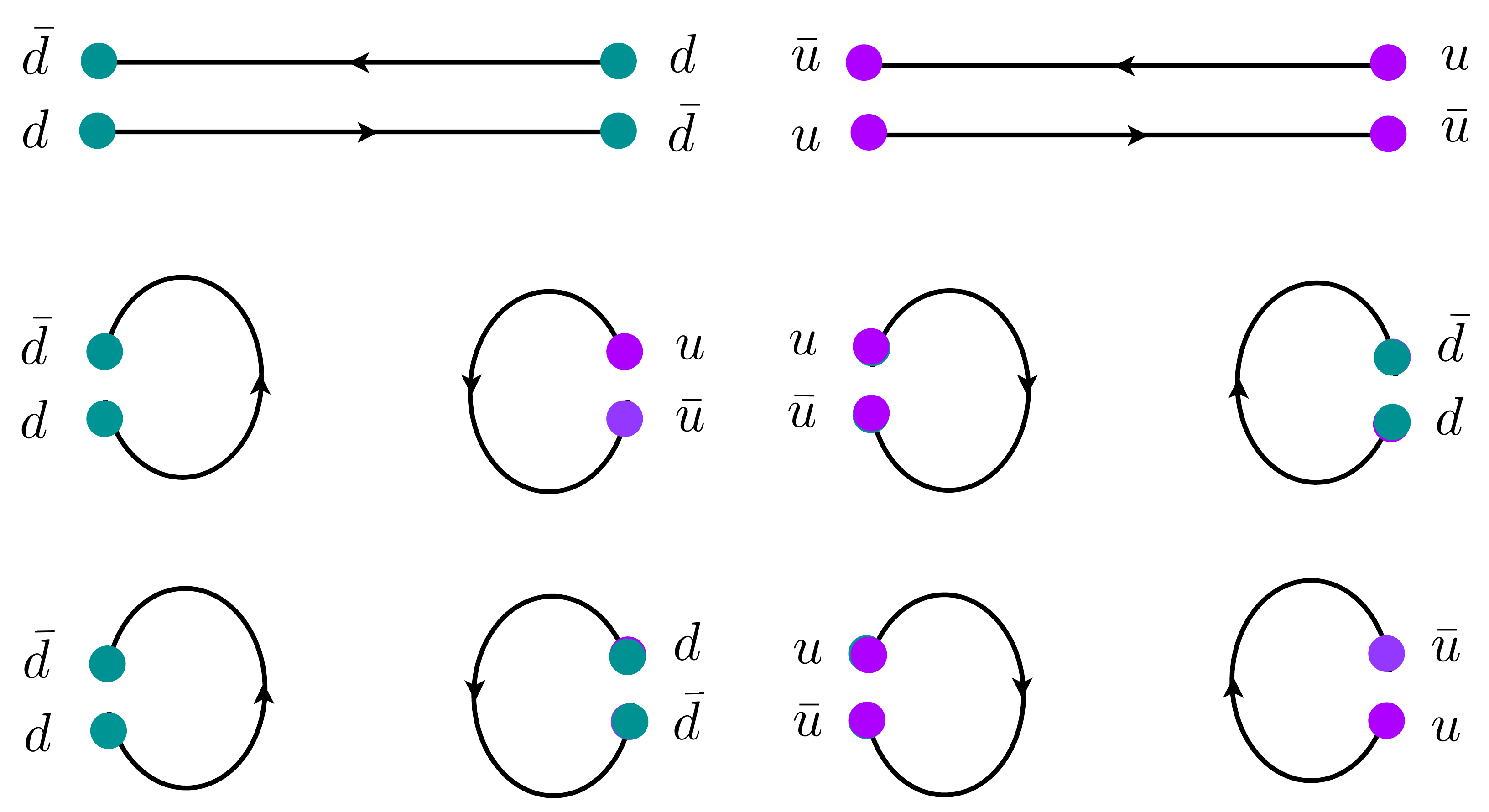}
\caption
{{\small The Wick contractions in the evaluation of the $\pi^0$ two-point correlation functions.}
}
\label{fig:Pion0-contraction}
\end{center}
\end{figure}
For flavor-singlet quantities, such as $\pi^0$, there are additional contributions to the correlation functions, namely the disconnected contributions, that put limitations on the calculation of such quantities with the current computational resources.\footnote{Some LQCD collaborations have started including the disconnected diagrams in their calculations, see Refs. \cite{Wagner:2012ay, Collins:2012mg, Alexandrou:2013cda, Abdel-Rehim:2013wlz, Dudek:2013yja, Alexandrou:2014yha, Bai:2014cva}.} Explicitly for the $\pi^0$ correlator with $\hat{O}=\frac{1}{\sqrt{2}}(\overline{u} \gamma^5 u - \overline{d} \gamma^5 d)$, we have
\begin{eqnarray}
\langle \hat{O}^{\pi^0}(n)\hat{O}^{\pi^0 \dagger}(0) \rangle_F &=&
-\frac{1}{2}{\rm Tr} \left[ \gamma^5 D_u^{-1}(n,0) \gamma^5 D_u^{-1}(0,n) \right]
\nonumber\\
&& + \frac{1}{2}{\rm Tr} \left[ \gamma^5 D_u^{-1}(n,n) \right] {\rm Tr} \left[ \gamma^5 D_u^{-1}(0,0) \right]
\nonumber\\
&& - \frac{1}{2}{\rm Tr} \left[ \gamma^5 D_u^{-1}(n,n) \right] {\rm Tr} \left[ \gamma^5 D_d^{-1}(0,0) \right] + \{ u \leftrightarrow d \},
\label{pio-contraction}
\end{eqnarray}
as depicted in Fig. \ref{fig:Pion0-contraction}. The second and third term, which contain the propagator from a single lattice point to itself, require evaluations of the \emph{all-to-all} propagators.\footnote{It must be noted that in the isospin limit, where the masses of $m_u$ and $m_d$ are set equal in the calculations, the disconnected contributions to the $\pi^0$ correlator vanish. This is the case for most of the lattice calculations that are currently performed. For isosinglet quantities such cancellation, even in the isospin limit, does not occur.} This introduces substantial extra cost in calculations as now instead of a column in the inverse Dirac operator matrix in position space, one needs to calculate the full matrix.

As the number of hadrons increases, the quark contractions to be performed become more involved, however the procedure described above remains the same. 
By taking advantage of various symmetries of multi-hadron systems and optimal choices of interpolators, the number of required contractions can be substantially reduced, see Refs. \cite{Shi:2011mr, Detmold:2012wc, Detmold:2010au, Doi:2012xd, Detmold:2012eu}. Due to the progress in the algorithms that perform contractions required in the evaluation of multi-baryon correlation functions \cite{Doi:2012xd, Detmold:2012eu}, obtaining the correlation functions of several nuclei up to $^{28}{\rm Si}$ are shown to be computationally plausible \cite{Detmold:2012eu}. Such developments gave rise to the first LQCD determination of the binding energies of the light nuclei and hypernuclei (up to atomic number 5) albeit at the heavy pion mass $m_{\pi} \approx 800~{\rm MeV}$ by the NPLQCD collaboration \cite{Beane:2012vq}, followed by the another determination of the binding of nuclei at a slightly lighter pion mass $m_{\pi} \approx 500~{\rm MeV}$ by Yamazaki, \emph{et al.} \cite{Yamazaki:2013rna}.

\emph{4) Extract masses and energies}: Let us first project the correlation function to a momentum $\bm{P}$,
\begin{eqnarray}
C(\mathbf{P};n_4) = \sum_{\mathbf{n}} e^{i \bm{P}.\bm{n}a} \langle \hat{O}(\mathbf{n},n_4)\hat{O}^{\dagger}(\mathbf{0},0) \rangle.
\label{corre-func-def-I}
\end{eqnarray}
Then upon inserting a complete set of states and using $\hat{O}(\mathbf{n},n_4)=e^{\hat{H}n_4}\hat{O}(\mathbf{n},0)e^{-\hat{H}n_4}$ (where Hamiltonian operator is defined through the lattice transfer matrix), the correlation function in the limit of large (Euclidean) time becomes
\begin{eqnarray}
C(\mathbf{P};n_4) &=& \sum_k \sum_{\mathbf{n}} e^{i \bm{P}.\bm{n}a} \langle 0  | \hat{O}(\mathbf{n},n_4)  | k \rangle \langle k  | \hat{O}^{\dagger}(\mathbf{0},0)  | 0 \rangle
\nonumber\\
&=&A_0 e^{-E(\bm{P}) |n_4| a}\left(1+\mathcal{O}(e^{-\Delta E(\bm{P}) |n_4| a})\right)
\label{corre-func-def-II},
\end{eqnarray}
where $E(\bm{P})$ is the lowest energy eigenvalue of the system and is related to the three-momentum through a (lattice) dispersion relation. $\Delta E(\bm{P})$ denotes the difference between the ground state energy and the first excited state energy. $A_0$ accounts for the overlap of the interpolator used onto the ground state.

A useful quantity which is commonly plotted is the effective mass/energy, defined as
\begin{eqnarray}
m_{eff}(n_t)=\ln \frac{C(\mathbf{P};n_t)}{C(\mathbf{P};n_t+1)}.
\label{m-eff}
\end{eqnarray}
As is clear, once the system approaches its ground state at large times, this quantity becomes constant. This defines a plateau region the the effective mass plot (EMP) as a function of time from which the ground state energy of the system can be read off. By using a larger basis of interpolating operators and by increasing the number of correlation function measurements, the excited state energies of the system can as well be extracted, see Refs. \cite{Beane:2005rj, Beane:2006mx, Beane:2006gf, Beane:2007es, Detmold:2008fn, Beane:2009py, Thomas:2011rh, Beane:2011sc, Basak:2005ir, Peardon:2009gh, Dudek:2010wm, Edwards:2011jj, Dudek:2012ag, Dudek:2012gj, Yamazaki:2009ua, Yamazaki:2012hi}.

\begin{figure}[b!]
\begin{centering}
\includegraphics[scale=0.445]{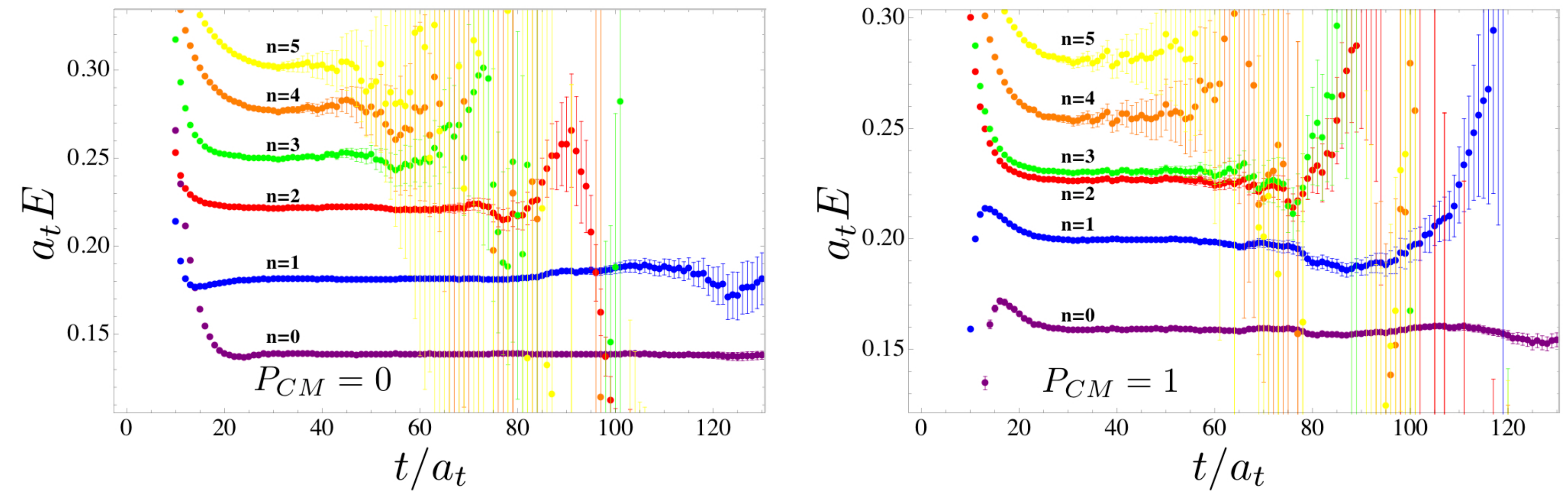}
\par\end{centering}
\caption{{\small The EMPs of the $\pi \pi$ system in the $I=2$ channel produced by the NPLQCD collaboration \cite{Beane:2011sc}. The plots correspond to two different values of the total CM momentum $|\bm{P}| \equiv P_{CM}=0,1$. Energy and time are made dimensionless using the temporal extent of the (anisotropic) lattice used in this calculation, $a_t \approx 0.035~{\rm fm}$. Different colors represent different energy levels labeled by index $n$. Figure is reproduced with the permission of the NPLQCD collaboration.}}
\label{fig:EMP-pipi}
\end{figure}
In Sec. \ref{subsec:ChiPT}, as an example of the interplay between LQCD calculations and the low-energy effective field theories, we presented the result of a LQCD determination of the scattering parameters of the $\pi\pi$ system in the $I=2$ channel by the NPLQCD collaboration \cite{Beane:2011sc}. Here we show the immediate output of this calculation which are the energy eigenvalues in Fig. \ref{fig:EMP-pipi} through EMPs. The plateau region can be clearly identified from the plots before the noise dominates the signal at later times. The calculations of the correlation functions have been done with various different total momenta $\bm{P}$ to increase the number of energy levels extracted. We will come back to this example in Sec. \ref{Two-body} and discuss a non-trivial step that led to the result presented earlier in this chapter.

\emph{5) Interpret the results}: The energy levels (masses) are the output of the calculations we discussed so far. The final step is to try to make sense of these extracted quantities in terms of physical quantities, i.e. those that correspond to the continuum infinite-volume limit of the calculations (and the physical light-quark masses when the calculations have not been performed with physical values of quark masses). Usually, available computational resources allow for multiple calculations with few several lattice spacings, volume sizes and quark masses and therefore extrapolations (interpolations) to the physical point are plausible. The rest of this thesis deals with situations where such extrapolation (interpolations) are not practical or when the determination of the physical quantities depends on calculations away from the physical scenarios, e.g. in determination of scattering parameters. We will not discuss the pion-mass interpolations further in the following, and will focus on the continuum and infinite-volume limits. Since the discussion of the finite-volume (FV) effects is rather extensive in the following chapters, we take the opportunity to introduce and motivate the FV formalism for LQCD in more details in the next section of this chapter.

\section{Infinite-volume Observables from a Finite-volume Formalism
\label{IV-intro}}

\subsection{Single-particle sector
\label{Sec:Single-FV}}
Given the finite extent of the volumes used in LQCD calculations, one should expect the masses extracted form the large-time behavior of lattice two-point correlation functions are not equal to their infinite-volume values. Qualitatively, this can be understood by noting that due to polarization effects, the exchanged particles can propagate to the boundaries of the volume and cause corrections to the mass that differ from those of the scenario when the boundary is in infinity. For the case of QED interactions, it is the photon that gives rise to polarization effects, and due to its zero mass, the FV corrections must scale as (inverse) powers of volume. The situation with the QCD interactions is different, as due to confinement, these are not the massless gluons that go around the volume, and they will not give rise to any significant volume effects. Instead the volume effects are dominantly due to the presence of the pGBs of the spontaneous breaking of the chiral symmetry (pions for the case of $SU(2)$ symmetry). Roughly speaking, by constraining a hadron to a finite volume, the pion could surrounding it is squeezed and the hadron mass is shifted. Here we discuss these corrections to the mass of hadrons through the example of nucleons' mass, to which we come back later in chapter \ref{chap:TBC}, where we apply different boundary conditions than periodic to explore its consequences. The discussion of the volume effects to the masses due to QED interactions will be delayed until chapter \ref{chap:EM}.

As rigorously proved by Martin L\"uscher in 1985 \cite{Luscher:1985dn} for a massive scalar field theory, the volume corrections to the mass of particles have a universal form, and fall off exponentially with volume with a rate that is set by the mass of the lightest particle that is exchanged in the theory. Since volume corrections are due to the IR manipulation of the system, these corrections, as described by L\"uscher, are of kinematic nature and the details of the interactions are not needed in obtaining these results -- a situation that continues to be the case for the two-body problem, see Sec. \ref{Two-body}. So we will consider the nucleon in the HB$\chi$PT and calculate the corrections to its mass due to enclosing it in a finite cubic volume with the PBCs. 

\begin{figure}[t!]
\begin{center}
\subfigure[]{  
\includegraphics[scale=0.460]{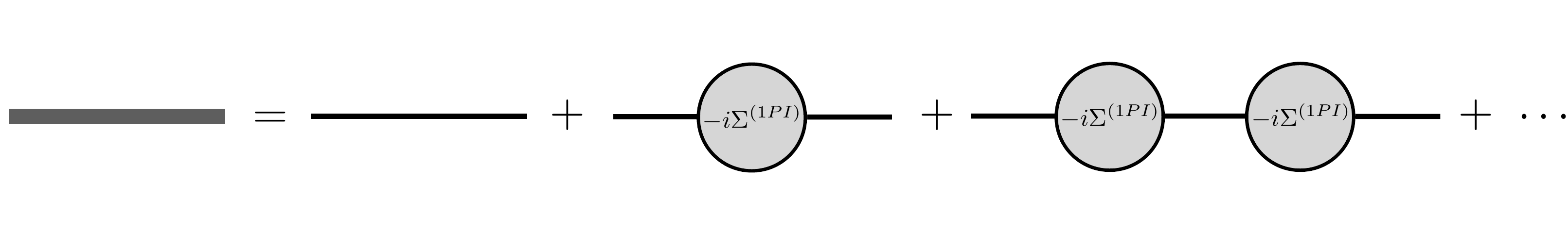}}
\subfigure[]{  
\includegraphics[scale=0.455]{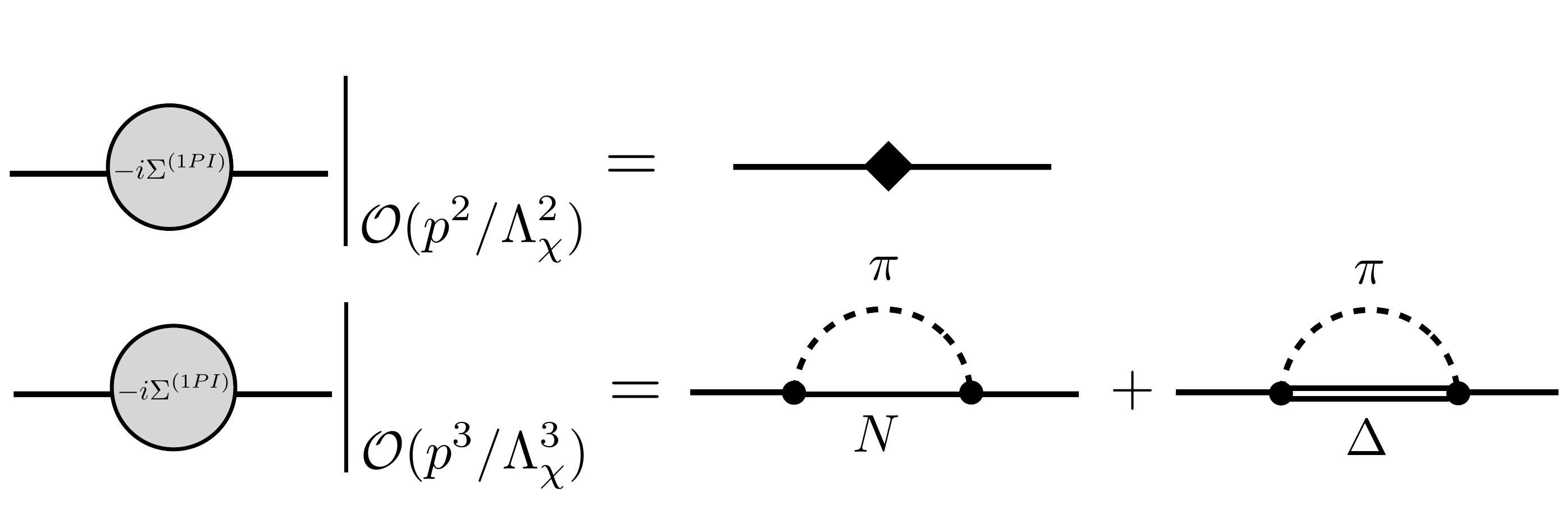}}
\caption
{{\small a) The 1PI self-energy diagrams contributing to the fully dressed nucleon propagator to all orders. The thick solid line denotes the full nucleon propagator. b) The leading contributions (the upper panel) to the 1PI self-energy diagram in HB$\chi$PT comes from an insertion of the quark mass matrix (the diamond) according to Eq. (\ref{L-BpGB-Mass}). The NLO contributions (the lower panel) arise from the pion loops where the possibility of the production of a delta resonance in the loop is taken into account. The black dots denote axial couplings. The solid line, solid-double line and dashed line denote bare nucleon, $\Delta$ resonance and pion propagators, respectively.}
}
\label{fig:Self-energy}
\end{center}
\end{figure}

In the heavy-baryon formalism (see Sec. \ref{subsec:ChiPT}), the mass of the nucleon with momentum $P_{\mu}=M_{N}^{(0)}v_{\mu}+l_{\mu}$ is obtained from the pole of the following fully dressed propagator 
\begin{align}
&\mathcal{D}_{N_l}=\frac{i}{P \cdot v-M_N^{(0)}+i \epsilon}\
\left[ 1-i \Sigma^{(1PI)} \frac{i}{P \cdot v-M_N^{(0)}+i \epsilon}+\left(-i \Sigma^{(1PI)} \frac{i}{P \cdot v-M_N^{(0)}}\right)^2+ \dots \right]
\nonumber\\
& ~~~~ = \frac{i}{P \cdot v-M_N^{(0)}-\Sigma^{(1PI)}+i \epsilon} 
~\equiv~ \frac{i Z_N}{P \cdot v-M_N+i \epsilon},
\label{S-N}
\end{align}
as depicted in Fig. \ref{fig:Self-energy}(a). $M_{N}^{(0)}$ is the nucleon bare mass and $\Sigma^{(1PI)}$ denotes the one-particle irreducible self energy of the nucleon. $\Sigma^{(1PI)}$ can be seen to depend on two scalar variables $v \cdot l$ and $l^2$, so by rewriting $l_{\mu}$ as $l_{\mu}=(M_N-M_N^{(0)})v_{\mu}+(P_{\mu}-M_N v_{\mu})$, it is easy to see that by requiring the on-shell condition $P.v=M_N$, the nucleon mass can be identified as
\begin{eqnarray}
M_N=M_N^{(0)}+\Sigma^{(1PI)}|_{v \cdot l =M_N-M_N^{(0)};~ l^2=(M_N-M_N^{(0)})^2}.
\label{MN-Sigma}
\end{eqnarray}

At LO in HB$\chi$PT, $\mathcal{O}({\frac{p^2}{\Lambda_{\chi}^2}})$, there is one contribution to the self-energy diagram, as shown in the upper panel of Fig. \ref{fig:Self-energy}(b).  It comes from an insertion of the light-quark mass matrix, arising from Lagrangian in Eq. (\ref{L-BpGB-Mass}). This contribution reads
\begin{eqnarray}
\Sigma^{(1PI)}_{LO}=-4 c_1m_{\pi}^2,
\label{LO-Sigma}
\end{eqnarray}
with $c_1=-0.93 \pm 0.10 ~{\rm GeV}^{-1}$ \cite{Bernard:1996gq}. At NLO in the chiral expansion, $\mathcal{O}({\frac{p^3}{\Lambda_{\chi}^3}})$, there are contributions from chiral loops as shown in the lower panel of Fig. \ref{fig:Self-energy}(b). Due to the small mass difference between the nucleon and the $\Delta$ resonance, $\Delta \approx 292~{\rm MeV}$, the contribution from this resonance to the self-energy of the nucleon must be taken into account at this order. We did not discusse the coupling of the baryon decuplets to the pGBs and to the baryon octets in Sec. \ref{subsec:ChiPT}, but it is straightforward to show that the only required vertex for this calculation comes from the following chirally invariant Lagrangian
\begin{eqnarray}
\mathcal{L}_{\Delta N}=g_{\Delta N} \overline{\Delta}^{abc,\nu} \mathcal{A}_{a,\nu}^d N_{b} \epsilon_{cd},
\label{L-Ndelta}
\end{eqnarray}
where the axial vector current $\mathcal{A}_{\nu}$ is defined in Eq. (\ref{Axial}). Then from this Lagrangian and that in Eq. (\ref{L-piN}) for the axial coupling of nucleons, it is easy to see that for the loop corrections, we have
\begin{eqnarray}
\Sigma^{(1PI)}_{NLO}=-i\frac{9 g_A^2}{2 f_{\pi}^2} \mathcal{I}(\infty,0)
-i\frac{4 g_{\Delta N}^2}{f_{\pi}^2} \mathcal{I}(\infty,\Delta),
\label{Sigma-NLO}
\end{eqnarray}
where
\begin{eqnarray}
\mathcal{I}(\infty,\Delta)=-\frac{1}{3} \int \frac{d^4 k}{(2 \pi)^4}
\frac{\bm{k}^2}{(k^0-\Delta+i \epsilon)(k^{02}-\bm{k}^2-m_{\pi}^2+i \epsilon)}.
\label{Ndelta-loop}
\end{eqnarray}
The integral is clearly UV divergent and must be renormalized. However, since we are interested in the FV corrections to the nucleon mass, we do not need to carry out this integration any further. The only observation to be made before moving on to the FV scenario is to note that the (renormalized) mass at this order is proportional to $m_{\pi}^3 \sim m_{q}^{3/2}$ (for $\Delta=0$ term) and is therefore non-analytic in the light-quark masses. The contribution from the $\Delta$-resonance introduces further nontrivial non-analytic corrections to the mass of the nucleon, see Refs. \cite{Jenkins:1990jv, Hemmert:1997ye, Bernard:1993nj, Beane:2002vq} for the discussion of baryon masses from (HB)$\chi$PT.

In a finite volume, the momentum modes are all discretized due to the PBCs, $\mathbf{k}=\frac{2\pi}{L}\mathbf{n},~\bm{n} \in \mathbb{Z}^3$. As a result the only difference between the FV and infinite-volume calculation arises from the loops where the integrals over momenta are replaced with sums \cite{AliKhan:2003cu, Beane:2004tw, Beane:2011pc},
\begin{eqnarray}
\mathcal{I}(L,\Delta)=-\frac{1}{3} \frac{1}{L^3} \sum_{\bm{k}} \int \frac{d k^0}{(2 \pi)^4}
\frac{\bm{k}^2}{(k^0-\Delta+i \epsilon)(k^{02}-\bm{k}^2-m_{\pi}^2+i \epsilon)}.
\label{Ndelta-loop-sum}
\end{eqnarray}

Note that we keep the temporal extent of the volume infinite for the discussion of FV effects. Since in practice LQCD calculations have a finite extent in the (imaginary) time direction, there will be contaminations to the extracted energies from the backward propagating states. Such \emph{thermal} effects must be dealt with separately but their effects can be shown numerically to be smaller that the (spatial) volume effects. Using the Poisson re-summation formula,
\begin{eqnarray}
\frac{1}{L^3} \sum_{\bm{k}} f(\mathbf{k})=\int \frac{d^3k}{(2 \pi)^3} f(\mathbf{k})+\sum_{\bm{m} \neq \bm{0}}\int \frac{d^3k}{(2 \pi)^3} f(\mathbf{k}) e^{i \bm{k}.\bm{m}L},
\label{Poisson}
\end{eqnarray}
where $\mathbf{m}$ is another triplet of integers, one can isolate the infinite-volume contribution to $\mathcal{I}(L,\Delta)$ in Eq. (\ref{Ndelta-loop}), which will be canceled out when taking the difference of the infinite-volume and FV masses. With the help of a useful identity, 
\begin{eqnarray}
\frac{1}{(\bm{k}^2+\mathcal{M}^2)^r}=\frac{1}{\Gamma(r)} \int_{0}^{\infty} ds s^{r-1} e^{-s(\bm{k}^2+\mathcal{M}^2)},
\label{Identity}
\end{eqnarray}
it then takes a few lines of algebra to show that \cite{Beane:2004tw},
\begin{eqnarray}
\delta_{L} M_{N} \equiv M_N(L)-M_N(\infty)= \frac{3 g_A^2}{8 \pi^2 f_{\pi}^2} \mathcal{K}(0)
+\frac{g_{\Delta N}^2}{3 \pi^2 f_{\pi}^2} \mathcal{K}(\Delta),
\label{deltaM}
\end{eqnarray}
where
\begin{eqnarray}
\mathcal{K}(0)= \frac{\pi}{2}m_{\pi}^2 \sum_{\mathbf{n}\neq \mathbf{0}} \frac{e^{-|\mathbf{n}| m_{\pi} L}}{|\mathbf{n}| L}
,
\end{eqnarray}
and
\begin{eqnarray}
\mathcal{K}(\Delta)
\ =\ 
\int_{0}^{\infty} d \lambda ~ \beta_{\Delta} ~ \sum_{\mathbf{n}\neq \mathbf{0}} 
\left[ 
\beta_{\Delta} K_0(\beta_{\Delta}|\mathbf{n}| L)
\ -\ 
\frac{1}{|\mathbf{n}| L} K_1(\beta_{\Delta}|\mathbf{n}| L)
\right]
.
\end{eqnarray}
$K_n(z)$ is the modified Bessel function of the second kind, and $\beta_{\Delta} = \lambda^2 + 2 \lambda \Delta + m_{\pi}^2$.\footnote{
Note that we have chosen to define the  $\mathcal{K}(\Delta)$ function with a negative sign compared to
Ref.~\cite{Beane:2004tw}.
} When expanded in the limit of large $L$, Eq.~(\ref{deltaM}) scales as $e^{-m_{\pi}L}/L$ at LO. Explicitly one obtains \cite{Beane:2004tw}
\begin{eqnarray}
\delta_{L} M_{N}^{asym} = \left[\frac{9 g_A^2 m_{\pi}^2}{8 \pi f_{\pi}^2}+
\frac{4g_{\Delta N}^2 m_{\pi}^{5/2}}{(2 \pi)^{3/2} f_{\pi}^2 \Delta}\frac{1}{L^{1/2}} \right]
\frac{1}{L}e^{-m_{\pi}L},
\label{deltaM}
\end{eqnarray}
As we already discussed, the exponential corrections of these types are general features of interacting theories with finite-range interactions. The reader can consult  Refs. \cite{Gasser:1986vb, Colangelo:2003hf, Colangelo:2005cg, Colangelo:2010ba, Beane:2011pc, Briceno:2013rwa} for the FV corrections to the masses of mesons and baryons.

\subsection{Two-particle sector
\label{Two-body}}
As mentioned, LQCD produces n-point correlation functions of Euclidean spacetime. Euclidean correlation functions with the \emph{reflection positivity} property can be Wick rotated back to Minkowski spacetime, as proved by Osterwalder and Schrader~\cite{Osterwalder:1973dx}. Therefore, if one was able to fully reconstruct the continuum correlation functions from the Euclidean lattice counterparts, such analytic continuation would not be formally problematic. However, lattice correlation functions are evaluated at a discrete set of spacetime points and are not exact. Maiani and Testa \cite{Maiani:1990ca} noted the Euclidian nature of LQCD calculations prohibits the determination of few-body scattering quantities from lattice correlation functions in the infinite-volume limit (unless at the kinematic threshold). However, LQCD correlation functions are evaluated in a \emph{finite volume}. In fact, it turns out that the scattering amplitudes of the infinite volume can be constructed from the spectrum of the interacting particles in a finite volume. 

The first realization of this statement goes back to 1957 when Huang and Yang \cite{Huang:1957im} considered a quantum mechanical two-body system in a finite volume interacting via a hard spherical potential and found out that the energy shift due to the interactions in a finite volume can be related to the two-body scattering length, $a$. The scattering length is defined  as $-1/a=\lim_{k^*\rightarrow0} k^* \cot \delta$, where $\delta$ is the scattering phase shift of the two-particle system and $k^*$ is the momentum of each particle in the CM frame. Thirty years later, Martin L\"uscher, motivated by LQCD applications, extended the Huang and Yang's relation to quantum field theory, and derived a non-perturbative relation between the two-body scattering amplitudes and the FV energy eigenvalues for scalar bosons with zero total momentum~\cite{Luscher:1986pf, Luscher:1990ux}. Various extensions of the L\"uscher relation that followed in subsequent years include generalization to boosted systems \cite{Rummukainen:1995vs, Kim:2005gf, Christ:2005gi}, asymmetric lattices \cite{Li:2003jn, Feng:2004ua, Detmold:2004qn}, systems with unequal masses \cite{Bour:2011ef,Davoudi:2011md, Fu:2011xz, Leskovec:2012gb}, two-body coupled channels \cite{He:2005ey, Liu:2005kr, Lage:2009zv, Bernard:2010fp, Hansen:2012tf, Hansen:2012bj, Briceno:2012yi, Li:2012bi, Briceno:2014oea, Li:2014wga}, nucleons with only S-wave interactions \cite{Beane:2003da}, systems with total spin $1/2$ (pion-nucleon scattering) \cite{Gockeler:2012yj}, with total spin $1$ (nucleon-nucleon scattering) \cite{Ishizuka:2009bx, Briceno:2013lba} with arbitrary spin \cite{Briceno:2014oea}, and calculations with twisted boundary conditions \cite{Bedaque:2004kc, Agadjanov:2013wqa, Briceno:2013hya}.\footnote{The most general form of the two-particle quantization condition incorporating all these extensions has been recently written down in Ref. \cite{Briceno:2014oea}.} Here we present a derivation of a form of the L\"uscher formula applicable to the case of (multi) coupled-channel scattering of scalar particles in the moving frame with arbitrary partial waves. We follow closely Ref. \cite{Briceno:2012yi}, however most of the details associated with generalizing to moving frames have been developed by Kim, $et\; al.$ \cite{Kim:2005gf}, which will be briefly reviewed here for completeness.\footnote{See Refs. \cite{Rummukainen:1995vs, Christ:2005gi} for alternative derivations of the moving frame generalization of the L\"uscher formula.} Generalization to the two-nucleon systems with both periodic and twisted boundary conditions, and their implications for the spectrum of the deuteron and the extraction of its properties will constitute the bulk of chapters \ref{chap:NN}, \ref{chap:deuteron}, \ref{chap:TBC} of this thesis.

Consider a system of multiple two-particle channels (spin 0) coupled via interactions of arbitrary strengths. Since we are interested in the IR modifications to the energy levels of the interacting system, the details of the interactions in the UV turned out to be immaterial for the discussions presented here. In fact, in (an effective) a field theory approach, one does not need to explicitly write down a Lagrangian for the system in obtaining the FV energy eigenvalues.\footnote{In chapter \ref{chap:NN}, we will present another derivation of the L\"uscher formula that starts from an effective Lagrangian using an auxiliary field method. We will see that by matching the parameters of the EFT to scattering amplitudes, the FV energy eigenvalues can be written in terms of these scattering parameters and the dependence on the LECs of the EFT are then eliminated.} The following observations enable us to take such a general approach:

\begin{enumerate}

\item Below the three-particle inelastic thresholds, a general interacting kernel for $2 \to 2$ processes can be replaced, under the following condition, by its infinite-volume counterpart up to exponential corrections in volume. The scale of the exponential suppressions is set by the mass of the particle that is produced, once inelastic thresholds are reached, and is given by $e^{-mL}$. To satisfy this property, all the s-channel contributions -- for which only two particles propagate inside the loops -- are separated from the rest of the contributions. We call such kernel the s-channel two-particle irriducible Bethe-Salpeter kernel, $\mathcal{K}$, see Fig. \ref{fig:twoparFV}(b).

\begin{figure}[t!] 
\begin{center}
\subfigure[]{
\includegraphics[scale=0.375]{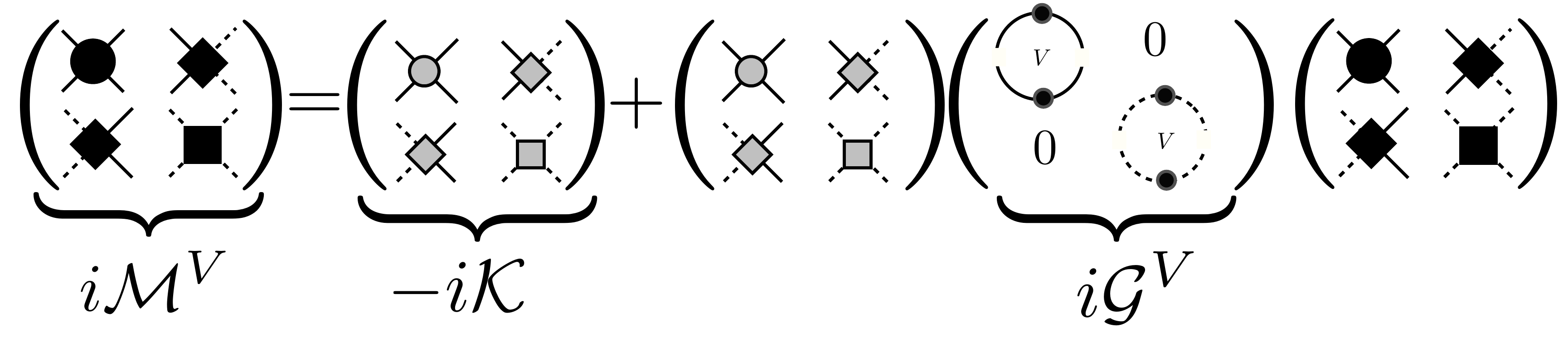}}
\subfigure[]{
\includegraphics[scale=0.23]{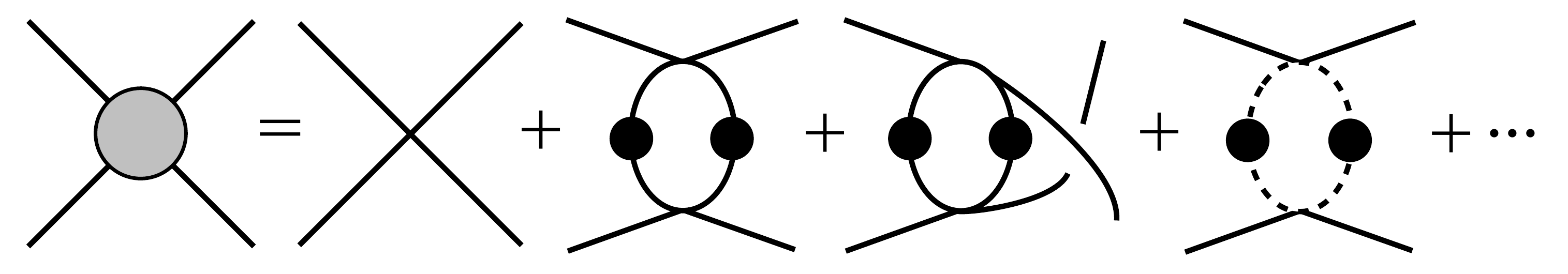}}
\subfigure[]{
\includegraphics[scale=0.23]{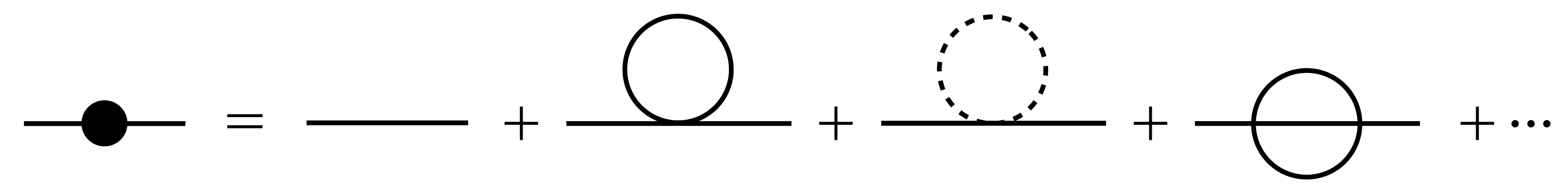}}
\caption{\small{ a) The fully-dressed FV two-particle propagator, $\mathcal{M}^V$ can be written in a self-consistent way in terms of the Bethe-Salpeter Kernel, $\mathcal{K}$ and the FV s-channel bubble $\mathcal{G}^V$. There are only two channels that are kinematically allowed, therefore the amplitude, the kernel and the FV two-particle propagator are $2 \times 2$ matrices. b) Shown is the $\mathcal{K}_{I,I}$-component of the kernel, which sums all s-channel two-particle irreducible diagrams for channel $I$. c) The fully dressed one-particle propagator is the sum of all one-particle irreducible diagrams and is denoted by a black dot on the propagator lines. The single particle propagator in channel I (II) is shown by the solid (dashed) lines. Note that we have chosen a $\lambda \phi^4$ theory to display the explicit contributions to  kernels and self energies, however the discussions of the FV formalism for any two-body coupled-channel systems as presented in the text is general. For this theory, the production of two particles is the first multi-particle inelastic threshold as there is no $2 \to 3$ couplings in the theory, e.g. interactions of the pGBs of the chiral symmetry breaking.}}
\label{fig:twoparFV}
\end{center}
\end{figure}

\item 
The fully-dressed propagators in a finite volume are exponentially close to their infinite-volume counterparts. This is again due to the fact that below the three-particle production thresholds, any loop correction to the single-particle propagator involves the production of an off-mass-shell particle, see Fig. \ref{fig:twoparFV}(c). We ignore such exponential corrections and replace the propagators with their infinite-volume counterparts in constructing the FV correlation functions (alternatively amplitudes).

\item
For any s-channel loop, the integral over the three-momenta is replaced by a sum over discrete momenta,
\begin{eqnarray}
\label{sum-int}
\int \frac{d^4 q}{(2\pi)^4} f(q) \to   \frac{1}{L^3} \sum_{\mathbf{q}=\frac{2\pi}{L} \mathbf{n}} \int \frac{dq^0}{2\pi} f(q),
\end{eqnarray}
where $\mathbf{n} \in \mathbb{Z}^3$. If the summand is not singular for any real value of $|\mathbf{q}|$ and falls off fast as $|\mathbf{q}| \to \infty$, the sum can be replaced with integral up to exponentially small corrections that we neglect in this formalism. For any s-channel diagram, see Eq. (\ref{loop3}), this however does not hold due to singularities of the two-particle propagator when the loop momenta coincides with the non-interacting momenta of each particle in the CM frame. This defines the on-shell condition due to which the particles can propagate far enough to encounter the boundary of the volume. This causes large power-law corrections to the FV spectrum as will be derived shortly.

\end{enumerate}

To proceed let us first define the kinematics of the problem of coupled-channel systems in a moving frame. If the total energy and momentum of the system in the laboratory frame (lattice frame) are $E$ and $\mathbf{P}$ respectively, then the total CM energy of the system is $E^*=\sqrt{E^2-\mathbf{P}^2}$, and can be written as  $E^*=E/ \gamma$ by introducing the relativistic $\gamma$ factor. For the $i^{th}$ channel, when the two particles each are having masses $m_{i,1}$ and $m_{i,2}$, the CM relative momentum of the particles, $k^*_i$, can be derived from
\begin{eqnarray}
\label{momentum}
k^{*2}_i=\frac{1}{4}\left(E^{*2}-2(m_{i,1}^2+m_{i,2}^2)+\frac{
(m_{i,1}^2-m_{i,2}^2)^2}{E^{*2}}\right),
\end{eqnarray}
which simplifies to $\frac{E^{*2}}{4}-m_{i}^2$ when $m_{i,1}=m_{i,2}=m_{i}$. 

For N coupled channels, the scattering amplitude for the $l^{th}$ partial wave can be written as a N-dimensional matrix. In order to have a fully relativistic result that holds for all possible energies below the three-particle production threshold, the scattering amplitude must include all possible diagrams, i.e. contributions from s-, t- and u-channels as well as self-energy corrections. Fig. \ref{fig:twoparFV}(a) depicts the FV analogue of the scattering amplitude, $\mathcal{M}^V$, for the special case of $N=2$ channels.\footnote{One should note that using the notion of FV scattering amplitude is merely for the mathematical convenience. As there is no asymptotic state by which one could define the scattering amplitude in a finite volume, one should in principle look at the pole locations of the two-body correlation functions. However, one can easily show that both correlation function and the so-called FV scattering amplitude have the same pole structure, so we use the latter for the sake of dealing with a simpler representation.} This amplitude is written in a self-consistent way in terms of the Bethe-Salpeter kernel, $\mathcal{K}$, which is the sum of all s-channel two-particle irreducible diagrams as defined above. We emphasize that for energies below the three-particle threshold, the intermediate particles in the kernel and the self-energy diagrams, Fig. \ref{fig:twoparFV}(b),(c), cannot go on-shell, and therefore these are exponentially close to their infinite-volume counterparts. As discussed, it is only in the s-channel diagrams that all intermediate particles can be simultaneously put on shell. 

By upgrading the kernel and the two-particle propagators to matrices in the space of open channels, Fig. \ref{fig:twoparFV}(a), it is straightforward to obtain a non-perturbative QC for the energy levels of the system. It is important to note that the channels only mix by off-diagonal terms in the kernel, which implies that in the absence of interactions a two-particle state $i$ continues to propagate as a two-particle state $i$. Now in the presence of momentum-dependent vertices a typical s-channel loop for channel $i$ can be written as 
\begin{eqnarray}
\label{loop3}
\left[iG^{V}(\mathbf{p}_a,\mathbf{p}_b)\right]_{ab}&\equiv&\frac{n_i}{L^3}\sum_{\mathbf{q}}\int\frac{dq^0}{2\pi}\frac{[\mathcal{K}(\mathbf{p}_a,\mathbf{q})]_{ai}~[\mathcal{K}(\mathbf{q},\mathbf{p}_b)]_{ib}}{[(q-P)^2-m_{i,1}^2+i\epsilon][q^2-m_{i,2}^2+i\epsilon]},
\end{eqnarray}
where the subscripts $a, i, b$ denote the initial, intermediate and final states, respectively, and $n_i$ is ${1}/{2}$ if the particles in the $i^{th}$ loop are identical and 1 otherwise. The sum over all intermediate states, and therefore index $i$ is assumed.
Since the FV corrections arise from the pole structure of the intermediate two-particle propagator, one would expect that the difference between this loop and its infinite-volume counterpart should depend on the on-shell momentum. The on-shell condition fixes the magnitude of the momentum running through the kernels but not its direction. Therefore it is convenient to decompose the product of the kernels into spherical harmonics. These depend not only on the directionality of the intermediate momentum but also on those of the incoming and outgoing momenta, $\mathbf{p}_a$ and $\mathbf{p}_b$. Further, one can represent the N two-body propagators as a diagonal matrix $\mathcal{G}={\rm diag}(\mathcal{G}_{1},\mathcal{G}_{2},\cdots, \mathcal{G}_{N})$ as depicted in Fig. \ref{fig:twoparFV}(a). These are infinite-dimensional matrices with with matrix elements \cite{Kim:2005gf}
\begin{eqnarray}
(\delta G^{V}_i)_{l,m;l',m'}\equiv
(G^{V}_i-G^{\infty}_i)_{l,m;l',m'}=
-i\left(\mathcal K \delta \mathcal G^{V}_i \mathcal K\right)_{l,m;l',m'},
\end{eqnarray}
where 
\begin{eqnarray}
\label{def0}
(\delta \mathcal{G}^{V}_{i})_{l_1,m_1;l_2,m_2}&=&i	\frac{k^*_in_i}{8\pi E^*}\left(\delta_{l_1,l_2}\delta_{m_1,m_2}+i\frac{4\pi}{k_i^*}\sum_{l,m}\frac{\sqrt{4\pi}}{k_i^{*l}}c^{\textbf{P}}_{lm}(k_i^{*2})\int d\Omega Y^*_{l_1m_1}Y^*_{lm}Y_{l_2m_2}	\right),\nonumber\\
\end{eqnarray}
and the function $c^{\textbf{P}}_{lm}$ is defined as\footnote{Note that our definition of the $c_{lm}^{\textbf{P}}$ function differs that of Ref. \cite{Kim:2005gf} by an overall sign.}
\begin{eqnarray}
\label{clm}
c^{\textbf{P}}_{lm}(x)=\frac{1}{\gamma}\left[\frac{1}{ L^3}\sum_{\textbf{q}}-\mathcal{P}\int\frac{d^3\mathbf{q}}{(2\pi)^3}\right]\frac{\sqrt{4\pi}Y_{lm}(\hat{q}^*)~q^{*l}}{{q}^{*2}-x} \ .
\end{eqnarray}
$\mathcal{P}$ in this relation denotes the principal value of the integral, and $\mathbf{q}^*={\gamma}^{-1}(\mathbf q_{||}-\alpha \mathbf P)+\mathbf q_{\perp}$, where $\mathbf q_{||}$ ($\mathbf q_{\perp}$) denotes the component of the momentum vector $\mathbf{q}$ that is parallel (perpendicular) to the boost vector $\mathbf{P}$ and $\alpha=\frac{1}{2}\left[1+\frac{m_1^2-m_2^2}{E^{*2}}\right]$~\cite{Davoudi:2011md, Fu:2011xz, Leskovec:2012gb}.\footnote{The kinematic function $c^{\textbf{P}}_{lm}(k_i^{*2})$ can also be written in terms of the three-dimensional Zeta function, $\mathcal{Z}^d_{lm}$,
 \begin{eqnarray}
 \nonumber
 	c^{\textbf{P}}_{lm}(k^{*2})=\frac{\sqrt{4\pi}}{\gamma L^3}\left(\frac{2\pi}{L}\right)^{l-2}\mathcal{Z}^\mathbf{d}_{lm}[1;(k^*L/2\pi)^2],\hspace{1cm} 
\mathcal{Z}^\mathbf{d}_{lm}[s;x^2]=\sum_{\mathbf r \in P_d}\frac{Y_{l,m}(\mathbf{r})}{(r^2-x^2)^s},
\end{eqnarray}
where the sum is performed over $P_d=\left\{\mathbf{r}\in \mathbb{R}^3\hspace{.1cm} | \hspace{.1cm}\mathbf{r}={\gamma}^{-1}(\mathbf m_{||}-\alpha \mathbf d)+\mathbf m_{\perp} \text{,}~ \mathbf{m} \in \mathbb{Z}^3\right\}$, $\mathbf d$ is the normalized boost vector $\mathbf d=\mathbf{P}L/2\pi$,  and $\alpha$ is defined above.}  This reduces to the NR value of $\alpha=\frac{m_1}{m_1+m_2}$ as is presented in Ref.~\cite{Bour:2011ef}. Note that this result is equivalent to the result obtained in Refs. \cite{Rummukainen:1995vs, Kim:2005gf, Christ:2005gi} for the boosted systems of particles with identical masses and with one channel. The only non-trivial piece in this relation is the momentum vectors $\mathbf{q}^*$ to be summed over in the energy quantization condition (QC). This is determined mainly from the on-shell kinematics of the two-particle states which depends on the boost vector, the masses of particles and the boundary conditions. Since we aim to present a general proof for the form of these momentum vectors with arbitrary twisted boundary conditions, we will delay the derivation until chapter \ref{chap:TBC}. The result in Eq. (\ref{clm}) can be recovered from that of presented in chapter \ref{chap:TBC} upon setting all the twist angles to zero.

The kernel, which is now not only a matrix in the channel space but also in the angular momentum space, is assured to reproduce the infinite-volume scattering amplitude matrix ($\mathcal{M}$) by solving the following matrix equation
\begin{eqnarray}
i\mathcal{M}&=&-i\mathcal{K}
-i\mathcal{K}\mathcal{G^{\infty}} \mathcal{K}
-i\mathcal{K}\mathcal{G^{\infty}} \mathcal{K}\mathcal{G^{\infty}}\mathcal{K}+\cdots=-i\mathcal{K}\frac{1}{1-\mathcal{G^{\infty}}\mathcal{K}}.
\label{MK-relation}
\end{eqnarray}
giving rise to
\begin{eqnarray}
\mathcal{K}=-\mathcal{M}\frac{1}{1-\mathcal{G^{\infty}}\mathcal{M}}.
\label{eq:kernel}
\end{eqnarray}
With this definition of the kernel, one can proceed to evaluate poles of the  N-channels FV scattering amplitude matrix by replacing the infinite-volume loops $\mathcal{G}^\infty$ with their FV $\mathcal{G}^V$ counterparts, 
\begin{eqnarray}
-i\mathcal{M}^{V}&=&-i\mathcal{K}-i\mathcal{K}\mathcal{G}^{V} \mathcal{K}-i\mathcal{G}^{V}\mathcal{K}\mathcal{G}^{V}\mathcal{K}+\cdots=-i\mathcal{K}\frac{1}{1-\mathcal{G}^{V}\mathcal{K}}
\nonumber\\
& = & -i\frac{1}{1-\mathcal{M}\mathcal{G}^{\infty}}\mathcal{M}\frac{1}{1+\delta \mathcal{G}^{V}\mathcal{M}}({1-\mathcal{M}\mathcal{G}^{\infty}}). 
\end{eqnarray} 
Finally arriving at the QC
\begin{eqnarray}
\label{det0}
\mathcal{R}e\left\{\det(\mathcal{M}^{-1}+\delta \mathcal{G}^{V})\right\}=\mathcal{R}e\left\{{\rm{det}}_{\rm{oc}}\left[\rm{det}_{\rm{pw}}\left[\mathcal{M}^{-1}+\delta \mathcal{G}^{V}\right]\right]\right\}=0,
\end{eqnarray} 
where the determinant $\rm{det}_{\rm{oc}}$ is over the N open channels and the determinant $\rm{det}_{\rm{pw}}$ is over the partial waves, and both $\mathcal{M}$ and $\delta \mathcal{G}^V$ functions are evaluated on the on-shell value of the momenta. This latter property, along with decomposing to partial waves, has enabled us to decouple the chain of loops in the expansion of the scattering amplitude, see Fig. \ref{fig:twoparFV}(a), and obtain an algebraic expansion (a geometric series) of the scattering amplitude matrix. We have taken the real part of the determinant in Eq. (\ref{det0}), but as it will be shown shortly, this determinant condition gives rise to only one single real condition for both single channel and two coupled-channel cases with $l_{max}=0$, so we omit the notion of the real part in the QC from now on. For a general proof of the reality of QC with any number of coupled channels see Refs. \cite{Hansen:2012tf, Hansen:2012bj}.

\subsubsection{Single-channel scattering} For N=1 the QC in Eq. (\ref{det0}) reproduces the L\"uscher formula \cite{Luscher:1986pf, Luscher:1990ux} when generalized to moving frames \cite{Rummukainen:1995vs, Kim:2005gf, Christ:2005gi}. In order to deduce this, let us write the relativistic single-channel scattering amplitude $\mathcal{M}_i$ as
\begin{eqnarray}
\label{def1}
(\mathcal{M}_{i})_{l_1,m_1;l_2,m_2}&=&\delta_{l_1,l_2}\delta_{m_1,m_2}\frac{8\pi E^*}{n_ik^*_i}\frac{e^{2i\delta^{(l)}_i(k^*_i)}-1}{2i},
\end{eqnarray}
where $\delta_i^{(l)}$ is the scattering phase shift in channel $i$ and in partial wave channel $l$. The Kronecker deltas that are introduced to insure the conservation of angular momentum between initial and final states scattering should not be confused with the phase shifts. 

Since the QC in Eq. (\ref{det0}), even for the case of single-channel scattering, is infinite dimensional, one should first perform a truncation in the angular momentum basis. Let us assume that the contributions from higher partial waves to the  scatterings are negligible (which is the case at low energies), so that one can truncate the determinant over the angular momentum at $l_{max}=0$. Then the QC for the S-wave scattering reads,
\begin{eqnarray}
k_i^*\cot {\delta_i^{(0)}}=4\pi c_{00}^{\mathbf{P}}(k^{*2}_i),
\label{Luscherl0}
\end{eqnarray}
It is convenient to introduce a pseudo-phase defined by ${k^*_i}\cot {\phi^\mathbf{P}_i}\equiv -4\pi{ c_{00}^\mathbf{P}}$, 
 to rewrite the QC as
\begin{eqnarray}
\label{quateq}
\cot {\delta_i}=-\cot {\phi^\mathbf{P}_i} \Rightarrow \delta_i+\phi^\mathbf{P}_i=n\pi,
\end{eqnarray}
where $n$ is an integer. In this form, the QC is manifestly real.

\begin{figure}[t!] 
\begin{center}
\includegraphics[scale=0.475]{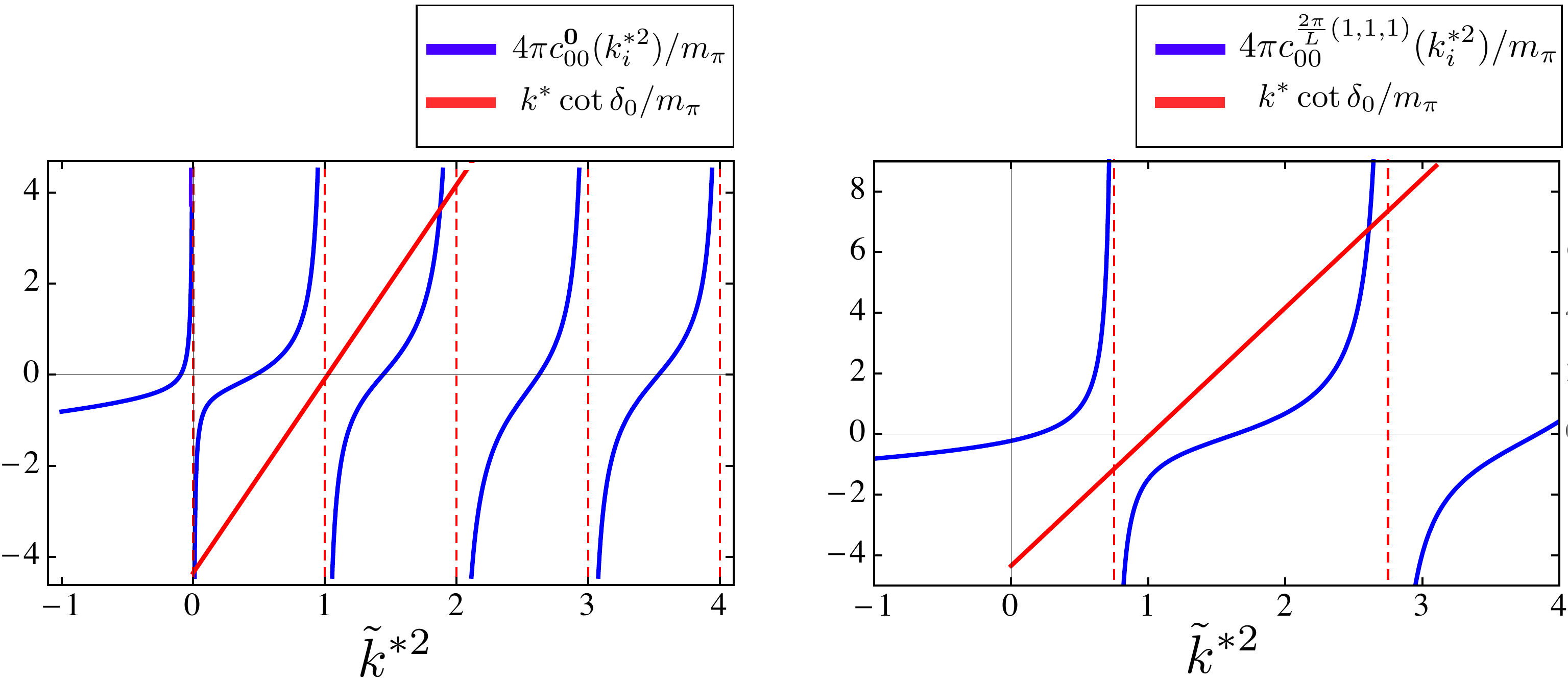}
\caption{\small{The FV function $4\pi c_{00}^{\mathbf{P}}$ (in blue) and the $k^* \cot \delta_0$ (in red) (both normalized by the pion mass) as a function of $\tilde{k}^{*2} \equiv \frac{k^{*2}L^2}{(2\pi)^2}$ for a chosen values of $m_{\pi}=390~{\rm MeV}$ and $L \approx 3.9~{\rm fm}$. The $k^* \cot \delta_0$ function is plotted using an ERE parametrization with the obtained values of $a_{0}^{(2)}$ and $r_{0}^{(2)}$ at this pion mass by the NPLQCD collaboration \cite{Beane:2011sc}.}}
\label{fig:kcot-c00}
\end{center}
\end{figure}

Here we discuss briefly how one implements the L\"uscher formula in LQCD studies of two-hadron systems.\footnote{We will continue discussing the NPLQCD study of $\pi\pi$ scattering in the $I=2$ channel \cite{Beane:2011sc}. For more examples of successful implementations of the L\"uscher formula in studies of two-hadron sector of QCD, including studies of resonances, the reader may consult the following references, \cite{Li:2007ey, Aoki:2007rd, Beane:2010hg, Beane:2011xf, Beane:2011sc, Beane:2011iw, Beane:2012ey,Yamazaki:2012hi, Lang:2012sv, Beane:2013br,  Pelissier:2011ib,  Lang:2011mn, Pelissier:2012pi, Ozaki:2012ce, Buchoff:2012ja, Dudek:2012xn, Dudek:2012gj, Lang:2014tia}.} In Sec. \ref{subsec:ChiPT} we saw that by inputting the S-wave scattering length and effective range of the $I=2$ $\pi\pi$ scattering into the chiral expansion of these parameters, the LECs of the $\chi$PT at NLO can be determined, and predictions at the physical values of light-quark masses are made possible, see Fig. \ref{fig:NPLQCD-pipi}. Furthermore, in Sec. \ref{LQCD}, we saw, through the same example of the $\pi\pi$ system in the $I=2$ channel, how the energy levels of the system can be extracted from the large-time dependence of the lattice correlation functions, see Fig. \ref{fig:EMP-pipi}. Now given the L\"uscher formula in Eq. (\ref{Luscherl0}), it is known how these energy eigenvalues turn into the corresponding phase shifts. Fig. \ref{fig:kcot-c00} demonstrates the relation between the $k^* \cot \delta_0$ function from Eq. (\ref{a-r-def}) and the FV function $c_{00}^{\mathbf{P}}$ evaluated numerically as a function of $\tilde{k}^{*2} \equiv \frac{k^{*2}L^2}{(2\pi)^2}$ for chosen values of $m_{\pi}=390~{\rm MeV}$ and $L \approx 3.9~{\rm fm}$. The point where the two functions intersect corresponds to an energy eigenvalue, which as seen differs for systems with different boosts. 
\begin{figure}[t!] 
\begin{center}
\includegraphics[scale=0.515]{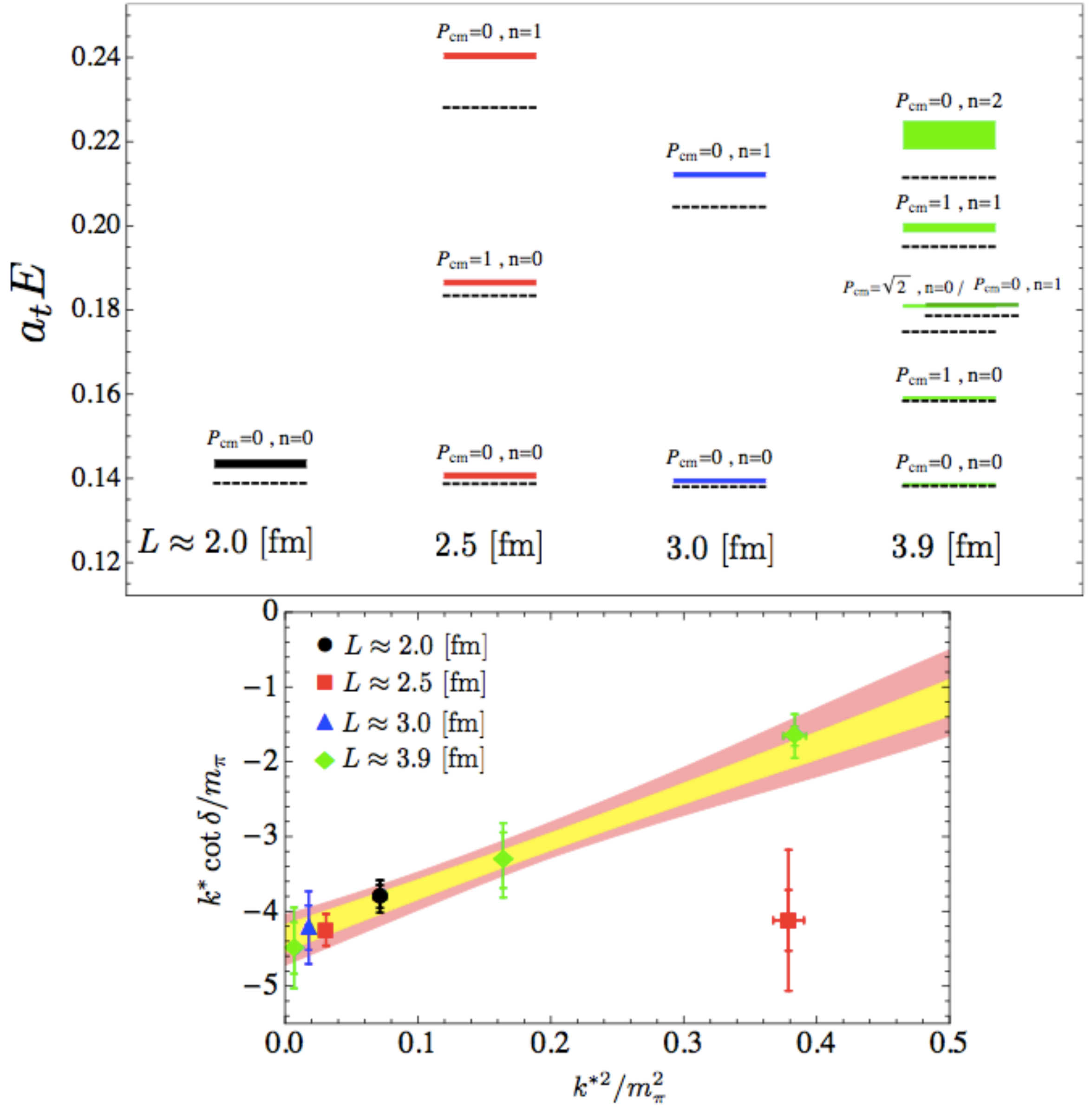}
\caption{\small{The upper panel depicts the NPLQCD extraction \cite{Beane:2011sc} of the lowest energy levels of the two-pion state in the $I=2$ channel, labeled by index $n$, for various total CM momenta $|\mathbf{P}|\equiv P_{cm}$. The energy is given in terms of the dimensionless quantity $a_tE,$ where $a_t \approx 0.035~{\rm fm}$ is the temporal extent of the (anisotropic) lattice used in this calculation. The dashed lines denote the location of the non-interacting energy levels. The lower panel contains the obtained S-wave phase shifts using the L\"uscher QC for several energy levels in the upper panel. The fit to the ERE with two parameters, $a_0^{(2)}$ and $r_0^{(2)}$, is shown where the shaded bands correspond to statistical (inner-yellow) and statistical and systematic added in quadrature (outer-pink). Figure is reproduced with the permission of the NPLQCD collaboration.}}
\label{fig:pipi}
\end{center}
\end{figure}
This clearly demonstrates that boosting the two-hadron system does not only give rise to a trivial shift in the total energy of the system, but also changes the location of the CM energies in a finite volume. This can be understood by noting that by boosting the two-body system in a finite cubic volume, the spatial extents of the volume as is seen in the rest frame of the system is changed, giving rise to a different FV symmetry groups than the unboosted case, see chapter \ref{chap:NN}. This is the reason why performing calculations with different CM is advantageous as it provides further energy inputs to the L\"uscher QC at a single volume, and puts better constraints on the extracted scattering parameters. When such procedure is carried out for the $\pi \pi$ scattering, the extracted energy levels as is shown in the upper panel of Fig. \ref{fig:pipi}, lead to several phase shift points that can then be fit using a two-parameter ERE in Eq. (\ref{a-r-def}) to obtain the values of $a_{0}^{(2)}$ and $r_{0}^{(2)}$. Note that although the L\"uscher formula can be used up to the inelastic threshold which is $k^{*2}=3m_{\pi}^2$ for this system, the range of validity of the extracted scattering length and the effective range is determined by the momentum at which the ERE breaks down, $k^{*2}=m_{\pi}^2$.

Although we have truncated the QC to $l_{max}=0$, as is seen from the definition of the FV matrix $\delta \mathcal{G}^{V}$ in Eq. (\ref{def0}), the reduced symmetry of the FV introduces off-diagonal terms in the FV matrix which couple different partial waves in the QC. For example, if the two equal-mass meson interpolating operator is in the $A_1^+$  irreducible representation (irrep) of the cubic group, the energy eigenstates of the system have overlap with the $l=0,4,6,\ldots$ angular momentum states at zero total momentum (see table \ref{table:irreps} for decomposition of the irreps of the cubic group in terms of the irreps of the rotational group), making the truncation at $l_{max}=0$ a rather reasonable approximation in the low-energy limit. When $\textbf{P}\neq 0$ but $\gamma \approx 1$, as will be discussed in details in chapter \ref{chap:NN}, the symmetry group is reduced, and at low energies the $l=0$ will mix with the $l=2$ partial wave as well as with higher partial waves.\footnote{A comprehensive study of the symmetry groups of the calculations with different boost vectors in the NR limit will be presented in chapter \ref{chap:NN}.} Recently, by analyzing the LQCD calculations of two-pion systems with several CM boosts, the $l=0,2$ phase shifts of $I=2$ $\pi\pi$ scattering have been simultaneously  extracted by Dudek, \emph{et al.} at $m_{\pi}=396~{\rm MeV}$ \cite{Dudek:2012gj}.

For two mesons with different masses, the symmetry group is even further reduced in the boosted frame, making the mixing to occur between $l=0$ and $l=1$ states as well as with higher angular momentum states \cite{Fu:2011xz}. An easy way to see the latter is to note that in contrast with the case of degenerate masses, the kinematic function $c_{lm}^{\textbf{P}}$ as defined in Eq. (\ref{clm}) is non-vanishing for odd $l$ when the masses are different. As a result even and odd angular momenta can mix in the QC. This however does not indicate that the spectrum of the system is not invariant under parity. As long as all  interactions between the particles are parity conserving, the spectrum of the system and its parity transformed counterpart are the same. The determinant condition, Eq. (\ref{det0}), guarantees this invariance: any  mechanism, for example, which takes an S-wave scattering state to an intermediate P-wave two-body state, would take it back to the final S-wave scattering state, and the system ends up in the same parity state.\footnote{We note that under parity $\mathcal{Z}^d_{lm}\rightarrow\left(-1\right)^{l}\mathcal{Z}^d_{lm}$. Also it can be seen that under the interchange of particles $\mathcal{Z}^d_{lm}\rightarrow\left(-1\right)^{l}\mathcal{Z}^d_{lm}$, so that for degenerate masses the $c_{lm}^{P}$ functions vanish for odd $l$. This is expected since the parity transformation in the CM frame is equivalent to the interchange of particles. However, as is explained above for the case of parity transformation, despite the fact that $\delta \mathcal{G}^{V}$ is not symmetric with respect to the particle masses, the QC is invariant under the interchange of the particles.}

\subsubsection{Two coupled channels} For the N=2 case, the expression for the scattering amplitude in Eq. (\ref{def1}) is modified, as it now depends on the mixing angle $\bar{\epsilon}$, and the scattering matrix is no longer diagonal in the channel basis. By labeling the off-diagonal terms as $\mathcal{M}_{I,II}$, and by choosing the ``barred" parameterization for the time-reversal invariant S-matrix  \cite{Stapp:1956mz}
\begin{eqnarray}
\label{smatrix2}
S_2=\begin{pmatrix} 
e^{i2\delta_I}\cos{2\overline{\epsilon}}&ie^{i(\delta_I+\delta_{II})}\sin{2\overline{\epsilon}}\\
ie^{i(\delta_I+\delta_{II})}\sin{2\overline{\epsilon}}&e^{i2\delta_{II}}\cos{2\overline{\epsilon}} \\
\end{pmatrix},
\end{eqnarray}
where the subscript $2$ on $S$ denotes the number of coupled channels, the scattering matrix elements can be written as
\begin{eqnarray}
\label{def}
(\mathcal{M}_{i,i})_{l_1,m_1;l_2,m_2}&=&\delta_{l_1,l_2}\delta_{m_1,m_2}\frac{8\pi E^*}{n_ik^*_i}\frac{\cos(2\bar{\epsilon})e^{2i\delta^{(l_1)}_i(k^*_i)}-1}{2i},\\ 
(\mathcal{M}_{I,II})_{l_1,m_1;l_2,m_2}&=&\delta_{l_1,l_2}\delta_{m_1,m_2}\frac{8\pi E^*}{\sqrt{n_In_{II}k^*_Ik^*_{II}}}\sin(2\bar{\epsilon})\frac{e^{i(\delta^{(l_1)}_I(k^*_I)+\delta^{(l_1)}_{II}(k^*_{II}))}}{2},
\end{eqnarray}
where the usual relativistic normalization of the states is used in evaluating the S-matrix elements.\footnote{Here we assume no physical partial-wave mixing occurs in either channels. This is of course not the case in the two-nucleon systems in the isosinglet channel. We revisit this formalism in chapter \ref{chap:NN} to incorporate such mixings.} From Eq. (\ref{det0}) one obtains \cite{Hansen:2012tf, Briceno:2012yi}
\begin{eqnarray}
\label{allorders}
\det\begin{pmatrix} 
1+\delta \mathcal{G}^{V}_I\mathcal{M}_{I,I}&\delta \mathcal{G}^{V}_I\mathcal{M}_{I,II}\\
\delta \mathcal{G}^{V}_{II}\mathcal{M} _{I,II}&1+\delta \mathcal{G}^{V}_{II}\mathcal{M}_{II,II}\\
\end{pmatrix}=0,
\end{eqnarray}
where the determinant is not only over the number of channels but also over angular momentum which is left implicit. 
For $l_{max}=0$ one can use the definition of the pseudo-phase to rewrite the QC in a manifestly real form,
 \begin{eqnarray}
\label{allorders2}
\cos{2\bar{\epsilon}}\cos{\left(\phi^\mathbf{P}_1+\delta_1-\phi^\mathbf{P}_2-\delta_2\right)}=\cos{\left(\phi^\mathbf{P}_1+\delta_1+\phi^\mathbf{P}_2+\delta_2\right)},
 \end{eqnarray} 
which is equivalent to the result given in Refs. \cite{He:2005ey, Liu:2005kr} in the CM frame. 
It is easy to see that in the $\bar \epsilon\rightarrow 0$ limit, one recovers the decoupled QCs for both channels $I$ and $II$, Eq. (\ref{quateq}). 

In order to understand the significance of such FV coupled-channels formalism, it is sufficient to note that the spectrum of QCD contains a wealth of resonances that sit above multi-particle thresholds. Since resonances are not isolated eigenstates of QCD Hamiltonian and can only be observed as resonances in multi-particle scattering amplitudes, an evaluation of the energy levels will not give direct insight into the mass and decay width of these resonances. Instead a L\"uscher-type methodology, by which the scattering phase shifts of the scattering states are evaluated at the calculated energy levels on the lattice, is required. Therefore, analyzing the FV spectra, in particular those of the excited spectra of QCD as produced by various LQCD collaborations (e.g., Refs.~\cite{Dudek:2009qf, Dudek:2010wm, Dudek:2011tt, Edwards:2011jj, Dudek:2011bn, Dudek:2013yja}), requires applying a multi-coupled channel formalism. This necessary step can provide some of the theoretical guidance for the forthcoming JLab GlueX experiment~\cite{Shepherd:2009zz, Zihlmann:2010zz, Somov:2011zz, Smith:2012ch} as well as other spectroscopy experiments worldwide.

Before concluding this section, let us emphasize that the formalism presented here and in chapters \ref{chap:NN}, \ref{chap:deuteron} and \ref{chap:TBC} is valid up to exponential corrections of the form $\mathcal{O}(e^{-m_{\pi}L})$. For the case of the nuclear force, these corrections are due to the finite (non-zero) range of interactions that is set by the pion mass, i.e. the lightest particle that is produced and mediated in the hadronic system due to strong interactions. In order for these corrections to be at  sub-percent level, the spatial extent of the volume must be chosen such that $m_{\pi}L \gtrsim 2\pi$. As the pion masses used in studies of multi-hadron systems approach their physical value, larger volumes must be used to insure these corrections will remain small.


\chapter{LATTICE OPERATORS AND RESTORATION OF ROTATIONAL SYMMETRY IN THE CONTINUUM LIMIT}
\label{chap:operators}

\begin{figure}[h!]
\begin{centering}
\includegraphics[scale=0.425]{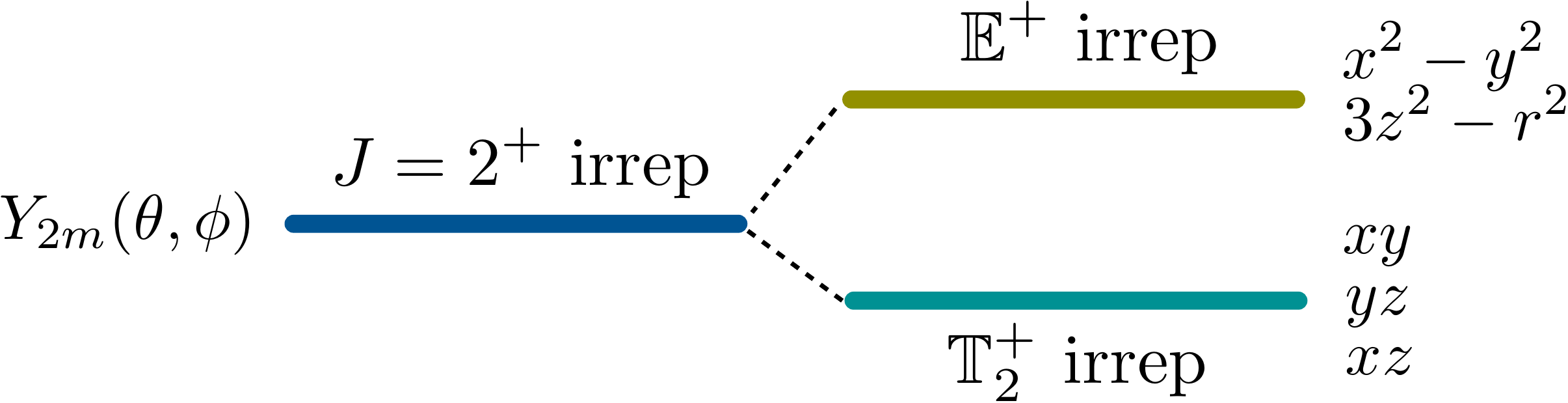}
\par\end{centering}
\caption{{\small The schematic level splitting due to the breakdown of the $SO(3)$ symmetry down to $O_h(3)$ symmetry due to a non-zero lattice spacing. The 5-fold degeneracy of the $J=2^+$ irrep of the $SO(3)$ group is split to a two-fold degeneracy ($\mathbb{E}^+$ irrep) and a three-fold degeneracy ($\mathbb{T}_2^+$ irrep). The basis functions of the corresponding irreps are given next to each level.}}
\label{fig:splitting}
\end{figure}

Efforts to reduce lattice artifacts and achieve a better behaved
theory in the continuum limit date back to early stages of development
of LQCD. Many that are part of the Symanzik improvement program
include a systematic modification of the action in such a
way to eliminate $\mathcal{O}\left(a^{n}\right)$ terms from physical
quantities calculated with  LQCD at each order in perturbation 
theory~\cite{Symanzik:1983dc,Symanzik:1983gh,Parisi:1985iv,WeiszI,WeiszII,Luscher,Curci,Hamber,Eguchi,Wetzel,Sheikholeslami}, or nonperturbatively.
However, as will be discussed, discretization effects are known to 
give rise to more subtle issues; the treatment of which turns out to be more
involved.
\begin{table}[b!]
\begin{centering}
\includegraphics[scale=0.875]{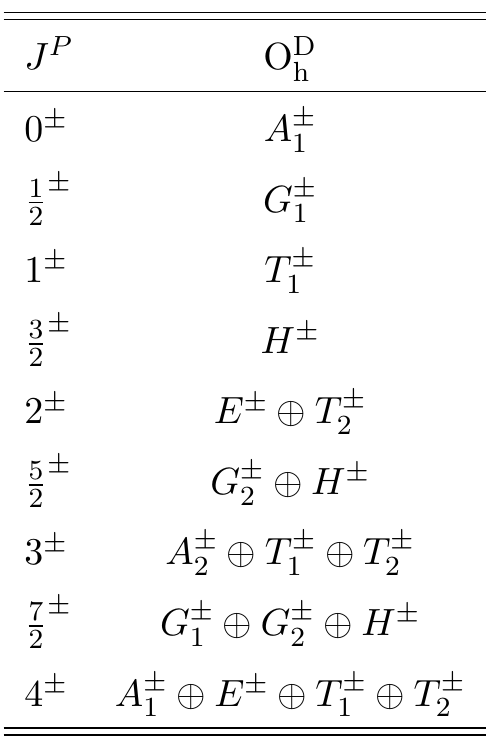}
\par\end{centering}
\caption{{\small The decomposition of the irreps of the SO(3) group up to $J=4$ in terms of the irreps of the $\mathrm{O^D_h}$ \cite{Luscher:1986pf, Luscher:1990ux, Mandula:1983wb, Johnson:1982yq, Berg, Basak:2005aq, Dresselhaus}.}}
\label{table:irreps}
\end{table}
LQCD is commonly formulated on a hyper-cubic grid, as a result the
full (Euclidean) Lorentz symmetry group of the continuum is
reduced to the discrete symmetry group of a hypercube. As
the (hyper) cubic group has only a finite number of irreducible representations
(irreps) compared to infinite number of irreps of the rotational group, a given irrep of the rotational group is not irreducible
under the (hyper) cubic group. 
Consequently, one can not assign
a well-defined angular momentum 
to a lattice state, which is generally a linear combination
of infinitely many different angular momentum states, see for example Table. \ref{table:irreps} for the decomposition of the irreps of the $SO(3)$ group up to $J=4$ in terms of the irreps of the $\mathrm{O^D_h}$ (the double cover of cubic $\mathrm{O_h}$ group). 
In principle, one can identify
the angular momentum of a corresponding continuum state in a lattice
calculation
from the degeneracies in the spectrum of
states belonging to different irreps of the cubic group as the lattice
spacing is reduced.\footnote{For some recent hadron spectroscopy works see Refs.
 \cite{BurchI,Gattringer,Petry,BurchII,Dudek:2009qf, Dudek:2010wm, Dudek:2011tt, Edwards:2011jj, Dudek:2011bn, Meinel, Dudek:2013yja} and for a review of baryon spectroscopy efforts see Ref. \cite{Lin}.} For example, a $J^P=2^+$ state can be identified if two (nearly) degenerate energy levels found in the spectra that are obtained using an $\mathbb{E}^+$ interpolator and a $\mathbb{T}_2^+$ interpolator for the state as $a \to 0$, see Fig. \ref{fig:splitting}. 
According to Table. \ref{table:irreps}, the $\mathbb{T}_2^+$ irrep, for example, has overlap with not only the $2^+$ irrep but also with $3^+,~4^+,~\dots$ irreps of the rotational group. In reality, there are error bars associated with these levels due systematic and statistical uncertainties. Therefore, as the density of degenerate states substantially increases with increasing
the angular momentum,
the identification of states with 
higher angular momentum becomes impossible with the current statistical precision.

The other issue is that the cubic symmetry of the lattice
allows the renormalization mixing of interpolating operators with
lower dimensional ones.  
The induced coefficients of the lower-dimensional operators
scale as inverse powers of the
lattice spacing, and hence diverge 
as the lattice spacing goes to zero. 
Although renormalization mixing of operators
is familiar from the continuum quantum field theory, it happens more
frequently in LQCD calculations as the reduced symmetry of the hyper-cube
is now less restrictive in preventing operators from mixing.
To obtain useful results for, as an example, the matrix elements of operators
from LQCD calculations,
non-perturbative
subtraction of the power divergences is required and generally introduces large statistical
uncertainties.

\begin{figure}[h!]
\begin{centering}
\includegraphics[scale=1.05]{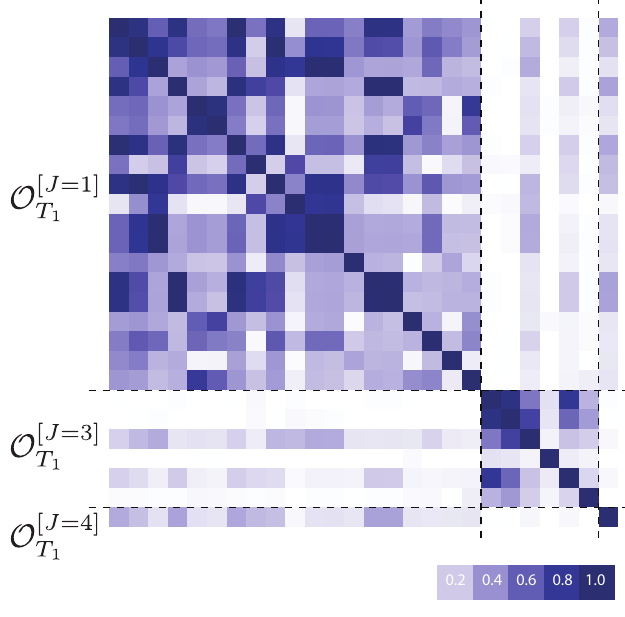}
\par\end{centering}
\caption{{\small The representative plot of the degree of the correlation among 26 different lattice operators, $\mathcal{O}^{[J]}_{\mathbb{T}_1^-}$, all transforming under $\mathbb{T}_1^-$ irrep of the cubic group, but are subduced from operators with three different angular momentum, $J=2,3,4$ as described in Ref. \cite{Dudek:2010wm}. The figure is reproduced with the permission of Jozef Dudek.}}
\label{fig:subduction}
\end{figure}

To overcome these obstacles, it has been proposed by Dudek,
{\it et al.}~\cite{DudekI,DudekII,Edwards} that by means of a novel construction of
interpolating operators, the excited states of several mesons and
baryons can be identified to high precision. 
The essence of this
method is that if one uses a set of cubically invariant local
operators which have already been subduced \cite{Basak} from a rotationally
invariant local operator with a definite angular momentum, $J$, while at the same
time smearing the gauge and quark fields over
the hadronic scale~\cite{Allton,Morningstar,Peardon}, the constructed
operator has maximum overlap onto a continuum state with angular momentum
$J$ if the lattice spacing is sufficiently small.  As an example, one can measure the correlation among operators belonging to a large set of constructed lattice operators that share the same transformation properties under a given irrep of the cubic group but are subduced from operators with different angular momentum. One such investigation is proposed in Ref. \cite{Dudek:2010wm} and is plotted in Fig. \ref{fig:subduction}. Interestingly, the correlation among operators subduced from different angular momentum representation of the continuum is minimal compared with the self correlations, confirming the effectiveness of their method in identification of the continuum states' angular momentum. 
The subduction is assumed to be responsible for retaining ``memory'' of the 
underlying angular momentum of the continuum operator, while 
the smearing is assumed to suppress mixing with operators of different angular momentum
-- by filtering contributions from ultraviolet (UV) modes. 
In another approach, states with  higher angular momentum  
in the glueball spectra
of $2+1$ dimensional $SU\left(2\right)$ gauge theories~\cite{Meyer,JohnsonRW}
are isolated by using glueball interpolating operators that are 
linear
combinations of Wilson loops which are rotated by arbitrary angles
in order to project out a particular angular momentum $J$ in the continuum.
In addition, 
the links are smeared, or blocked, in order to be smooth over  physical
length scales  rather than just in the UV~\cite{Teper}.
So by monitoring the angular content of the glueball wavefunction
in the continuum limit with a probe with definite $J$, 
the $0^{-}/4^{-}$ puzzle in the glueball
spectroscopy has been tackled.
The prominent feature of these works
is that the recovery of rotational symmetry for sufficiently small
lattice spacings is qualitatively emergent from their numerical results.

The same issue occurs in LQCD calculations of higher moments
of hadron structure functions, the extraction
of which requires the matrix elements of local operators
between hadronic states.
Although 
Lorentz invariance forbids
twist-2 operators with different $J$
from mixing in the continuum, generally 
they can mix in LQCD calculations with power-divergent mixing 
coefficients~\cite{Capitani,Beccarini}.
The power-divergent  mixing problem associated with the  lower
moments can be avoided  by several means as described, for example, in Refs. ~\cite{Beccarini,GockelerI,GockelerII,GockelerIII,GockelerIV,GockelerVI,GockelerV,MartinelliI,MartinelliII,MartinelliIII,GockelerVI}. 
In addition to these approaches,
two methods~\cite{Dawson,Detmold} have been suggested
that highlight the idea of approaching the continuum properties
of the hadronic matrix elements by suppressing the contributions from the UV, 
and in that sense resemble the idea of operator smearing in the proposals described above. 
In LQCD calculations of non-leptonic K-decay, 
Dawson {\it et al.}~\cite{Dawson}
suggested that point-splitting the hadronic currents by a distance
larger than the lattice spacing, but smaller than the
QCD scale, results in an operator product expansion of
the currents with the coefficients of lower dimensional operators
scaling with inverse powers of the point-splitting distance,
as opposed to the inverse lattice spacing.
This  considerably reduces  the numerical issues
introduced by  the operator mixing. 
In a different, but still physically equivalent approach, Detmold and Lin~\cite{Detmold}
showed that in LQCD calculations of matrix elements of the Compton
scattering tensor, 
the introduction of a fictitious, non-dynamical, heavy
quark coupled to physical light quarks 
removes the power divergences of the mixing
coefficients.
This technique enables the extraction of matrix elements
of higher spin twist-2 operators. 
The essence of this method is that the
heavy quark propagator acts as a smearing function in the momentum space,
suppressing contributions from the high-energy modes, provided
that its mass is much smaller than the inverse lattice spacing.

Encouraged by the results of these 
numerical non-perturbative investigations, 
we aim to quantify the recovery of rotational symmetry 
with analytical, perturbative calculations  in
$\lambda\phi^4$ theory and in QCD. 
In order to
achieve this goal, we first define a composite operator on the lattice
which has a well-defined angular momentum in the continuum limit and
is smeared over a finite physical region, and show how the non-continuum
contributions to the multipole expansion of the operator scales as
the lattice spacing is reduced toward the continuum. 
Tree-level contributions to matrix elements that violate
rotational symmetry,
either by the lattice operator
matching onto continuum operators with the ``wrong'' angular
momentum,
or matching onto continuum operators that explicitly 
violate rotational symmetry,
scale as $\mathcal{O}\left(a^{2}\right)$
as $a\rightarrow 0$. 
This includes the (naively) power-divergent contributions from
lower-dimension operators.
In order to make definitive statements about the size of 
violations to rotational symmetry, it must be ensured that the 
tree-level scalings are not ruined by quantum
fluctuations. 
This is  demonstrated by a perturbative calculation of the two-point function in 
$\lambda\phi^{4}$ scalar field theory with an insertion of such an operator.
It is confirmed that quantum
corrections at any order in perturbation theory do not alter the observed
classical scalings of non-continuum contributions.

After gaining experience with this operator in scalar field theory, 
the generalization to gauge theories is straightforward.
Special attention must be paid to the gauge links that appear in the
definition of gauge-invariant operator(s) that are the analogue of those
considered in the scalar field theory.
Also, it is well known that the perturbative expansion of operators used in
LQCD  are not well-behaved due to the presence of tadpole
diagrams~\cite{Lepage}. 
Naively, tadpoles make enhanced contributions to the matrix elements of the
operators we consider, and that tadpole improvement of the gauge links  and
smearing of the gluon fields 
are
crucial to the suppression of violations of rotational symmetry. 
After discussing the continuum behavior of the QCD operator(s), and their potential mixings, which violate rotational invariance at ${\cal O}(a^2)$,
we determine the renormalization of the operator(s) on the lattice
at one-loop order. 
The leading rotational invariance violating contributions to the renormalized
lattice 
operator are suppressed by ${\cal O}(\alpha_{s} a^2)$,
provided that the gauge
fields 
are also smeared over a physical region similar to the matter fields. 
This means that the leading rotational invariance violating operators
introduced by the quantum loops make subleading contributions 
compared to tree-level, ${\cal O} (a^2)$.
The loop contributions that scale as ${\cal O}(\alpha_{s} a)$ 
do not violate rotational symmetry, and hence 
are absorbed into the operator $Z$-factor.

\section{Operators in Scalar Field Theory
\label{sec:Classical}}
\begin{figure}[t!]
\begin{centering}
\includegraphics[scale=0.40]{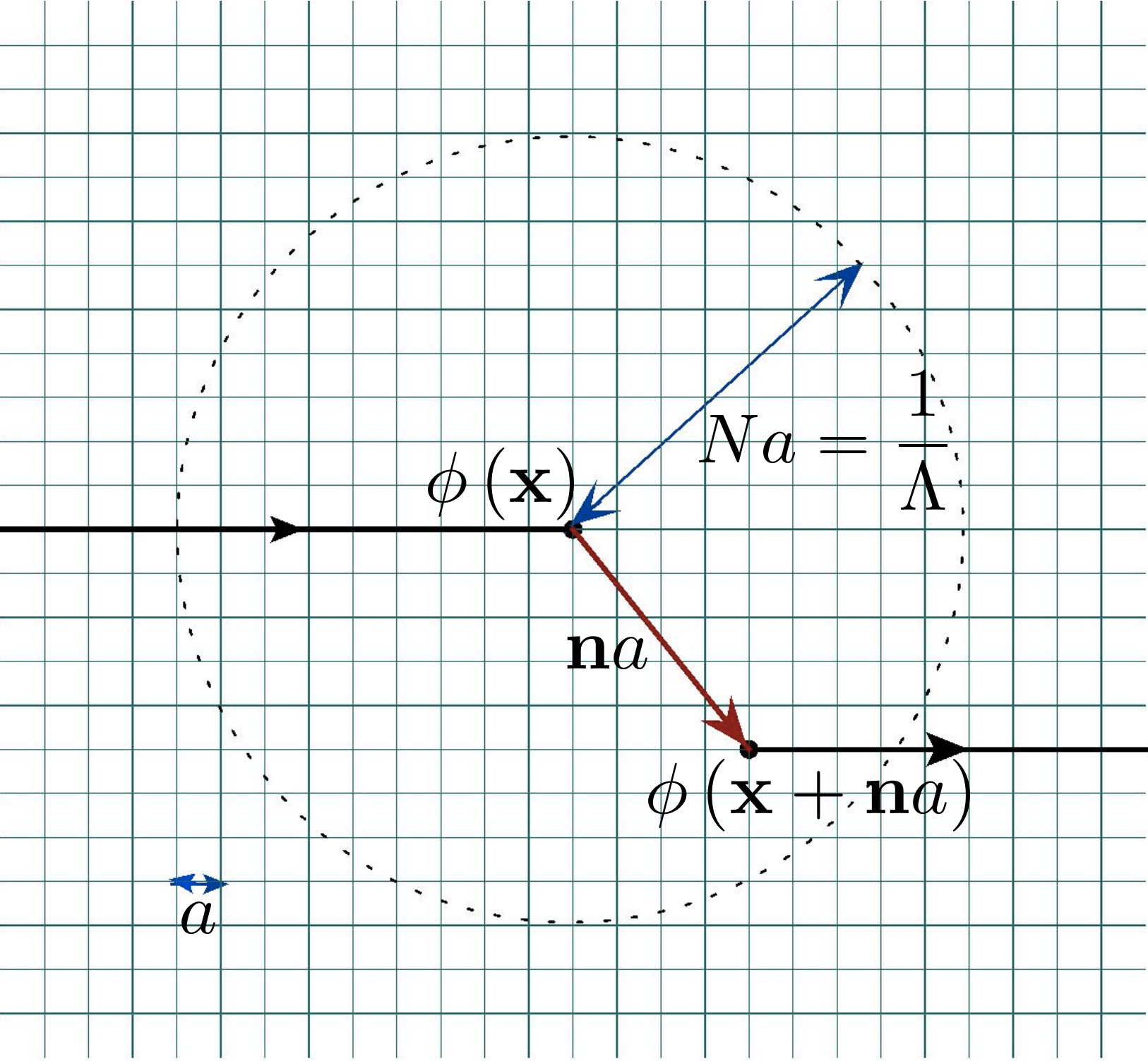}
\par\end{centering}
\caption{{\small 
A contribution to the lattice operator defined in 
Eq.~(\ref{eq:1}), with $\left|\mathbf{n}\right|\leq N$. 
All the points inside the three-dimensional spherical shell
$\left|\mathbf{n}a\right|=Na$ are included in the operator. 
The two length scales defining the operator, 
the lattice spacing, $a$, 
and the operator size, $Na=1/\Lambda$, 
are shown.}}
\label{fig:operator}
\end{figure}

\noindent
The goal is to construct a bilinear operator of the scalar fields
on a cubic lattice which has certain properties. First of all, as
it was discussed earlier, it has to be smeared over a finite region
of space. 
This physical region should be large compared to the lattice
spacing, 
and, for our purposes,  small compared to the typical length scale of
the system  to allow for a perturbative analysis.
The spatial extent of the operator can be identified with its renormalization scale.
Secondly, 
it is required to transform as a spherical tensor with well-defined angular
momentum in the continuum limit.
An operator that satisfies these conditions is~\footnote{
This corresponds to one particular choice of radial structure of the operator.
However, the results of the calculations  and the physics conclusions presented
in this work do not change qualititively when other smooth radial structures are
employed, such as a Gaussian or exponential.
} 
\begin{equation}
\hat{\theta}_{L,M}\left(\mathbf{x};a,N\right)
\ =\ 
\frac{3}{4\pi
  N^{3}}
\sum_{\mathbf{n}}^{\left|\mathbf{n}\right|\leq N}\phi\left(\mathbf{x}\right)
\phi\left(\mathbf{x}+\mathbf{n}a\right)
\ Y_{L,M}\left(\hat{\mathbf{n}}\right)
,
\label{eq:1}
\end{equation}
where $\mathbf{n}$ denotes a triplet of integers, and  
it is normalized
by the spatial volume of the region over which it is distributed.
$\phi({\bf x})$ is the scaler field operator, 
$N$ is the maximum number of lattice sites in the radial direction, and 
$Y_{LM}\left(\hat{\mathbf{n}}\right)$ is a spherical harmonic evaluated at the
angles defined by the unit vector in the direction of  $\mathbf{n}$, 
$\hat{\mathbf{n}}$, as shown in Fig.~\ref{fig:operator}.
This operator can also be written in a multipole expansion about its center as
\begin{equation}
\hat{\theta}_{L,M}\left(\mathbf{x};a,N\right)=\frac{3}{4\pi
  N^{3}}
\sum_{\mathbf{n}}^{\left|\mathbf{n}\right|\leq N}\sum_{k}\frac{1}{k!}\ 
\phi\left(\mathbf{x}\right)\left(a\mathbf{n}\cdot\mathbf{\nabla}\right)^{k}
\phi\left(\mathbf{x}\right)
\ Y_{L,M}\left(\hat{\mathbf{n}}\right)
,
\label{eq:2}
\end{equation}
where the gradient operator acts on the ${\bf x}$ variable,
$\mathbf{\nabla}\equiv \mathbf{\nabla}_{\bf x}$.
Although the operator $\hat{\theta}_{L,M}\left(\mathbf{x};a,N\right)$
is labeled  by its angular momentum in the continuum limit, from the right hand side of
Eq.~(\ref{eq:2}), it is clear that it is a linear combination
of an infinite number of operators with angular momentum compatible with its parity. 
To be more specific, consider the $M=0$
component of the operator expanded in a derivative  operator basis,
\begin{equation}
\hat{\theta}_{L,0}\left(\mathbf{x};a,N\right)=
\sum_{L^{\prime},d}\frac{C_{L0;L^{\prime}0}^{\left(d\right)}
\left(N\right)}{\Lambda^{d}}\mathcal{O}_{z^{L^{\prime}}}^{\left(d\right)}
\left(\mathbf{x};a\right)
,
\label{eq:3}
\end{equation}
where $\mathcal{O}_{z^{L^{\prime}}}^{\left(d\right)}\left(\mathbf{x};a\right)$
are defined in Appendix~\ref{app:operators}.
The operator subscript denotes  that there are $L^{\prime}$ 
free indices in the derivative operator, while $d$ denotes the total number of derivatives. 
As is discussed in the
Appendix~\ref{app:operators}, 
there are operators in this basis which are not rotationally
invariant but only cubically invariant. 
$C_{L0;L^{\prime}0}^{\left(d\right)}\left(N\right)$
are coefficients of each operator in the expansion whose values are
determined by matching Eq. (\ref{eq:2}) with Eq.~(\ref{eq:3}). 
Finally $\Lambda=1/(Na)$ is the momentum scale of the smeared
operator which is kept fixed as the lattice spacing is varied. 
Therefore, as the lattice spacing decreases,
more point shells (shells of integer triplets) 
are included in the sum in Eq.~(\ref{eq:2}).
The convergence of this derivative expansion is guaranteed
as the scale $\Lambda$ is set to be much larger than the typical
momentum encountered by the operator.

\subsection{Classical scalar field theory}
\label{sec:Clasphi}

In order for the operator to recover its continuum limit as the lattice
spacing vanishes, 
the coefficients $C_{L0;L^{\prime}0}^{\left(d\right)}$
should have certain properties. 
First of all, those associated with
the operators with $L\neq L^{\prime}$ as well as the rotational invariance
violating operators, should vanish as $a\rightarrow0$.
Also the
coefficients of rotational invariant operators with $L=L^{\prime}$
should reach a finite value in this limit. 
These properties will be
shown to be the case in a formal way shortly, but in order to get
a general idea of the classical scaling of the operators and the size
of mixing coefficients, we first work out a particular example. 
Consider
the operator $ $$\hat{\theta}_{3,0}\left(\mathbf{x};a,N\right)$
expanded out up to five derivative operators,
\begin{eqnarray}
\hat{\theta}_{3,0}\left(\mathbf{x};a,N\right)
&=&
\frac{C_{30;10}^{\left(1\right)}\left(N\right)}{\Lambda}\mathcal{O}_{z}^{\left(1\right)}
\left(\mathbf{x};a\right)
+
\frac{C_{30;10}^{\left(3\right)}\left(N\right)}{\Lambda^{3}}\mathcal{O}_{z}^{\left(3\right)}
\left(\mathbf{x};a\right)
+ 
\frac{C_{30;10}^{\left(5\right)}\left(N\right)}{\Lambda^{5}}\mathcal{O}_{z}^{\left(5\right)}
\left(\mathbf{x};a\right)
+
\nonumber\\
&&\frac{C_{30;10}^{\left(5;RV\right)}\left(N\right)}{\Lambda^{5}}\mathcal{O}_{z}^{\left(5;RV\right)}
\left(\mathbf{x};a\right)
+
\frac{C_{30;30}^{\left(3\right)}\left(N\right)}{\Lambda^{3}}\mathcal{O}_{zzz}^{\left(3\right)}
\left(\mathbf{x};a\right)
+
\nonumber\\
&&\frac{C_{30;30}^{\left(5\right)}\left(N\right)}{\Lambda^{5}}\mathcal{O}_{zzz}^{\left(5\right)}
\left(\mathbf{x};a\right)
+
\frac{C_{30;50}^{\left(5\right)}\left(N\right)}{\Lambda^{5}}\mathcal{O}_{zzzzz}^{\left(5\right)}
\left(\mathbf{x};a\right)
+
\mathcal{O}\left(\frac{\nabla_{z}^{7}}{\Lambda^{7}}\right),
\label{eq:4}
\end{eqnarray}
where the superscript RV denotes the rotational invariance violating operator and its corresponding coefficient in the above expansion.
\begin{figure}[!ht]
\begin{centering}
\includegraphics[scale=0.8]{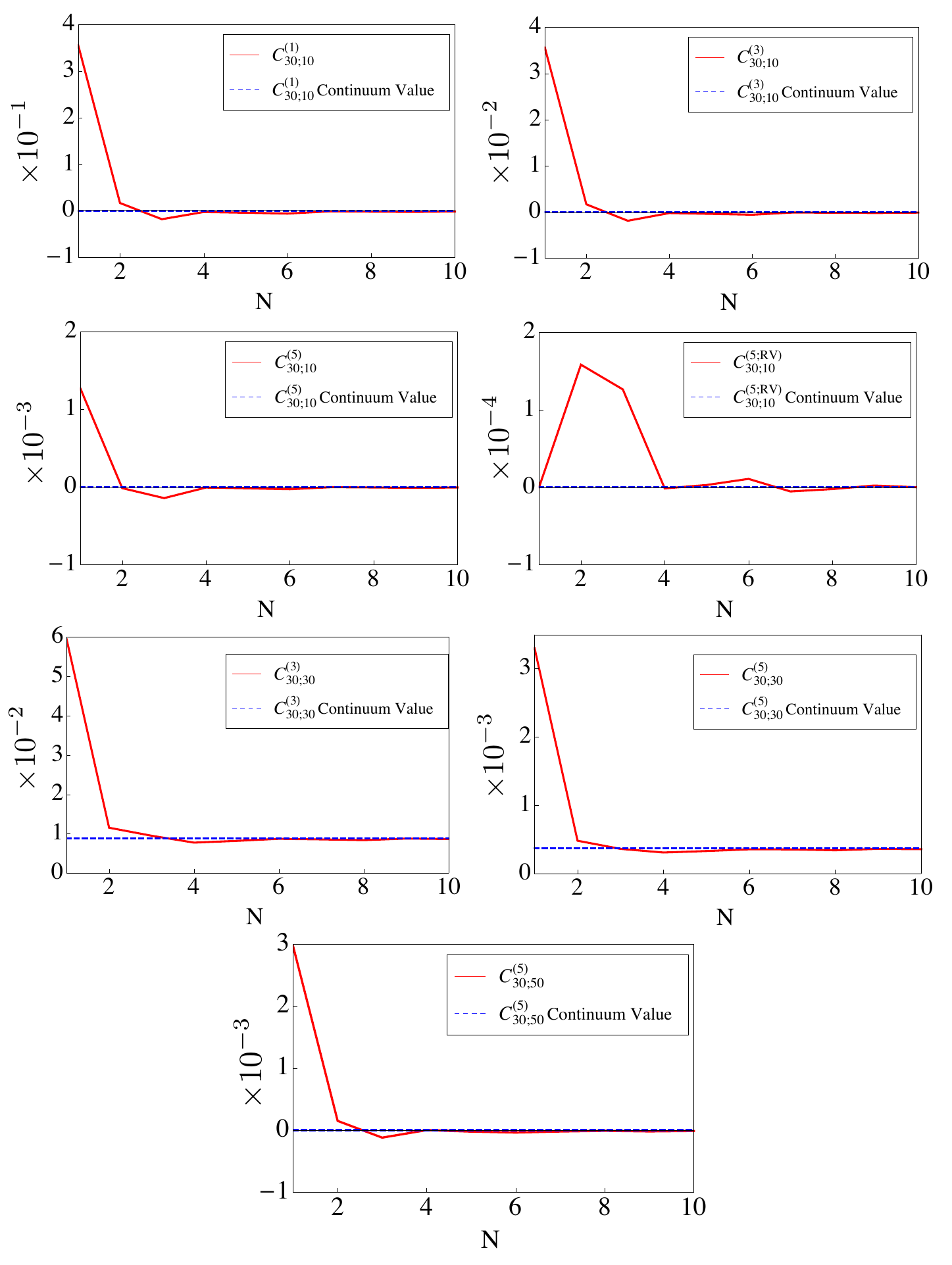}
\par\end{centering}
\caption{{\small 
The tree-level values of the 
coefficients $C^{(d)}_{30;L^\prime 0}$  appearing in
Eq.~(\ref{eq:4})
as a function of the largest $n$-shell included in the summation in Eq.~(\ref{eq:1}).
}}
\label{fig:TheCs}
\end{figure}
\begin{figure}[!ht]
\begin{centering}
\includegraphics[scale=0.4]{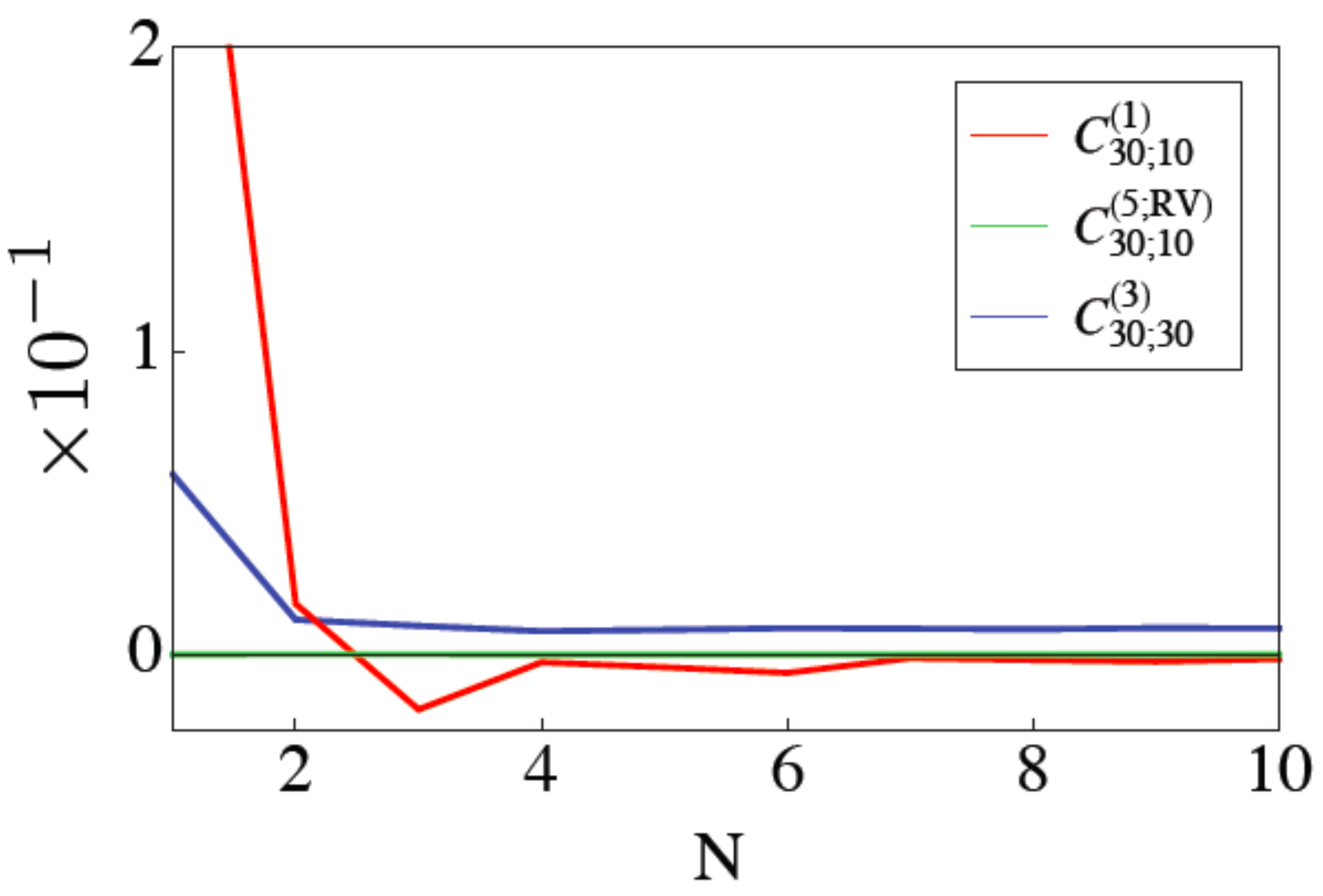}
\par\end{centering}
\caption{{\small 
A comparison between the tree-level coefficients $C^{(d)}_{30;L^\prime 0}$ to illustrate
the relative rates of convergence to the continuum limit.
}}
\label{fig:TheComp}
\end{figure}

The numerical values of the coefficients in Eq.~(\ref{eq:4}), at the classical level,
as a
function of the maximum shell included in the sum in Eq.~(\ref{eq:2})
are shown in Fig.~\ref{fig:TheCs} and Fig.~\ref{fig:TheComp}.
From these plots it is clear that while the coefficients
$C_{30;30}^{\left(3\right)}$ and $C_{30;30}^{\left(5\right)}$ reach
a finite value for large N, the coefficients of lower and higher angular
momentum operators,  as well as the rotational invariance violating
operator, approach zero. 
To find the values of the 
leading order (LO) 
coefficients in this limit, as well as to see how the non-leading
contributions scale with $N=1/(\Lambda a)$, one can apply the Poisson
re-summation formula to the right hand side of Eq.~(\ref{eq:2}),
\begin{equation}
\hat{\theta}_{L,M}\left(\mathbf{x};a,N\right)
 =\ 
\frac{3}{4\pi N^{3}}\sum_{k}\frac{a^{k}}{k!}
\sum_{\mathbf{p}}\int d^{3}y\ 
\theta\left(N-y\right)\ 
e^{i2\pi\mathbf{p}\cdot\mathbf{y}}\ 
\phi\left(\mathbf{x}\right)\left(\mathbf{y}\cdot\mathbf{\nabla}\right)^{k}
\phi\left(\mathbf{x}\right)\ 
Y_{L,M}\left(\hat{\mathbf{y}}\right)
,
\label{eq:5}
\end{equation}
where $\mathbf{p}$ is another triplet of integers, and the
$\mathbf{p}$ summation is unbounded. 
The continuum values of the coefficients
obtained in the $N\rightarrow\infty$ limit, 
corresponding to the $\mathbf{p}=0$ term in Eq.~(\ref{eq:5}),
are
\begin{eqnarray}
C_{30;30}^{\left(d\right)}
& = & 
{15\over 4}\ 
\sqrt{7\over\pi}\ 
{ d^2-1 \over (d+4)!}
 \qquad {\rm with}\qquad d=3,5,...
,
\label{eq:6}
\end{eqnarray}
while the other coefficients 
in Eq.~(\ref{eq:4})
vanish in this limit as expected.
The LO  corrections to these continuum values can be calculated
as following. 
The deviation of $C_{30;30}^{\left(3\right)}$ from
its continuum value can be found from 
\begin{eqnarray}
I_{30}
& \sim\ & 
\frac{3}{4\pi}\frac{\left(Na\right)^{3}}{3!}
\sum_{\mathbf{p}\neq 0}
\int_{0}^{1}dy\ y^{2}\ d\Omega_{\hat{\mathbf{y}}}\ 
e^{i2\pi N\mathbf{p}\cdot\mathbf{y}}
\ \phi\left(\mathbf{x}\right)\ 
\ 
\left(\hat{\mathbf{y}}\cdot\mathbf{\nabla}\right)^{3}
\phi\left(\mathbf{x}\right)\ 
\ Y_{3,0}\left(\hat{\mathbf{y}}\right)
,
\label{eq:7}
\end{eqnarray}
where $\mathbf{\nabla}=\nabla_{z}\hat{e}_{z}$ and the $y$-variable  in Eq.~(\ref{eq:7})
is redefined to lie between $0$ and $1$,
and it is straightforward to show that
\begin{equation}
\delta C_{30;30}^{\left(3\right)}
\ =\ 
\frac{1}{N^{2}}
\ \frac{1}{32\pi^{2}}
\ \sqrt{\frac{7}{\pi}}\ 
\sum_{\mathbf{p}\neq0}
\frac{\cos\left(2\pi
    N\left|\mathbf{p}\right|\right)}{\left|\mathbf{p}\right|^{8}}
\left(-\frac{3}{2}\left|\mathbf{p}\right|^{6}+15\left|\mathbf{p}\right|^{2}p_z^4-\frac{25}{2}p_{z}^{6}\right)
.
\label{eq:8}
\end{equation}
It is interesting to note that, after trading $N$ for $1/(a \Lambda)$,
the finite lattice spacing corrections are not monotonic in $a$,
but exhibit oscillatory behavior, which is clearly evident in Fig.~\ref{fig:TheCs}.

The  deviation of $C_{30;10}^{\left(1\right)}$ from its continuum value of zero
follows similarly, and is found to scale as $\sim 1/N^2$, 
\begin{equation}
\delta C_{30;10}^{\left(1\right)}
\ =\ \frac{1}{N^{2}}
\ \frac{3}{16\pi^{2}}\sqrt{\frac{7}{\pi}}
\ \sum_{\mathbf{p}\neq0}
\ \frac{\cos\left(2\pi
    N\left|\mathbf{p}\right|\right)}{\left|\mathbf{p}\right|^{6}}
\ \left(\left|\mathbf{p}\right|^{4}-5p_{z}^{4}\right)
.
\label{eq:9}
\end{equation}
As in the case of the operator that conserves angular momentum in the continuum
limit,
the sub-leading correction (and in this case the first non-zero contribution) 
to the coefficient is suppressed by $1/N^{2}$. 
This can be shown to be the case for all the sub-leading
contributions to the coefficients $C_{LM;L^{\prime}M^{\prime}}^{\left(d\right)}$ as follows.
As is evident from Eq.~(\ref{eq:5}), the integrals
that are required
in calculating deviations from the continuum
values have the general form
\begin{equation}
I^{i_1...i_k}
\ \sim\ 
\frac{3}{4\pi}\frac{\left(Na\right)^{k}}{k!}
\ \sum_{\mathbf{p}\neq0}
\ \int_{0}^{1}dy\ y^{2+k}\ \int d\Omega_{\hat{\mathbf{y}}}
\ e^{i2\pi N\mathbf{p}\cdot \mathbf{y}}\ \hat{\mathbf{y}}^{i_{1}}\
\hat{\mathbf{y}}^{i_{2}}...\hat{\mathbf{y}}^{i_{k}}
\ Y_{LM}\left(\Omega_{\hat{\mathbf{y}}}\right)
,
\label{eq:10}
\end{equation}
which can be written as
\begin{eqnarray}
I^{i_1...i_k}
&  \sim & 
\frac{3}{4\pi}
\ \frac{\left(Na\right)^{k}}{k!}\frac{1}{\left(i2\pi N\right)^{k}}
\ \sum_{\mathbf{p}\neq0}\frac{\partial}{\partial
  p_{i_{1}}}...\frac{\partial}{\partial p_{i_{k}}}
\ \int_{0}^{1}dy\ y^{2+k}\ \int d\Omega_{\hat{\mathbf{y}}}
\ e^{i2\pi N\mathbf{p}\cdot\mathbf{y}}
\ Y_{LM}\left(\Omega_{\hat{\mathbf{y}}}\right)
\nonumber\\
& \sim & 
\frac{3}{4\pi}\frac{\left(Na\right)^{k}}{k!}\ 
\frac{4\pi i^{L}}{\left(i2\pi
    N\right)^{k}}\sum_{\mathbf{p}\neq0}
\ \frac{\partial}{\partial
  p_{i_{1}}}...\frac{\partial}{\partial
  p_{i_{k}}}
\ Y_{LM}\left(\Omega_{\hat{\mathbf{p}}}\right)
\ \int_{0}^{1}dy\ y^{2+k}\ j_{L}\left(2\pi N\left|\mathbf{p}\right|y\right)
.
\nonumber\\
\label{eq:11}
\end{eqnarray}
The y integration over the Bessel function gives rise to
either 
$-\frac{\cos\left(2\pi N\left|\mathbf{p}\right|\right)}{\left(2\pi N\left|\mathbf{p}\right|\right)^{2}}$
or 
$-\frac{\sin\left(2\pi N\left|\mathbf{p}\right|\right)}{\left(2\pi N\left|\mathbf{p}\right|\right)^{2}}$,
up to higher orders in $1/N$, 
depending on whether $L$ is even or odd. 
Thus the LO contribution from
Eq.~(\ref{eq:11}) in the  large $N$ limit 
is obtained by acting 
on the numerator
with the $p$ derivatives,  producing  $k$ powers
of $N$,  multiplying the $1/N^{2}$ from the denominator. 
Therefore, Eq.~(\ref{eq:11}) scales as
\begin{equation}
I^{i_1...i_k}\ \sim\ 
\left(Na\right)^{k}\frac{1}{N^{k}}\frac{N^{k}}{N^{2}}
\ \sim\ {1\over\Lambda^k}\ {1\over N^2}
,
\label{eq:12}
\end{equation}
and, 
in general,
the deviation of any coefficient from its  continuum value  
is suppressed by $1/N^{2}=\Lambda^{2}a^{2}$. 
This result implies that in calculating the matrix element of $L=3$ operator,
one has a derivative expansion of the form
\begin{eqnarray}
\Lambda^{3}\hat{\theta}_{3,0}\left(\mathbf{x};a,N\right)
& = & 
\alpha_{1}\ 
\frac{\Lambda^{2}}{N^{2}}\mathcal{O}_{z}^{\left(1\right)}\left(\mathbf{x};a\right)
\ +\ 
\alpha_{2}\ 
\frac{1}{N^{2}}\mathcal{O}_{z}^{\left(3\right)}\left(\mathbf{x};a\right)
\ +\ 
\alpha_{3}\ 
\frac{1}{\Lambda^{2}N^{2}}\mathcal{O}_{z}^{\left(5\right)}\left(\mathbf{x};a\right)
\nonumber\\
& + & 
\alpha_{4}\ 
\frac{1}{\Lambda^{2}N^{2}}\mathcal{O}_{z}^{\left(5;RV\right)}\left(\mathbf{x};a\right)
\ +\ 
\alpha_{5}\ 
\mathcal{O}_{zzz}^{\left(3\right)}\left(\mathbf{x};a\right)
 \ +\ 
\alpha_{6}\ 
\frac{1}{\Lambda^{2}}\mathcal{O}_{zzz}^{\left(5\right)}\left(\mathbf{x};a\right)
\nonumber\\
& + & 
\alpha_{7}\ 
\frac{1}{\Lambda^{2}N^{2}}\mathcal{O}_{zzzzz}^{\left(5\right)}\left(\mathbf{x};a\right)
 \ +\ 
\mathcal{O}\left({\nabla_{z}^{7}\over \Lambda^{4}}\right)
,
\label{eq:13}
\end{eqnarray}
where the mixing with $L\neq3$ operators
(with coefficients $\alpha_{1,2,3,7,...}$),
as well as the
operator with broken rotational symmetry
(with coefficient $\alpha_4$), 
vanish in the  large $N$ limit, 
while
the coefficients of $L=3$ operators 
(with coefficients $\alpha_{5,6,...}$),
are fixed by the 
scale of the operator, $\Lambda$. 
It is clear that for $N=1$ and $\Lambda=1/a$, where no smearing
is performed, the problem with divergent coefficients of the lower
dimensional operators is obvious, as,  for example, the coefficient of
$\mathcal{O}_{z}^{\left(1\right)}\left(\mathbf{x};a\right)$ diverges
as $1/a^{2}$ as $a\rightarrow0$, as is well known.

The fact that all the sub-leading contributions to the classical operator
are suppressed at least by $1/N^{2}$ regardless of $L$ and $L^{\prime}$
can be understood as follows. 
In the classical limit, where the short
distance fluctuations of the operator are negligible, the operator
does not probe the distances of the order of lattice spacing when
$a\rightarrow 0$. 
The angular resolution of the operator is
dictated by the solid angle discretization of the physical region
over which the operator is smeared, and therefore is proportional to
$1/N^{2}$. 
The question to answer is whether the 
quantum fluctuations modify this general result.

Before proceeding with the quantum loop calculations, 
it is advantageous to transform the operator into 
momentum-space to simplify loop integrals.
This can be done
easily by noting that for zero momentum insertion,
the operator acting on the field with momentum
$\mathbf{k}$ is
\begin{equation}
\hat{\tilde{\theta}}_{LM}\left(\mathbf{k};a,N\right)
\ =\ 
\frac{3}{4\pi N^{3}}\ 
\sum_{\mathbf{n}}^{|\mathbf{n}|\leq N}
\ e^{i\mathbf{k}\cdot\mathbf{n}a}
\ Y_{LM}\left(\mathbf{n}\right)
\ \tilde{\phi}\left(\mathbf{k}\right)
\ \tilde{\phi}\left(-\mathbf{k}\right)
,
\label{eq:14}
\end{equation}
which, after using the partial-wave expansion of
$e^{i\mathbf{k}\cdot\mathbf{n}a}$
and the exponential term resulting from the Poisson relation,
can be written as
\begin{eqnarray}
\hat{\tilde{\theta}}_{LM}\left(\mathbf{k};a,N\right)
& = &
6\sqrt{\pi}\ 
 \tilde{\phi}\left(\mathbf{k}\right)
 \tilde{\phi}\left(-\mathbf{k}\right)
\ \sum_{\mathbf{p}}\sum_{L_{1},M_{1},L_{2},M_{2}}i^{L_{1}+L_{2}}
\ \sqrt{\frac{\left(2L_{1}+1\right)\left(2L_{2}+1\right)}{2L+1}}
\nonumber\\
&&
\times
\ \left\langle L_{1}0;L_{2}0\left|L0\right.\right\rangle 
\ \left\langle L_{1}M_{1};L_{2}M_{2}\left|LM\right.\right\rangle 
Y_{L_{1}M_{1}}\left(\Omega_{\hat{\mathbf{k}}}\right)
\ Y_{L_{2}M_{2}}\left(\Omega_{\hat{\mathbf{p}}}\right)
\nonumber\\
&&
\times
\ \int_{0}^{1}dy\ y^{2}
\ j_{L_{1}}\left(aN\left|\mathbf{k}\right|y\right)
\ j_{L_{2}}\left(2\pi N\left|\mathbf{p}\right|y\right)
 .
\label{eq:15}
\end{eqnarray}
Although this form seems to be somewhat more complicated than in
position space, 
it turns out that it is advantageous
to work in momentum space when dealing with higher angular
momenta, as well as for $M\ne 0$.
Further, the dimensionless parameters 
$|{\bf k}|/\Lambda$ and $N$ that define the physics of such systems
are now explicit.
It is straightforward to show this form recovers 
the values of the
leading and sub-leading coefficients
given in Eqs.~(\ref{eq:8}) and (\ref{eq:9}),
and it is worth mentioning how they emerge from Eq.~(\ref{eq:15}).
For a non-zero value of $|{\bf p}|$ and $N=\infty$, the spherical
Bessel function $j_{L_{2}}\left(2\pi N\left|\mathbf{p}\right|y\right)$ vanishes
for any value of $L_2$.  However, for large values of $N$ but $|{\bf p}|=0$ the
only non-zero contribution is from $L_2=0$, and thus $L_1=L$, leaving a
straightforward integration over a single spherical Bessel function
$j_{L}\left(aN\left|\mathbf{k}\right|y\right)$ to obtain the continuum limit
given in Eq.~(\ref{eq:6}).
Extracting the subleading contributions and the violations of rotational
symmetry 
is somewhat more involved, and we provide an explicit
example in Appendix~\ref{app:RIviolation}.

\subsection{Quantum corrections in $\lambda\phi^{4}$
\label{sec:Scalar}}
\noindent 
In order to determine the impact of quantum fluctuations on the matrix elements
of $\hat{\theta}_{L,M}$,
defined in Eq.~(\ref{eq:1}), we consider loop contributions in
$\lambda\phi^{4}$ theory.
Beside its simplicity which enables us to develop tools in performing 
the analogous calculations in Lattice QCD, 
this theory corresponds to some interesting condensed matter systems.
For example, three dimensional O(N) models,  which describe important
critical phenomena in nature, have a corresponding $\lambda\phi^{4}$
field theory formulation. 
As pointed out in Refs.~\cite{CampostriniI,CampostriniII},
anisotropy in space either due to the symmetries of the physical system,
or due to an underlying lattice formulation, 
will result in the presence of irrelevant operators in the effective
Hamiltonian which are not rotationally invariant,
and 
introduce deviations of two-point functions
from their rotationally invariant scaling law near the fixed point. 
However,
as the rotationally invariant fixed point of the theory is approached,
the anisotropic deviations vanish like $1/\xi^{\rho}$ where $\xi^{2}$
is the second moment correlation length derived from the two-point function, 
and $\rho$ is a critical
exponent which is related to the critical effective dimension of the
leading irrelevant operator breaking rotational invariance. 
It has
been shown that in the large N approximation
of $O\left(N\right)$ models, $\rho\simeq2$ for cubic-like lattices. 
In the following, it
will be shown that, 
by inserting
$\hat{\theta}_{L,M}$ defined in
Eq.~(\ref{eq:1}) into the two-point function,
the same scaling law emerges when approaching the rotational-invariant
continuum limit of $\lambda\phi^{4}$ theory.
\begin{figure}[h!]
\begin{centering}
\includegraphics[scale=0.15]{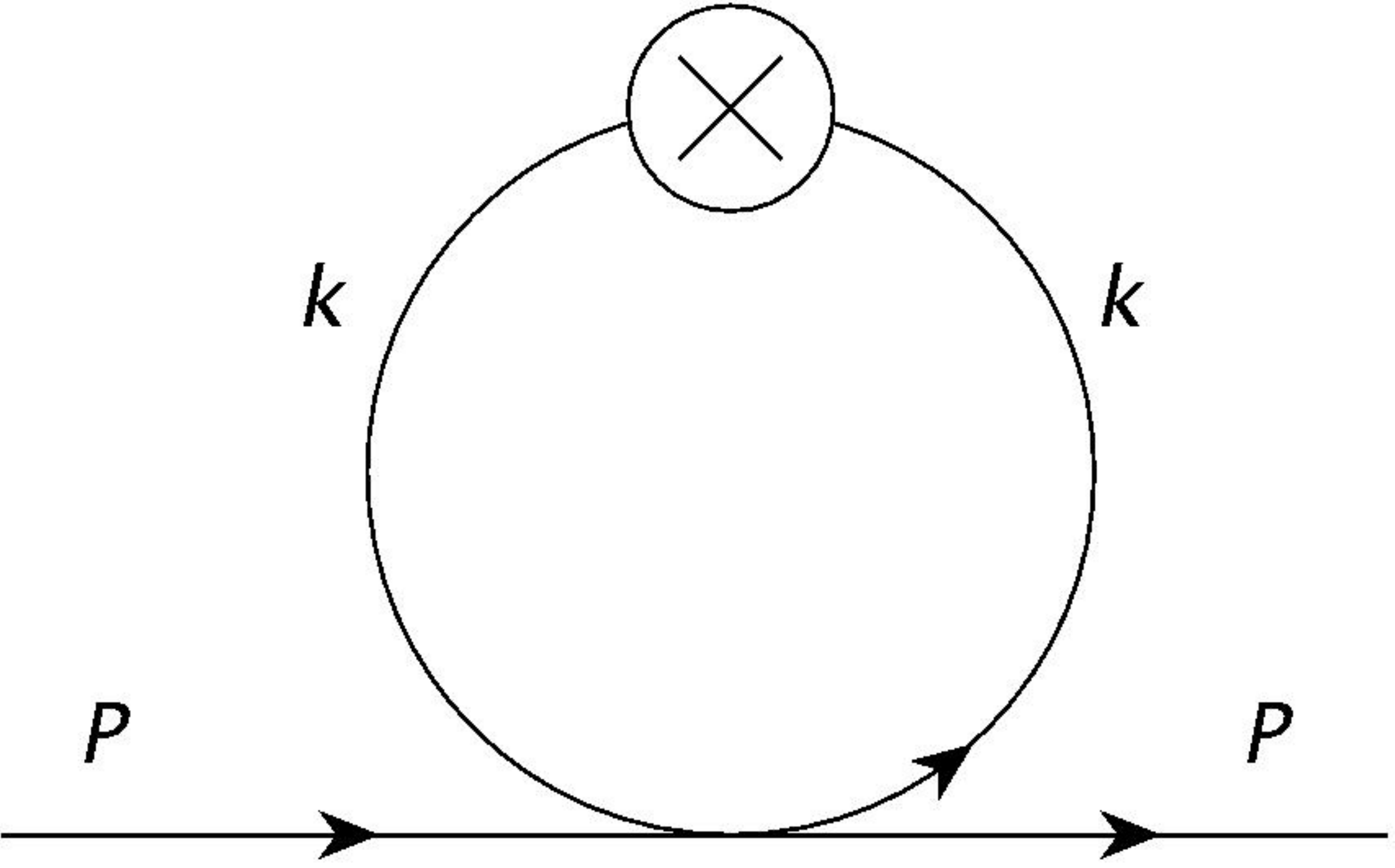}
\par\end{centering}

\caption{{\small One-loop correction to the two-point function with an insertion
of $\hat{\theta}_{L,M}$ in $\lambda\phi^{4}$}}
\label{fig:onelooplf4}
\end{figure}

At tree level, the contributions to the two-point function 
from an insertion of 
$\hat{\theta}_{L,M}$
at zero momentum transfer
has been already discussed in section \ref{sec:Clasphi}.
At one-loop order, there is only one diagram with an insertion of $\hat{\theta}_{L,M}$
that contributes to the two-point function, as shown in
Fig.~\ref{fig:onelooplf4}. 
This diagram introduces corrections only to the $L=0$ matrix element as there are
no free indices associated with the loop. 
The lattice integral associated with this one-loop diagram is 
\begin{equation}
J_{LM}
\ =\ 
\frac{3\lambda}{4\pi
  N^{3}}\sum_{\mathbf{n}}^{|\mathbf{n}|\leq N}
\ \int_{-\frac{\pi}{a}}^{\frac{\pi}{a}}\frac{d^{4}k}{\left(2\pi\right)^{4}}
\ \frac{e^{i\mathbf{k}\cdot\mathbf{n}a}}{\left(\hat{\mathbf{k}}^{2}+m^{2}\right)^{2}}
\ Y_{LM}\left(\Omega_{\mathbf{n}}\right)
,
\label{eq:17}
\end{equation}
where 
$\hat{\mathbf{k}}^{2}= {4\over a^2}\sum\limits_{\mu} \sin^{2}\left({ k_{\mu}a\over
    2}\right)$, 
$\lambda$ is the coupling constant and $m$
is the $\phi$ mass.
The three-momentum integration can be
evaluated by noting that the region of integration
can be split into two parts: region I where $0\leq\left|\mathbf{k}\right|\leq\pi/a$
and therefore is rotationally symmetric, and region II where $\pi/a\leq\left|\mathbf{k}\right|\leq\sqrt{3}\pi/a$
which consists of disconnected angular parts. Also as the three-momentum
integration is UV convergent, a small $a$ expansion of the integrand
can be performed. 
Using Eq.~(\ref{eq:15}), the contribution from
region I to the $\mathbf{p}=0$ term in the Poisson sum is
\begin{eqnarray}
J_{LM}^{\left(I\right)}\left(\mathbf{p}=0\right)
& = & 
\frac{3\lambda}{\left(2\pi\right)^{4}}i^{L}
\int_{-\frac{\pi}{a}}^{\frac{\pi}{a}}dk_{4}\int_{0}^{\frac{\pi}{a}}dkk^{2}
\int d\Omega_{\hat{\mathbf{k}}}\frac{1}{\left(\hat{\mathbf{k}}^{2}+m^{2}\right)^{2}}
\nonumber\\
&&\qquad\qquad  \times
\left[\ 
\int_{0}^{1}dy\ y^{2}\ j_{L}\left(aN\left|\mathbf{k}\right|y\right)\ \right]\ 
Y_{LM}\left(\Omega_{\mathbf{k}}\right)
\nonumber\\
& = & 
\frac{3\lambda}{16\pi^4}i^{L}
\left[J_{LM}^{LO}+J_{LM}^{NLO}+\mathcal{O}\left(1/N^{4}\right)\right]
,
\label{eq:18}
\end{eqnarray}
where
\begin{eqnarray}
J_{LM}^{LO}
& = & 
2\sqrt{\pi}
\delta_{L,0}\delta_{M,0}
\int_{-\frac{\pi}{\Lambda a}}^{\frac{\pi}{\Lambda a}}\ 
dq_{4}\int_{0}^{\frac{\pi}{\Lambda
    a}}dqq^{2}\frac{1}{\left[q^{2}+q_{4}^{2}+m^{2}/\Lambda^{2}
\right]^{2}}
\ \int_{0}^{1}dy\ y^{2}\ j_0\left(qy\right)
,
\nonumber
\end{eqnarray}
\begin{eqnarray}
J_{LM}^{NLO}
& = & 
\frac{1}{N^{2}}\int_{-\frac{\pi}{\Lambda a}}^{\frac{\pi}{\Lambda a}}
dq_{4}\int_{0}^{\frac{\pi}{\Lambda
    a}}dqq^{2}\frac{q^{4}}{\left[q^{2}+q_{4}^{2}+m^{2}/\Lambda^{2}\right]^{3}}
\nonumber\\
&& 
\qquad
\times
\left[
\frac{6\sqrt{\pi}}{5}\ \delta_{L,0}\ \delta_{M,0}\
\int_{0}^{1}dyy^{2}j_{0}\left(qy\right)
\right.\nonumber\\
&& \left.
\qquad
+\ \delta_{L,4}\left(\frac{2}{3}\sqrt{\frac{2\pi}{35}}\delta_{M,-4}
+\frac{4\sqrt{\pi}}{15}\delta_{M,0}
+\frac{2}{3}\sqrt{\frac{2\pi}{35}}\delta_{M,4}\right)\ \int_{0}^{1}dy\ y^{2}\ 
j_{4}\left(qy\right)\right]
\ ,
\nonumber\\
\label{eq:19}
\end{eqnarray}
with 
$q=\left|\mathbf{k}\right|/\Lambda$ and $q_{4}=k_{4}/\Lambda$.
The LO integral,  $J_{LM}^{LO}$, is convergent, while the NLO contribution, 
$J_{LM}^{NLO}$, while not convergent, is not divergent, but is of the 
form $\sin\left(N\pi\right)/N^2$.
This implies that they depend on the ratio of the two mass scales, $\Lambda$ and $m$, 
but without inverse powers of $a$. 
So as $a\rightarrow 0$, 
the LO $L=0$ operator makes an unsuppressed contribution to the $L=0$ matrix
element, while the contributions to this matrix element 
from the NLO rotational-symmetry violating $L=0$ and $L=4$ operators 
are suppressed by $1/N^{2}$.

A simple argument shows that contributions from integration region II, for
which  $\pi/a\leq\left|\mathbf{k}\right|\leq\sqrt{3}\pi/a$, 
are also
suppressed by $1/N^{2}$. 
After defining a new momentum variable 
$l_{\mu}=k_{\mu}a$ and $l^{2}=l_{1}^{2}+l_{2}^{2}+l_{3}^{2}$, the
$\mathbf{p}=0$ term of the Poisson sum in region II is
\begin{eqnarray}
J_{LM}^{\left(II\right)}\left(\mathbf{p}=0\right)
&  = & 
\frac{3\lambda}{16\pi^4}i^{L}
\int_{-\pi}^{\pi}dl_{4}\int_{\pi}^{\sqrt{3}\pi}dl\ l^{2}\ 
\int_{f\left(\Omega_{\mathbf{l}}\right)}d\Omega_{\mathbf{l}}
\nonumber\\
&&
\qquad\qquad
\frac{Y_{LM}\left(\Omega_{\mathbf{l}}\right)}{\left(4\sum_{\mu}\sin^{2}\left(l_{\mu}/2\right)
+a^{2}m^{2}\right)^{2}}
\int_{0}^{1}dy\ y^{2}\ j_{L}\left(Nly\right)
,
\label{eq:20}
\end{eqnarray}
where $f\left(\Omega_{\mathbf{l}}\right)$ identifies the angular
region of integration, and whose parametric form does not matter for
this discussion. This region still exhibits cubic symmetry, and gives
rise to contribution to the $L=0,4,6,8,...$ operators. On the other
hand, the three-momentum integration is entirely located in the UV
as $a\rightarrow0$, and thus
\begin{equation}
\sin^{2}\left(l_{1}/2\right)+\sin^{2}\left(l_{2}/2\right)+\sin^{2}\left(l_{3}/2\right)
+\sin^{2}\left(l_{4}/2\right)\geq 1
.
\label{eq:21}
\end{equation}
Also, integration over the Bessel function brings in a factor of 
$\ -\cos\left(Nl\right)/(N^{2}l^{2})$, up to higher orders in $1/N$. 
So the integrand does not have any
singularities  in region II of the 
integration, and is bounded.  
As a result,
\begin{equation}
\left|J_{LM}^{\left(II\right)}\left(\mathbf{p}=0\right)\right|
\ \leq\ 
\frac{1}{N^{2}}
\frac{3\lambda}{(4\pi)^4}
\ \int_{-\pi}^{\pi}dl_{4}
\ \int_{\pi}^{\sqrt{3}\pi}dl
\ \int_{f\left(\Omega_{\mathbf{l}}\right)}d\Omega_{\mathbf{l}}
\ Y_{LM}\left(\Omega_{\mathbf{l}}\right)
,
\label{eq:22}
\end{equation}
and consequently 
$J_{LM}^{\left(II\right)}\left(\mathbf{p}=0\right)$ 
itself is suppressed by $1/N^{2}$. 
This completes the discussion of the $\mathbf{p}=0$
term in the Poisson sum,
corresponding to a zero-momentum insertion of the continuum operator into the
loop diagram.
It then remains to determine the scaling of the  $\mathbf{p}\neq 0$ terms in the summation
in the large $N$ limit. 
The integral arising from the  $\mathbf{p}\neq 0$ terms
is, up to  numerical factors,
\begin{eqnarray}
{\cal I}_{\mathbf{p}\neq 0}
& \sim & 
\lambda\sum_{\mathbf{p}\neq 0}\int_{-\frac{\pi}{a}}^{\frac{\pi}{a}}\frac{d^{4}k}{\left(2\pi\right)^{4}}
\frac{1}{\left(\hat{\mathbf{k}}^{2}+m^{2}\right)^{2}}\ 
Y_{L_{1}M_{1}}\left(\Omega_{\hat{\mathbf{k}}}\right)\ 
Y_{L_{2}M_{2}}\left(\Omega_{\hat{\mathbf{p}}}\right)
\nonumber\\ 
&& \qquad\qquad \qquad \qquad
\times\ 
\int_{0}^{1}dy\ y^{2}\ 
j_{L_{1}}\left(Na\left|\mathbf{k}\right|y\right)\ 
j_{L_{2}}\left(2\pi  N\left|\mathbf{p}\right|y\right)
.
\label{eq:23}
\end{eqnarray}
This integral is finite in UV, and  integrand can be expanded in powers of $a$,
giving  a leading contribution of 
\begin{eqnarray}
{\cal I}_{\mathbf{p}\neq 0}
& \sim & 
\lambda\sum_{\mathbf{p}\neq 0}\int_{-\frac{\pi}{\Lambda a}}^{\frac{\pi}{\Lambda
    a}}\frac{d^{3}q\ dq_{4}}{\left(2\pi\right)^{4}}\frac{1}{\left(q^{2}+q_{4}^{2}+m^{2}/\Lambda^{2}\right)^{2}}
\ Y_{L_{1}M_{1}}\left(\Omega_{\hat{\mathbf{q}}}\right)
\ Y_{L_{2}M_{2}}\left(\Omega_{\hat{\mathbf{p}}}\right)
\nonumber\\ 
&& \qquad \qquad \qquad \qquad \qquad \qquad
\times\ 
\int_{0}^{1}dy\ y^{2}\ 
j_{L_{1}}\left(qy\right)
\ j_{L_{2}}\left(2\pi N\left|\mathbf{p}\right|y\right)
.
\label{eq:24}
\end{eqnarray}
A non-zero angular integration requires that $L_1=0$,
and the integral 
is suppressed at least by a factor of $1/N^{2}$ as
integration over the Bessel functions introduces a factor 
of $1/\left(2\pi N\left|\mathbf{p}\right|\right)^{2}$
up to a numerical coefficient and a bounded trigonometric function
at leading order in $1/N$. 
The next order term in the small $a$ expansion of the integrand
can be easily shown to bring in an additional  factor of $1/N^{2}$.
So one can see that the $\mathbf{p}\neq 0$ terms in the Poisson summation,
which give rise to non-continuum contributions to the two-point function
at one loop, are always suppressed by at least a factor of $1/N^{2}$.

The result of the one-loop calculation is promising: all the sub-leading
contributions that break rotational symmetry
are suppressed by $1/N^{2}$
compared to the leading $L=0$ continuum operator contribution to
the two-point function. 
A little investigation shows that this scaling
also holds to higher orders in $\lambda\phi^{4}$ theory. Suppose
that the operator is inserted into a  propagator inside an n-loop diagram
contributing to the two-point function. 
Considering the 
continuum part of the operator first, the leading term in the small $a$
expansion of the integrand gives rise to $2n$ propagators,
while the integration measure contributes $4n$ powers of momentum.
Although this appears to be logarithmically divergent, 
the spherical Bessel function contributes a factor of inverse three-momentum 
and either a sine or cosine of the three-momentum, rendering the diagram finite.
The same argument applies to the NLO term in the small $a$ expansion of
the integrand, resulting  in a $1/N^{2}$ suppression of the 
breaking of rotational invariance.
Insertion of the non-continuum operator in loop diagrams are also suppressed by
$1/N^2$ for similar reasons.

The interpretation of finite-size scaling results presented in 
Refs.~\cite{CampostriniI,CampostriniII}
in terms of what has been observed in this section is now straightforward.
Near the critical point, the correlation length is the only relevant
physical scale in the problem, and tends to infinity. 
So as the critical point is approached, 
one does not probe the underlying lattice structure
as the correlation length becomes much larger than the lattice spacing, 
and extends over an
increasing number of point shells. 
In comparison,
inserting an operator which only probes distances of the order of
a physical scale that is much larger than the lattice spacing, 
resembles the physics near a rotational-invariant fixed point, and the same scaling
law for the non-rotational invariant operators is expected (in the
same theory) as the lattice spacing goes to zero.

\section{Operators in QCD
\label{sec:QCD}}
\noindent 
The necessity of introducing a gauge link to connect the fermionic
fields in a gauge-invariant way, makes the discussion of the operator
and its renormalization more involved in gauge theories. 
The reason
is two-folded: firstly as is well known, perturbative
LQCD is ill-behaved as a result of non-vanishing tadpoles which diverge
in the  UV, making the small coupling series expansion of the operators
slowly convergent. The other difficulty is that as the operator is
smeared over many lattice sites, the links are necessarily extended links.
Thus, to analytically investigate the deviations
from a rotational invariant path, working with a well-defined
path on the grid is crucial. In this section, the strategies to deal
with these problems are discussed, and the scaling laws of different
operator contributions to the two-point function in 
QCD with an insertion of the smeared operator are deduced.

In position space, perhaps the simplest gauge-invariant smeared operator of
quark bilinears is
\begin{equation}
\hat{\theta}_{L,M}\left(\mathbf{x};a,N\right)
\ =\ 
\frac{3}{4\pi N^{3}}\sum_{\mathbf{n}}^{\left|\mathbf{n}\right|\leq
  N}\overline{\psi}
\left(\mathbf{x}\right)U\left(\mathbf{x},\mathbf{x}+\mathbf{n}a\right)\psi\left(\mathbf{x}+\mathbf{n}a\right)
\ Y_{L,M}\left(\hat{\mathbf{n}}\right)
,
\label{eq:25}
\end{equation}
with
\begin{equation}
U\left(\mathbf{x},\mathbf{x}+\mathbf{n}a\right)
\ =\ 
e^{ig\int_{\mathbf{x}}^{\mathbf{x}+\mathbf{n}a}\mathbf{A}\left(z\right)\cdot d\mathbf{z}}
\ =\ 
1+ig\int_{\mathbf{x}}^{\mathbf{x}+\mathbf{n}a}\mathbf{A}\left(z\right)\cdot d\mathbf{z}
+\mathcal{O}\left(g^{2}\right)
,
\label{eq:26}
\end{equation}
where the actual path defining $U$ will be considered subsequently.
As the fermion operator is a spin singlet, $S=0$, 
the total angular momentum of this operator in the continuum is $J=L$.
One could also consider operators of the form
\begin{equation}
\hat{\theta}_{JL,M}^{\mu}\left(\mathbf{x};a,N\right)
\ =\ 
\frac{3}{4\pi N^{3}}
\sum_{\mathbf{n}}^{\left|\mathbf{n}\right|\leq   N}
\overline{\psi}
\left(\mathbf{x}\right)\ \gamma^\mu \ 
U\left(\mathbf{x},\mathbf{x}+\mathbf{n}a\right)
\psi\left(\mathbf{x}+\mathbf{n}a\right)
\ Y_{L,M}\left(\hat{\mathbf{n}}\right)
,
\label{eq:25b}
\end{equation}
which can be used to form operators with $J=L+1, L, L-1$.  
It is clear that
the set of operators with angular momentum $J$ will mix under renormalization,
but the vector nature of QCD precludes mixing between the
$\overline{\psi}\psi$ and $\overline{\psi}\gamma^\mu \psi$ operators in the
chiral limit.
However to capture the main features
of operator mixing in the continuum limit of LQCD, it suffices to
work with the simplest operator, in Eq.~(\ref{eq:25}). At tree-level, the contributions of this operator away from the continuum limit
scale in the same way as in the scalar theory, 
with contributions that violate rotational invariance 
suppressed by $\sim 1/N^2$.

\begin{figure}[h!]
\begin{centering}
\includegraphics[scale=0.115]{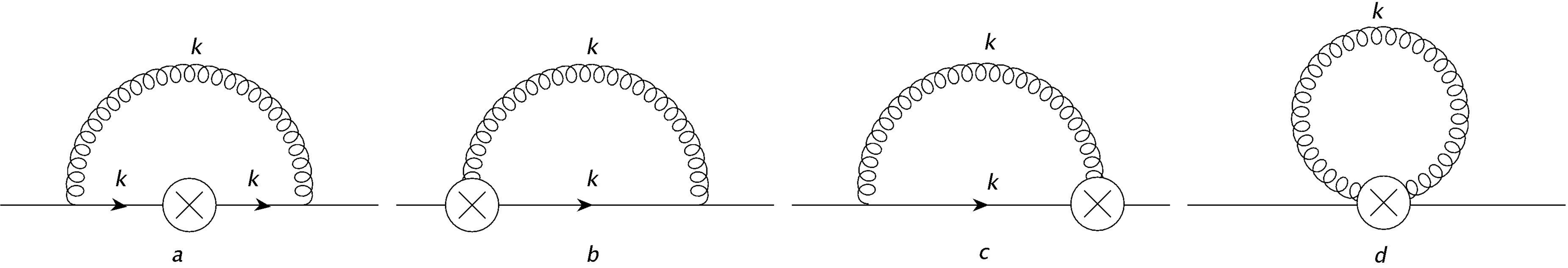}
\par\end{centering}
\caption{{\small 
One-loop QCD corrections to the fermionic two-point function
with an insertion of 
$\hat{\theta}_{L,M}$, given in Eq.~(\ref{eq:25}),
at zero external momentum}}
\label{fig:qcdOneLoop}
\end{figure}
%

\subsection{Continuum operator and its renormalization}
Let us first discuss the one-loop renormalization of the operator
in the continuum. There are four one-loop diagrams contributing to
the operator renormalization as shown in Fig.~\ref{fig:qcdOneLoop}. 
The diagram in
Fig.~\ref{fig:qcdOneLoop}(a) 
results from inserting the leading order term in the small
coupling expansion of the operator in the loop. At zero external momentum
this diagram is
\begin{eqnarray}
\Gamma^{(a)} & \sim & 
-T^{a}T^{a}\frac{3ig^{2}}{4\pi}\int_{0}^{1}dyy^{2}
\int d\Omega_{\mathbf{y}}\int\frac{d^{4}k}{\left(2\pi\right)^{4}}
\frac{\gamma_\alpha
\left(ik_{\mu}\gamma^{\mu}+m\right)^{2}
\gamma^\alpha
}{\left(k^{2}+m^{2}\right)^{2}k^{2}}
e^{iNa\mathbf{k}\cdot\mathbf{y}}Y_{LM}\left(\Omega_{\mathbf{y}}\right)
,
\nonumber\\
\label{eq:27}
\end{eqnarray}
which is clearly convergent in the UV. 
Also it contains $L=0$
as well as $L=1$ operator as can be seen from the angular
part of the integral
\begin{eqnarray}
&&
\sum_{L^{\prime},M^{\prime}}\int d\Omega_{\mathbf{y}}d\Omega_{\mathbf{k}}
\left[f_{1}\left(k^{2},m,k_{4}\right)+f_{2}\left(k^{2},m,k_{4}\right)\mathbf{k}\cdot\vec{\mathbf{\gamma}}\right]
\ Y_{L^{\prime}M^{\prime}}\left(\Omega_{\mathbf{k}}\right)
\ Y^{*}_{L^{\prime}M^{\prime}}\left(\Omega_{\mathbf{y}}\right)
\ Y_{LM}\left(\Omega_{\mathbf{y}}\right)
\nonumber\\
&&
\ =\ 
\sqrt{4\pi}f_{1}\left(k^{2},k_{4},m\right)\delta_{L,0}\delta_{M,0}+
\nonumber\\
&& \qquad
\sqrt{\frac{4\pi}{3}}f_{2}\left(k^{2},k_{4},m\right)\left|\mathbf{k}\right|
\delta_{L,1}\left[\gamma_{1}\left(\frac{\delta_{M,-1}-\delta_{M,1}}{\sqrt{2}}\right)
+i\gamma_{2}\left(\frac{\delta_{M,-1}+\delta_{M,1}}{\sqrt{2}}\right)+\gamma_{3}\delta_{M,0}
\right]
,
\nonumber\\
&&
\label{eq:28}
\end{eqnarray}
where $f_{1}$ and $f_{2}$ are some functions of their arguments.
One can check however that as $m/\Lambda\rightarrow0$ (the chiral
limit), the contribution to the $L=1$ operator is suppressed by the
quark mass.

The diagrams in Fig.~\ref{fig:qcdOneLoop}(b) comes from the next term in the expansion
of Eq. (\ref{eq:26}). It is straightforward to show that the Feynman
rule for the one-gluon vertex with zero momentum insertion into the
operator is
\begin{eqnarray}
V_g^\lambda = 
\frac{3}{4\pi N^{3}}\sum_{\mathbf{n}}^{\left|\mathbf{n}\right|\leq  N}
 g a n^\lambda
\frac{1}{\left(\mathbf{p}-\mathbf{p^{\prime}}\right)\cdot\mathbf{n}a}
\left(e^{i\left(\mathbf{k}+\mathbf{p^{\prime}}\right)\cdot \mathbf{n}a}
-e^{i\mathbf{p^{\prime}\cdot}\mathbf{n}a}\right)\delta^{4}\left(p-p^{\prime}-k\right)
Y_{L,M}\left(\hat{\mathbf{n}}\right),
\label{eq:29}
\end{eqnarray}
where the radial path between points $\mathbf{x}$ and $\mathbf{x}+\mathbf{n}a$
is taken in evaluating the link integral, $p$ and $p^{\prime}$ are
the momenta of incoming and outgoing fermions respectively, 
$\lambda$ is the Lorentz-index of the gluon field,
and $k$
is the momentum of the gluon coming out of the vertex. Note that in
principle, any path between points $\mathbf{x}$ and $\mathbf{x}+\mathbf{n}a$
can be taken in the above calculation, but if one is interested in
deviations of the renormalized lattice operator from the rotational
invariance compared to the continuum operator, a path between two
points should be chosen in the continuum in such a way that it respects
rotational invariance explicitly. 
Any path other than the radial path, on the other hand, is equivalent to
infinite many other paths resulting
from rotated versions of the original path around the radial path.
To reveal rotational invariance at the level of the
continuum operator, an averaging over these infinite copies of the
path is needed, and this makes the calculation of the link more involved.

Now at zero external momentum, using expression (\ref{eq:29}) with
$p=0$, the contribution from the second and third diagrams in
Fig.~\ref{fig:qcdOneLoop}b is
\begin{eqnarray}
\Gamma^{(b,c)}
& \sim & 
- T^{a}T^{a}
\frac{3 g^2}{2\pi}\int_{0}^{1}dyy^{2}\int d\Omega_{\mathbf{y}}
\int\frac{d^{4}k}{\left(2\pi\right)^{4}}
\frac{
i {\bf k}\cdot {\bf y} 
+m \mathbf{y}\cdot\vec{\gamma}
}{\left(k^{2}+m^{2}\right)k^{2}}
\nonumber\\
&&
\qquad \qquad
\qquad \qquad
\times\frac{1}{\mathbf{k}.\mathbf{y}}\left(e^{iNa\mathbf{k}\cdot\mathbf{y}}-1\right)
\ Y_{LM}\left(\Omega_{\mathbf{y}}\right)
.
\label{eq:30}
\end{eqnarray}
As is evident, because of a non-oscillatory contribution to the operator,
there is a logarithmically divergent piece from the above integration
contributing to the $L=0$ operator, which along with the logarithmic
divergent  contribution  
from wavefunction renormalization, contributes
to the anomalous dimension of the operator. Also the angular integration
of the above expression: 
\begin{align}
&\int d\Omega_{\mathbf{y}}d\Omega_{\mathbf{k}}
\left[1+\frac{\mathbf{y}\cdot\vec{\gamma}}{i\mathbf{k}\cdot\mathbf{y}} m \right]
\left(e^{iNa\mathbf{k}\cdot\mathbf{y}}-1\right)
\ Y_{LM}\left(\Omega_{\mathbf{y}}\right)&
\nonumber\\
&
\qquad \qquad
=\int d\Omega_{\mathbf{y}}
\left[g_{1}\left(Nay\left|\mathbf{k}\right|\right)
+g_{2}\left(Nay\left|\mathbf{k}\right|\right) m \mathbf{y}\cdot\vec{\gamma}\right]
\ Y_{LM}\left(\Omega_{\mathbf{y}}\right)
,
\label{eq:31}
\end{align}
indicates that as before, in addition to $L=0$ operator, an $L=1$
contribution is present which is finite at UV, and can be shown to
vanish for $m/\Lambda\rightarrow0$. $g_{1}$ and $g_{2}$ are some
functions of their arguments whose explicit forms do not matter for
this discussion.

\begin{figure}[h!]
\begin{centering}
\includegraphics[scale=0.15]{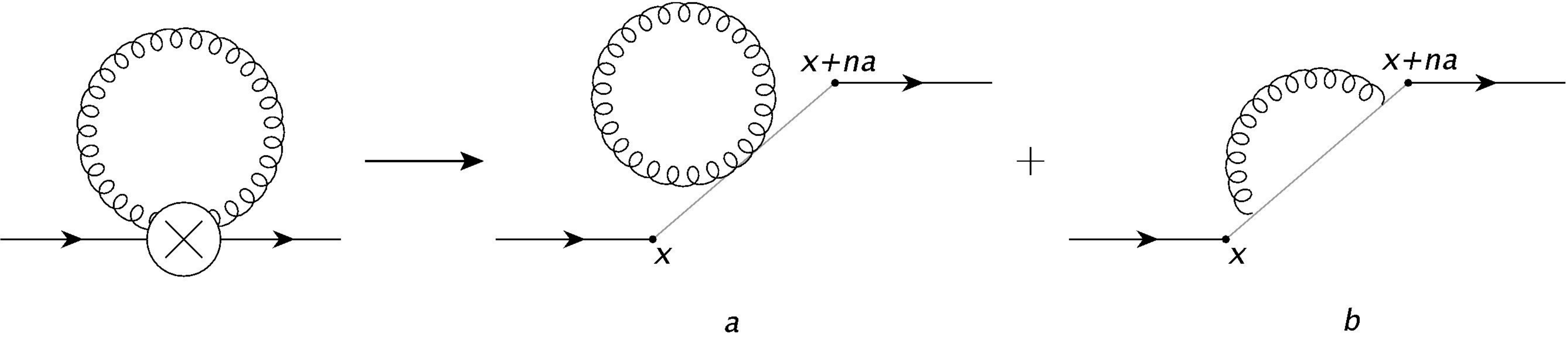}
\par\end{centering}
\caption{{\small The tadpole contribution consists of the conventional tadpole
diagram (a), which vanishes when using a mass-independent regulator in the
continuum (such as dimensional  regularization), 
as well as the diagram
shown in (b) which is of the order of $\alpha_{s}/\left|\mathbf{\Delta x}\right|^{2}$,
where $\mathbf{\Delta x}$ is the distance between two gluon vertices.}}
\label{fig:tadpoles}
\end{figure}
The last diagram in Fig.~\ref{fig:qcdOneLoop} corresponds to the $\mathcal{O}\left(g^{2}\right)$
term in the small coupling expansion of the gauge link. 
It contains
the tadpole of the continuum theory whose value depends in general
on the regularization scheme.
For example, by using a hard momentum cutoff which is matched easily
with the lattice regularization, it diverges quadratically. However,
it is not hard to see that in dimensional regularization which respects
the full rotational symmetry of the continuum, it vanishes in $d=4$,
therefore it does not contribute to the renormalization of the continuum
operator. 
But the fourth diagram in Fig.~\ref{fig:qcdOneLoop} does not only include the
conventional tadpoles, Fig.~\ref{fig:tadpoles}(a), it also contains the diagram where
a gluon is emitted by the Wilson line inside the operator and then
absorbed at another point on the Wilson line, Fig.~\ref{fig:tadpoles}(b) as a consequence
of the matter fields being separated by a distance $\mathbf{n}a$.
It is straightforward to show this diagram is convergent, and scales
by $\alpha_{s}/\left|\mathbf{\Delta x}\right|^{2}$ where $\mathbf{\Delta x}$
is the distance between two gluon vertices and $\alpha_{s}$ is evaluated
at the energy scale of the order of $1/\left|\mathbf{\Delta x}\right|$.
This completes the qualitative discussion of the operator renormalization
and mixing at one-loop order in the continuum.

\subsection{Lattice operator and its renormalization}
Let us start the discussion of the lattice operator by assuming that its
definition is still given by Eq.~(\ref{eq:25}). However, this can
be shown to be a naive definition of the operator on the lattice.
The reason is implicit in the discussion of tadpoles given above.
Although  tadpoles are absent from the operator renormalization
in the continuum,
on the lattice, they are non-vanishing, and result in large
renormalizations, as can be seen  in perturbative lattice QCD calculations. 
As was suggested
long ago by Lepage and Mackenzie \cite{Lepage}, to make the perturbative
expansion of the lattice quantities well-behaved, and to define an
appropriate connection between the lattice operators and their continuum
counterparts, one can remove tadpoles from the expansion of the lattice
operators in a non-perturbative manner by dividing the gauge link
by its expectation value in a smooth gauge,
\begin{equation}
U\left(x,x+a\hat{\mu}\right)
\rightarrow
\frac{1}{u_{0}}U\left(x,x+a\hat{\mu}\right)
,
\label{eq:32}
\end{equation}
where a simpler, gauge invariant choice of $u_{0}$ uses the measured
value of the plaquette in the simulation,
 $u_{0}\equiv\left\langle \frac{1}{3}{\rm Tr}\left(U_{plaq}\right)\right\rangle
 ^{1/4}$.
There remains still another issue regarding the tadpole contributions
to the smeared operator which is not fully taken care of by the simple
single-link improvement procedure explained above. The operator introduced in
Eq.~(\ref{eq:25}) is smeared over several lattice sites, and as a
result includes extended links. As will be explained shortly, in
spite of $\mathcal{O}\left(\alpha_{s}\right)$ corrections due to
tadpoles from a single link, 
there is an $\mathcal{O}\left(N\alpha_{s}\right)$
enhancement due to the tadpoles from the extended link with length $\sim Na$. 
So although a non-perturbative tadpole improvement
could introduce non-negligible  statistical errors, this improvement is crucial,
otherwise the relation between the lattice smeared operator and
the corresponding continuum operator is somewhat obscure.
\begin{figure}[t]
\begin{centering}
\includegraphics[scale=0.15]{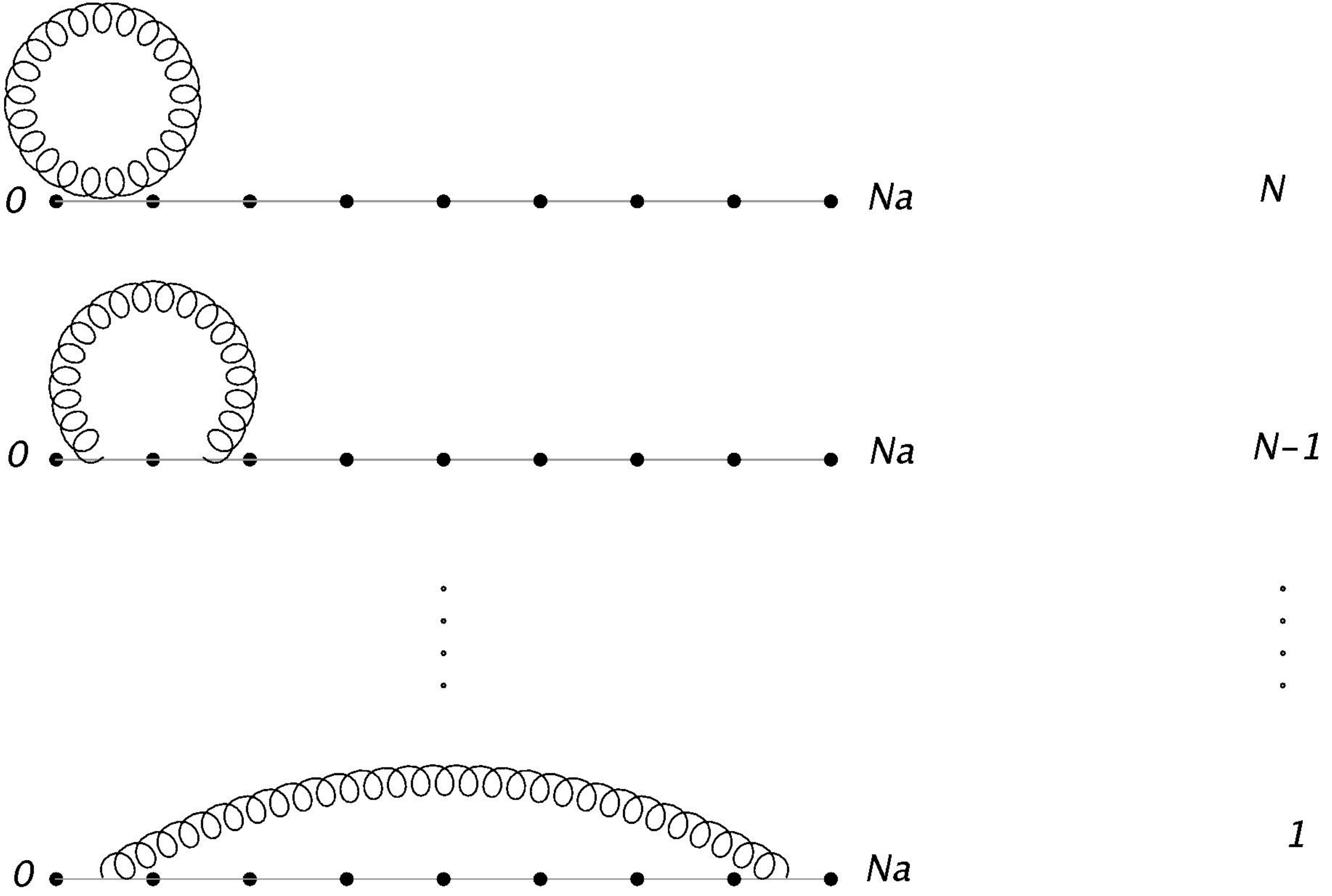}
\par\end{centering}
\caption{{\small Tadpole diagrams contributing to the smeared operator at one-loop
order. Shown in the right are the number of diagrams of each type.}}
\label{fig:tadpoleextended}
\end{figure}

The reason for the $\mathcal{O}\left(N\alpha_{s}\right)$ enhancement 
of tadpoles from the extended links
can be illustrated by working out a particular example. Suppose that
the link is extended between points $\mathbf{x}$ and $\mathbf{x}+Na\hat{e}_{1}$
entirely along the $1$ axis. 
Then in order to make a tadpole, not
only can each gauge field be contracted with the other gauge field belonging to
the same elementary link, but also it can be contracted with a gauge field
from one of the remaining $N-1$ elementary links 
(see Fig.~\ref{fig:tadpoleextended}). 
Note that each diagram in Fig.~\ref{fig:tadpoleextended} 
comes with a multiplicity of $N-m$, where $m$ is the number of
links between the contracted gluonic vertices.
At LO  in $a$, the corresponding contribution from the extended tadpole (ET) is of the form
\begin{equation}
\Gamma^{(ET)}
\sim
 \alpha_{s}a^{2}\int_{-\frac{\pi}{a}}^{\frac{\pi}{a}}d^{4}k
\ \frac{e^{imak_{1}}}{k_{1}^{2}+k_{2}^{2}+k_{3}^{2}+k_{4}^{2}}
\ \sim\ 
\frac{\alpha_{s}}{m^{2}}
,
\label{eq:33}
\end{equation}
from which
the contribution from all the diagrams in Fig.~\ref{fig:tadpoleextended} can be obtained,
\begin{equation}
\sum_{m=1}^{N-1}(N-m)\frac{\alpha_{s}}{m^{2}}
\ =\ 
\mathcal{O}\left(N\alpha_{s}\right)
.
\label{eq:34}
\end{equation}
Note that the $m=0$ term, corresponding to the first diagram in
fig~\ref{fig:tadpoleextended}, 
has been excluded from the above sum 
as it is just the single link tadpole contribution.
Given that there are N single links, the total contribution from single link
tadpoles is  $\mathcal{O}\left(N\alpha_{s}\right)$ as well.

Another issue with the extended links is the fact that without
tadpole improvement, breakdown of rotational symmetry occurs
at $\mathcal{O}\left(N\alpha_{s}\right)$. 
The reason is that without tadpole improvement of the extended links, 
contributions from the different 
$A_{1}$ irreps in a given point shell are normalized differently.
For example, there are more tadpole diagrams at 
$\mathcal{O}\left(g^{2}\right)$ contributing to an extended 
link between points $\left(0,0,0\right)$ and $\left(2,2,1\right)$ (six single links)
than to an extended link between points $\left(0,0,0\right)$ and
$\left(3,0,0\right)$
(three single links)
although both points belong to the same point shell  
(i.e. have the same separation in position space). 
This fact magnifies the necessity of tadpole improvement as well as providing  
a prescription for an appropriate improvement of an extended link. 
As the
expectation value of a link belonging to a given $A_{1}$ irrep in
a given shell is in general different from the expectation value of
the link belonging to another $A_{1}$ irrep in the same shell, one
needs to redefine the link in a given irrep by dividing it by its
expectation value in the same irrep,
\begin{equation}
U_{A_{1}^{i}}\left(x,x+a\mathbf{n}\right)
\ \rightarrow
\ \frac{1}{u_{A_{1}^{i}}}U_{A_{1}^{i}}\left(x,x+a\mathbf{n}\right)
,
\label{eq:35}
\end{equation}
where $u_{A_{1}^{i}}=\left\langle U_{A_{1}^{i}}\left(x,x+a\mathbf{n}\right)\right\rangle $,
and the $A_{1}^{i}$'s are different $A_{1}$ irreps belonging to the
$n^{2}$-shell. 
With this prescription for tadpole improvement of the extended links, the
renormalized operator is 
assured
to be safe from large rotational invariance breaking effects of the
order of $\mathcal{O}\left(N\alpha_{s}\right)$. 
With this new definition
of the gauge link, Eq. (\ref{eq:25}) is now a well-defined lattice
operator with an appropriate continuum limit which can be used
in our subsequent analysis.

As the cancellation of the tadpole diagram is assured by the new definition
of the operator, there are only  three one-loop diagrams  that contribute
to the renormalization of the lattice operator. 
The first
diagram in Fig.~\ref{fig:qcdOneLoop} corresponds to the following loop integral at zero
external momentum for Wilson fermions,
\begin{eqnarray}
\Gamma^{(a)}
& \sim & 
\left(ig\right)^{2}T^{a}T^{a}\frac{3}{4\pi N^{3}}\sum_{\mathbf{n}}
 \int_{-\frac{\pi}{a}}^{\frac{\pi}{a}}\frac{d^{4}k}{\left(2\pi\right)^{4}}
 e^{i\mathbf{k}\cdot\mathbf{n}a}
 \left[\gamma_{\rho}\cos\left(\frac{k_{\rho}a}{2}\right)-ir\sin\left(\frac{k_{\rho}a}{2}\right)\right]
\nonumber\\
&&
\times
\ \left(\frac{-i\sum_{\mu}\gamma_{\mu}\frac{\sin\left(k_{\mu}a\right)}{a}+M\left(k\right)}
{\sum_{\mu}\frac{\sin^{2}\left(k_{\mu}a\right)}{a^{2}}+M\left(k\right)^{2}}\right)^{2}
\left[\gamma^{\rho}\cos\left(\frac{k^{\rho}a}{2}\right)-ir\sin\left(\frac{k^{\rho}a}{2}\right)\right]
\nonumber\\
&&
\times\ 
\frac{i}{\frac{4}{a^{2}}\sum_{\nu}\sin^{2}\left(\frac{k_{\nu}a}{2}\right)}
\ Y_{LM}\left(\Omega_{\mathbf{n}}\right)
 , 
\label{eq:36}
\end{eqnarray}
where $M\left(k\right)\equiv M+2r/a\sum\limits_{\mu}\sin^{2}\left(k_{\mu}a/2\right)$,
and $r$ is the Wilson parameter. Clearly at LO in
the lattice spacing, one recovers the corresponding diagram with the
insertion of the continuum operator, Eq.~(\ref{eq:27}), and 
so it contributes to both the $L=0$ and $L=1$ operators. 
Note that although the integration region
is not rotationally symmetric like the continuum integral, the convergence
of integral at UV ensures that the contributions from non-rotationally
symmetric integration region II, defined in section~\ref{sec:Scalar},
are suppressed by additional  powers of $1/N$ compared to the rotational
invariant region I:
\begin{eqnarray}
\delta\Gamma^{(a)}
& \sim & 
-ig^{2}T^{a}T^{a}\frac{3
    i^{L}}{16\pi^{4}}
\int_{-\pi}^{\pi}dl_{4}\int_{\pi}^{\sqrt{3}\pi}dl\ l^{2}\ 
\int_{f\left(\Omega_{\mathbf{l}}\right)}d\Omega_{\mathbf{l}}\ 
\frac{\left(il_{\mu}\gamma^{\mu}+ma\right)^{2}}{\left(l^{2}+m^{2}a^{2}\right)^{2}l^{2}}
\nonumber\\
&&
\qquad
\qquad
\times\ Y_{LM}\left(\Omega_{\mathbf{l}}\right)\left[\int_{0}^{1}dyy^{2}j_{L}\left(Nly\right)\right]
,
\label{eq:37}
\end{eqnarray}
where: $l_{\mu}=k_{\mu}a$ and $l^{2}=l_{1}^{2}+l_{2}^{2}+l_{3}^{2}$.
The integrand is clearly convergent, and the integration region is
entirely in the UV, and  so the only dependence on $a=1/(\Lambda N)$ comes from
the integration over the Bessel function, giving a LO contribution 
proportional to $1/N^{2}$. 
However,
the first sub-leading contribution from this
diagram scales as  $\sim \alpha_{s}/N$ for Wilson fermions instead
of $\sim \alpha_{s}/N^{2}$. 
The reason is that the small $a$
expansion of the integrand in Eq.~(\ref{eq:36}) includes terms at
$\mathcal{O}\left(a\right)$ which is proportional to the Wilson
parameter. 
The integrand
scales as $\sim 1/k^{3}$ multiplied by the spherical Bessel function
in the UV which still gives rise
to a convergent four-momentum integration for any value of $L$,
\begin{equation}
\delta\Gamma^{(a,r)}
\sim a\int d^{4}k\frac{1}{k^{3}}\left[\int_{0}^{1}dy\ y^{2}
\ j_{L}\left(Naky\right)\right]\sim a\Lambda=\frac{1}{N}
.
\label{eq:38}
\end{equation}
These contributions are rotational invariant, and will be included in the
renormalization 
$Z$-factor of the operator when matching the lattice operator with its
continuum counterpart. 
Further,
the integrals that appear at $\mathcal{O}\left(a^{2}\right)$
in an expansion of Eq.~(\ref{eq:36}) are also convergent,
and the terms containing rotational invariance breaking
contributions are suppressed by $1/N^{2}$. This completes discussion
of the first one-loop diagram of Fig.~\ref{fig:qcdOneLoop}.

The second diagram contains the one-gluon vertex operator, and requires
evaluating a line  integral over the path on the grid defining the extended
link. 
As was pointed
out in the discussion of the path in the continuum, in general any
path can be chosen in evaluating the operator both in the continuum
or on the lattice, but requiring the recovery of rotational symmetry
at the level of the operator means that the extended link has to exhibit
rotational symmetry in the continuum limit. 
As already discussed, the simplest rotational
invariant path in the continuum is the radial path between the points,
so it makes sense to try to construct a path on the grid which remains
as close as possible to the radial path between points $x$ and $x+\mathbf{n}a$. One might expect though that
choosing a path in continuum which is the same as its lattice counterpart
is a more legitimate choice. One example of such a path is an $L$-shaped
path. However, it is not hard to verify that the $L$-shaped link does not
restore rotational invariance in the continuum limit as 
the continuum path explicitly breaks rotational symmetry.
So the problem of evaluating the one-gluon vertex of the smeared operator
is reduced to finding the closest path to the straight line on the
grid. 
In a lattice calculation, one can, in principle, construct an algorithm
which finds a path on the three-dimensional grid in such a way that
the area between the path and the rotational invariant radial path
is a minimum. 
One such algorithm has already been used in Ref.~\cite{Meyer}
to construct a path that follows the straight line between sites A and B as
closely as possible, by forming a diagonal link at each step which has
the maximum projection onto the vector $\overrightarrow{AB}$. 
By
this construction of ``super links'', the authors have been able
to form arbitrary (approximate) rotations of the Wilson loops, therefore
constructing glueball operators which project onto a definite spin $J$
in the continuum limit. 
However, 
the analytic form of the super link has not been given.
In appendix \ref{app:linksongrid}, a method to evaluate the link on such a path
is illustrated with a small number of examples. 
For the following discussion
however, a particular example has been considered which encapsulates
the essential features of the recovery of the rotational path, and
gives us an idea how to deal with the general case.

\begin{figure}
\begin{centering}
\includegraphics[scale=0.16]{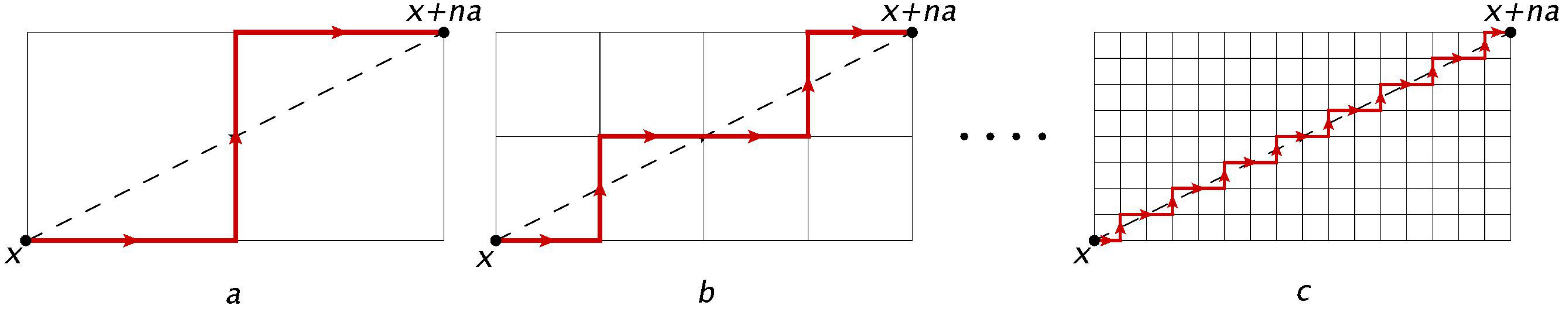}
\par\end{centering}
\caption{{\small a) The link between points $x$ and $x+\mathbf{n}a$ for $\mathbf{n}=\left(2,1,0\right)$
which remains as close as possible to the diagonal link, b) The link
between the same points for $\mathbf{n}=2\left(2,1,0\right)$ which
consists of two separate links of part a) with the lattice spacing
being halved, c) The link for $\mathbf{n}=2^{K}\left(2,1,0\right)$
which consists of $2^{K}$ separate links of part a) with the  lattice
spacing divided by $2^{K}$.}}
\label{fig:links}
\end{figure}
Suppose that the link connects points $x$ and $x+\mathbf{n}a$ on
a cubic lattice where
$\mathbf{n}=\frac{a_{0}}{a}\left(Q,1,0\right)$,
and $a_{0}=2^{K}a$. As usual $a$ denotes the lattice spacing, and
$Q$ is an arbitrary integer. The continuum limit is recovered when
the integer $K$ tends  to infinity for a finite value of $a_{0}$.
Then as is shown in appendix \ref{app:linksongrid}, for a path which is symmetric under
reflection about its midpoint and remains as close as possible to
the vector $\mathbf{n}a$ (see Fig.~\ref{fig:links}), the $\mathcal{O}\left(g\right)$
term in the momentum-space expansion of the link has the following
form
\begin{eqnarray}
U^{\left(1g\right)}\left(q\right)
 & = &  
ig\frac{a_{0}}{2^{K}}e^{i\mathbf{q}\cdot\mathbf{n}a/2}
\frac{\sin\left(\frac{\mathbf{q}\cdot\mathbf{n}a}{2}\right)}{
\sin\left(\frac{\mathbf{q}\cdot\mathbf{n}a}{2^{K+1}}\right)}
\left[A_{y}\left(q\right)
\right.
\nonumber\\
&&
\left.
\qquad\qquad
+2A_{x}\left(q\right)\frac{\sin\left(Qq_{x}a_{0}/2^{K+2}\right)}{
\sin\left(q_{x}a_{0}/2^{K+1}\right)}\cos\left(\frac{Qq_{x}a_{0}}{2^{K+2}}+\frac{q_{y}a_{0}}{2^{K+1}}\right)
\right]
.
\label{eq:39}
\end{eqnarray}
As $K\rightarrow\infty$ limit which corresponds to $a\rightarrow 0$, one
obtains
\begin{eqnarray}
U^{\left(1g\right)}\left(q\right)
 & = &  
2ige^{i\mathbf{q}\cdot\mathbf{n}a/2}\frac{\sin\left(
\frac{\mathbf{q}\cdot\mathbf{n}a}{2}\right)}{\mathbf{q}\cdot\mathbf{n}a}
\left[
\mathbf{A}\cdot\mathbf{n}a+\frac{a^{2}}{24}
\left(q_{x}Q+q_{y}\right)^{2}\mathbf{A}\cdot\mathbf{n}a
\right.
\nonumber\\
&& 
\left.
\qquad\qquad
-\frac{a^{2}}{24}QA_{x}a_{0}\left(q_{x}^{2}\left(Q^{2}-1\right)
+3Qq_{x}q_{y}+3q_{y}^{2}\right)
+\mathcal{O}\left(a^{4}\right)
\right]
 ,
\label{eq:40}
\end{eqnarray}
recovering the  continuum link, given in  Eq.~(\ref{eq:29}),
and contains broken rotational invariance contributions which are
suppressed by $\sim {\cal O}(a^{2})$. This
scaling has been shown in appendix \ref{app:linksongrid} to hold for vectors $\mathbf{n}$
of the forms: $\frac{a_{0}}{a}\left(Q,1,1\right)$, $\frac{a_{0}}{a}\left(Q,Q,1\right)$
and $\frac{a_{0}}{a}\left(Q,Q,Q\right)$ as well.

Let us now examine
how the insertion of this contribution from the operator  modifies the scaling
of the rotational invariance violating operators at one-loop.
The contribution from the second diagram in Fig.~\ref{fig:qcdOneLoop} with
the insertion of this vertex can be calculated order by order in small
$a$ by expanding the vertices and propagators as before. 
At the LO one gets
\begin{align}
&\Gamma^{(b)}
\sim 
-ig^{2}T^{a}T^{a}\frac{3}{4\pi
  N^{3}}\sum_{\mathbf{n}}\int_{-\frac{\pi}{a}}^{\frac{\pi}{a}}
\frac{d^{4}k}{\left(2\pi\right)^{4}}\frac{ik_{\mu}\gamma^{\mu}+m}{\left(k^{2}+m^{2}\right)k^{2}}
\frac{e^{i\mathbf{k}\cdot\mathbf{n}a}-1}{i\mathbf{k}\cdot\mathbf{n}a}
\ Y_{LM}\left(\Omega_{\mathbf{n}}\right)\times
\nonumber\\
& ~~
\left[a\mathbf{n}\cdot\vec{\gamma}
+\frac{a^{2}}{24}\left(k_{x}Q+k_{y}\right)^{2}a\mathbf{n}\cdot\vec{\gamma}
-\frac{a^{2}}{24}Q\left(k_{x}^{2}\left(Q^{2}-1\right)+3Qk_{x}k_{y}+3k_{y}^{2}\right)\gamma_{x}a_{0}\right]
.
\qquad
\label{eq:41}
\end{align}
Clearly, after adding the contribution from the third diagram in Fig.~\ref{fig:qcdOneLoop}, the LO contribution from the above expression,
the first term in the bracket of Eq.~(\ref{eq:41}),
 recovers
the results obtained previously for the insertion of the continuum
operator, up to suppressed contributions from the integration region
II, as discussed before. Therefore this term contributes to the $L=0$
operator with a logarithmically divergent coefficient, which along
with the wavefunction renormalization contributes to the anomalous
dimension of the lattice operator. Note that the wavefunction renormalization
gives rise to a logarithmically divergent contribution to the $L=0$
operator at LO  in the lattice spacing, recovering the continuum result,
and the sub-leading contributions are suppressed at least by $a=1/(N\Lambda)$
for Wilson fermions. This term also contains and $L=1$ operator which
is proportional to $m$, and vanishes in the chiral limit.

The second  term in the bracket of  Eq.~(\ref{eq:41}) 
is ${\cal O}(a^2)$, and can be written as
\begin{eqnarray}
\delta\Gamma^{(b,c),2}
& =  & \
-i{g^2 a^2\over 8\pi N^3}\ 
T^{a}T^{a}
\
\sum_{\mathbf{n}}
\int_{-\frac{\pi}{a}}^{\frac{\pi}{a}}\frac{d^{4}k}{\left(2\pi\right)^{4}}
\left[1+\frac{m}{i\mathbf{k}\cdot\mathbf{n}a}a\mathbf{n}\cdot\vec{\gamma}\right]
\frac{e^{i\mathbf{k}\cdot\mathbf{n}a}-1}{\left(k^{2}+m^{2}\right)k^{2}}
\nonumber\\
&&
\qquad
\qquad
\qquad
\qquad
\qquad
\qquad
\times\left(k_{x}Q+k_{y}\right)^{2}Y_{LM}\left(\Omega_{\mathbf{n}}\right)
\nonumber\\
& \sim & \mathcal{O}\left(g^{2}a^{0}\right)
.
\label{eq:42}
\end{eqnarray}
This scaling arises as a result of the UV divergence of the non-oscillatory
contribution to the integral and is entirely a UV effect.  For this term there
is no dependence upon ${\bf n}$ and as such the factor of $N^{-3}$ is canceled
by a corresponding $N^3$ from the sum.
Terms proportional
to the mass are convergent in the UV, and as such are suppressed by $a^{2}$
in the continuum limit.

The last term in the above expression Eq. (\ref{eq:41}) contains
rotational breaking contributions. 
It is multiplied by an explicit factor of 
$a^{2}$, 
but as seen in the previous term, 
the power divergence of the non-oscillatory part of the integral gives rise to 
an overall scaling of $\mathcal{O}\left(g^{2}\right)$. 
This completes the discussion
of the one-loop corrections to the lattice operator for the specific displacement
vector $\mathbf{n}a$ used above. 
It is also straightforward to check
the obtained scaling of different terms for other choices of the vector
$\mathbf{n}a$. 
In general, sub-leading contributions to the continuum link
are ${\cal O}(a^{2})$, and
so by dimensional analysis it 
has an associated factor of momentum squared. 
On the other hand, it always contains a non-oscillatory
term, and  as a result, the non-continuum
contributions  and the violations of rotational symmetry 
scale as $\mathcal{O}\left(\alpha_{s}\right)$.

Given the discussion of the previous paragraphs, we naively conclude that the
rotational symmetry breaking scales as $\sim {\cal O}(\alpha_s)$ in the continuum limit.
It is the one-gluon vertex associated with the smeared-operator that is
dominating this behavior, with the contributions from other diagrams
scaling as $\sim \alpha_s/N$ for Wilson fermions (Eq.~(\ref{eq:36}) and Eq.~(\ref{eq:37}))
and $\alpha_s/N^2$ from the other loop
diagrams compared with $\sim 1/N^2$ from the tree-level matching.
However, this scaling can be further improved by smearing the gauge-field.
The ${\cal O}(\alpha_s)$ contributions are due to the  explicit factor of $a^2$ being
compensated by a quadratic loop divergence, $\left(\pi/a\right)^2$, rendering a
suppression by only the coupling in the continuum limit, analogous to the
impact of tadpole diagrams. 
However, by smearing the gluon field over a volume of radius 
$1/\Lambda_g = a N_g$~\footnote{We have
distinguished the smearing radius of the operator, $N$, from  the smearing
radius 
of the gluons, $N_g$, but in principle they could be set equal.},
the offending diagrams in Fig.~\ref{fig:qcdOneLoop} scale as 
\begin{eqnarray}
\delta\Gamma^{(b,c),2,3}
& \sim  & \alpha_s\ a^2\ \Lambda_g^2
\ \sim\ {\alpha_s \over N_g^2}
,
\label{eq:42b}
\end{eqnarray}
due to the suppression of the high momentum modes in the gluon propagator.

The natural question to ask here is what is the scale of the coupling
in this process? Note that the bare coupling constant of lattice QCD
suffers from large renormalization as discussed before, so a better-behaved
weak coupling expansion of the lattice quantities uses a renormalized
coupling constant as the expansion parameter. As is suggested by Lepage
and Mackenzie \cite{Lepage}, one first fixes the renormalization
scheme by determining the renormalized coupling $\alpha_{s}^{ren}\left(k^{*}\right)$
from a physical quantity such as the heavy quark potential. Then the
scale of the coupling is set by the typical momentum of the gluon
in a given process. In the case considered above, the energy scale
of the strong coupling constant is dictated by the scale of the
gluon smearing region as the dominant contribution to the integral comes from
this region of the 
integration: $k^{*}\sim\pi/(N_g a)$. A better estimate
of the scale can be obtained by the method explained in Ref.~\cite{Lepage},
but since we are interested in the continuum limit where $a\rightarrow0$,
this is already a reliable estimation of the momentum scale of the
running coupling.

The analysis in QCD 
is more complex at one-loop level than in the scalar theory
due to the presence of the gauge-link required to render the
operator gauge-invariant.  
We have found that the contributions from the operator defined in
Eq.~(\ref{eq:25}) scale in the same way as those in the scalar theory,
with the violation of rotational symmetry  suppressed by factors of $\sim
1/N^2$,  but 
both tadpole improvement of the extended links and
smearing of the gauge-field is required.  
Our analysis of Wilson fermions reveals the contributions to matrix elements
that violate rotational invariance in the
continuum limit at the one-loop level are suppressed by factors of 
$\sim \alpha_s/N^2$ and $\sim \alpha_s/N_g^2$, and thus for a smearing defined in physical units,
deviations from rotational invariance scale as  ${\cal O}(a^2)$.  
Contributions that scale as $\sim \alpha_s/N$ and are proportional to the
Wilson parameter, conserve angular momentum and can be absorbed by the operator Z-factor.
Most importantly, as in the scalar theory, there are no mixings with lower dimension operators that
diverge as inverse powers of the lattice spacing.

\section*{Summary and discussions}

In this chapter, a mechanism for the restoration of rotational symmetry
in the continuum limit of lattice field theories is considered. The
essence of this approach is to construct an appropriate operator
on the cubic lattice which has maximum overlap onto the states with
definite angular momentum in the continuum. In analogy to the operator
smearing proposals given in Refs.~\cite{DudekI,DudekII,Edwards} and 
Refs.~\cite{Meyer,JohnsonRW},
the operator is constructed on multiple lattice sites.
Using spherical harmonics in the definition of the operator
is key to having  the leading contributions
to the classical operator be those with the desired angular momentum.
The
sizes of the contributions are controlled by the scale of the smearing of the operator, 
with sub-leading
contributions to both lower and higher dimensional operators
that violate rotational symmetry
being suppressed by $1/N^{2}$ - reflective of the pixelation of the
operator and fields. 
The  $\lambda\phi^{4}$ scalar field theory
is shown to preserve this universal scaling of the leading non-rotationally
invariant contributions at all orders in perturbation theory, compatible
with the finite-size scaling results of $\lambda\phi^{4}$-type theories
near their rotational invariant fixed points~\cite{CampostriniI,CampostriniII}.
The same can be shown to be true in $g\phi^3$ scalar field theory.

Gauge invariance somewhat complicates the construction and analysis of analogous
operators in QCD.  Although the tree-level lattice operator in QCD exhibits the
same scaling 
properties as the scalar operator, extended gauge links connecting the quark fields generate gluonic
interactions that contribute to loop diagrams that are power-law divergent.
Such contributions are either eliminated by tadpole improvement of the extended
links, or are suppressed by smearing of the gauge field.  
We find that it is the physical length scales
and continuum renormalization scale that dictate the size of matrix elements.
The leading non-continuum corrections from the one-loop diagrams preserve
angular momentum, scaling as $\sim \alpha_s a$ for Wilson fermions, and can be
absorbed by the operator $Z$-factor. 
In contrast, contributions that violate rotational symmetry are suppressed by $\alpha_s a^2$ as $a\rightarrow 0$. 
While we have chosen a specific form for the smeared operator, we expect that
the results, in particular the scaling of the violations to rotational symmetry,
are general features of a smeared operator with any (smooth) profile. Also, it is worth mentioning that although the calculations preformed in this work, and the subsequent conclusions, relate operators and matrix elements in $O_h(3)$ to those in $SO(3)$, the methodology and results are expected to hold in relations between $O_h(4)$ and $SO(4)$.  Instead of working with operators formed with spherical harmonics to recover $SO(3)$ invariance, one would work with operators formed with hyper-spherical harmonics to recover $SO(4)$ symmetry.

We conclude the chapter by discussing the practicality of this result
for the current LQCD calculations as well as its connection to the recovery of IR rotational invariance in the lattice theories:
\begin{figure}
\begin{centering}
\includegraphics[scale=0.4]{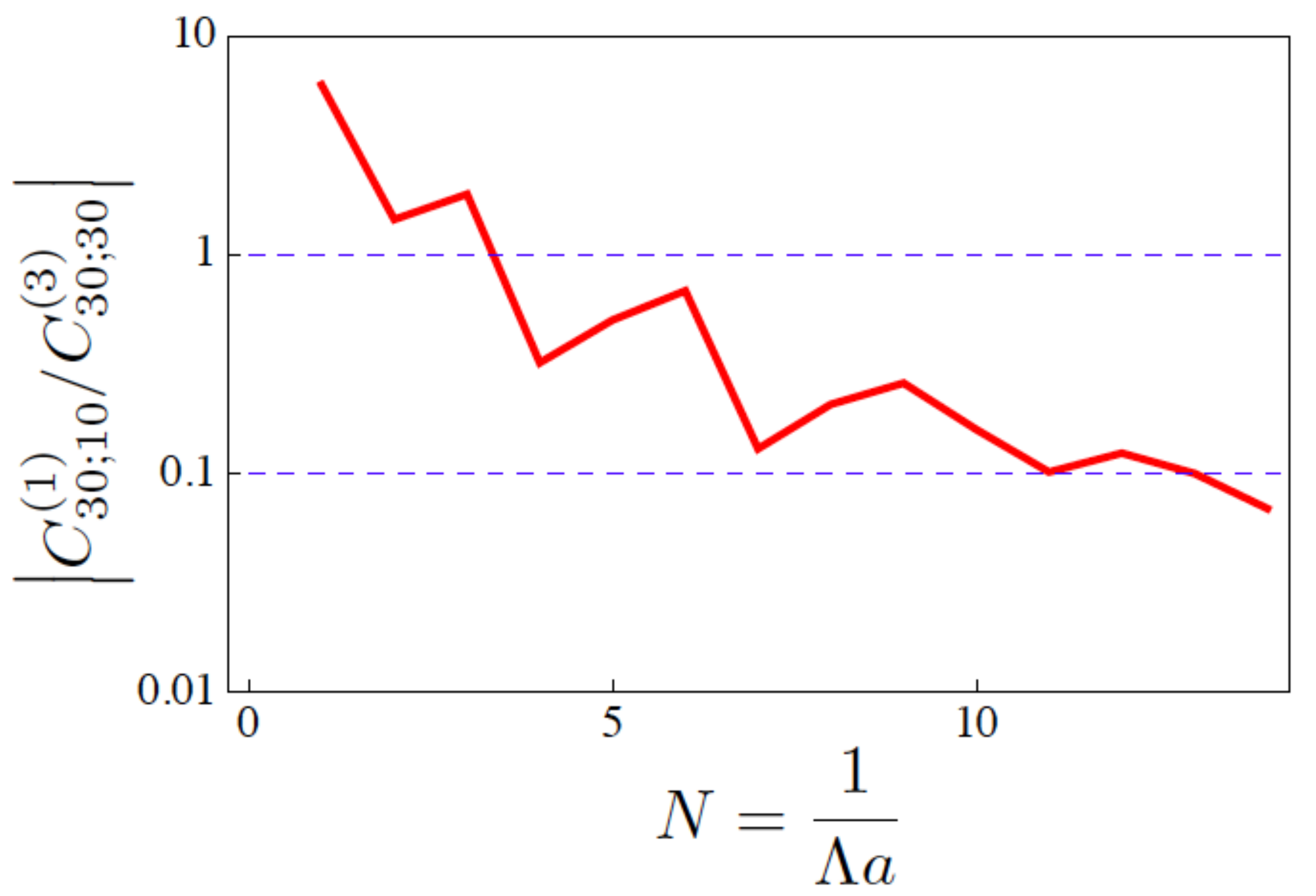}
\par\end{centering}
\caption{{\small 
The absolute value of the ratio of the tree-level coefficient,
    $C_{30;10}^{(1)}$,  of 
a lowest dimension operator  with $L=1$ 
to the tree-level coefficient, $C_{30;30}^{(3)}$, of the lowest dimension operator with
angular momentum, $L=3$, resulting from the $L=3$ operator in Eq. (\ref{eq:1}),
as a function of the number of included point shells.
}}
\label{fig:ratioplot}
\end{figure}
\begin{itemize}
\item
It is important to understand and to quantify 
the violation of angular momentum conservation in the
states and matrix elements calculated using Lattice QCD
with the lattice spacings currently employed. 
One interesting result is 
that by using the tadpole-improved operator extended over several lattice sites
and built from the smeared gauge links, 
the quantum corrections introduce non-continuum corrections to
the tree-level results that are suppressed by at least $\alpha_s$, i.e. they do
not introduce power-divergent contributions.
As an example, 
suppose that a lattice calculation aims to determine a matrix element of an
operator with $L=3$. 
Then, as is demonstrated in Fig. (\ref{fig:ratioplot}), 
the coefficient of the lower dimensional derivative operator with $L=1$ is almost $10$
times larger  than the coefficient of the $L=3$ derivative operator 
when the operator is defined  over one lattice site, $N=1$. 
The computational time required to accurately perform the subtraction of the
$L=1$ contribution is significant for  a smearing scale of, say, $\Lambda\sim 2~{\rm GeV}$.
Fortunately, by halving the lattice spacing and smearing the operator over just two point shells ($N=2$), 
the contamination from the lower dimensional operator is reduced by
a factor of $\sim 3$, requiring a factor of $\sim 10$ less computational resources
to accurately perform the subtraction at the same level of precision.
Further, by smearing the operator over ten point shells, 
the contamination from the lower dimensional operator is reduced to 
$\sim 1\%$ of its value at $N=1$.
Given that the lattice spacing associated with  $\Lambda = 2~{\rm GeV}$
is $a\sim 0.1~{\rm fm}$ for $N=1$, to be able to smear out to  the $N=2$ shell requires a
lattice spacing of $a\sim 0.05~{\rm fm}$, pushing the limits of current lattice
generation.  To smear out to the $N=10$ shell would require  a lattice spacing
of  $a\sim 0.01~{\rm fm}$ which is currently impractical.

\item The restoration of rotational invariance as discussed in this chapter
regards only the UV asymptote of the lattice theories: as one reaches
a good pixelation of a region of space where the lattice operator
probes, the identification of eigenstates of the angular momentum
operator becomes possible. In the other words, the more point shells
included in the lattice operator, the larger  overlap the operator
has onto a definite angular momentum state. However, the full recovery
of rotational invariance in the lattice theories requires the suppression
of rotational symmetry breaking contributions to the physical quantities
not only as a result of short-distance discretization effects, but
also as a result of boundary effects of the finite cubic lattice in the
IR regime of the theories. The finite size of the lattice imposes
(anti-)periodic boundary conditions on the lattice wavefunctions which
enforces the lattice momenta to be discretized, 
${\bf p}=\frac{2\pi {\bf n} }{L}$,
where $L$ is the spatial extent of the lattice 
and ${\bf n}$ is a vector of
integers. The IR rotational invariant theory is achieved as the
lattice becomes infinitely large, 
corresponding to a large number of point shells in the momentum space. 
However, beyond this intuitive picture, one
needs to examine in a quantitative way how this recovery takes place
in the large-volume limits of the lattice theories in the same way
as it was discussed for small lattice spacing limit of the theories.
One quantitative explanation of this IR recovery, has been given 
in Ref.~\cite{Luu:2011ep} in the context of the 
extraction of phase shifts in higher
partialwaves from the energies of scattering particles in a finite volume
using L\"uschers method.
The idea is that as one includes higher momentum shells, the number
of occurrence (multiplicity) of any given irrep of the cubic group increases. 
As a result, for a fixed energy in the large volume limit,
linear combinations of different states of a given irrep can be formed
which can be shown to be energy eigenstates; and 
the energy shift of each combination due to interactions is suppressed 
in all but one partialwave in the infinite-volume limit. 
So, although each state has an
overlap onto infinitely many angular momentum states, the high multiplicity
of a given irrep in a large momentum shell generates energy eigenstates
which 
dominantly overlap onto states of definite angular momentum,
and the mixing with other angular momentum states becomes insignificant in the
large volume limit.
This picture also helps to better understand the mechanism of the
UV rotational invariance recovery due to the operator smearing. 
It
is the high multiplicity of the irreps in large (position-space) 
shells that  is responsible
for projecting out a definite angular momentum eigenstate. 
These large
shells are obtained by reducing the pixelation of the lattice by taking
$a\rightarrow 0$ in position space,  or increasing the size of the
lattice by taking $L\rightarrow\infty$ in momentum space -- both are required in
order to recover rotational invariance from calculations performed on a lattice.
\end{itemize}


\chapter{TWO-NUCLEON SYSTEMS FROM A FINITE-VOLUME FORMALISM}
{\label{chap:NN}}

Despite tight empirical constraints on the two-body nuclear force, the investigation of the two-nucleon sector within LQCD is still warranted.  Understanding the energy dependence of the scattering phase shifts of two-body hadronic states, for example, is essential in obtaining physical matrix elements of current operators in the two-body sector \cite{Detmold:2004qn, Meyer:2012wk, Briceno:2012yi, Bernard:2012bi, Meyer:2013dxa}.  Additionally, as LQCD calculations are currently done at unphysical pion masses, a rigorous study of three (multi)-nucleon systems from LQCD requires not only the knowledge of two-nucleon phase shifts,  but also their pion mass dependence as shown in Refs. \cite{Briceno:2012rv, Kreuzer:2008bi, Kreuzer:2009jp, Kreuzer:2010ti, Kreuzer:2012sr}. The LQCD determination of the scattering parameters of two-nucleon systems at unphysical pion masses by itself is an interesting problem as it reveals the dependence of the two-body nuclear force on the masses of quarks in nature. Progress in this direction will have striking impact on our understanding of some of the most fundamental questions regarding the nuclear fine tunings in nature and the anthropic view of the Universe. As discussed in Refs. \cite{Epelbaum:2012iu, Epelbaum:2013wla, Bedaque:2010hr}, the survivability of Carbon-Oxygen based life is related to the variation of the inverse scattering lengths of NN scattering in the isosinglet and isotriplet channels, and a precise LQCD determination of these parameters will put tighter constraints on this quantity. In fact, for the first time LQCD has started addressing the question of  naturalness of the NN interactions. A nice example is the extraction of the $S$-wave NN scattering length and effective range by the NPLQCD collaboration \cite{Beane:2013br} at a pion mass of $m_{\pi}\approx 800~{\rm MeV}$ which has enabled them to study the variation of the NN scattering parameters and the corresponding bound-state energies with respect to the light-quark masses, see Fig. \ref{a-to-r}.
\begin{figure}[h]
\begin{center}
\label{a-to-r}
\includegraphics[scale=0.695]{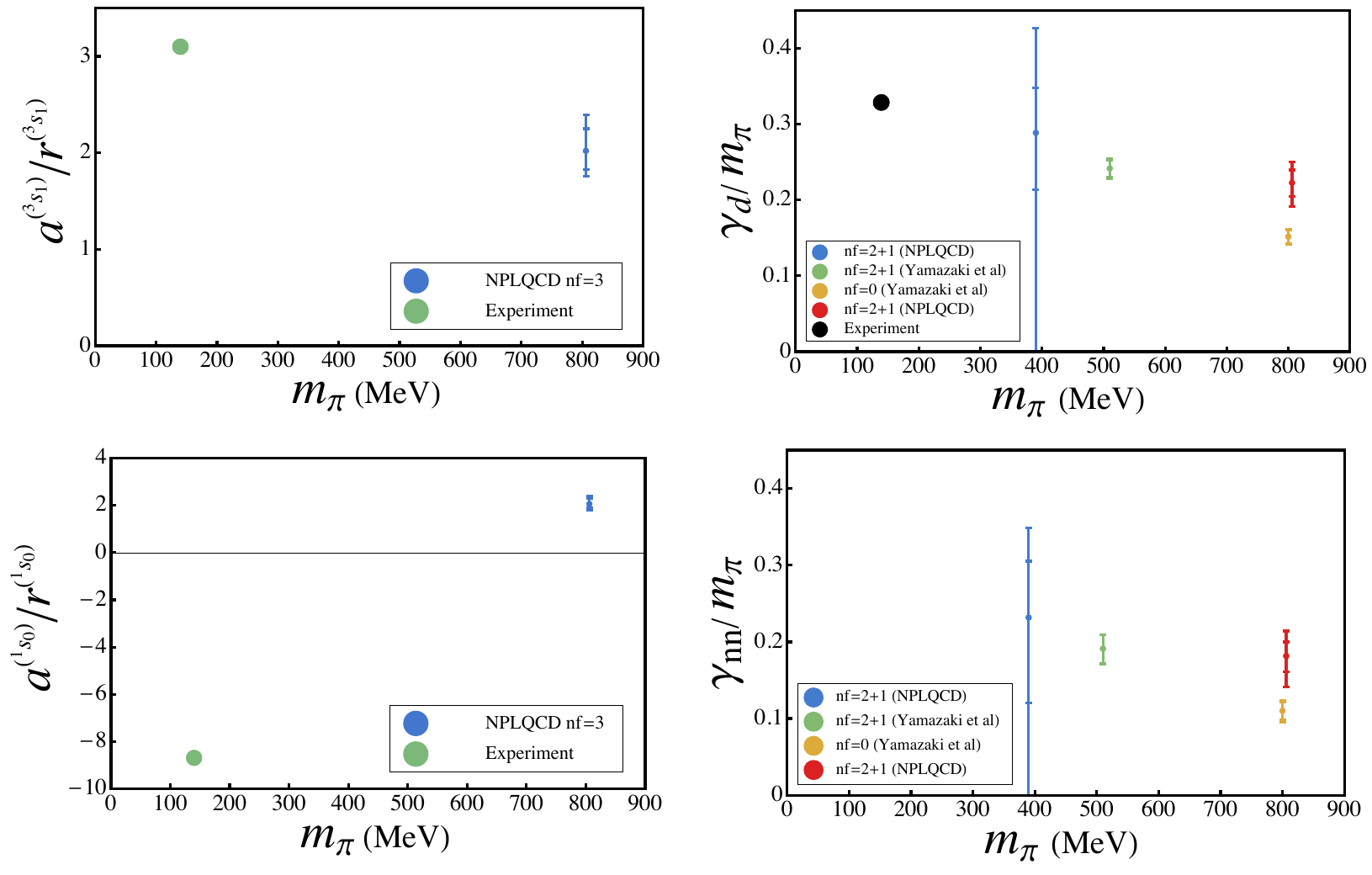}
\caption{{\small The left panel represents the ratio of the two-nucleon scattering length, $a$, to the effective range, $r$, in the ${^3}S_1$ (top) and ${^1}S_0$ (bottom) channels at the physical point as well at the $SU(3)$ symmetric point with $m_{\pi}\approx 800~{\rm MeV}$ \cite{Beane:2013br}. As can be inferred from the plots, the NN interactions remain unnatural over a wide range of pion masses. The right panel represents the plots of the binding energy as a function of pion mass. These indicate that the size of the deuteron and the $nn$ bound state remain large compared with the range of interactions at heavier pion masses. The figure is reproduced with the permission of the NPLQCD collaboration.}}
\label{a-to-r}
\end{center}
\end{figure}
To appropriately utilize these LQCD calculations, in particular for the realistic NN systems with physical partial-wave mixings -- which occurs due to the action of non-central forces in nuclear systems, the FV formalisms and their associated QCs must be developed.
 
In this chapter, we derive and present the generalization of the L\"uscher formula for nucleon-nucleon (NN) scattering valid below the inelastic threshold for all spin and isospin channels in both positive and negative parity sectors. This formula is derived using the auxiliary field (dimer) formalism in the language of a NR effective field theory EFT for NN interactions. As introduce in chapter \ref{chap:intro}, a S-wave dimer field -- that sums all $2\rightarrow2$ interactions non-perturbatively \cite{Kaplan:1996nv, Beane:2000fi} -- significantly simplifies the diagrammatic representation of multi-nucleon scattering amplitudes \cite{Bedaque:1997qi, Bedaque:1998mb, Bedaque:1998kg, Bedaque:1998km, Gabbiani:1999yv, Bedaque:1999vb, Bedaque:1999ve, Bedaque:2000ft}. However, to account for scattering in higher angular momentum channels, this dimer field must be generalized to arbitrary partial waves. This is particularly important when such an auxiliary field is used in constructing a FV formalism for three-body scattering processes. As is pointed out in Ref. \cite{Briceno:2012rv}, the leading systematics of the results presented in Refs. \cite{Briceno:2012rv,Kreuzer:2008bi, Kreuzer:2009jp, Kreuzer:2010ti, Kreuzer:2012sr}, for the relation between three-body scattering amplitude and the FV spectrum of the three-particle system, arises from the FV-induced mixing between S-wave and D-wave scattering modes of the two-particle sub-system.  A S-wave dimer field therefore does not incorporate possible mixings in the FV formalism and will not give rise to a full quantization condition in arbitrary partial waves in both two-body and three-body systems. To address this defect, we generalize the dimer field to higher partial waves  and utilize the result to derive the generalized L\"uscher formula for the two-body boosted systems within both scalar and nucleon sectors.

As discussed in chapter \ref{chap:intro}, performing LQCD calculations for systems with different CM momenta gives access to more energy levels at a given volume and provides additional QCs for the energy eigenvalues of the system in terms of scattering parameters. 
 Although the master formula that will be derived is self-contained and incorporates all the necessary details to be implemented in practice, deducing the relations, or QCs, among phase shifts in different partial waves and the energy levels of a specific LQCD calculation requires multiple nontrivial steps. The corresponding procedure is sometimes called the reduction of the L\"uscher formula. The difficulty associated with this procedure is due to the fact that LQCD calculations are performed in a finite periodic cubic volume (for calculations at rest). As a result, the degeneracy of energy eigenvalues of the system in such calculations is determined according to the irreps of the cubic group. Since the phase shifts are characterized according to the irreps of the SO(3) rotational group, the energy eigenvalues of the system in a given irrep of the cubic group in general depend on the phase shifts of more than one partial-wave channel. Performing  LQCD calculations of energy levels in different irreps of the cubic group would provide multiple QCs depending on different linear combinations of the scattering phase shifts, leading to better constraints on these quantities. Therefore it is necessary to identify all the QCs satisfied by a given scattering parameter in a partial-wave channel. While L\"uscher's original work presents the reduction of the master formula to a QC for the cubic $A_1$ irrep, Ref. \cite{Luu:2011ep} provides the full quantization conditions for the energy eigenvalues of different irreps of the cubic group, in both positive and negative parity sectors for orbital angular momentum $l\leq6$ as well as $l=9$ in the scalar sector. For scattering involving a spin-$\frac{1}{2}$ particle and a scalar particle, the L\"uscher formula can be generalized such that the energy eigenvalues of the meson-baryon system in a given irrep of the double-cover of the cubic group is related to the corresponding phase shifts \cite{Bernard:2008ax}. This generalization has been also presented for NN scattering, where due to the the possibility of physical mixing among different partial-wave channels, more complexities arise.\footnote{The L\"uscher formula to study two-nucleon systems were first presented in Ref. \cite{Beane:2003da}, although due to constraining the calculation to the S-wave scattering, the complexity of the two-nucleon systems has not been dealt with. The only previous attempt to address this problem, including the spin, isospin and angular momentum degrees of freedom, is the work by N. Ishizuka \cite{Ishizuka:2009bx}, where the quantization conditions for energy eigenvalues of a two-nucleon system at rest in the positive and negative parity isosinglet channels were obtained for $J\leq 4$.}

By investigating the symmetry groups of the boosted systems along one and two Cartesian axes as well as that of the unboosted system, we have identified all the QCs satisfied by the phase shifts and mixing parameters in channels with total angular momentum $J\leq4$; ignoring scattering in partial-wave channels with $l\geq4$. Different QCs correspond to different irreps of the cubic ($O$), tetragonal ($D_{4}$) and orthorhombic ($D_{2}$) point groups that represent the symmetry group of systems with CM momentum $\mathbf{P}=0$, $\mathbf{P}=\frac{2\pi}{L}(0,0,1)$ and $\mathbf{P}=\frac{2\pi}{L}(1,1,0)$ respectively, where $L$ denotes the spatial extent of the cubic volume. As will be discussed later, these QCs can be also utilized for boost vectors of the form $\frac{2\pi}{L}(2n_1,2n_2,2n_3)$, $\frac{2\pi}{L}(2n_1,2n_2,2n_3+1)$ and $\frac{2\pi}{L}(2n_1+1,2n_2+1,2n_3)$ and all cubic rotations of these vectors where $n_1,n_2,n_3$ are integers.  Although the master formula presented in this article in the limit of zero CM momentum has been already derived in Ref. \cite{Ishizuka:2009bx} for NN systems using a relativistic quantum field theory approach, the full classifications of different QCs for all the spin and isospin channels and for two non-zero CM momenta were first obtained through completion of this thesis and are already presented in Refs. \cite{Briceno:2013lba, Briceno:2013rwa}. These lengthy relations are tabulated in a Mathematica notebook supplemented to the published paper \cite{BDLsupp} and we refrain to repeat those in this thesis. However, the procedure of deducing these relations will be presented through an example in Sec. \ref{app:red-example}. Strategies for deducing such QC for more general cases will be briefly discussed in Sec. \ref{red-syst}. These relations make the implementation of the generalized L\"uscher formula for NN systems straightforward for future LQCD calculations of the NN system.

 \section{Finite Volume Formalism with the Auxiliary Field Method \label{sec: dimer}}

The goal of this section is to extend L\"usher's formula \cite{Luscher:1986pf, Luscher:1986pf} to the case of two nucleons within the context of a NR EFT, using an auxiliary field method. Although there has been several derivations for the L\"uscher formula (an example of which presented in Sec. \ref{IV-intro}), the formalism that will be presented here makes the study of two-baryon systems with arbitrary quantum numbers straightforward. Additionally the methodology developed here can be used to generalize the FV formalism presented in Ref. \cite{Briceno:2012rv} to three hadrons with arbitrary partial waves. In order to be able to incorporate the specific features of the two-nucleon systems in the formalism, it is instructive to start with developing a general dimer formalism for scalar particles. However, such formalism by itself is valuable in studies of multi-meson systems in a finite volume, see for example, Refs. \cite{Briceno:2012rv,Kreuzer:2008bi, Kreuzer:2009jp, Kreuzer:2012sr}.
  
 \subsection{Two-boson systems \label{sec: Scalar}}
 
\begin{figure}[t]
\begin{center}
\subfigure[]{
\includegraphics[scale=0.415]{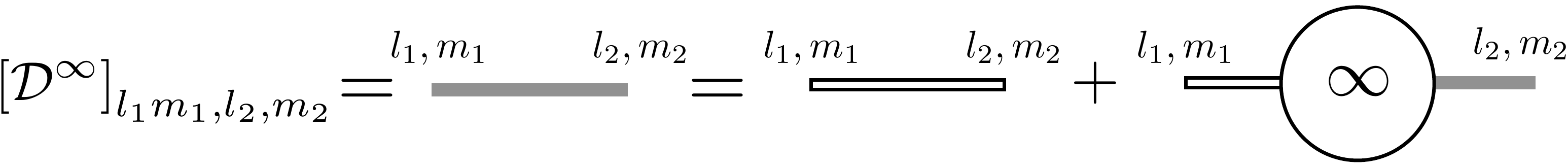}}
\subfigure[]{
\includegraphics[scale=0.415]{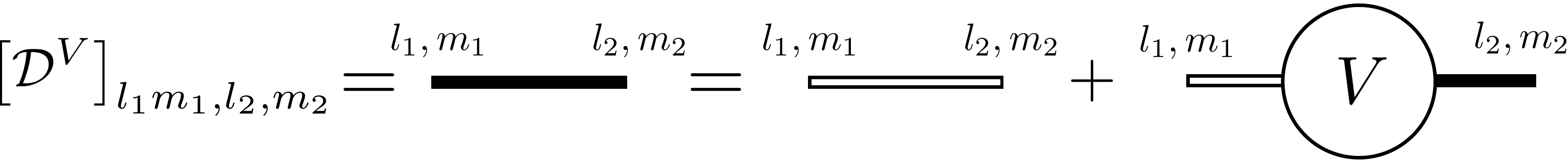}}
\caption{{\small a) Diagrammatic equation satisfied by the matrix elements of the full dimer propagator in a) infinite volume and b) finite volume. The grey (black) band represents the full infinite (finite) volume propagator, $\mathcal{D}^{\infty}$ ($\mathcal{D}^V$), while the double lines represent the bare propagator, $\mathcal{D}^B$.}}\label{fig:dimer}
\end{center}
\end{figure}

Consider two identical bosons with mass $M$ that interact in a partial-wave channel $(l,m)$ via a short-range interaction that can be effectively described by derivative couplings to the fields. Let $\phi_{k}$ and  $d_{lm,P}$ denote the interpolating operators that annihilate a boson with NR four-momentum $k$, and a dimer (with quantum numbers of two bosons) with NR four-momentum $P$ and angular momentum $(l,m)$, respectively. Then if $P^\mu=(E,\mathbf{P})$ denotes the NR four-momentum of the system, one can write a Galilean-invariant action that describes such system in the infinite volume in terms of a Lagrange density in the momentum space,
\begin{eqnarray}
\label{action}
{S}&=&\int\frac{d^{4}P}{(2\pi)^{4}}\left[\phi_{P}^{\dagger}(E-\frac{\textbf{P}^{2}}{2M})\phi_{P}-\sum_{l,m}d_{lm,P}^{\dagger}\left(E-\frac{\textbf{P}^{2}}{4M}-\Delta_{l}+\sum_{n=2}^{\infty}c_{n,l}(E-\frac{\textbf{P}^{2}}{4M})^{n}\right)d_{lm,P}\right]
\nonumber\\
&~& \qquad \qquad  -\int\frac{d^{4}P}{(2\pi)^{4}}~\frac{d^{4}k}{(2\pi)^{4}}\sum_{l,m}~\frac{g_{2,l}}{2}\left[d_{lm,P}^{\dagger}~\sqrt{4\pi}~Y_{lm}(\hat{\textbf{k}}^{*})~|\mathbf{k}^{*}|^{l}\phi_{{k}}\phi_{P-{k}}+h.c.\right],
\end{eqnarray}
where ${\textbf{k}}^*=\textbf{k}-\textbf{P}/2$ denotes the relative momentum of two bosons in the interaction term. Note that the interactions between bosons in partial-wave channel $(l,m)$ is mediated by a corresponding dimer field, $d_{lm}$. As is evident, upon integrating out such auxiliary field, one recovers the four-boson interaction term in a Lagrangian with only $\phi$-field degrees of freedom. Since this is a theory of identical bosons, all couplings of the dimer field to a two-boson state with an odd partial wave vanish. Eq. (\ref{action}) clearly reduces to the S-wave result of Refs. \cite{Kaplan:1996nv, Beane:2000fi, Griesshammer:2004pe}. This action can be easily generalized for systems involving distinguishable scalar bosons (e.g. for P-wave scattering see Ref. \cite{Braaten:2011vf}). As usual, the LECs $\{\Delta_{l},c_{l,n}, g_{2,l}\}$ in the effective Lagrangian must be tuned to reproduce the ERE of the $l^{th}$-partial wave,
\begin{eqnarray}
k^{*2l+1}\cot\delta_{d}^{(l)}=-\frac{1}{a_l}+\frac{r_{l}k^{*2}}{2}+\sum_{n=2}^\infty\frac{\rho_{n,l}}{n!}~(k^{*2})^{n},
\label{NN-ERE}
\end{eqnarray}
where $k^*\equiv |\mathbf{k}^*|=\sqrt{ME-\frac{\mathbf{P}^2}{4}}$ is the relative \emph{on-shell} momentum of the bosons in the CM frame. $\delta_d^{(l)}$ is the phase shift in the $l^{th}$-partial wave, and $\{a_l,r_l, \rho_{n,l}\}$ are the corresponding scattering length, effective range and all higher order shape parameters, respectively. The fully dressed dimer propagator can be obtained by summing up the self-energy bubble diagrams to all orders, Fig. \ref{fig:dimer}(a), the result of which is the following
\begin{eqnarray}
\mathcal{D}^{\infty}(E,\mathbf{P})=\frac{1}{(\mathcal{D}^{B})^{-1}-I^{\infty}(E,\mathbf{P})},
\label{D-infinity}
\end{eqnarray}
where $\mathcal{D}^B$ denotes the bare dimer propagator, 
\begin{eqnarray}
\left[\mathcal{D}^{B}(E,\mathbf{P})\right]_{l_1m_1,l_2m_2}=\frac{-i~\delta_{l_1l_2}\delta_{m_1m_2}}{E-\frac{\mathbf{P}^2}{4M}-\Delta_{l}+\sum_{n=2}^{\infty}c_{n,l}(E-\frac{\textbf{P}^{2}}{4M})^{n}+i\epsilon},
\label{D-bare}
\end{eqnarray}
and $I^{\infty}$ denotes the value of the bubble diagram evaluated using the power divergence subtraction (PDS) scheme \cite{Kaplan:1998tg, Kaplan:1998we, Beane:2003da},
\begin{eqnarray}
\left[I^{\infty}(E,\mathbf{P})\right]_{l_1m_1,l_2,m_2}=\frac{iM}{8\pi}g_{2,l_1}^2k^{*2l_1}(\mu+ik^*)\delta_{l_1l_2}\delta_{m_1m_2},
\label{I-infinity}
\end{eqnarray}
 where $\mu$ is the renormalization scale. By requiring the full dimer propagator, $\mathcal{D}^{\infty}$, in the infinite volume to reproduce the full scattering amplitude in any given partial wave,
\begin{eqnarray} 
\mathcal{M}^{\infty}_{l_1m_1,l_2m_2}&=&-[~g~\mathcal{D}^{\infty}(E,\mathbf{P})~g~]_{l_1m_1,l_2m_2}=
\frac{8\pi}{M}~
\frac{1}{k^{*}\cot{\delta^{(l_1)}_d}-ik^{*}}\delta_{l_1l_2}\delta_{m_1m_2},
\end{eqnarray}
one arrives at
\begin{eqnarray}
g_{2,l}^2=\frac{16\pi}{M^2r_{l}} ~\text{for}~l~\text{even}, ~~ \Delta_l=\frac{2}{Mr_l}\left(\frac{1}{a_l}-\mu k^{*2l}\right), ~~ c_{n,l}=\frac{2}{Mr_l}\frac{\rho_{n,l}M^n}{n!}.
\label{g2l}
\end{eqnarray}

In the finite volume, the two-boson system can still be described by the action in Eq. (\ref{action}) except the periodic boundary conditions constrain the momenta to be discretized. In particular, the integral over three-vector momenta in Eq. (\ref{action}) is replaced by a sum over discrete momenta, $P=\frac{2\pi}{L}\mathbf{n}$, where $\mathbf{n}$ is an integer triplet. Then it is straightforward to evaluate the corresponding bubble diagram in the finite volume,
\begin{eqnarray}
\left[I^{V}\right]_{l_1m_1,l_2,m_2}&=&\frac{iM}{8\pi}g_{2,l_1}g_{2,l_2}k^{*l_1+l_2}
\nonumber\\
&~&\times\left[\mu~\delta_{l_1l_2}\delta_{m_1m_2}+\sum_{l,m}\frac{(4\pi)^{3/2}}{k^{*l}}c^{{P}}_{lm}(k^{*2})\int d\Omega~Y^*_{l_1,m_1}Y^*_{l,m}Y_{l_2,m_2}\right],
\label{I-V}
\end{eqnarray}
where as defined in Sec. \ref{IV-intro},
\begin{eqnarray}
\label{clm}
c^{{\mathbf{P}}}_{lm}(x)=\left[\frac{1}{L^3}\sum_{\textbf{q}}-\mathcal{P}\int\frac{d^3\mathbf{q}}{(2\pi)^3}\right]{q}^{*l}\frac{\sqrt{4\pi}Y_{lm}(\hat{\mathbf{q}^*})}{{\mathbf{q}^*}^{2}-x} \ ,
\end{eqnarray}
$\mathbf{q}^*=\mathbf q- \mathbf P/2$, and $\mathcal{P}$ denotes the principal value of the integral. The full dimer propagator, $\mathcal{D}^V$, can then be obtained by summing up the infinite series of bubble diagrams in Fig. \ref{fig:dimer}(b), where the LEC of the theory are matched with the the physical quantities according to Eq. (\ref{g2l}),
\begin{eqnarray}
\mathcal{D}^{V}(E,\mathbf{P})=\frac{1}{(\mathcal{D}^{B})^{-1}-(\mathcal{D}^{B})^{-1}I^{V}(E,\mathbf{P})\mathcal{D}^{B}}.
\label{D-finite}
\end{eqnarray}
Note that, just like $\mathcal{D}^\infty$ in Eq.~(\ref{D-infinity}), $\mathcal{D}^V$ is a matrix in the angular momentum space. The poles of the FV dimer propagator give the spectrum of two-boson system in a finite volume in terms of the scattering parameters. These energy eigenvalues satisfy the following determinant condition
\begin{eqnarray}
\det \left[k^*\cot \delta-\mathcal{F}^{FV}\right]=0,
\label{FullQCboson}
\end{eqnarray}
where both $\cot \delta$ and $\mathcal{F}^{FV}$ are matrices in the angular momentum space,
\begin{eqnarray}
\cot \delta \equiv \cot (\delta_{l_1}) \delta_{l_1l_2}\delta_{m_1m_2},
\label{cot}
\end{eqnarray}
\begin{eqnarray}
\left[\mathcal{F}^{FV}\right]_{l_1m_1,l_2m_2}=\sum_{l,m}\frac{(4\pi)^{3/2}}{k^{*l}}c^{\mathbf{P}}_{lm}(k^{*2})\int d\Omega~Y^*_{l_1,m_1}Y^*_{l,m}Y_{l_2,m_2}.
\label{F}
\end{eqnarray}

In Eq.~(\ref{cot}) the Kronecker deltas that are constraining the $(l,m)$ quantum numbers should not be confused with the phase shift $\delta_{l_1}$. This quantization condition agrees with the NR limit of the results presented in Refs. \cite{Rummukainen:1995vs, Kim:2005gf, Christ:2005gi} for the generalization of the L\"uscher formula to the boosted systems, and upon truncating the angular momentum sum to $l_{max}=0$, reduces to the S-wave result of Ref. \cite{Briceno:2012rv} where an S-wave dimer field is used to derive the L\"uscher formula. This derivation shows that upon incorporating higher partial waves in the construction of the dimer Lagrangian, as well as accounting for higher order terms in the EFR expansion, all the two-body physics is fully encapsulated in this formalism. As a result the systematic errors of those FV multi-particle calculations that have used a S-wave dimer field up to next-to-leading order in ERE (see Refs. \cite{Briceno:2012rv, Kreuzer:2008bi, Kreuzer:2009jp, Kreuzer:2010ti, Kreuzer:2012sr}), can be easily avoided.

\subsection{Two-nucleon systems \label{sec: Nuclear}}
Due to spin and isospin degrees of freedom, the two-nucleon system exhibits some specific features. In particular, the anti-symmetricity of the two-nucleon state constrains the allowed spin and isospin channels for a given parity state. Additionally, any spin-triplet two-nucleon state is an admixture of two different orbital-angular momentum states. For example, as is well known, the two-nucleon state in the deuteron channel with $J^{P}=1^{+}$ is an admixture of S-wave and D-wave states. In general, a positive parity two-nucleon state with total angular momentum $J$ is a linear combination of states with the following orbital angular momentum $L$ and total spin $S$\footnote{The $L$ that is introduced here and elsewhere as the partial-wave label of quantities should not be confused with the spatial extent of the lattice $L$ that appears in the definition of the $c_{lm}^{\mathbf{P}}$ functions.}
\begin{eqnarray}
\left(L=J\mp\frac{1}{2}(1-(-1)^J), S=\frac{1}{2}(1-(-1)^J)\right),
\label{positive}
\end{eqnarray}
while in the negative parity sector, the states that are being mixed have\footnote{Note, however, that for a $J$-even state in the first case and a $J$-odd state in the second case, there is only one angular momentum state allowed and no mixing occurs.}
\begin{eqnarray}
\left(L=J\mp\frac{1}{2}(1+(-1)^J), S=\frac{1}{2}(1+(-1)^J)\right).
\label{negative}
\end{eqnarray}
Table (\ref{JP}) shows the allowed spin and angular momentum of NN states in both isosinglet and isotriplet channels with $J\leq3$.
 \begin{center}
\begin{table}[h]
\includegraphics[scale=0.9]{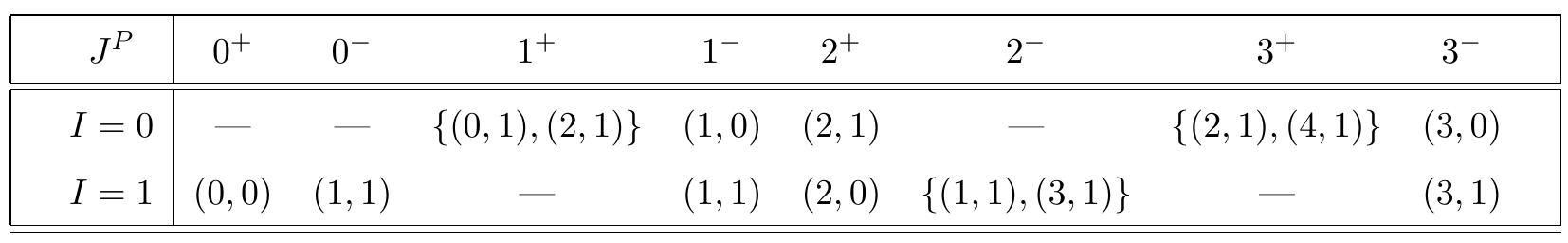}
\caption{{\small The allowed spin and angular momentum states, $(L,S)$, for NN-states with $J\leq3$ assuming an exact isospin symmetry. Note that depending on the parity of the states, the partial-wave mixing can occur in either the isosinglet or isotriplet channels.}\label{JP}}
\end{table}
\end{center}

In order to write the most general Lagrangian describing NN scattering in all spin, isospin and angular momentum channels, let us introduce an operator that creates a NN-state with total four-momentum $P$ and the relative momentum ${\textbf{k}}^*={\textbf{k}}-\frac{{\textbf{P}}}{2}$ in an arbitrary partial wave $(L,M_L)$ in the following way
\begin{eqnarray}
|NN;P,k^*\rangle _{LM_L,SM_S,IM_I}=\mathcal{N}_L\int d\Omega_{\textbf{k}^*}~Y^*_{LM_L}(\hat{\textbf{k}}^{*})k^{*L}\left[N^T_{P-k}~\hat{\mathcal{P}}_{(SM_S,IM_I)}~N_k\right]^\dag|0\rangle,
\end{eqnarray}
where $k^*=\left|\mathbf{k}^*\right|$. $\hat{\mathcal{P}}_{(SM_S,IM_I)}$ is an operator which projects onto a two-nucleon state with spin $(S,M_S)$ and isospin $(I,M_I)$, and $\mathcal{N}_L$ is a normalization factor. By requiring such state to have a non-zero norm, and given the anti-commutating nature of nucleon fields, one can infer that for positive parity states the operator $\hat{\mathcal{P}}_{(SM_S,IM_I)}$ must be necessarily antisymmetric, while for negative parity states it must be symmetric. Since this operator is a direct product of two projection operators in the space of spin and isospin, these requirements can be fulfilled by constructing the corresponding operators using the appropriate combinations of Pauli matrices, $\sigma_j$ ($\tau _j$), that act on the spin (isospin) components of the nucleon field. To proceed with such construction, let us define the following operators
\begin{eqnarray}
\alpha_j^I=\tau_y\tau_j,\hspace{1cm}
\alpha_j^S=\sigma_y\sigma_j,\hspace{1cm}
\beta^I=\tau_y,\hspace{1cm}
\beta^S=\sigma_y. 
\end{eqnarray}
Note that the matrices that are named as $\alpha$ are symmetric while those that are named as $\beta$ are antisymmetric. Superscript $I$ ($S$) implies that the operator is acting on the spin (isospin) space, and index $j=1,2,3$ stands for the Cartesian components of the operators. Alternatively one can form linear combinations of $\alpha^S_j$ ($\alpha^I_j$) that transform as a rank one spherical tensor.\footnote{A Cartesian vector $\mathbf{r}$ can be brought into a spherical vector according to 
\begin{eqnarray}
\label{spherical}
r^{(0)}\equiv r_{z},\hspace{1cm}
r^{(\pm 1)}\equiv\mp \frac{\left(r_{x} \pm ir_{y}\right)}{\sqrt{2}}.
\nonumber
\end{eqnarray}
} Using these matrices, it is straightforward to see that an antisymmetric $\hat{\mathcal{P}}_{(SM_S,IM_I)}$ can have one of the following forms 
\begin{eqnarray}
{\hat{\mathcal{P}}}_{(00,1M_I)} \equiv \frac{\alpha^{(M_I)}_I\otimes \beta_S}{\sqrt{8}},\hspace{.25cm}
{\hat{\mathcal{P}}}_{(1M_S,00)} \equiv \frac{\beta_I\otimes \alpha^{(M_s)}_S}{\sqrt{8}},
\end{eqnarray}
which can project onto two-nucleon states with $\left(S=0,I=1\right)$ and $(S=1,I=0)$ respectively. Note that these are the conventional isotriplet and isosinglet projection operators in the positive parity sector that are used frequently in literature \cite{Savage:1998ae, Chen:1999tn}.
On the other hand, a symmetric $\hat{\mathcal{P}}_{(SM_S,IM_I)}$ can project onto two-nucleon states with $(S=0,I=0)$ and $(S=1,I=1)$ and should have one of the following forms,
\begin{eqnarray}
{\hat{\mathcal{P}}}_{(00,00)} \equiv \frac{\beta_I\otimes \beta_S}{\sqrt{8}},\hspace{.25cm}
{\hat{\mathcal{P}}}_{(1M_S,1M_I)} \equiv \frac{\alpha^{(M_I)}_I\otimes \alpha^{(M_S)}_S}{\sqrt{8}},
\end{eqnarray}
respectively.

As it is the total angular momentum $J$ that is conserved in a two-nucleon scattering process, as opposed to the orbital angular momentum $L$, it is convenient to project a two-nucleon state in the $|LM_L,SM_S\rangle$ basis into a state in the $|JM_J,LS\rangle$ basis using the Clebsch-Gordan coefficients,
\begin{eqnarray} 
|NN;P,k^*\rangle _{JM_J,LS,IM_I}&=&\sum_{M_L,M_S}\langle JM_J|LM_L,SM_S\rangle~ |NN;P,k^*\rangle _{LM_L,SM_S,IM_I}.
\label{NNstate}
\end{eqnarray} 

Note that isospin remains a conserved quantum number up to small isospin breaking effects that we ignore for the nucleon systems.
In order to describe NN interactions, we introduce an auxiliary dimer filed, similar to the scalar theory.\footnote{The S-wave dimer field in the nuclear sector is commonly referred to as a di-baryon field.} This field, that will be labeled $d^{LS}_{JM_J,IM_I;P}$, has the quantum numbers of two-nucleon states with total angular momentum $(J,M_J)$ and isospin quantum number of ${(I,M_I)}$ with orbital angular momentum $L$ and spin $S$. Now the action corresponding to the Lagrangian density of free nucleon and dimer fields in the momentum space can be written as
\begin{eqnarray} 
&&S_{kinetic}=\int\frac{d^4P}{(2\pi)^4}\left[N^\dag_P(E-\frac{\textbf{P}^2}{2m})N_P \right .
\nonumber\\
&&\left . -\sum_{\substack{J,M_J, I,M_I}}\sum_{L,S}~\left(d^{LS}_{JM_J,IM_I;P}\right)^\dag\left(E-\frac{\textbf{P}^2}{4m}-\Delta^{LS}_{JI}+\sum_{n=2}^{\infty} c^{LS}_{JI,n}(E-\frac{\textbf{P}^2}{4m})^{n}\right)d^{LS}_{JM_J,IM_I;P}\right].
\nonumber\\
\label{Skin}
\end{eqnarray} 

In order to write the interaction Lagrangian, one should note that, while the total angular momentum, parity, isospin and spin are conserved in a strongly interacting NN process, the orbital angular momentum can change due to the action of tensor forces in nuclear physics. This is easy to implement in this formalism, as the two-nucleon states that are formed, Eq. (\ref{NNstate}), are compatible with the symmetries of the two-nucleon states. The interacting part of the action that does not mix angular momentum states, $S_{int,1}$, can then be written as
\begin{eqnarray} 
S_{int,1}&=&-\int\frac{d^4P}{(2\pi)^4}~\frac{d^4k}{(2\pi)^4}
\sum_{\substack{J,M_J, I,M_I}} ~ \sum_{L,M_L,S,M_S}
~{g^{LS}_{JI}}~
\langle JM_J|LM_L,SM_S\rangle
\nonumber\\
&~& \qquad \qquad \times \left[\left(d^{LS}_{JM_J,IM_I;P}\right)^\dag~\sqrt{4\pi}~Y_{LM_L}(\hat{\textbf{k}}^*)~{k}^{*L}~N^T_{k}~\hat{\mathcal{P}}_{(SM_S,IM_I)}~N_{P-k}+h.c. \right],
\nonumber\\
\label{S1}
\end{eqnarray} 
where ${g^{LS}_{JI}}$ denotes the coupling of a dimer field to the two-nucleon state with quantum numbers $\{J,I,L,S\}$. Note that the interactions must be azimuthally symmetric and so the reason the couplings are independent of azimuthal quantum numbers. Eqs. (\ref{positive}, \ref{negative}) now guide us to write the most general form of the interacting part of the action that is not diagonal in the angular momentum space, ${S}_{int,2}$, as follows
\begin{eqnarray} 
{S}_{int,2}&=&-\int\frac{d^4P}{(2\pi)^4}\frac{d^4k}{(2\pi)^4}
\sum_{\substack{J,M_J, I,M_I}} ~ \sum_{L,M_L,L',M_L',S,M_S} 
\nonumber\\
&~& \qquad \qquad {h_{JI}}~ \delta_{I,\frac{1+(-1)^J}{2}}~\delta_{S,1}(\delta_{L,J+1}\delta_{L',J-1}+\delta_{L,J-1}\delta_{L',J+1})\langle JM_J|L'M_L',SM_S\rangle~
\nonumber\\
&~& \qquad \qquad \times \left[\left(d^{LS}_{JM_J,IM_I;P}\right)^\dag~\sqrt{4\pi}~Y_{L'M_{L}'}(\hat{\textbf{k}}^*)~{k}^{*{L'}}~N^T_{k}~\hat{\mathcal{P}}_{(SM_S,IM_I)}~N_{P-{k}}+h.c.\right].
\nonumber\\
\label{S2}
\end{eqnarray} 
Note that in this interacting term, spin, isospin and the initial and final angular momenta are all fixed for any given total angular momentum $J$. As a result we have only specified the $(JI)$ quantum numbers corresponding to coupling $h$. As in the scalar case, all the LECs of this effective Lagrangian, $\{\Delta^{LS}_{JI},c^{LS}_{JI,n}, g^{LS}_{JI}, h_{JI}\}$, can be tuned to reproduce the \textit{low-energy} expansion of the scattering amplitudes in the $J^{th}$ angular momentum channel with a given spin and isospin. As discussed in Sec. \ref{sec: Scalar}), in the scalar sector the LECs can be easily determined in terms of the ERE parameters and the renormalization scale. For coupled-channel systems, obtaining the LECs in terms of the scattering parameters requires solving a set of coupled equations. The tuning of the LECs is only an intermediate step in obtaining the relationship between the FV spectrum and the scattering amplitude, which can be easily circumvented by introducing the Bethe-Salpeter kernel.  

\begin{figure}[t!]
\begin{centering}
\includegraphics[scale=0.415]{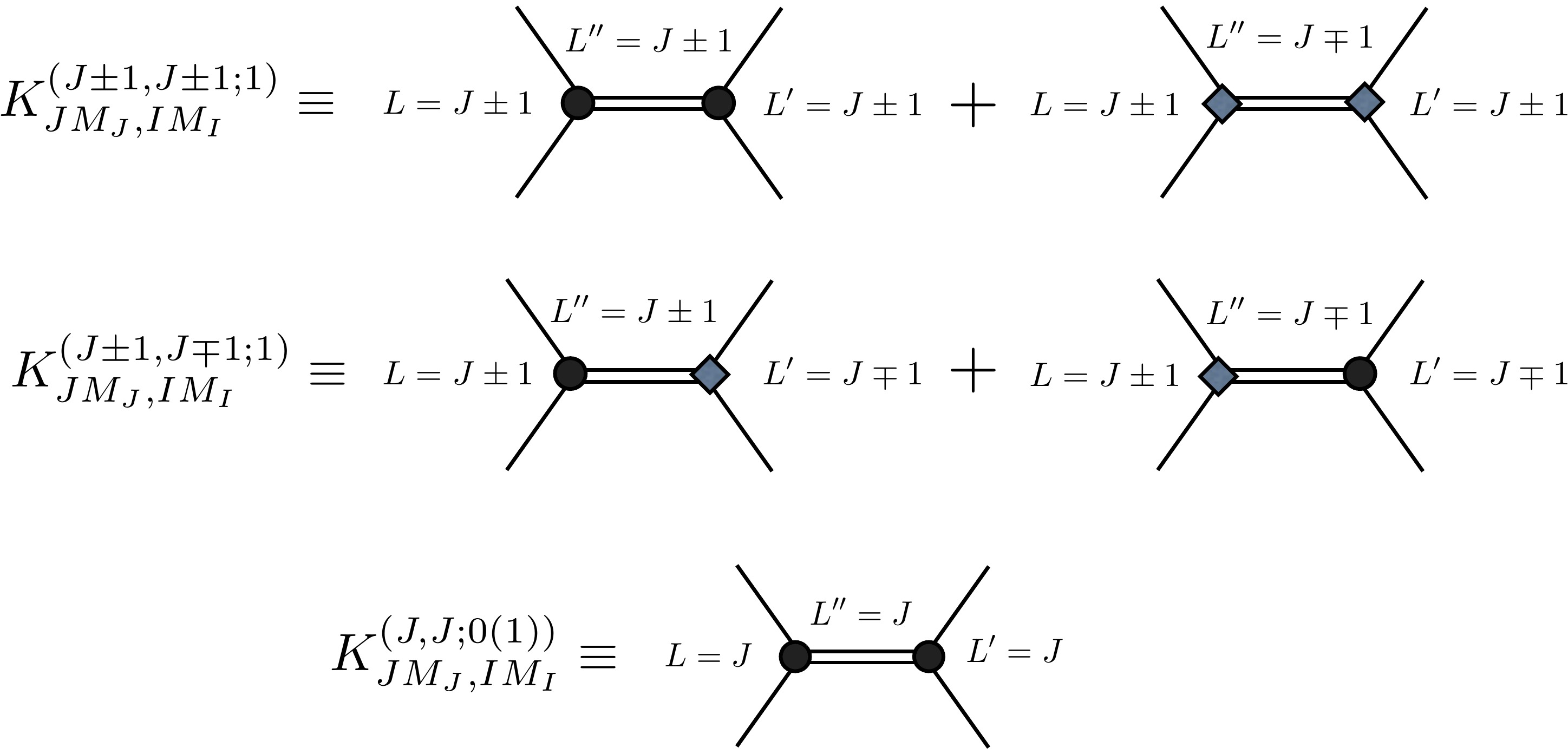}
\par
\caption{{\small The leading $2\rightarrow2$ transition amplitudes in the sector with total angular momentum $J$, Eq. (\ref{KernelJ}). The superscripts in the kernels denote the initial angular momentum, $L$, final angular momentum, $L'$ and the conserved spin of the channels, $S$, respectively. The black dot represents the interaction vertex that conserves the partial wave of the channel, and whose strength is parametrized by the coupling $g^{LS}_{JI}$, Eq. (\ref{S1}). The grey diamond denotes the vertex that mixes partial waves, and whose strength is given by $h_{JI}$, Eq. (\ref{S2}). the double lines are the bare propagators corresponding to a dimer field with angular momentum $L''$.}}\label{fig: Kernels}
\end{centering}
\end{figure}

Let us encapsulate the leading $2\rightarrow2$ transition amplitude between a two-nucleon state with $(JM_J,IM_I,LS)$ quantum numbers and a two-nucleon state with $(JM_J,IM_I,L'S')$ quantum numbers in the Bethe-Salpeter kernel, $K$. Since total angular momentum, spin and isospin are conserved in each $2\rightarrow 2$ transition, the kernel can be fully specified by $K_{JM_J;IM_I}^{(LL';S)}$. Since $J$ is conserved, the full kernel in the space of total angular momentum can be expressed as a block-diagonal matrix. In fact, it is straightforward to see that for each $J$-sector, the corresponding subblock of the full matrix has the following form
\begin{eqnarray}
\left(\begin{array}{cccc}
K_{JM_{J};IM_{I}}^{(J-1,J-1;1)} & 0 & 0 & K_{JM_{J};IM_{I}}^{(J-1,J+1;1)}\\
0 & K_{JM_{J};IM_{I}}^{(J,J;0)} & 0 & 0\\
0 & 0 & K_{JM_{J};I'M_{I'}}^{(J,J;1)} & 0\\
K_{JM_{J};IM_{I}}^{(J+1,J-1;1)} & 0 & 0 & K_{JM_{J};IM_{I}}^{(J+1,J+1;1)}
\end{array}\right).
\label{KernelJ}
\end{eqnarray}
We keep in mind that for any given $J$, $I$, $L$ and $S$, there are $(2J+1)^2\times(2I+1)^2$ elements accounting for different values of $M_J$ and $M_I$ quantum numbers. We also note that the value of the isospin is fixed for each transition kernel. Explicitly, one finds that $I=\frac{1+(-1)^J}{2}$ and $I'=\frac{1+(-1)^{J+1}}{2}$.\footnote{There is no $(I=0,S=0)$ channel for scattering in an even $J$ sector. Also there is no $(I=1,S=0)$ channel for scattering in an odd $J$ sector.} For the special case of $J=0$, the corresponding sub-sector is 
\begin{eqnarray}
\left(\begin{array}{cc}
K_{00;1M_I}^{(0,0;0)} & 0\\
0 & K_{00;1M_{I}}^{(1,1;1)}
\end{array}\right).
\label{KernelJ0}
\end{eqnarray}
These kernels, that correspond to leading transitions in all spin and isospin channels, are depicted in Fig. \ref{fig: Kernels}. Although one can read off the Feynman rules corresponding to these kernels from the  Lagrangian, Eqs. (\ref{Skin}, \ref{S1}, \ref{S2}), the FV energy eigenvalues can be determined without having to reference to the explicit form of these kernels, as will become evident shortly.

The scattering amplitude can be calculated by summing up all the $2\rightarrow2$ diagrams which can be obtained by any number of insertions of the transition kernels and the two-particle propagator loops. It can be easily seen that the infinite-volume two-particle loops, $\mathcal{G}^{\infty}$, are diagonal in total angular momentum, spin, isospin and orbital angular momentum. It is easy to show that $\mathcal{G}^{\infty}=2~I^{\infty}$, where $I^{\infty}$ is the infinite-volume loop for two identical bosons, Eq. (\ref{I-infinity}), hence the overall factor of two.
As a result, the scattering amplitude can be expressed as
\begin{eqnarray}
\mathcal{M}^{\infty}=-\mathcal{K}\frac{1}{1-\mathcal{G}^{\infty}\mathcal{K}},
\end{eqnarray}
where $\mathcal{K}$ is a matrix whose $J^{th}$-subblock is given by Eq. (\ref{KernelJ}). Since $\mathcal{G}^{\infty}$ is diagonal, the $J^{th}$-subblock of the infinite-volume scattering amplitude reads 
\begin{eqnarray}
\left(\begin{array}{cccc}
\mathcal{M}_{JM_{J};IM_{I}}^{(J-1,J-1;1)} & 0 & 0 & \mathcal{M}_{JM_{J};IM_{I}}^{(J-1,J+1;1)}\\
0 & \mathcal{M}_{JM_{J};IM_{I}}^{(J,J;0)} & 0 & 0\\
0 & 0 & \mathcal{M}_{JM_{J};I'M_{I'}}^{(J,J;1)} & 0\\
\mathcal{M}_{JM_{J};IM_{I}}^{(J+1,J-1;1)} & 0 & 0 & \mathcal{M}_{JM_{J};IM_{I}}^{(J+1,J+1;1)}
\end{array}\right),
\label{amplitude}
\end{eqnarray}
for any non-zero $J$ and
\begin{eqnarray}
\left(\begin{array}{cc}
\mathcal{M}_{00;1M_I}^{(0,0;0)} & 0\\
0 & \mathcal{M}_{00;1M_{I}}^{(1,1;1)}
\end{array}\right),
\label{amplitude}
\end{eqnarray}
for $J=0$. As is conventional, the scattering amplitude in channels with no partial-wave mixing can be parametrized by a scattering phase shift, $\delta_{JI}^{LS}$, according to
\begin{eqnarray}
\mathcal{M}_{JM_{J};IM_{I}}^{(JJ;S)}=\frac{4\pi}{Mk^*}\frac{e^{2i\delta_{JI}^{LS}}-1}{2i}\delta_{L,J}=\frac{4\pi}{Mk^*}\frac{1}{\cot{\delta_{JI}^{LS}}-i}\delta_{L,J},
\label{M-single}
\end{eqnarray}
while in channels where there is a mixing between the partial waves, it can be characterized by two phase shifts and one mixing angle, $\bar{\epsilon}_J$, \cite{Stapp:1956mz}\footnote{We take a different parametrization for the S-matrix when studying the deuteron spectrum for reasons that will be discussed in chapter \ref{chap:deuteron}.}
\begin{eqnarray}
\mathcal{M}_{JM_{J};IM_{I}}^{(J\pm1,J\pm1;S)}=\frac{4\pi}{Mk^*}\frac{\cos{2\bar{\epsilon}_J}e^{2i\delta_{JI}^{LS}}-1}{2i}\delta_{L,J\pm1},
\label{M-coupled1}
\\
\mathcal{M}_{JM_{J};IM_{I}}^{(J\pm1,J\mp1;S)}=\frac{4\pi}{Mk^*}\sin{2\bar{\epsilon}_J}\frac{e^{i(\delta_{JI}^{LS}+\delta_{JI}^{L'S})}}{2}\delta_{L,J\pm1}\delta_{L',J\mp1}.
\label{M-coupled2}
\end{eqnarray}
These relations are independent of $M_J$ and $M_I$ as the scatterings are azimuthally symmetric. We emphasize again that Kronecker deltas used to specify the $L$ quantum numbers should not be confused with the phase shifts. Note that for each $J$ sector, there is only one mixing parameter and as result no further labeling other than the $J$ label is necessary in case of $\bar{\epsilon}_J$.

The FV kernels are equal to the infinite-volume kernels (up to exponentially suppressed terms in volume below the pion production, and in particular the $J^{th}$-subblock of such kernel is given by Eq. (\ref{KernelJ}). As in the scalar case, the only difference between the finite volume and infinite volume shows up in the s-channel bubble diagrams, where the two particles running in the loops can go on shell and give rise to power-law volume corrections. It is straightforward to show that the two-nucleon propagator in the finite volume, $\mathcal{G}^V$, can be written as
\begin{eqnarray}
\mathcal{G}^V=\mathcal{G}^{\infty}+\delta\mathcal{G}^V,
\label{M-infinity}
\end{eqnarray}
where $\delta\mathcal{G}^V$ is a matrix in the $(JM_J,IM_I,LS)$ basis whose matrix elements are given by
\begin{align}
&\left[\delta\mathcal{G}^V\right]_{JM_J,IM_I,LS;J'M_J',I'M_I',L'S'}=\frac{iMk^*}{4\pi}\delta_{II'}\delta_{M_IM_I'}\delta_{SS'}\left[\delta_{JJ'}\delta_{M_JM_J'}\delta_{LL'} +i\sum_{l,m}\frac{(4\pi)^{3/2}}{k^{*l+1}}c_{lm}^{\mathbf{P}}(k^{*2}) \right.
\nonumber\\
& ~~~ \qquad \qquad \qquad ~ \left .  \times \sum_{M_L,M_L',M_S}\langle JM_J|LM_L,SM_S\rangle \langle L'M_L',SM_S|J'M_J'\rangle \int d\Omega~Y^*_{L,M_L}Y^*_{l,m}Y_{L',M_L'}\right],
\nonumber\\
\label{deltaG}
\end{align}
and, as is evident, is \emph{neither} diagonal in the $J$-basis nor in the $L$-basis. The kinematic function $c_{lm}^{\mathbf{P}}(k^{*2})$ is defined in Eq. (\ref{clm}) and is evaluated at the on-shell relative momentum of two nucleons in the CM frame. The full FV two-nucleon scattering amplitude can be evaluated by summing up all $2\rightarrow2$ FV diagrams,
\begin{eqnarray}
\mathcal{M}^{V}=-\mathcal{K}\frac{1}{1-\mathcal{G}^{V}\mathcal{K}}=\frac{1}{(\mathcal{M}^{\infty})^{-1}+\delta\mathcal{G}^{V}},
\label{M-V}
\end{eqnarray}
where in the second equality the kernel is eliminated in favor of $\mathcal{M}^{\infty}$ and $\mathcal{G}^{\infty}$ using Eq. (\ref{M-infinity}). The energy eigenvalues of the two-nucleon system arise from the poles of $\mathcal{M}^V$ which satisfy the following determinant condition
\begin{eqnarray}
\det\left[{(\mathcal{M}^{\infty})^{-1}+\delta\mathcal{G}^{V}}\right]=0.
\label{NNQC}
\end{eqnarray}
This quantization condition clearly reduces to Eq. (\ref{FullQCboson}) for two-boson systems when setting $S=0$\footnote{The symmetry factor in both the scattering amplitude and the FV function cancel out in the determinant condition, leaving the FV QC in Eq. \ref{FullQCboson}, insensitive to the distinguishability of the particles.}, and is in agreement with the result of Ref. \cite{Bernard:2008ax} for meson-baryon scattering after setting $S=1/2$. This result also extends the result of Ref. \cite{Ishizuka:2009bx} for two-nucleon systems to moving frames.

Although both $(\mathcal{M}^{\infty})^{-1}$ and $\delta\mathcal{G}^{V}$ are complex, for each partial wave, there occurs an intricate cancellation of their imaginary parts.  One could, on the other hand, work with the kernel $\mathcal{K}$ and finite-volume propagator $\mathcal{G}^V$, both manifestly real expressions, to define the QC.  From Eq. (\ref{M-V}) it is evident that the finite volume QC can be equivalently written as $\det[\mathcal{K}^{-1}-\mathcal{G}^{V}]=0$,
which has yet no reference to the scattering amplitude. However Eq. (\ref{NNQC}) is comprised of renormalization-scale \emph{independent} quantities, whereas $\mathcal{K}$ and $\mathcal{G}^V$ are both scale \emph{dependent} (only their difference is scale independent).  For this reason and to make connection with infinite-volume observables easier, we utilize Eq. \ref{NNQC}. Since the cancellation of imaginary terms is not trivial in channels with partial-wave mixing, we explicitly show how this cancellation occurs at the end of this section.

It is important to note that in deriving this result we have only assumed that the FV kernels are exponentially close to their infinite volume counterparts. Therefore the result obtained is valid for energies up to the inelastic threshold. In the nuclear sector this corresponds to the pion production threshold, $E^*=m_\pi$, which is well above the t-channel cut defined by $E^*_{cut}\equiv m_\pi^2/4m_N$ ($\sim 5$~MeV at the physical point). For energies above $E^*_{cut}$, the dimer formalism written in Eqs.~(\ref{Skin}-\ref{S2}) will get corrections from coupling of nucleons to pions and the LECs appearing in the action will get $m_\pi$-dependent corrections \cite{Weinberg:1990rz, Jenkins:1990jv, Kaplan:1998tg, Kaplan:1998we, Beane:2001bc}. All such corrections due to the dynamical pions above the t-channel cut, including pion exchange diagrams, can be still embedded in the interacting kernels in both infinite volume and finite volume. Below the pion production threshold, these corrections in the FV kernels are still exponentially close to their infinite volume counterparts, making the results presented in this section valid beyond the t-channel cut.

The determinant of the QC is defined in the basis of $(JM_J,IM_I,LS)$ quantum numbers and is over an infinite dimensional matrix. To be practical, this determinant should be truncated in the space of total angular momentum and orbital angular momentum. Such truncation is justified since in the low-momentum limit the scattering phase shift of higher partial waves $L$  scales as $k^{*2L+1}$. In the next section, by truncating the partial waves to $L\leq3$, we unfold this determinant condition further, and present strategies to reduce this master formula to separate QCs for energy eigenvalues in different irreps of the corresponding symmetry group of the two-nucleon system. The first trivial reduction in the QC clearly takes place among different spin/isospin channels. In particular, it is straightforward to see that the QC in Eq. (\ref{NNQC}) does not mix $(S=0,I=1)$, $(S=1,I=0)$, $(S=0,I=0)$ and $(S=1,I=1)$ sectors, and automatically breaks into four independent determinant conditions that correspond to different spin-isospin sectors,
\begin{eqnarray}
\textrm{Det}\left[{(\mathcal{M}^{\infty})^{-1}+\delta\mathcal{G}^{V}}\right]=\prod_{I=0}^{1}\prod_{S=0}^{1} \det \left[(\mathcal{M}^{\infty}_{(I,S)})^{-1}+\delta \mathcal{G}^{V}_{(I,S)} \right]=0 .
\label{NNQC-IS}
\end{eqnarray}
This is due to the fact that each J-sub block of the scattering amplitude matrix can be separated into three independent sectors as following
\begin{align}
&\mathcal{M}^{\infty}_{(I,1)}\equiv\left(\begin{array}{ccc}
\mathcal{M}_{J;I}^{(J-1,J-1;1)} &  & \mathcal{M}_{J;I}^{(J-1,J+1;1)}\\
\\
\mathcal{M}_{J;I}^{(J+1,J-1;1)} &  & \mathcal{M}_{J;I}^{(J+1,J+1;1)}
\end{array}\right), ~\mathcal{M}^{\infty}_{(I,0)}\equiv\begin{array}{c}
\mathcal{M}_{J;I}^{(J,J;0)}\end{array}, ~\mathcal{M}^{\infty}_{(I',1)}\equiv
\begin{array}{c}
\mathcal{M}_{J;I'}^{(J,J;1)}\end{array},
\nonumber\\
\label{amplitude-IS}
\end{align}
where $I$ and $I'$ are defined after Eq. (\ref{KernelJ}). Since the $M_J$ and $M_I$ indices are being suppressed, one should keep in mind that each block is still a $(2J+1)^2\times(2I+1)^2$ diagonal matrix. If $J$ is even, these amplitudes describe scattering in the negative parity isotriplet, positive parity isotriplet and positive parity isosinglet channels, respectively. For an odd $J$, these amplitudes correspond to scattering in the positive parity isosinglet, negative parity isosinglet and negative parity isotriplet channels, respectively.
Due to the reduced symmetry of the FV, $\delta \mathcal{G}^V$ has off-diagonal terms in the basis of total angular momentum $J$. So although the QC in Eq. (\ref{NNQC}) fully breaks down in the $(I,S)$-basis, it remains coupled in the $(J,L)$-basis. In order to further reduce the determinant conditions in Eq. (\ref{NNQC-IS}), the symmetries of the FV functions must be studied in more detail. This will be the topic of the next section of this chapter.

Before moving on to such symmetry considerations, let us conclude this section by showing that the master QC derived in this section is real as it must be. It can be verified that, e.g. for systems at rest, the only imaginary part of the FV matrix $\delta \mathcal{G}^V$ shows up in the diagonal elements of this matrix.\footnote{This is not always the case as for example, the $\delta \mathcal{G}^V$ matrix for the $\mathbf{d}=(1,1,0)$ boost contains off-diagonal complex elements as well. For all of those case, we have verified that although the elements of the matrix $(\mathcal{M}^{\infty})^{-1}+\delta\mathcal{G}^{V}$ are complex, the determinant of the matrix remains real, see Refs. \cite{Briceno:2013lba, BDLsupp, Briceno:2013rwa}.} So in general, for the angular momentum channels $J$ where there is no coupling between different partial waves, the inverse scattering amplitude matrix has only diagonal elements, whose imaginary part exactly cancels that of the $\delta \mathcal{G}^V$ matrix, see Eq. (\ref{M-single}). Explicitly,
\begin{eqnarray}
\Im [(\mathcal{M}^{LL;S}_{JM_J;IM_I})^{-1}+\delta \mathcal{G}^{V,(LL;S)}_{JM_J,JM_J;IM_I}]=-\frac{iMk^*}{4\pi}+\frac{iMk^*}{4\pi}=0.
\label{Im-diagonal}
\end{eqnarray}
For the angular momentum channels where there are off-diagonal terms due to the partial-wave mixing, one can still write the inverse of the scattering amplitude in that sector, Eqs. (\ref{M-coupled1}, \ref{M-coupled2}), as following
\begin{small}
\begin{align}
& (\mathcal{M}^{LL';1})^{-1}=
\left( \begin{array}{cc}
-\frac{Mk^*}{4\pi}
\frac{\cos{2\epsilon}~\sin({\delta '-\delta})
+\sin({\delta '+\delta})}{\cos({\delta '+\delta})-\cos({\delta '-\delta})\cos({2\epsilon})}-\frac{iMk^*}{4\pi}&
\frac{Mk^*}{2\pi}
\frac{\cos(\epsilon)\sin(\epsilon)}{\cos(\delta '+\delta)-\cos(\delta '-\delta)\cos(2\epsilon)}
\\
\frac{Mk^*}{2\pi}
\frac{\cos({\epsilon})\sin({\epsilon})}{\cos({\delta '+\delta})-\cos({\delta '-\delta})\cos({2\epsilon})}
&
-\frac{Mk^*}{4\pi}
\frac{\cos(2\epsilon)~\sin({\delta-\delta '})
+\sin(\delta '+\delta)}{\cos(\delta '+\delta)-\cos(\delta '-\delta)\cos(2\epsilon)}-\frac{iMk^*}{4\pi}
\\
\end{array} \right),
\nonumber\\
\label{Minverse-coupled}
\end{align}
\end{small}
where $L=J \pm 1$ ($L'=J \mp 1$) and $\delta$ ($\delta '$) denotes the phase shift corresponding to the $L$ ($L'$) partial wave. As is seen the off-diagonal elements of this matrix are real. Given that the FV function $\delta \mathcal{G}^V$ has real off-diagonal terms in this case, these terms in the QC lead to a real off-diagonal element. For the diagonal elements, the imaginary part of the inverse scattering amplitude is isolated and has the same form as the imaginary part of the $\delta \mathcal{G}^V$ matrix, so a similar cancellation as that given in Eq. (\ref{Im-diagonal}) occurs in this case as well.

\section{Symmetry Considerations and Quantization Conditions \label{sec: Reduction}}
Although it is convenient to think of the determinant condition, Eq. (\ref{NNQC}), as a determinant in the $J$ basis, one should expect that for zero CM momentum, this equation splits into $5$ independent QCs corresponding to the $5$ irreps of the cubic group (see table (\ref{groups})). Furthermore, the degeneracy of the energy eigenvalues will reflect the dimension of the corresponding irrep. In general, the FV matrix $\delta \mathcal{G}^{V}$, Eq. (\ref{deltaG}), although being sparse, mixes states corresponding to different irreps of the cubic group. As a result, at least a partial block diagonalization of this matrix is necessary to unfold different irreps that are present due to the decomposition of a given total angular momentum $J$. When the two-particle system is boosted, the symmetry group of the system is no longer cubic, and the reduction of the determinant condition, Eq. (\ref{NNQC}), takes place according to the irreps of the corresponding point group, Table. (\ref{groups}). In the following section, this reduction procedure and the method of block diagonalization will be briefly discussed.
\begin{table} [h]
\begin{centering}
\includegraphics[scale=0.965]{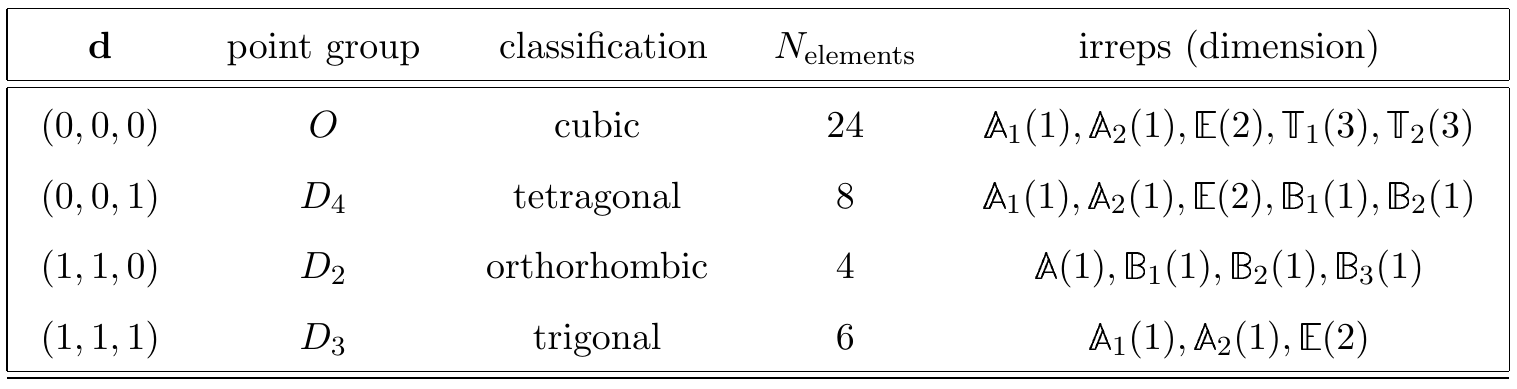}
\caption{{\small The classification of the point groups corresponding to the symmetry groups of the FV calculations for NN systems with the exact isospin symmetry with three selected lowest boost vectors.}\label{groups}}
\par\end{centering}
\end{table}

In order to calculate matrix elements of the FV matrix $\delta \mathcal{G}^{V}$, one can take advantage of the symmetries of the $c_{lm}^{\mathbf{P}}$ functions as defined in Eq. (\ref{clm}). The relations between non-zero $c_{lm}^{\mathbf{P}}$s for any given angular momentum $l$ can be easily deduced from the transformation properties of these functions under symmetry operations of the corresponding point groups
\begin{eqnarray}
c^{\mathbf{P}}_{lm}=\sum_{m'=-l}^{l}\mathcal{D}^{(l)}_{mm'}(R_{\mathcal{X}})~c^{\mathbf{P}}_{lm'},
\label{clm-trans}
\end{eqnarray}
where $R_{\mathcal{X}}$ is the rotation matrix corresponding to each symmetry operation $\mathcal{X}$ of the group, and $\mathcal{D}^{(l)}_{mm'}$ denotes the matrix elements of the Wigner $\mathcal{D}$-matrix \cite{Luscher:1990ux}. Besides these transformations, one can see that $c^\mathbf{P}_{lm}$s are invariant under inversion as can be easily verified from Eq. (\ref{clm}) for an arbitrary boost, and as a result all $c^\mathbf{P}_{lm}$s with an odd $l$ vanish.\footnote{For systems with non-equal masses, this is no longer true when the system is boosted. Since the parity is broken for such systems, even and odd partial waves mix with each other in the QCs, see Refs. \cite{Bour:2011ef, Davoudi:2011md, Fu:2011xz, Leskovec:2012gb}.} Table (\ref{nonzero-clm}) contains all such relations for non-vanishing $c^\mathbf{P}_{lm}$s up to $l=6$ for $\mathbf{d}=(0,0,0)$, $\mathbf{d}=(0,0,1)$ and $\mathbf{d}=(1,1,0)$ boost vectors.\footnote{A closer look at nonrelativistic $c_{lm}^{\mathbf{P}}$ functions shows that all $c_{2,\pm2}^{\mathbf{(1,1,0)}}$ and $c_{4,\pm2}^{\mathbf{(1,1,0)}}$ vanish. This extra symmetry of NR systems with equal masses significantly simplifies the QCs presented in appendix C for boost vector $(1,1,0)$. In this limit, the QCs corresponding to the boost vector $(1,1,1)$ are equivalent to the ones obtained for the system at rest.}
\begin{table} [h]
\label{tab:param3}
\begin{centering}
\begin{tabular}{|c|c|c|}
\hline 
\textbf{d}=(0,0,0) & \textbf{d}=(0,0,1) & \textbf{d}=(1,1,0)\tabularnewline
\hline 
\hline 
$~c_{00}^{P}~$ & $~c_{00}^{P}~$ & $~c_{00}^{P}~$\tabularnewline
$~c_{40}^{P}~$ & $~c_{20}^{P}~$ & $~c_{20}^{P}~$\tabularnewline
$~c_{44}^{P}=c_{4,-4}^{P}=\sqrt{\frac{5}{14}}c_{40}^{P}~$ & $~c_{40}^{P}~$ & $~c_{22}^{P}=-c_{2,-2}^{P}~$\tabularnewline
$~c_{60}^{P}~$ & $~c_{44}^{P}=c_{4,-4}^{P}~$ & $~c_{40}^{P}~$\tabularnewline
$~c_{64}^{P}=c_{6,-4}^{P}=-\sqrt{\frac{7}{2}}c_{60}^{P}~$ & $~c_{60}^{P}~$ & $~c_{42}^{P}=-c_{4,-2}^{P}~$\tabularnewline
 & $~c_{64}^{P}=c_{6,-4}^{P}~$ & $~c_{44}^{P}=c_{4,-4}^{P}~$\tabularnewline
 &  & $~c_{60}^{P}~$\tabularnewline
 &  & $~c_{62}^{P}=-c_{6,-2}^{P}~$\tabularnewline
 &  & $~c_{64}^{P}=c_{6,-4}^{P}~$\tabularnewline
\hline 
\end{tabular}
\par\end{centering}
\caption{The nonzero $c_{lm}^P$s up to $l=6$ for three different boost vectors $\mathbf{d}$ and for two particles with equal masses..}
\label{nonzero-clm}
\end{table}

An important point regarding the $c^\mathbf{P}_{lm}$ functions is that they explicitly depend on the direction of the boost vector. In other words, $c^\mathbf{P}_{lm}$s that correspond to different boost vectors with the same magnitude $|\mathbf{d}|=n$  are not equal. As a result the corresponding set of non-zero $c^\mathbf{P}_{lm}$s as well as the relations among them, for permutations of the components of $(0,0,1)$ and $(1,1,0)$ boost vectors are different from those that are listed in Table (\ref{nonzero-clm}). Although this difference in general results in different $\delta \mathcal{G}^{V}$ matrices, e. g. for $(1,0,0)$, $(0,1,0)$ and $(0,0,1)$ boost vectors, as is shown in appendix \ref{app:invariant}, the master equation (\ref{NNQC}) is invariant under a  $\mathbf{P}\rightarrow \mathbf{P}'$ transformation when $\mathbf{P}$ and $\mathbf{P}'$ are related by a cubic rotation and  $|\mathbf{P}|=|\mathbf{P}'|$. The reason is that there exists a unitary transformation that relates $\delta \mathcal{G}^{V,\mathbf{P}}$ to $\delta \mathcal{G}^{V,\mathbf{P}'}$, leaving the determinant condition invariant. Since the relations among $c^\mathbf{P}_{lm}$ are simpler when one assumes boost vectors that are special with respect to the z-axis, we have presented in Refs. \cite{Briceno:2013lba, BDLsupp, Briceno:2013rwa} the QCs corresponding to $\mathbf{d}=(0,0,1)$ and $\mathbf{d}=(1,1,0)$ boost vectors only. One can still use the deduced QCs to extract the scattering parameters of the NN system from the energy eigenvalues of lattice calculations with other permutations of these boost vectors. It is however crucial to input the boost vectors that are specified in this paper when calculating the $c^\mathbf{P}_{lm}$ functions in the QCs (instead of the boost vectors that are used in the lattice calculation). In order to increase statistics and the precision of results, one should perform the lattice calculation with all possible boost vectors of a given magnitude that belong to the same $A_1$ irrep of the cubic group,\footnote{In higher momentum shells, there occurs multiple $A_1$ irreps of the cubic group. This indicates that there are classes of momentum vector that do not transform into each other via a symmetry operation of the cubic group, e. g. $(2,2,1)$ and $(0,0,3)$ vectors in the $\mathbf{n}^2=9$ shell. However, as is discussed, another property of the $c_{lm}^{\mathbf{P}}$ functions for non-relativistic degenerate masses indicates that the value of the FV function is the same for these two boost vectors as they are both of the form $(2n_1,2n_2,2n_3+1)$ with $n_i \in \mathbb{Z}$.} and use the average energy eigenvalues in the QCs presented to determine the scattering parameters; keeping in mind that $c^\mathbf{P}_{lm}$ functions have to be evaluated at the boost vectors considered in Refs. \cite{Briceno:2013lba, BDLsupp, Briceno:2013rwa}.

The other fact that should be pointed out is that due to the symmetries of the $c^\mathbf{P}_{lm}$ function for equal masses, the system at rest with $\mathbf{d}=(0,0,0)$ exhibits the same symmetry transformation as that of the $(2n_1,2n_2,2n_3)$ boost where $n_1,n_2,n_3$ are integers. Similarly, the symmetry group of the calculations with $(0,0,1)$ ($(1,1,0)$) boost is the same as that of $(2n_1,2n_2,2n_3+1)$ ($(2n_1+1,2n_2+1,2n_3)$) boosts. As a result, the quantization conditions presented in Refs. \cite{Briceno:2013lba, BDLsupp, Briceno:2013rwa} can be used with these boost vectors as well. It is worth mentioning that for relativistic two-particle systems with degenerate masses, the above statement is no longer true. This is due to the fact that the boost vector dependence of the relativistic $c_{lm}^{\mathbf{P}}$ function is different from that of the NR counterpart, leading to more distinct point group symmetries for different boosts \cite{Rummukainen:1995vs, Kim:2005gf, Christ:2005gi}.

Back to our main goal, we aim to break the master equations (\ref{NNQC-IS}) into separate QCs corresponding to each irrep of the symmetry group of the problem. In fact, from the transformation law of the $\delta \mathcal{G}^{V}$ function under a symmetry operation of the group, 
\begin{align}
&\left[\delta\mathcal{G}^V\right]_{JM_J,LS;J'M_J',L'S}=\sum_{\bar{M}_J=-J}^{J}\sum_{\bar{M}_J'=-J'}^{J'}\mathcal{D}^{(J)}_{M_J,\bar{M}_J}(R_{\mathcal{X}})\left[\delta\mathcal{G}^V\right]_{J\bar{M}_J,LS;J'\bar{M}_J',L'S}\mathcal{D}^{(J')}_{\bar{M}_J',M_J'}(R_{\mathcal{X}}^{-1}),
\nonumber\\
\label{dG-trans}
\end{align}
one can deduce that there is a unitary transformation which brings the matrix $\delta \mathcal{G}^{V}$ to a block-diagonal form. Note that we have suppressed the isospin quantum numbers as $\delta\mathcal{G}^V$ is diagonal in the isospin basis. Each of these blocks then can be identified by a given irrep of the symmetry group of the problem. Such transformation eventually breaks the determinant conditions (\ref{NNQC-IS}) to separate determinant conditions corresponding to each irrep of the point group of the system. Explicitly in each spin and isospin sector,
\begin{eqnarray}
\det \left[(\mathcal{M}^{\infty}_{(I,S)})^{-1}+\delta \mathcal{G}^{V}_{(I,S)} \right]=\prod_{\Gamma^i}\det\left[(\mathcal{M}^{\infty-1}_{(I,S)})_{\Gamma^i}+\delta \mathcal{G}^{V,\Gamma^i}_{(I,S)} \right]^{N(\Gamma^i)}=0 .
\label{NNQC-irrep}
\end{eqnarray}
where $\Gamma^i$ denotes each irrep of the corresponding group and $N(\Gamma^i)$ is the dimensionality of each irrep. The dimensionality of each of these smaller determinant conditions is  given by the multiplicity of each irrep in the decomposition of angular momentum channels that are being included in the scattering problem. As is seen in Refs. \cite{Briceno:2013lba, BDLsupp, Briceno:2013rwa}, although from the master quantization condition, for some of the NN channels with $J\leq4$ and $l\leq 3$, one has to deal with a determinant of $30 \times 30$ matrices, upon such reduction of the master equation, one arrives at QCs that require taking the determinant of at most $9 \times 9$ matrices. We demonstrate this procedure in more detail for one example and refer the reader to a complete list of 49 deduced QCs that we presented in Refs. \cite{Briceno:2013lba, BDLsupp, Briceno:2013rwa}.

\subsection{Reduction procedure for positive-parity isosinglet channel with $\mathbf{P}=\mathbf{0}$ \label{app:red-example}}
\noindent Consider the NN system in the positive parity isosinglet channel where the ground state in the infinite volume is known to be a shallow bound state, the deuteron, whose wave-function is an admixture of both S-wave and D-wave. In order to obtain the phase shifts and mixing parameter in this channel from the energy eigenvalues of the two-nucleon system at rest from a LQCD calculation, one must first construct sources and sinks that transform according to a given irrep of the cubic group, e.g. $T_1$ when $\textbf{P}=0$. The extracted energies then needs to be put in the determinant condition for this channel in the corresponding irrep of the cubic group, Eq. (\ref{NNQC-irrep}), and subsequently solve for the scattering parameters. If one assumes the contributions from scattering channels with $J>4$ and $l\geq4$ to be negligible, the scattering amplitude matrix in the LHS of Eq. (\ref{NNQC-irrep}) can be written as
\begin{eqnarray}
\mathcal{M}^{\infty}_{(0,1)}=\left(\begin{array}{cccc}
\mathcal{M}_{1;0}^{(0,0;1)} & \mathcal{M}_{1;0}^{(0,2;1)} & 0 & 0\\
\mathcal{M}_{1;0}^{(2,0;1)} & \mathcal{M}_{1;0}^{(2,2;1)} & 0 & 0\\
0 & 0 & \mathcal{M}_{2;0}^{(2,2;1)} & 0\\
0 & 0 & 0 & \mathcal{M}_{3;0}^{(2,2;1)}
\end{array}\right),
\end{eqnarray}
where each element, $ \mathcal{M}_{J;I}^{(L,L';S)}$, is a diagonal $(2J+1)^2\times(2I+1)^2$ dimensional matrix. As a result, this is an $18\times18$ matrix which is parametrized by two phase shifts and one mixing angle in the $J=1$ channel, and two D-wave phase shifts in the $J=2$ and $J=3$ channels. Although there is a mixing between D-wave and G-wave channels in the $J=3$ sector, due to the assumption of a negligible $G$-wave scattering, the scattering amplitude in this channel is truncated to the D-wave.

The elements of the FV matrix $\delta \mathcal{G}^V$ in the LHS of Eq. (\ref{NNQC-irrep}) for this channel can be evaluated from Eq. (\ref{deltaG}). The result reads
\begin{eqnarray}
\delta \mathcal{G}^V_{(0,1)}=\left(\begin{array}{cccc}
\delta\mathcal{G}{}_{1,1;0}^{V,(0,0;1)} & \delta\mathcal{G}_{1,1;0}^{V,(0,2;1)} & \delta\mathcal{G}_{12;0}^{V,(0,2;1)} & \delta\mathcal{G}_{1,3;0}^{V,(0,2;1)}\\
\\
\delta\mathcal{G}_{1,1;0}^{V,(2,0;1)} & \delta\mathcal{G}_{1,1;0}^{V,(2,2;1)} & \delta\mathcal{G}_{1,2;0}^{V,(2,2;1)} & \delta\mathcal{G}_{1,3;0}^{V,(2,2;1)}\\
\\
\delta\mathcal{G}_{2,1;0}^{V,(2,0;1)} & \delta\mathcal{G}_{2,1;0}^{V,(2,2;1)} & \delta\mathcal{G}_{2,2;0}^{V,(2,2;1)} & \delta\mathcal{G}_{2,3;0}^{V,(2,2;1)}\\
\\
\delta\mathcal{G}_{3,1;0}^{V,(2,0;1)} & \delta\mathcal{G}_{3,1;0}^{V,(2,2;1)} & \delta\mathcal{G}_{3,2;0}^{V,(2,2;1)} & \delta\mathcal{G}_{3,3;0}^{V,(2,2;1)}
\end{array}\right),
\end{eqnarray}
where each element still represents a matrix $\delta\mathcal{G}_{J,J';I}^{V,(L,L';S)}$
in the $|J,M_J\rangle$ basis and whose explicit forms are as following\footnote{We will drop the superscript $\mathbf{P}$ on the $c_{lm}$s in this example as they are evaluated for $\mathbf{P}=0$.}
\begin{eqnarray}
\delta\mathcal{G}_{1,1;0}^{V,(0,0;1)}&=&\delta\mathcal{G}_{1,1;0}^{V,(2,2;1)}=M(-c_{00}+\frac{i k^*}{4 \pi })~\mathbf{I}_3,
\end{eqnarray}
\begin{eqnarray}
 \delta\mathcal{G}_{1,3;0}^{V,(2,2;1)}&=&\left[\delta\mathcal{G}_{3,1;0}^{V,(2,2;1)}\right]^T=\frac{M}{k^{*4}}c_{40}\left(
\begin{array}{ccccccc}
 0 & 0 & -\frac{3  }{7} & 0 & 0 & 0 & -\frac{\sqrt{15}}{7} \\
 0 & 0 & 0 & \frac{2 \sqrt{6}}{7} & 0 & 0 & 0 \\
 -\frac{\sqrt{15}}{7} & 0 & 0 & 0 & -\frac{3}{7} & 0 & 0
\end{array}
\right),
\end{eqnarray}
\begin{eqnarray}
\delta \mathcal{G}^{V,(2,2;1)}_{(2,2;0)}&=& M(-c_{00}+\frac{i k^*}{4 \pi})~\mathbf{I}_5+\frac{M}{k^{*4}}c_{40}\left(\begin{array}{ccccc}
 \frac{2 }{21} & 0 & 0 & 0 & \frac{10 }{21} \\
 0 &-\frac{8 }{21} & 0 & 0 & 0 \\
 0 & 0 & \frac{4 }{7} & 0 & 0 \\
 0 & 0 & 0 &-\frac{8 }{21} & 0 \\
 \frac{10 }{21} & 0 & 0 & 0 & \frac{2 }{21}
\end{array}\right),
\end{eqnarray}
\begin{eqnarray}
\delta \mathcal{G}^{V,(2,2;1)}_{(2,3;0)}&=&\left[\delta \mathcal{G}^{V,(2,2;1)}_{(3,2;0)}\right]^T=\frac{M}{k^{*4}}c_{40}\left(
\begin{array}{ccccccc}
 0 & \frac{5 \sqrt{2}}{21} & 0 & 0 & 0 & \frac{5 \sqrt{2}}{21} & 0 \\
 0 & 0 & -\frac{5 \sqrt{5}}{21} & 0 & 0 & 0 & \frac{5}{7 \sqrt{3}} \\
 0 & 0 & 0 & 0 & 0 & 0 & 0 \\
 -\frac{5}{7 \sqrt{3}} & 0 & 0 & 0 & \frac{5 \sqrt{5}}{21} & 0 & 0 \\
 0 & -\frac{5 \sqrt{2}}{21} & 0 & 0 & 0 & -\frac{5 \sqrt{2}}{21} & 0
\end{array}
\right),
\nonumber\\
\end{eqnarray}
\begin{eqnarray}
\delta \mathcal{G}^{V,(2,2;1)}_{(3,3;0)}&=&M(-c_{00}+\frac{i k^*}{4 \pi})~\mathbf{I}_7-
\frac{M}{k^{*4}}c_{40}\left(\begin{array}{ccccccc}
 \frac{1}{7} & 0 & 0 & 0 & \frac{\sqrt{\frac{5}{3}}}{7} & 0 & 0 \\
 0 & -\frac{1}{3} & 0 & 0 & 0 & \frac{5}{21} & 0 \\
 0 & 0 & \frac{1}{21} & 0 & 0 & 0 & \frac{\sqrt{\frac{5}{3}}}{7} \\
 0 & 0 & 0 & \frac{2}{7} & 0 & 0 & 0 \\
 \frac{\sqrt{\frac{5}{3}}}{7} & 0 & 0 & 0 & \frac{1}{21} & 0 & 0 \\
 0 & \frac{5}{21} & 0 & 0 & 0 & -\frac{1}{3} & 0 \\
 0 & 0 & \frac{\sqrt{\frac{5}{3}}}{7} & 0 & 0 & 0 & \frac{1}{7}
\end{array}\right),
\nonumber\\
\end{eqnarray}
where $\mathbf{I}_n$ is the $n \times n$ identity matrix, and the rest of the blocks are zero. As is suggested in Ref. \cite{Luu:2011ep}, a unitary matrix, that can bring the $\delta \mathcal{G}^V$ matrix into a block-diagonalized form, can be found by diagonalizing the blocks that are located on the diagonal of the $\delta \mathcal{G}^V$ matrix, $\delta \mathcal{G}^{V,(L,L';1)}_{(J,J;0)}$. Then a unitary matrix can be found easily based on the method described in Ref. \cite{Luu:2011ep} which brings $\delta \mathcal{G}^V$ to a (partially) block-diagonal form. One finds
\begin{eqnarray}
S=\left(\begin{array}{ccc}
S_{11} & 0 & 0\\
0 & S_{22} & 0\\
0 & 0 & S_{33}
\end{array}\right),
\end{eqnarray}
where the zero elements denote subblocks of appropriate dimension with all elements equal to zero, and the nontrivial blocks are the following matrices
\begin{small}
\begin{align}
&S_{11}=\mathbf{I}_6,
~S_{22}=\left(\begin{array}{ccccc}
0 & 0 & 0 & 1 & 0\\
0 & 1 & 0 & 0 & 0\\
0 & 0 & 1 & 0 & 0\\
-\frac{1}{\sqrt{2}} & 0 & 0 & 0 & \frac{1}{\sqrt{2}}\\
\frac{1}{\sqrt{2}} & 0 & 0 & 0 & \frac{1}{\sqrt{2}}
\end{array}\right), 
~S_{33}=\left(\begin{array}{ccccccc}
0 & 0 & \sqrt{\frac{3}{8}} & 0 & 0 & 0 & \sqrt{\frac{5}{8}}\\
\sqrt{\frac{5}{8}} & 0 & 0 & 0 & \sqrt{\frac{3}{8}} & 0 & 0\\
0 & 0 & 0 & 1 & 0 & 0 & 0\\
0 & 0 & -\sqrt{\frac{5}{8}} & 0 & 0 & 0 & \sqrt{\frac{3}{8}}\\
0 & \frac{1}{\sqrt{2}} & 0 & 0 & 0 & \frac{1}{\sqrt{2}} & 0\\
-\sqrt{\frac{3}{8}} & 0 & 0 & 0 & \sqrt{\frac{5}{8}} & 0 & 0\\
0 & -\frac{1}{\sqrt{2}} & 0 & 0 & 0 & \frac{1}{\sqrt{2}} & 0
\end{array}\right).
\nonumber\\
\end{align}
\end{small}
The resultant (partially) block-diagonalized matrix can then be obtained by,
\begin{small}
\begin{eqnarray}
&& S[(\mathcal{M}^{\infty}_{(0,1)})^{-1}+\delta \mathcal{G}^{V}_{(0,1)}]S^T=
\nonumber\\
&& \left(\begin{array}{cccccccccccccccccc}
x_{1} & 0 & 0 & y_{1} & 0 & 0 & 0 & 0 & 0 & 0 & 0 & 0 & 0 & 0 & 0 & 0 & 0 & 0\\
0 & x_{1} & 0 & 0 & y_{1} & 0 & 0 & 0 & 0 & 0 & 0 & 0 & 0 & 0 & 0 & 0 & 0 & 0\\
0 & 0 & x_{1} & 0 & 0 & y_{1} & 0 & 0 & 0 & 0 & 0 & 0 & 0 & 0 & 0 & 0 & 0 & 0\\
y_{1} & 0 & 0 & x_{2} & 0 & 0 & 0 & 0 & 0 & 0 & 0 & -y_{2} & 0 & 0 & 0 & 0 & 0 & 0\\
0 & y_{1} & 0 & 0 & x_{2} & 0 & 0 & 0 & 0 & 0 & 0 & 0 & 0 & y_{2} & 0 & 0 & 0 & 0\\
0 & 0 & y_{1} & 0 & 0 & x_{2} & 0 & 0 & 0 & 0 & 0 & 0 & -y_{2} & 0 & 0 & 0 & 0 & 0\\
0 & 0 & 0 & 0 & 0 & 0 & x_{3} & 0 & 0 & 0 & 0 & 0 & 0 & 0 & 0 & 0 & y_{3} & 0\\
0 & 0 & 0 & 0 & 0 & 0 & 0 & x_{3} & 0 & 0 & 0 & 0 & 0 & 0 & y_{3} & 0 & 0 & 0\\
0 & 0 & 0 & 0 & 0 & 0 & 0 & 0 & x_{4} & 0 & 0 & 0 & 0 & 0 & 0 & 0 & 0 & 0\\
0 & 0 & 0 & 0 & 0 & 0 & 0 & 0 & 0 & x_{3} & 0 & 0 & 0 & 0 & 0 & -y_{3} & 0 & 0\\
0 & 0 & 0 & 0 & 0 & 0 & 0 & 0 & 0 & 0 & x_{4} & 0 & 0 & 0 & 0 & 0 & 0 & 0\\
0 & 0 & 0 & -y_{2} & 0 & 0 & 0 & 0 & 0 & 0 & 0 & x_{5} & 0 & 0 & 0 & 0 & 0 & 0\\
0 & 0 & 0 & 0 & 0 & -y_{2} & 0 & 0 & 0 & 0 & 0 & 0 & x_{5} & 0 & 0 & 0 & 0 & 0\\
0 & 0 & 0 & 0 & y_{2} & 0 & 0 & 0 & 0 & 0 & 0 & 0 & 0 & x_{5} & 0 & 0 & 0 & 0\\
0 & 0 & 0 & 0 & 0 & 0 & 0 & y_{3} & 0 & 0 & 0 & 0 & 0 & 0 & x_{6} & 0 & 0 & 0\\
0 & 0 & 0 & 0 & 0 & 0 & 0 & 0 & 0 & -y_{3} & 0 & 0 & 0 & 0 & 0 & x_{6} & 0 & 0\\
0 & 0 & 0 & 0 & 0 & 0 & y_{3} & 0 & 0 & 0 & 0 & 0 & 0 & 0 & 0 & 0 & x_{6} & 0\\
0 & 0 & 0 & 0 & 0 & 0 & 0 & 0 & 0 & 0 & 0 & 0 & 0 & 0 & 0 & 0 & 0 & x_{7}
\end{array}\right),
\nonumber\\
\label{BD-form}
\end{eqnarray}
\end{small}
where
\begin{small}
\begin{align}
&x_{1}=-M c_{00}+\frac{iMk^*}{4\pi}+\frac{\mathcal{M}_{1;0}^{(2,2;1)}}{\det(\mathcal{M}^{SD})},~
x_{2}=-M c_{00}+\frac{iMk^*}{4\pi}+\frac{\mathcal{M}_{1;0}^{(0,0;1)}}{\det(\mathcal{M}^{SD})},
\nonumber\\
&x_{3}=-M c_{00}-\frac{8}{21}\frac{M}{k^{*4}}c_{40}+\frac{iMk^*}{4\pi}+\frac{1}{\mathcal{M}^{(22;1)}_{2;0}}, ~
x_{4}=-M c_{00}+\frac{4}{7}\frac{M}{k^{*4}}c_{40}+\frac{iMk^*}{4\pi}+\frac{1}{\mathcal{M}^{(22;1)}_{2;0}},
\nonumber\\
&x_{5}=-M c_{00}-\frac{2}{7}\frac{M}{k^{*4}}c_{40}+\frac{iMk^*}{4\pi}+\frac{1}{\mathcal{M}^{(22;1)}_{3;0}},~
x_{6}=-M c_{00}+\frac{2}{21}\frac{M}{k^{*4}}c_{40}+\frac{iMk^*}{4\pi}+\frac{1}{\mathcal{M}^{(22;1)}_{3;0}},
\nonumber\\
&x_{7}=-M c_{00}+\frac{4}{7}\frac{M}{k^{*4}}c_{40}+\frac{iMk^*}{4\pi}+\frac{1}{\mathcal{M}^{(22;1)}_{3;0}},
y_{1}=-\frac{\mathcal{M}_{1;0}^{(0,2;1)}}{\det(\mathcal{M}^{SD})},
y_{2}=\frac{2\sqrt{6}}{7}\frac{M}{k^{*4}}c_{40},
y_{3}=\frac{10\sqrt{2}}{21}\frac{M}{k^{*4}}c_{40},
\nonumber\\
\end{align}
\end{small}
and $\det(\mathcal{M}^{SD})$ in these relations denotes the determinant of the $J=1$ subblock of the scattering amplitude, $\det(\mathcal{M}^{SD})=\mathcal{M}_{1;0}^{(0,0;1)} \mathcal{M}_{1;0}^{(2,2;1)}- (\mathcal{M}_{1;0}^{(0,2;1)})^2$. This matrix can now clearly be broken to 4 independent blocks corresponding to 4 irreps of the cubic group. The degeneracy of the diagonal elements of this matrix, as well as the coupling between different rows and columns, indicate which irrep of the cubic group each block corresponds to. According to table \ref{irreps}, the one-dimensional irrep $A_2$ only occurs in the decomposition of $J=3$ angular momentum. As is seen from Eq. (\ref{BD-form}), the element $x_7$ belongs to the $J=3$ sector and has a one-fold degeneracy. Also it does not mix with other angular momentum channels, therefore it must correspond to the $A_2$ irrep. So the one-dimensional QC corresponding to the $A_2$ irrep is
\begin{eqnarray}
A_2: ~ \frac{1}{\mathcal{M}^{(22;1)}_{3;0}}-M c_{00}+\frac{4}{7}\frac{M}{k^{*4}}c_{40}+\frac{iMk^*}{4\pi}=0.
\label{A2}
\end{eqnarray}

The QC corresponding to the two-dimensional irrep $E$ can be also deduced easily as it only has overlap with the $J=2$ channel. Clearly the element corresponding to this irrep is $x_4$ with two-fold degeneracy and the corresponding QC reads
\begin{eqnarray}
E: ~ \frac{1}{\mathcal{M}^{(22;1)}_{2;0}}-M c_{00}+\frac{4}{7}\frac{M}{k^{*4}}c_{40}+\frac{iMk^*}{4\pi}=0.
\label{E}
\end{eqnarray}

The three-dimensional irrep $T_2$ appears in the decomposition of both $J=2$ and $J=3$ angular momentum, and as is seen from Eq. (\ref{BD-form}) mixes the $x_3$, $x_6$ and $y_3$ elements through the following QC
\begin{align}
&T_2: ~ \det\left(\begin{array}{cc}
\frac{1}{\mathcal{M}^{(22;1)}_{2;0}}-M c_{00}-\frac{8}{21}\frac{M}{k^{*4}}c_{40}+\frac{iMk^*}{4\pi} & \frac{10\sqrt{2}}{21}\frac{M}{k^{*4}}c_{40} \\
\frac{10\sqrt{2}}{21}\frac{M}{k^{*4}}c_{40} & \frac{1}{\mathcal{M}^{(22;1)}_{3;0}}-M c_{00}+\frac{2}{21}\frac{M}{k^{*4}}c_{40}+\frac{iMk^*}{4\pi}
\end{array}\right)=0.
\nonumber\\
\label{T2}
\end{align}
As is clear, the energy eigenvalues in this irrep have a three-fold degeneracy (there are three copies of this QCs) that is consistent with the dimensionality of the irrep. The remaining irrep is $T_1$ which is a three-dimensional irrep and contribute to both $J=1$ and $J=3$ channels. As there are two $J=1$ sectors corresponding to S-wave and D-wave scatterings, the QC must be the determinant of a $3 \times 3$ matrix. This is in fact the case by looking closely at the (partially) block-diagonalized matrix in Eq. (\ref{BD-form}). One finds explicitly
\begin{small}
\begin{align}
&T_1: \det\left(\begin{array}{ccc}
\frac{\mathcal{M}_{1;0}^{(2,2;1)}}{\det(\mathcal{M}^{SD})}-M c_{00}+\frac{iMk^*}{4\pi} & -\frac{\mathcal{M}_{1;0}^{(0,2;1)}}{\det(\mathcal{M}^{SD})} & 0\\
-\frac{\mathcal{M}_{1;0}^{(0,2;1)}}{\det(\mathcal{M}^{SD})} & \frac{\mathcal{M}_{1;0}^{(0,0;1)}}{\det(\mathcal{M}^{SD})}-M c_{00}+\frac{iMk^*}{4\pi} & -\frac{2\sqrt{6}}{7}\frac{M}{k^{*4}}c_{40}\\
0 & -\frac{2\sqrt{6}}{7}\frac{M}{k^{*4}}c_{40} & \frac{1}{\mathcal{M}^{(22;1)}_{3;0}}-M c_{00}-\frac{2}{7}\frac{M}{k^{*4}}c_{40}+\frac{iMk^*}{4\pi}
\end{array}\right)
\nonumber\\
& \qquad \qquad \qquad \qquad \qquad \qquad
\qquad \qquad  \qquad  \qquad \qquad \qquad \qquad \qquad 
\qquad \qquad \qquad \qquad ~ =0.
\label{T1}
\end{align}
\end{small}
Again there is a three-fold degeneracy for the energy-eigenvalues as there are three copies of this QC for this irrep. This is an important QC as it gives access to the mixing angle between S and D partial waves.

As is clear now, the QC for $A_2$ irrep, Eq. (\ref{A2}), by its own determines the phase shift in the $J=3$ channel, which can then be used in Eq. (\ref{T1}) for the $T_1$ irrep to determine the phase shifts and mixing angle in the $J=1$ channel. Eq. (\ref{E}) for the $E$ irrep gives access to the phase shift in the $J=2$ channel, and finally Eq. (\ref{T2}) provides another relation for the phase shifts in the $J=2$ and $J=3$ channels. In practice, one needs multiple energy levels in order to be able to reliably extract these parameters from the QCs presented. This is specially a challenging task when it comes to the determination of the scattering parameters in the channels with physical mixing, e. g. S-D mixing, since there are at least three unknown parameters to be determined from the QC, e. g. see Eq. (\ref{T1}). By doing the LQCD calculations of the boosted two-nucleon system, one will attain more energy levels that will correspond to another set of QCs. These QCs then provide a set of equations that the same scattering parameters satisfy, and therefore better constraints can be put on these quantities. These are the set of QCs we have tabulated in Refs. \cite{Briceno:2013lba, Briceno:2013rwa}.

\subsection{The systematic procedure for the reduction of the master quantization condition
\label{red-syst}}
\begin{center}
\begin{table}[h!]
\includegraphics[scale=0.525]{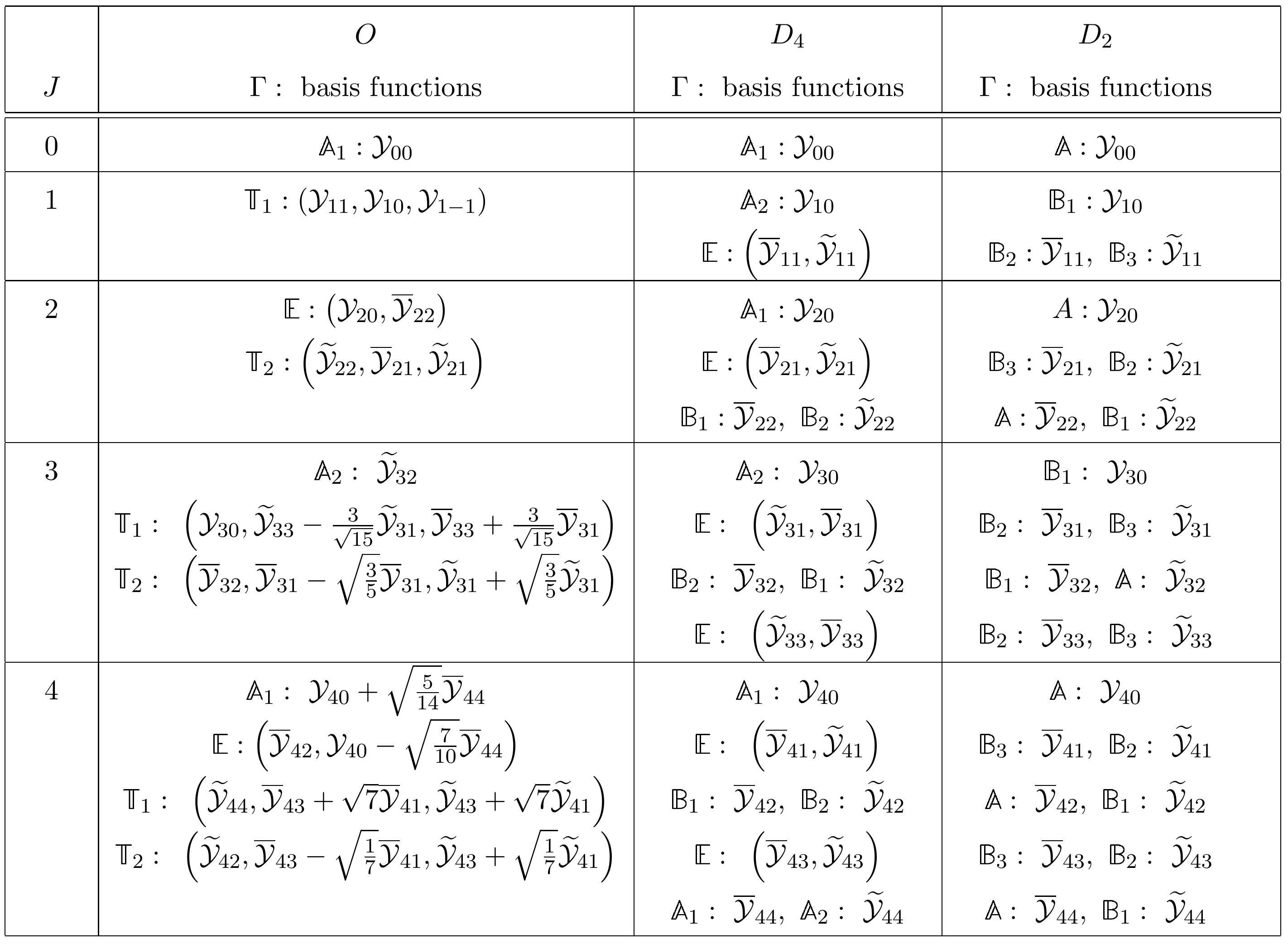}
\caption{{\small The decomposition of the irreps of the rotational group up to $J=4$ in terms of the irreps of the cubic ($O$), tetragonal ($D_4$) and orthorhombic ($D_2$) groups, see Refs. \cite{Luscher:1990ux, Feng:2004ua, Dresselhaus}. The corresponding basis functions  of each irrep are also given in the table in terms of the $SO(3)$ functions $\mathcal{Y}_{lm}$, where $\overline{\mathcal{Y}}_{lm}\equiv\mathcal{Y}_{lm}+\mathcal{Y}_{l-m}$ and $\widetilde{\mathcal{Y}}_{lm}\equiv \mathcal{Y}_{lm}-\mathcal{Y}_{l-m}$. These basis functions become useful in reduction of the full determinant condition, Eq. (\ref{NNQC}) into separate QCs corresponding to each irrep of the point group considered.
}\label{irreps}}
\end{table}
\end{center}
When there are multiple occurrences of a given irrep in each angular momentum $J$ (see table \ref{irreps}), the procedure of block diagonalization becomes more cumbersome, and a systematic procedure must be taken which is based on the knowledge of the basis functions corresponding to each occurrence of any given irrep. Such basis functions for the irreps of the point groups considered in this paper are tabulated in table \ref{irreps}. These basis functions correspond to each occurrence of the irreps in the decomposition of the angular momentum states into the irreps of the $O$, $D_4$ and $D_2$ point groups up to $J=4$. Each J-block (for any given $L$) of the desired unitary matrix $S$ can be constructed from the corresponding eigenvectors of each irrep and will immediately bring the FV matrix $\delta{G}^V$ to a block-diagonalized form. It is however important to keep in mind that the these basis functions are defined with respect to the boost vector of the system (helicity basis), and must be transformed to the $| J,M_J \rangle$ basis by proper rotation matrices as the $\delta \mathcal{G}^V$ and $\mathcal{M}$ matrices in Eqs. (\ref{amplitude}) and (\ref{deltaG}) are given in this latter basis.

For two particles with equal masses in the NR limit, the parity remains a good quantum number even for non-zero CM momentum. This is manifested in vanishing $c_{lm}^{\mathbf{P}}$s with an odd $l$ which ensures no mixing between different parity sectors will occur. This greatly simplifies the reduction of the QCs with a definite point group symmetry. In the next chapter, we encounter a situation where the FV calculation with particular BCs on the fields can suffer from such mixing between different parity sectors. Another example is the boosted two-particle system with non-equal masses (both relativistic and NR) \cite{Davoudi:2011md, Fu:2011xz, Leskovec:2012gb} as discussed in Sec. \ref{Two-body}. This implies that the upcoming FV studies of NN systems with physical values of $u$ and $d$ quark masses will suffer from a more severe FV-induced partial-wave mixings than that of the current calculations with the exact isospin symmetry.

\section*{Summary and Discussions \label{sec: S&M}}

The auxiliary field formalism was extended to arbitrary partial waves in both the scalar and nuclear sectors in infinite and finite volumes. Such a formalism can be used to derive a master equation that relates the FV two-nucleon energies and the scattering parameters of the two-nucleon systems with arbitrary spin, isospin and angular momentum. This master equation, Eq.~(\ref{NNQC}), that is the extension of the L\"uscher formalism to two-nucleon systems,  is valid up to inelastic thresholds and is general for any non-relativistic CM boost. 

The QC is a determinant over an infinite-dimensional matrix in the basis of angular momentum, and in practice it is necessary to truncate the number of partial waves that contribute to the scattering. By taking advantage of symmetries within the problem, we show how the master equation can be reduced to finite-size blocks that relate particular partial-wave channels (and their mixing) to different spin-isospin channels and different irreps of the corresponding point group of the system. By truncating the matrices at $J\leq 4$ and $l\leq 3$, this procedure requires block diagonalizing matrices as large as $30\times30$. The resulting QCs are determinant conditions involving matrices that are at most $9\times9$, and are therefore practical to be used in future LQCD calculations of NN systems. We have provided one explicit example of this reduction for the scattering in the positive-parity isosinglet channel for zero CM momentum.  All other QCs for different CM boosts, parity, isospin, spin, and angular momentum are enumerated in Refs. \cite{Briceno:2013lba, BDLsupp, Briceno:2013rwa}. These form $49$ independent QCs for four different spin and isospin channels giving access to all 16 phase shifts and mixing parameters in these channels. Given the fact that NN-systems couple different partial waves, in order to reliably extract scattering parameters from LQCD calculations, these calculations must be necessarily performed  in multiple boosts and various irreps of the corresponding symmetry group.
 
As is extensively discussed in this chapter, the QCs obtained in this chapter (see Refs. \cite{Briceno:2013lba, Briceno:2013rwa}) can be used for a more general set of boost vectors. Since the symmetry group of the two equal-mass problem depends on the evenness and oddness of the components of the boost vectors, the QCs presented for different irrepsof the corresponding symmetry groups in case of $(0,0,0)$, $(0,0,1)$ and $(1,1,0)$ boosts, can be equally used for the $(2n_1,2n_2,2n_3)$, $(2n_1,2n_2,2n_3+1)$ and $(2n_1+1,2n_2+1,2n_3)$ boost vectors $\forall~n_i\in \mathbb{Z}$. 
 The other generality of the QCs with regard to the boost vectors is that upon a cubic rotations of the CM boost vectors, the QCs remain unchanged although the FV functions will be different. As a result, one may use an average of the energy levels extracted from the NN correlation functions for all different boosts with the same magnitude, belonging to the same $A_1$ irrep of the cubic group, in the QCs of Refs. \cite{Briceno:2013lba, BDLsupp, Briceno:2013rwa} to improve the statistics of the calculation. However, it should be noted that in evaluating the FV functions ($c_{lm}^{\mathbf{P}}s$) only the boost vectors considered in this paper must be used. In summary, one can extract the desired scattering parameters of the NN-system by performing the following steps:
\begin{enumerate}
\item For a given irrep $\Gamma$, evaluate the $NN$ correlation function with all possible boost vectors with magnitude ${d}$ that are related to each other via a cubic rotation,

{\small $\{C_{NN}^{\Gamma,\textbf{d}_1},\ldots,C_{NN}^{\Gamma,\textbf{d}_{N_d}}\}$}.
\item Average the value of the correlation functions over all boost vectors used in the previous step, $C_{NN}^{\Gamma,{d}}=\sum_{i}^{N_d}C_{NN}^{\Gamma,\textbf{d}_i}/N_d$.
\item Obtain the non-relativistic finite volume energy, $E_{NR}^\Gamma=E_{NN}^\Gamma-2m_N$, from the asymptotic behavior of the correlation function and therefore obtain the value of the relative momentum $k^*$ from $k^*=\sqrt{M_NE-(\pi\mathbf{d})^2/L^2}$.
\item Determine scattering parameters from the QCs of Refs. \cite{Briceno:2013lba, BDLsupp, Briceno:2013rwa}:

\begin{enumerate}
\item Use $\mathbf{d}=(0,0,0)$ if $\mathbf{d}$ is a permutation of $(2n_1,2n_2,2n_3)$,
\item Use $\mathbf{d}=(0,0,1)$ if $\mathbf{d}$ is a permutation of $(2n_1,2n_2,2n_3+1)$,
\item Use $\mathbf{d}=(1,1,0)$ if $\mathbf{d}$ is a permutation of $(2n_1+1,2n_2+1,2n_3)$.
\end{enumerate}
\end{enumerate}

In the following chapter, we will use the QCs derived here, coupled with empirical two-nucleon phase shifts, to provide estimates for the energy levels expected to be seen in future LQCD calculations at the physical pion mass. This will allow for an estimation of the precision that is needed for such calculations specially with regard to the deuteron state. Clearly the impact of this formalism on our understanding of the nature of nuclear forces depends upon best implementing this formalism in the upcoming LQCD calculations of the NN systems.


\chapter{DEUTERON AND ITS PROPERTIES FROM A FINITE-VOLUME FORMALISM}
{\label{chap:deuteron}}


The lightest nucleus, the deuteron, played an important historical role in 
understanding the form of the nuclear forces and 
the developments that led to the modern phenomenological nuclear potentials,
e.g. Refs.~\cite{Wiringa:1994wb,Machleidt:2000ge}.
While challenging for LQCD calculations, postdicting the properties of the
deuteron, and other light nuclei, is a critical part of the verification of
LQCD technology
that is required in order to trust predictions of quantities
for which there is little or no experimental guidance.
In nature, the deuteron,
with  total angular momentum and parity of $J^\pi=1^+$,
is the only bound state of a neutron and proton,
bound by $B_d^{\infty}= 2.224644(34)~{\rm MeV}$.
While predominantly S-wave, the non-central components  of the nuclear forces
(the tensor force) induce a D-wave component,
and the $J^\pi=1^+$ two-nucleon (NN) sector 
that contains the deuteron
is a 
$\siii$-$\diii$ coupled-channels system.
An important consequence of the nonconservation
of orbital angular momentum
is that the deuteron is not spherical,
and possesses a non-zero quadrupole moment
(the experimentally measured value of the electric quadrupole moment of the  
deuteron is $Q_d=0.2859(3)~{\rm fm}^2$~\cite{PhysRevA.20.381}).
The S-matrix for this coupled-channels system 
can be parameterized by two phase shifts 
and one mixing angle,
with the mixing angle manifesting itself in the 
asymptotic $D/S$ ratio of the deuteron wavefunction, 
$\eta = 0.02713(6)$~\cite{PhysRev.93.1387, PhysRevC.47.473, deSwart:1995ui}.
A direct calculation of the three scattering 
parameters from QCD, at both physical and unphysical light-quark masses,
would provide important insights into the tensor components of the nuclear forces.

Corrections to the binding energy of a bound state, such as
the deuteron, depend exponentially upon the volume,
and are dictated by its size,
and also by the range of the nuclear forces.
With the assumption of a purely S-wave deuteron, 
the leading order (LO) volume corrections 
have been determined
for a deuteron at rest in a cubic volume of spatial extent $L$ and with the
fields subject to periodic BCs in the spatial directions~\cite{Luscher:1986pf, Luscher:1990ux, Beane:2003da}.
They are found to 
scale as ${1\over L}e^{-\kappa_d^{\infty} L}$, where
$\kappa_d^{\infty}$ is the 
infinite-volume
deuteron binding momentum
(in the non-relativistic limit, $\kappa_d^{\infty}=\sqrt{M B_d^{\infty}}$, with $M$ being the nucleon mass).
Volume corrections beyond LO have been determined, and extended to
systems that are moving in the volume~\cite{Konig:2011nz, Bour:2011ef,Davoudi:2011md}.
As $\eta$, $Q_d$, and other observables
dictated  by the tensor interactions, are small at the physical light-quark masses, 
FV analyses of existing LQCD calculations~\cite{Beane:2006mx, Beane:2012vq,  Beane:2013br,
  Yamazaki:2012hi} using 
L\"uscher's method~\cite{Luscher:1986pf, Luscher:1990ux} 
have taken the deuteron to be purely S-wave, neglecting the D-wave admixture,
even at unphysical pion masses, introducing a
systematic uncertainty into these 
analyses.~\footnote{Recent lattice effective field theory (EFT) calculations include
the effects of higher partial waves and mixing~\cite{Lee:2008fa, Bour:2012hn}, 
and thus are able to calculate matrix
elements of non-spherical quantities like $Q_d$ up to a given order in the
low-energy EFT, but their FV
analyses treat the deuteron as a S-wave~\cite{Lee:2008fa, Bour:2012hn}.
}
Although the mixing between the S-wave and D-wave is known to be small at the
physical light-quark masses, its contribution to the 
calculated FV binding energies must be determined in order to
address this systematic uncertainty.
Further, it is not known if the mixing
between these channels remains small at unphysical quark masses.
As the central and tensor components of the nuclear forces have
different forms, their contribution to the FV effects will, in
general, differ.
The contributions from the tensor interactions 
are found to be relatively enhanced for certain CM boosts in modest
volumes due to the reduced spatial symmetry of the system.
Most importantly, 
extracting the  S-D mixing angle at the deuteron binding energy,
in addition to  the S-wave scattering parameters,
requires a complete coupled-channels analysis of the FV 
spectrum.

In this chapter, we utilize the formalism presented in the previous chapter for NN systems with arbitrary CM momenta, spin, angular momentum and isospin, to explore how the S-D mixing angle 
at the deuteron binding energy, along with the binding energy itself,
can be optimally extracted from LQCD calculations performed in cubic
volumes with fields subject to 
periodic BCs (PBCs) in the spatial directions.
Using the phase shifts and mixing angles generated by phenomenological 
NN potentials that are fit to NN scattering data~\cite{NIJMEGEN},
the expected FV energy spectra in the 
positive-parity isoscalar channels
are determined  at the physical pion mass (we assume exact
isospin symmetry throughout). 
It is found that correlation functions of 
boosted NN systems will play a key role in extracting the S-D mixing angle in future LQCD calculations. 
The FV energy shifts of the ground state of different irreps of the symmetry groups 
associated with momenta $\mathbf{P}=\frac{2\pi}L(0,0,1)$ 
and $\frac{2\pi}L(1,1,0)$,
are found to have enhanced sensitivity to the mixing angle in modest volumes
and to
depend both on its magnitude and sign.
A feature of the FV spectra, 
with practical implications for future LQCD calculations,
is that the contribution 
to the energy splittings from  channels with $J>1$, 
made possible by the  reduced symmetry of the volume,
are negligible for $L\gtrsim 10~\text{fm}$ as the phase shifts in
those channels are small at low energies.
As the generation of multiple ensembles of 
gauge-field configurations at the physical light-quark masses
will require significant computational resources on capability-computing platforms,
we have investigated the viability of precision determinations of the deuteron
binding energy and scattering parameters 
from one lattice volume using 
the 
six bound-state energies
associated with 
CM momenta $|\mathbf{P}|\leq\frac{2\pi}L\sqrt{3}$. 
We have also considered extracting the asymptotic D/S ratio from the
behavior of the deuteron FV wavefunction and its
relation to the S-D mixing angle.
\section{Deuteron and the Finite-volume Spectrum
\label{sec:DeutFV}
}
\noindent
The spectra of energy eigenvalues of two nucleons in the isoscalar
channel with positive parity in a cubic volume subject to  PBCs are dictated by
the S-matrix elements in this sector, including those defining the 
$\siii$-$\diii$ coupled channels that contain the deuteron, as discussed in detail in the previous chapter. For the deuteron at rest as well as boosts vectors $(0,0,1)$, $(1,1,0)$ and $(1,1,1)$, the required QCs for the NN system in the positive-parity isoscalar channel are given in Appendix~\ref{app: QC} \cite{Briceno:2013lba, BDLsupp, Briceno:2013rwa, Briceno:2013bda}. 
Although the ultimate goal is to utilize these QCs in the analysis of the NN
spectra extracted from LQCD calculations, 
they can be used, in combination with the experimental NN scattering data, 
to predict the FV  spectra at the physical light-quark masses,
providing important guidance for future LQCD calculations. 
While for scattering states, 
the phase shifts and mixing angle from 
phenomenological  analyses of the experimental data~\cite{PhysRevC.48.792,
  PhysRevC.49.2950, PhysRevC.54.2851, PhysRevC.54.2869}
can be used in the QCs,
for bound states, however, it is necessary to use fit functions 
of the correct form to be continued to negative energies. 
Here we choose a different parametrization for the $J=1$ S-matrix than what we used in chapter \ref{chap:NN}. It is the Blatt-Biedenharn (BB) parameterization~\cite{Blatt:1952zza,PhysRev.93.1387}
\begin{eqnarray}
S_{(J=1)}=\left( \begin{array}{cc}
\cos\epsilon_1&-\sin\epsilon_1\\
\sin\epsilon_1&\cos\epsilon_1\\
\end{array} \right)
\left( \begin{array}{cc}
e^{2i\delta_{1\alpha}}&0\\
0&e^{2i\delta_{1\beta}}\\
\end{array} \right)
\left( \begin{array}{cc}
\cos\epsilon_1&\sin\epsilon_1\\
-\sin\epsilon_1&\cos\epsilon_1\\
\end{array} \right),
\label{eq:BBSmatrix}
\end{eqnarray}
\begin{figure}[b!]
\begin{center}  
\subfigure[]{
\includegraphics[scale=0.135]{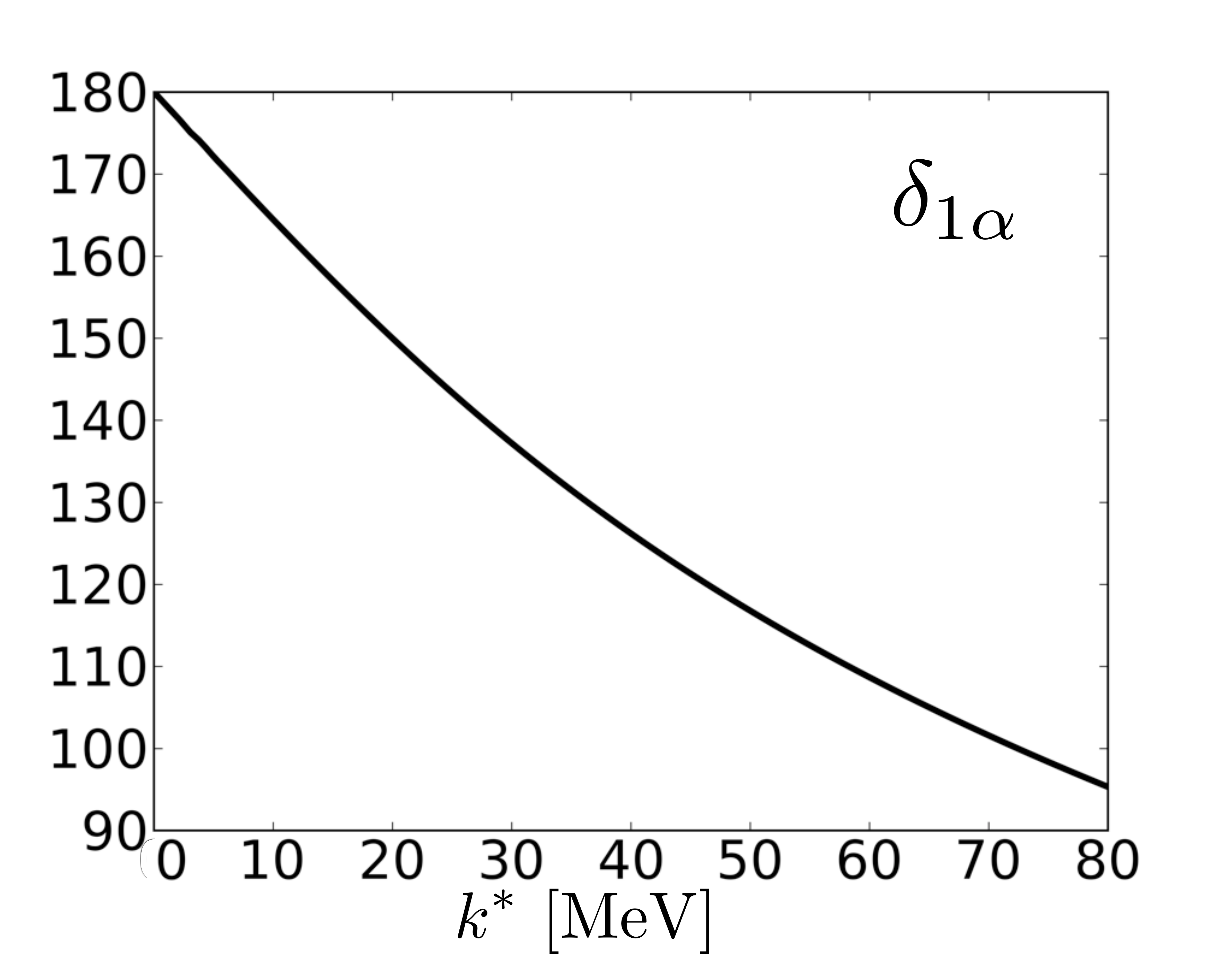}}
\subfigure[]{
\includegraphics[scale=0.135]{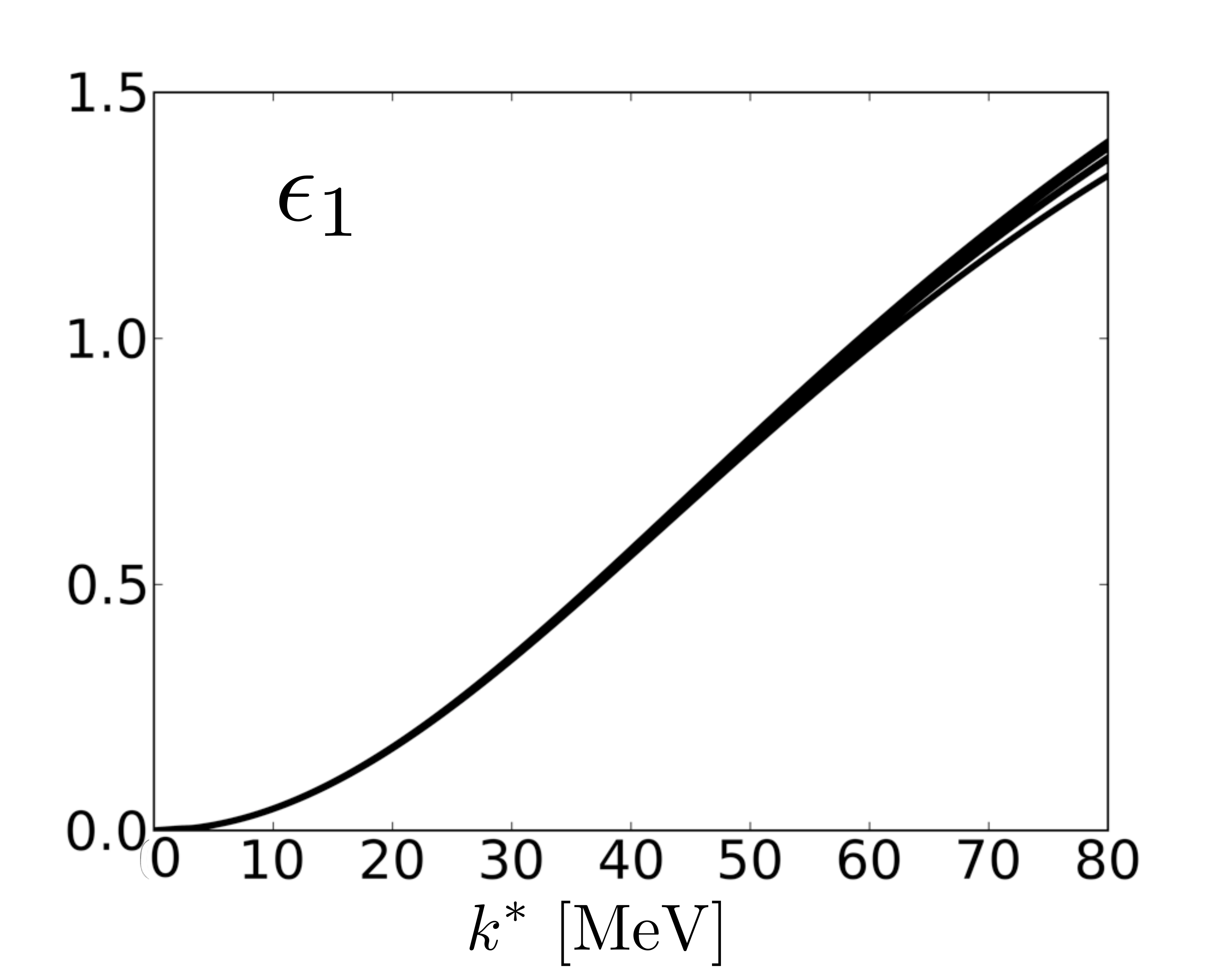}}
\subfigure[]{
\includegraphics[scale=0.135]{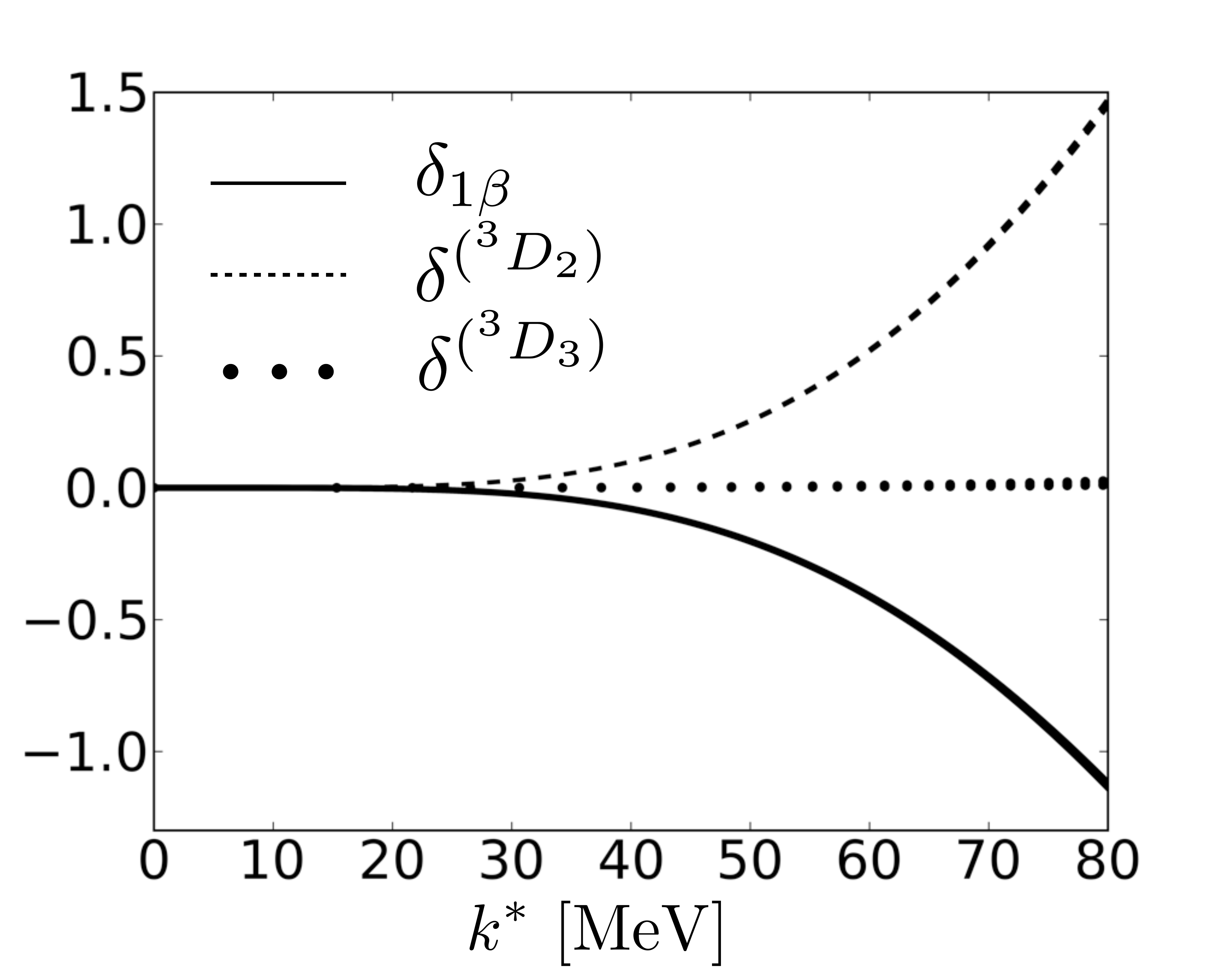}}
\subfigure[]{
\includegraphics[scale=0.135]{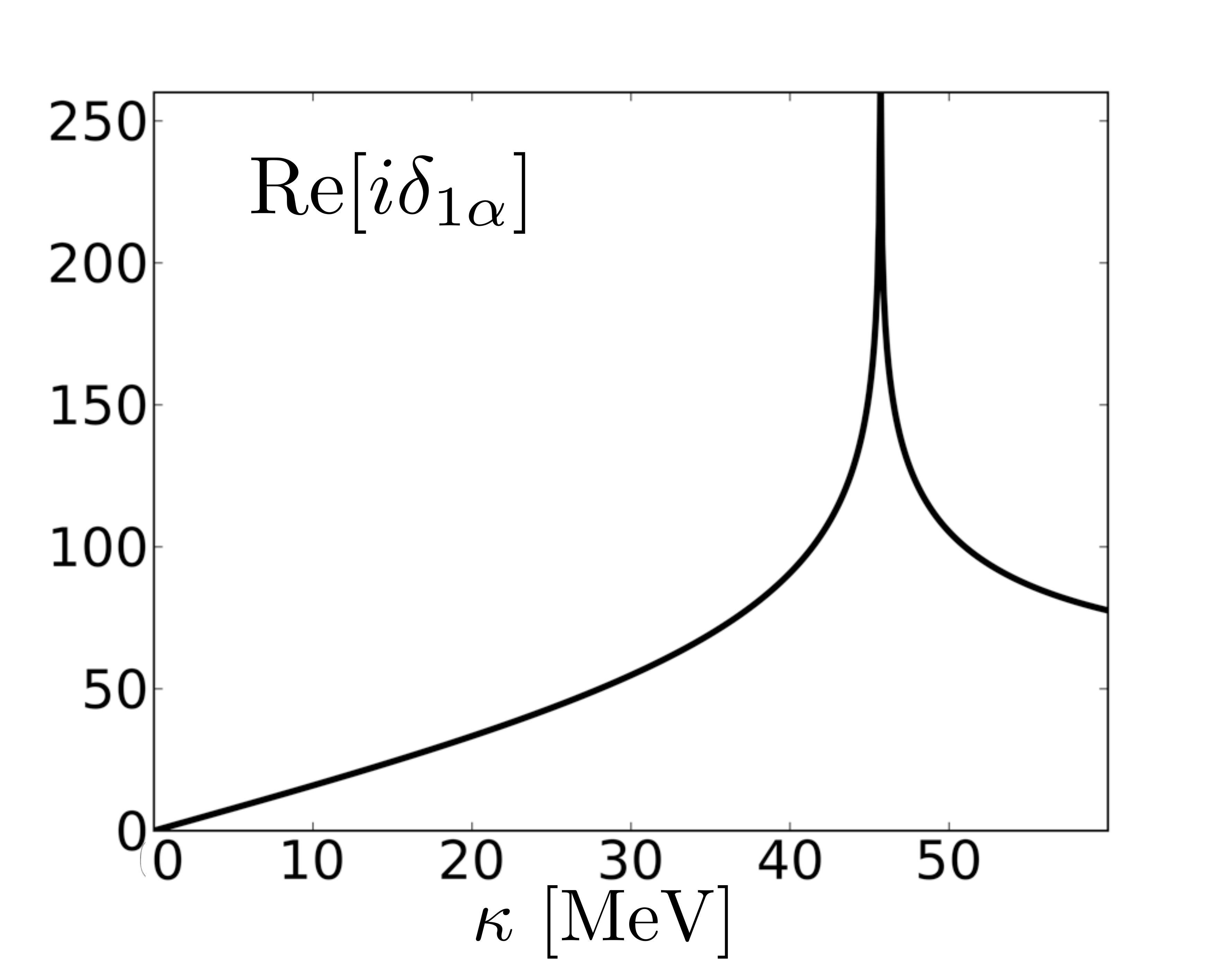}}
\subfigure[]{
\includegraphics[scale=0.135]{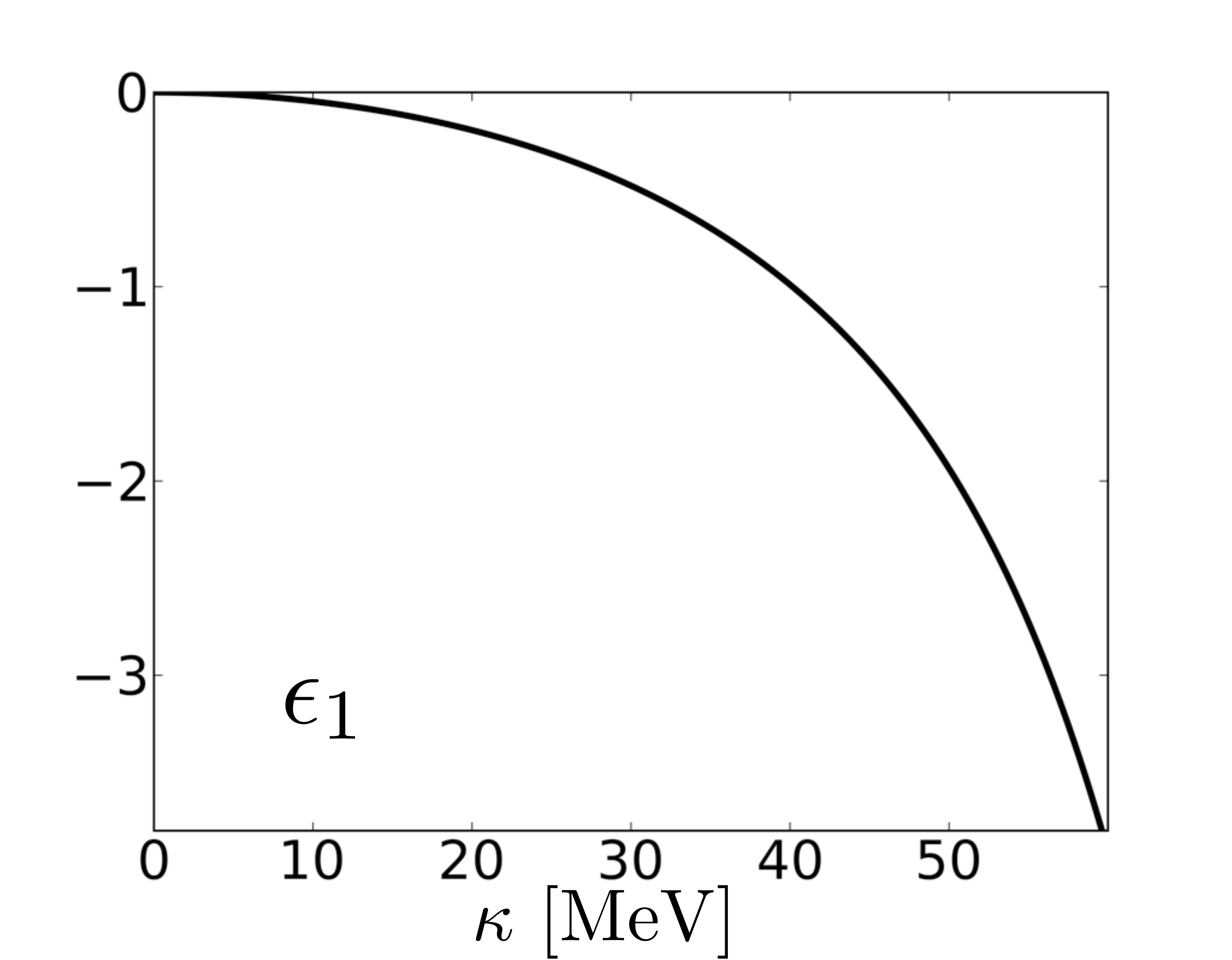}}
\subfigure[]{
\includegraphics[scale=0.135]{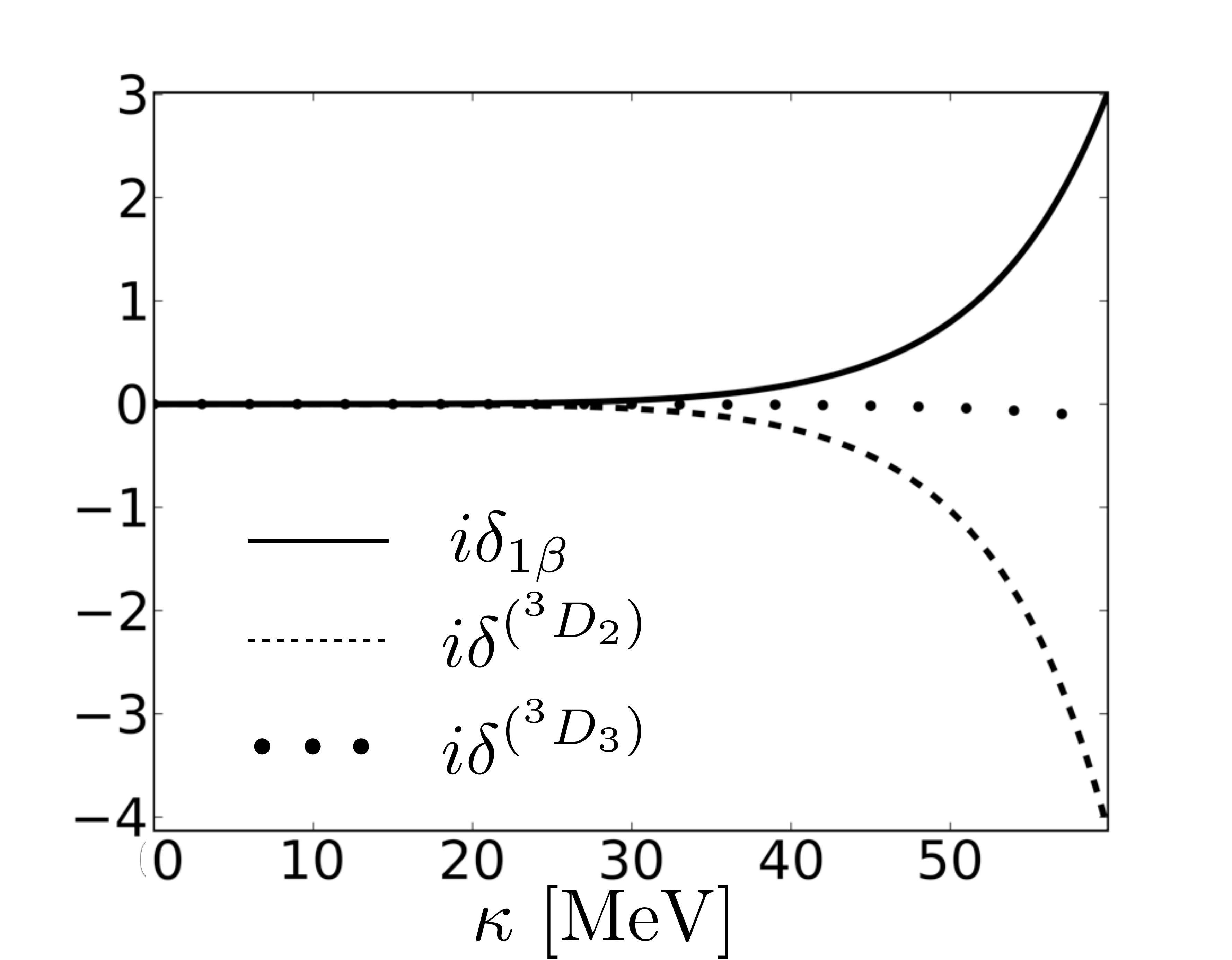}}
\caption{
{\small Fits to the experimental values of 
a) the $\alpha$-wave phase shift, 
b) the mixing angle and 
c) the $J=1$ $\beta$-wave and $J=2,3$ D-wave  phase shifts 
(in  degrees), in the Blatt-Biedenharn (BB) parameterization~\cite{Blatt:1952zza}, 
 as a function of
  momentum of each nucleon in the CM frame, $k^*$, 
based on six different phase shifts analyses~\protect\cite{PhysRevC.48.792,
    PhysRevC.49.2950, PhysRevC.54.2851, PhysRevC.54.2869}. 
d) The $\alpha$-wave phase shift, 
e) the mixing angle and 
f)  the $J=1$ $\beta$-wave and $J=2,3$ D-wave phase shifts 
(in degrees) as a function of $\kappa=-ik^*$
obtained from the fit functions. 
}}
\label{fig:Fits}
\end{center}
\end{figure}
whose mixing angle, $\epsilon_1$, when evaluated at the deuteron binding
energy, is directly related to the asymptotic D/S
ratio in the deuteron wavefunction. 
$\delta_{1\alpha}$ and $\delta_{1\beta}$ are the scattering phase shifts
corresponding to two eigenstates of the S-matrix; 
the so called ``$\alpha$'' and ``$\beta$'' waves respectively. 
At low energies, the $\alpha$-wave is predominantly S-wave with a small
admixture of the D-wave, 
while the $\beta$-wave is predominantly D-wave with a small admixture of the
S-wave. 
The location of the deuteron pole is determined
by one condition 
on the $\alpha$-wave phase shift, $\cot \delta_{1\alpha}|_{k^*=i\kappa}=i$.  
In addition, $\epsilon_1$ in this parameterization is an analytic function of energy near
the deuteron pole 
(in contrast with $\bar{\epsilon}$ in the barred parameterization
\cite{Stapp:1956mz}).  
With a truncation of $l_{max}=2$ imposed upon the scattering amplitude matrix,
the scattering parameters required for the analysis of the FV spectra
are $\delta_{1\alpha}$, $\delta_{1\beta}$, $\delta^{(^3D_2)}$,
$\delta^{(^3D_3)}$ and $\epsilon_1$. 
Fits to six different phase-shift analyses 
(PWA93~\cite{PhysRevC.48.792}, Nijm93~\cite{PhysRevC.49.2950}, Nijm1~\cite{PhysRevC.49.2950}, 
Nijm2~\cite{PhysRevC.49.2950}, Reid93~\cite{PhysRevC.49.2950} and 
ESC96~\cite{PhysRevC.54.2851, PhysRevC.54.2869}) 
obtained from Ref.~\cite{NIJMEGEN} are shown in
Fig.~\ref{fig:Fits}(a-c).~\footnote{
The $\alpha$-wave was fit by  a pole term and a
polynomial, while the other parameters were fit with polynomials alone.  
The order of the polynomial for each parameter was determined by the goodness
of fit to phenomenological model data below the t-channel cut.
} 
In order to obtain the scattering parameters at negative energies,
the fit functions are
continued to imaginary momenta, $k^*\rightarrow i\kappa$. 
Fig.~\ref{fig:Fits}(d-f)  shows the phase shifts and the mixing angle as a
function of $\kappa$ 
below the t-channel cut
(which approximately corresponds to the 
positive-energy fitting range). 
$\epsilon_1$ is observed  to be 
positive for positive energies, and becomes
negative when continued to negative energies
(see Fig.~\ref{fig:Fits}). 
The slight difference between phenomenological models 
gives rise to a small  ``uncertainty band'' for each of the parameters.

\begin{figure}[h!]
\begin{center}
\subfigure[]{
\label{T1spec}
\includegraphics[scale=0.215]{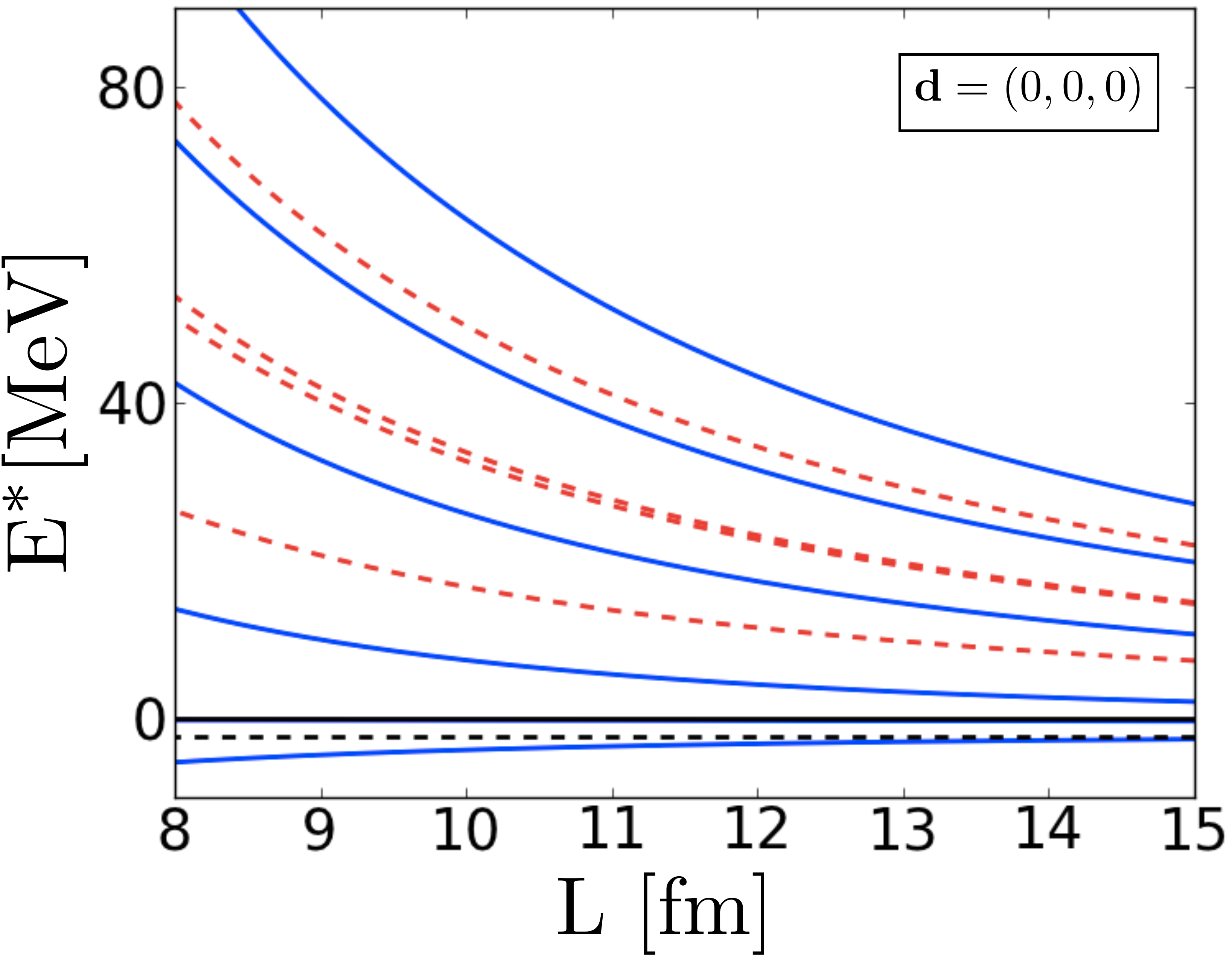}}
\subfigure[]{
\label{T1qtilde}
\includegraphics[scale=0.2175]{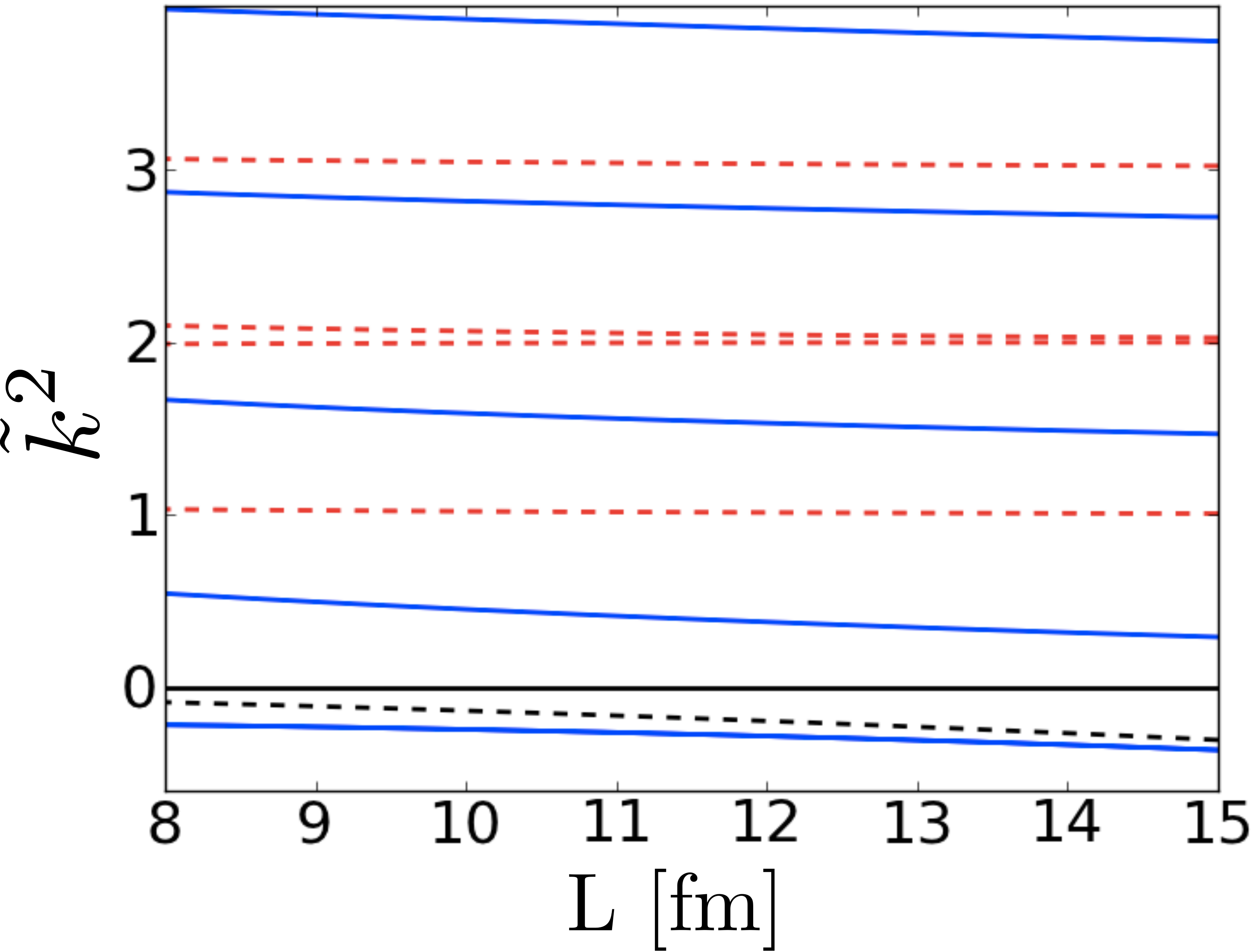}}
\caption
{{\small a) 
The nine lowest energy eigenvalues satisfying the QC for the
$\mathbb{T}_1$-irrep of the cubic group, 
Eq.~(\ref{I000T1}), and 
b) the dimensionless quantity $\tilde{k}^2=ME^*{L}^2/4\pi^2$ as a
function of $L$.
The blue-solid lines correspond to states that are predominantly S-wave,
while the red-dashed lines represent states that are predominantly D-wave. 
The black-dashed line shows the infinite-volume binding energy of the
deuteron.} 
}
\label{T1specfull}
\end{center}
\end{figure}

For the NN system at rest in the 
positive-parity isoscalar channel,
the only irrep of the cubic group that 
has overlap with the $J=1$ sector is $\mathbb{T}_1$,
which also has overlap with the $J=3$ and higher channels. 
Using the scattering parameters of the $J=1$ and $J=3$  
channels, the nine lowest $\mathbb{T}_1$ energy levels (including the
bound-state level)
are shown in  Fig.~\ref{T1spec}  as a function of 
$L$. In the limit that $\epsilon_1$ vanishes, the $\mathbb{T}_1$ QC,
given in Eq.~(\ref{I000T1}), can be written as a product of two 
independent QCs. 
One of these QCs depends only on  
$\delta_{1\alpha}\rightarrow \delta^{(^3S_1)}$, while the other depends 
on $\delta_{1\beta}\rightarrow \delta^{(^3D_1)}$ and $\delta^{(^3D_3)}$. 
By comparing the $\mathbb{T}_1$ spectrum with that
obtained for $\epsilon_1=0$, the $\mathbb{T}_1$
states can be classified as predominantly S-wave 
or predominantly D-wave states. 
The dimensionless quantity
$\tilde{k}^2=ME^*{L}^2/4\pi^2$ as a 
function of volume is shown in Fig. \ref{T1qtilde}, 
from which it is clear that the predominantly
D-wave energy levels remain close to 
the non-interacting energies, corresponding to
$\tilde{k}^2=1,2,3,4,5,6,8,\ldots$,
consistent
with the fact that both the mixing angle and the D-wave phase shifts are
small at low energies, as seen in Fig.~\ref{fig:Fits}. 
The states that are predominantly
S-wave are negatively shifted in energy 
compared with the non-interacting  states due to the attraction of the NN interactions.

\begin{figure}[ht!]
\begin{center}  
\subfigure[]{
\label{deut_cub-I}
\includegraphics[scale=0.210]{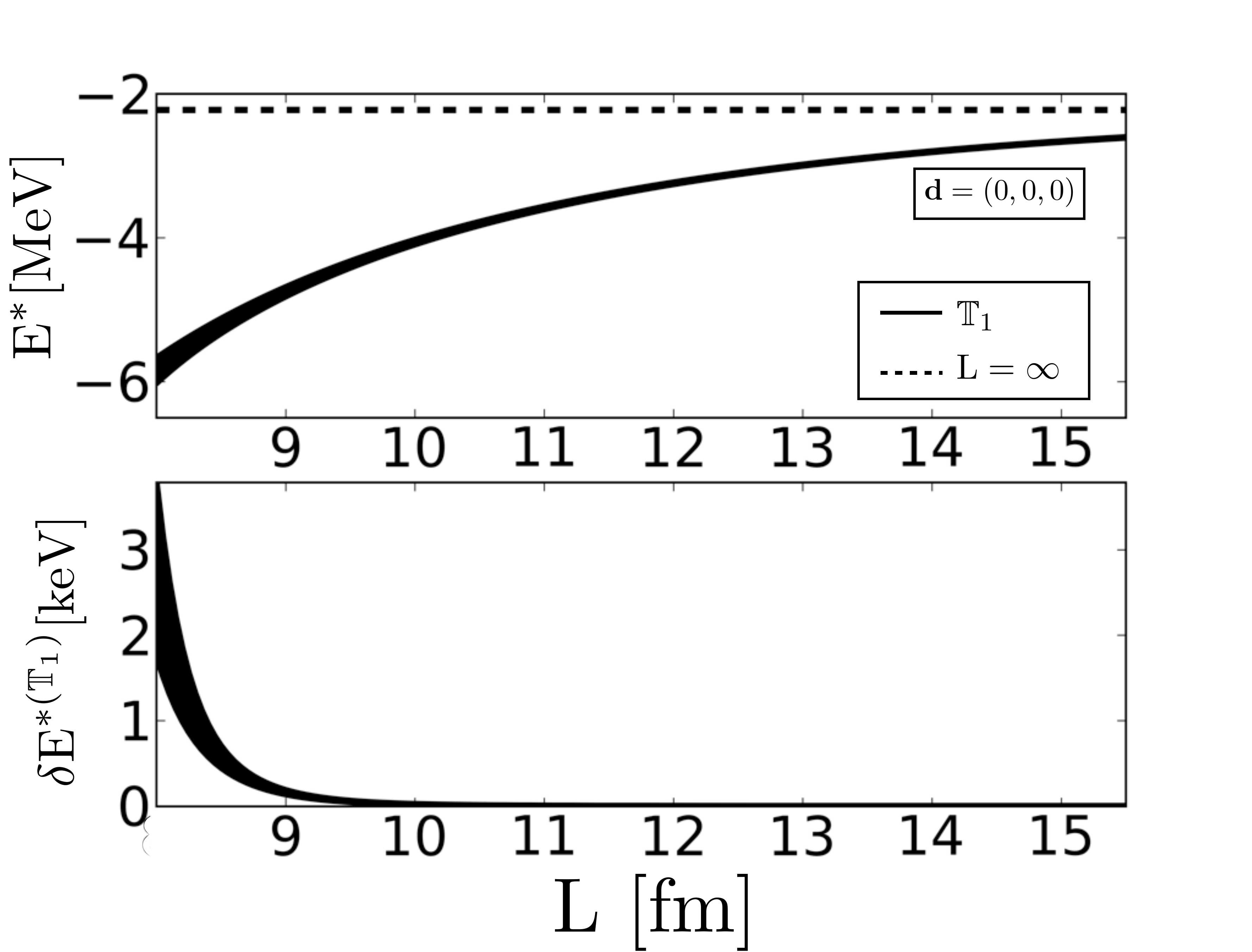}}
\subfigure[]{
\label{deut_trig}
\includegraphics[scale=0.2]{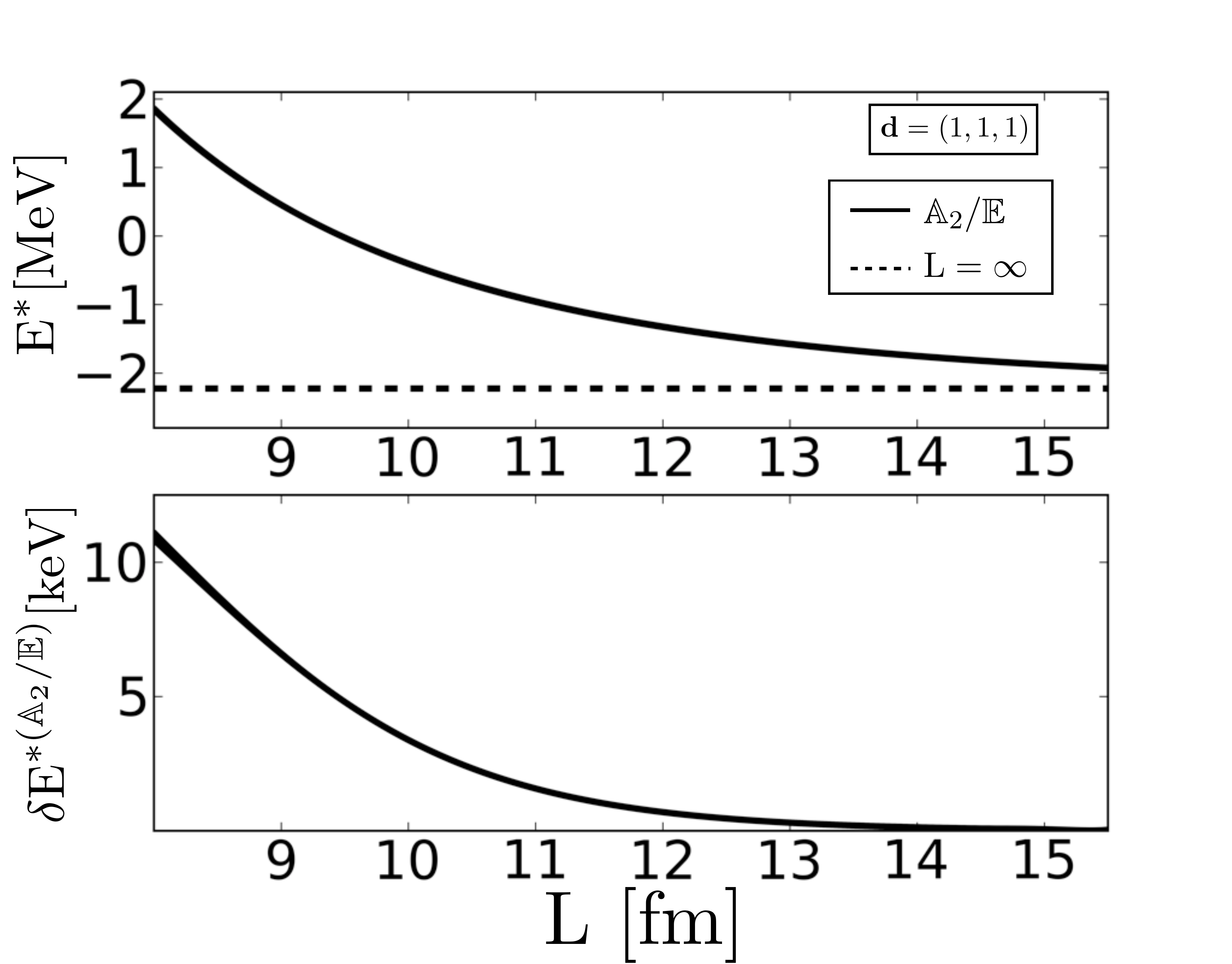}}
\caption{
{\small a) The upper panel shows the energy of two nucleons at rest in the 
positive-parity isoscalar channel  as a function of $L$
extracted from the $\mathbb{T}_1$ QC given in  Eq.~(\ref{I000T1}). 
The uncertainty band is
associated with fits to different phenomenological analyses of the
experimental data,
and the dashed line denotes the infinite-volume deuteron binding
energy. 
The lower panel shows the contribution of the mixing angle to the 
energy,
$\delta {\rm E}^{*(\mathbb{T}_1)}= {\rm E}^{*(\mathbb{T}_1)}-{\rm E}^{*(\mathbb{T}_1)}(\epsilon_1=0)$.
b) The same quantities as in (a) for the NN
system with
${\bf d}=(1,1,1)$ obtained from the  $\mathbb{A}_2/\mathbb{E}$ QCs, Eqs.~(\ref{I111A2},
\ref{I111E}).
}}
\label{deut_cub}
\end{center}
\end{figure}

Focusing on the deuteron, it is important to quantify the effect of the
mixing between the S-wave and D-wave on the energy of the deuteron in a finite volume. 
The upper panel of Fig. \ref{deut_cub-I} provides 
a closer look at the binding energy of the deuteron as function of $L$ 
extracted from the $\mathbb{T}_1$ QC given in Eq.~(\ref{I000T1}). 
While in larger volumes the uncertainties in the predictions due to the fits to experimental data 
are a few keV, 
in smaller volumes the uncertainties increase
because the fit functions are valid only below the t-channel cut 
and are not expected to describe the data above the cut. 
It is interesting  to examine the difference between the bound-state energy
obtained from the full $\mathbb{T}_1$ QC 
and that  obtained  with $\epsilon_1=0$, 
$\delta
{\rm E}^{*(\mathbb{T}_1)}={\rm E}^{*(\mathbb{T}_1)}-{\rm E}^{*(\mathbb{T}_1)}(\epsilon_1=0)$.
This quantity, shown in the lower panel of Fig.~\ref{deut_cub-I}, does not
exceed a few $\rm{keV}$ in smaller volumes, $L \lesssim9~\rm{fm}$, 
and is significantly smaller in larger volumes,
demonstrating that the spectrum of $\mathbb{T}_1$ irrep is quite  insensitive
to the small mixing angle 
in the $\siii$-$\diii$ coupled channels.
Therefore, a determination of the mixing angle from the spectrum of two nucleons at rest
will be challenging for LQCD calculations. 
The spectra in the $\mathbb{A}_2/\mathbb{E}$ irreps of the trigonal group for
$\mathbf{d}=(1,1,1)$ 
exhibit the same feature, as shown in Fig. \ref{deut_trig}. 
By investigating the QCs in Eqs.~(\ref{I000T1}, \ref{I111A2}, \ref{I111E}), 
it is straightforward to show that the difference between the bound-state
energy 
extracted from the full QCs 
(including physical and FV-induced mixing between S-waves and D-waves) 
and from the uncoupled QC is proportional to ${\sin^2 \epsilon_1}$,
and is further suppressed by FV corrections and the 
small  $\beta$-wave and D-wave
phase shifts.

The boost vectors $\mathbf{d}=(0,0,1)$ and $(1,1,0)$
distinguish the $z$-axis from the 
other two axes, and result in an asymmetric volume as viewed
in the rest frame of the deuteron. 
In terms of the periodic images of the deuteron, 
images that are located in the $z$-direction with opposite signs compared with the
images in the $x$- and $y$-directions~\cite{Bour:2011ef,Davoudi:2011md} 
result in the quadrupole-type shape
modifications to the 
deuteron, as will be elaborated  on in Sec.~\ref{sec:wavefunction}. 
As a result, the energy of the deuteron, as well as its
shape-related quantities such as its quadrupole moment, 
will be affected more by the finite extent of the volume (compared with the
systems with $\mathbf{d}=(0,0,0)$ and $(1,1,1)$).

\begin{figure}[ht!]
\begin{center}  
\includegraphics[scale=0.30]{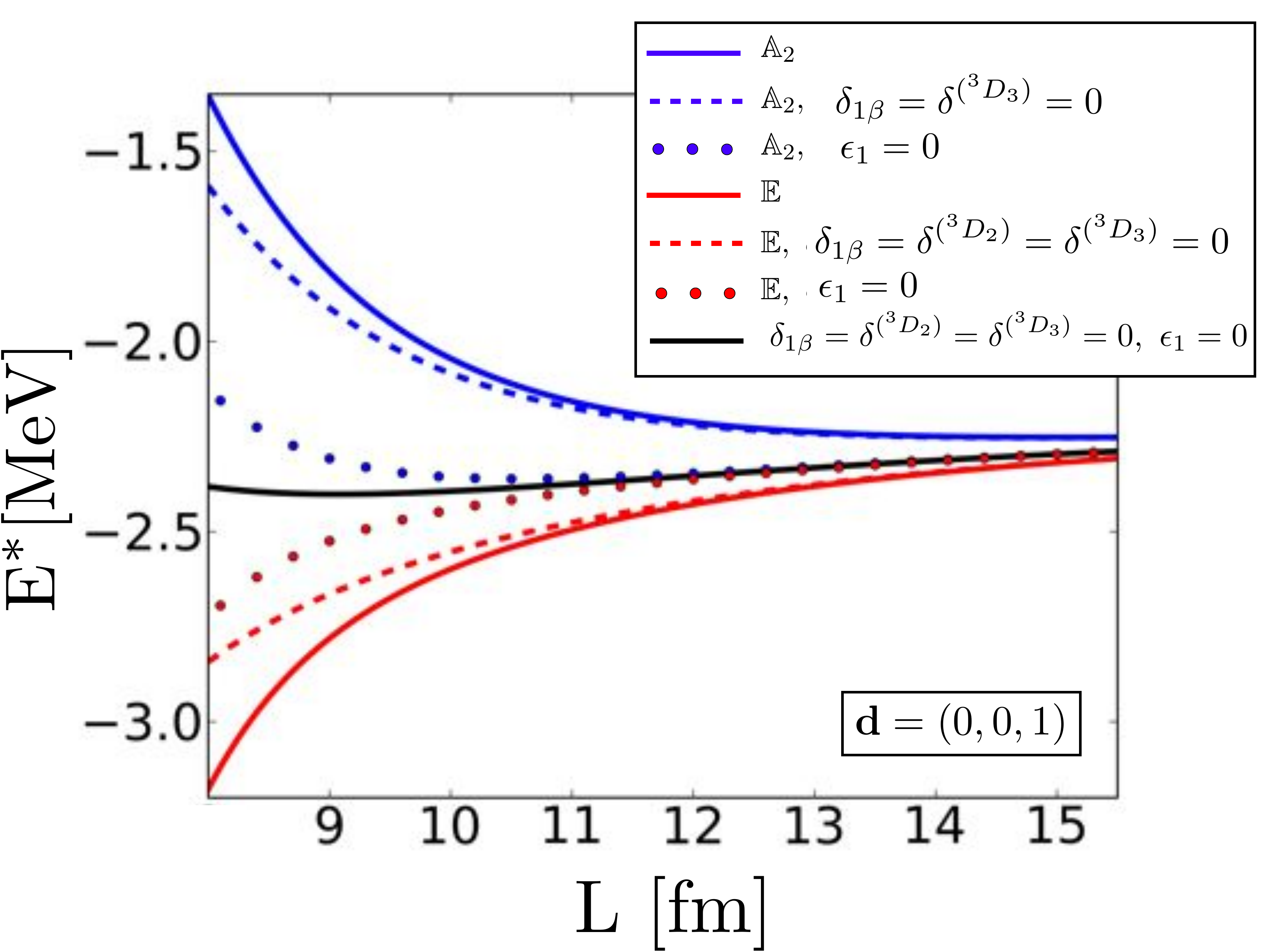}
\caption{{\small The energy of two nucleons in the
 positive-parity
isoscalar channel with $\mathbf{d}=(0,0,1)$ as a function of
  $L$, 
extracted from the $\mathbb{A}_2$ (blue) and $\mathbb{E}$ (red) QCs given in 
Eq.~(\ref{I001A2}) and Eqs.~(\ref{I001E}), respectively. 
The systematic uncertainties  associated with fitting 
different phenomenological analyses of the experimental data  are included. 
}}
\label{deut_tet}
\end{center}
\end{figure}
As is clear from the QCs for $\mathbf{d}=(0,0,1)$ systems, given in Eqs.~(\ref{I001A2}, \ref{I001E}), 
there are two
irreps of the tetragonal group, $\mathbb{A}_2$ 
and $\mathbb{E}$,  that have overlap with the $J=1$ channel. 
These irreps represent states with $M_J=0$ for the $\mathbb{A}_2$ 
irrep and $M_J=\pm 1$ for the $\mathbb{E}$ irreps,
see Table~\ref{irreps-deuteron}. 
The bound-state energies of  these two irreps  are shown in  Fig.~\ref{deut_tet}
as a function of $L$. 
For comparison, the energy of the bound state with  ${\bf d}=(0,0,1)$
in the limit of vanishing mixing angle and D-wave phase shifts is also shown 
(black-solid curve) in Fig. \ref{deut_tet}. 
The energy of the bound states obtained in  both the $\mathbb{A}_2$ irrep
(blue-solid curve) and the $\mathbb{E}$ irrep (red-solid curve) 
deviate substantially from the energy of the purely S-wave bound state for modest
volumes, $L\lesssim14~\rm{fm}$.~\footnote{LQCD calculations at the physical
  pion mass require volumes with $L \gtrsim 9~\rm{fm}$ so that the
  systematic uncertainties associated with
the finite range of the nuclear forces  are below the percent level.} 
These deviations are such that the energy gap between the systems in the two
irreps is 
$\sim 80\%$ of the
infinite-volume deuteron binding energy 
at $L=8~\rm{fm}$, decreasing to $\sim 5\%$  for $L=14~\rm{fm}$. 
This gap is largely due to the mixing between S-wave and D-wave in
the infinite volume, as verified by evaluating the bound-state energy in the
$\mathbb{A}_2$ and $\mathbb{E}$ irrep in the limit where $\epsilon_1=0$ 
(the blue and red dotted curves in Fig. \ref{deut_tet}, respectively.)
Another feature of the $\textbf{d}=(0,0,1)$ FV bound-state energy is that
the contribution from the 
 $\beta$-wave and D-wave
states cannot be neglected for $L\lesssim 10~\rm{fm}$. 
The blue (red) dashed curve in Fig.~\ref{deut_tet} 
results from the $\mathbb{A}_2$ ($\mathbb{E}$) QC in this limit. 
The D-wave states in the $J=2$ and $J=3$ channels mix with the $J=1$ $\alpha$-
and $\beta$-waves due to the reduced symmetry of the system, 
and as a result they, and the $\beta$-wave state, contribute to the energy of the
predominantly S-wave bound state in the finite volume.

\begin{table}[t!]
\begin{centering}
\begin{tabular}{|c|c|c|c|c|c|}
\hline
J &$O$&${D}_{4}$&${D}_{2}$& ${D}_{3}$
 \\\hline\hline
1
&$\mathbb{T}_1:(\mathcal{Y}_{11},\mathcal{Y}_{10},\mathcal{Y}_{1-1})$
&$\mathbb{A}_2:\mathcal{Y}_{10}$
&$\mathbb{B}_1:\mathcal{Y}_{10}$
& $\mathbb{A}_2:\mathcal{Y}_{10}$ \\
&
&$\mathbb{E}:\left(\overline{\mathcal{Y}}_{11},\widetilde{\mathcal{Y}}_{11}\right)$
&$\mathbb{B}_2:\overline{\mathcal{Y}}_{11},~\mathbb{B}_3:\widetilde{\mathcal{Y}}_{11}$
& $\mathbb{E}:\left(\overline{\mathcal{Y}}_{11},\widetilde{\mathcal{Y}}_{11}\right)$ \\
\hline
\end{tabular}
\caption{
{\small Decomposition of the $J=1$ irrep of the rotational group in terms of
  the irreps of the cubic ($O$), 
tetragonal ($D_4$), orthorhombic ($D_2$) and trigonal ($D_3$) groups, see
Refs. \cite{Luscher:1990ux, Feng:2004ua, Dresselhaus}. 
The corresponding basis functions  of each irrep are also shown
in terms of the SO(3) functions
$\mathcal{Y}_{lm}$,
where 
$\overline{\mathcal{Y}}_{lm}\equiv \mathcal{Y}_{lm}+\mathcal{Y}_{l-m}$ and 
$\widetilde{\mathcal{Y}}_{lm}\equiv \mathcal{Y}_{lm}-\mathcal{Y}_{l-m}$.
}}
\label{irreps-deuteron}
\par\end{centering}
\end{table}
%

\begin{figure}[!ht]
\begin{center}  
\includegraphics[scale=0.30]{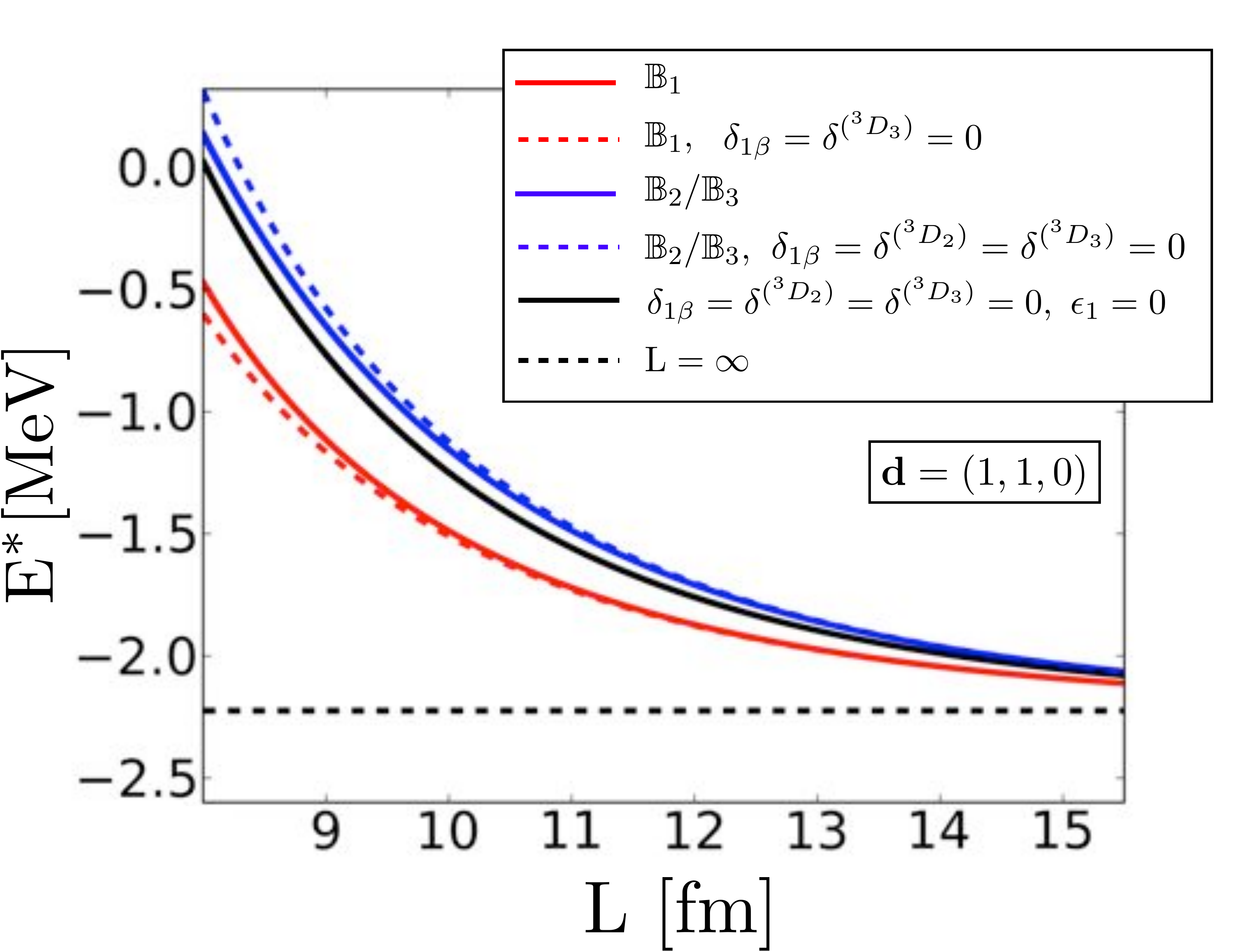}
\caption{
The energy of two nucleons in the
positive-parity
isoscalar channel with $\mathbf{d}=(1,1,0)$ as a function of
  $L$, 
extracted from the $\mathbb{B}_1$ (red) and $\mathbb{B}_2/\mathbb{B}_3$ (blue)
QCs given in Eqs.~(\ref{I110B1}) and (\ref{I110B2}, \ref{I110B3}), 
respectively. 
The systematic uncertainties associated with fitting to different
phenomenological analyses of the experimental 
data are included. 
}
\label{deut_ort}
\end{center}
\end{figure}
The FV energy eigenvalues 
for the NN system in the positive-parity
isoscalar channel with ${\bf d}=(1,1,0)$ 
can be obtained from QCs in 
Eqs.~(\ref{I110B1}) and (\ref{I110B2}, \ref{I110B3}), corresponding to
$\mathbb{B}_1$ and $\mathbb{B}_2/\mathbb{B}_3$ irreps of 
the orthorhombic group, respectively. 
These irreps represent states with $M_J^{\prime}=0$ for the $\mathbb{B}_1$ irrep and 
$M_J^{\prime}=\pm 1$ for the $\mathbb{B}_2/\mathbb{B}_3$ irreps (where $M_J^\prime$ now is the projection of total angular momentum
along the twist direction),
see Table~\ref{irreps}. 
The bound-state energies of these systems are shown in
Fig.~\ref{deut_ort}, and are found to  
deviate noticeably from the purely  S-wave limit (black-solid curve in
Fig.~\ref{deut_ort}), 
however the deviation is not as large as the 
case of ${\bf d}=(0,0,1)$. 
The energy gap between the systems in the two irreps is $\sim 30\%$ of the
 infinite-volume deuteron binding energy 
at $L=8~\rm{fm}$, decreasing  to $\sim 5\%$  for $L=14~\rm{fm}$. 
Eliminating the  $\beta$-wave and $J=2,3$ D-wave interactions, leads to the dashed curves in
Fig.~\ref{deut_ort}, indicating the 
negligible effect that they have on the bound-state energy in
these irreps.

To understand the large FV energy shifts from the purely  $\alpha$-wave
estimates for ${\bf d}=(0,0,1)$ and $(1,1,0)$ systems compared 
with $(0,0,0)$ and $(1,1,1)$ systems, 
it is instructive to examine the QCs given in Appendix~\ref{app: QC} in the limit
where the  $\beta$-wave and D-wave phase shifts vanish. 
This is a reasonable approximation  for volumes with $L \gtrsim 10~\rm{fm}$,  as
illustrated in Fig.~\ref{deut_tet} and Fig.~ \ref{deut_ort}.
It is straightforward to show that in this limit, 
the QC of the system with $\mathbf{d}=(0,0,0)$ reduces to a purely $\alpha$-wave condition
\begin{align}
&\mathbb{T}_1:\hspace{.1cm}k^*\cot\delta_{1\alpha}-4 \pi
c_{00}^{(0,0,0)}(k^{*2}; L)=0
.
\label{appr-T1}
\end{align}
The QCs for a system with $\mathbf{d}=(0,0,1)$ are
\begin{align}
&\mathbb{A}_2:\hspace{.1cm}
k^*\cot\delta_{1\alpha}
-4 \pi  c_{00}^{(0,0,1)}(k^{*2}; L)
\ =\ 
-{1\over\sqrt{5}}
\frac{4\pi}{k^{*2}}\ 
c_{20}^{(0,0,1)}(k^{*2}; L)
\ 
(\sqrt{2}\sin2\epsilon_1-\sin^2\epsilon_1)
,
\\
&\mathbb{E}:\hspace{.1cm}
 k^* \cot \delta_{1\alpha}
-4 \pi  c_{00}^{(0,0,1)}(k^{*2}; L)
\ =\ 
+{1\over 2\sqrt{5}}
\frac{4\pi}{k^{*2}}\ 
c_{20}^{(0,0,1)}(k^{*2}; L)
\ 
(\sqrt{2}\sin2\epsilon_1-\sin^2\epsilon_1)
,
\label{appr-E}
\end{align}
which include corrections to the $\alpha$-wave limit that scale with  $\sin \epsilon_1$
at LO. 
This is the origin of  the large deviations of these energy eigenvalues from the purely S-wave values. 
The same feature is seen in the systems with $\mathbf{d}=(1,1,0)$, where the QCs reduce to
\begin{align}
&\mathbb{B}_1:
 k^*\cot\delta_{1\alpha}
-4 \pi  c_{00}^{(1,1,0)}(k^{*2}; L)
\ =\ 
-{1\over\sqrt{5}}
\frac{4\pi}{k^{*2}}\ 
c_{20}^{(1,1,0)}(k^{*2}; L)
\ 
(\sqrt{2}\sin2\epsilon_1-\sin^2\epsilon_1)
,
\\
&\mathbb{B}_2/\mathbb{B}_3:
 k^* \cot \delta_{1\alpha}
-4 \pi  c_{00}^{(1,1,0)}(k^{*2}; L)
\ =\ 
+{1\over 2\sqrt{5}}
\frac{4\pi}{k^{*2}}\ 
c_{20}^{(1,1,0)}(k^{*2}; L)
\ 
(\sqrt{2}\sin2\epsilon_1-\sin^2\epsilon_1)
.
\label{appr-B2}
\end{align}
Similarly, the QC with $\mathbf{d}=(1,1,1)$ in this limit is
\begin{align}
&\mathbb{A}_2/\mathbb{E}:\hspace{.1cm}k^*\cot\delta_{1\alpha}-4 \pi
c_{00}^{(1,1,1)}(k^{*2}; L)=0
.
\label{appr-EA2}
\end{align}

The LO corrections to the QCs 
in Eqs.~(\ref{appr-T1}-\ref{appr-EA2}) 
are not only suppressed by
the 
$J=1$ $\beta$-wave and $J=2,3$ D-wave
phase shifts, but also by FV corrections 
that are further exponentially suppressed compared with the leading FV corrections. 
It is straightforward to show that the leading neglected terms in the QCs
presented above are 
$\sim \frac{1}Le^{-2\kappa L}\tan{\delta_{1\beta}}$ and 
$\frac{1}Le^{-2\kappa L}\tan{\delta_{D_{J=2,3}}}$,
while the
FV contributions to the approximate relations given in 
Eqs.~(\ref{appr-T1}-\ref{appr-EA2})  are  $\sim {1\over L}e^{-\kappa L}$. 
In Appendix \ref{app: clm}, the explicit volume dependence of $c^{\mathbf{d}}_{LM}$ functions
are given 
for the case of $k^{*2}=-\kappa^2<0$. 
These explicit forms are useful in obtaining the leading exponential corrections to the
QCs. 
We emphasize that the smaller volumes considered have $\kappa L =2-2.5$, 
and therefore it is not a good approximation to replace the $c^{\mathbf{d}}_{LM}$
functions with their leading exponential terms,
and the complete form of these functions should be used in analyzing the FV spectra.

\begin{figure}[!ht]
\begin{center}  
\subfigure[]{
\label{spin-ave-001}
\includegraphics[scale=0.20]{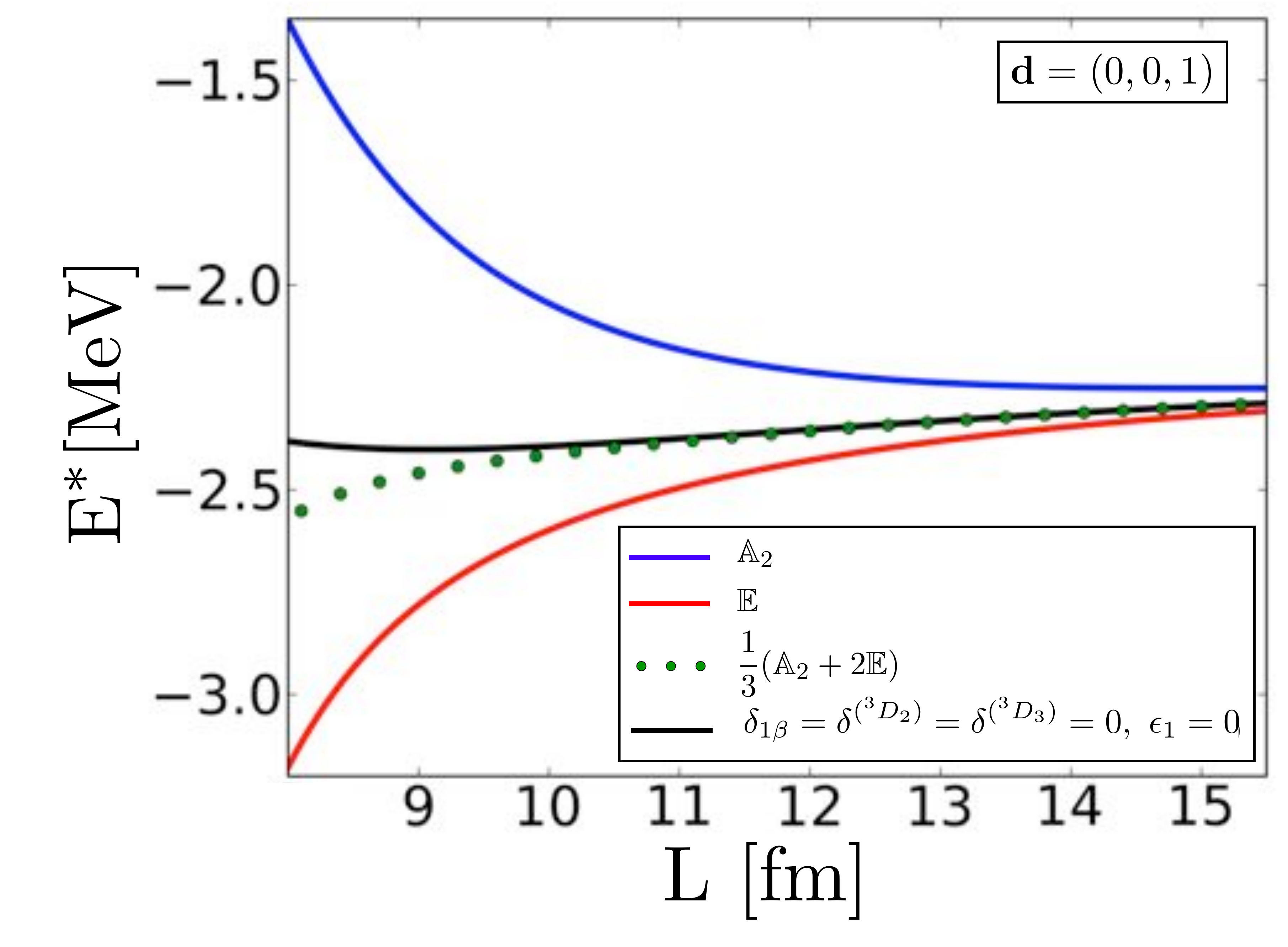}}
\subfigure[]{
\label{spin-ave-110}
\includegraphics[scale=0.20]{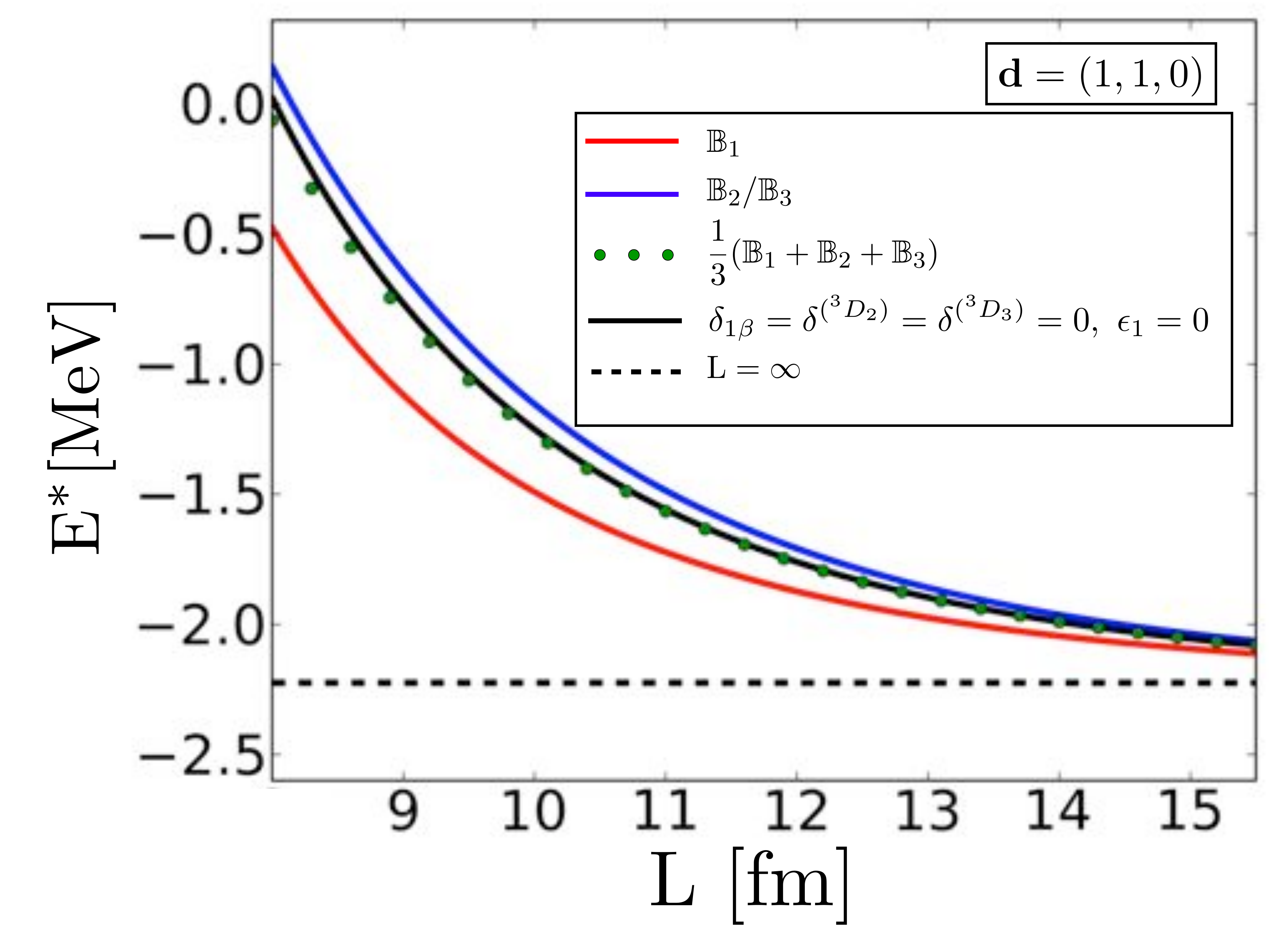}}
\caption{
{\small a) The dotted curve shows the $M_J^{\prime}$-averaged quantity
$\frac{1}{3}(E^{*(\mathbb{A}_2)}+2E^{*(\mathbb{E})})$ 
as function of $L$, while the solid curves show the energy of the state  
in the $\mathbb{A}_2$ (blue) and $\mathbb{E}$ (red) irreps of the tetragonal
group, 
as well as that of the state with $\epsilon_1=0$ (black). 
b) The dotted curve shows the $M_J^{\prime}$-averaged quantity
$\frac{1}{3}(E^{*(\mathbb{B}_1)}+E^{*(\mathbb{B}_2)}+E^{*(\mathbb{B}_3)})$
as function of $L$, while the solid curves show the energy of the state 
in the $\mathbb{B}_1$ (red) and $\mathbb{B}_2/\mathbb{B}_3$ (blue) irreps of
the orthorhombic group, 
as well as that of the state with $\epsilon_1=0$ (black).}}
\label{spin-ave}
\end{center}
\end{figure}
In the limit of vanishing 
$J=1$ $\beta$-wave and $J=2,3$ D-wave
phase shifts,
the QCs show that the energy shift of each pair of irreps of the systems 
with $\mathbf{d}=(0,0,1)$ and $(1,1,0)$ differ in sign. 
It is also the case that the $M_J^{\prime}$-averaged energies are approximately the
same as the purely S-wave case. 
In fact, as illustrated in Fig. \ref{spin-ave-001}, the energy level
corresponding to 
$\frac{1}{3}(E^{*(\mathbb{A}_2)}+2E^{*(\mathbb{E})})$ quickly
converges to the S-wave energy with ${\bf d}=(0,0,1)$. 
Similarly, the $M_J^{\prime}$-averaged quantity
$\frac{1}{3}(E^{*(\mathbb{B}_1)}+E^{*(\mathbb{B}_2)}+E^{*(\mathbb{B}_3)})$
almost coincides with the S-wave state with ${\bf d}=(1,1,0)$,
Fig. \ref{spin-ave-110}. 
This is to be expected, as $M_J^{\prime}$-averaging is equivalent to averaging
over the orientations of the image systems, suppressing  the anisotropy induced by
the boost phases in the FV corrections, Eqs.~(\ref{c00-exp})-(\ref{c40-exp}).
These expressions also demonstrate that, unlike the case of degenerate, scalar
coupled-channels systems \cite{Berkowitz:2012xq,Oset:2012bf}, 
the NN spectra (with spin degrees of freedom) depend on the sign of
$\epsilon_1$. 
Of course, this sensitivity to the sign of $\epsilon_1$ can be deduced
from the full QCs in Eqs.~(\ref{I000T1}-\ref{I111E}). 
Upon fixing the phase  convention of the angular momentum states, 
both the magnitude and sign of the mixing angle  can be extracted from 
FV calculations, 
as will be discussed in more detail in Section~\ref{sec:extraction}.

\section{Extracting the Scattering Parameters From Synthetic Data
\label{sec:extraction}
}
\noindent
Given the features of the energy spectra associated with different boosts, 
it is interesting  to consider how well the scattering parameters
can be extracted from future LQCD calculations at the physical pion mass.
With the truncations we have imposed, 
the full QCs for the FV  states that have overlap with the
$\siii$-$\diii$ 
coupled channels depend on four scattering phase shifts and the $J=1$ mixing angle.
As discussed in Sec.~\ref{sec:DeutFV}, for bound states these are equivalent to
QCs that depend solely on $\delta_{1\alpha}$
and $\epsilon_1$  up to corrections of
$\sim \frac{1}Le^{-2\kappa L}\tan{\delta_{1\beta}}$ and 
$\frac{1}Le^{-2\kappa L}\tan{\delta_{D_{J=2,3}}}$,
as given in Eqs.~(\ref{appr-T1}-\ref{appr-EA2}). 
By considering the boosts with $|\textbf{d}|\leq\sqrt{3}$, six
independent bound-state energies that asymptote to 
the physical deuteron energy can be obtained. 
For a single volume, these give six different constraints on
$\delta_{1\alpha}$ and $\epsilon_1$ for energies in the 
vicinity of the deuteron pole. 
Therefore, by parameterizing the momentum dependence of these two
parameters, and requiring them 
to simultaneously satisfy Eqs.~(\ref{appr-T1}-\ref{appr-EA2}), 
their low-energy behavior can be extracted.

Using the fact that the $\alpha$-wave is dominantly S-wave with $\epsilon_1$
and $\delta^{(\diii)}$
small, we use the  effective range expansion (ERE) of the inverse 
S-wave scattering amplitude,
which is valid below the t-channel cut,
to parameterize~\cite{PhysRev.93.1387}
\begin{eqnarray}
\label{eq:ERE}
k^*\cot\delta_{1\alpha}&=&-\frac{1}{a^{(^3S_1)}}+\frac{1}{2}r^{(^3S_1)}k^{*2}+ \dots
,
\\
\epsilon_1&=&h_1~k^{*2}+ \dots~
.
\end{eqnarray}
Therefore, up to $\mathcal{O}(k^{*2})$, 
the three  parameters, denoted by $a^{(^3S_1)}$, $r^{(^3S_1)}$ and $h_1$,
 can be over-constrained by the bound-state spectra in 
a single volume. 
To illustrate this point, we fit the six independent energies 
to ``synthetic data''
using the approximated QCs, Eqs.~(\ref{appr-T1}-\ref{appr-EA2}). 
The precision with which  $\{a^{(^3S_1)},r^{(^3S_1)},h_1\}$ can be extracted 
depends on the precision and correlation of the energies
determined in LQCD calculations. 
Therefore, we consider four possible scenarios,
corresponding  to the energies being extracted 
from a given LQCD calculation
with $1\%$ and $10\%$ precision, and with uncertainties that are
uncorrelated or fully correlated with each other. 
It is likely that the energies of these irreps will be determined in 
LQCD calculations on the same ensembles of gauge-field configurations, 
and consequently they are likely to
be highly correlated - a feature that has been
exploited 
extensively in the past when determining energy differences.

\begin{figure}[t!]
\begin{center} 
\subfigure[]{

\label{a_corr}
\includegraphics[scale=0.175]{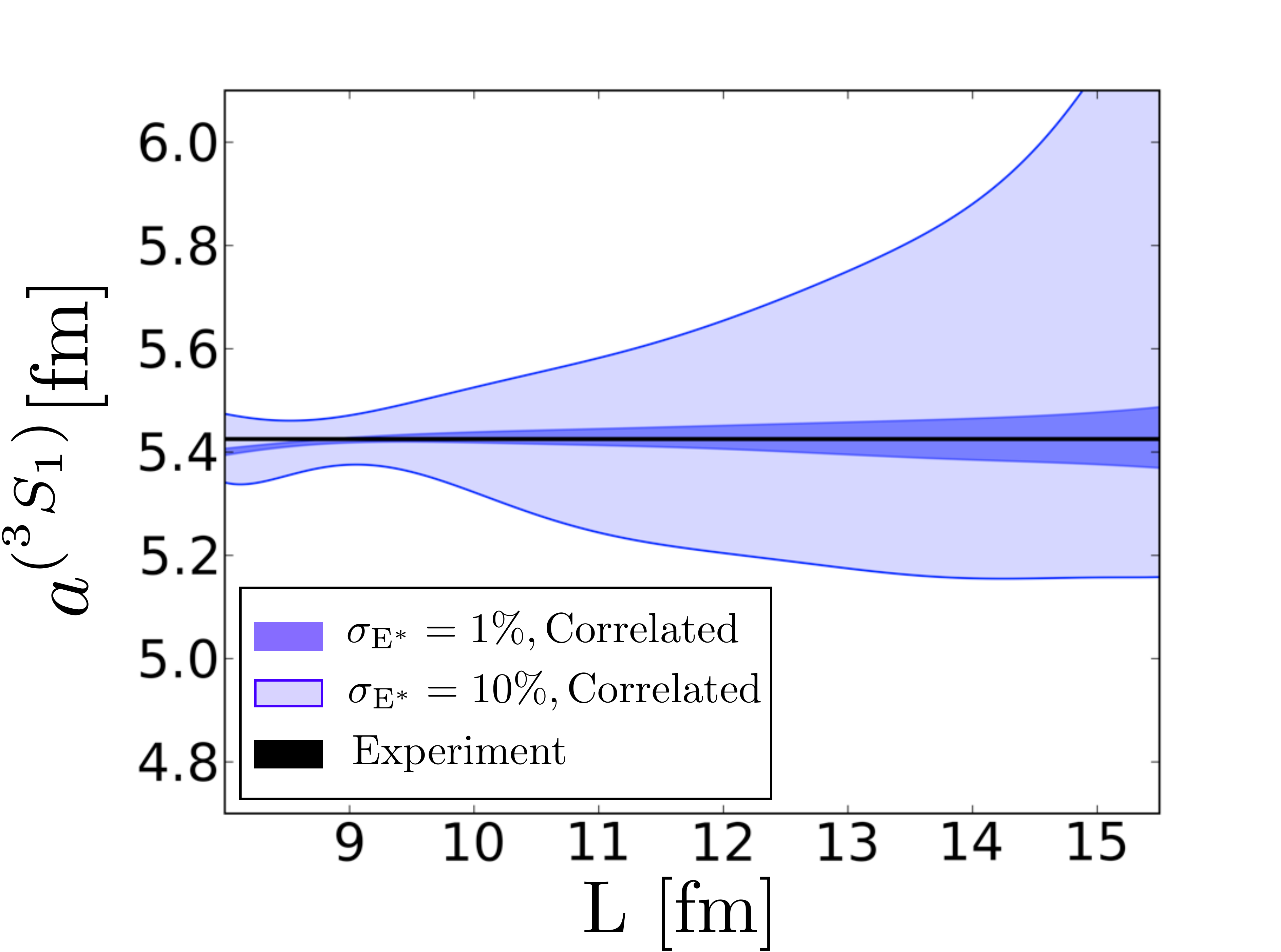}}
\subfigure[]{
\label{a_uncorr}
\includegraphics[scale=0.175]{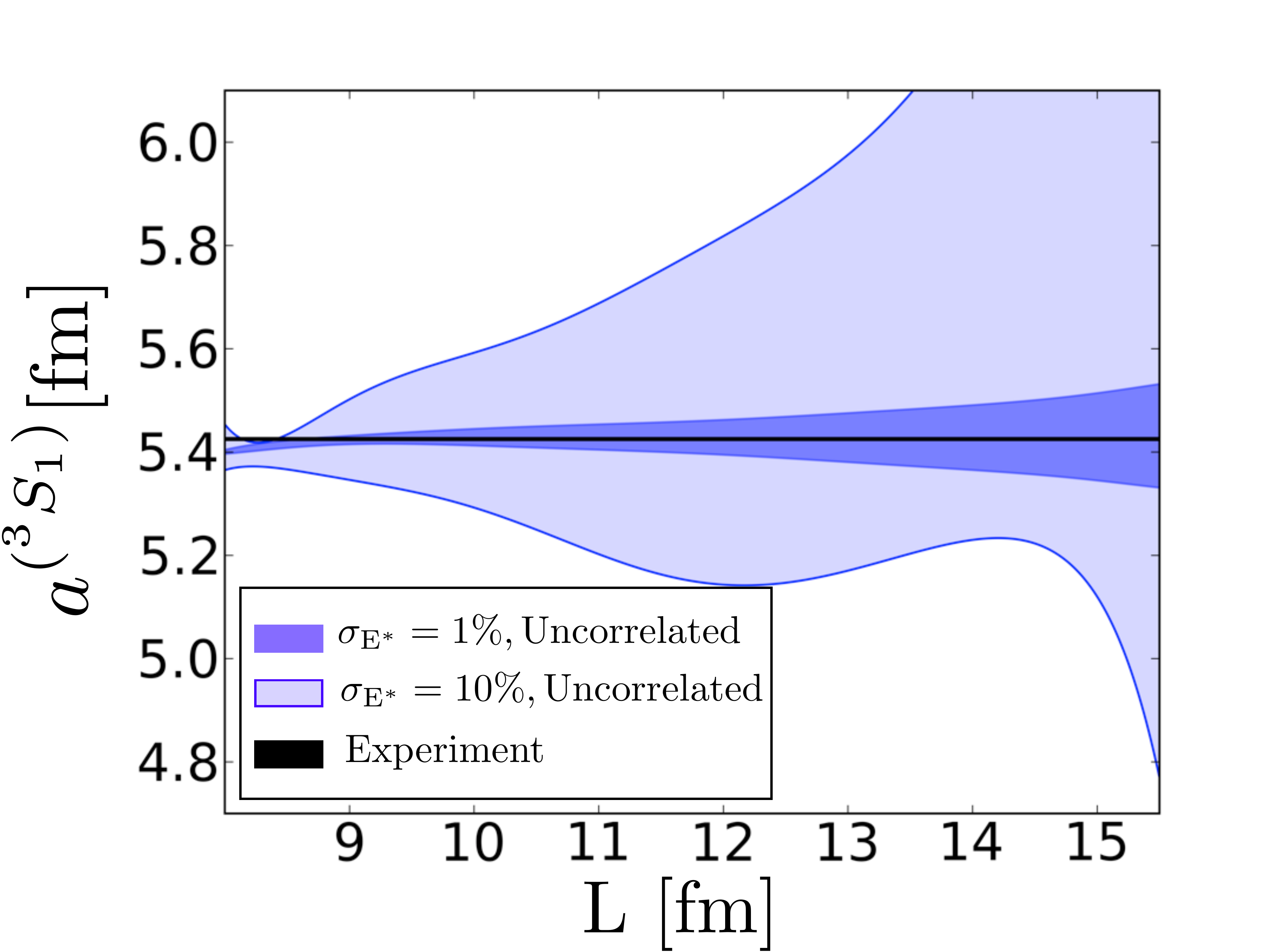}}
\subfigure[]{
\label{r_corr}
\includegraphics[scale=0.175]{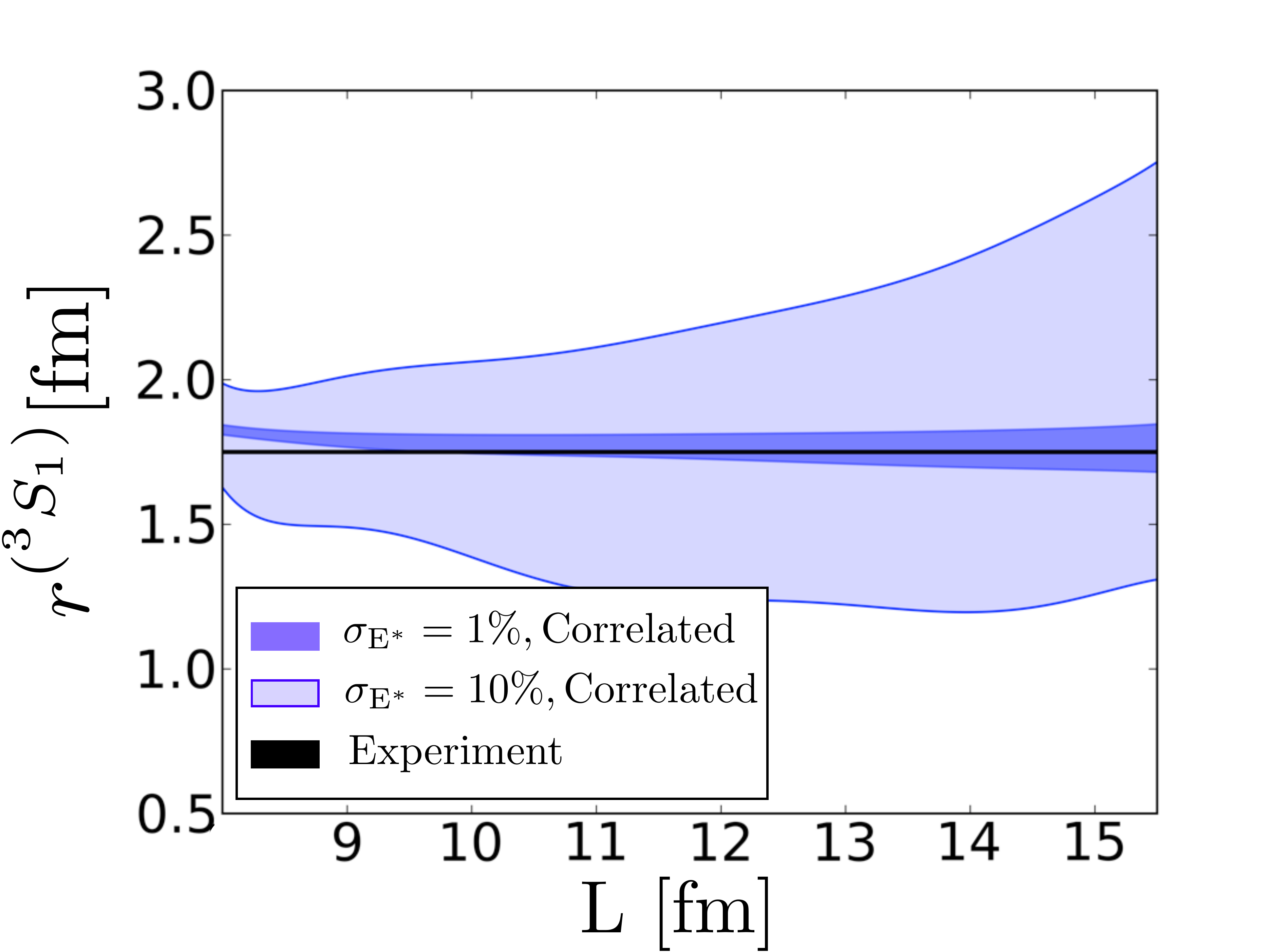}}
\subfigure[]{
\label{r_uncorr}
\includegraphics[scale=0.175]{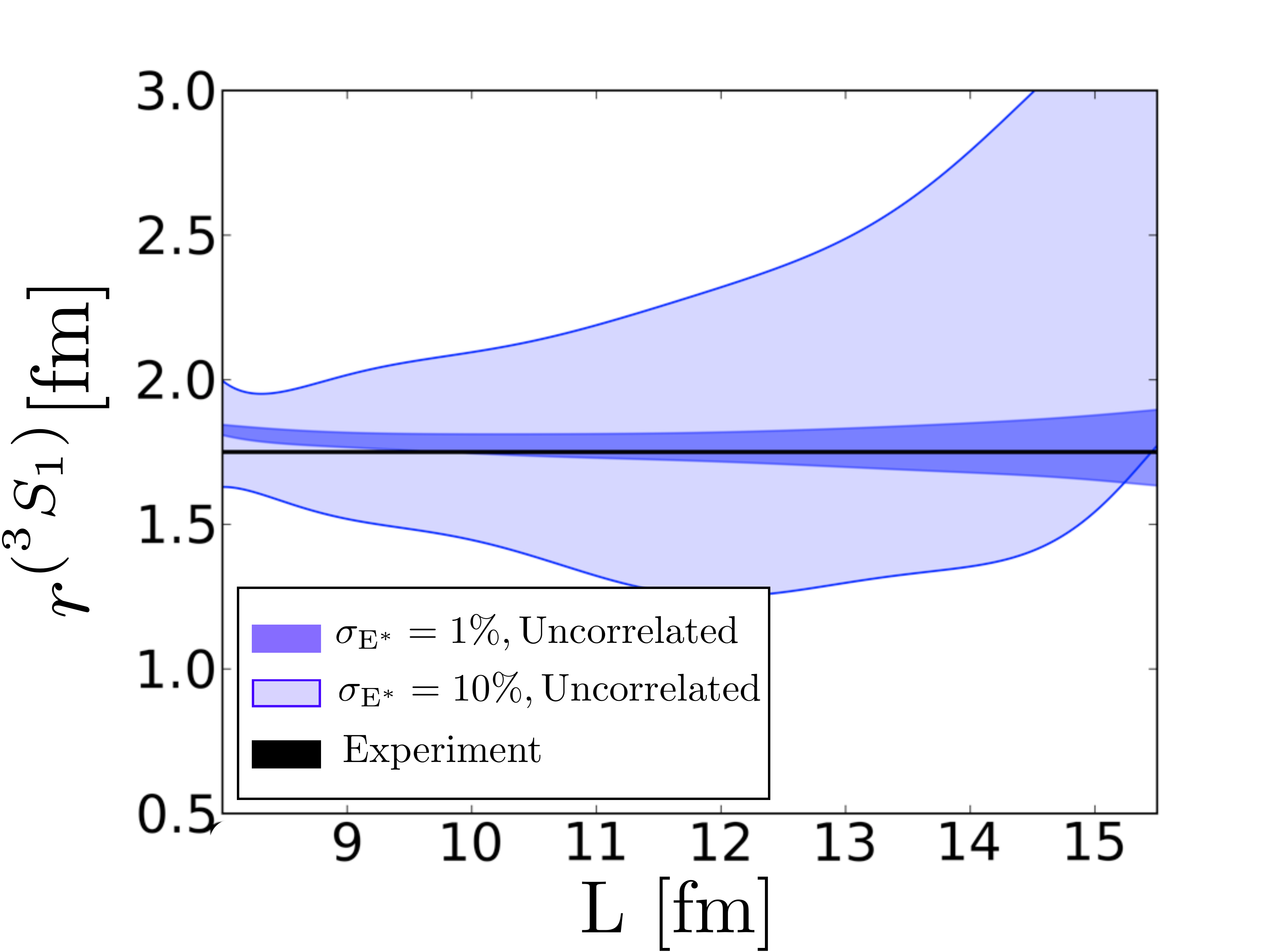}} 
\caption{
{\small The  values of $\{a^{(^3S_1)}, r^{(^3S_1)}\}$ obtained by
  fitting the six independent bound-state energies with $|\textbf{d}|\leq\sqrt{3}$
  (depicted in Figs.~\ref{deut_cub}, \ref{deut_tet}, \ref{deut_ort}),
generated from synthetic LQCD calculations,
using the approximate QCs in Eqs.~(\ref{appr-T1}-\ref{appr-EA2}), as discussed
in the text. 
The black lines denote
the experimental value of these quantities determined by fitting the scattering
parameters obtained from Ref.~\cite{NIJMEGEN}. 
The dark (light) inner (outer) band is the 
$1\sigma$ band corresponding to the energies being  determined with 
1\% (10\%) precision. 
}}
\label{fig:fakedata1}
\end{center}
\end{figure}

\begin{figure}[t!]
\begin{center}  
\subfigure[]{
\label{Bd_corr}
\includegraphics[scale=0.175]{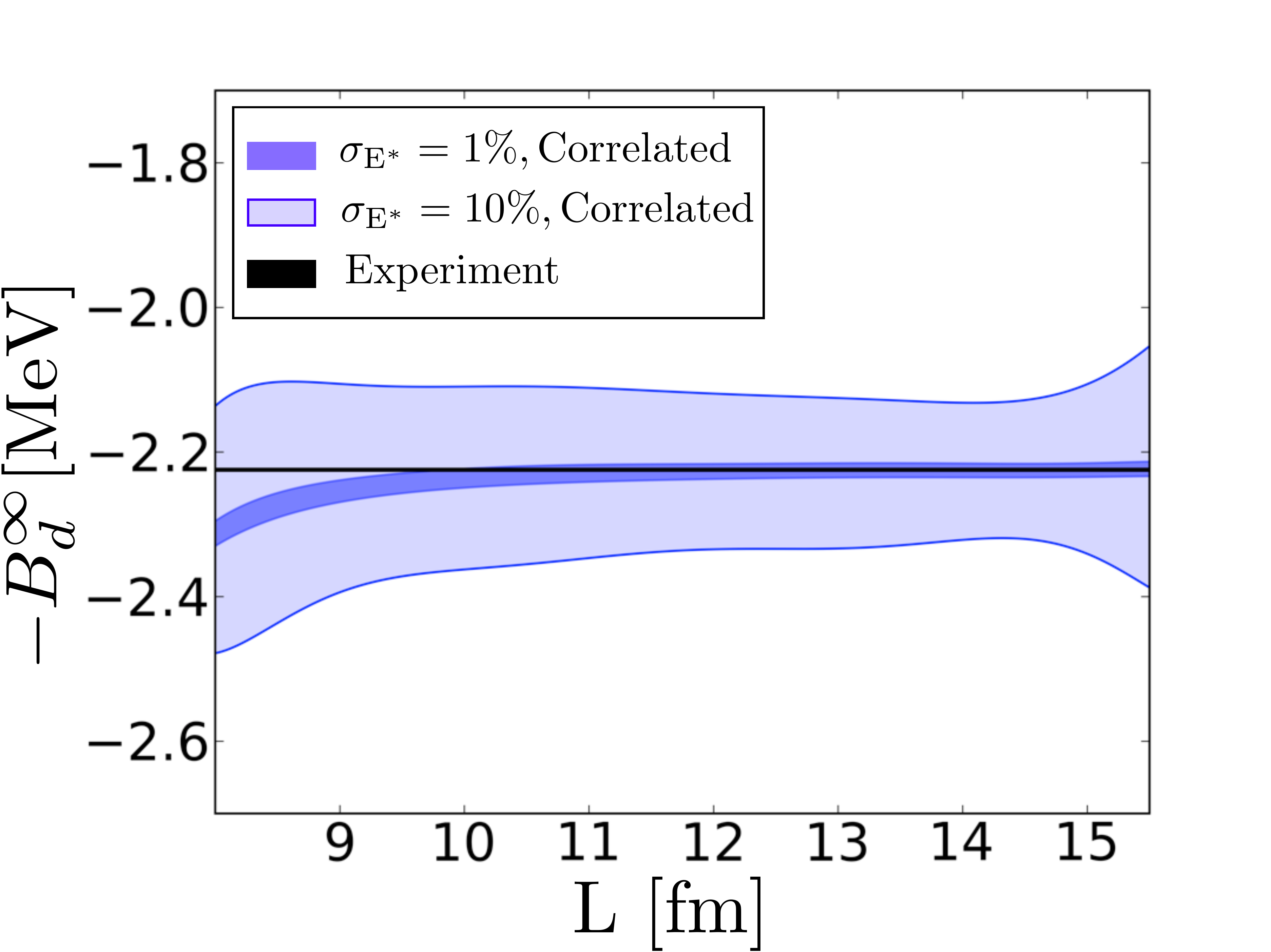}}
\subfigure[]{
\label{Bd_uncorr}
\includegraphics[scale=0.175]{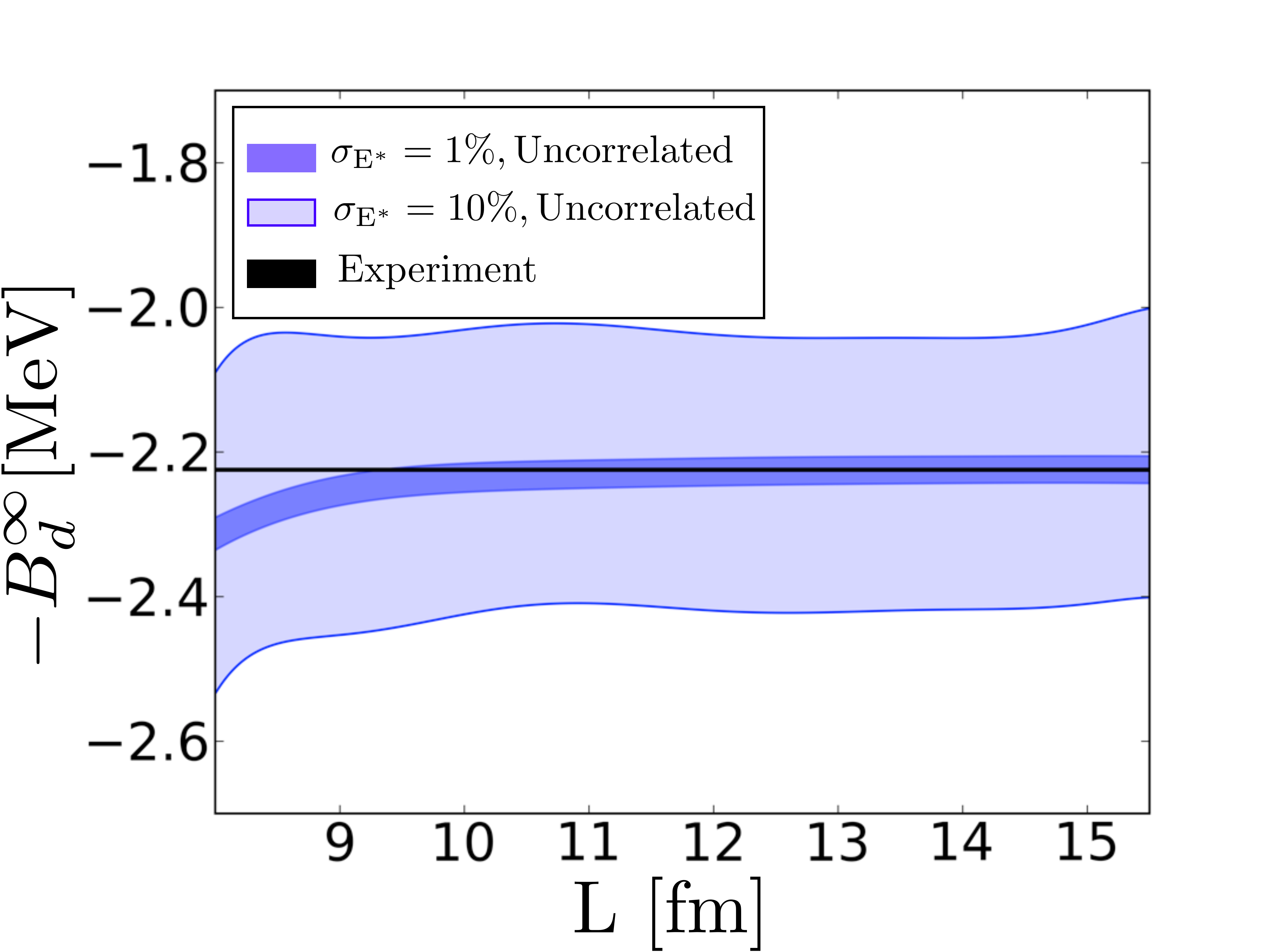}}
\subfigure[]{
\label{e1_corr}
\includegraphics[scale=0.175]{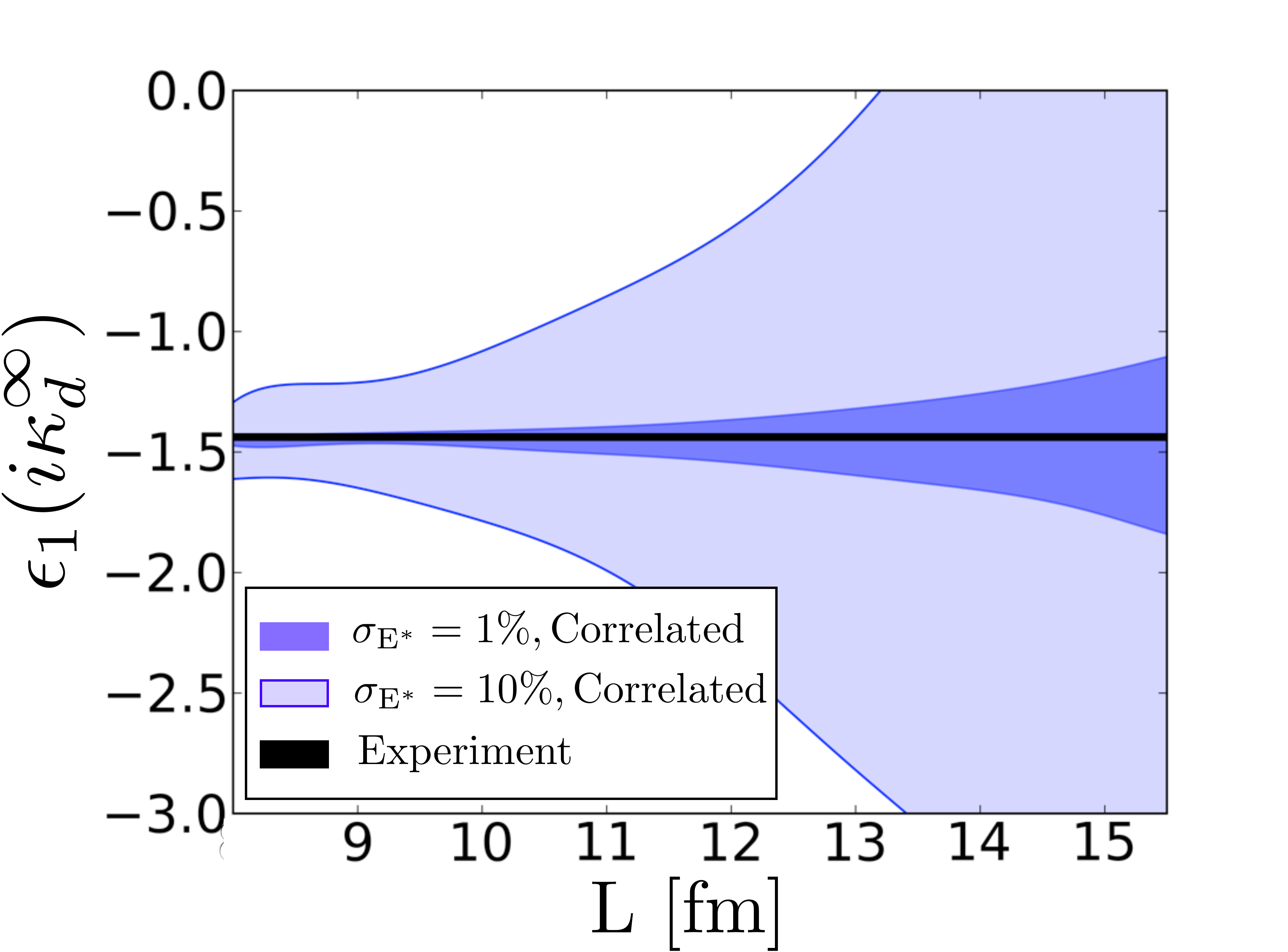}}
\subfigure[]{
\label{e1_uncorr}
\includegraphics[scale=0.175]{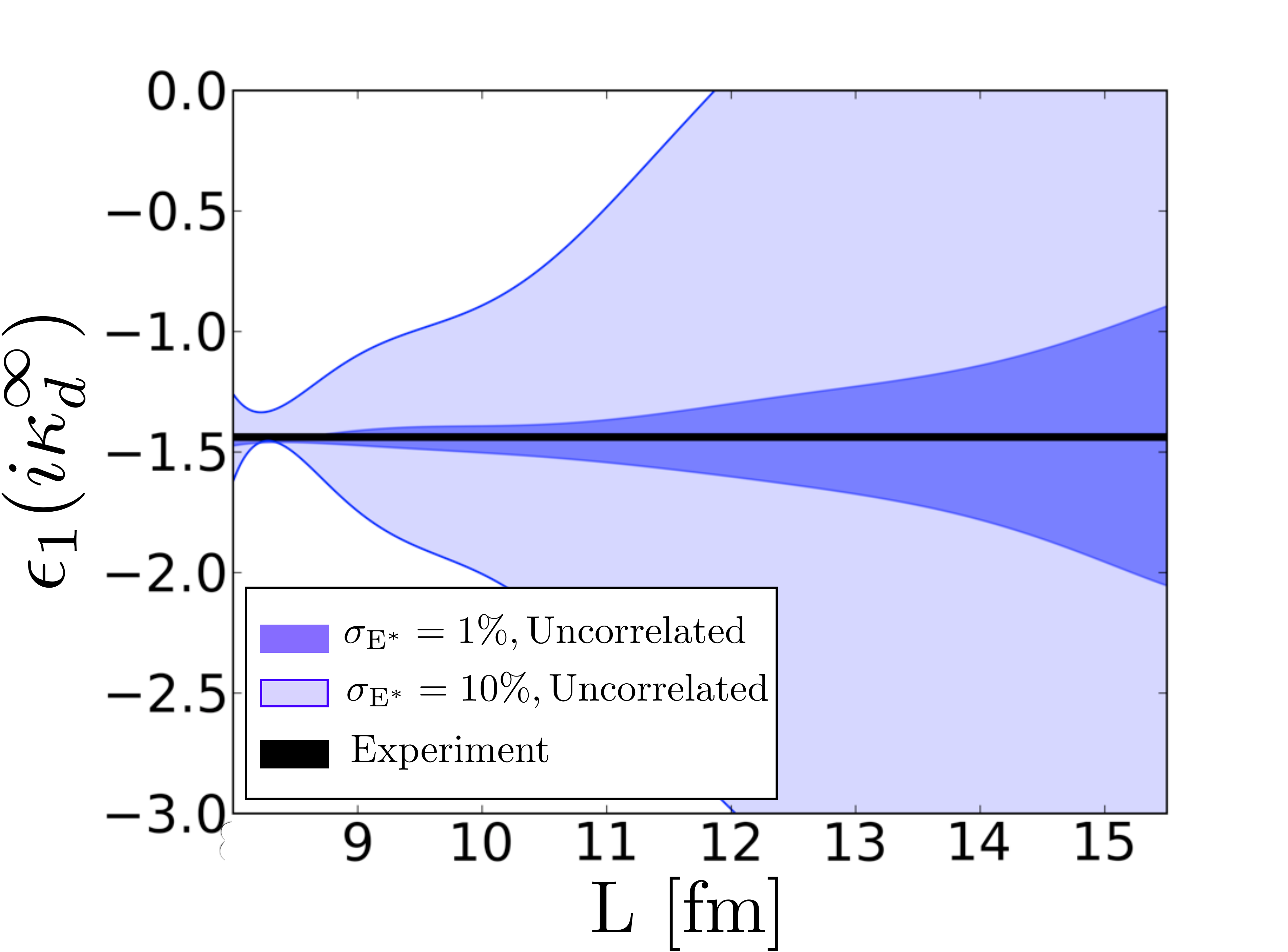}}
\caption{
{\small The values of
  $\{-B_d^{\infty},\epsilon_1(\mathit{i}\kappa_d^{\infty})\}$ obtained by fitting the
  six independent 
bound-state energies with $|\textbf{d}|\leq\sqrt{3}$ (depicted in
Figs.~\ref{deut_cub}, \ref{deut_tet}, \ref{deut_ort}),
generated from synthetic LQCD calculations,
using the approximate QCs in
Eqs.~(\ref{appr-T1}-\ref{appr-EA2}). 
$\epsilon_1$ is in degrees and ${B_d^\infty}(\kappa_d^\infty)$ 
denotes the infinite-volume deuteron binding energy (momentum). 
The black lines denote
the experimental value of these quantities determined by fitting the scattering
parameters obtained from Ref.~\cite{NIJMEGEN}. 
The dark (light) inner (outer) band is the 
$1\sigma$ band corresponding to the energies being  determined with 
1\% (10\%) precision. 
}}
\label{fig:fakedata2}
\end{center}
\end{figure}

Using the QCs,
the ground-state energy in each irrep is determined for a given lattice volume.
The level of precision of such a future LQCD calculation is introduced  by selecting a
modified energy for each ground state from a Gaussian distribution with the
true energy for its mean and the precision level multiplied by the mean
for its standard deviation.
This generates one set of uncorrelated ``synthetic LQCD calculations''.
To generate fully correlated ``synthetic LQCD calculations'', the same
fluctuation (appropriately scaled) 
is chosen for each energy.~\footnote{Partially-correlated 
``synthetic LQCD calculations'' 
can be generated by forming a weighted average of the uncorrelated and
fully-correlated calculations.}
These synthetic data are then taken to be the results of a possible future LQCD
calculation and analyzed accordingly to extract the scattering parameters \footnote{A similar analysis has been carried out in Ref. \cite{Beane:2010em} where the S-wave scattering length, effective range and the deuteron binding energy are extracted from ``synthetic LQCD calculations'', but using a purely S-wave quantization condition.}.
The values of $\{a^{(^3S_1)},
  r^{(^3S_1)},-B_d^{\infty},\epsilon_1(\mathit{i}\kappa_d^{\infty})\}$ 
extracted from an analysis of the synthetic data
are shown in Figs.~\ref{fig:fakedata1},~\ref{fig:fakedata2} for both
correlated and uncorrelated energies. 
Since for $L\lesssim 10~\rm{fm}$ the contribution of the D-wave phase shifts to the
bound-state spectrum is not negligible, the mean values of the scattering
parameters extracted using the approximated 
QCs deviate from their experimental values. 
This is most noticeable when the binding energies are determined at the 1\%
level of precision, 
where the S-matrix parameters and predicted $B_d^\infty$ can deviate by
$\sim 3\sigma$ from the experimental values for this range of volumes. 
For $10~\rm{fm}$$ < L < $$14~\rm{fm}$, one can see that these quantities can be extracted
with high accuracy using this method,
but it is important to note that the precision with which $\{a^{(^3S_1)},
  r^{(^3S_1)},\epsilon_1(\mathit{i}\kappa_d^{\infty)}\}$ 
can be extracted decreases as a function of increasing volume. 
The reason is that the bound-state energy in each irrep asymptotes to the physical deuteron binding energy
in the infinite-volume limit.
In this limit, sensitivity to $\epsilon_1$ is lost and the $\alpha$-wave phase shift
is determined  at a single energy, the deuteron pole. 
Therefore, for sufficiently large volumes one cannot  independently  resolve $a^{(^3S_1)}$
and $r^{(^3S_1)}$. 
This analysis of synthetic data reinforces the fact that the FV spectrum not
only depends on the magnitude of $\epsilon_1$ but also its sign. 
As discussed in Sect.~\ref{sec:DeutFV}, this sensitivity can be deduced
from the full QCs in Eqs.(\ref{I000T1}-\ref{I111E}), 
but it is most evident from the approximated
QCs in Eqs.~(\ref{appr-T1}-\ref{appr-EA2}).

In performing this analysis, we have benefited from two important pieces of
{\it apriori} knowledge at the physical light-quark masses. 
First is that in the volumes of interest, the bound-state energy in each
irrep falls within the radius of convergence of the 
ERE, $|{\rm E}^*|<m_\pi^2/4M$. 
For unphysical light-quark masses, the S-matrix elements could in principle
change in such a way that this need not be the case
and pionful EFTs would be required 
to extract the scattering parameters from the FV spectrum. 
Second is that the D-wave phase shifts are naturally small. 
Again, since the dependence of these phase shifts on the light-quark masses 
can only be estimated, further investigation would be required.
To improve upon this analysis,  
the J=1 $\beta$-wave and  $J=2,3$ D-wave phase shifts
would have to be extracted from the scattering states.
As is evident from Fig.~\ref{T1specfull}, states that have a strong dependence
on the D-wave phase shifts will, in general, 
lie above the t-channel cut. 
In principle, one could attempt to extract them by fitting the FV bound-state
energies for $L\leq 10~{\rm fm}$ with the full QCs. 
In practice, this will be 
challenging as eight scattering parameters appear in the ERE at
the order at which the 
$J=1$ $\beta$-wave and $J=2,3$ D-wave
phase shifts first contribute.
This is also formally problematic since for small volumes, $m_\pi L \lsim
2\pi $, 
finite range effects are no longer negligible. 
Although these finite range effects have been estimated for two nucleons in a  S-wave~\cite{Sato:2007ms}, 
they remain to be examined for the general NN  system.

\section{The Finite-volume Deuteron Wavefunction and the Asymptotic D/S
  Ratio \label{sec:wavefunction}}
\noindent
The S-matrix dictates the asymptotic behavior of the NN wavefunction, 
and as a result the IR distortions of the wavefunction inflicted by
the boundaries of the lattice volume have a direct connection to the parameters
of the scattering matrix, as exploited by L\"uscher. 
Outside the range of the nuclear forces,   the FV wavefunction of the NN system 
is obtained from the solution of the
Helmholtz equation in a cubic volume with the 
PBCs~\cite{Luscher:1986pf,Luscher:1990ux, Rummukainen:1995vs, Ishizuka:2009bx}. 
By choosing the amplitude of the $l=0$ and $l=2$ components of the 
FV wavefunction to recover the asymptotic D/S ratio of the infinite-volume
deuteron,
it is straightforward to show~\cite{Luscher:1986pf,Luscher:1990ux} 
that the unnormalized FV deuteron wavefunctions
associated with the approximate QCs in Eqs.~(\ref{appr-T1}-\ref{appr-EA2})
are
\begin{eqnarray}
\psi^{V,{\bf d}}_{1,M_J} (\mathbf{r};\kappa)
\ =\ 
\psi^{\infty}_{1,M_J} (\mathbf{r};\kappa)
\ +\ 
\sum_{\mathbf{n} \neq \mathbf{0}} e^{i \pi \mathbf{n} \cdot \mathbf{d}} \ 
\psi^{\infty}_{1,M_J} (\mathbf{r}+\mathbf{n}L;\kappa)
,
\label{psi-V}
\end{eqnarray}
with $r = |{\bf r}|>R$,
where  $\mathbf{r}$ denotes the relative displacement 
of the two nucleons, 
and $R >>  1/m_\pi$ is the approximate range of the nuclear interactions.
The subscripts on the wavefunction refer to the $J=1,~M_J=0,\pm1$ 
quantum numbers of the state and 
$\mathbf{n}$ is an integer triplet. 
In order for Eq.~(\ref{psi-V}) to be an energy eigenstate of the Hamiltonian, 
$E^*=-{\kappa^2}/{M}$ has to be an energy eigenvalue of the NN  system in
the finite volume,
obtained from the  QCs in Eqs.~(\ref{appr-T1}-\ref{appr-EA2}). 
$\psi^{\infty}_{1,M_J}(\mathbf{r})$ is the asymptotic infinite-volume wavefunction of the deuteron,
\begin{eqnarray}
\psi^{\infty}_{1,M_J} (\mathbf{r};\kappa)
 \ =\ 
\mathcal{A}_S 
\ \left(\ 
\frac{e^{-\kappa r}}{r}
 \mathcal{Y}_{1M_J;01}(\hat{\mathbf{r}})
\ +\ 
\eta~\frac{e^{-\kappa r}}{r} (1+\frac{3}{\kappa
  r}+\frac{3}{\kappa^2r^2}) 
\mathcal{Y}_{1M_J;21}(\hat{\mathbf{r}})
\ \right)
.
\label{psi-inf}
\end{eqnarray}
with $\mathcal{Y}_{JM_J;L1}$ being the well-known spin-orbital functions,
\begin{eqnarray}
\mathcal{Y}_{JM_J;L1}(\hat{\mathbf{r}})
\ =\ 
\sum_{M_L,M_S}\left\langle L M_L 1 M_S|J M_J\right\rangle  \
Y_{L M_L}(\hat{\mathbf{r}})\ \mathcal{\chi}_{1 M_S}
,
\label{Y-def}
\end{eqnarray}
where $\mathcal{\chi}_{1 M_S}$ is the spin wavefunction of the deuteron. 
$\eta$ is the deuteron asymptotic D/S ratio which is
related to the mixing angle via
$\eta=-\tan{\epsilon_1}|_{k^*=i\kappa^\infty_d}$~\cite{Blatt:1952zza}. 
As is well known from the effective range theory~\cite{PhysRev.76.38,
  PhysRev.77.647}, 
the short-distance contribution to the \textit{outer} quantities of the
deuteron, 
such as  the quadrupole moment,
can be approximately taken into account by requiring the
normalization  of the asymptotic wavefunction of the deuteron,
obtained from the residue of the S-matrix at the
deuteron pole, 
to be approximately 
$|\mathcal{A}_S|^2\approx{2\kappa}/({1-\kappa  r^{(^3S_1)}})$. 
Corrections to this normalization arise at 
$\mathcal{O}\left({\kappa^3}/{R}^2,\kappa\eta^2\right)$,
at the same order the 
$J=1$ $\beta$-wave and $J=2,3$ D-waves
contribute.
In writing the FV wavefunction in Eq.~(\ref{psi-V}) 
contributions from  these
waves have been neglected.

\begin{figure}[h!]
\begin{center}  
\subfigure[]{
\label{WF-T1-L10}
\includegraphics[scale=0.215]{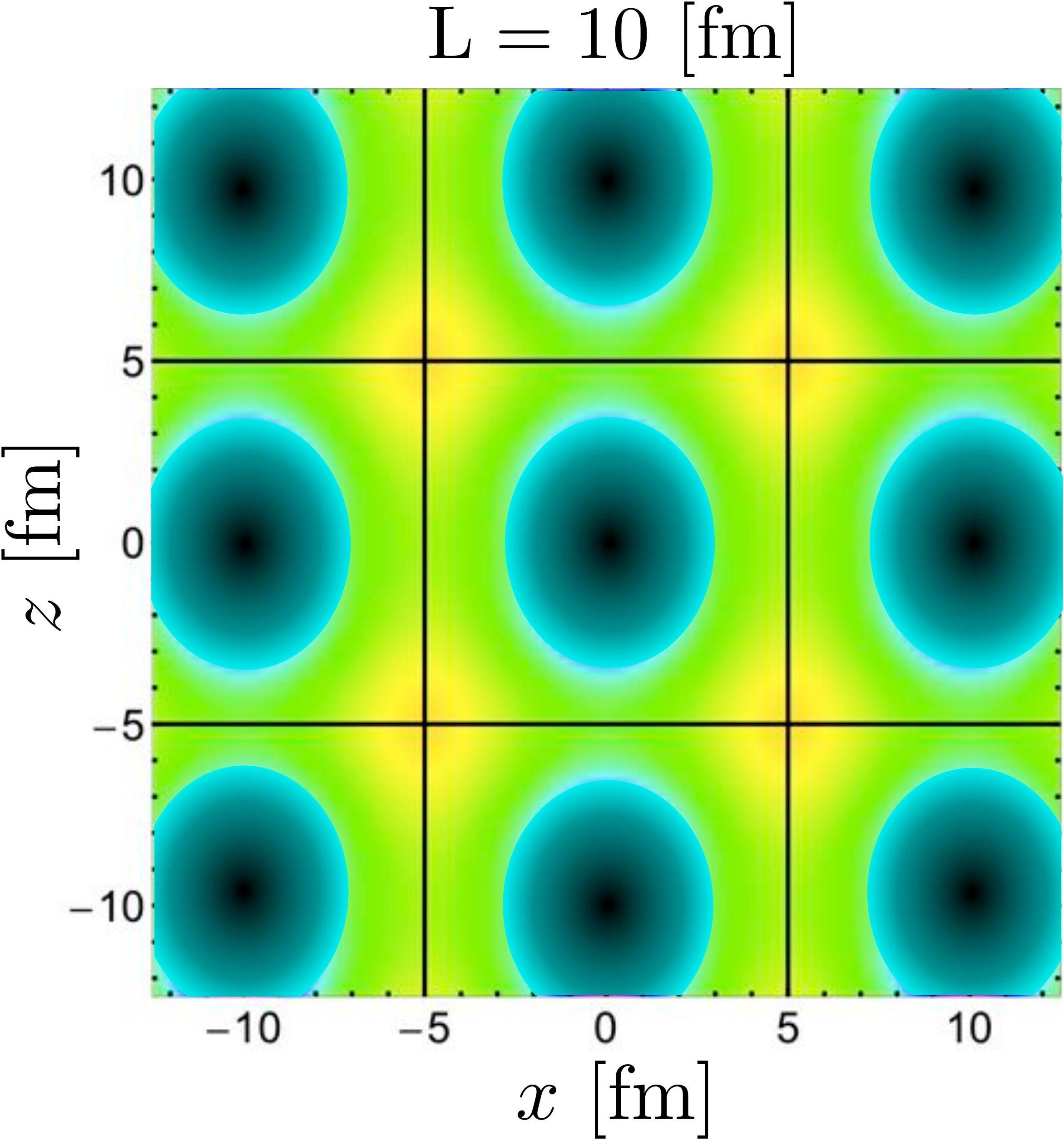}}
\subfigure[]{
\label{WF-T1-L15}
\includegraphics[scale=0.215]{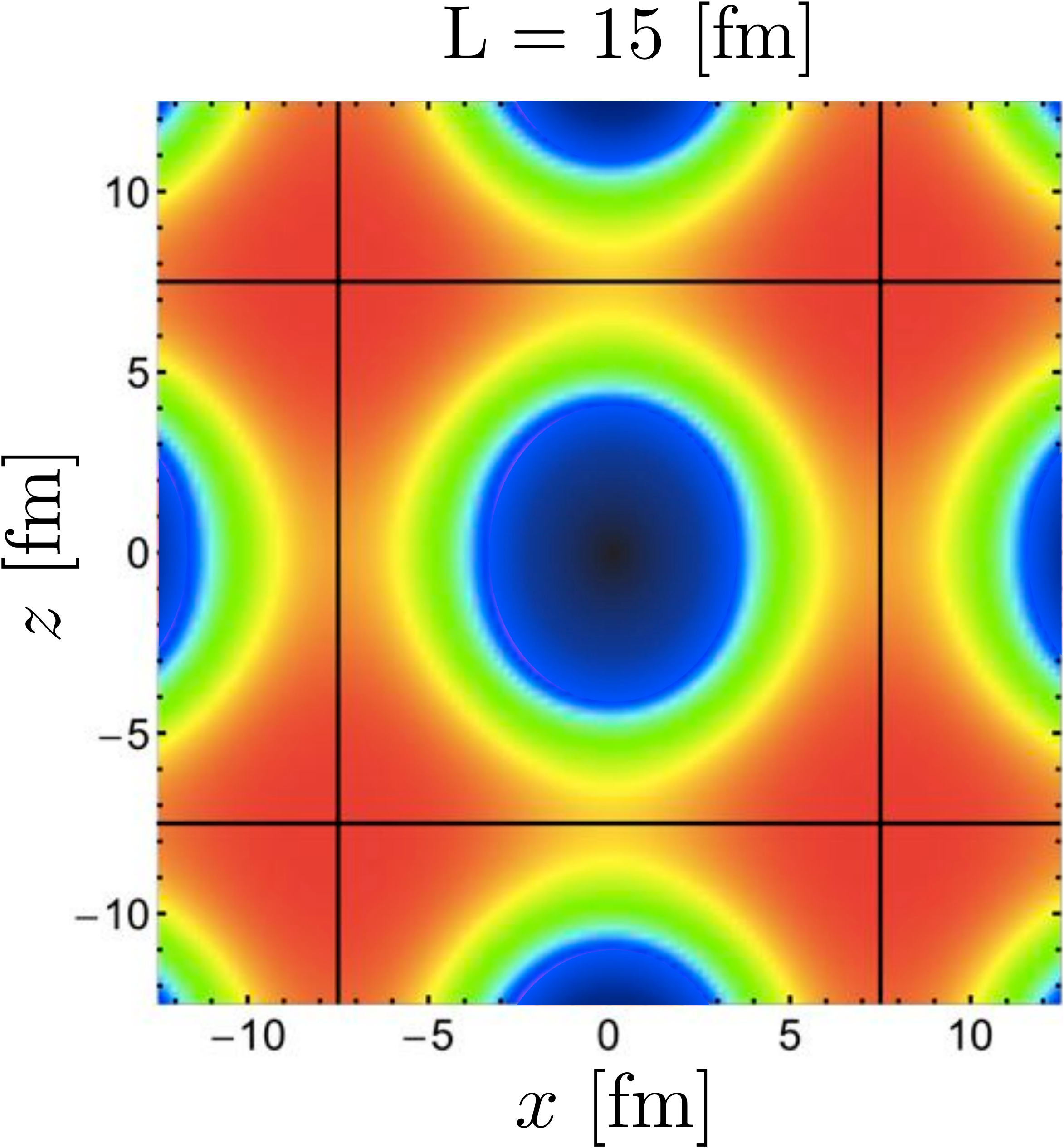}}
\subfigure[]{
\label{WF-T1-20}
\includegraphics[scale=0.215]{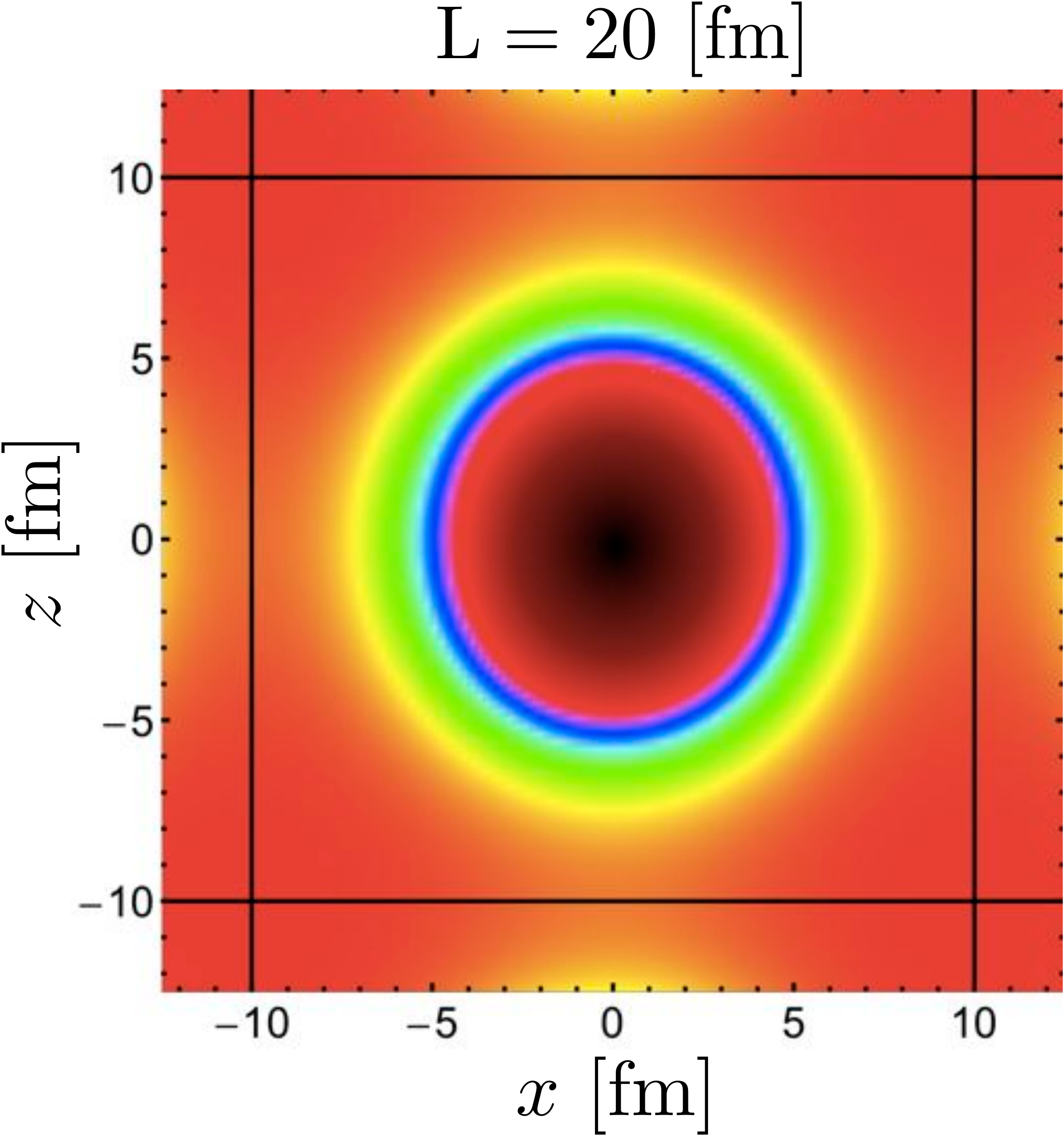}}
\subfigure[]{
\label{WF-T1-30}
\includegraphics[scale=0.215]{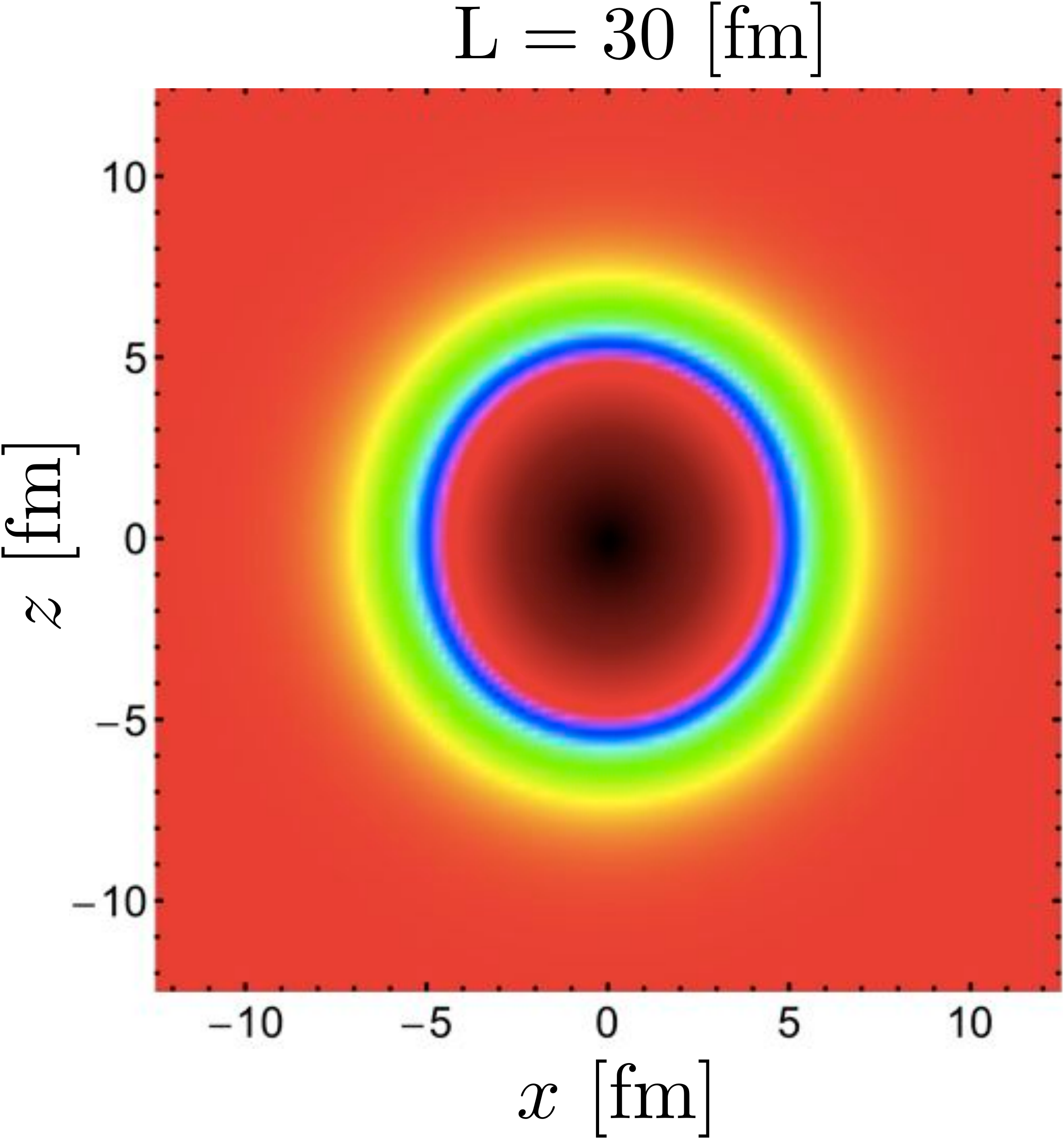}}
\caption{{\small The mass density  in the $xz$-plane from the $\mathbb{T}_1$ FV deuteron wavefunction  at rest
  for 
$L=10,15,20$, and $30~{\rm fm}$.
}}
\label{WF-T1}
\end{center}
\end{figure}
An important feature of the FV wavefunction in Eq. (\ref{psi-V}) is the
contribution from partial waves 
other than $l=0$ and $l=2$, 
which results from the cubic distribution of the periodic images. 
While there are also FV corrections to the $l=0$ component of the wavefunction, 
the FV corrections to the $l=2$ component are enhanced  for systems with
${\bf d}=(0,0,1)$ and $(1,1,0)$. 
By forming appropriate linear combinations of the $\psi^{V,{\bf d}}_{1,M_J}$ 
that transform according to a given irrep of the cubic, tetragonal, 
orthorhombic and trigonal point groups (see Table \ref{irreps}),
wavefunctions for the systems with ${\bf d}=(0,0,0)$, $(0,0,1)$, $(1,1,0)$ and
$(1,1,1)$
can be obtained.
The 
mass density in the $xz$-plane from 
the FV wavefunction  of the deuteron 
at rest in the volume, obtained from the $\mathbb{T}_1$ irrep of the
  cubic group 
is shown in Fig.~\ref{WF-T1}
for $L=10,15,20$, and $30~{\rm fm}$,
and for the boosted systems in  Figs.~\ref{WF-A2}-\ref{WF-A2E} of Appendix \ref{sec:Wavefunc}. 
As the interior region of the wavefunctions cannot be deduced from its asymptotic
behavior alone, it is ``masked'' 
in  Fig.~\ref{WF-T1} and Figs.~\ref{WF-A2}-\ref{WF-A2E} by a 
shaded disk.
Although the deuteron wavefunction exhibits its slight prolate shape 
(with respect to its spin axis) 
at large volumes, 
it is substantially deformed in smaller volumes, 
such that the deuteron can no longer be thought as a compact bound state within
the lattice volume. 
When the system is at rest, the FV deuteron 
is more prolate than the infinite-volume deuteron. 
When the deuteron is boosted along the $z$-axis with ${\bf d}=(0,0,1)$, 
the distortion of the wavefunction 
is large, 
and in fact, for a significant  
range of volumes ($L \lesssim 30~\rm{fm}$), the FV effects
give rise to an oblate (as opposed to prolate) deuteron in the 
$\mathbb{E}$ irrep, 
Fig. \ref{WF-E}, and a more prolate shape in the 
$\mathbb{A}_2$ irrep, 
Fig. \ref{WF-A2}. 
For ${\bf d}=(1,1,0)$, 
the system remains prolate for the deuteron in the 
$\mathbb{B}_2/\mathbb{B}_3$ irreps,
Fig. \ref{WF-B2B3}, 
while it becomes  
oblate in the $\mathbb{B}_1$ irrep, Fig. \ref{WF-B1}, for volumes up to $L \sim
30~\rm{fm}$. 

\begin{figure}[h!]
\begin{center}  
\subfigure[]{
\label{NDS-T1}
\includegraphics[scale=0.215]{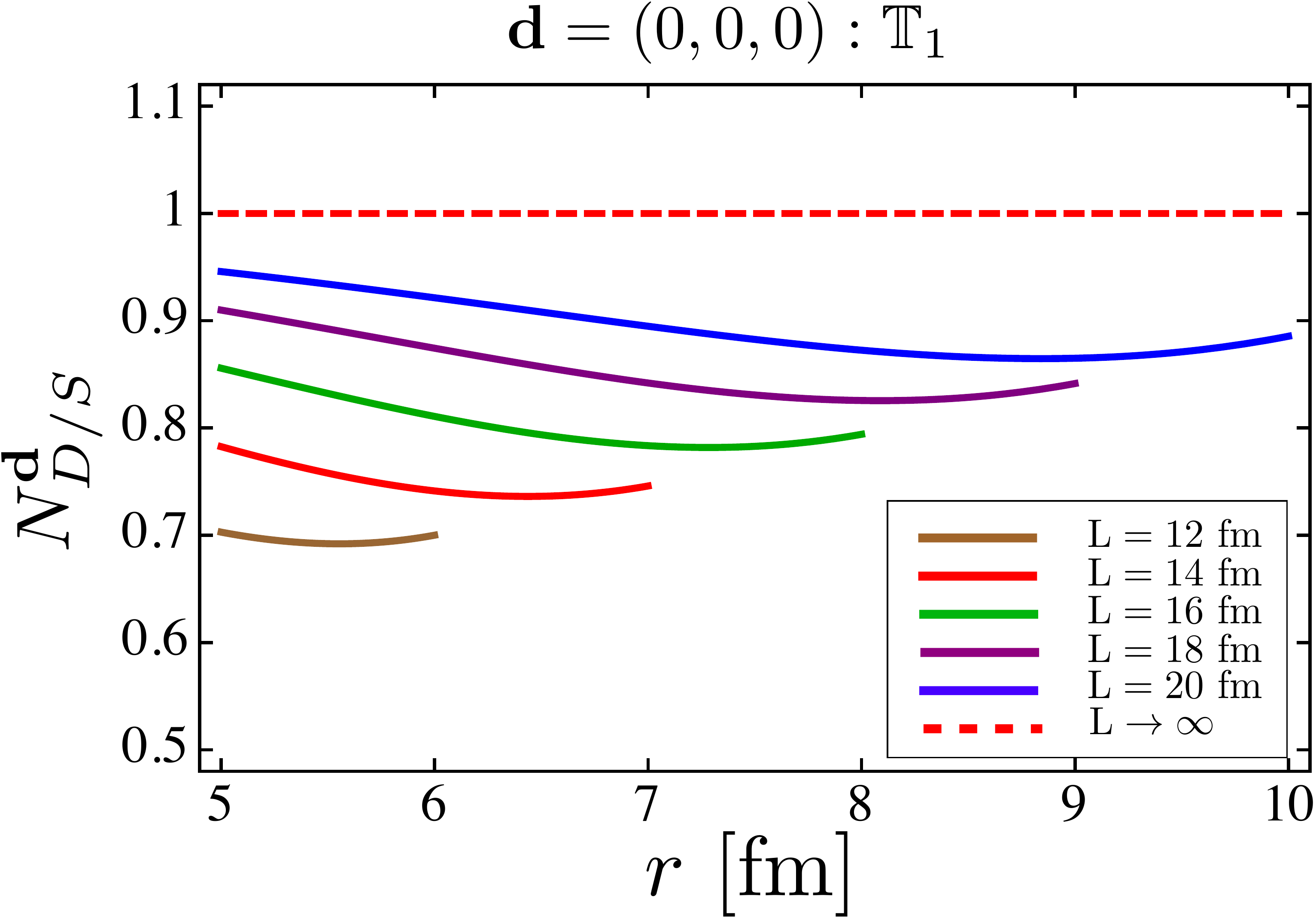}}
\subfigure[]{
\label{NDS-E}
\includegraphics[scale=0.215]{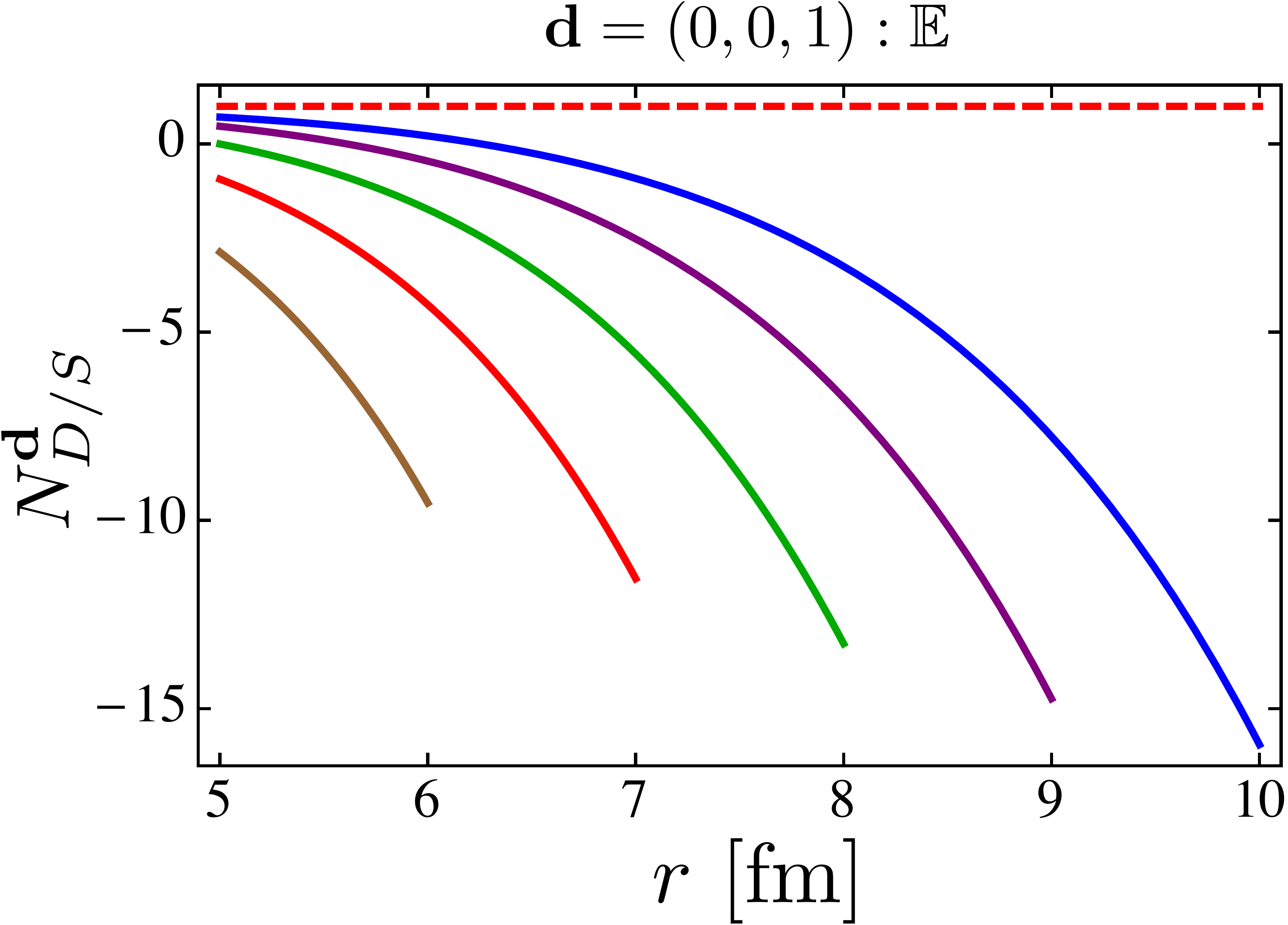}}
\subfigure[]{
\label{NDS-B2B3}
\includegraphics[scale=0.215]{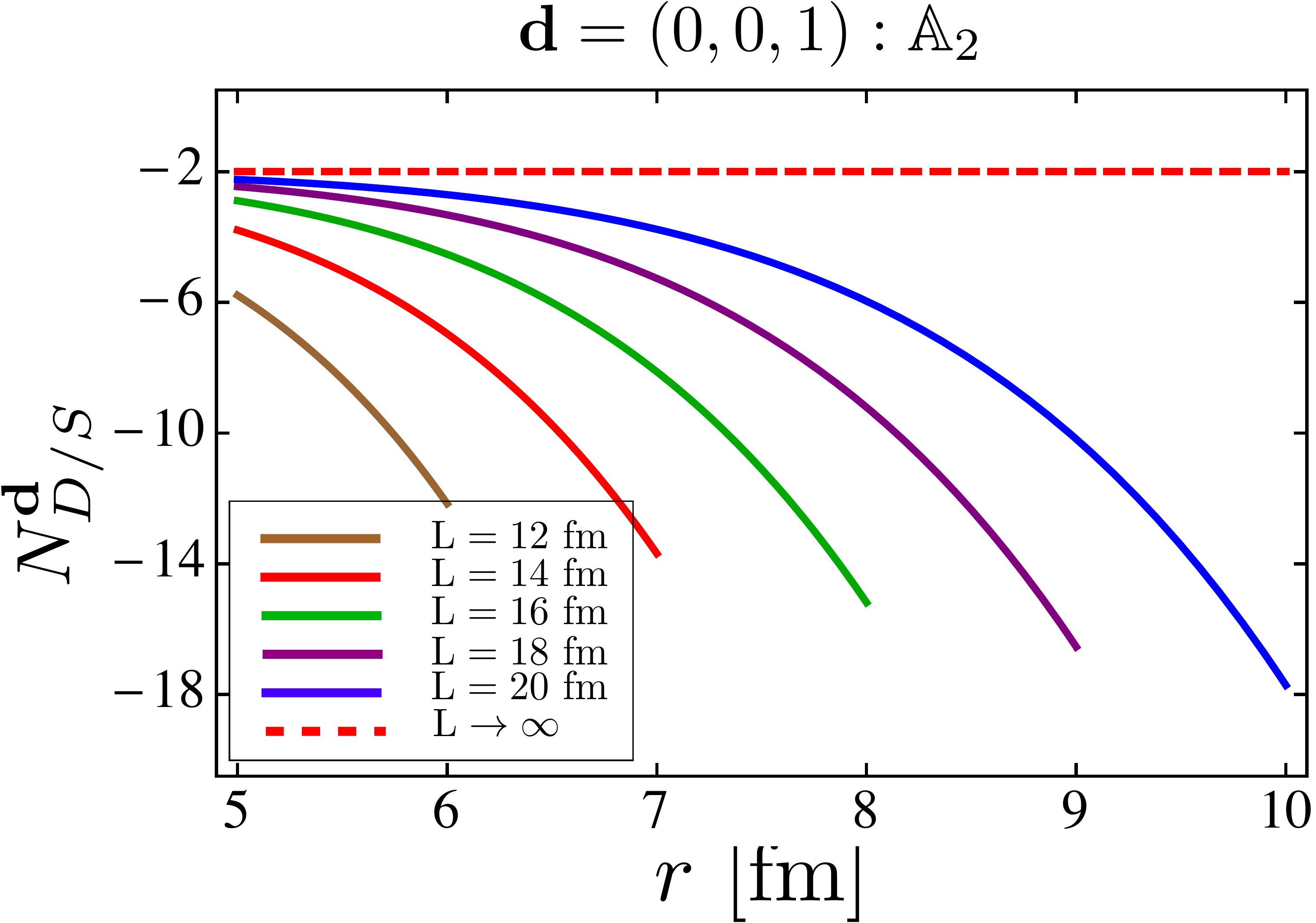}}
\subfigure[]{
\label{NDS-A2E}
\includegraphics[scale=0.215]{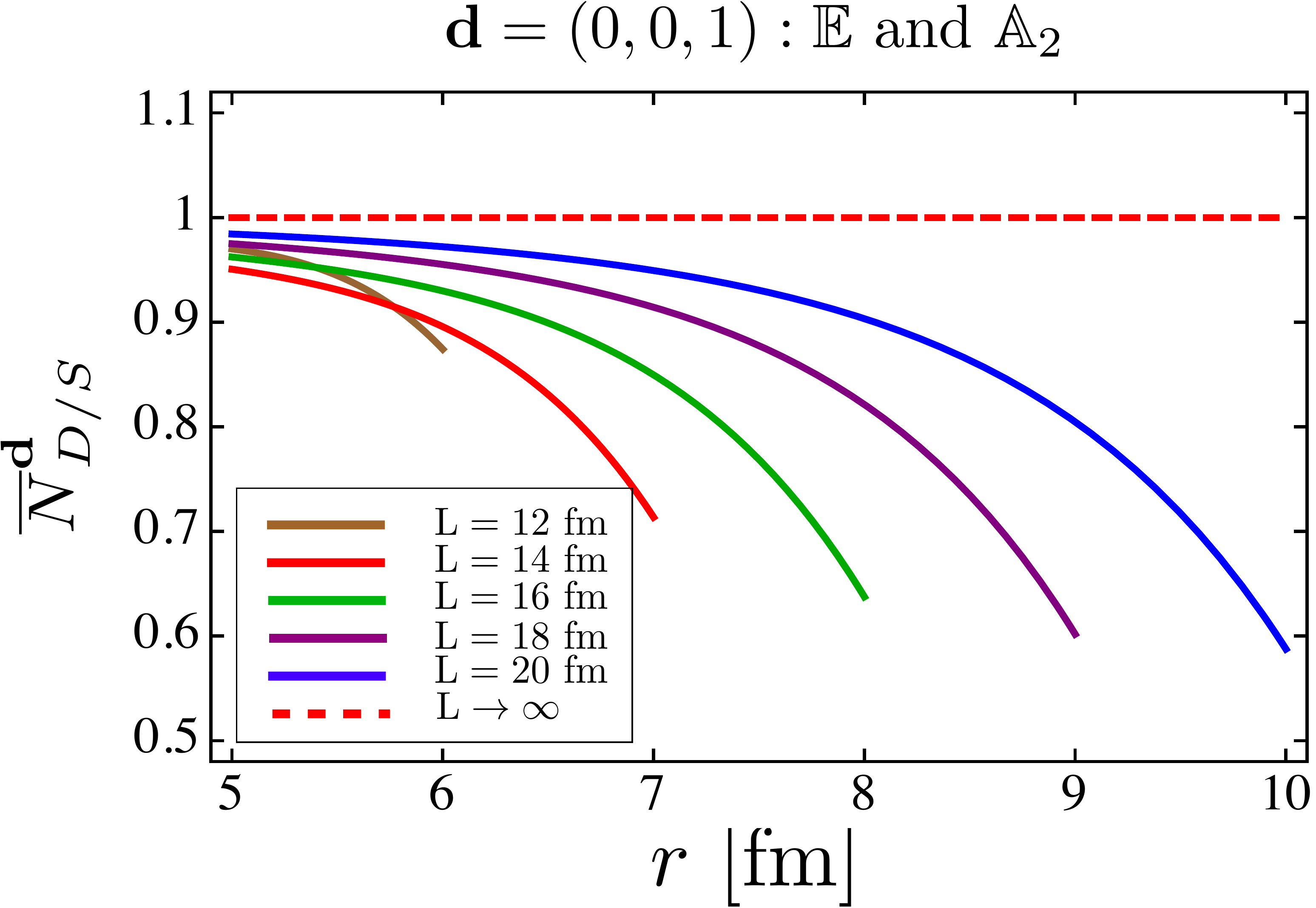}}
\caption{
{\small The normalized D/S ratio of the deuteron wavefunction with $M_J=M_S=1$, defined in 
Eq.~(\protect\ref{ratio}) and Eq.~(\protect\ref{ratiob})
in the 
a) $\mathbb{T}_1$, 
b) $\mathbb{E}$ and 
c) with $M_J=M_S=0$ in the $\mathbb{A}_2$ irrep,
along with 
d) the difference of the D/S ratios in  the  $\mathbb{E}$ and  $\mathbb{A}_2$ irreps, defined
in Eq.~(\protect\ref{eq:A2Ediff}). 
The red-dashed lines show 
the infinite-volume value.}}
\label{fig:NDS}
\end{center}
\end{figure}

Although the normalization factor $\mathcal{A}_S$ corrects  for
the fact that the complete wavefunction is not given by the asymptotic form
given in Eq.~(\ref{psi-inf}) for  $|{\bf r}|\lesssim r^{(^3S_1)}/2$ in infinite volume, it gives rise to a
normalization ambiguity in the FV. 
On the other hand, the asymptotic D/S ratio is protected by the
S-matrix, and 
can be directly extracted from the long-distance tail of the lattice
wavefunctions.~\footnote{
The energy-dependent ``potentials'' generated by HALQCD
  and used to compute scattering parameters, including $\epsilon_1$ (at
unphysical light-quark masses)~\cite{Murano:2013xxa},
are expected to reproduce the predictions of QCD only at the energy eigenvalues of
their LQCD calculations.  
Hence, if they had found a bound deuteron,
their prediction for $\epsilon_1$ would be expected to be correct at the
calculated deuteron binding energy.
} 
It is evident from Eq.~(\ref{psi-inf}) that the ratio
\begin{eqnarray}
N_{D/S}^{{\bf d}; M_J, M_S}(r;\kappa) \equiv
\frac{\psi_{D;M_J,M_S}^{V,{\bf d}}(r;\kappa)}{\eta~\chi(r;\kappa)~\psi_{S;M_J,M_S}^{V,{\bf d}}(r;\kappa)}
,
\label{ratio}
\end{eqnarray}
with $\chi(r;\kappa)=\sqrt{\frac{1}{10}}(1+\frac{3}{\kappa
  r}+\frac{3}{\kappa^2r^2})$, 
is unity  for the $M_J=M_S=1$ component of the infinite-volume deuteron wavefunction
(and is equal to $-2$ for the $M_J=M_S=0$ component),
where
$\psi_{S;M_J,M_S}^{V,{\bf d}}$  and $\psi_{D;M_J,M_S}^{V,{\bf d}}$ 
are 
\begin{eqnarray}
\psi_{S;M_J,M_S}^{V,{\bf d}}(r;\kappa)
& = &  
\int d\Omega_{\hat{\mathbf{r}}} \ 
\psi^{V,{\bf d}}_{1,M_J}(\mathbf{r},\kappa)
\big|_{M_S}\ 
\ Y_{00}(\hat{\mathbf{r}})
,
\nonumber\\
\psi_{D;M_J,M_S}^{V,{\bf d}}(r;\kappa)
& = & 
\int d\Omega_{\hat{\mathbf{r}}}\  
\psi^{V,{\bf d}}_{1,M_J}(\mathbf{r},\kappa) \big|_{M_S}\ 
\ Y_{20}(\hat{\mathbf{r}})
,
\label{ratiob}
\end{eqnarray}
with $r\le L/2$.
By evaluating the FV wavefunction 
in different irreps with $|\mathbf{d}|\leq \sqrt{3}$, 
this ratio can be determined in the FV, as is shown in 
Fig.~\ref{fig:NDS}. 
Not only does it exhibit strong dependence on the volume, 
but also varies dramatically as a function of $r$. 
This is due to the fact that the periodic images give rise to exponentially
growing contributions 
to the FV wavefunction in $r$. 
For the FV deuteron at rest and with $\mathbf{d}=(1,1,1)$, 
a sufficiently small $r$ gives rise to a $N_{D/S}^{{\bf d};M_J,M_S}$ that is not severely
distorted by volume effects even in small volumes.
In contrast,  
this ratio deviates significantly from its infinite-volume value for systems
with $\mathbf{d}=(0,0,1)$ and $(1,1,0)$ even in large volumes ($L \lesssim
20~\rm{fm}$). 
This feature is understood by noting that 
while  the leading correction to $N_{D/S}^{{\bf d};M_J,M_S}$ 
is $\sim \eta~e^{-\kappa L}$ for systems with ${\bf d}=(0,0,0)$ and
$(1,1,1)$, 
they are $\sim e^{-\kappa L}$ for systems
with ${\bf d}=(0,0,1)$ and $(1,1,0)$. 
The periodic images
of the wavefunction with 
the latter boosts are quadrupole distributed, and consequently modify the
$l=2$ component of the 
wavefunction by contributions that are not suppressed by $\eta$.
However, for these systems, there are two irreps that receive similar FV
corrections to their ratios,
which can be largely removed
by forming differences, 
e.g. for the system with ${\bf d}=(0,0,1)$,
\begin{eqnarray}
\overline{N}_{D/S}^{(0,0,1)}
& = & 
{1\over 3}\left(
N_{D/S}^{(0,0,1);1,1}
\ -\ 
N_{D/S}^{(0,0,1);0,0}
 \right)
\ =\ 
{1\over 3}\left(
N_{D/S}^{(0,0,1);\mathbb{E}}
\ -\ 
N_{D/S}^{(0,0,1);\mathbb{A}_2}
 \right)
,
\label{eq:A2Ediff}
\end{eqnarray}
as shown in 
Fig.~\ref{fig:NDS}.  A similar improvement is found for systems with 
${\bf d}=(1,1,0)$.
It is also worth noting that the contributions to the wavefunction from higher
partial waves, $l\geq2$, can be added to $\psi^{V,{\bf d}}$ with 
coefficients that depend on their corresponding phase shifts, and therefore are
small  under the assumption of low-energy scattering \cite{Luscher:1990ux,
Rummukainen:1995vs, Ishizuka:2009bx}. 
However, the partial-wave decomposition of the FV wavefunction in Eq.~(\ref{psi-V})
contains contributions with $l\geq2$. 
In the limit where the corresponding phase shifts vanish, the wavefunction, in contrast to the spectrum,  
remains  sensitive to these contributions,
resulting in  the larger FV modifications  of
quantities   compared with their spectral analogues.

An extraction of $\eta$ is possible by taking sufficiently large volumes such that a large NN
separation can be 
achieved without approaching the boundaries of the volume. 
While 
the wavefunctions corresponding to the deuteron at rest or with
$\mathbf{d}=(1,1,1)$ 
provide an opportunity to extract $\eta$
with an accuracy of $\sim 15-20\%$ in volumes of  $L\sim 14~\rm{fm}$,
combinations of the ratios obtained from the two irreps 
in both the systems with $\mathbf{d}=(0,0,1)$ and $(1,1,0)$
will provide for a $\sim 10\%$ determination  in volumes of  $L\sim 12~\rm{fm}$, 
as shown in  Fig. \ref{fig:NDS}.
As it is possible that the uncertainties in the
extraction of $\eta$ can be systematically reduced,
those due to the neglect of the
$J=1$ $\beta$-wave and $J=2,3$ D-wave
phase shifts, as well as higher order terms in the ERE, 
deserve further investigation.

\section*{Summary and Conclusion 
\label{sec:conclusion}}
\noindent
A Lattice QCD calculation of the deuteron and its properties would be a
theoretical milestone on the path toward  calculating quantities of importance in low-energy nuclear physics 
from quantum chromodynamics without uncontrolled approximations or assumptions.
While there is no formal impediment to calculating the deuteron binding energy
to arbitrary precision when sufficient computational resources become
available, determining its  properties and interactions presents a
challenge that has largely remained unexplored~\cite{Detmold:2004qn,Meyer:2012wk,Briceno:2013lba}. 
Using the NN formalism developed in Ref.~\cite{Briceno:2013lba}, 
we have explored  the 
FV energy 
spectra of states that
have an overlap with the $\siii$-$\diii$ coupled-channels system in which
 the deuteron resides.
Although the full FV QCs associated with the 
$^3S_1$-$^3D_1$ coupled channels depend on interactions in all 
positive-parity isoscalar channels, 
a low-energy expansion
depends only on four scattering
phases and one mixing angle.
Further, for the deuteron, these truncated QCs can be further simplified to
depend only upon one phase shift and the mixing angle,
with corrections suppressed by 
$\sim\frac{1}Le^{-2\kappa L} \tan{\delta_i}$
where $\delta_i$ denotes the 
$J=1$ $\beta$-wave and $J=2,3$ D-wave
phase shifts
which are all small at the deuteron binding energy.
We have demonstrated that the infinite-volume deuteron binding energy and
leading scattering parameters,
including the mixing angle, $\epsilon_1$, 
that dictate the low-energy behavior of the scattering amplitudes,
can  be (in principle) determined with precision
from the bound-state spectra of deuterons, both at rest and in motion,
in a single modest volume, with $L=10$-$14~\rm{fm}$.
Calculations in a second lattice volume would reduce
the systematic uncertainties introduced by truncating the QCs.

We have investigated the feasibility of extracting $\epsilon_1$ from the
asymptotic D/S ratio of the deuteron FV wavefunction using the
periodic images associated with the $\alpha$-wavefunction. 
As the amplitude of the $J=1$ $\beta$-wave and the $J=2,3$ D-wave components of
the wavefunction are not constrained by the infinite-volume deuteron wavefunction, 
the analysis is limited by an imposed truncation of the ERE, 
which is at the same level of approximation as the approximate QCs. 
The systematic uncertainties introduced by this
truncation 
are currently unknown, but will be 
suppressed by the small phase shifts in those channels in addition to being 
exponentially suppressed with $L$.
This is in contrast to the extraction from the FV spectra where the
systematic uncertainties have been determined to be small. 
With this approximation, it is estimated  that 
volumes with $L \gsim 12~\rm{fm}$ are required to extract $\epsilon_1$  
with $\sim 10\%$ level of accuracy from the asymptotic form of the
wavefunctions.


\chapter{FINITE-VOLUME FORMALISM WITH TWISTED BOUNDARY CONDITIONS}
{\label{chap:TBC}}


LQCD calculations are commonly performed with PBCs imposed upon the quark fields in the spatial directions,
constraining the quark momentum modes in the volume to satisfy
$\mathbf{p}=\frac{2\pi}{L}\mathbf{n}$ with $\mathbf{n}$ being an integer triplet. 
PBCs are a subset of a larger class of BCs called twisted BCs (TBCs).
 TBCs~\cite{PhysRevLett.7.46} are those that require the quark fields to acquire 
a  phase $\theta$ at the boundary, 
$\psi(\mathbf{x}+\mathbf{n}L )=e^{i{\rm{\theta}} \cdot \mathbf{n}}\psi(\mathbf{x})$,
where $0<\theta_i<2\pi$ is the twist angle in the $i^{\rm th}$ Cartesian direction. 
Bedaque~\cite{Bedaque:2004kc}  introduced this idea to the LQCD community, and showed  that 
TBCs are equivalent to having a $U(1)$ background gauge field in the QCD Lagrangian 
with the quarks subject to PBCs. By choosing this constant background field to be, e.g., $\mathbf{A}=\frac{\theta_z}{L}\hat{e}_z$, the quark wavefunction will acquire a non-vanishing phase $\frac{\theta_z}{L}$ when wraps around the boundary of the volume in the $z$ direction due to the Aharonov-Bohm effect, despite the magnetic field strength being zero on the lattice. Alternatively, a quark field redefinition, $\psi (\mathbf{x})\rightarrow \widetilde{\psi}(\mathbf{x})= e^{i \frac{\theta_z}{L}z} \psi(\mathbf{x})$, will eliminate this background field provided that the new field $\widetilde{\psi}$ satisfies the TBCs, $\widetilde{\psi}(x,y,z+L)=e^{i\theta_z}\widetilde{\psi}(x,y,z)$. So the lattice gauge field configurations can be generated with these choices of BCs. The benefit of such BCs is that an arbitrary momentum can be selected for a (non-interacting) hadron by a judicious choice
of the twist angles of its valence quarks, 
$\mathbf{p}=\frac{2\pi}{L}\mathbf{n}+\frac{\bm{\phi}}{L}$,
where $\bm{\phi}$ is the sum of the twists of the valence quarks, again with 
$0<\phi_i<2\pi$, and $\mathbf{n}$ is an integer triplet. 
TBCs have been  shown to 
be useful in LQCD calculations of 
 the low-momentum transfer behavior of form factors required in determining 
hadron radii and moments, circumventing the need for large-volume lattices~\cite{Tiburzi:2005hg,Jiang:2006gna,Boyle:2007wg,Simula:2007fa,Boyle:2008yd,Aoki:2008gv,Boyle:2012nb, Brandt:2013mb}.
They have also been speculated to be helpful 
in calculations of $K\rightarrow \pi\pi$ decays
by bringing the initial and final FV states  closer in energy~\cite{deDivitiis:2004rf, Sachrajda:2004mi}.

In addition to performing calculations with a particular twist,
by averaging the results of calculations over twist angles, the discrete sum over momentum modes becomes 
an integral over momenta, 
\begin{eqnarray}
\int \ {d^3\bm{\phi}\over (2\pi)^3}\ 
\frac{1}{L^3}\ \sum_{\mathbf{n} \in \mathbb{Z}^3}
& \equiv & 
\int \frac{d^3\mathbf{p}}{(2\pi)^3}
.
\end{eqnarray}
Although the volume dependence of most quantities   is non linear due to interactions, such averaging can eliminate significant FV effects. 
This was first examined in the context of condensed-matter physics where, for example, the finite-size effects in the finite-cluster calculations 
of correlated electron systems are shown to be reduced by the boundary condition integration technique \cite{Gros, PhysRevB.53.6865}. 
This technique is implemented in quantum Monte Carlo QMC algorithms of many-body  systems, and results in faster convergence of energies to the thermodynamic limit~\cite{PhysRevE.64.016702}. 

In this chapter, we discuss the advantages of using TBCs to reduce the FV modifications to the mass of hadrons and to the 
binding energy of two-hadron bound states, such as the deuteron. 
In particular, we consider the FV effects resulting 
from averaging the results obtained from PBC and anti-PBCs (APBCs), 
from a specific choice of the twist angle,  \emph{i}-PBCs, 
and from averaging over twist angles.
For the two-nucleon systems, the volume improvement is explored both analytically and numerically with the use of the 
 developed FV formalism for NN systems (see chapter \ref{chap:NN}), that is generalized to systems with TBCs. 
As was first noted by Bedaque and Chen~\cite{Bedaque:2004ax},  the need to generate new gauge field configurations 
with fully twisted BCs can be circumvented by  imposing TBCs on the valence quarks only, which defines partial twisting. 
Partial twisting gives rise to corrections beyond full twisting that scale as $e^{-m_{\pi}L}/L$, 
and can be neglected for sufficiently large volumes compared to the FV effects from the size of weakly  bound states. 
Although the validity of partial twisting makes it feasible to achieve an approximate twist-averaged result in LQCD calculations, 
this remains a computationally expensive technique.
We demonstrate that certain hadronic twist angles can result in an exponentially-improved convergence to the infinite-volume 
limit of certain quantities, with an accuracy that is comparable  to the  twist-averaged mean.
Further, we speculate that  similar improvements are also present in arbitrary n-body systems.

As discussed extensively in the previous chapters, in several situations, given the Euclidean nature of lattice correlation functions, it is desirable to keep the volume finite as the extraction of physical 
quantities relies on non-vanishing FV effects. 
 For example, as we saw, the ability to extract the $S$-$D$ mixing parameter, $\epsilon_1$,  and consequently the D/S ratio of the deuteron from LQCD calculations, 
depends upon the  FV modifications to the binding energy when the deuteron is boosted in particular directions 
within the lattice volume \cite{Briceno:2013bda}. 
The use of TBCs will further enhance the effectiveness of such calculations.
By appropriate choices of the twist angles of each hadron, different CM energies can be accessed in a 
single lattice volume,  further constraining the scattering parameters with the use of L\"uscher's method 
(see e.g. Refs.~\cite{Bernard:2010fp, Doring:2011vk, Doring:2012eu, Ozaki:2012ce} for demonstrations of this technique in 
studying hadronic resonances). 
Due to the possibility of partial twisting in NN scattering, 
these extra energy levels can be obtained without having to generate additional ensembles of
gauge-field  configurations, in analogy with the boosted calculations
(this technique has recently been used to calculate $J/\psi$-$\phi$ scattering~\cite{Ozaki:2012ce}).
Of course, the spectra of energy eigenvalues determined with a range of twist angles allow for fits to 
parametrizations of the S-matrix elements, which can then be used to predict infinite-volume quantities, such as binding 
energies~\cite{Beane:2010em,Prelovsek:2013sxa}.
TBCs provide a way to reduce the systematic uncertainties  that are currently present in analyses of 
coupled-channels systems by providing the ability to control, at some level, 
the location of eigenstates.

\section{Nucleon Mass
\label{sec:Single}
}
\noindent
If the up and down quarks have distinct twist angles,  the charged pions, the proton and the neutron will acquire 
net twist angles denoted as 
${\bm{\phi}}^{\pi^{+}}=-\bm{\phi}^{\pi^{-}}$, $\bm{\phi}^{p}$ and $\bm{\phi}^{n}$, respectively, 
while the flavor-singlet mesons, such as $\pi^0$, will remain untwisted, $\bm{\phi}^{\pi^{0}}=\mathbf{0}$.
The optimal set of quark twists depends upon the desired observable, and 
an appropriate choice  can yield a relation between the twists of different hadrons, or
leave a hadron untwisted.

In chapter chapter \ref{chap:intro} we calculated the FV corrections to the mass of nucleon $M_N$, in a cubic volume with PBCs imposed on the quark fields, 
at one-loop order in two-flavor baryon $\chi$PT with the inclusion of $\Delta$ resonance, see Eq. (\ref{deltaM}). The masses of the proton and neutron in a FV
at the one-loop level with TBCs hare also calculated using the HB$\chi$PT~\cite{Jiang:2008ja}.\footnote{The FV corrections to meson masses, decay constants and semileptonic form factors 
with both the TBCs and the partially-TBCs have been calculated at LO in $\chi$PT in Ref. \cite{Sachrajda:2004mi}.
}
We use Poisson re-summation formula to 
factor the dependences on the twist angles as pure phases and put the expressions for the masses into a 
simple form.
The proton mass is found to be 
\footnote{Since nucleons with non-zero twists are not at rest, these expressions represent the corrections to their rest energy. The kinetic energy, $E_{K}^{p(n)}=(\phi^{p(n)})^2/2M_NL^2$, is however subleading at this order in HB$\chi$PT and these expressions can be considered as corrections to the mass of the nucleons.}
\begin{eqnarray}
\delta_{L} M_{p}=\frac{3g_A^2}{8 \pi^2 f_{\pi}^2} \mathcal{K}^{p}(0;\bm{\phi}^{\pi}) + \frac{g_{\Delta N}^2}{3 \pi^2 f_{\pi}^2} 
\mathcal{K}^{p}(\Delta;\bm{\phi}^{\pi})
,
\label{eq:MnucVol}
\end{eqnarray}
where
\begin{eqnarray}
\mathcal{K}^{p}(0;\bm{\phi}^{\pi})= \frac{\pi}{3}m_{\pi}^2 \sum_{\mathbf{n}\neq \mathbf{0}} 
\frac{e^{-|\mathbf{n}| m_{\pi} L}}{|\mathbf{n}| L} (\frac{1}{2}+e^{-i\mathbf{n}\cdot\bm{\phi}^{\pi^+}})
,
\label{eq:NNpisum}
\end{eqnarray}
and
\begin{eqnarray}
\mathcal{K}^{p}(\Delta;\bm{\phi}^{\pi})
\ =\ \frac{1}{2}
\int_{0}^{\infty} d \lambda ~ \beta_{\Delta}  
&&\sum_{\mathbf{n}\neq \mathbf{0}} 
\left[ 
\beta_{\Delta} 
K_0(\beta_{\Delta}|\mathbf{n}| L)
\ -\ 
\frac{1}{|\mathbf{n}| L} K_1(\beta_{\Delta}|\mathbf{n}| L)
\right]
\nonumber\\
&&\qquad \qquad \qquad 
~ \times (e^{-i\mathbf{n}\cdot\bm{\phi}^{\pi^-}}+\frac{2}{3}+\frac{1}{3}e^{-i\mathbf{n}\cdot\bm{\phi}^{\pi^+}})
,
\label{eq:NDpisum}
\end{eqnarray}
and the neutron mass can be found from these expressions by  the substitutions 
$p\rightarrow n$ and $\pi^+\leftrightarrow\pi^-$.
It is convenient to consider the periodic images associated with the nucleon 
having their contributions modified by the appropriate phase factor 
due to the TBCs.

After twist averaging (over the twists of the pion field, see Appendix~\ref{app: TI}), 
the leading FV corrections to the mass of both the proton 
and the neutron arising from Eq.~(\ref{eq:MnucVol})
are $1/3$ of their value when calculated with PBCs, Eq.~(\ref{deltaM}).\footnote{If the twist of the up and down 
quarks is the same, $\bm{\phi}^{\pi^{\pm}}$ vanishes and no volume improvement will be obtained by averaging.} 
Of course, calculations at  multiple twist angles need not be performed  to estimate the twist-averaged value, and 
special twist angles can be selected based upon the symmetries of the integer 
sums in Eqs.~(\ref{eq:NNpisum}) and (\ref{eq:NDpisum}).
 In particular, it is notable that the leading volume effects of the form 
$e^{-m_\pi L}/L$, $e^{-\sqrt{2}m_\pi L}/L$ and $e^{-\sqrt{3}m_\pi L}/L$, 
can be reduced by a factor of three
with \emph{i}-PBCs,  by setting the pion twist angle to 
$\bm{\phi}^{\pi^+}=(\frac{\pi}{2},\frac{\pi}{2},\frac{\pi}{2})$.
Averaging  the masses  calculated with PBCs and APBCs also reduces the 
leading contribution by a factor of three.
The leading volume dependence can be eliminated completely by choosing
$\bm{\phi}^{\pi^+}=(\frac{4 \pi}{3},\frac{4\pi}{3},\frac{4\pi}{3})$, leaving  volume corrections to the nucleon mass
of the form $\sim e^{-\sqrt{2} m_\pi L}/L$.
It is likely that optimal twists exist for other single nucleon properties, such as matrix elements of the isovector axial current, $g_A$.

For arbitrary quark twists, the proton and neutron have, in general,  
different phase spaces as the  momentum modes that exist in the FV differ.
As an example, while quark twists can be chosen to keep the proton at rest in the volume
and allow for averaging over the charged pion twists, 
${\bm\phi}^{(d)} = -2{\bm\phi}^{(u)}$, 
in general the neutron will have non-zero momentum.\footnote{
Such non-trivial phase spaces somewhat complicate the analysis of LQCD calculations of multi-baryon systems.
}

\section{Two Baryons and Twisted Quantization Condition}
\noindent
The S-wave NN energy QC was generalized to systems with TBCs at rest in Ref.~\cite{Bedaque:2004kc}, 
and to more general two-hadron systems in Ref.~\cite{Agadjanov:2013kja}. 
L\"uscher's energy QC~\cite{Luscher:1986pf, Luscher:1990ux},
which determines the form of the FV corrections,
is dictated by the on-shell two-particle states within the volume.
Once the kinematic constraints on the momentum modes of the two-particle states in the FV are determined, the corresponding 
QC can be determined in a straightforward manner. 
Explicitly, the QC is once again of the form
\begin{eqnarray}
\det\left[{(\mathcal{M}^{\infty})^{-1}+\delta\mathcal{G}^{V}}\right]\ =\ 0
,
\label{NNQC}
\end{eqnarray}
where $\mathcal{M}^{\infty}$ is the infinite-volume scattering amplitude matrix evaluated at the on-shell momentum of each particle 
in the CM frame, $k^*$.
It is convenient to express the QC in the $\left|JM_J(LS)\right\rangle$ basis, where $J$ is the total angular momentum,  $M_J$ is the eigenvalue of the $\hat
J_z$ operator, and $L$ and $S$ are the orbital angular momentum and the total spin of the system, respectively.
The matrix elements of $\delta\mathcal{G}^V$ in this basis 
are very similar to Eq. (\ref{deltaG}) for the case of NN scattering with PBCs,
\begin{align}
& \left[\delta\mathcal{G}^V\right]_{JM_J,LS;J'M_J',L'S'}=i	\eta \frac{k^*}{8\pi E^*}
\delta_{SS'}\left[\delta_{JJ'}\delta_{M_JM_J'}\delta_{LL'} +i\sum_{l,m}\frac{(4\pi)^{3/2}}{k^{*l+1}}
c_{lm}^{\mathbf{d},\bm{\phi}_1,\bm{\phi}_2}(k^{*2};L) \right.
\nonumber\\
& \qquad \qquad \qquad \left .  \times \sum_{M_L,M_L',M_S}\langle JM_J|LM_L,SM_S\rangle \langle L'M_L',SM_S|J'M_J'\rangle 
\int d\Omega~Y^*_{L M_L}Y^*_{l m}Y_{L' M_L'}\right],
\nonumber\\
\label{deltaG-TBC}
\end{align}
where $\eta=1/2$ for identical particles and $\eta=1$ otherwise,
and $\langle JM_J|LM_L,SM_S\rangle$ are Clebsch-Gordan coefficients. 
$E^*$ is the total (relativistic) CM energy of the system, $E^*=\sqrt{k^{*2}+m_1^2}+\sqrt{k^{*2}+m_2^2}$ where
$m_1$ and $m_2$ are the masses of the particles, 
and $\bm{\phi}_1$  and $\bm{\phi}_2$ are their respective twist angles. 
The total momentum of the system is  $\mathbf{P}=\frac{2\pi}{L}\mathbf{d}+\frac{\bm{\phi}_1+\bm{\phi}_2}{L}$ 
with $\mathbf{d}\in \mathbb{Z}^3$. 
 The only difference between this equation and Eq. (\ref{deltaG}), besides the relativistic kinematics used, is in the $c_{lm}^{\mathbf{d},\bm{\phi}_1,\bm{\phi}_2}(k^{*2};L)$ function who carries the volume dependence and the dependence on the BCs in the QC. Explicitly 
 \begin{eqnarray}
c^{\textbf{d},\bm{\phi}_1,\bm{\phi}_2}_{lm}(k^{*2};L)
\ =\ \frac{\sqrt{4\pi}}{\gamma L^3}\left(\frac{2\pi}{L}\right)^{l-2}
\mathcal{Z}^{\mathbf{d},\bm{\phi}_1,\bm{\phi}_2}_{lm}[1;(k^*L/2\pi)^2]
,
\label{clm}
\end{eqnarray}
with
\begin{eqnarray}
\mathcal{Z}^{\mathbf{d},\bm{\phi}_1,\bm{\phi}_2}_{lm}[s;x^2]
\ =\ \sum_{\mathbf r \in \mathcal{P}_{\mathbf{d},\bm{\phi}_1,\bm{\phi}_2}}
\frac{ |{\bf r}|^l \ Y_{l m}(\mathbf{r})}{(\mathbf{r}^2-x^2)^s}
.
\label{Zlm}
\end{eqnarray}
$\gamma=E/E^*$ where E is the total energy of the system in the rest frame of the volume (the lab frame), 
$E^2=\mathbf{P}^2+E^{*2}$. 
The sum in Eq.~(\ref{Zlm}) is performed over the momentum vectors $\mathbf{r}$ that belong to the 
set $\mathcal{P}_{\mathbf{d},\bm{\phi}_1,\bm{\phi}_2}$, 
which remains to be determined.  
 
Consider the two-hadron wavefunction in the lab frame~\cite{Rummukainen:1995vs, Davoudi:2011md} 
that is subject to the TBCs,
\begin{eqnarray}
\psi_{{\rm Lab}}(\mathbf{x}_1+L \mathbf{n}_1,\mathbf{x}_2+L \mathbf{n}_2)
\ =\ 
e^{i{\bm{\phi}}_1 \cdot \mathbf{n}_1+i{\bm{\phi}}_2 \cdot \mathbf{n}_2}
\ \psi_{{\rm Lab}}(\mathbf{x}_1,\mathbf{x}_2)
,
\label{WF-BC}
\end{eqnarray}
where $\mathbf{x}_1$ and $\mathbf{x}_2$ denote the position of the hadrons, 
and $\mathbf{n}_1,\mathbf{n}_2\in\mathbb{Z}^3$. 
As the total momentum of the system is conserved, the wavefunction can be written as an eigenfunction of 
the total momentum $P=(E,\mathbf{P})$. 
In the lab frame, the equal-time wavefunction of the system is
\begin{eqnarray}
\psi_{{\rm Lab}}(x_1,x_2) 
\ =\  e^{-iE X^0+i\mathbf{P} \cdot \mathbf{X}}
\ \varphi_{{\rm Lab}}(0,\mathbf{x}_1-\mathbf{x}_2)
,
\end{eqnarray}
where the position of the CM is $X$, and
\begin{eqnarray}
X & = & \alpha x_1+(1-\alpha) x_2
, \ \ \ \ 
\alpha=\frac{1}{2}\left(1+\frac{m_1^2-m_2^2}{E^{*2}}\right)
,
\end{eqnarray}
for  systems with unequal masses~\cite{Davoudi:2011md}.  
Since the CM wavefunction is independent of the relative time coordinate~\cite{Rummukainen:1995vs},  
$\varphi_{{\rm Lab}}(0,\mathbf{\mathbf{x}_1-\mathbf{x}_2})=\varphi_{{\rm CM}}(\hat{\gamma} (\mathbf{x}_1-\mathbf{x}_2))$,
where the boosted relative position vector is
$\hat{\gamma} \mathbf{x}=\gamma \mathbf{x}_{\Vert}+ \mathbf{x}_{\bot}$, 
with $\mathbf{x}_{\Vert}$ ($\mathbf{x}_{\bot}$) being the component of $\mathbf{x}$ that is 
parallel (perpendicular) to $\mathbf{P}$. 
By expressing $\psi_{{\rm Lab}}$ in  Eq.~(\ref{WF-BC}) in terms of $\varphi_{CM}$, it straightforwardly follows that
\begin{eqnarray}
e^{i\alpha\mathbf{P}\cdot(\mathbf{n}_1-\mathbf{n}_2)L+i\mathbf{P}\cdot\mathbf{n}_2L}
\ \varphi_{{\rm CM}}(\mathbf{y}^*+\hat{\gamma}(\mathbf{n}_1-\mathbf{n}_2)L)
\ =\ 
e^{i{\bm{\phi}}_1 \cdot \mathbf{n}_1+i{\bm{\phi}}_2 \cdot \mathbf{n}_2}
\ \varphi_{{\rm CM}}(\mathbf{y}^*)
,
\end{eqnarray}
where $\mathbf{y}^*=\mathbf{x}_1^*-\mathbf{x}_2^*$ is the relative coordinate of two hadrons in the CM frame. 
By Fourier transforming this relation, and using the form of the total momentum $\mathbf{P}$ from above, 
the relative momenta allowed in the FV energy QC are constrained to be 
\begin{eqnarray}
\mathbf{r}
\ =\ 
\frac{1}{L}\ 
\hat{\gamma}^{-1}
\ \left[2\pi(\mathbf{n}-\alpha\mathbf{d})-(\alpha-\frac{1}{2})(\bm{\phi}_1+\bm{\phi}_2)+\frac{1}{2}(\bm{\phi}_1-\bm{\phi}_2)\right]
,
\label{r-TBC}
\end{eqnarray}
where $\mathbf{n}\in\mathbb{Z}^3$ is the three-vector that is summed over in Eq.~(\ref{Zlm}).
These results encapsulate those of Refs.~\cite{Rummukainen:1995vs, Davoudi:2011md, Fu:2011xz, Leskovec:2012gb, Bour:2011ef} 
when the PBCs are imposed, 
i.e.,  when $\bm{\phi}_1=\bm{\phi}_2=\mathbf{0}$. 
It also recovers two limiting cases that are considered in Ref.~\cite{Agadjanov:2013kja} for the use of TBCs in the scalar sector of QCD. 
It should be noted that for particles with equal masses, $\alpha=1/2$, 
the set of allowed momentum vectors reduces to
\begin{eqnarray}
\mathbf{r}
\ =\ 
\frac{1}{L}\ \hat{\gamma}^{-1}
\ \left[2\pi(\mathbf{n}-\frac{1}{2}\mathbf{d})+\frac{1}{2}(\bm{\phi}_1-\bm{\phi}_2)\right]
.
\label{eq-mass}
\end{eqnarray}
It is important to note that for two identical hadrons, when $\bm{\phi}_1=\bm{\phi}_2=\bm{\phi}$, the FV spectra show no non-trivial 
dependence on the twist other than a  shift in the total energy of the system, 
$E^2=(\frac{2\pi}{L}\mathbf{d}+\frac{\bm{\phi}}{L})^2+E^{*2}$. 
As a result, twisting will not provide additional constraints on the scattering amplitude in, 
for instance, the $\si$ nn or pp channels.
This is also the case for the FV studies of NN scattering in the $\siii$-$\diii$ coupled channels if the same twist is 
imposed on the up and down quarks.\footnote{This result differs somewhat from the conclusion of Ref.~\cite{Bedaque:2004kc}.}

\section{Deuteron Binding and Volume Improvement}
\noindent
In the previous chapter, we obtained the expected energy spectra of two nucleons with spin $S=1$ in a FV subject to PBCs 
and with a range of CM momenta from the 
experimentally measured phase shifts and mixing angles~\cite{Briceno:2013bda}.
In particular, the dependence of the bound-state spectra on the non-zero mixing angle between S and D waves, $\epsilon_1$, 
was determined. 
As seen from Eq.~(\ref{eq-mass}), the 
effects of the 
twist angles $\frac{1}{2\pi}(\bm{\phi}_1~-~\bm{\phi}_2)=(0,0,1),(1,1,0),(1,1,1)$ 
on the CM spectra
are the same as those of  (untwisted) boost vectors, ${\bf d}$,
considered in chapter \ref{chap:deuteron}.
Therefore,  different TBCs can provide additional CM energies in a single volume, 
similar to boosted calculations, 
which can be used to better constrain  scattering parameters and the S-matrix. 
However, twisting may be a more powerful tool as it 
provides access to a continuum of momenta.

If imposing TBCs on the quark fields would require the generation of new ensembles of gauge-field configurations,
it would likely not be optimal to expend  large computational resources  on multiple twisted calculations.
However,  PBCs can be retained on the sea quarks  and 
TBCs can be imposed only in the valence sector~\cite{Bedaque:2004ax}. 
The reason for this is that there are no disconnected diagrams  associated with the NN interactions.\footnote{
As  recently demonstrated, disconnected diagrams will not hinder the use of partially-TBCs in studies of the scalar 
sector of QCD either~\cite{Agadjanov:2013kja}. 
The graded symmetry of ``partially-quenched'' QCD  results in  cancellations among 
contributions from intermediate non-valence mesons.} 
\begin{figure}[h!]
\begin{center}
\subfigure[]{
\includegraphics[scale=0.8]{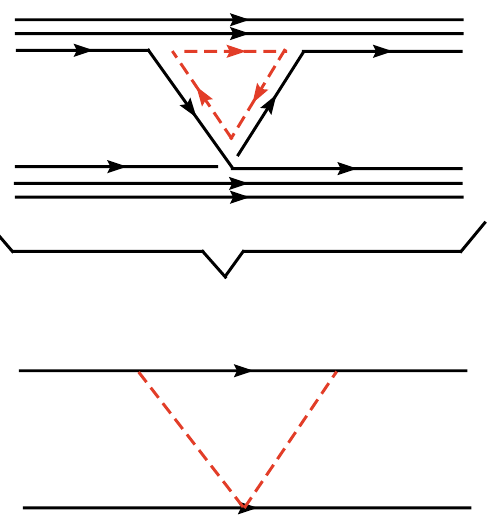}}
\subfigure[]{
\includegraphics[scale=0.8]{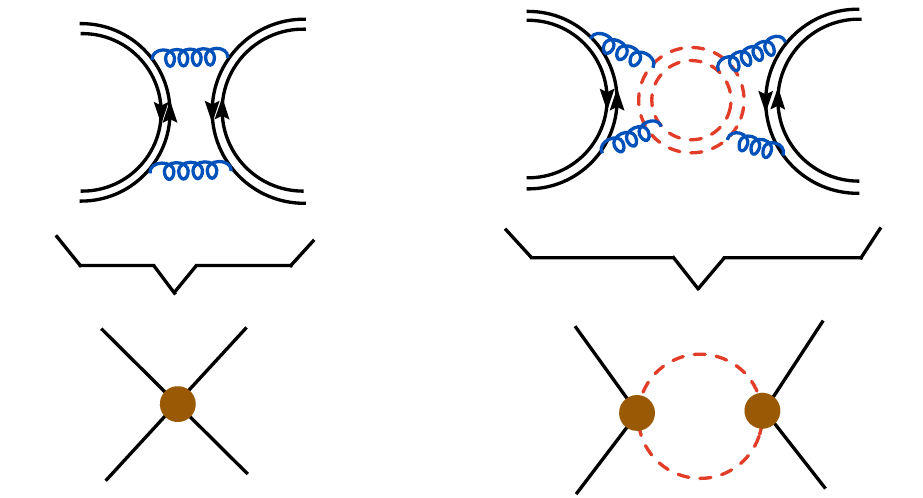}}
\caption{
{\small a) A sea-quark (red dashed lines) contribution to the NN scattering in QCD can be described by an EFT at low energies where the bosons exchanged (red dashed lines) contain a a sea quark. No sea contribution to the NN scattering can occur in the s channel. b) The $\pi \pi$ scattering in $I=0$ channel is an example of a disconnected process where due to the intermediate sea quark creation and annihilation loops, the corresponding EFT description will necessary get contribution from s-channel diagrams whose intermediate mesons contain the sea quarks \cite{Bedaque:2004ax}. Figure is reproduced with the permission of Paulo Bedaque.}
}
\label{fig:sea}
\end{center}
\end{figure}
At the level of the low-energy EFT, this indicates that there are no intermediate s-channel diagrams in which a nucleon or meson 
containing a sea quark can go on-shell. 
Such off-mass-shell hadrons modify the NN interactions by  $\sim e^{-m_{\pi}L}/L$, and do not invalidate the use 
of the QC in  Eq.~(\ref{NNQC}) with the partially-TBCs as long as the calculations are performed
in sufficiently large volumes,  
$L \gtrsim 9 ~ {\rm fm}$.

One significant advantage of imposing TBCs is the improvement in the volume dependence of the deuteron binding energy. 
Although  the formalism presented in the previous chapters can be used to fit to various scattering parameters~\cite{Briceno:2013bda} 
(and consequently determine the deuteron binding energy), 
we will show that with a judicious choice of twist angles, the extracted energies in future LQCD calculations should be close to 
the infinite-volume values,  even in volumes as small as $\sim (9~{\rm fm})^3$.

As discussed in the previous section, the CM energy of the np system is sensitive to TBCs only if 
different twists are imposed upon the up and down quarks. 
This means that, even if exact isospin symmetry is assumed, the proton and the neutron will have different phase spaces
due to the different BCs. 
By relaxing the interchangeability constraint on the np state, as required by the different phase spaces, 
the NN positive-parity channels will mix with the negative-parity channels. 
This admixture of parity eigenstates is entirely a FV effect induced by the boundary conditions, 
and does not require parity violation in the interactions, 
manifesting itself in  non-vanishing $c_{lm}^{\mathbf{d},\bm{\phi}_1,\bm{\phi}_2}$ functions 
for odd values of $l$.  
As such, the spin of the NN system is preserved.

The procedure to obtain the expected spectra of the deuteron is similar to our method in chapter \ref{chap:deuteron} and is as follows. The QC in Eq.~(\ref{NNQC}) depends on S-matrix elements in all partial waves, however it can be truncated
to include only channels with $L \leq 2$
(requiring $J \leq 3$) 
because of the reducing size of the  low-energy phase shifts in the higher channels. 
For arbitrary twist angles, the truncated QC can be represented by a $27 \times 27$ matrix in the $\left|JM_J(LS)\right\rangle$ basis, the eigenvalues of which dictate the energy eigenvalues.
Fits to the experimentally known phase shifts and 
mixing parameters~\cite{NIJMEGEN, PhysRevC.48.792, PhysRevC.49.2950, PhysRevC.54.2851, PhysRevC.54.2869} 
are used to extrapolate to negative 
energies~\cite{Briceno:2013bda} to provide the inputs into the truncated QC,
from which the deuteron spectra in a cubic volume with TBCs are predicted. 
The scattering parameters entering the  analysis are 
$\delta_{1\alpha},~\epsilon_1,~\delta_{1\beta},~\delta^{({^3}P_0)},~\delta^{({^3}P_1)},~\delta^{({^3}P_2)},
~\delta^{({^3}D_2)}$ and $\delta^{({^3}D_3)}$, 
where the Blatt-Biedenharn (BB) parameterization~\cite{Blatt:1952zza} is used in the $J=1$ sector. 
The twist angles explored in this work are 
$\bm{\phi}^p=-\bm{\phi}^n \equiv \bm{\phi}=(0,0,0)$ (PBCs), 
$(\pi,\pi,\pi)$ (APBCs)
and $(\frac{\pi}{2},\frac{\pi}{2},\frac{\pi}{2})$ (\emph{i}-PBCs).
 At the level of the quarks, this implies that the twist angles of the (valence) up and down quarks 
are ${\bm\phi}^u =  -{\bm\phi}^d = {\bm\phi}$.
We also set $\mathbf{d}=\mathbf{0}$ in Eq.~(\ref{r-TBC}) so that the np system is at rest in the lab frame. 
The reason for this choice of twist angles is  that they (directly or indirectly) give rise a significant cancellation of the 
leading FV corrections to the masses of the nucleons, as shown in Sec.~\ref{sec:Single}. 
The number of eigenvalues of 
${(\mathcal{M}^{\infty})^{-1}+\delta\mathcal{G}^{V}}$, 
and their degeneracies, reflect the spatial-symmetry group of the FV. 
Calculations with $\bm{\phi}=\bm{0}$ respect the cubic ($O_h$) symmetry, while 
for $\bm{\phi}=(\frac{\pi}{2},\frac{\pi}{2},\frac{\pi}{2})$ the symmetry group is reduced to the $C_{3v}$ point 
group.~\footnote{
There is a correspondence between the FV spatial symmetry in twisted calculations with  
arbitrary twists
$\bm{\phi}^p \neq \bm{\phi}^n$ 
and the FV symmetry in (boosted) NN calculations with PBCs when isospin breaking is considered. 
For example, the point symmetry group corresponding to twisted calculations with $\bm{\phi}^p = -\bm{\phi}^n=(0,0,\frac{\pi}{2})$ 
and that of the physical np system with $\mathbf{P}=\frac{2\pi}L(0,0,1)$ with PBCs are both $C_{4v}$. 
} 
However, for $\bm{\phi}=(\pi,\pi,\pi)$ 
the system has inversion symmetry, and respects the $D_{3h}$ point symmetry~\cite{Dresselhaus}. 
By examining the transformation properties of the 
$c_{lm}^{\mathbf{d},\bm{\phi}_1,\bm{\phi}_2}$ 
functions under the symmetry operations of these groups, certain relations are found for any given $l$. 
These relations, 
as well as the eigenvectors of the FV matrices,
which are tabulated elsewhere~\cite{Luscher:1990ux, Rummukainen:1995vs, Briceno:2013lba, Gockeler:2012yj,Thomas:2011rh,Dudek:2012gj},
can be used to block diagonalize the $27 \times 27$ matrix representation of the  QCs, 
where each block corresponds to an irrep of the point-group symmetry of the system. 
For the selected twist angles, the QCs of the irreps of the corresponding point 
groups that have overlap with the deuteron are given in Appendix~\ref{app: QC-TBC} .

\begin{figure}[h]
\begin{centering}
\includegraphics[scale=0.40]{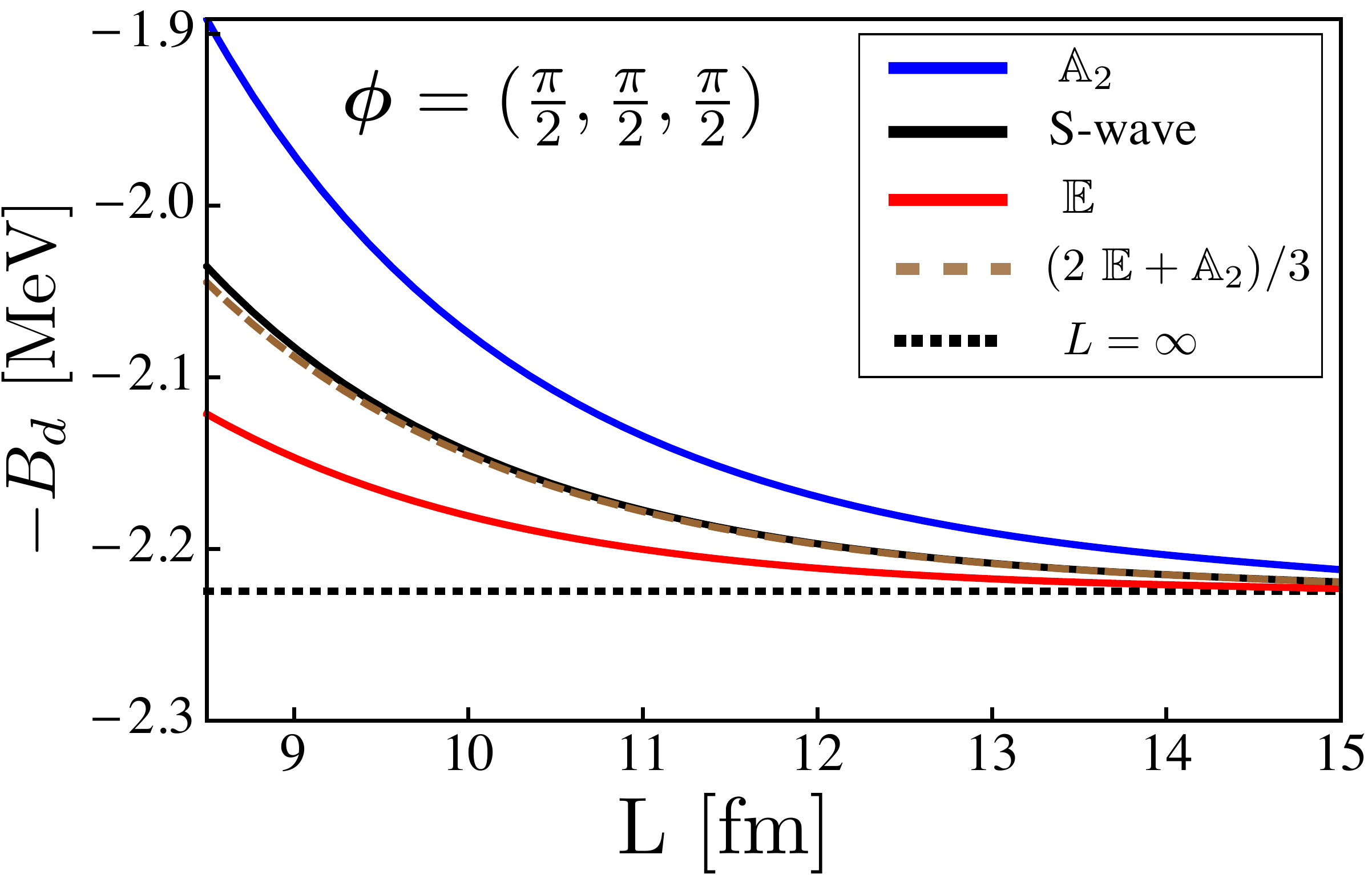}
\caption{{\small
The deuteron binding energy as a function of $L$ 
using \emph{i}-PBCs
($\bm{\phi}^p=-\bm{\phi}^n \equiv \bm{\phi}=(\frac{\pi}{2},\frac{\pi}{2},\frac{\pi}{2})$). 
The blue curve corresponds to the $\mathbb{A}_2$ irrep of the $C_{3v}$ group, 
while the red curve corresponds to the $\mathbb{E}$ irrep. 
The brown-dashed curve corresponds to the weighted average of the $\mathbb{A}_2$ and 
$\mathbb{E}$ irreps, $-\frac{1}{3}(2B_d^{(\mathbb{E})}+B_d^{(\mathbb{A}_2)})$,
while the black-solid curve corresponds to the S-wave limit.
The infinite-volume deuteron binding energy is shown  by the black-dotted line.
}}
\label{fig:A2-E}
\par\end{centering}
\end{figure}
\begin{figure}[h]
\begin{centering}
\subfigure[]{
\includegraphics[scale=0.3]{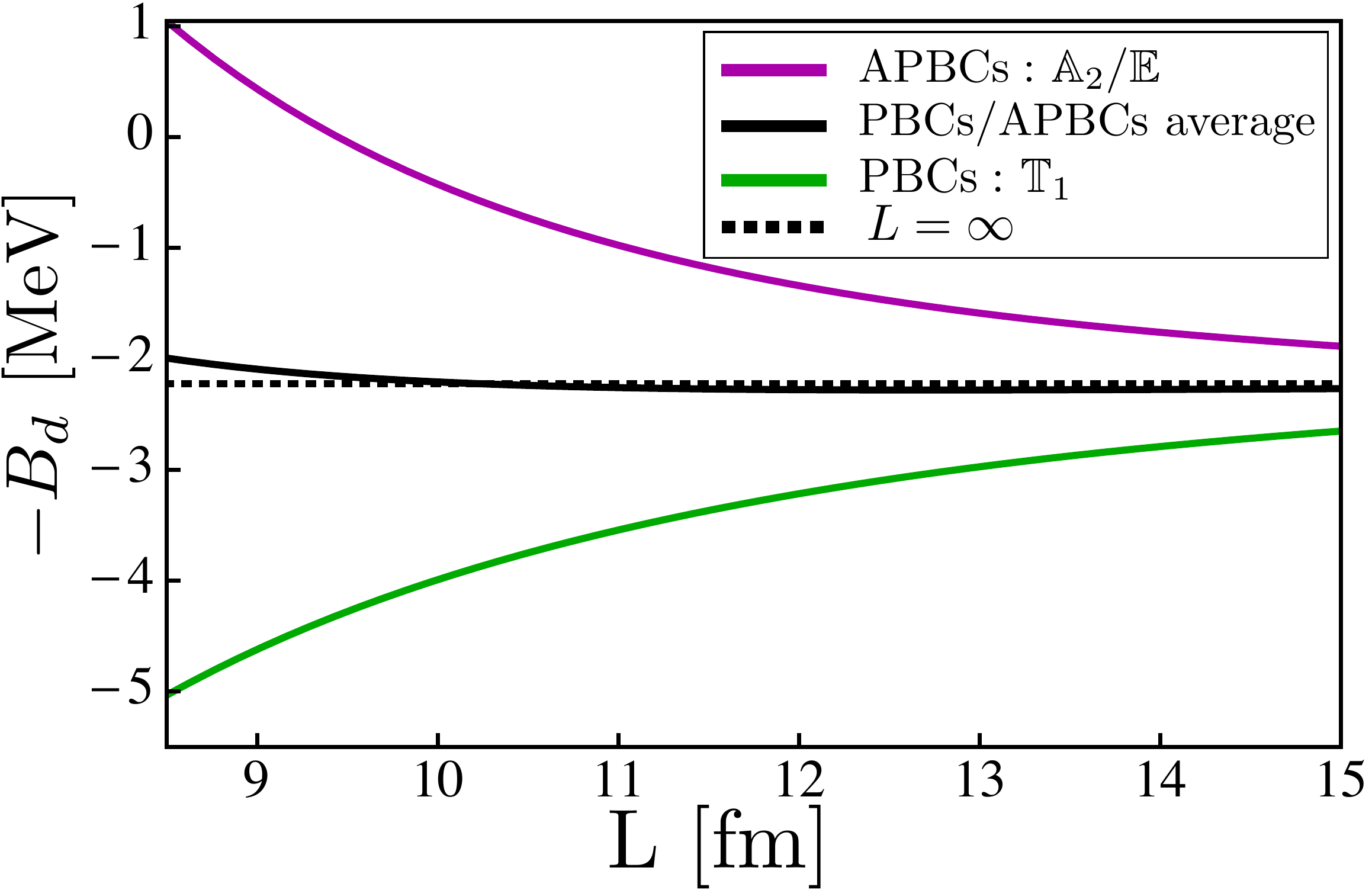}}
\subfigure[]{
\includegraphics[scale=0.3]{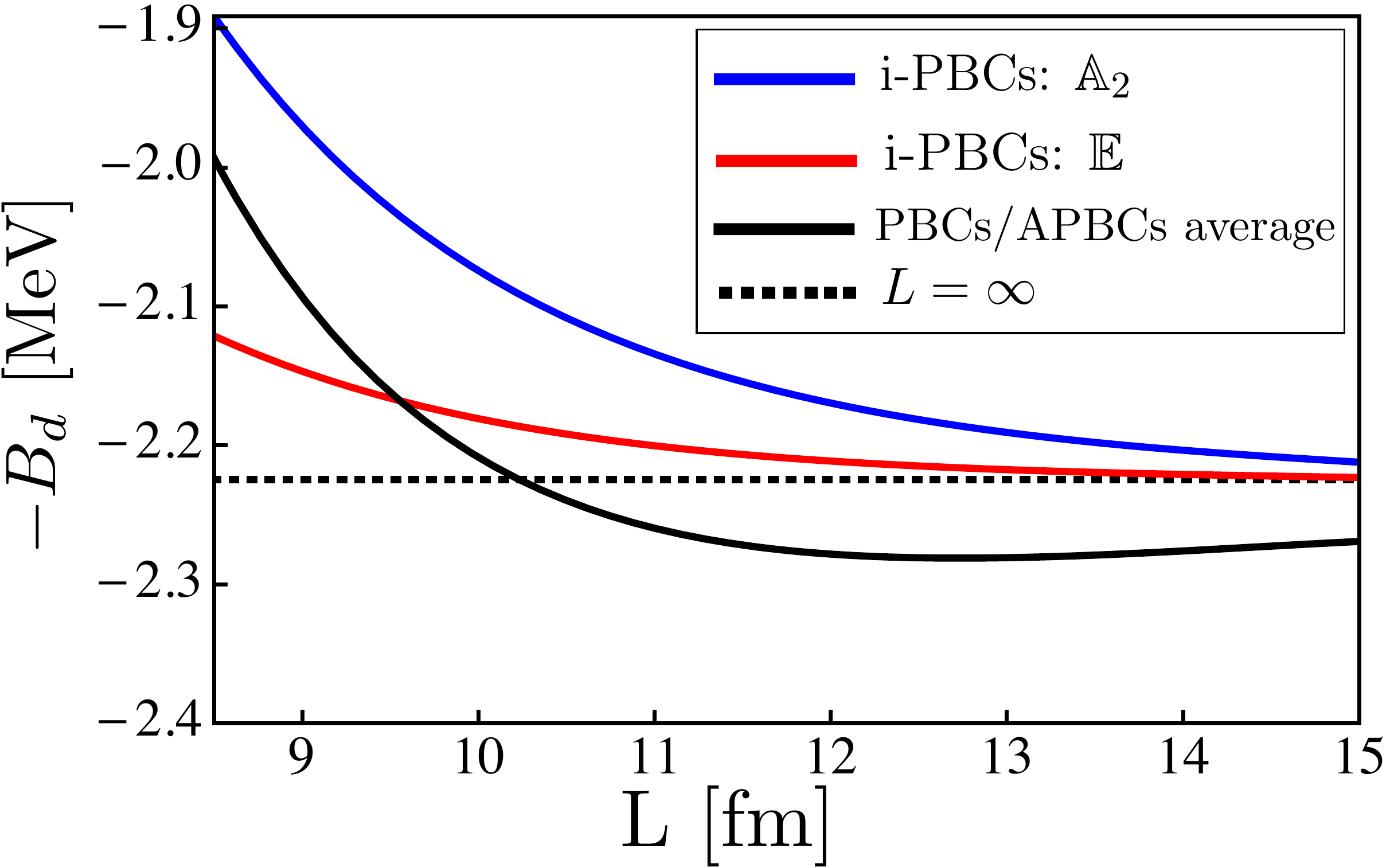}}
\caption{{\small
a) The deuteron binding energy as a function of $L$ 
from  PBCs (green curve) and  from APBCs (purple curve). 
The black-solid curve represents the average of these energies. 
b) A closer look at the average in part (a) compared with  energies obtained with \emph{i}-PBCs,
$\mathbb{A}_2$ (blue curve) and $\mathbb{E}$ (red curve).
}}
\label{fig:PBC-APBC}
\par\end{centering}
\end{figure}

For \emph{i}-PBCs, there are two irreps of the $C_{3v}$ group, namely the one-dimensional irrep $\mathbb{A}_1$ 
and the two-dimensional irrep $\mathbb{E}$, that have overlap with the $^3{S_1}$-${^3}{D_1}$ coupled channels. 
Fig.~\ref{fig:A2-E} shows the binding energy (the CM energy minus the rest masses of the nucleons), $-B_d=E^*-M_p-M_n$, 
as a function of $L$ corresponding to $\mathbb{A}_2$ irrep (blue curve) and $\mathbb{E}$ irrep (red curve), 
obtained from the QCs in Eqs. (\ref{pi2pi2pi2A2}) and (\ref{pi2pi2pi2E}).  
Even at $L\sim9~{\rm fm}$, the deuteron binding energies extracted from both irreps are close to the infinite-volume value. 
In particular, calculations in the $\mathbb{E}$ irrep of the $C_{3v}$ group provide a few percent-level accurate 
determination of the deuteron binding energy 
in this volume. 
The black-solid curve in Fig.~\ref{fig:A2-E} 
represents the S-wave limit of the interactions, 
when the S-D mixing parameter and all phase shifts except that in the S-wave are set equal to zero. 
The $M_J^\prime$-averaged binding energy, 
$-\frac{1}{3}(2B_d^{(\mathbb{E})}+B_d^{(\mathbb{A}_2)})$, 
converges to this S-wave limit, as shown 
in Fig. \ref{fig:A2-E}
(the $\mathbb{A}_2$ irrep contains the $M_J^\prime=0$ state 
while $\mathbb{E}$ contains the $M_J^\prime=\pm 1$ states,
where as mentioned before, $M_J^\prime$ is the projection of total angular momentum
along the twist direction). 
In order to appreciate the significance of calculations performed with the $\bm{\phi}=(\frac{\pi}{2},\frac{\pi}{2},\frac{\pi}{2})$ twist angles, 
it is helpful to recall the deuteron binding energy obtained in calculations with PBCs. 
For PBCs, 
we remind that the only irrep of the cubic group that has overlap with the $\siii$-$\diii$ coupled channels is the 
three-dimensional irrep $\mathbb{T}_1$, Eq.~(\ref{000T1}), 
and the corresponding binding energy is shown in Fig. \ref{fig:PBC-APBC}(a) (green curve). 
As was already seen in the previous chapter, the binding energy deviates significantly from its infinite-volume value, 
such that the FV deuteron is approximately twice as bound as the infinite-volume deuteron at $L=9~{\rm fm}$. 
For APBCs, two irreps of the $D_{3h}$ group overlap with the deuteron channel, 
$\mathbb{A}_2$ and $\mathbb{E}$ (Eqs. (\ref{pipipiA2},\ref{pipipiE})), 
and yield degenerate binding energies as shown in Fig. \ref{fig:PBC-APBC}(a) (purple curves). 
As seen in Fig. \ref{fig:PBC-APBC}(a), 
the deuteron becomes unbound over a range of volumes 
and  asymptotes slowly to the infinite-volume limit. 
However, in analogy with the nucleon masses, 
the volume dependence of the deuteron binding energy 
is significantly reduced 
by averaging the results obtained with PBCs and APBCs,
as shown in Fig. \ref{fig:PBC-APBC}(a) (black-solid curve). 
Fig. \ref{fig:PBC-APBC}(b) provides a magnified  view of this averaged quantity (black-solid curve),
where the two energy levels associated  with \emph{i}-PBCs are shown for comparison.

In order to understand the observed volume improvements,  
consider the volume scaling of the full QC assuming that the phase shifts beyond the $\alpha$-wave 
are small.
In this limit, for a general set of twist angles and boosts, the QC  collapses to 
\begin{eqnarray}
&&
\!\!\!\!\!\!\!\!\!
\det\left[
\left(
k^*\cot\delta_{1\alpha} - 4\pi c_{00} 
\right)
\left(
\begin{array}{ccc}
1&0&0\\ 0&1&0\\ 0&0&1
\end{array}
\right)
\right.\nonumber\\
&&
\left.
- 
{2\pi\over \sqrt{5} k^{*2}}\left(\sqrt{2}\sin 2\epsilon_1 - \sin^2\epsilon_1 \right)
\left(
\begin{array}{ccc}
c_{20}&\sqrt{3} c_{21}&\sqrt{6} c_{22}\\
-\sqrt{3} c_{2-1}&-2c_{20}&-\sqrt{3} c_{21}\\
\sqrt{6} c_{2-2}&\sqrt{3} c_{2-1}&c_{20}
\end{array}
\right)
\ \right]  \ = \  0
,
\label{eq:QC3by3}
\end{eqnarray}
which depends upon the $\alpha$-wave phase shift and the mixing parameter, $\epsilon_1$.
Shorthand notation has been used for convenience, $c_{lm} = c_{lm}^{{\bf d},{\bm\phi}_1,{\bm\phi}_2}(k^{*2}; L)$.
For generic twist angles,
deviations between the energy eigenvalues resulting from this truncated QC and the full QC scale as 
$\sim \tan\delta_i \ e^{-2 \kappa L}/(\kappa L^2)$,
where $\delta_i$ denotes phase shifts beyond the $\alpha$-wave
(see Appendix~\ref{app:TwistC} for  expansions of the $c_{lm}^{{\bf d},{\bm\phi}_1,{\bm\phi}_2}$ functions).
For \emph{i}-PBCs, the leading corrections are from the P-waves, as can be seen from the expansions of the $c_{lm}$ in Table~\ref{app:TwistC}.
By neglecting the small mixing between the S-wave and D-waves in Eq.~(\ref{eq:QC3by3}), 
the QC  dictated by  S-wave interactions is~\footnote{
In the limit where $\epsilon_1=0$, the $J=1$ $\alpha$-wave is entirely S-wave, 
while the $\beta$-wave is entirely D-wave. 
This approximation neglects FV effects of the form $\epsilon_1 e^{- \kappa L}/L$.
}
\begin{eqnarray}
k^*\cot\delta^{{(^3S_1)}}|_{k^*=i\kappa}+\kappa=\sum_{\mathbf{n}\neq\mathbf{0}}
e^{i  (\alpha-\frac{1}{2}) \mathbf{n} \cdot (\bm{\phi}^p+\bm{\phi}^n)} e^{-i \frac{1}{2} \mathbf{n} \cdot (\bm{\phi}^p-\bm{\phi}^n)}
e^{i 2\pi \alpha \mathbf{n} \cdot \mathbf{d}}
~\frac{e^{-|\hat{\gamma}\mathbf{n}| \kappa L}}{|\hat{\gamma}\mathbf{n}|L}
.
\label{S-QC}
\end{eqnarray}
The volume dependence of the deuteron binding momentum, $\kappa$, originates from the right-hand side of this equation.
For ${\bf d}={\bf 0}$,  the $c_{2m}$ functions vanish  for both PBCs and APBCs, leading to Eq.~(\ref{S-QC}) without further approximation.
For the twist angles $\bm{\phi}^p=-\bm{\phi}^n \equiv \bm{\phi}=(\frac{\pi}{2},\frac{\pi}{2},\frac{\pi}{2})$ and boost $\mathbf{d}=\mathbf{0}$,
the first few terms in the summation on the right-hand side of Eq.~(\ref{S-QC}) ($\mathbf{n}^2\leq 3$)
vanish, leaving the leading volume corrections to scale as $\sim e^{-2\kappa L}/L$.
A lesser 
cancellation occurs in the average of binding energies obtained with PBCs and APBCs, giving rise to deviations from the infinite-volume energy by terms that scale as $\sim e^{-\sqrt{2}\kappa L}/L$.

The result of Monte Carlo twist averaging of the deuteron binding energy 
can be ascertained from the behavior of the two extreme contributions, the PBC and APBC results.
While the average binding energy obtained from $N$ randomly selected sets of twist angles scales as 
$B_d^{(\infty)} +  {\cal O}\left( e^{-2\kappa L}/L \right)$, 
the standard deviation of the mean scales as $\sim e^{-\kappa L}/(\sqrt{N} L)$, 
giving rise to a signal-to-noise ratio in the binding energy that scales as 
$\sim  \sqrt{N} \ B_d^{(\infty)}\  L\ e^{\kappa L}$, which even for $L\sim 14~{\rm fm}$ allows only for a poor extraction, as 
can be deduced from Fig. \ref{fig:PBC-APBC}(a).
It is clear that such a method is inferior to that of pair-wise averaging, such as from PBCs and APBCs, or choosing special twists, such as 
\emph{i}-PBCs.

We have restricted ourselves to the scenarios where the net twist angles in each 
Cartesian direction (the lattice axes) are the same. 
One reason for this is that systems with arbitrary twists give rise to three distinct, but nearby, energy eigenvalues 
associated with combinations of each of the three $M_J$-states of the deuteron - a sub-optimal system to analyze
in LQCD calculations.
Another reason is that a twist of $\frac{\pi}{2}$  in each direction is optimal in minimizing the FV effects in 
both the two-body binding energies and the single-baryon masses.
Further, averaging the results of calculations with PBCs and APBCs also eliminates the leading FV corrections to both quantities.
We re-emphasize that ultimately, one wants to extract as many scattering parameters as feasible from calculations 
in a single volume, requiring calculations with multiple boosts of the CM as well as multiple arbitrary twists, in order  
to maximize the inputs to the energy QCs. 
In general, with arbitrary twist angles,  $\bm{\phi}=(\phi_x,\phi_y,\phi_z)$,
the $27 \times 27$ matrix representation of the QC matrix cannot be block diagonalized and it has $27$ distinct eigenvalues.
The truncation to the $3\times 3$ matrices given in Eq.~(\ref{eq:QC3by3}) remains valid, as do the estimates of the truncation errors, but this truncated QC will provide three distinct energy eigenvalues.

While not the focus of this work, it is worth reminding ourselves about the behavior of the positive-energy states in the FV,
such as the higher states associated with the $\siii$-$\diii$ coupled channel or those associated with the $\si$ $np$ channel,
as described in Eq.~(\ref{NNQC}).
For an arbitrary twist, the non-interacting energy levels in the FV are determined by integer triplets and the twist angles.
Interactions will produce deviations from these non-interacting levels, that become smaller
as the lattice volume increases, scaling with $\sim \tan\delta(k^*)/(M L^2)$.
As discussed previously, as there is no underlying symmetry for arbitrary twists, the eigenstates will, in general, be non degenerate.

\section*{Summary and Conclusions}
\noindent
Twisted boundary conditions have been successfully used in numerical calculations of important observables, 
both in nuclear and particle physics with Lattice QCD, 
as well as in others areas such as condensed-matter physics.
They provide a means with which to select the phase space of particles  in a given finite volume, beyond that allowed by periodic or anti-periodic boundary conditions.
In LQCD calculations, TBCs have been used to resolve the threshold region required in the evaluation of 
transition matrix elements without requiring large lattice volumes \cite{Tiburzi:2005hg, Jiang:2006gna, Boyle:2007wg, Simula:2007fa, Boyle:2008yd, Aoki:2008gv, Boyle:2012nb, Brandt:2013mb,Boyle:2013gsa}. 
They can also be
used in calculations of elastic $2 \rightarrow 2$ processes by providing a better sampling of 
CM kinematics in a single volume, allowing for 
better constraints on scattering parameters~\cite{Bernard:2010fp, Doring:2011vk, Doring:2012eu, Ozaki:2012ce}. 
In this chapter, we have explored the use of TBCs in calculating the mass of single baryons, and in determining the binding of two-hadron 
systems in a FV, with a focus on the deuteron.  
In particular, we have used experimentally known scattering data to determine the location of the lowest-lying 
FV states that have overlap with the deuteron for a selection of twist angles, and combinations thereof.
We have formally found that twisting provides an effective way of exponentially reducing the impact of the finite lattice volume on the calculation of 
two-body binding energies.  Pair-wise combining results obtained with particular twists, such as 
PBCs and APBCs, 
can eliminate the leading volume dependence.  
The same is true for twist averaging, but the uncertainty resulting from a finite number of randomly selected twists can be large.
Importantly, we have 
determined that the \emph{i}-PBCs, with 
$\bm{\phi}=(\frac{\pi}{2},\frac{\pi}{2},\frac{\pi}{2})$,
eliminate the first three FV corrections to the dominant S-wave contribution to 
the two-hadron binding energies, suppressing such effects from 
$\mathcal{O}\left(e^{-\kappa L}/L\right)$
to
$\mathcal{O}\left(e^{- 2 \kappa L}/L\right)$, while also reducing the FV modifications to the nucleon mass, of the form
$\mathcal{O}\left(e^{- m_\pi L}/L\right)$, by a factor of three.
This translates into at least an order of magnitude improvement in the accuracy of the deuteron binding energy 
extracted from  LQCD correlation functions in volumes as small as $\sim (9~{\rm fm})^3$. 
As partially-TBCs modify the nuclear forces by terms of order $\mathcal{O}\left(e^{-m_{\pi}L}/L\right)$,
such calculations of the deuteron and other bound states can be performed without the need for 
multiple ensembles of gauge-field configurations, significantly reducing the required computational resources.

Given the generalized L\"uscher FV formalism for NN systems \cite{Briceno:2013lba} with TBCs, 
not only can the binding energy of the deuteron be obtained from the upcoming LQCD calculations, 
but the relevant scattering parameters, including the S-D mixing parameter, can be well constrained. 
While giving different  twists to the up and down quarks modifies the neutron and proton phase 
space in different ways that allows for a parametric reduction in volume effects to the deuteron binding energy, 
and control on the location of the positive-energy scattering states,
it does not change the CM phase space in the neutron-neutron or proton-proton systems.  
Therefore, it is  not a useful tool in 
refining calculations of scattering parameters in these channels.

Inspired by the volume improvement seen in the QMC calculations of few and many-body systems with 
twist-averaged BCs~\cite{PhysRevE.64.016702, PhysRevLett.73.1959, PhysRevB.51.10591, PhysRevB.53.1814, Wilcox:1999ux}, 
and  studies of Dirichlet BCs and PBCs in QMC and Density-Functional Theory, e.g. Refs.~\cite{Bulgac:2013mz,Erler:2012qd},
and considering the twist-phase modifications to the images associated with a given system,
we speculate that the FV modifications to the spectrum of three-nucleon and multi-nucleon systems can  be   reduced 
by TBCs.  The magnitude of the improvement  
will depend upon the inter-particle forces being short ranged compared to the extent of the system. 
Due to the complexity of such systems, particularly in a FV~\cite{Polejaeva:2012ut, Briceno:2012rv,Hansen:2013dla}, a definitive 
conclusion can only be arrived at upon further investigation.


\chapter{FINITE-VOLUME QED EFFECTS IN THE SINGLE-PARTICLE SECTOR}
\label{chap:EM}

Lattice QCD 
has matured to the point where basic properties of the light hadrons are being calculated at the physical 
pion mass~ \cite{Aoki:2009ix, Durr:2010vn, Arthur:2012opa, Aoki:2012st, Durr:2013goa}.
In some instances, the up- and down-quark masses and QED have been included in an effort to 
precisely postdict the observed isospin splittings in the 
spectrum of hadrons~\cite{Blum:2007cy,Basak:2008na,Blum:2010ym,Portelli:2010yn,Portelli:2012pn,Aoki:2012st,deDivitiis:2013xla,Borsanyi:2013lga,Drury:2013sfa}.
While naively appearing to be a simple extension of pure LQCD calculations,
there are subtleties associated with including
QED. 
In particular, Gauss's law and Ampere's law cannot be satisfied 
when the electromagnetic gauge field is subject to PBCs~\cite{Hilf1983412,Duncan:1996xy,Hayakawa:2008an}. 
However, a uniform background charge density can be introduced to circumvent this problem and  restore these laws. 
This is equivalent to removing the zero modes of the photon in a FV calculation,
which does not change the infinite-volume value of calculated quantities.

One-loop level calculations in $\chi$PT
and partially-quenched $\chi$PT ($PQ\chi$PT) have been performed~\cite{Hayakawa:2008an}
to determine the leading FV modifications to the mass of  mesons induced by constraining QED to a cubic volume subject to 
PBCs.\footnote{
Vector dominance~\cite{PhysRevLett.62.1343}
has been previously used to model the low-momentum contributions to the
FV electromagnetic mass splittings of the pseudo-scalar 
mesons, see Refs. \cite{Duncan:1996xy, Blum:2007cy}.
}
Due to the photon being massless, the FV QED 
corrections to the mass of the $\pi^+$ are predicted to be an expansion in powers of the volume, 
and have been determined to be of the form 
$\delta m_{\pi^+}\sim 1/L + 2/(m_{\pi^+} L^2) + \cdots $,
where $L$ is the spatial extent of the cubic volume.
As the spatial extents of present-day gauge-field configurations at the physical pion mass are not large, with $m_\pi L\lsim 4$,
the exponentially suppressed strong interaction FV effects,  ${\cal O}\left( e^{- m_\pi L}\right)$,  
are not negligible for precision studies of hadrons, and 
when QED is included, the power-law corrections, although suppressed by $\alpha_e$, are expected to be important, 
particularly in mass splittings.

In this chapter, we return to the issue of calculating  FV QED effects, and show that non-relativistic effective field theories (NREFTs)
provide a straightforward way to calculate such corrections to the properties of hadrons.  
With these EFTs, the FV mass shift of  
mesons, baryons and nuclei are calculated 
out to ${\cal O}\left(1/L^4\right)$ in the 
$1/L$ expansion, 
including contributions from their charge radii, magnetic moments and polarizabilities.
The NREFTs have the advantage that the coefficients of  operators coupling to the electromagnetic field 
are directly related,
order by order in the $\alpha_e$,
to the electromagnetic moments of the hadrons (in the continuum limit), 
as opposed to a 
perturbative estimate thereof (as is the case in $\chi$PT).
For protons and neutrons, the NREFT is the well-established 
non-relativistic QED (NRQED)~\cite{Isgur:1989vq,Isgur:1989ed,Jenkins:1990jv,Jenkins:1991ne,Thacker:1990bm, Labelle:1992hd,Manohar:1997qy,Luke:1997ys, Hill:2011wy},  
modified to include the finite extent of the charge and current densities~\cite{Chen:1999tn}.
Including multi-nucleon interactions, this framework has been 
used extensively to describe the low-energy behavior of nucleons and nuclear interactions, $\nopi$,
along with their interactions with electromagnetic fields~\cite{Kaplan:1998tg, Kaplan:1998we, Chen:1999tn, Butler:1999sv, Butler:2000zp, Butler:2001jj},
and is straightforwardly generalized to hadrons and nuclei with arbitrary angular momentum.
LQCD calculations performed with background electromagnetic fields  are currently making use of these NREFTs 
to extract  the properties of hadrons, including magnetic moments and polarizabilities \cite{Martinelli:1982cb, Fiebig:1988en, Bernard:1982yu, Lee:2005ds, Christensen:2004ca, Lee:2005dq, Engelhardt:2007ub, Detmold:2009dx, Alexandru:2009id, Detmold:2010ts, Primer:2013pva, Lee:2013lxa}.

\section{Finite-Volume QED}
\noindent
The issues complicating the inclusion of QED in FV calculations 
with PBCs are well 
known,
the most glaring of which is the inability to preserve Gauss's law~ \cite{Duncan:1996xy,Blum:2007cy,Hayakawa:2008an}, 
which relates the electric flux penetrating 
any closed surface to the charge enclosed by the surface,  and Ampere's Law, which relates the integral of the 
magnetic field around a closed loop to the current penetrating the loop.
An obvious way to see the problem is to consider the electric field along the axes of the cubic volume 
(particularly at the surface) associated with a point charge at the center.
Restating the discussions of Ref.~\cite{Hayakawa:2008an}, the variation of the QED action is,
for a fermion of charge $e Q$,
\begin{eqnarray}
\delta S & = & 
\int\ d^4x\ 
\left[ \ 
\partial_\mu F^{\mu\nu} (x)
\ -\ 
e\ Q\ \overline{\psi} (x)\gamma^\nu \psi  (x)
\ \right]
\ \delta \left(A_\nu (x)\right)
\nonumber\\
& = & 
\int  dt\ 
{1\over L^3}\ \sum_{\bf q}\ 
\delta\left(\tilde A_\nu (t,{\bf q})\right)
\int_{L^3}\ d^3 {\bf x}\ 
e^{i{\bf q}\cdot  {\bf x}}\ 
\left[\ 
\partial_\mu F^{\mu\nu} (t,{\bf x})
\ -\ e\ Q\ \overline{\psi}  (t,{\bf x})\gamma^\nu \psi  (t,{\bf x})
\ \right],
\nonumber\\
\label{eq:gauss}
\end{eqnarray}
where 
$\tilde A_\nu (t,{\bf q})$ is the spatial Fourier transform of $A_\nu (t,{\bf x})$, and
$e=|e|$ is the magnitude of the electronic charge.
For simplicity, here and in what follows, 
we assume the time direction of the FV to be infinite~\footnote{
In practice, there are thermal effects in LQCD calculations due to the finite extent of the time direction.}
 while the spatial directions are of length $L$.
Eq. (\ref{eq:gauss}) leads to 
$\partial_\mu F^{\mu\nu} = e Q \overline{\psi} \gamma^\nu \psi $ 
for 
$\delta S=0$ and hence Gauss's Law and Ampere's Law. 
This can be modified to $\partial_\mu F^{\mu\nu} = e Q \overline{\psi} \gamma^\nu \psi  + b^\nu$ simply by omitting the 
spatial zero modes of 
$A_\mu$, i.e.  $\tilde A_\nu (t,{\bf 0}) = 0$, 
or more generally by setting $\delta \tilde A_\nu (t,{\bf 0}) = 0$,
where $b^\nu$ is some uniform background charge 
distribution~\cite{Portelli:2010yn}.~\footnote{
The introduction of a uniformly charged background is a technique that has 
been used extensively to include electromagnetic interactions into calculations of many-body systems, 
such as nuclear matter and condensed matter, see for example Ref.~\cite{2000physics..12024C}.
} 
This readily eliminates the relation between the electric flux penetrating a closed surface and the inserted charge,
and the analogous relation between the magnetic field and 
current.~\footnote{For a discussion about including QED with C-PBCs (anti-PBCs), see Ref.~\cite{Kronfeld:1992ae}.}
Ensuring this constraint is preserved under gauge transformations, 
$A_\mu (t,{\bf x})\rightarrow A^\prime_\mu (t,{\bf x})  = A_\mu (t,{\bf x}) + \partial_\mu \Lambda(t,{\bf x})$, where 
$\Lambda$ is a periodic function in the spatial volume,
requires
$ \partial_0 \tilde\Lambda(t,{\bf 0})=0$, where $\tilde\Lambda (t,{\bf q})$ is the Fourier transform of $\Lambda (t,{\bf x})$.
Modes with ${\bf q}\ne {\bf 0}$ are subject to the standard gauge-fixing conditions, and in LQCD calculations it is
sometimes convenient to work in Coulomb gauge,
${\bm\nabla}\cdot {\bf A}=0$.
This is because of the asymmetry between the spatial and temporal directions that is present in most ensembles of gauge field configurations,
along with the fact that the photon fields are generated in momentum space as opposed to position space.\footnote{The generation of gauge-field configurations in the non-compact formulation of lattice QED is usually performed in momentum space. This, first of all, makes the exclusion of the zero  modes of the QED gauge field easy. 
Secondly, the lattice gauge condition in momentum space provides a  linear relation among modes and one  of the degrees of freedom of $A_i$ can be eliminated in favor of the other two, see Ref. \cite{Blum:2007cy}.}

In infinite volume, the Coulomb potential energy between charges $eQ$ is well known to be 
$U(r) = \frac{\alpha_e Q^2}{r}$, where $\alpha_e=e^2/4\pi$ is the QED fine-structure constant,
while in a cubic spatial volume 
with the zero modes removed, it is
\begin{eqnarray} 
U( {\bf r} ,L) 
& = &  
{\alpha_e Q^2 \over \pi L}
\sum_{{\bf n}\ne {\bf 0}}
{1\over |{\bf n}|^2}
e^{i 2\pi {\bf n}\cdot {\bf r}\over L}
\nonumber\\
& = & 
{\alpha_e Q^2\over \pi L}
\left[-1 + 
\sum_{{\bf n}\ne {\bf 0}}
{e^{-|{\bf n}|^2}\over |{\bf n}|^2}
e^{i 2\pi {\bf n}\cdot {\bf r}\over L}
+
\sum_{\bf p}\ \int_0^1\ dt\ \left({\pi\over t}\right)^{3/2}
e^{ -  {\pi^2 |{\bf p}-{\bf r}/L |^2\over t}  }
\ \right],
\label{eq:Vgreen}
\end{eqnarray}
where 
${\bf n}$ and $\mathbf{p}$ are triplets of integers.
The latter,
exponentially accelerated,  expression in Eq.~(\ref{eq:Vgreen}) is obtained from the former using the Poisson summation formula.
\begin{figure}[t]
  \centering
     \includegraphics[scale=0.55]{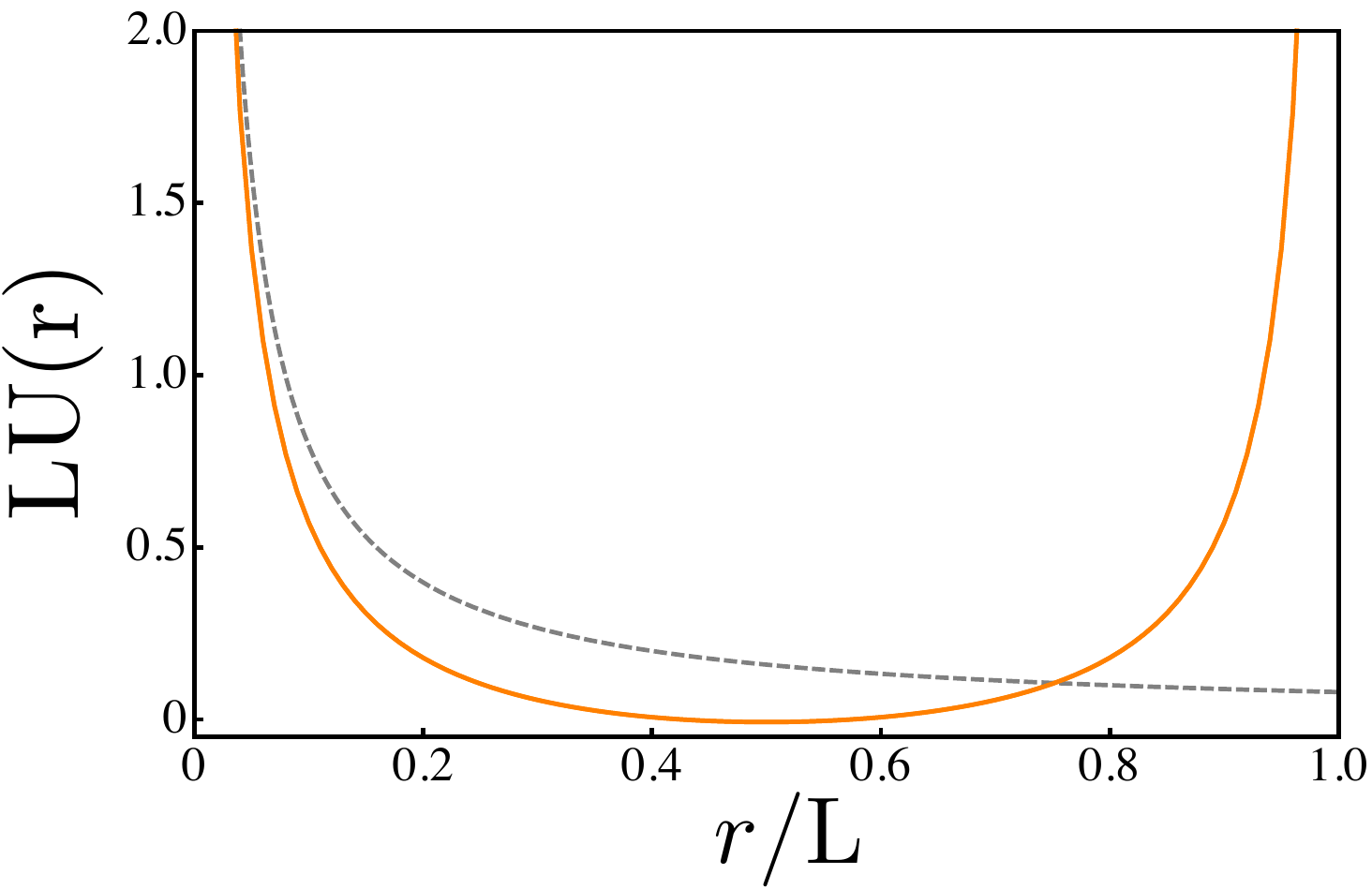}     
     \caption{{\small The FV potential energy between two charges with $Qe=1$, along one of the axes of a cubic volume of spatial extent $L$ (solid orange curve), 
     obtained from Eq.~(\protect\ref{eq:Vgreen}), and the corresponding infinite-volume Coulomb potential energy (dashed gray curve).
      } }
  \label{fig:pot}
\end{figure}
The  FV potential energy between two charges with $Qe  = 1$, and the
corresponding infinite-volume Coulomb potential energy are shown in Fig.~\ref{fig:pot}.

In the next sections, we construct non-relativistic EFTs to allow for order-by-order calculations of the FV QED modifications
to the energy of hadrons in 
the continuum limit of
LQCD calculations, going beyond the first two orders in the $1/L$ expansion that have 
been determined previously.
While these EFTs permit calculations to any given precision, including quantum fluctuations, some of 
the results that will be presented can be determined simply without the EFTs; 
a demonstration of which is the self-energy of a uniformly charged, rigid and fixed,  sphere in a FV. 
In this textbook case, the self-energy can be determined directly by integrating the interaction 
between infinitesimal volumes of the charge density, as governed by the modified Coulomb potential, Eq.~(\ref{eq:Vgreen}), 
over the sphere of radius $R$.  It is straightforward to show that the self-energy can be written in an expansion of $R/L$,
\begin{eqnarray}
\label{ChargedSphere}
U^{\rm sphere} (R,L) & = & 
{3\over 5} {(Qe)^2\over 4 \pi R}\ 
\ +\ 
{(Qe)^2\over 8\pi L}\ c_1
\ +\ 
{(Qe)^2\over 10 L} \left({R\over L}\right)^2\ 
 +\ \cdots,
\end{eqnarray}
where $c_1 =-2.83729$~\cite{Luscher:1986pf, Hasenfratz:1989pk, Luscher:1990ux}. 
The leading contribution is the well-known result for a uniformly charged sphere,
while the second term, the 
LO FV correction, is independent of the structure of the charge distribution. 
This suggests that it is also valid for a point particle; a result that proves to be valid 
for the corrections to the masses of single particles calculated with $\chi$PT and with the 
NREFTs presented in this work. 
It is simply the modification to the Coulomb self-energy of a point charge. 
The third term can be written as ${(Qe)^2} \langle r^2\rangle /  6 L^3$, 
where $ \langle r^2\rangle=\frac{3}{5}R^2$ is the mean-squared radius of the sphere, and 
reproduces the charge-radius contributions determined with the NREFTs, 
as will be shown in the next section.

\section{ Scalar NRQED for Mesons and $J=0$ Nuclei}
\noindent
LQCD calculations including QED have been largely 
focused on the masses of the pions and kaons in an effort to extract the values of electromagnetic counterterms of $\chi$PT, thus
we begin by considering the FV corrections to the masses of scalar hadrons.
In the limit where the volume of space is much larger than that of the hadron, keeping in mind that only the zero modes are being excluded from the photon fields, 
the FV corrections to the mass of the hadron will have a power-law dependence upon $L$, and  vanish as $L\rightarrow\infty$.
As the modifications to the self-energy arise from the infrared behavior of the theory, 
low-energy EFT provides a tool 
with which to systematically determine the FV effects in an expansion in one or more small parameters.

Using the methods developed to describe heavy-quark and heavy-hadron 
systems~\cite{Isgur:1989vq,Isgur:1989ed,Jenkins:1990jv,Jenkins:1991ne,Thacker:1990bm,Labelle:1992hd,Manohar:1997qy,Luke:1997ys,Chen:1999tn,Beane:2007es,Lee:2013lxa},
the Lagrange density describing the low-energy dynamics of a charged composite scalar particle, $\phi$, with charge $eQ$ 
can be written as an expansion in $1/m_{\phi}$ and in the  scale of compositeness,
\begin{eqnarray}
{\cal L}_\phi
& = & 
\phi^\dagger \left[\ 
iD_0
\ +\ {|{\bf D}|^2\over 2 m_\phi}
\ + \ { |{\bf D}|^4 \over 8 m_\phi^3}\ 
\ +\   {e \langle r^2\rangle_\phi\over 6}\ {\bm\nabla}\cdot {\bf E}
\ +\ 2 \pi \tilde\alpha_E^{(\phi)}  |{\bf E}|^2
\ +\ 2\pi \tilde\beta_M^{(\phi)}  |{\bf B}|^2
\right.\nonumber\\
&&\left.
\qquad \qquad 
\ +\ i  e  c_M\ { \{ D^i , ({\bm\nabla}\times {\bf B})^i \} \over  8 m_\phi^3}
\ +\  \cdots
\ \right] \phi,
\label{eq:scalarLag}
\end{eqnarray}
where $m_{\phi}$ is the mass of the particle, the covariant derivative is $D_\mu = \partial_\mu + i e \hat Q A_\mu$ with $\hat Q$ the charge operator.
$\langle r^2\rangle_\phi$ is the mean-squared charge radius of the $\phi$,
and we have performed the standard field redefinition to the NR normalization of states,
$\phi\rightarrow \phi/\sqrt{2 m_\phi}$.
The remaining coefficients of operators involving the 
electric, ${\bf E}$, and magnetic, ${\bf B}$, fields,
have been determined by matching this EFT to  scalar QED, to yield
\begin{eqnarray}
\tilde\alpha_E^{(\phi)}  & = & 
\alpha_E^{(\phi)}  -  {\alpha_e Q\over 3 m_\phi} \langle r^2\rangle_\phi 
\ \ ,\ \ 
\tilde\beta_M^{(\phi)}  \ = \
\beta_M^{(\phi)}  
\ \ ,\ \ 
c_M\ =\  {2\over 3} m_\phi^2 \langle r^2\rangle_\phi,
\label{eq:scalarpols}
\end{eqnarray}
where $\alpha_E^{(\phi)} , \beta_M^{(\phi)}$ are the electric and magnetic polarizabilities of the 
$\phi$.~\footnote{
The presence of a charge-radius dependent term in the coefficient of the 
electric polarizability  indicates a subtlety in using this EFT  to describe hadrons in a 
background electric field~\cite{Lee:2013lxa}. 
Such contributions can be cancelled by including redundant operators in the EFT Lagrange density 
when matching to S-matrix elements. 
Since a classical uniform electric field modifies the equations of motion, such operators must  be retained in the 
Lagrange density and their coefficients matched directly to Green functions. 
}
These coefficients will be modified at higher orders in perturbation theory, 
starting at $\mathcal{O}(\alpha_e)$.
They will also be modified by terms that are exponentially suppressed by  compositeness length scales, e.g. 
$\sim e^{-m_\pi L}$ for QCD.
The ellipses denote terms that are higher order in 
derivatives acting on the  fields, with  coefficients dictated by the mass and compositeness scale 
-- the chiral symmetry breaking scale, $\Lambda_\chi$, for mesons and baryons.
For  one-body observables,  terms beyond 
$\phi^\dagger i \partial_0 \phi$ are treated in perturbation theory, providing  a systematic expansion in $1/L$.

\begin{figure}[!ht]
  \centering
     \includegraphics[scale=0.230]{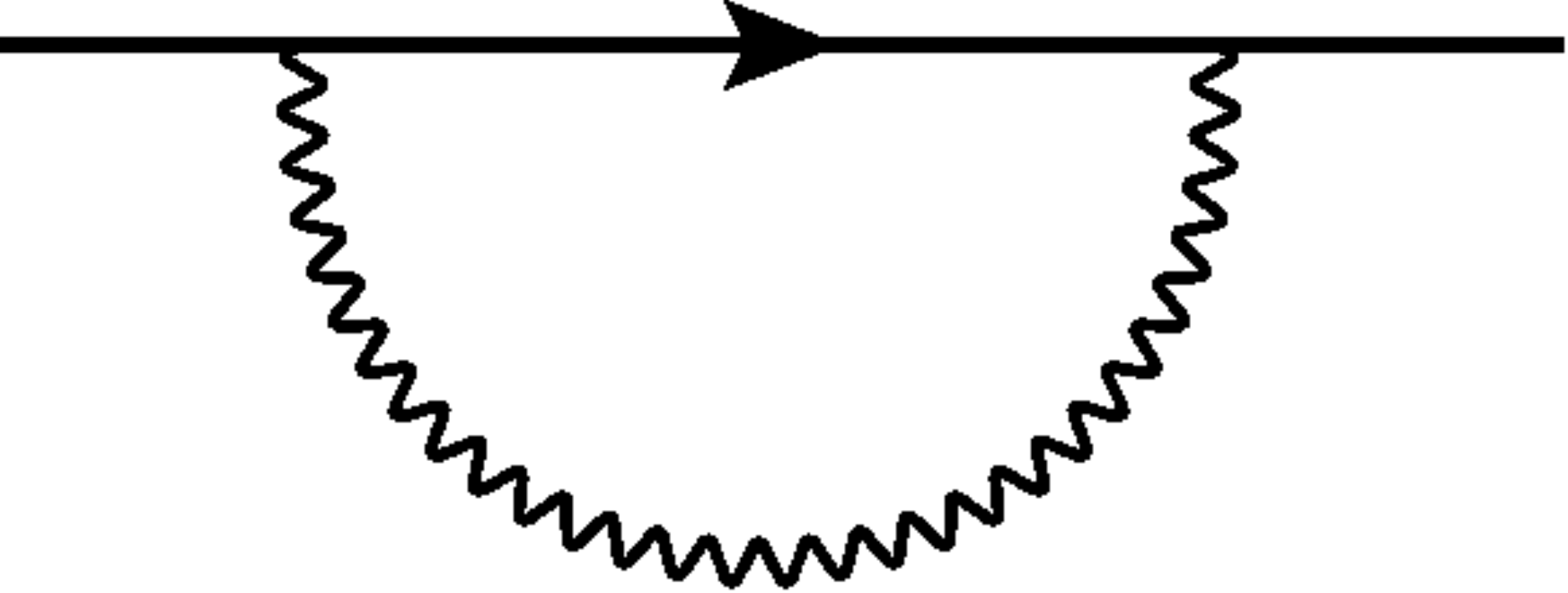}     
     \caption{{\small The one-loop diagram providing the LO, $\mathcal{O}(\alpha_e/L)$, 
     FV correction to the mass of a charged scalar particle. 
     The solid straight line denotes a scalar particle, while the wavy line denotes a photon.
     }}
  \label{fig:LO}
\end{figure}
The LO, $\mathcal{O}(\alpha_e/L)$, 
correction to the 
mass of a charged scalar particle in FV, $\delta m_\phi$,
is from  the one-loop diagram shown in 
Fig.~\ref{fig:LO}.
While most simply calculated in Coulomb gauge, the diagram can be calculated in any gauge and,
in agreement with previous determinations~\cite{Hayakawa:2008an},   is
\begin{eqnarray}
\delta m_\phi^{({\rm LO})}
& = & 
 {\alpha_e Q^2\over 2\pi L}\ 
 \hat{\sum_{ {\bf n}\ne {\bf 0}}}\ 
 {1\over |{\bf n}|^2}
\ =\ 
{\alpha_e Q^2\over 2 L}\ c_1,
\label{eq:scalarLO}
\end{eqnarray}
with $c_1 = -2.83729$. 
The sum, $\hat{\sum}$, represents the difference between the sum over the FV modes and the 
infinite-volume integral, e.g.
\begin{eqnarray}
{1\over L^3}\hat{\sum_{{\bf k}\ne {\bf 0}}}\ 
f({\bf k})
&\equiv &
{1\over L^3}\sum_{{\bf k}\ne {\bf 0}}\ f({\bf k})
\ -\ 
\int {d^3{\bf k}\over (2\pi)^3}\ f({\bf k})
 ,
\label{eq:FVsumint}
\end{eqnarray}
for an arbitrary function $f({\bf k})$,
and is therefore finite.
This shift is a power law in $1/L$ as expected, and provides a reduction in the mass of the hadron.
As the infinite-volume Coulomb interaction increases the mass, and the FV result is obtained 
from the modes that satisfy the PBCs (minus the zero modes), the sign of the correction is also expected.
The result in Eq.~(\ref{eq:scalarLO}) is nothing more than the difference between the FV and infinite-volume 
contribution to the Coulomb self-energy of a charged point particle, as seen from 
Eq.~(\ref{eq:Vgreen}), $U({\bf 0},L)/2$.

\begin{figure}[!ht]
\begin{center}
\subfigure[]{
\includegraphics[scale=0.24]{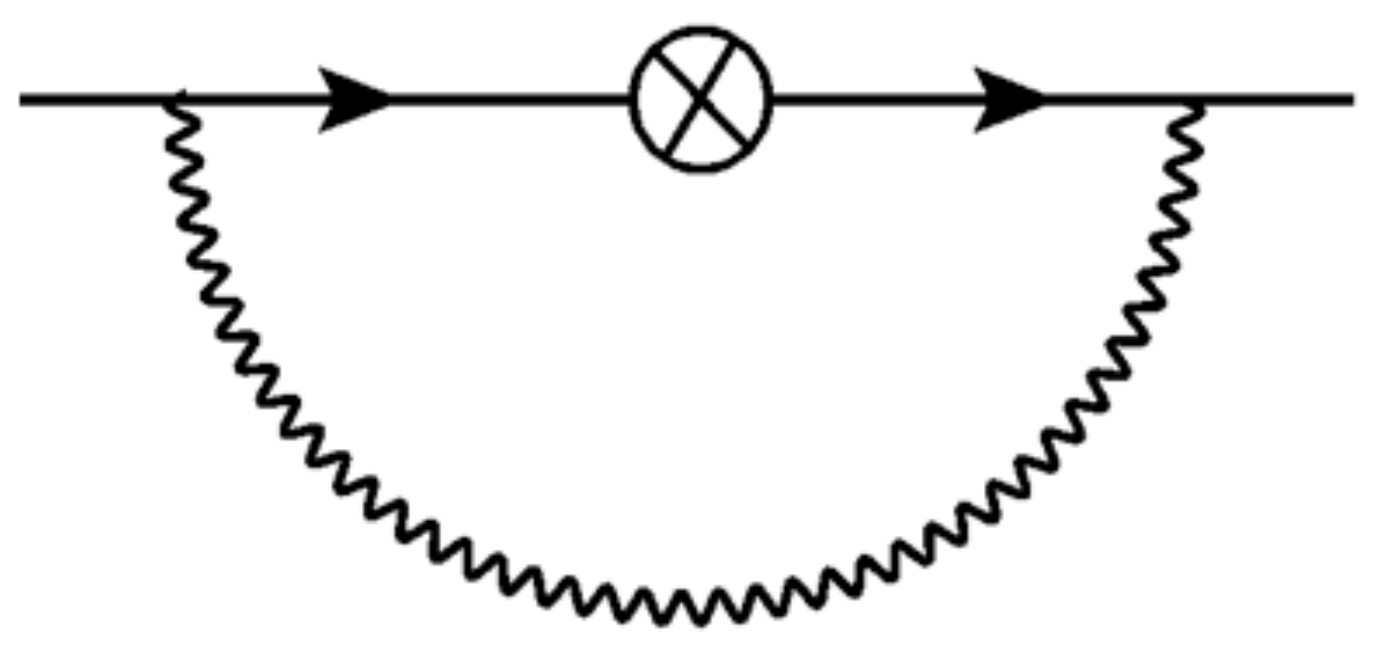}}
\subfigure[]{
\includegraphics[scale=0.22]{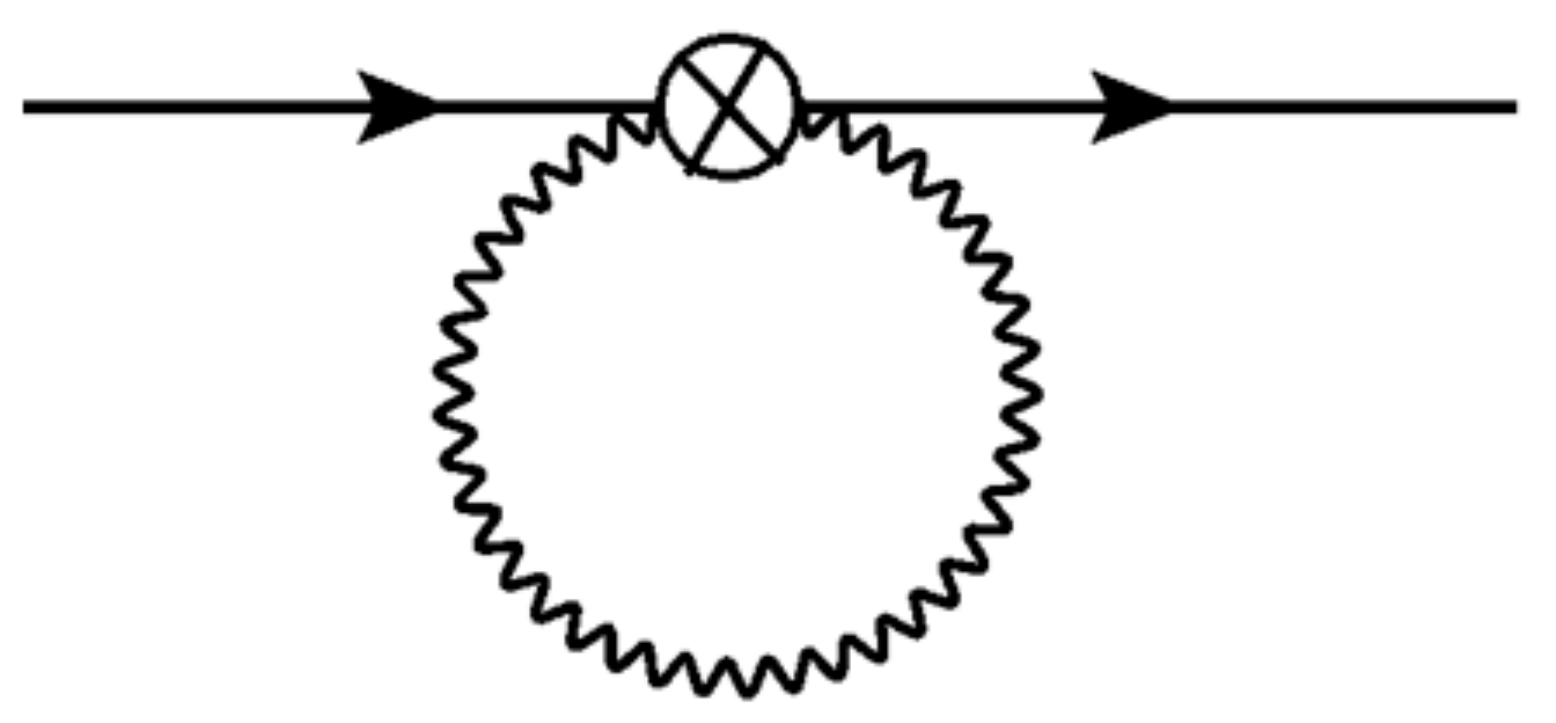}}
\subfigure[]{
\includegraphics[scale=0.24]{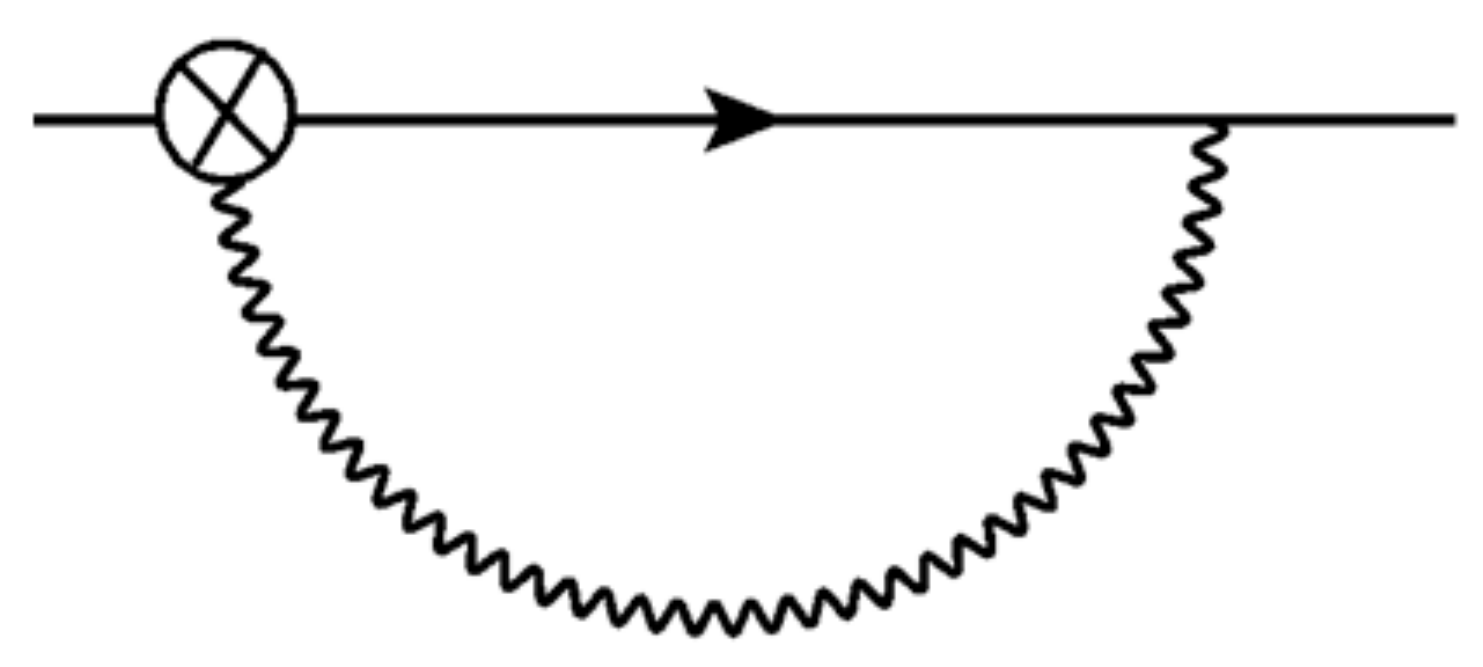}}
\subfigure[]{
\includegraphics[scale=0.24]{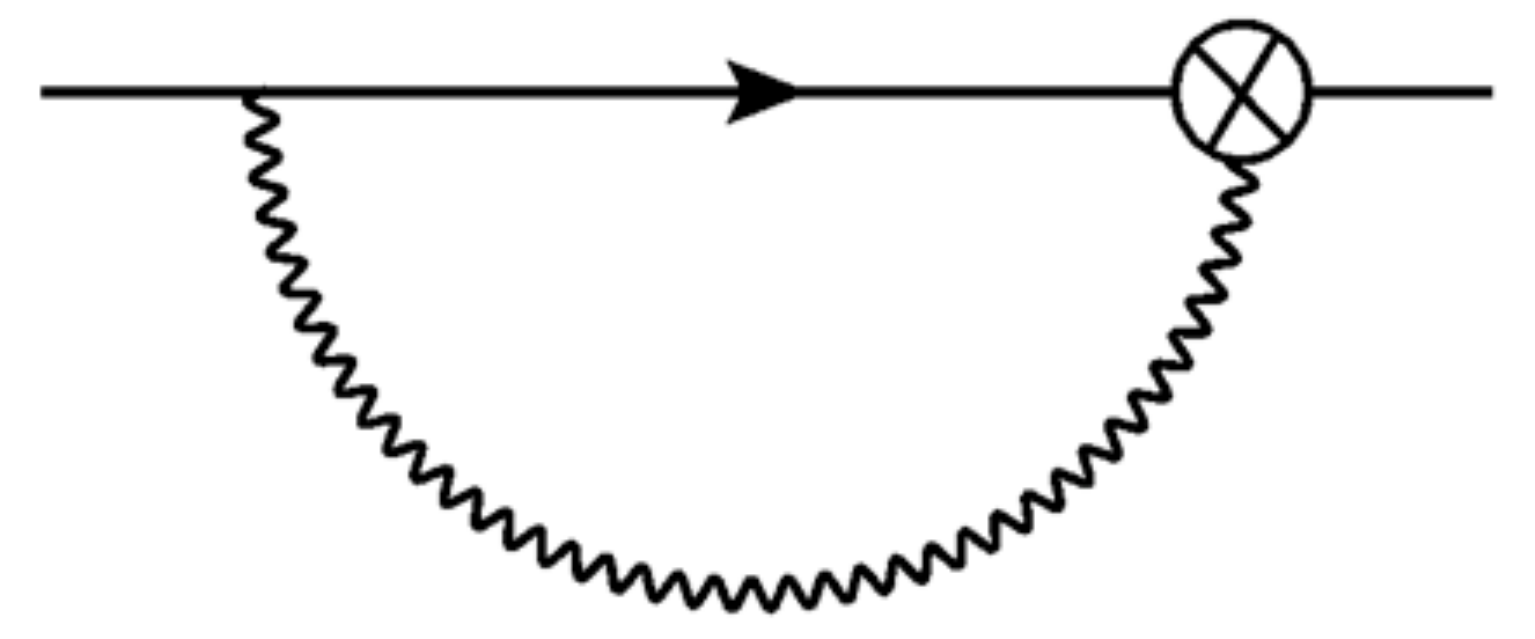}}
\caption{{\small Diagrams contributing at NLO,
$\mathcal{O}(\alpha_e/m_{\phi}L^2)$, in the ${1/ L}$ expansion.
The crossed circle denotes an insertion of  the $|{\bf D}|^2/2 m_\phi$ 
      operator in the scalar QED Lagrange density, Eq.~(\protect\ref{eq:scalarLag}).
}}
\label{fig:mall}
\end{center}
\end{figure}
The next-to-LO (NLO) contribution,  ${\cal O}\left(\alpha_e/L^2\right)$, 
arises from a single insertion of the 
$|{\bf D}|^2/2 m_\phi$ operator in Eq.~(\protect\ref{eq:scalarLag}) into the one-loop diagrams shown in Fig.~\ref{fig:mall}.
The contribution from each of these diagrams depends upon the choice of gauge, however the sum is gauge independent,~\footnote{
The sums appearing at LO and NLO are 
\begin{eqnarray}
 \hat{\sum_{ {\bf n}\ne {\bf 0}}}\  {1\over |{\bf n}|} & = & c_1
 \ \ ,\ \ 
  \hat{\sum_{ {\bf n}\ne {\bf 0}}}\  {1\over |{\bf n}|^2} \ = \ \pi\  c_1
  .
  \nonumber
\end{eqnarray}
} 
\begin{eqnarray}
\delta m_\phi^{({\rm NLO})}
& = & 
 {\alpha_e Q^2\over m_\phi L^2 }\ 
 \hat{\sum_{{\bf n}\ne {\bf 0}}}\ 
 {1\over |{\bf n}|}
\ =\ 
{\alpha_e Q^2\over m_\phi L^2 }\ c_1
  .
\label{eq:scalarNLO}
\end{eqnarray}
This NLO recoil correction agrees with previous calculations~\cite{Hayakawa:2008an,deDivitiis:2013xla},
and is the highest order in the $1/L$ expansion 
to which these FV effects have been previously 
determined.~\footnote{
The $\mathcal{O}(\alpha_e)$ calculations of Ref.~\cite{Hayakawa:2008an} at NLO in $\chi$PT and PQ$\chi$PT  do not 
include the full contributions from the meson charge radius and polarizabilities, but are perturbatively close. 
This is in contrast to the NREFT calculations presented in this work
where the low-energy coefficients are matched to these quantities 
order by order in  $ \alpha_e$,
and provide the  result at any given order in  $1/L$  as an expansion in $ \alpha_e$.
}

\begin{figure}[!ht]
\begin{center}
\subfigure[]{
\includegraphics[scale=0.21]{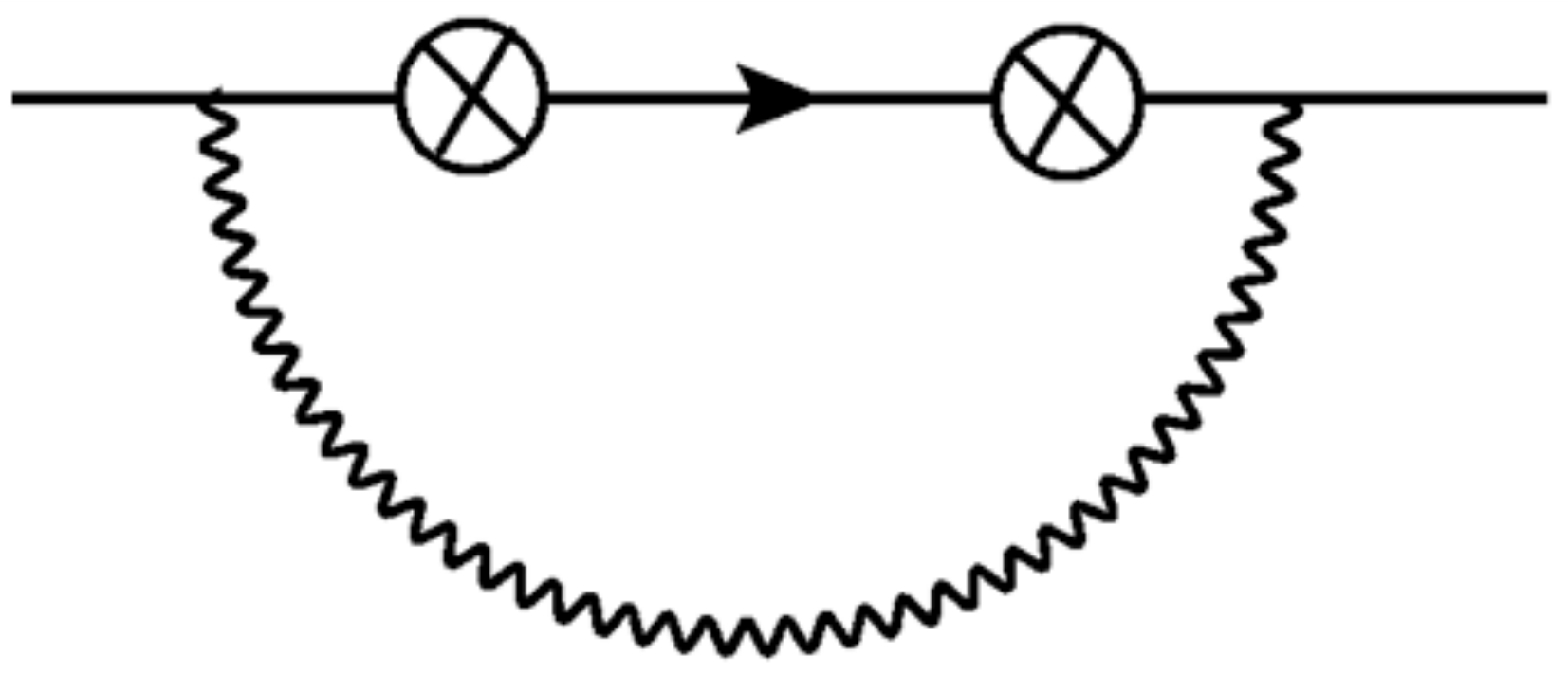}}
\subfigure[]{
\includegraphics[scale=0.21]{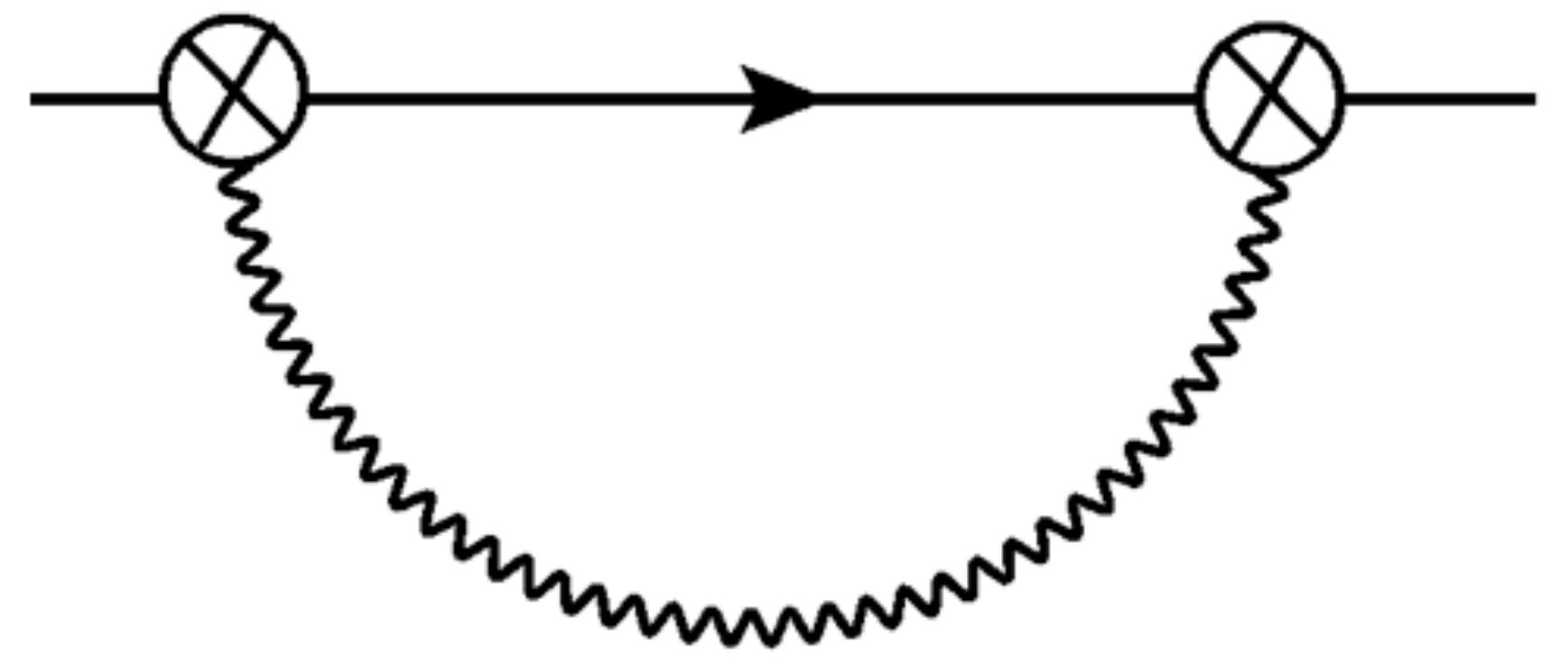}}
\subfigure[]{
\includegraphics[scale=0.21]{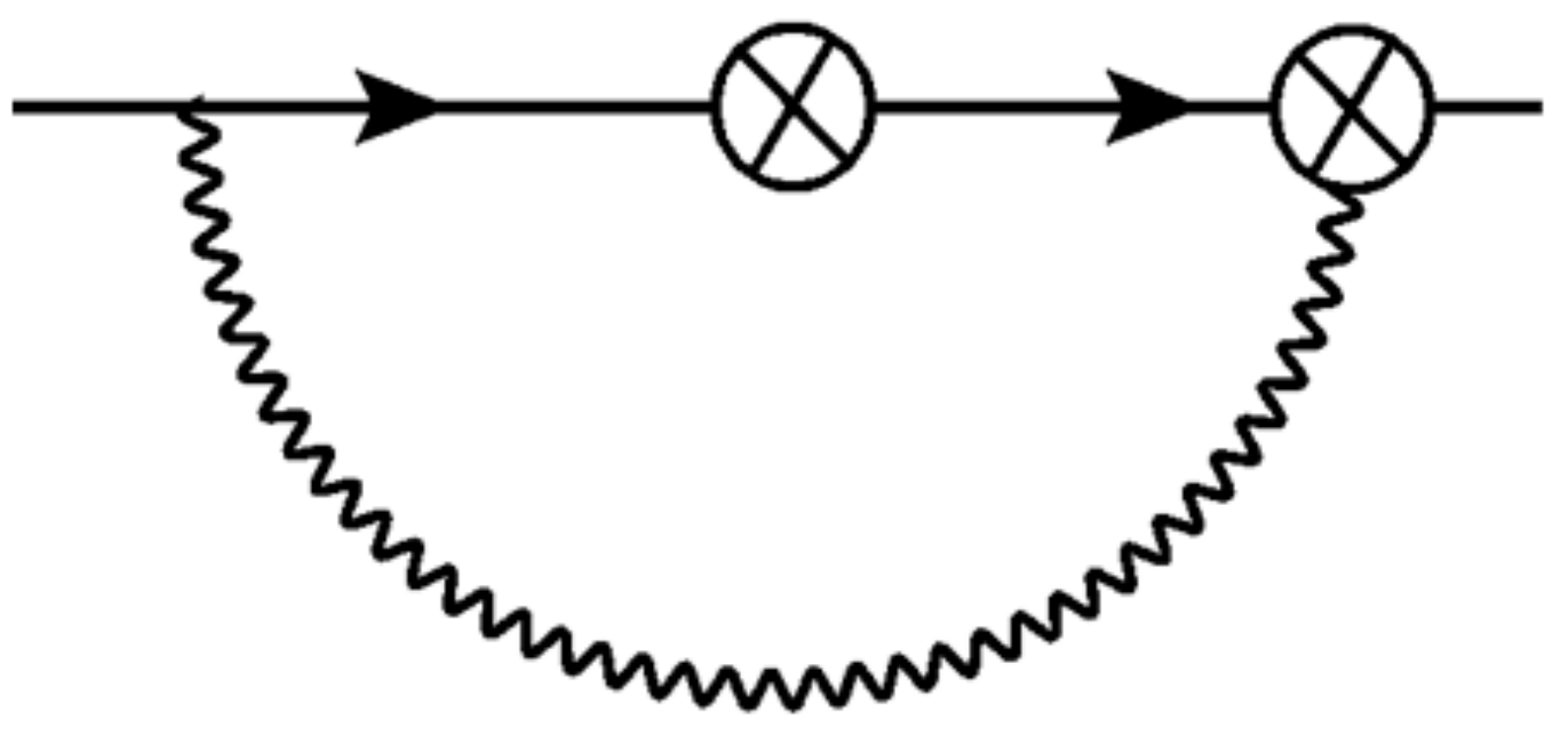}}
\subfigure[]{
\includegraphics[scale=0.21]{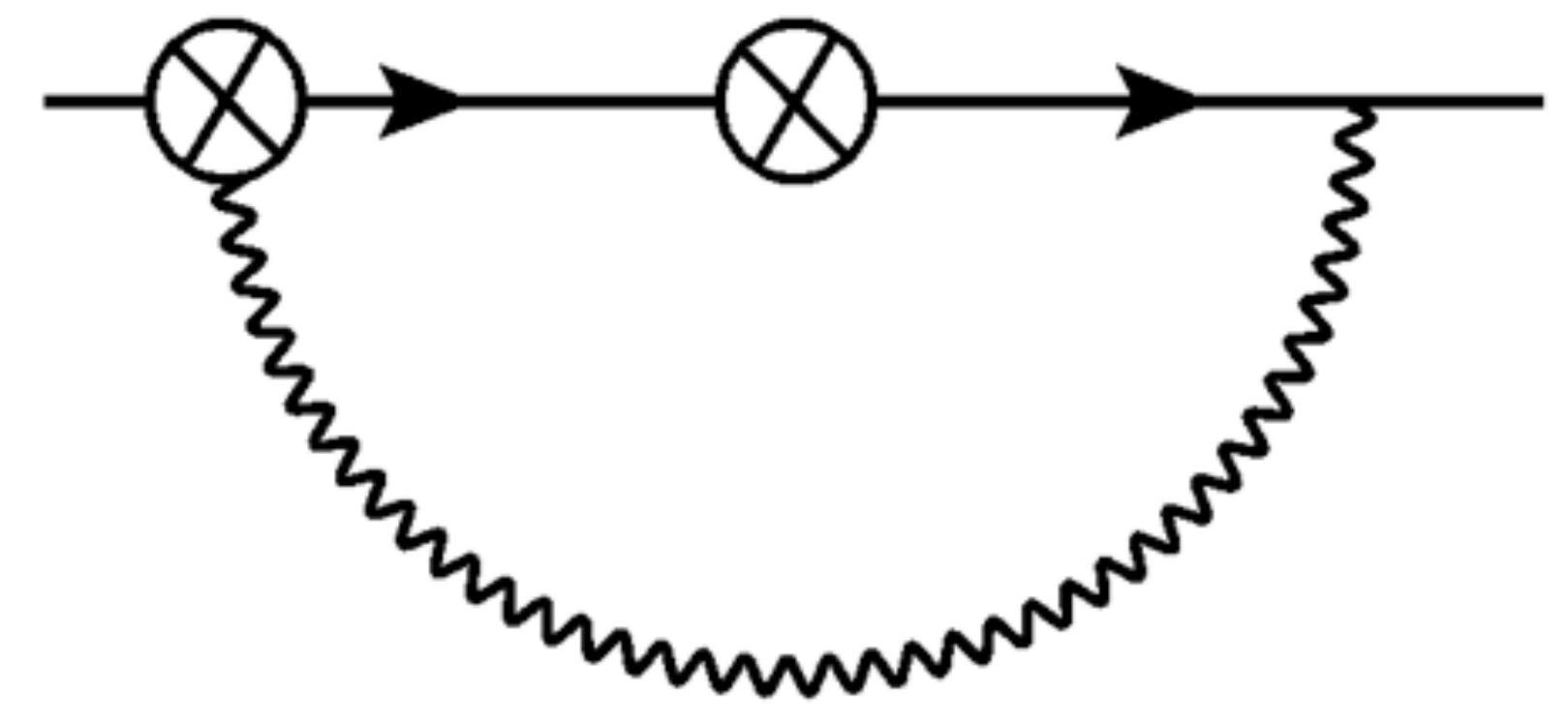}}
\subfigure[]{
\includegraphics[scale=0.185]{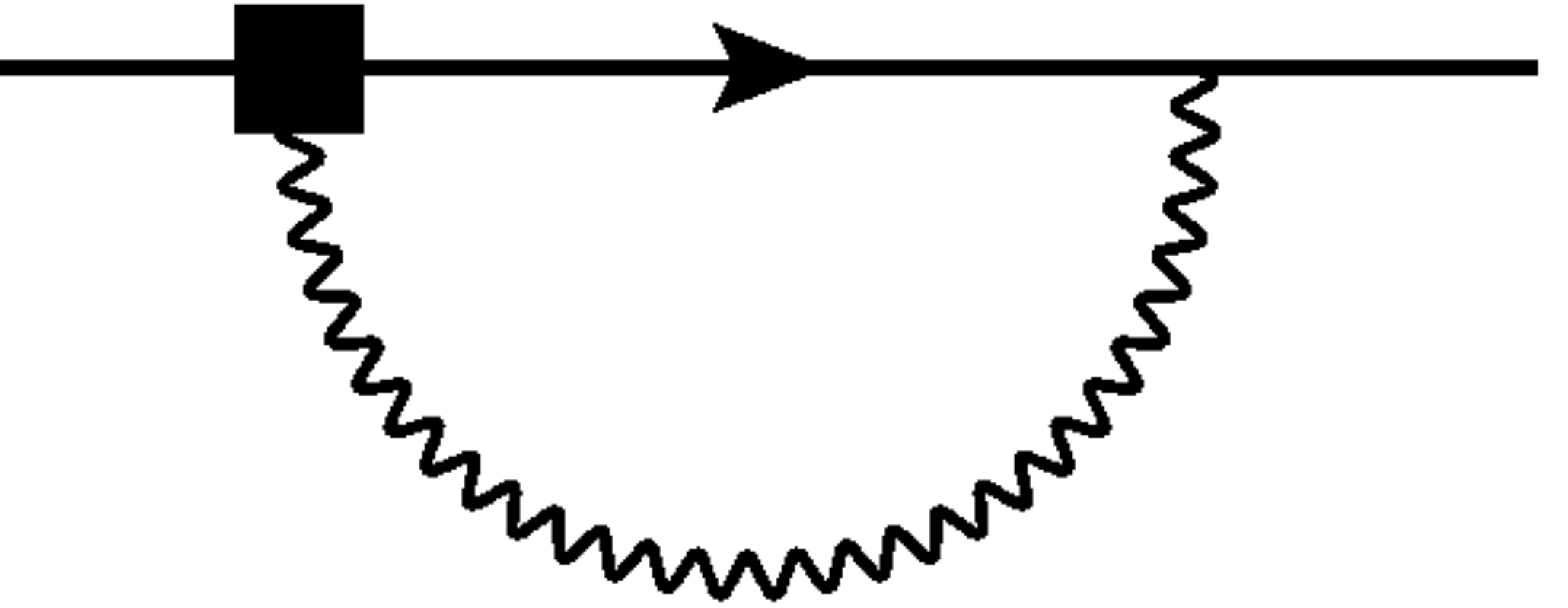}}
\subfigure[]{
\includegraphics[scale=0.185]{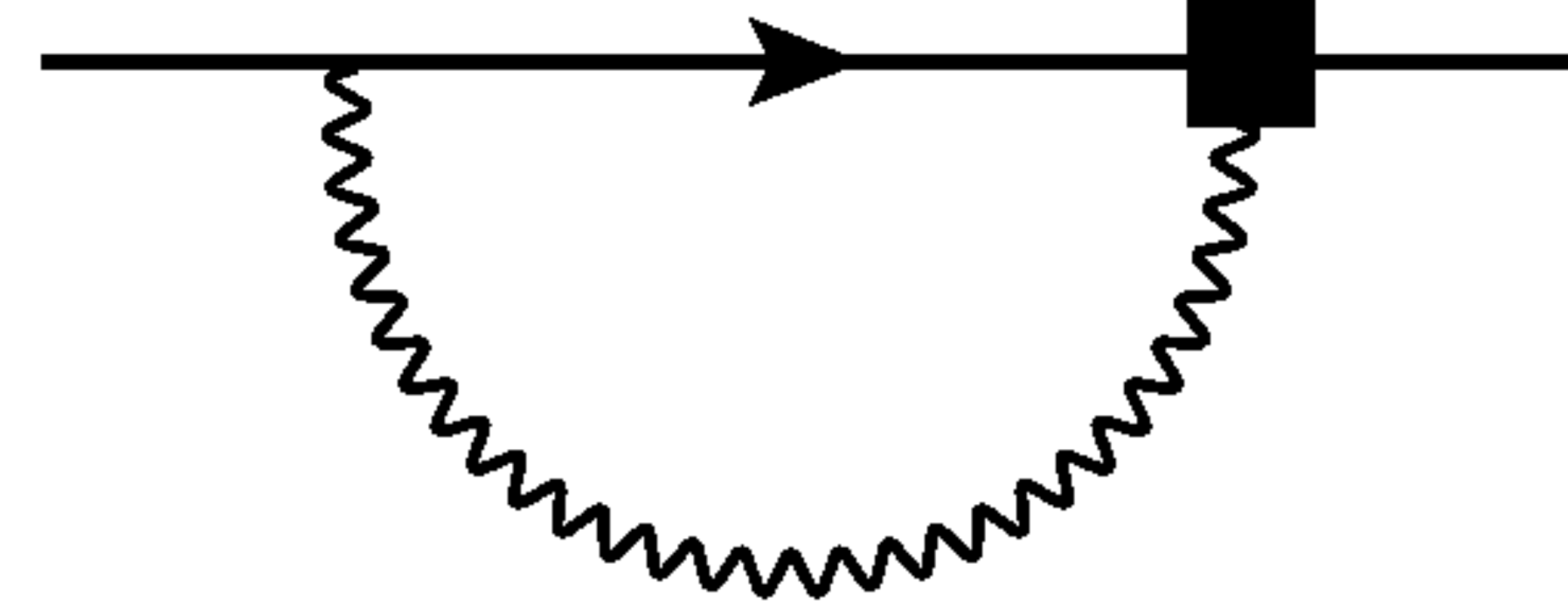}}
\caption{{\small 
(a-d) One-loop diagrams giving rise to the recoil corrections of $\mathcal{O}(\alpha_e/m_\phi^2 L^3)$. 
The crossed circle denotes an insertion of the $|{\bf D}|^2/2 m_\phi$ operator. 
(e,f) One-loop diagrams providing the leading contribution from the charge radius of the scalar hadron, 
$\sim \alpha_e \langle r^2\rangle_\phi/L^3$. 
The solid square denotes an insertion of the charge-radius 
operator in the scalar Lagrange density, Eq.~(\protect\ref{eq:scalarLag}).
}}
\label{fig:m2}
\end{center}
\end{figure}
At next-to-next-to-LO
(N$^2$LO), ${\cal O}\left(\alpha_e/L^3\right)$, 
there are potentially two contributions - one is a recoil correction of the form $\sim  \alpha_e /m_\phi^2 L^3$ 
and one is from the charge radius, $\sim \alpha_e  \langle r^2\rangle_\phi/L^3$.
An evaluation of the one-loop diagrams giving rise to the recoil contributions, Fig.~\ref{fig:m2}(a-d), shows that
while individual diagrams are generally non-zero for a given gauge, their sum vanishes in 
any gauge.  
Therefore, there are no contributions of the form $\alpha_e/m_\phi^2 L^3$ to the mass of $\phi$.
In contrast, the leading contribution from the charge radius of the scalar particle, resulting from the one-loop diagrams shown in Fig.~\ref{fig:m2}(e,f)
gives a contribution of the form
\begin{eqnarray}
\delta m^{({\rm N^2LO})}_\phi & = & 
-{2\pi\alpha_e  Q\over 3 L^3}\  \langle r^2 \rangle_\phi \ 
\hat{\sum_{{\bf n}\ne {\bf 0}}} ~ 1
\ =\ 
+{2\pi\alpha_e Q\over 3 L^3}\  \langle r^2 \rangle_\phi ,
\label{eq:CRscalar}
\end{eqnarray}
where  $\hat{\sum\limits_{\bf n} }~1 = 0$.

\begin{figure}[t!]
\begin{center}
\subfigure[]{
\includegraphics[scale=0.22]{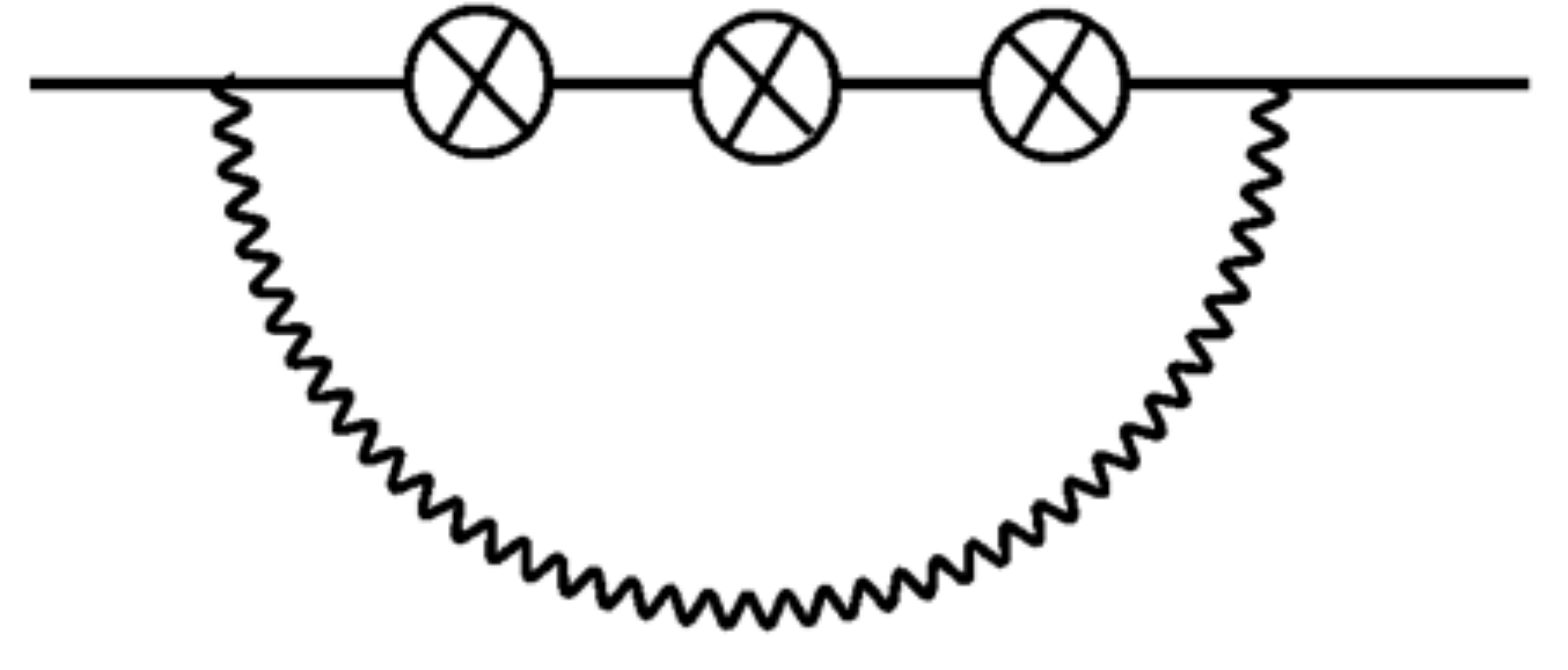}}
\subfigure[]{
\includegraphics[scale=0.22]{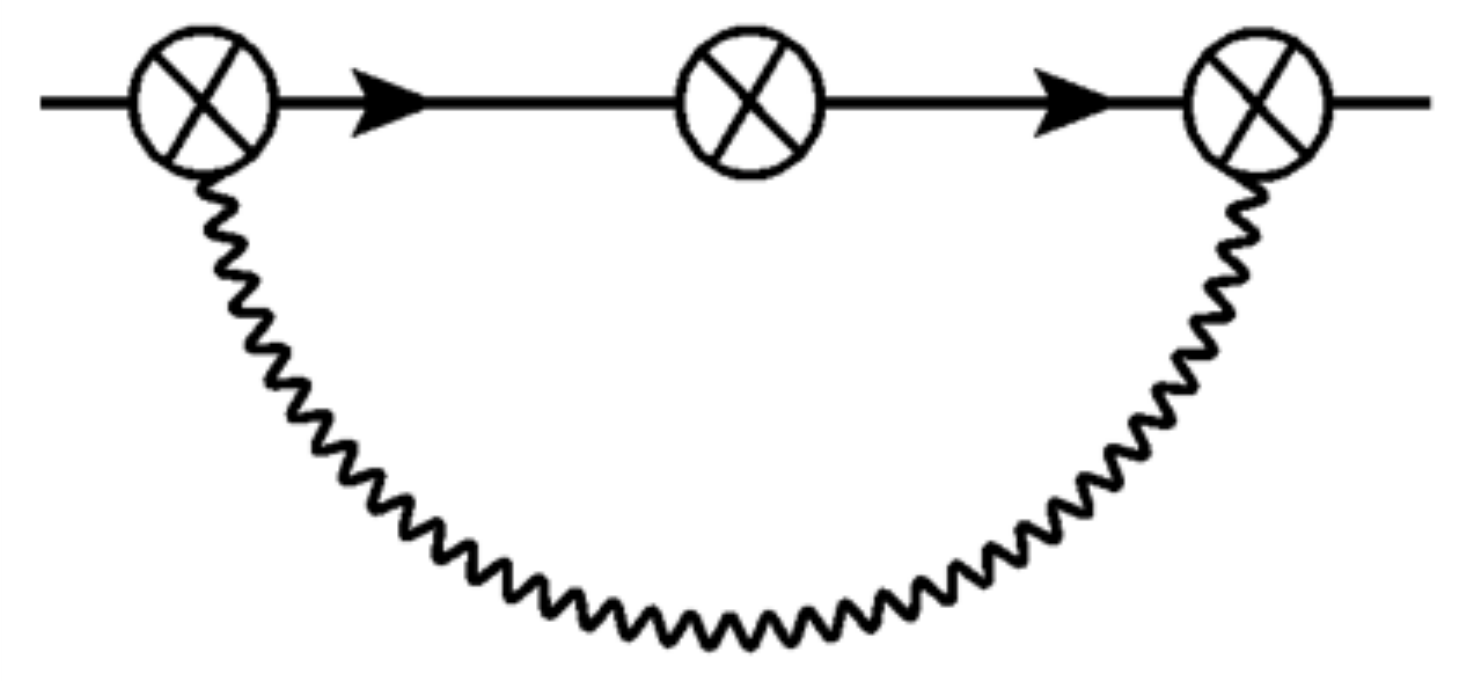}}
\subfigure[]{
\includegraphics[scale=0.22]{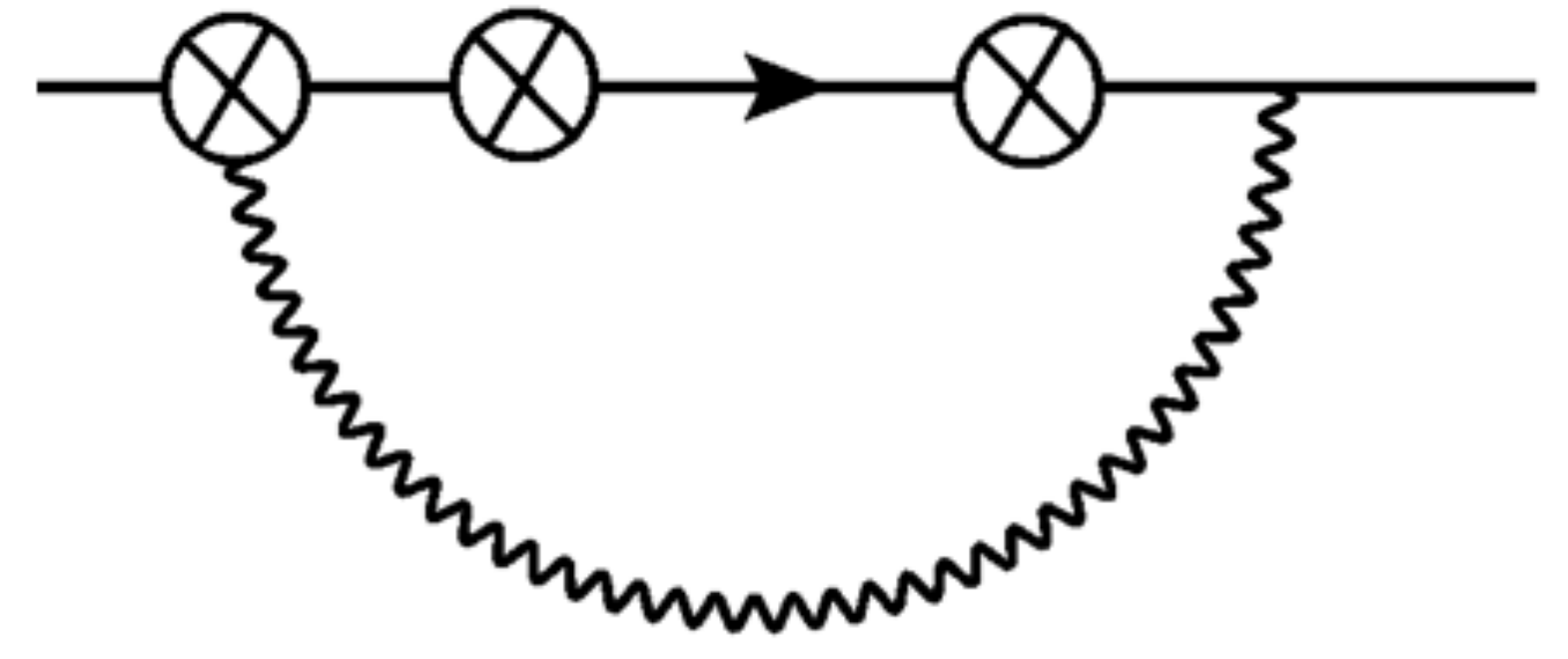}}
\subfigure[]{
\includegraphics[scale=0.22]{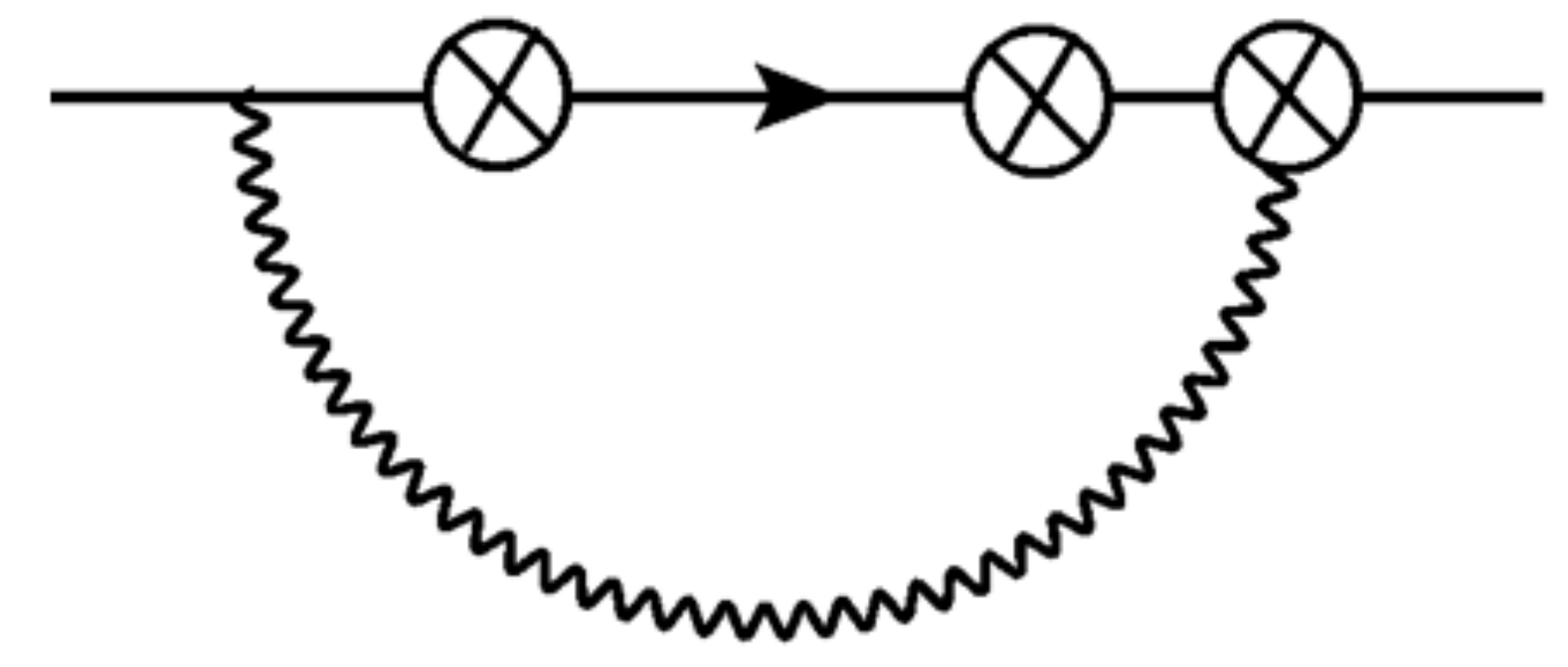}}
\subfigure[]{
\includegraphics[scale=0.22]{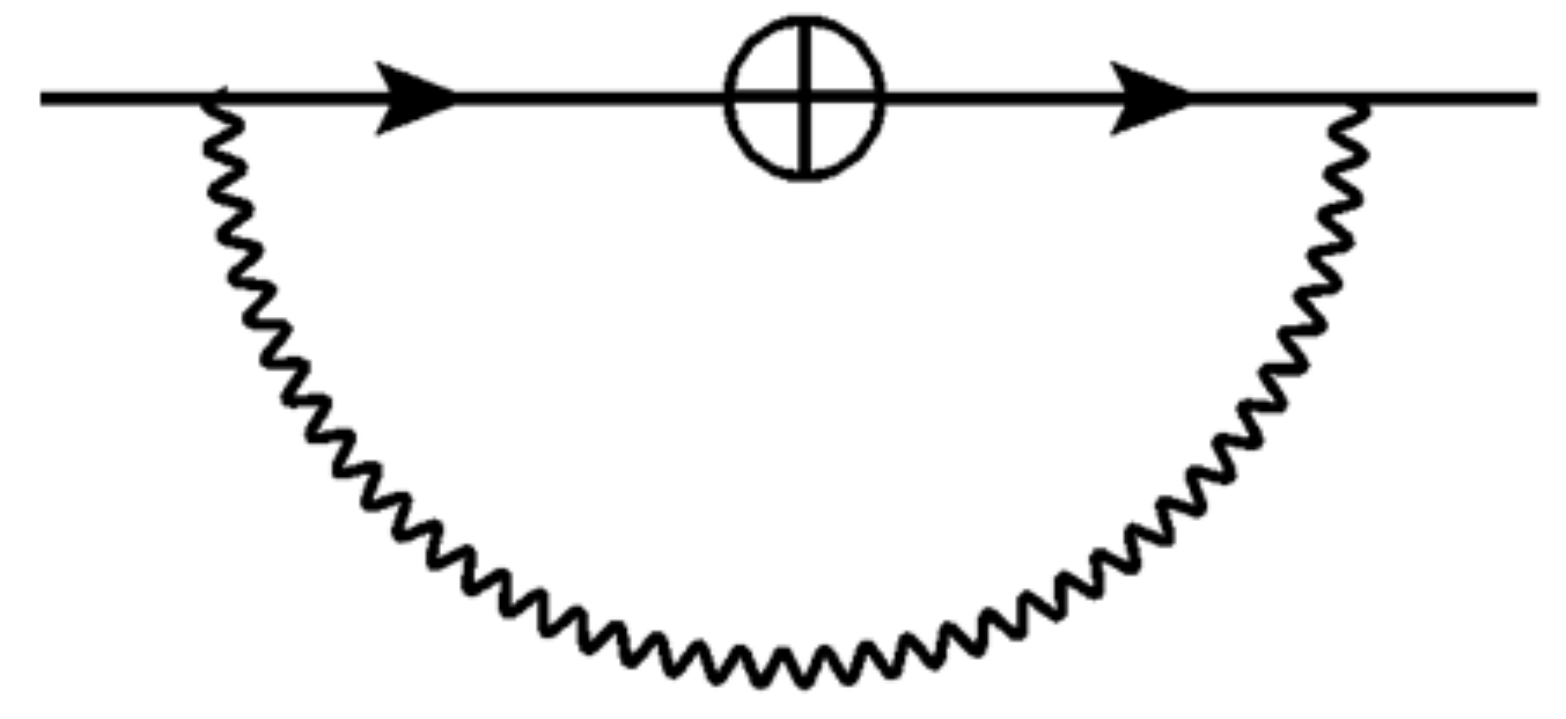}}
\subfigure[]{
\includegraphics[scale=0.21]{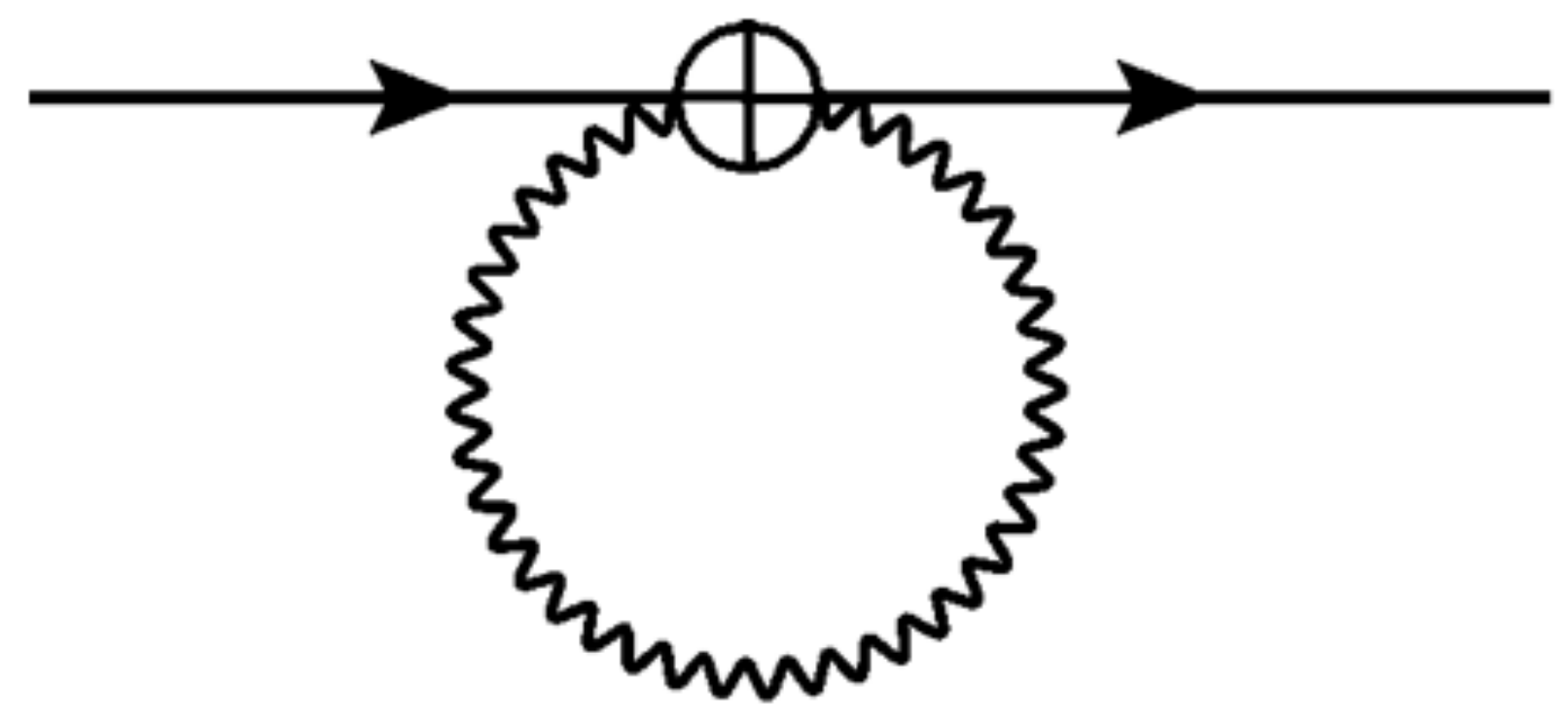}}
\subfigure[]{
\includegraphics[scale=0.14]{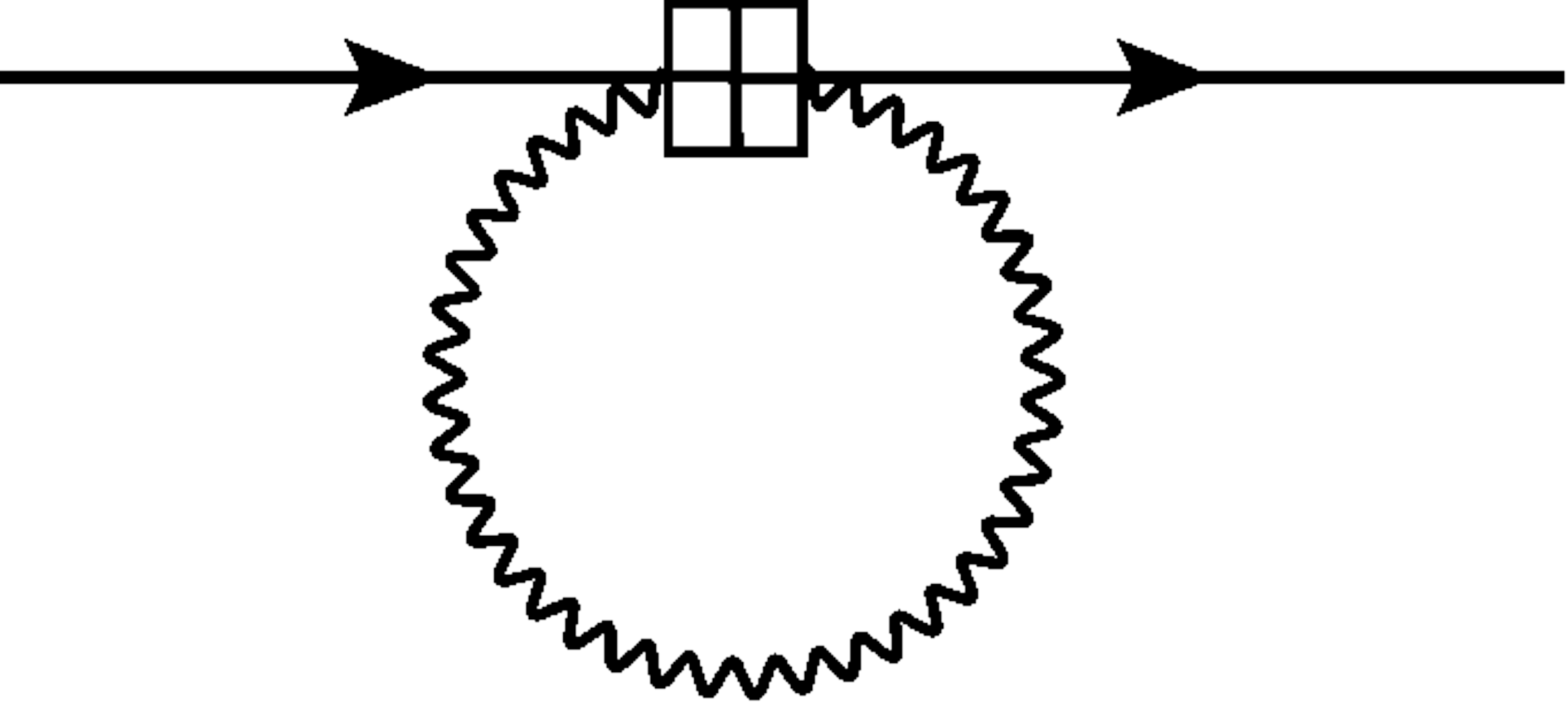}}
\caption{{\small 
One-loop diagrams contributing to the FV corrections to the mass of a scalar hadron at 
N$^3$LO, ${\cal O}\left(1/L^4\right)$.
Diagrams (a-d) involve  three insertions of the $|{\bf D}|^2/2 m_\phi$ operator (crossed circles) 
in the scalar QED Lagrange density in Eq.~(\protect\ref{eq:scalarLag}), 
while (e,f) involve one insertion of the $|{\bf D}|^4/8 m_\phi^3$ operator (the sun cross), 
giving a $\mathcal{O}(\alpha_e/m_\phi^3 L^4)$ correction. 
Diagram (g) involves an  insertion of $\tilde{\alpha}_E^{(\phi)} \ |{\bf E}|^2$ and $\tilde{\beta}_M^{(\phi)}\  |{\bf B}|^2$, 
operators (crossed square), 
contributing terms of the form 
$\sim (\alpha_E+\beta_M)/L^4$ and $\sim \alpha_e\langle r^2 \rangle_\phi/m_{\phi}L^4)$.
A diagram analogous to (g) provides the leading contribution from the $c_M$ operator at   $\mathcal{O}(\alpha_e/m_\phi L^4)$.
}}
\label{fig:m3A}
\end{center}
\end{figure}
At N$^3$LO, ${\cal O}\left(\alpha_e/L^4\right)$, 
there are potentially three contributions: 
recoil corrections, $\sim \alpha_e/m_\phi^3 L^4$, 
contributions from the electric and magnetic polarizability operators, 
$\sim \tilde\alpha_E^{(\phi)}/L^4$ , $\tilde\beta_M^{(\phi)} /L^4$, 
and contributions from the $c_M$ operator, Eq. (\ref{eq:scalarLag}). 
There are two distinct sets of recoil corrections at this order. 
One set is from diagrams involving three insertions of the $|{\bf D}|^2/2 m_\phi$ operator, 
as shown in Fig.~\ref{fig:m3A}(a-d), 
and the other is from a single insertion of the $|{\bf D}|^4/8 m_\phi^3$ operator, shown in Fig.~\ref{fig:m3A}(e,f). 
The sum of diagrams contributing to each set vanishes, and so there are no contributions of the form $\alpha_e/m_\phi^3 L^4$.
The other contributions, that include the electric and magnetic polarizabilities, arise from the one-loop diagrams shown in Fig.~\ref{fig:m3A}(g). A straightforward evaluation yields  a mass shift of 
\begin{eqnarray}
\delta m_\phi^{({\rm N^3LO}; \tilde{\alpha},\tilde{\beta})} 
&& = 
- {4\pi^2\over L^4}\ \left( \tilde{\alpha}_E^{(\phi)} + \tilde{\beta}_M^{(\phi)} \right)\ 
\hat{\sum_{{\bf n}\ne {\bf 0}}}\ |{\bf n}|\nonumber\\
&& = 
- {4\pi^2\over L^4}\ \left( \alpha_E^{(\phi)} + \beta_M^{(\phi)} \right)\ c_{-1}
+{4\pi^2 \alpha_e Q \over 3 m_\phi L^4} \ 
\langle r^2 \rangle_\phi \ c_{-1},
\end{eqnarray}
where the regularized sum is the same that contributing to the energy density  associated with the Casimir effect,
and is
$c_{-1} = -0.266596$ \cite{Hasenfratz:1989pk}.
A similar calculation yields the contribution from the $c_M$ operator, 
\begin{eqnarray}
\delta m_\phi ^{({\rm N^3LO}; c_M)} 
& = & 
+{4\pi^2 \alpha_e Q \over 3 m_\phi L^4} \ 
\langle r^2 \rangle_\phi \ c_{-1}
.
\end{eqnarray}

\begin{figure}[!ht]
  \centering
     \includegraphics[scale=0.335]{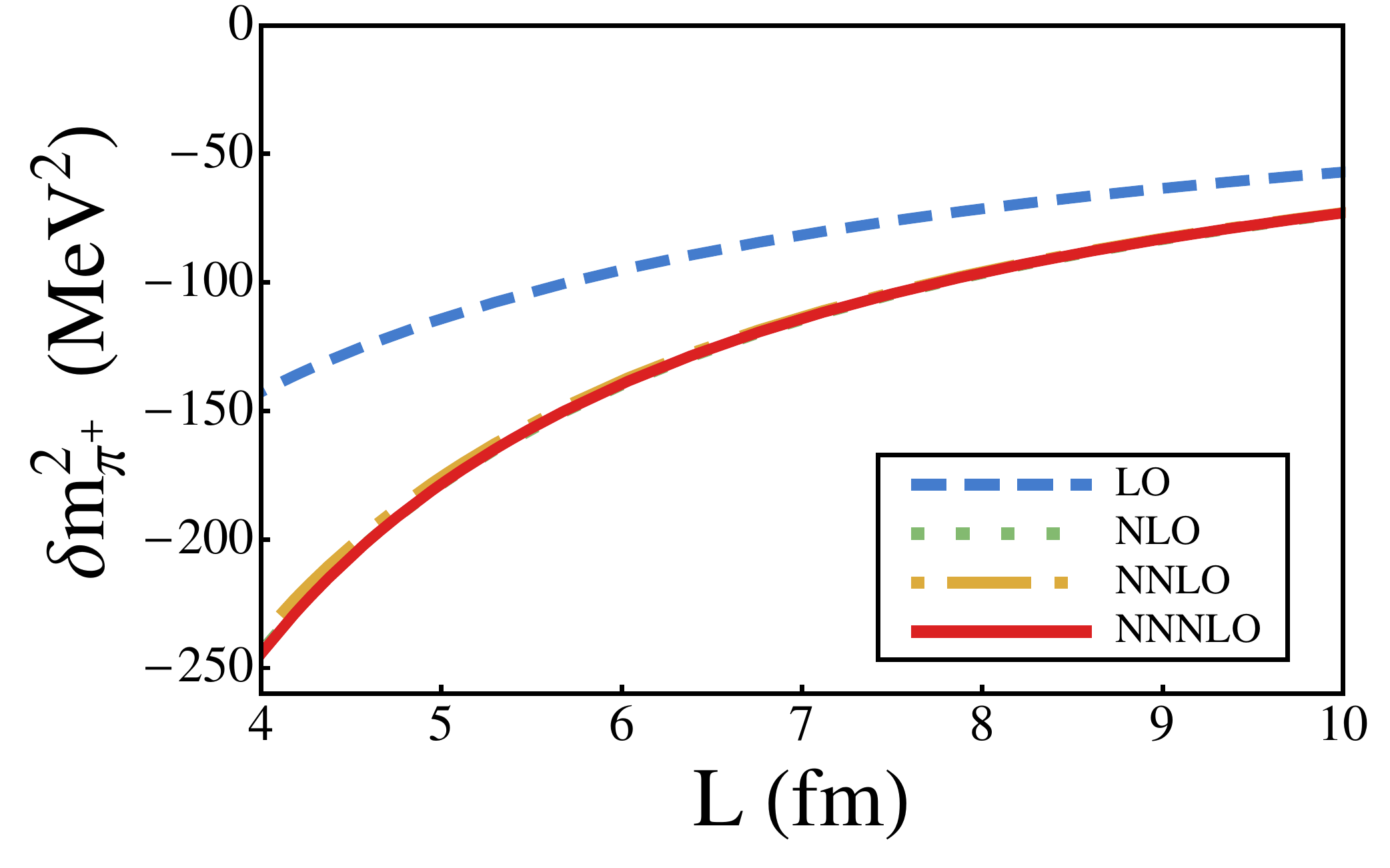}     \qquad
     \includegraphics[scale=0.335]{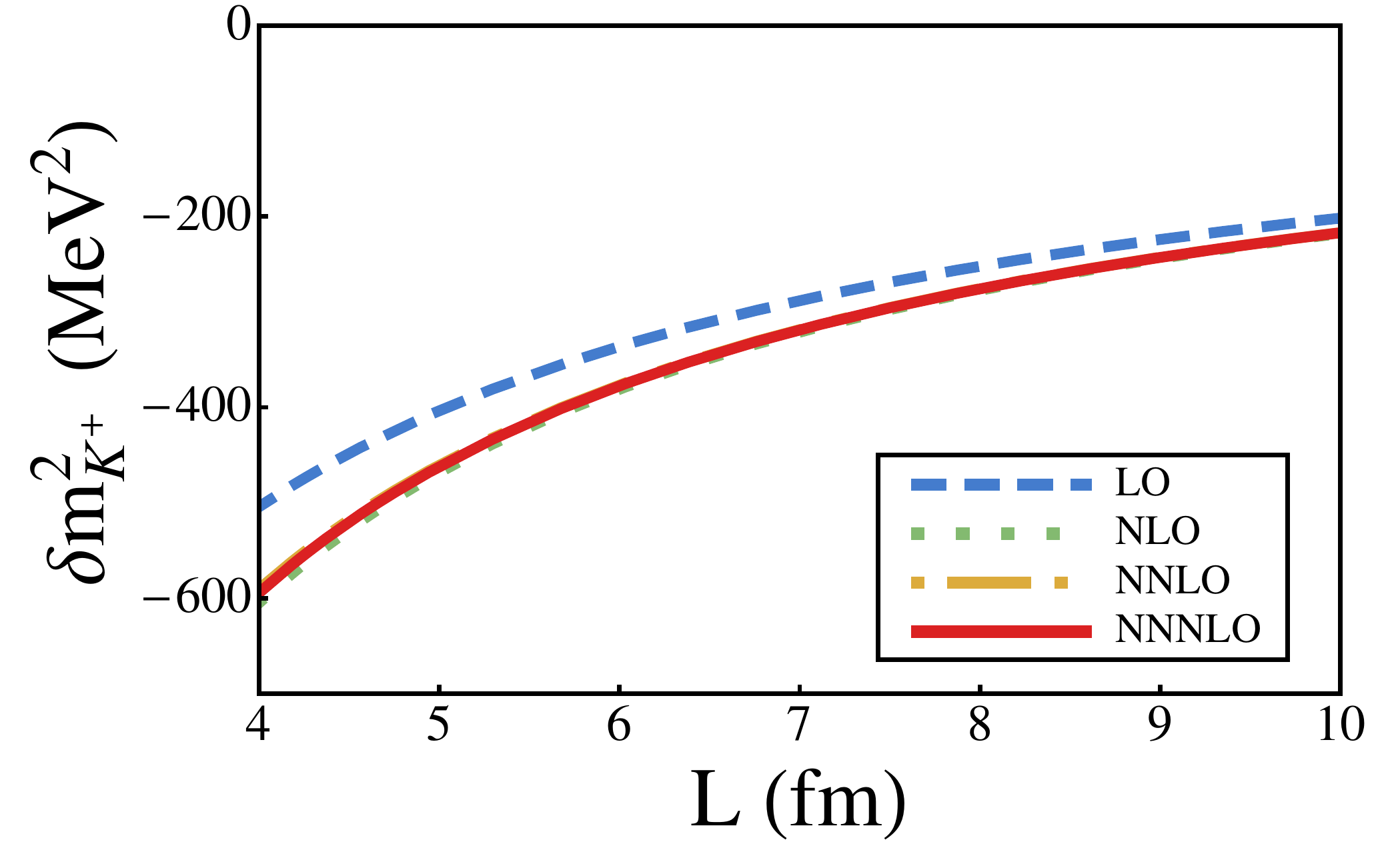}     
     \caption{{\small The FV QED correction to  the mass squared of a charged pion (left panel) and kaon (right panel) 
     at rest in a FV at the physical pion mass.
     The leading contribution is due to  their electric charge, and scales as $1/L$.
     The $1-\sigma$ uncertainty bands associated with each order in the expansion are determined from the uncertainties in the experimental and theoretical inputs.
    }}
  \label{fig:chargedkaonpionmsquare}
\end{figure}

\begin{figure}[!ht]
  \centering
     \includegraphics[scale=0.335]{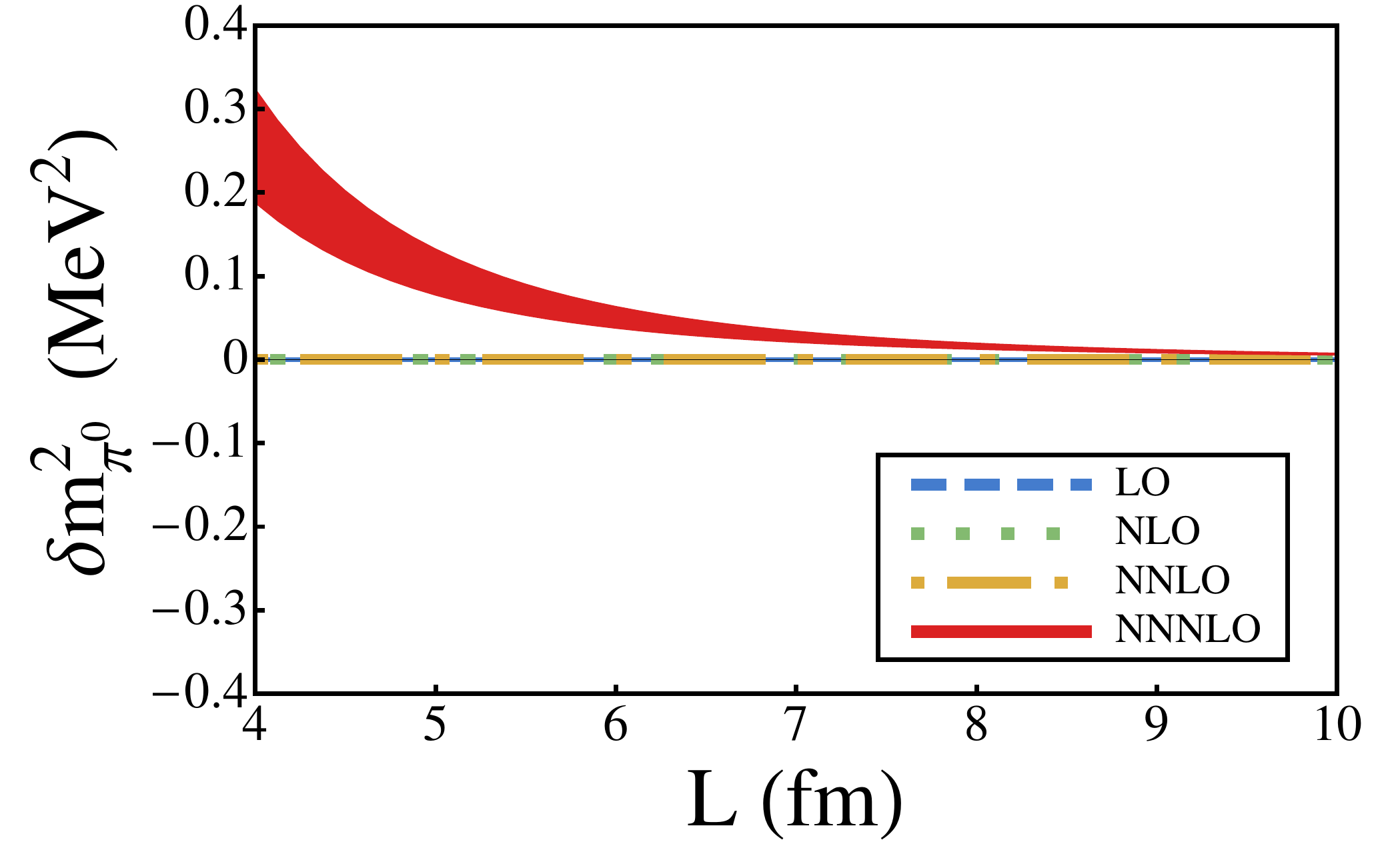}     \qquad
     \includegraphics[scale=0.335]{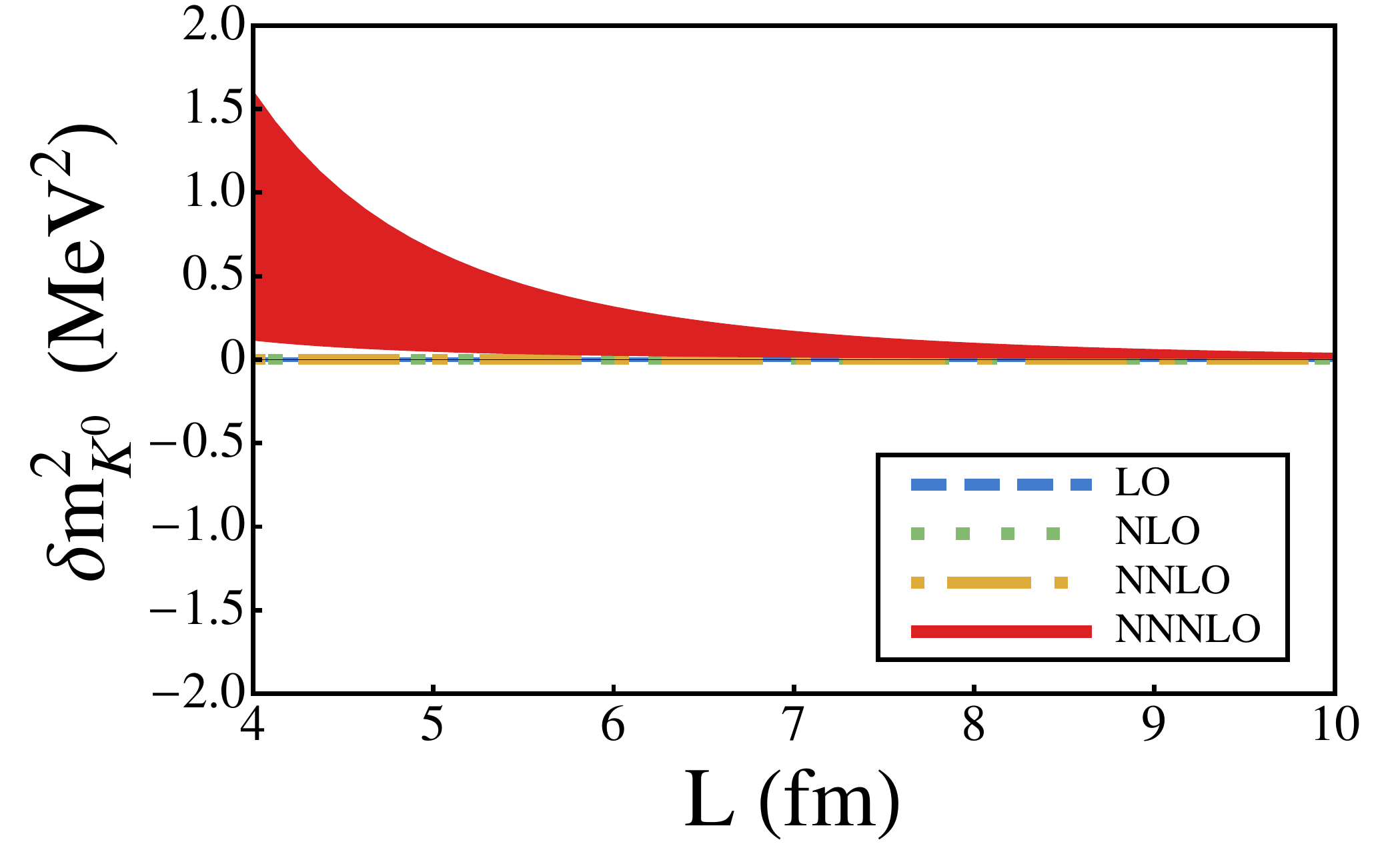}     
     \caption{{\small The FV QED correction to  the mass squared of a neutral pion (left panel) and kaon (right panel) at rest in a FV
     at the physical pion mass.
     The leading contributions are from their polarizabilities, and scale as $1/L^4$.
     The $1-\sigma$ uncertainty bands associated with each order in the expansion are determined from the uncertainties in the experimental and theoretical inputs.
    }}
  \label{fig:neutralkaonpionmsquare}
\end{figure}

Collecting  the contributions up to N$^3$LO,
the mass shift of a composite scalar particle in the $1/L$ expansion is
\begin{eqnarray}
\delta m_\phi & = & 
{\alpha_e Q^2\over 2 L} c_1 
\left( 1 + {2\over m_\phi L} \right)
+ {2\pi \alpha_e  Q\over 3 L^3}  \left(1+ {4\pi\over m_\phi L} c_{-1} \right) \langle r^2 \rangle_\phi 
-  {4\pi^2\over L^4}\ \left( \alpha_E^{(\phi)} + \beta_M^{(\phi)} \right) c_{-1}
.
\nonumber\\
\end{eqnarray}
Therefore, for the charged and neutral pions, the mass shifts are 
\begin{eqnarray}
\delta m_{\pi^+}& = & 
{\alpha_e \over 2 L} c_1 
\left( 1 + {2\over m_{\pi^+} L} \right)
+ {2\pi \alpha_e \over 3 L^3} \left(1+ {4\pi\over m_{\pi^+} L} c_{-1} \right) \langle r^2 \rangle_{\pi^+}
-  {4\pi^2\over L^4} \left( \alpha_E^{(\pi^+)} + \beta_M^{(\pi^+)} \right) c_{-1},
\nonumber\\
\delta m_{\pi^0}& = & 
\ -\  {4\pi^2\over L^4}\ \left( \alpha_E^{(\pi^0)} + \beta_M^{(\pi^0)} \right)\ c_{-1}
,
\end{eqnarray}
where  potential complications due to the electromagnetic decay of the $\pi^0$ via the anomaly  have been neglected .
The shifts of the charged and neutral kaons have the same form, with $m_{\pi^{\pm ,0}} \rightarrow m_{K^{\pm ,0}}$,
$ \langle r^2 \rangle_{\pi^+} \rightarrow \langle r^2 \rangle_{K^+}$, 
$\alpha_E^{(\pi^{\pm , 0})}\rightarrow \alpha_E^{(K^{\pm , 0})}$
and 
$\beta_E^{(\pi^{\pm , 0})}\rightarrow \beta_E^{(K^{\pm , 0})}$. With the experimental constraints on the charge radii 
and polarizabilities of the pions and kaons, 
numerical estimates of the FV corrections can be performed at N$^3$LO.
The LO and NLO contributions are dictated by  only  the charge and mass of the meson.
The N$^2$LO contribution depends upon the charge and charge radius, which, for the charged mesons, are known experimentally to 
be~\cite{Beringer:1900zz},
\begin{eqnarray}
\sqrt{ \langle r^2 \rangle}_{\pi^+} 
& = & 
0.672\pm 0.008~{\rm fm}
\ \ ,\ \ 
\sqrt{ \langle r^2 \rangle}_{K^+} 
\ = \ 
0.560\pm 0.031~{\rm fm}
.
\end{eqnarray}
The N$^3$LO contribution from the electric and magnetic polarizabilities of the mesons depends upon their sum.
The Baldin sum rule determines the charged pion combination, while the result of a two-loop $\chi$PT calculation is used for the neutral pion 
combination~\cite{Holstein:2013kia},
\begin{align}
& \alpha_E^{(\pi^+)} + \beta_M^{(\pi^+)} &  = & 
 \left(0.39\pm 0.04\right)\times 10^{-4}~{\rm fm}^3
 \ ,\ 
 \alpha_E^{(\pi^0)} + \beta_M^{(\pi^0)}  =  
 \left(1.1\pm 0.3\right)\times 10^{-4}~{\rm fm}^3
 .
\end{align}
Unfortunately, little is known about the polarizabilities of the kaons, and so 
naive dimensional analysis is used to provide an estimate 
of their contribution~\cite{Holstein:2013kia}, 
$ \alpha_E^{(K^+)} + \beta_M^{(K^+)}$, $ \alpha_E^{(K^0)} + \beta_M^{(K^0)} = \left(1\pm 1\right)\times 10^{-4}~{\rm fm}^3$.
With these values, along with their experimentally measured masses, 
the expected FV corrections to the charged meson masses are shown in 
Fig.~\ref{fig:chargedkaonpionmsquare} and to the neutral meson masses in 
Fig.~\ref{fig:neutralkaonpionmsquare}.~\footnote{
When comparing with  previous results one should note that the squared mass shift of the $\pi^+$, as an example, due to  
FV QED is 
\begin{eqnarray}
\delta m_{\pi^+}^2 
& = & 
\left(
m_{\pi^+} + \delta m_{\pi^+} 
\right)^2 - m_{\pi^+}^2 
\ =\ 2 m_{\pi^+}  \ \delta m_{\pi^+} \ +\ {\cal O}(\alpha_e^2)
,
\nonumber
\end{eqnarray}
As is evident, the leading contribution to the mass squared scales as $1/L$, 
contrary to a recent suggestion in the literature~\cite{Portelli:2012pn} of $1/L^2$. 
Note that the quantity shown in 
Fig.~\ref{fig:chargedkaonpionmsquare} and Fig.~\ref{fig:neutralkaonpionmsquare} 
is $\delta m_\phi^2$ as opposed to $\delta m_\phi$,
as it is this that enters into the determination of the light-quark masses from LQCD calculations.
}

In a volume with $L=4~{\rm fm}$, the FV QED mass shift of a charged meson  is approximately $0.5~{\rm MeV}$.
Figure~\ref{fig:chargedkaonpionmsquare} shows that for volumes with $L\gsim 4~{\rm fm}$,
the meson charge is responsible for essentially all of the FV modifications, with their compositeness 
making only a  small contribution, i.e. the differences between the NLO and N$^2$LO mass shifts are small.
For the neutral mesons, the  contribution from the polarizabilities is very small, but with substantial uncertainty.
It is worth re-emphasizing that in forming these estimates of the QED power-law corrections, 
 exponential corrections  of the form $e^{-m_\pi L}$ have been neglected.

\section{NRQED for the Baryons and $J={1\over 2}$ Nuclei}
\begin{figure}[!ht]
\begin{center}
\includegraphics[scale=0.175]{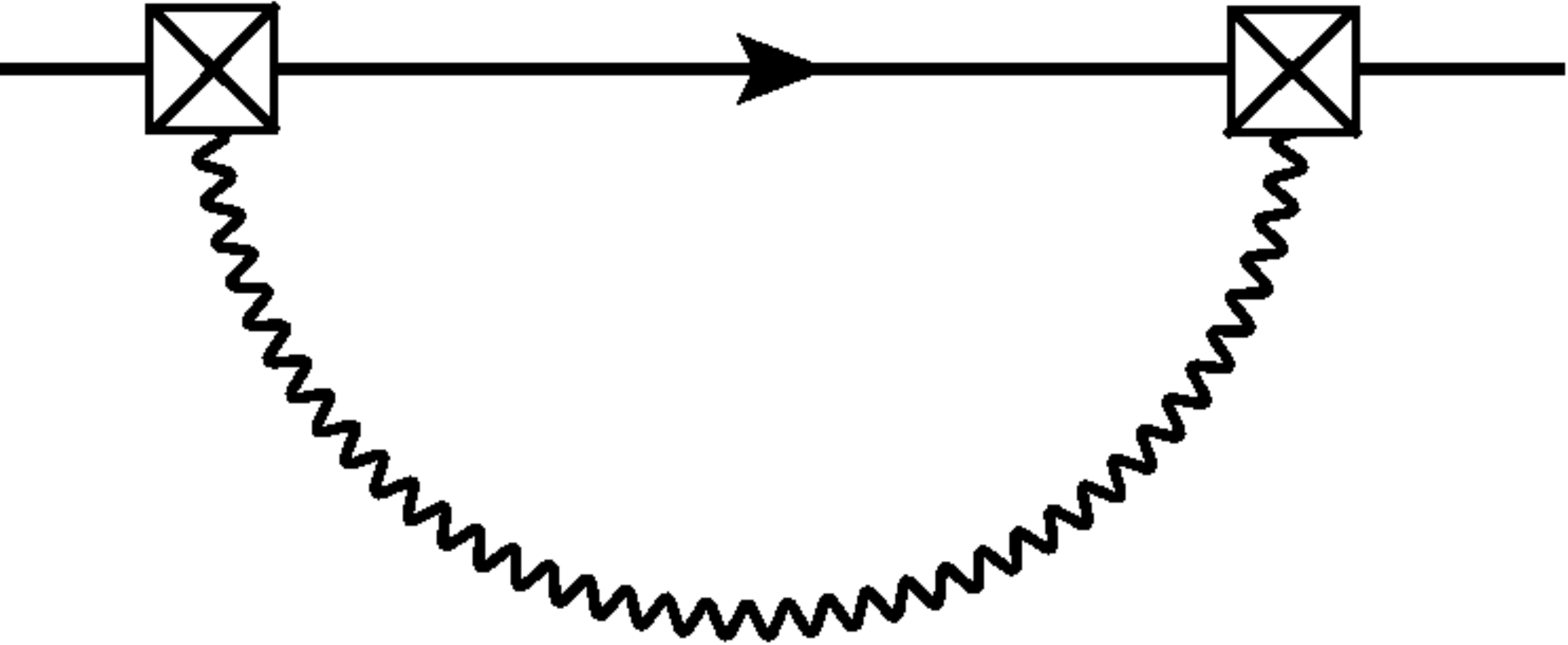}
\caption{{\small The N$^2$LO, ${\cal O}\left( \alpha_e/M_\psi^2 L^3 \right)$, 
 FV QED correction to the mass of a baryon from its magnetic moment. 
 The crossed square denotes an insertion of the magnetic moment operator given in Eq.~(\protect\ref{eq:baryonL}).}}
\label{fig:magmagloop-NNLO}
\end{center}
\end{figure}

\begin{figure}[!ht]
\begin{center}
\subfigure[]{
\includegraphics[scale=0.275]{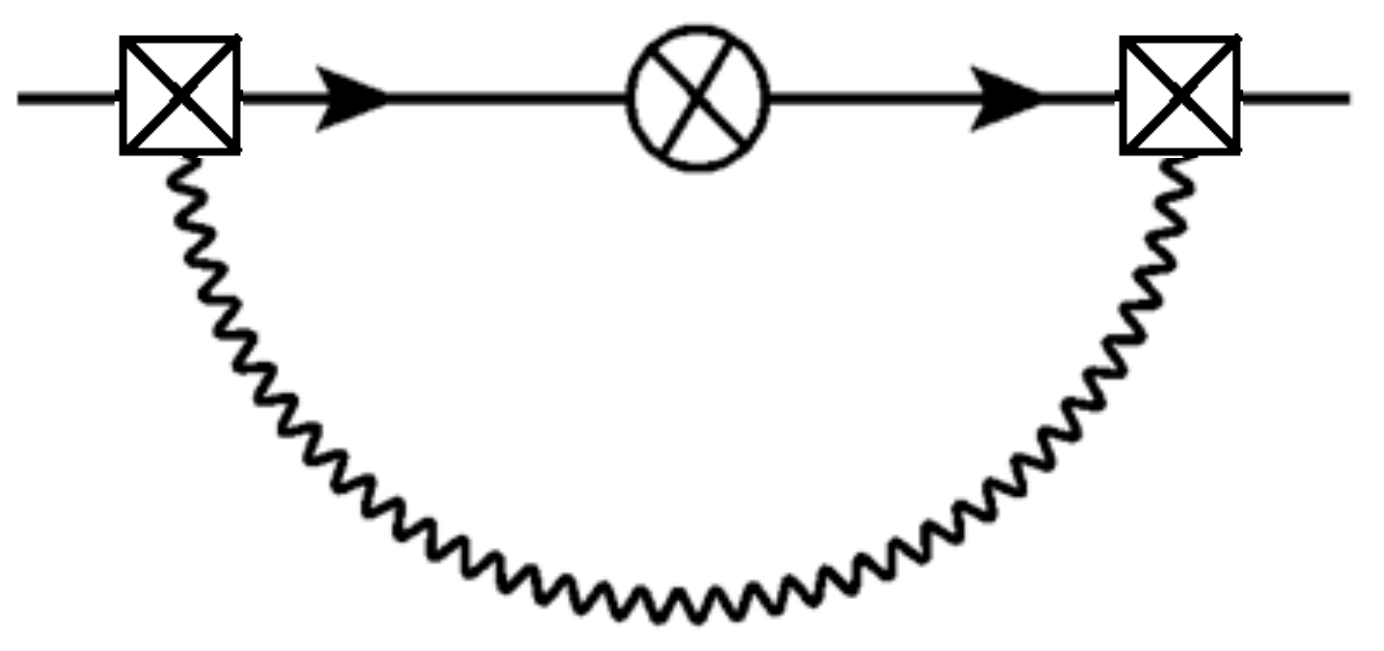}}
\subfigure[]{
\includegraphics[scale=0.31]{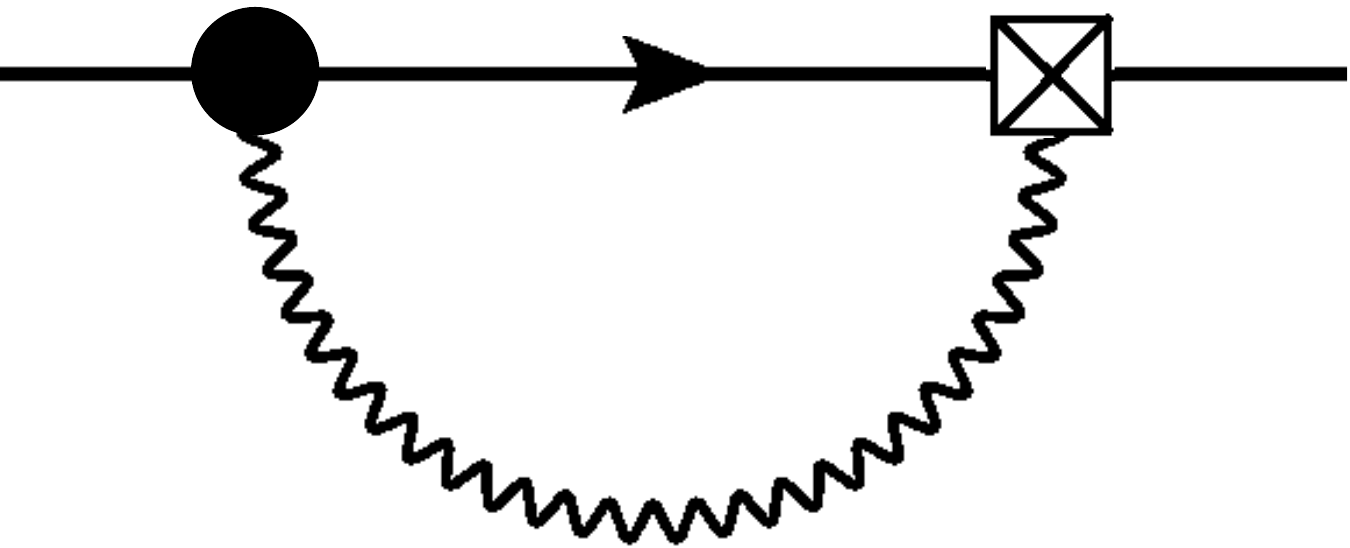}}
\subfigure[]{
\includegraphics[scale=0.31]{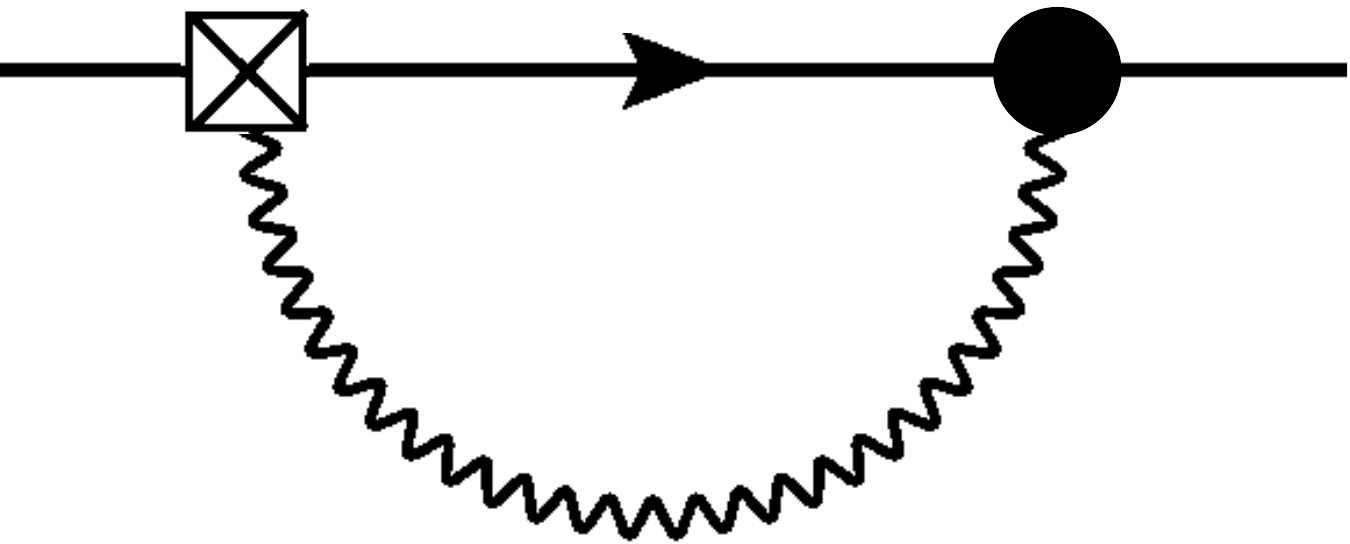}}
\caption{{\small  a) The N$^3$LO, ${\cal O}\left( \alpha_e/M_\psi^3 L^4 \right)$
 FV QED correction to the mass of a baryon from its magnetic moment. 
 The crossed square denotes an insertion of the magnetic moment operator given in Eq.~(\protect\ref{eq:baryonL}) while the crossed circle denotes an insertion from the $|{\bf D}|^2/2 m_\psi$ operator. b) Other non-vanishing contributions at this order arise from insertions of the $c_F$ and $c_S$ operators as given in Eq. (\protect\ref{eq:baryonL}). The black circles denote insertions of the $c_S$ operator.}}
\label{fig:magmagloop-NNNLO}
\end{center}
\end{figure}

\begin{figure}[t]
  \centering
     \includegraphics[scale=0.335]{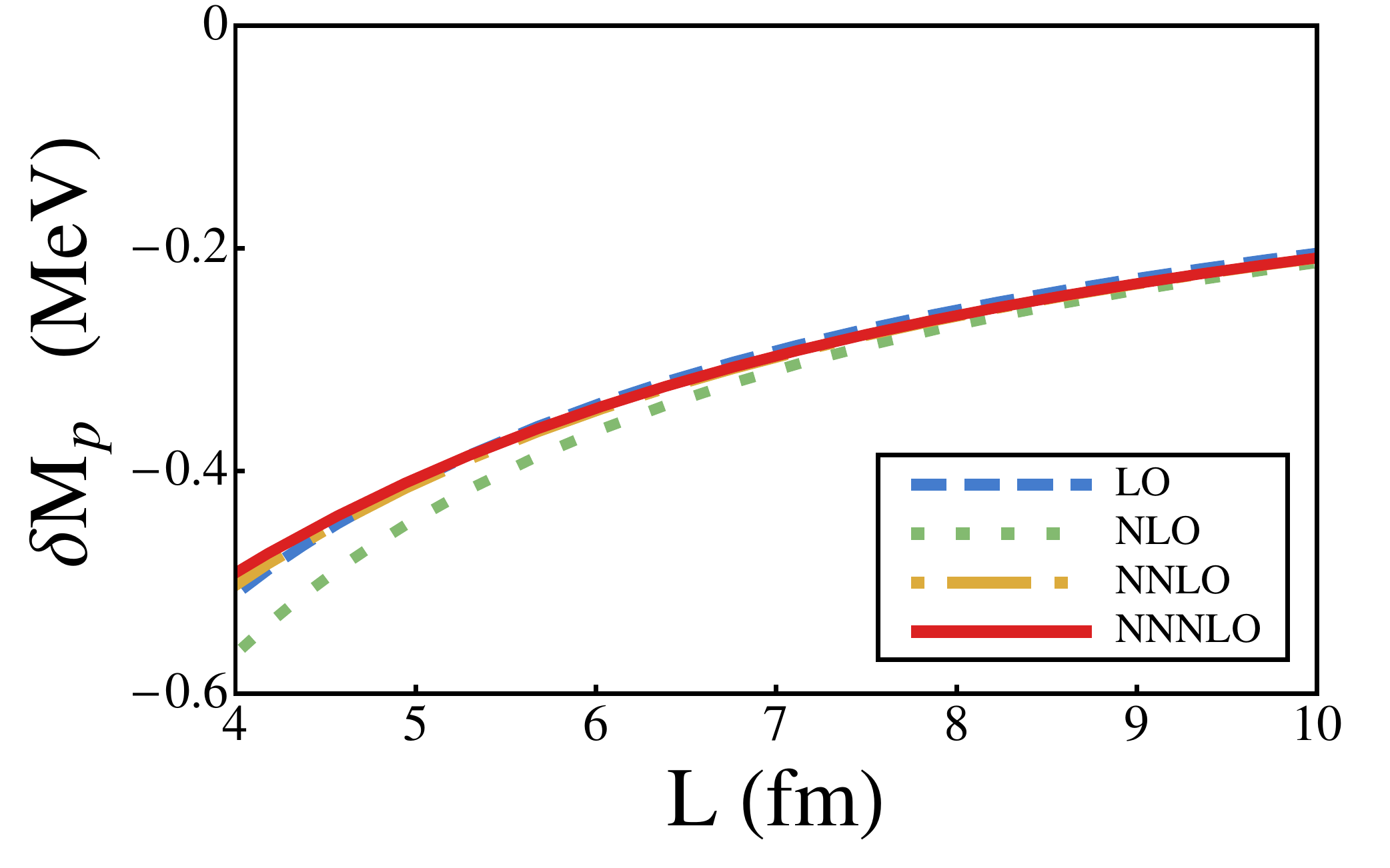}     \qquad
     \includegraphics[scale=0.335]{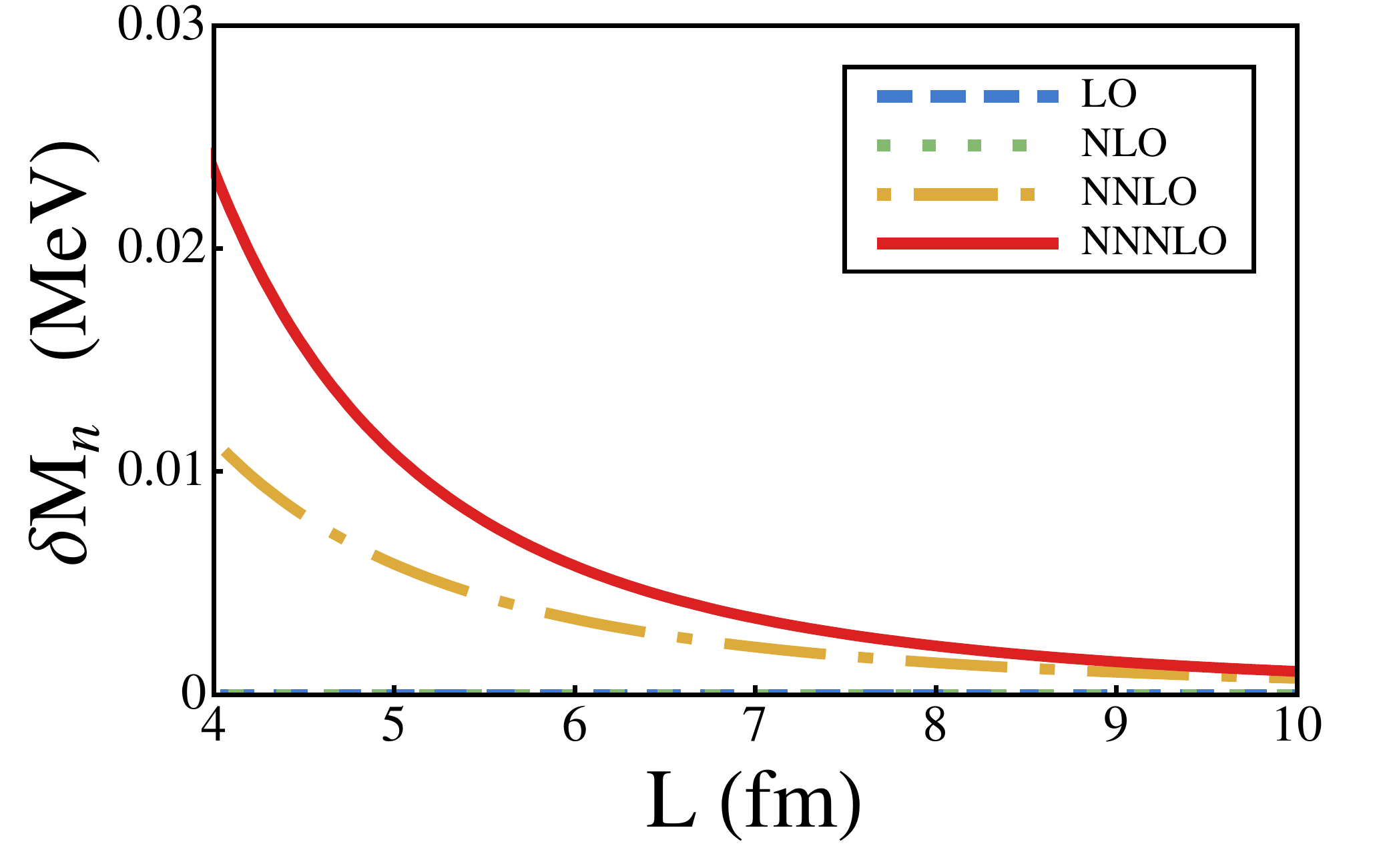}     
     \caption{{\small The FV QED correction to the mass of the proton (left panel) and neutron (right panel)
     at rest in a FV at the physical pion mass.
     The leading contribution to the proton mass shift is due to its electric charge, 
     and scales as $1/L$, while the leading contribution to the neutron mass shift is due to 
     its magnetic moment, and scales as $1/L^3$.
     The $1-\sigma$ uncertainty bands associated with each order in the expansion are determined from the uncertainties in the experimental and theoretical inputs.
    }}
  \label{fig:protonneutron}
\end{figure}

\noindent
In the case of baryons 
and $J={1\over 2}$ nuclei,
the method for determining the 
FV QED corrections is analogous to that 
for the mesons, described in the previous section, but modified to include the effects of spin and the 
reduction from a four-component  to a two-component spinor.
The low-energy EFT describing the interactions between the nucleons and the electromagnetic field is NRQED, but enhanced 
to include the compositeness of the nucleon.  
A nice review of NRQED, including the contributions from the non point-like structure of the nucleon, 
can be found in Ref.~\cite{Hill:2012rh}, and the relevant terms in the NRQED Lagrange density for 
a N$^3$LO calculation
are~\cite{Isgur:1989vq,Isgur:1989ed,Jenkins:1990jv,Jenkins:1991ne,Thacker:1990bm,Labelle:1992hd,Manohar:1997qy,Luke:1997ys,Chen:1999tn,Beane:2007es,Lee:2013lxa,Hill:2012rh}
\begin{eqnarray}
{\cal L}_\psi
& = & 
\psi^\dagger \left[
iD_0
\ +\ {|{\bf D}|^2\over 2 M_\psi}
\ + \ { |{\bf D}|^4 \over 8 M_\psi^3}\ 
\ +\ c_F {e\over 2M_\psi} {\bm\sigma}\cdot {\bf B}
\ +\  c_D {e\over 8 M_\psi^2} {\bm\nabla}\cdot {\bf E}
\right.\nonumber\\
&&\left. \qquad \qquad
\ +\ i c_S {e\over 8 M_\psi^2}\ {\bm\sigma}\cdot\left( {\bf D}\times {\bf E} -  {\bf E} \times {\bf D} \right)
\ +\ 2 \pi \tilde\alpha_E^{(\psi)}  |{\bf E}|^2
\ +\ 2\pi \tilde\beta_M^{(\psi)}  |{\bf B}|^2
\right.\nonumber\\
&&\left. \qquad \qquad
\ +\ e\ c_{W_1}\ {\{  {\bf D}^2 , {\bm\sigma}\cdot {\bf B} \} \over 8 M_\psi^3}
\ -\  e\ c_{W_2}\ { D^i {\bm\sigma}\cdot {\bf B} D^i\over 4 M_\psi^3}
\right.\nonumber\\
&&\left. \qquad \qquad
\ +\ e\ c_{p^\prime p}\ { {\bm\sigma}\cdot {\bf D} {\bf B}\cdot {\bf D} +  {\bf B}\cdot {\bf D} {\bm\sigma}\cdot {\bf D} \over  8 M_\psi^3}
\ +\ i e \ c_M\ { \{ D^i , ({\bm\nabla}\times {\bf B})^i \} \over  8 M_\psi^3}
\ +\  \cdots
 \right] \psi
,
\nonumber\\
\label{eq:baryonL}
\end{eqnarray}
where $c_F = Q + \kappa_\psi + {\cal O}(\alpha_e)$ is the coefficient of the magnetic-moment interaction, 
with $\kappa_\psi$ 
related to the anomalous magnetic moment of $\psi$,
$c_D =  Q + {4\over 3} M_\psi^2 \langle r^2\rangle_\psi + {\cal O}(\alpha_e)$ contains the leading 
charge-radius contribution, $c_S=2c_F-Q$ is the coefficient of the spin-orbit interaction and 
$c_M = (c_D-c_F)/2$.
The coefficients of the $|{\bf E}|^2$ and $|{\bf B}|^2$ terms contain the polarizabilities,
$1/M_\psi$
and $1/M_\psi^{3}$ corrections,
\begin{eqnarray}
\tilde\alpha_E^{(\psi)}  & = & 
\alpha_E^{(\psi)}  - {\alpha_e \over 4 M_\psi^3}\left(Q^2+\kappa_\psi^2\right) - {\alpha_e Q\over 3 M_\psi} \langle r^2\rangle_\psi
\ \ ,\ \ 
\tilde\beta_M^{(\psi)}  \ =  
\beta_M^{(\psi)}  +  {\alpha_e Q^2 \over 4 M_\psi^3}
.
\label{eq:pols}
\end{eqnarray}
The operators with coefficients $c_{W_1}$, $c_{W_2}$ and $c_{p^\prime p}$, 
given in Ref.~\cite{Hill:2012rh},
do not contribute to the FV corrections at this order.
The ellipses denote terms that are higher orders in $1/M_\psi$ and $1/\Lambda_\chi$.
Two insertions of the magnetic-moment operator provide its leading contribution, 
as shown in Fig.~\ref{fig:magmagloop-NNLO}, giving rise to $\mathcal{O}(\alpha_e/L^3)$ corrections to the mass of spin-$\frac{1}{2}$ particles. Although a single insertion of the $c_S$ operator seems to contribute at N$^2$LO, a straightforward calculation shows that this contribution is vanishing. At N$^3$LO, in addition to the operators contributing to the scalar case, one needs to take into account a diagram with two insertions of the magnetic-moment operator and one insertion of the $|{\bf D}|^2/2 m_\psi$ operator, plus diagrams with insertions of the $c_F$ and $c_S$ operators, as shown in Fig. \ref{fig:magmagloop-NNNLO}. 
Without replicating the detail presented in the previous section,
the sum of the contributions to the FV self-energy modification of a composite fermion, up to N$^3$LO,  is\footnote{As first noticed by the authors of Ref. \cite{Borsanyi:2014jba}, performing a non-relativistic expansion of the QED self-energy diagram for a point-like particle, although reproduces the result obtained via a NREFT at LO and NLO, naively turns out to be a factor of two bigger than the NNLO (and all higher orders) result presented in this chapter for both scalar and spinor QED. The source of discrepancy appears to be due to separating the range of (scalar) QED momentum summation to IR and UV modes where only in the IR part of the sum an expansion of the summand in $1/m$ is legitimate. The regulated UV sum must be evaluated as well keeping in mind that there exists an ambiguity in defining such a separation scale with $\Lambda \ll m$. The NREFT avoids such issues by appropriately incorporating all the UV contributions in a systematic expansion in the local operators with coefficients that are already matched to reproduce the (scalar) QED on-shell amplitudes. The LO and NLO contributions in the (scalar) QED calculation do not arise from any expansion in $1/m$ and as a result consistently reproduce our results.}
\begin{eqnarray}
\delta M_\psi && = 
{\alpha_e Q^2\over 2 L} c_1 
\left( 1 + {2\over M_\psi L} \right)
\ +\ {2\pi \alpha_e Q\over 3 L^3} \langle r^2 \rangle_\psi 
\ +\ {\pi \alpha_e\over M_\psi^2 L^3}\ \left[\ {1\over 2} Q^2\ + \ (Q+\kappa_{\psi})^2 \right]
\nonumber\\
&& -  
{4\pi^2\over L^4} \left( \tilde\alpha_E^{(\psi)} + \tilde\beta_M^{(\psi)} \right) c_{-1}
+  {\pi^2 \alpha_e Q  \over M_\psi^3 L^4}\ \left( {4\over 3} M_\psi^2  \langle r^2 \rangle_\psi  - \kappa_\psi \right) c_{-1}-\frac{\alpha_e \pi^2}{M_{\psi}^3 L^4}\kappa_{\psi}(Q+\kappa_{\psi})c_{-1}.
\nonumber
\\
\end{eqnarray}
Therefore, for the proton and neutron, the FV QED mass shifts are
\begin{eqnarray}
\delta M_p & = & 
{\alpha_e \over 2 L} c_1 
\left( 1 + {2\over M_p L} \right)
+ {2 \pi \alpha_e  \over 3 L^3} \left( 1 + {4\pi\over M_p L} c_{-1} \right) \langle r^2 \rangle_p
+ {\pi \alpha_e\over M_p^2 L^3}\ \left( {1\over 2} + (1+\kappa_p)^2 \right)
\nonumber\\
& &
-  {4\pi^2\over L^4}\left( \alpha^{(p)}_E + \beta^{(p)}_M \right) c_{-1}
\ -\  {2\pi^2 \alpha_e  \kappa_p \over  M_p^3 L^4}\  c_{-1},
\nonumber\\
\delta M_n & = & 
\kappa_n^2\ {\pi \alpha_e \over  M_n^2 L^3}
\ -\  {4\pi^2\over L^4} \left( \alpha^{(n)}_E + \beta^{(n)}_M \right) c_{-1},
\end{eqnarray}
where the anomalous magnetic moments of the proton and neutron give
$\kappa_p = 1.792847356(23)$ and 
$\kappa_n = -1.9130427(5) M_n/M_p $, respectively~\cite{Beringer:1900zz}.
One of the N$^2$LO contributions to the proton  FV QED correction depends upon its charge radius, which is known experimentally to 
be,
$\langle r^2 \rangle_{p}  = 0.768\pm 0.012~{\rm fm}^2 $~\cite{Beringer:1900zz}.
Further, part of the N$^3$LO contribution depends upon the electric and magnetic polarizabilities, 
which are constrained by
the Baldin sum rule,~\cite{Holstein:2013kia}
\begin{align}
 &\alpha_E^{(p)} + \beta_M^{(p)} &  = & 
 \left(13.69\pm 0.14\right)\times 10^{-4}~{\rm fm}^3
 \ ,\ 
 \alpha_E^{(n)} + \beta_M^{(n)}  =  
 \left(15.2\pm 0.5\right)\times 10^{-4}~{\rm fm}^3
 .
\end{align}
With these values for the properties of the proton and neutron, along with their experimentally measured masses, 
the expected FV modifications to their masses are shown in Fig.~\ref{fig:protonneutron}.

The   proton FV QED corrections are consistent with those of the charged scalar mesons.
However, the neutron corrections, while very small, of the order of a few keVs, exhibit more structure.  
The N$^2$LO contribution from the magnetic moment increases the mass in  FV, scaling as $1/M_n^2 L^3$, similar to the 
polarizabilities which make a positive contribution and scale as $1/L^4$ (N$^3$LO).  
Note that the polarizabilities of the nucleon are dominated by the
response of the pion cloud, while the magnetic moments are dominated by physics at the 
chiral symmetry breaking scale.  
Further the magnetic-moment contributions are suppressed by two powers of the nucleon mass.

There is an interesting difference between the meson and baryon FV modifications.  
As the nucleon mass is approximately seven times the pion mass, and twice the kaon mass, the recoil corrections 
are suppressed compared with those of the mesons. 
Further, the nucleons are significantly ``softer'' than the mesons, as evidenced by their polarizabilities.
However, the NLO recoil corrections to the proton mass 
are of approximately the same size as the N$^2$LO structure contributions, as seen in 
Fig.~\ref{fig:protonneutron}.

\section{Nuclei: Deuteron and Helium 4}
\noindent
As mentioned in chapter \ref{chap:intro}, a small number of LQCD collaborations have been calculating the binding of light nuclei and 
hypernuclei at unphysical light-quark masses in the isospin limit  and without 
QED~ \cite{Beane:2009py,Yamazaki:2009ua,Beane:2010hg,Inoue:2010es,Inoue:2011pg,Beane:2011iw,Yamazaki:2011nd,Yamazaki:2012hi,Yamazaki:2012fn,Beane:2012vq}.
However, it is known that 
as the atomic number of a nucleus increases, 
the Coulomb energy increases with the square of its charge, and   significantly reduces the 
binding of  large  nuclei.

\begin{figure}[!tt]
  \centering
     \includegraphics[scale=0.35]{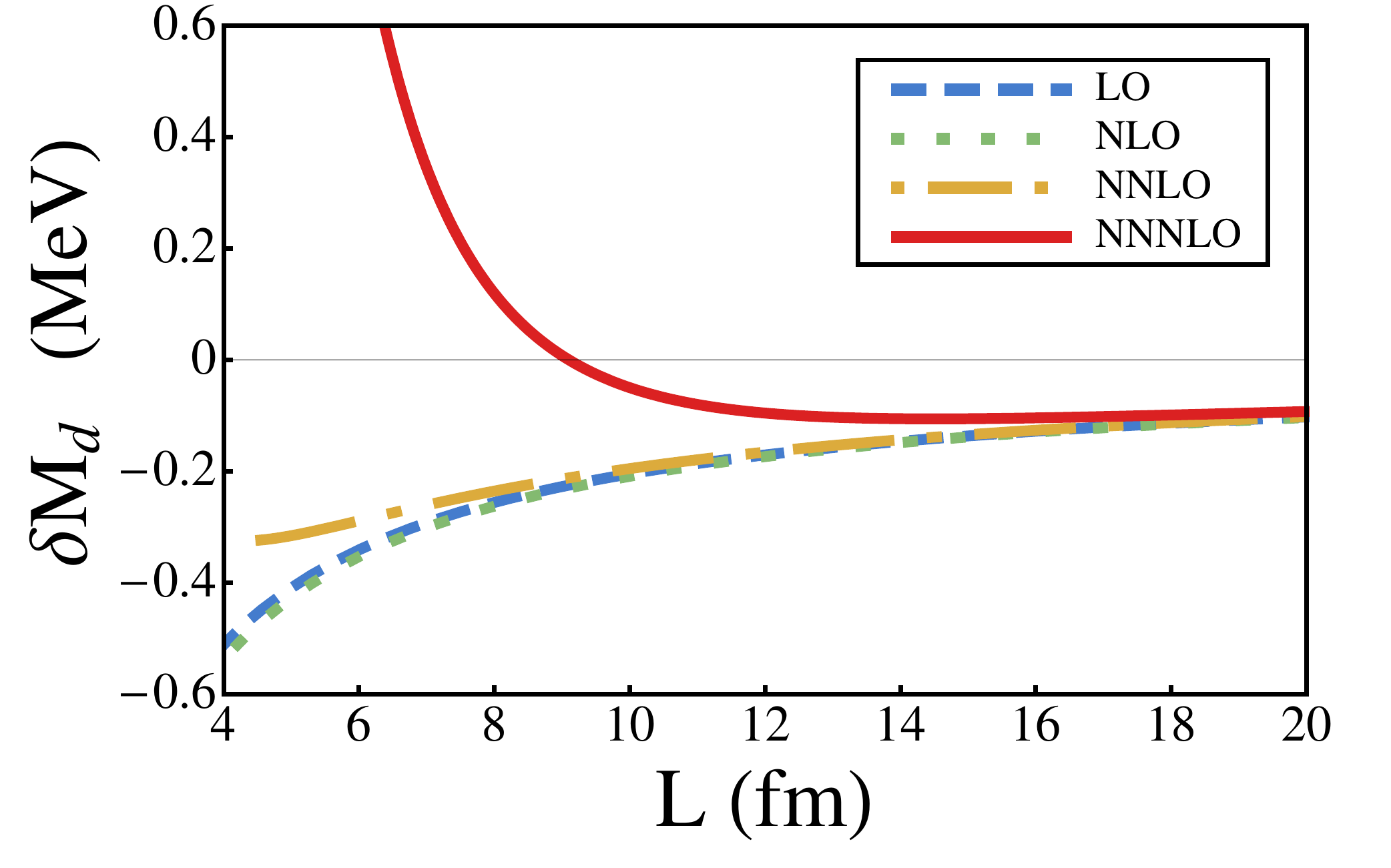}   \ \  
     \includegraphics[scale=0.35]{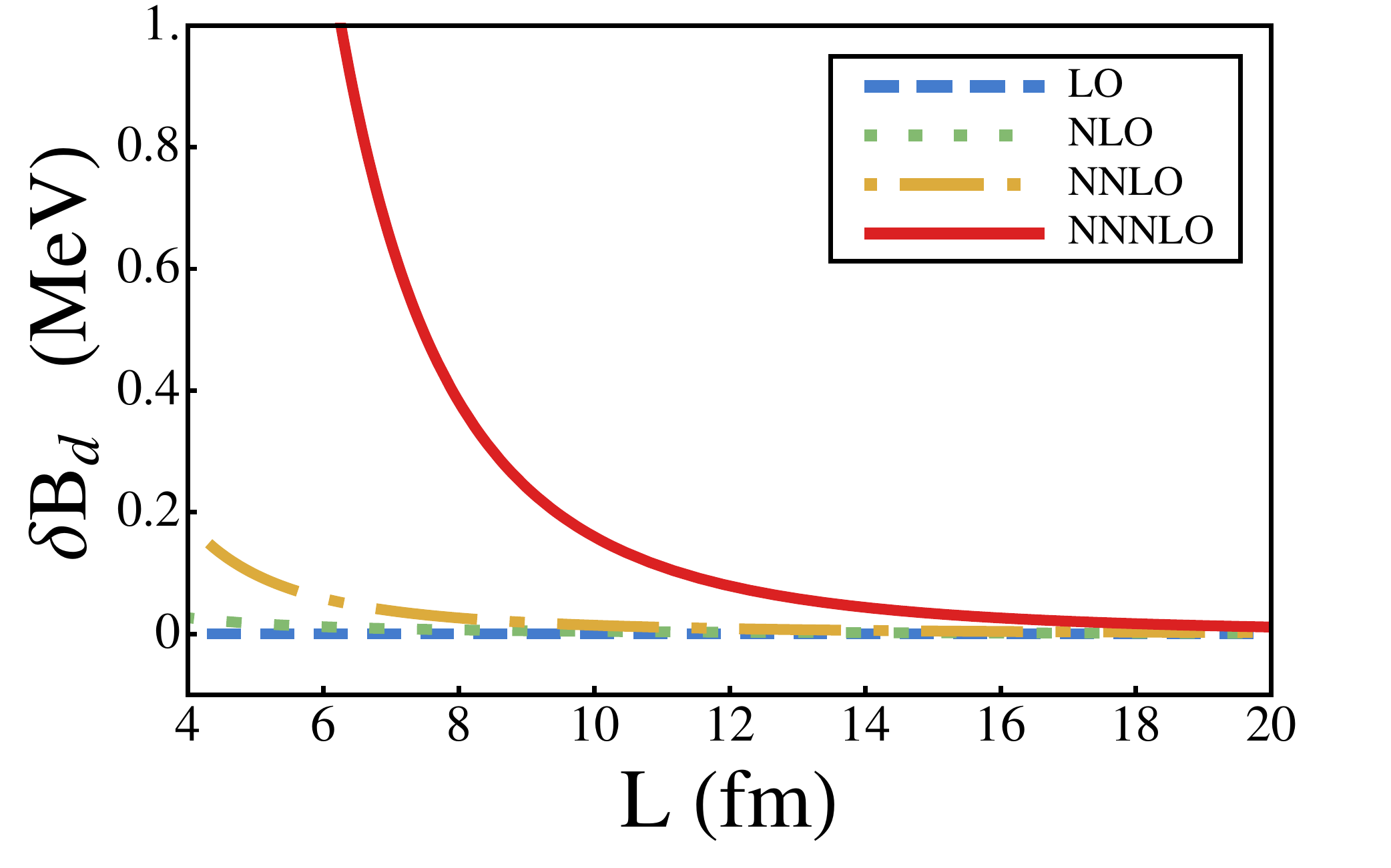}    
     \caption{{\small The left panel shows the FV QED correction to the mass  of the deuteron at rest  in a FV at the physical pion mass.
     The leading contribution is from its electric charges, and scales as $1/L$.
     The right panel shows the FV QED correction to the deuteron binding energy for which the  $1/L$ contributions cancel.
     The $1-\sigma$ uncertainty bands associated with each order in the expansion are determined from the uncertainties in the experimental and theoretical inputs.
         }}
  \label{fig:deut}
\end{figure}

A NREFT for  vector QED shares the features of the NREFTs for scalars and fermions that are relevant for the current analysis. 
One difference is in the magnetic moment contribution,  and another is the contribution from the quadrupole interaction.
The FV corrections to the deuteron mass and binding energy, $\delta {\rm B}_d$, are shown in Fig.~\ref{fig:deut}, where the
experimentally determined charge radius, magnetic moment and polarizabilities have been used.
Due to the large size of the deuteron, and its large polarizability, the $1/L$ expansion converges slowly in  modest volumes, and it
appears that $L\gsim 12~{\rm fm}$ is required for a reliable determination of  the QED FV effects. The QED FV corrections to the deuteron binding energy are seen to be significantly smaller than its total energy in large volumes, 
largely because the leading contribution to the deuteron and to the proton cancel.
Deuteron also possesses a quadrupole moment which contributes to  the FV QED effects 
at ${\cal O}\left(1/L^5\right)$ through two insertions.

\begin{figure}[!ht]
  \centering
     \includegraphics[scale=0.335]{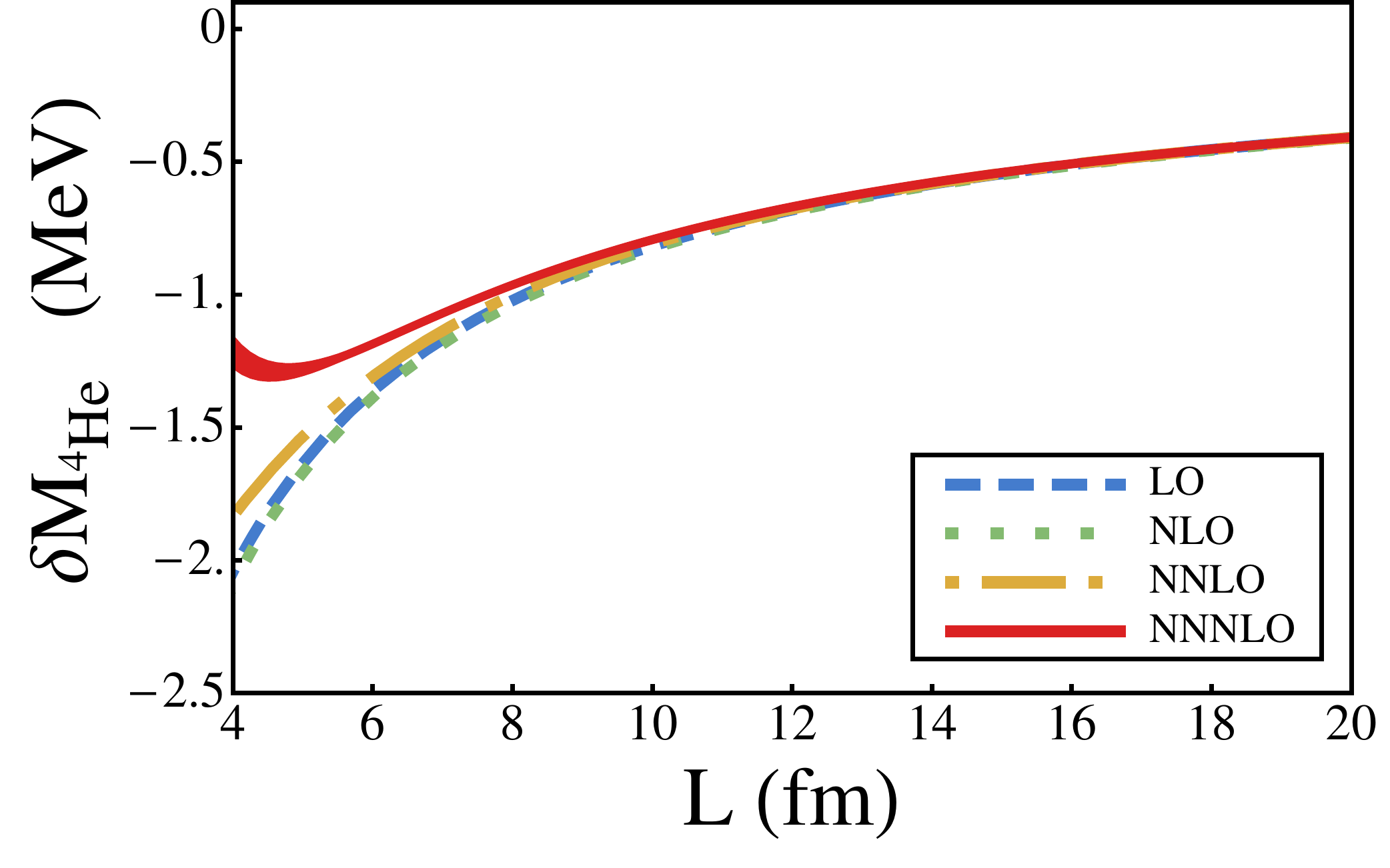}     \ \ 
     \includegraphics[scale=0.335]{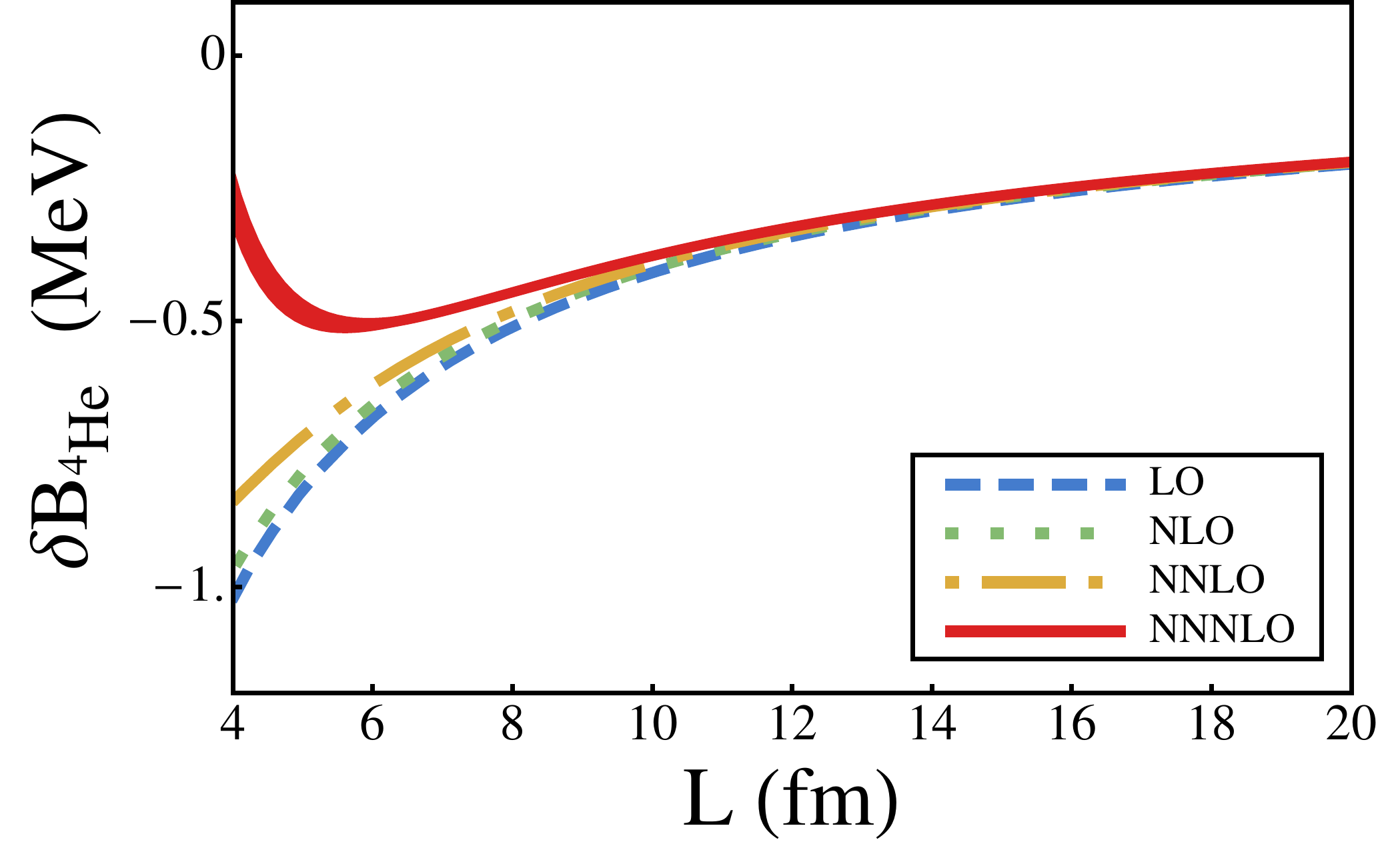}    
     \caption{{\small The left panel shows the FV QED correction to the mass  of  $^4$He at rest in a FV
     at the physical pion  mass.
     The leading contribution is from its electric charge, and scales as $1/L$.
     The right panel shows the FV QED correction to  the  $^4$He binding energy.
     The uncertainty bands associated with each order in the expansion are determined from the uncertainties in the experimental and theoretical inputs.
      }}
  \label{fig:he4}
\end{figure}

The NREFTs used to study the FV contributions to the mass of the pions in the previous section also apply to the $^4$He nucleus,
and the FV corrections to  the mass of $^4$He and its binding energy, $\delta {\rm B}_{^4{\rm He}}$,  are shown in Fig.~\ref{fig:he4}.
Unlike the deuteron, the leading FV corrections to  $^4$He
do not cancel in the binding energy due to the interactions between the two protons, but are reduced by a factor of two.
%

\section*{Summary and discussions}
\noindent
For Lattice QCD calculations performed in 
volumes that are much larger than the inverse pion mass, 
the finite-volume electromagnetic corrections to hadron masses
 can be calculated systematically using a NREFT.
The leading two orders in the $1/L$ expansion for mesons
have been previously calculated using chiral perturbation theory, and depend only upon their electric charge and  mass.  
We have shown that these two orders are universal FV QED corrections to the mass of charged particles.
Higher orders in the expansion are determined by 
recoil corrections and by the structure of the hadron, 
such as its electromagnetic multipole moments and polarizabilities, which we calculate using a NREFT.  
One advantage enjoyed by the NREFT  
is that the coefficients of the operators in the Lagrange density are directly related to the structure of the hadron, 
order by order in $\alpha_e$,
as opposed to being perturbative approximations as computed, for instance, in $\chi$PT.
For the mesons and baryons, the FV QED effects associated with their structure, beyond their charge, 
are  found to be small even in modest  lattice volumes.   
For nuclei, as long as the volume is large enough so that the non-QED effects are exponentially small, 
dictated by the nuclear radius,
their charge dominates the FV QED corrections, with only small modifications due to the structure of the nucleus. 

The results that we have presented in the previous sections have 
assumed a continuous spacetime, and have not yet considered the impact of a finite lattice spacing.
With the inclusion of QED, there are two distinct sources of lattice spacing artifacts that will modify the FV QED  corrections we have considered.
The coefficients of each of the higher dimension operators in the NREFTs will receive lattice spacing corrections, and 
for an ${\cal O}\left(a\right)$-improved action ($a$ is the lattice spacing) 
they are a polynomial in powers of $a$ of the form 
$d_i\sim d_{i0} +  d_{i2} a^2 + d_{i3} a^3 + ...$.
The coefficients $d_{ij}$ are determined by the strong interaction dynamics and the particular discretizations used in a given calculation.
In addition, the electromagnetic interaction will be modified in analogy with the strong sector, giving rise to further lattice spacing artifacts in the matching conditions between the full and the NR theories, and also in the value of one-loop diagrams.\footnote{The lattice artifacts will depend upon whether the compact or non-compact formulation of QED is employed - the former inducing
non-linearities in the electromagnetic field which vanish in the continuum limit.
The discussions we present in this section apply to both the compact and non-compact formulations.
}. For an improved action, the naive expectation is that such correction will first appear,
beyond the trivial correction from the modified hadron mass in the NLO term,
at
$\alpha_e a^2/L^3$ in the $1/L$ expansion.
They are a N$^2$LO contribution arising from
modifications to the one-loop  Coulomb self-energy diagram.
This is the same order as  contributions from the charge radius, recoil corrections and the magnetic moment, which are found to make a small contribution
to the mass shift in modest lattice volumes.  
As the lattice spacing is small compared to the size of the proton and the inverse mass of the proton, 
these lattice artifacts are expected to provide  a small modification to the N$^2$LO terms we have determined.
In addition, there are operators in the Symanzik action~\cite{Symanzik:1983dc,Symanzik:1983gh,Parisi:1985iv} 
that violate Lorentz symmetry as the calculations are performed on an underlying hypercubic grid.
Such operators require the contraction of at least four Lorentz vectors in order to form a hypercubically-invariant, but Lorentz-violating, operator,
for instance three derivatives and one electromagnetic field, or four derivatives. 
We have discussed the suppression of Lorentz-violating contributions at small lattice spacings, along with smearing, 
in chapter \ref{chap:operators}.

A second artifact  arises from the lattice volume.
The NREFTs are constructed as an expansion in derivatives acting on fields near their classical trajectory.  
As emphasized by Tiburzi~\cite{Tiburzi:2007ep} and others, this leads to modifications in calculated 
matrix elements because derivatives are approximated by finite differences in lattice calculations.  
For large momenta, this is a small effect because of the large density of states, but at low momenta,
particularly near zero, this can be a non-negligible effect that must be accounted for.  
This leads to a complication in determining, for instance,  magnetic moments from the forward limit of a form factor, relevant to the
discussion in the previous section.
However, this does not impact  the present calculations of FV QED corrections to the masses of the mesons, baryons and nuclei.


\chapter{CONCLUSION AND OUTLOOK}
{\label{chap:conclusion}}
Lattice quantum chromodynamics (LQCD) will soon become the primary method in rigorous studies of single- and multi-hadron sectors of QCD. It is truly \emph{ab initio} meaning that its only parameters are those of standard model; the theory that is confirmed to be the true underlying theory of particles and interactions at short distances (below the energy scale where any new physics arises). Progress in LQCD calculations of hadronic spectrum and structure, as well as hadronic interactions and resonances have been significant in recent years. As the computational resources become available in the upcoming years, the systematic uncertainties of these calculations that are associated with discretization and finite volume of spacetime, as well as the input of unphysical light-quark masses will be reduced/eliminated. The topics presented in this thesis include several improvements in our formal understanding of some of these uncertainties, but most importantly they provide new proposals for accessing physical quantities, e.g. those of two-nucleon systems such as scattering parameters, that otherwise would not have been directly accessible through numerical lattice QCD calculations of hadronic (Euclidean) correlation functions. Here we summarize the major lessons to be learned from the findings of the chapters presented and refer the reader to the the summary sections of the corresponding chapters for details. 

\begin{itemize}

\item
The singularities observed in taking the continuum limit of correlations functions of higher spin/twist operators is an artifact of probing the short-distance physics of a process whose typical scale extends over a physical region rather than just a lattice site. This is analogue of the idea that physical theories are effective description of physics at a given energy scale where the effect of short distance physics and its underlying symmetries become irrelevant with probes that are smeared over the corresponding distance of the low-energy scale. We have explicitly shown how such mechanism works, through examples in scalar and gauge field theories, where by increasing the resolution of the calculation but keeping fixed the physical region which the operators can probe, the continuum limit of operators with well-defined angular momentum are smoothly approached. Interestingly the coefficients of the``wrong'' angular momentum operators or those of the rotational breaking operators scale at worse with $a^2$, even at presence of quantum effects, consistent with the solid-angle resolution of the physical region that the operator is smeared over. Although already confirmed numerically, our analytical study provides a rigorous understanding of such operator improvement scenarios, and ensures that there is formally no issue with the restoration of rotational symmetry in the continuum limit of lattice field theories.

\item
Larger volumes are not necessarily what LQCD calculations should always ultimately aim for. As realized by Maiani and Testa and signified by L\"uscher's pioneering works in the case of scattering amplitudes, the finite volume of the system of interacting particles is what saves us from running into inaccessibility of amplitudes from the Euclidean correlation functions when the infinite-volume limit is taken. The D/S ratio of the deuteron is a prominent example where, upon using the proper finite-volume (FV) formalism for spin systems and examining the FV symmetries of moving frame calculations, the effect of such a tiny quantity can be artificially enhanced in a FV calculation of the energy levels, giving presumably the only plausible path to a precise first-principle determination of this quantity from QCD. By aiming only for large volumes, such opportunities, in particular with regard to scattering states, are lost. The other side of the story is that by tuning the volumes in the moderate range, the breakdown of rotational symmetry introduces non-negligible systematics in the calculations that must be identified, and whose size must be well quantified. We investigated an example of this again for the deuteron system, where by analyzing the expected energy levels in the upcoming LQCD calculations, we show that the FV-induced partial-wave mixing with higher J-states becomes significant in small to moderate volumes. However a knowledge of the underlying formalism of such mixings helps the lattice practitioner to identify these errors and correct for them systematically.

\item
For the case of masses and binding energies, for which the calculated quantities on the lattice can be directly interpreted as the infinite-volume value through a straightforward extrapolation in volume, novel ideas can result in major volume improvements. This can save significant computational resources by eliminating the need to enlarge the volumes in generating the gauge-field configurations. We have explored this idea through modifying the (valence) quark boundary conditions, and have proved that an order of magnitude volume improvement in the upcoming extracted binding energies of two-nucleon systems is expected by an optimized choice of (twisted) boundary conditions. This is an encouraging result and urges the exploration of similar ideas for other important quantities -- those that are already under investigations by LQCD, such as nucleon structure quantities, e.g., $g_A$, or  the binding energies of multi-nucleon systems.

\item
FV effects are the consequence of the manipulation of the system in the IR and therefore the details of the short-range theory used to describe the interactions within the finite volume is irrelevant. As long as the interactions are consistent with the symmetries of the system and are not infinite in range, a low-energy effective theory should suffice to correctly identify the volume effects. We have taken this idea, along with the elimination of the zero mode of the photon in a FV calculation, to estimate volume corrections expected to arise in the masses of hadrons due to quantum electrodynamics (QED) interactions. At each order in the QED coupling and the $1/L$ expansion ($L$ being the spatial extent of the volume), the coefficients of the correction terms are exact to all order in strong interaction effects as they are matched directly to the experimental values of the mass, charge and electromagnetic (EM) radii and moments of particles. The generality of this approach has enabled us to extend it readily to the case of nuclei with spin $0$, $1/2$ and $1$, that is going to be useful in extrapolating extracted masses and binding energies of these systems to the infinite-volume limit once QED interactions are included in these calculations. Nonetheless, quantification of errors associated with the long range of photons has not yet been properly done in most of EM processes in a finite volume. An important example is the LQCD+QED calculation of the hadronic contributions to the muon $g-2$ where an understanding of the expected form of volume corrections is crucial in estimating the uncertainties of the current calculations.

\end{itemize}

The topics presented in this thesis mainly concerns the formal developments in the single- and two-nucleon sectors. We conclude this thesis by briefly discussing the path forward with regard to multi-nucleon(hadron) calculations. There are reasons to speculate that LQCD calculations will only slowly go beyond a direct calculation of the properties of the few-nucleon systems, given the scale of the computational resources available at present and in near future. We should note however that significant efforts are underway on both theoretical and computational sides to improve the situation. These reasons can be put into two categories:

\begin{enumerate}
\item
\emph{Computational issues due to the poor signal to noise ratio}: The poor signal/noise (sign) problem that is inherent in performing LQCD calculation with finite baryon density (chemical potential) continues to pose a challenge, despite much activities at elucidating this problem \cite{Lepage89, MJSsign, Lee:2011sm, Endres:2011jm, Endres:2011mm, Grabowska:2012ik, Detmold:2014hla}. In particular, this has been the main reason why the binding energies of light nuclei have only been calculated via LQCD for pion masses as heavy as $500~{\rm MeV}$ and for systems with no more than four (hyper)nucleons \cite{Beane:2012vq, Yamazaki:2009ua}. Novel theoretical ideas and/or significantly larger computational resources can change the situation. This is going to be an active field of research and investigation in the upcoming years among experts in both nuclear physics and lattice QCD.

\item
\emph{Lack of a finite-volume formalism for multi-particle systems}: Despite the FV formalism (L\"uscher method) for two-hadron sector, the FV formalism for few-body systems either does not yet exist or has not reached the same level of maturity as the two-body case. The first necessary step in being able to properly understand few-body systems would require having a non-perturbative, model-independent framework for the \emph{three-body} sector. There has been much activity in deriving a formal result for the energy quantization conditions of the systems of three bosons in relation to scattering amplitudes \cite{Roca:2012rx,  Polejaeva:2012ut,Briceno:2012rv, Hansen:2013dla}, however non of these has yet been implemented in practice. This is due to the non-algebraic form of these quantization conditions -- in contrast to the L\"uscher two-body formula -- which makes the practicality of these formalisms less promising. Having such formalism however seems crucial as otherwise not much information can be gained about multi-nucleon interactions -- no matter how precisely the energy levels are obtained in future LQCD calculations, see Fig. \ref{fig:Multi-NPLQCD}.

\begin{figure}[h!]
\begin{centering}
\includegraphics[scale=0.45]{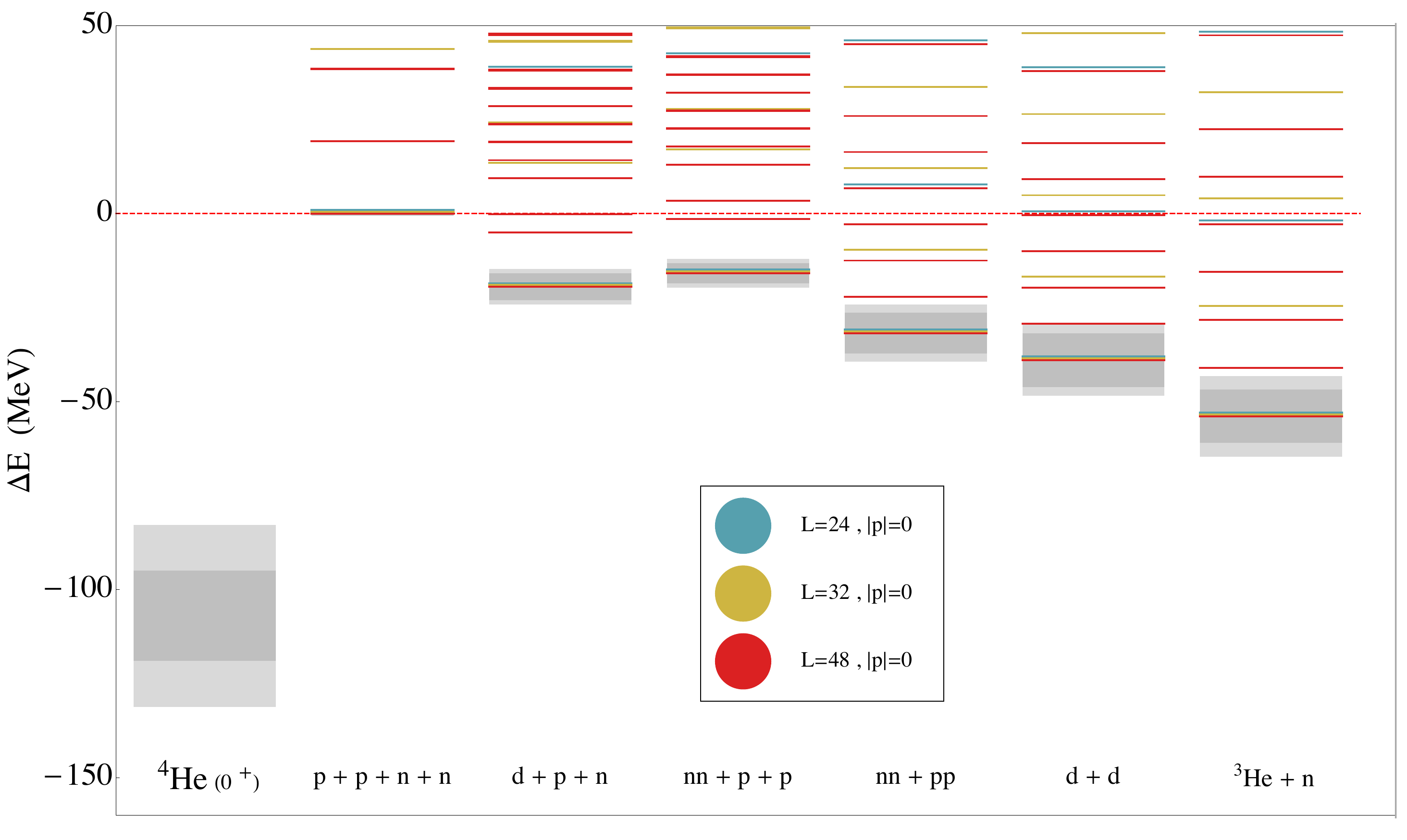}
\caption{{\small The location of energy levels of the four-nucleon sector with quantum numbers of $^4He$ can be estimated based on the non-interacting energy levels of $1+1+1+1$, $2+1+1$, $2+2$ and $3+1$ systems. Different colors denote different lattice volumes corresponding to spatial extents of $3.4~{\rm fm}$ (green), $4.5~{\rm fm}$ (orange) and $6.7~{\rm fm}$ (red). The total momentum of the system of particles is zero. The calculated binding energy of $^4He$ at the pion mass of $\approx 800~{\rm MeV}$ is shown \cite{Beane:2012vq}. Even if the computational challenges associated with resolving the dense excited states are overcome, there exists no FV formalism at the moment to extract the corresponding scattering parameters in these channels. Figure is reproduced with the permission of the NPLQCD collaboration.
}}
\label{fig:Multi-NPLQCD}
\par\end{centering}
\end{figure}
\end{enumerate}

It seems that the optimal way to proceed is to take advantage of effective field theories (EFTs) describing few-nucleon systems at low energies. As we already mentioned, experimental data can be used to constrain the value of low-energy constants (LECs) of an EFT.\footnote{EFTs such as $\chi$PT have already been used extensively to interpolate LQCD calculations to the physical pion mass and to get a systematic control over the size of the volume corrections in several quantities such as masses and magnetic moments, see chapter \ref{chap:intro}.} However, there are cases where there is little or no experimental data available to constrain EFTs. In the case of $\pi\pi$ scattering mentioned in chapter \ref{chap:intro}, these LECs were known from the wealth of experimental data, but, for example, in the case of nucleon-hyperon (nucleons with one or more u and d quarks replaced by the strange quark) interactions, these LECs are not known \cite{Beane:2012ey}. In other words, the result of LQCD calculations of few-body systems complement the experimental input into EFT calculations.  This direction most likely to be the avenue for implementing the results of first principle LQCD calculations to make systematic EFT-based predictions for multi-hadronic systems in the upcoming years (see a  review of this topic in Ref. \cite{Briceno:2014}).

It is notable that by matching to EFTs, the seemingly complicated L\"uscher-type  formalism of the multi-particle systems -- by which the scattering parameters are directly evaluated instead of EFT interaction kernels --  will not be essential. The limitation of this method is in lowering the range of validity of the FV formalism to the range of validity of EFTs; as opposed to that of the next particle production threshold (e.g. four-body inelastic thresholds in case of three-body FV formalism). However, using EFTs is far more realistic in terms of their implementation as signified by the recent work of Barnea, \emph{et al.} \cite{Barnea:2013uqa}. Is this work such matching is carried out to the LQCD binding energies of few-nucleon systems at heavy pion masses, and is moving toward making predictions for larger systems of nucleons with solely LQCD inputs \cite{Beane:2012vq}.


\printendnotes

%
%
\bibliographystyle{plain}
\bibliography{uwthesis}
%

%
\appendix
\raggedbottom\sloppy
 
\chapter{Details on the construction and properties of the smeared operator}

\section{Operator Basis}
\label{app:operators}

\noindent
In this appendix, a basis for composite local operators used in chapter \ref{chap:operators} is presented. 
Any local operator that is  bilinear in the scalar field 
with $L$ spatial indices, and that is invariant under cubic transformations,
can be written as
\begin{equation}
\mathcal{O}_{i_{1}i_{2}...i_{L}}^{\left(d\right)}\left(\mathbf{x}\right)
\ =\ 
\phi^{\dagger}\left(\mathbf{x}\right)
\ Q_{i_{1}i_{2}...i_{L}}^{\left(d\right)}
\ \phi\left(\mathbf{x}\right)
,
\label{eq:45}
\end{equation}
where $Q_{i_{1}i_{2}...i_{L}}^{\left(d\right)}$ is a homogeneous
function of the operator ${\nabla}_{i}$, 
and degree $d$ ($d\geq L$) is defined to be the number of ${\nabla}$s. 
Their forms are determined by the symmetric traceless tensor of rank $L$ 
that  respect cubic symmetry constructed from $d$ ${\nabla}$s.
The operators composed of fewer than seven derivatives and 
with no spatial indices  
are
\begin{eqnarray}
\mathcal{O}^{\left(0\right)}\left(\mathbf{x}\right)
& = & 
\phi^{\dagger}\left(\mathbf{x}\right)\phi\left(\mathbf{x}\right)
\nonumber\\
\mathcal{O}^{\left(2\right)}\left(\mathbf{x}\right)
&  = & 
\phi^{\dagger}\left(\mathbf{x}\right)\mathbf{\nabla}^{2}\phi\left(\mathbf{x}\right)
\nonumber\\
\mathcal{O}^{\left(4\right)}\left(\mathbf{x}\right)
& = & 
\phi^{\dagger}\left(\mathbf{x}\right)\left(\mathbf{\nabla}^{2}\right)^{2}\phi\left(\mathbf{x}\right)
\nonumber\\
\mathcal{O}^{\left(4,RV\right)}\left(\mathbf{x}\right)
& = &
\phi^{\dagger}\left(\mathbf{x}\right)\ \sum_{j}\nabla_{j}^{4}\ \phi\left(\mathbf{x}\right)
\nonumber\\
\mathcal{O}^{\left(6\right)}\left(\mathbf{x}\right)
& = & 
\phi^{\dagger}\left(\mathbf{x}\right)\left(\mathbf{\nabla}^{2}\right)^{3}\phi\left(\mathbf{x}\right)
\nonumber\\
\mathcal{O}^{\left(6,RV;1\right)}\left(\mathbf{x}\right)
& = & 
\phi^{\dagger}\left(\mathbf{x}\right) \mathbf{\nabla}^{2} \
\sum_{j}\nabla_{j}^{4}\ \phi\left(\mathbf{x}\right)
\nonumber\\
\mathcal{O}^{\left(6,RV;2\right)}\left(\mathbf{x}\right)
& = &
\phi^{\dagger}\left(\mathbf{x}\right)\ \sum_{j}\nabla_{j}^{6}\ \phi\left(\mathbf{x}\right).
\label{eq:46}
\end{eqnarray}
Except for three of these operators which explicitly
break the rotational symmetry, 
they transform as $L=0$ under rotations.

The operators with one spatial index with up to six derivatives
are
\begin{eqnarray}
\mathcal{O}_{i}^{\left(1\right)}\left(\mathbf{x}\right)
& = & 
\phi^{\dagger}\left(\mathbf{x}\right)\mathbf{\nabla}_{i}\phi\left(\mathbf{x}\right)
\nonumber\\
\mathcal{O}_{i}^{\left(3\right)}\left(\mathbf{x}\right)
& = &
\phi^{\dagger}\left(\mathbf{x}\right)\mathbf{\nabla}^{2}\mathbf{\nabla}_{i}\phi\left(\mathbf{x}\right)
\nonumber\\
\mathcal{O}_{i}^{\left(5\right)}\left(\mathbf{x}\right)
 & = & 
\phi^{\dagger}\left(\mathbf{x}\right)\left(\mathbf{\nabla}^{2}\right)^{2}\mathbf{\nabla}_{i}\phi\left(\mathbf{x}\right)
\nonumber\\
\mathcal{O}_{i}^{\left(5,RV\right)}\left(\mathbf{x}\right)
 & = & \phi^{\dagger}\left(\mathbf{x}\right)\sum_{j}\nabla_{j}^{4}\mathbf{\nabla}_{i}\phi\left(\mathbf{x}\right).
\label{eq:47}
\end{eqnarray}
There is one operator which breaks rotational invariance, and
the rest  transform as $L=1$ under rotations.

The operators with two  spatial index with up to six derivatives
are
\begin{eqnarray}
\mathcal{O}_{ij}^{\left(2\right)}\left(\mathbf{x}\right)
 & = & 
\phi^{\dagger}\left(\mathbf{x}\right)
\left[\mathbf{\nabla}_{i}\mathbf{\nabla}_{j}-\frac{1}{3}\delta_{ij}\mathbf{\nabla}^{2}\right]
\phi\left(\mathbf{x}\right)
\nonumber\\
\mathcal{O}_{ij}^{\left(4\right)}\left(\mathbf{x}\right)
& = &
\phi^{\dagger}\left(\mathbf{x}\right)
\mathbf{\nabla}^{2}\left[\mathbf{\nabla}_{i}\mathbf{\nabla}_{j}-\frac{1}{3}\delta_{ij}\mathbf{\nabla}^{2}\right]
\phi\left(\mathbf{x}\right)
\nonumber\\
\mathcal{O}_{ij}^{\left(6\right)}\left(\mathbf{x}\right)
& = &
\phi^{\dagger}\left(\mathbf{x}\right)
\left(\mathbf{\nabla}^{2}\right)^{2}\left[\mathbf{\nabla}_{i}\mathbf{\nabla}_{j}
-\frac{1}{3}\delta_{ij}\mathbf{\nabla}^{2}\right]
\phi\left(\mathbf{x}\right)
\nonumber\\
\mathcal{O}_{ij}^{\left(6,RV\right)}\left(\mathbf{x}\right)
 & = & 
\phi^{\dagger}\left(\mathbf{x}\right)
\sum_{k}\nabla_{k}^{4}\left[\mathbf{\nabla}_{i}\mathbf{\nabla}_{j}
-\frac{1}{3}\delta_{ij}\mathbf{\nabla}^{2}\right]
\phi\left(\mathbf{x}\right).
\label{eq:48}
\end{eqnarray}
There is one operator which breaks rotational invariance, and
the rest  transform as $L=2$ under rotations.

Operators with three, four and five spatial indices which have $L=3$,
$L=4$ and $L=5$ respectively are listed below. There is no operator
which breaks rotational invariance up to six derivatives:

\begin{eqnarray}
\mathcal{O}_{ijk}^{\left(3\right)}\left(\mathbf{x}\right)
 & = & 
\phi^{\dagger}\left(\mathbf{x}\right)
\left[\mathbf{\nabla}_{i}\mathbf{\nabla}_{j}\mathbf{\nabla}_{k}
-\frac{1}{5}\mathbf{\nabla}^{2}
\left(\delta_{ij}\mathbf{\nabla}_{k}+\delta_{jk}\mathbf{\nabla}_{i}+\delta_{ki}\mathbf{\nabla}_{j}\right)\right]
\phi\left(\mathbf{x}\right)
\nonumber\\
\mathcal{O}_{ijk}^{\left(5\right)}\left(\mathbf{x}\right)
& = & 
\phi^{\dagger}\left(\mathbf{x}\right)
\mathbf{\nabla}^{2}\left[\mathbf{\nabla}_{i}\mathbf{\nabla}_{j}\mathbf{\nabla}_{k}
-\frac{1}{5}\mathbf{\nabla}^{2}
\left(\delta_{ij}\mathbf{\nabla}_{k}+\delta_{jk}\mathbf{\nabla}_{i}+\delta_{ki}\mathbf{\nabla}_{j}\right)\right]
\phi\left(\mathbf{x}\right),
\label{eq:49}
\end{eqnarray}
\begin{eqnarray}
\mathcal{O}_{ijkl}^{\left(4\right)}\left(\mathbf{x}\right)
& = &  
\phi^{\dagger}\left(\mathbf{x}\right)
\left[
\mathbf{\nabla}_{i}\mathbf{\nabla}_{j}\mathbf{\nabla}_{k}\mathbf{\nabla}_{l}
\phantom{\frac{1}{7}}
\right.\nonumber\\
&&\left.
\qquad
\ -\
\frac{1}{7}\mathbf{\nabla}^{2}\left(\delta_{ij}\mathbf{\nabla}_{k}\mathbf{\nabla}_{l}
+\delta_{ik}\mathbf{\nabla}_{j}\mathbf{\nabla}_{l}
+\delta_{il}\mathbf{\nabla}_{k}\mathbf{\nabla}_{j}+\delta_{jk}\mathbf{\nabla}_{i}\mathbf{\nabla}_{l}
+\delta_{jl}\mathbf{\nabla}_{i}\mathbf{\nabla}_{k}+\delta_{kl}\mathbf{\nabla}_{i}\mathbf{\nabla}_{j}\right)
\right.\nonumber\\
&&\left.
\qquad
\ +\
\frac{1}{35}\left(\mathbf{\nabla}^{2}\right)^{2}\left(\delta_{ij}\delta_{kl}
+\delta_{ik}\delta_{jl}+\delta_{il}\delta_{jk}\right)
\right]
\phi\left(\mathbf{x}\right)
\nonumber\\
\mathcal{O}_{ijkl}^{\left(6\right)}\left(\mathbf{x}\right)
& = & 
\phi^{\dagger}\left(\mathbf{x}\right)
\mathbf{\nabla}^{2}
\left[
\mathbf{\nabla}_{i}\mathbf{\nabla}_{j}\mathbf{\nabla}_{k}\mathbf{\nabla}_{l}
\phantom{\frac{1}{7}}
\right.\nonumber\\
&&\left.
\qquad
\ -\
\frac{1}{7}\mathbf{\nabla}^{2}\left(\delta_{ij}\mathbf{\nabla}_{k}\mathbf{\nabla}_{l}
+\delta_{ik}\mathbf{\nabla}_{j}\mathbf{\nabla}_{l}+\delta_{il}\mathbf{\nabla}_{k}\mathbf{\nabla}_{j}
+\delta_{jk}\mathbf{\nabla}_{i}\mathbf{\nabla}_{l}+\delta_{jl}\mathbf{\nabla}_{i}\mathbf{\nabla}_{k}
+\delta_{kl}\mathbf{\nabla}_{i}\mathbf{\nabla}_{j}\right)
\right. \nonumber\\
&& \left.
\qquad
\ +\ 
\frac{1}{35}\left(\mathbf{\nabla}^{2}\right)^{2}\left(\delta_{ij}\delta_{kl}+\delta_{ik}\delta_{jl}
+\delta_{il}\delta_{jk}\right)
\right]
\phi\left(\mathbf{x}\right),
\label{eq:50}
\end{eqnarray}

\begin{eqnarray}
\mathcal{O}_{ijklm}^{\left(5\right)}\left(\mathbf{x}\right)
& = & \phi^{\dagger}\left(\mathbf{x}\right)
\left[
\mathbf{\nabla}_{i}\mathbf{\nabla}_{j}\mathbf{\nabla}_{k}\mathbf{\nabla}_{l}\mathbf{\nabla}_{m}
\right.
\nonumber\\
&& \left.
-\frac{1}{7}\mathbf{\nabla}^{2}
\left(
\delta_{ij}\mathbf{\nabla}_{k}\mathbf{\nabla}_{l}\mathbf{\nabla}_{m}
+\delta_{ik}\mathbf{\nabla}_{j}\mathbf{\nabla}_{l}\mathbf{\nabla}_{m}
+\delta_{il}\mathbf{\nabla}_{k}\mathbf{\nabla}_{j}\mathbf{\nabla}_{m}
+\delta_{im}\mathbf{\nabla}_{k}\mathbf{\nabla}_{l}\mathbf{\nabla}_{j}
\right.\right.
\nonumber\\
&& \left.
\qquad\qquad
+\delta_{jk}\mathbf{\nabla}_{i}\mathbf{\nabla}_{l}\mathbf{\nabla}_{m}
+\delta_{jl}\mathbf{\nabla}_{k}\mathbf{\nabla}_{i}\mathbf{\nabla}_{m}
+\delta_{jm}\mathbf{\nabla}_{k}\mathbf{\nabla}_{i}\mathbf{\nabla}_{l}
+\delta_{kl}\mathbf{\nabla}_{i}\mathbf{\nabla}_{j}\mathbf{\nabla}_{m}
\right.\nonumber\\
&&\left.
\qquad\qquad
+\delta_{km}\mathbf{\nabla}_{i}\mathbf{\nabla}_{j}\mathbf{\nabla}_{l}
+\delta_{lm}\mathbf{\nabla}_{i}\mathbf{\nabla}_{j}\mathbf{\nabla}_{k}
\right)
\nonumber\\
&& 
\ +\ \frac{1}{63}
\left(\mathbf{\nabla}^{2}\right)^{2}
\left[
\left(
\delta_{ij}\delta_{kl}+\delta_{ik}\delta_{jl}+\delta_{il}\delta_{jk}
\right)\mathbf{\nabla}_{m}
+
\left(
\delta_{ij}\delta_{km}+\delta_{ik}\delta_{jm}+\delta_{im}\delta_{jk}\right)\mathbf{\nabla}_{l}
\right.\nonumber\\
&&
\left.
\qquad\qquad
+
\left(\delta_{ij}\delta_{ml}+\delta_{im}\delta_{jl}+\delta_{il}\delta_{jm}\right)\mathbf{\nabla}_{k}
+
\left(\delta_{im}\delta_{kl}+\delta_{ik}\delta_{ml}+\delta_{il}\delta_{mk}\right)\mathbf{\nabla}_{j}
\right.\nonumber\\
&&
\left.\left.
\qquad\qquad
+
\left(\delta_{mj}\delta_{kl}+\delta_{mk}\delta_{jl}+\delta_{ml}\delta_{jk}\right)\mathbf{\nabla}_{i}
\right]
\right]
\phi\left(\mathbf{x}\right).
\label{eq:51}
\end{eqnarray}

Note that as demonstrated in Eq.~(\ref{eq:46}), there can be more than one operator
that breaks rotational invariance at a given order
in derivative expansion. 
To arrive at a notation that is general and useful,
one can use the fact that any cubically invariant polynomial of a 
three-vector $\mathbf{V}$, can be expanded in terms of only three  cubically
invariant structures,
\begin{equation}
\sum_{k}V_{k}^{2}\ ,\ 
\sum_{k}V_{k}^{4}\ ,\ 
\sum_{k}V_{k}^{6}.
\label{eq:52}
\end{equation}
The number of times each structure appears in a derivative
operator, as well as the number of free indices,
uniquely specify the operator.
For example,  with nine derivatives and one spatial
index, one can make four independent   operators,
\begin{eqnarray}
\mathcal{O}_{i}^{\left(4,0,0\right)}\left(\mathbf{x}\right)
 & = & 
\phi^{\dagger}\left(\mathbf{x}\right)\left(\mathbf{\nabla}^{2}\right)^{4}\mathbf{\nabla}_{i}\phi\left(\mathbf{x}\right)
\nonumber\\
\mathcal{O}_{i}^{\left(2,1,0\right)}\left(\mathbf{x}\right)
& = & 
\phi^{\dagger}\left(\mathbf{x}\right)\left(\mathbf{\nabla}^{2}\right)^{2}
\left(\sum_{k}\nabla_{k}^{4}\right)\mathbf{\nabla}_{i}\phi\left(\mathbf{x}\right)
\nonumber\\
\mathcal{O}_{i}^{\left(1,0,1\right)}\left(\mathbf{x}\right)
& = &
\phi^{\dagger}\left(\mathbf{x}\right)\left(\mathbf{\nabla}^{2}\right)\left(\sum_{j}\nabla_{j}^{6}\right)\mathbf{\nabla}_{i}\phi\left(\mathbf{x}\right)
\nonumber\\
\mathcal{O}_{i}^{\left(0,2,0\right)}\left(\mathbf{x}\right)
& = & 
\phi^{\dagger}\left(\mathbf{x}\right)\left(\sum_{k}\nabla_{k}^{4}\right)^{2}\mathbf{\nabla}_{i}\phi\left(\mathbf{x}\right),
\label{eq:53}
\end{eqnarray}
and generally,
\begin{eqnarray}
\mathcal{O}_{i}^{\left(m,n,p\right)}\left(\mathbf{x}\right)
& = & 
\left(\mathbf{\nabla}^{2}\right)^{m}
\left(\sum_{k}\nabla_{k}^{4}\right)^n
\left(\sum_{k}\nabla_{k}^{6}\right)^p
\mathbf{\nabla}_{i}
\phi\left(\mathbf{x}\right).
\label{eq:53b}
\end{eqnarray}
It is then obvious that $d=2m+4n+6p+L$
gives the total number of derivatives in the operator, where $L$ is
the number of free indices. For $n=p=0$, the operator is rotationally
invariant with angular momentum $L$.

\section{An Example for Rotational Invariance Violating Coefficients}
\label{app:RIviolation}

In this appendix, an explicit derivation of a rotational invariance
violating coefficient in both coordinate-space, and momentum-space
formalism, introduced in section~\ref{sec:Classical}, 
is presented.
Consider the position space operator 
$\hat{\theta}_{00}^{\left(4\right)}\left(\mathbf{x};a,N\right)$
where superscript indicates that only operators with
four derivatives are retained in the expansion of $\hat{\theta}_{00}$.
The goal is to derive the LO correction to the continuum
values of coefficients $C_{00,00}^{\left(4\right)}$ 
and $C_{00,00}^{\left(4;RV\right)}$:
\begin{eqnarray}
\hat{\theta}_{00}^{\left(4\right)}\left(\mathbf{x};a,N\right)
& = & 
\phi\left(\mathbf{x}\right)\left[\left(Na\right)^{4}C_{00,00}^{\left(4\right)}\left(\nabla^{2}\right)^{2}
+\left(Na\right)^{4}C_{00,00}^{\left(4;RV\right)}
\left(\nabla_{x}^{4}+\nabla_{y}^{4}+\nabla_{z}^{4}\right)\right]
\phi\left(\mathbf{x}\right)
\nonumber\\
& = & 
\frac{3}{4\pi}\frac{\left(aN\right)^{4}}{4!}
\ \sum_{\mathbf{P}}
\ \int_{0}^{1}dy\ y^{6}\ \int
d\Omega_{\mathbf{y}}\ 
e^{i2\pi  N\mathbf{p}\cdot\mathbf{y}}
\phi^{\dagger}\left(\mathbf{x}\right)
\left(\hat{\mathbf{y}} \cdot \mathbf{\nabla}\right)^{4}
\phi\left(\mathbf{x}\right)Y_{00}\left(\Omega_{\mathbf{y}}
\right).
\nonumber\\
\label{eq:54}
\end{eqnarray}
The $y$ integration is
\begin{eqnarray}
\int_{0}^{1}dy\ y^{6}\ 
&& \int d\Omega_{\mathbf{y}}\ e^{i2\pi N\mathbf{p}\cdot\mathbf{y}}y^{i}y^{j}y^{k}y^{l}
\ =\ 
\alpha\left(p^{i}p^{j}p^{k}p^{l}\right) 
\ +\ \gamma\left(\delta^{ij}\delta^{kl}+\delta^{ik}\delta^{jl}+\delta^{il}\delta^{jk}\right)
\nonumber\\
& & +\beta\left(p^{i}p^{j}\delta^{kl}+p^{i}p^{k}\delta^{jl}
+p^{i}p^{l}\delta^{jk}+p^{k}p^{l}\delta^{ij}+p^{j}p^{l}\delta^{ik}+p^{j}p^{k}\delta^{il}
\right),
\label{eq:55}
\end{eqnarray}
and the coefficients $\alpha$, $\beta$ and $\gamma$ can be determined.
It is easy to see that coefficient $\alpha$ 
makes the dominant contribution 
in the large $N$ limit. 
Using
\begin{equation}
\sum_{\mathbf{p}}f\left(p^{2}\right)\left(\mathbf{p}.\mathbf{A}\right)^{4}=\sum_{\mathbf{p}}f\left(p^{2}\right)\left(\rho\left|\mathbf{A}\right|^{4}+\sigma\sum_{j}\left(A^{j}\right)^{4}\right)
,
\label{eq:56}
\end{equation}
for any rotational invariant function $f$ of the vector $\mathbf{p}$,
with 
\begin{eqnarray}
\rho & = & 
\frac{1}{2}\left(\left|\mathbf{p}\right|^{4}-3p_{z}^{4}\right)
\ \ ,\ \ 
\sigma\ =\ 
\frac{1}{2}\left(5p_{z}^{4}-\left|\mathbf{p}\right|^{4}\right)
,
\label{eq:sigrho}
\end{eqnarray}
one finds that the deviations of $C_{00,00}^{\left(4\right)}$
and $C_{00,00}^{\left(4;RV\right)}$ from their continuum values are
\begin{eqnarray}
\delta
C_{00,00}^{\left(4\right)}
& = & 
\frac{1}{96\sqrt{\pi}}
\ \sum_{\mathbf{p}\neq\mathbf{0}}
\left(-\frac{3\cos\left(2\pi
      N\left|\mathbf{p}\right|\right)}{4\pi^{2}
\left|\mathbf{p}\right|^{6}N^{2}}\right)\left(-3p_{z}^{4}+\left|\mathbf{p}\right|^{4}\right)
\nonumber\\
\delta C_{00,00}^{\left(4;RV\right)}
& = & \frac{1}{96\sqrt{\pi}}\sum_{\mathbf{p}\neq\mathbf{0}}
\left(-\frac{3\cos\left(2\pi N\left|\mathbf{p}\right|\right)}{4\pi^{2}
\left|\mathbf{p}\right|^{6}N^{2}}\right)\left(5p_{z}^{4}-\left|\mathbf{p}\right|^{4}\right).
\label{eq:57}
\end{eqnarray}

The emergence of rotational invariance violating coefficients from
the momentum-space construction  is somewhat less obvious. 
From Eq.~(\ref{eq:14}) and Eq.~(\ref{eq:15})
the operator $\hat{\tilde{\theta}}_{00}^{\left(4\right)}\left(\mathbf{k};a,N\right)$
can be written as
\begin{eqnarray}
\hat{\tilde{\theta}}_{00}^{\left(4\right)}\left(\mathbf{k};a,N\right)
& = & 
\tilde{\phi}\left(\mathbf{k}\right)
\tilde{\phi}\left(-\mathbf{k}\right)
\left[\left(Na\right)^{4}C_{00,00}^{\left(4\right)}\left|\mathbf{k}\right|^{4}
+\left(Na\right)^{4}C_{00,00}^{\left(4;RV\right)}(k_{x}^{4}+k_{y}^{4}+k_{z}^{4})\right]
\nonumber\\
& = & \tilde{\phi}\left(\mathbf{k}\right)
\tilde{\phi}\left(-\mathbf{k}\right)
6\sqrt{\pi}\ 
\sum_{\mathbf{p}}\sum_{L_{1},M_{1},L_{2},M_{2}}i^{L_{1}+L_{2}}\sqrt{\frac{\left(2L_{1}+1\right)
\left(2L_{2}+1\right)}{2L+1}}
\nonumber\\
&&
\times
\left\langle L_{1}0;L_{2}0\left|00\right.\right\rangle 
\left\langle L_{1}M_{1};L_{2}M_{2}\left|00\right.\right\rangle
\
Y_{L_{1}M_{1}}\left(\Omega_{\hat{k}}\right)Y_{L_{2}M_{2}}\left(\Omega_{\hat{p}}\right)
\nonumber\\
&&
\times
 \int_{0}^{1}dy\ y^{2}\ j_{L_{1}}\left(aN\left|\mathbf{k}\right|y\right)
\ j_{L_{2}}\left(2\pi
  N\left|\mathbf{p}\right|y\right)
\Bigl\lvert_{k^4}
,
\label{eq:58}
\end{eqnarray}
where only the terms of order $k^4$ are retained from the integral.
As such,
only $L_1=4$ with 
$Y_{4\pm4}\left(\Omega_{\hat{p}}\right)$ and
$Y_{40}\left(\Omega_{\hat{p}}\right)$, and 
$L_1=0$ with $Y_{00}\left(\Omega_{\hat{p}}\right)$, contribute to the sum.
This reduces the relation to 
\begin{eqnarray}
\hat{\tilde{\theta}}_{00}^{\left(4\right)}\left(\mathbf{k};a,N\right)
& = & 
6\sqrt{\pi}\ 
\sum_{\mathbf{p}}\left\{
\   Y_{00}\left(\Omega_{\hat{k}}\right)Y_{00}\left(\Omega_{\hat{p}}\right)
\ \int_{0}^{1}dy\ y^{2}\ 
\frac{\left(aN\left|\mathbf{k}\right|y\right)^{4}}{120}j_{0}\left(2\pi N\left|\mathbf{p}\right|y\right)
\right.
\nonumber\\
& + &
9\left[
\left\langle 40;40\left|00\right.\right\rangle ^{2}
Y_{40}\left(\Omega_{\hat{k}}\right)Y_{40}
\left(\Omega_{\hat{p}}\right)+\left\langle 40;40\left|00\right.\right\rangle 
\left\langle 44;4-4\left|00\right.\right\rangle
\right.
\nonumber\\
&&
\left.\times Y_{44}\left(\Omega_{\hat{k}}\right)
Y_{4-4}\left(\Omega_{\hat{p}}\right) \left.+\left\langle 40;40\left|00\right.\right\rangle \left\langle
      4-4;44\left|00\right.\right\rangle
    Y_{4-4}\left(\Omega_{\hat{k}}\right)Y_{44}\left(\Omega_{\hat{p}}\right)
\right]
\right.
\nonumber\\
& & \left. 
\qquad\qquad
\times
\int_{0}^{1}dy\ y^{2}\ 
\frac{\left(aN\left|\mathbf{k}\right|y\right)^{4}}{945}
\ j_{4}\left(2\pi N\left|\mathbf{p}\right|y\right)
\ \right\}.
\label{eq:60}
\end{eqnarray}
Using the relations
\begin{eqnarray}
\sum_{\mathbf{p}}f\left(p^{2}\right)\left(\left|\mathbf{p}\right|^{4}
Y_{40}\left(\Omega_{\hat{p}}\right)\right)
& = & 
\frac{21}{16}\sqrt{\frac{1}{\pi}}\sum_{\mathbf{p}}f\left(p^{2}\right)
\left(5p_{z}^{4}-\left|\mathbf{p}\right|^{4}\right)
,
\nonumber\\
\sum_{\mathbf{p}}f\left(p^{2}\right)\left(\left|\mathbf{p}\right|^{4} 
Y_{4\pm4}\left(\Omega_{\hat{p}}\right)\right)
& = & 
\frac{3}{16}\sqrt{\frac{35}{2\pi}}\sum_{\mathbf{p}}f\left(p^{2}\right)
\left(5p_{z}^{4}-\left|\mathbf{p}\right|^{4}\right)
,
\label{eq:62}
\end{eqnarray}
and keeping the LO term in $1/N$ from the y integration
gives
\begin{eqnarray}
&& \hat{\tilde{\theta}}_{00}^{\left(4\right)}\left(\mathbf{k};a,N\right)
\ = \ 
3\left(aN\left|\mathbf{k}\right|\right)^{4}
\sum_{\mathbf{p}\neq\mathbf{0}}\left(-\frac{\cos\left(2\pi
      N\left|\mathbf{p}\right|\right)}{4\pi\left|\mathbf{p}\right|^{2}N^{2}}\right)\left\{ \frac{1}{120}Y_{00}\left(\Omega_{\hat{k}}\right)
\right.
\nonumber\\
&& \left.
\qquad
+\frac{\sqrt{4\pi}}{945}\left(5p_{z}^{4}-\left|\mathbf{p}\right|^{4}\right)
\left[\frac{21}{16}\sqrt{\frac{1}{\pi}}
Y_{40}\left(\Omega_{\hat{k}}\right)+\frac{3}{16}
\sqrt{\frac{35}{2\pi}}\left(Y_{4-4}
\left(\Omega_{\hat{k}}\right)
+Y_{44}\left(\Omega_{\hat{k}}\right)\right)\right]\right\}.
\nonumber\\
\label{eq:63}
\end{eqnarray}
Finally, we use the relation
\begin{equation}
\frac{k_{x}^{4}+k_{y}^{4}+k_{z}^{4}}{\left|\mathbf{k}\right|^{4}}
\ =\ 
\frac{6\sqrt{\pi}}{5}Y_{00}\left(\Omega_{\hat{k}}\right)
+\frac{4\sqrt{\pi}}{15}Y_{40}\left(\Omega_{\hat{k}}\right)
+\frac{2}{3}\sqrt{\frac{2\pi}{35}}\left(Y_{4-4}\left(\Omega_{\hat{k}}\right)
+Y_{44}\left(\Omega_{\hat{k}}\right)\right)
,
\label{eq:64}
\end{equation}
to identify the
coefficients $\delta C_{00,00}^{\left(4\right)}$ and $\delta C_{00,00}^{\left(4;RV\right)}$
from Eq.~(\ref{eq:63})
\begin{eqnarray}
\delta C_{00,00}^{\left(4\right)}
& = & 
\frac{1}{96\sqrt{\pi}}\sum_{\mathbf{p}\neq\mathbf{0}}
\left(-\frac{3\cos\left(2\pi
      N\left|\mathbf{p}\right|\right)}{4\pi^{2}\left|\mathbf{p}\right|^{6}N^{2}}\right)
\left(-3p_{z}^{4}+\left|\mathbf{p}\right|^{4}\right),
\nonumber\\
\delta C_{00,00}^{\left(4;RV\right)}
& = & \frac{1}{96\sqrt{\pi}}\sum_{\mathbf{p}\neq\mathbf{0}}
\left(-\frac{3\cos\left(2\pi N\left|\mathbf{p}\right|\right)}{4\pi^{2}
\left|\mathbf{p}\right|^{6}N^{2}}\right)\left(5p_{z}^{4}-\left|\mathbf{p}\right|^{4}
\right),
\label{eq:65}
\end{eqnarray}
which recovers the position-space results given in Eq.~(\ref{eq:57}).

\section{Matrix Elements for Non-Zero External Momentum}
\label{app:nonzerop}
The loop calculations presented in chapter \ref{chap:operators} 
have  been performed for vanishing 
external momentum, therefore only the quantum corrections to the $L=0$ operator
have been considered.
\begin{figure}[!ht]
\begin{centering}
\includegraphics[scale=0.13]{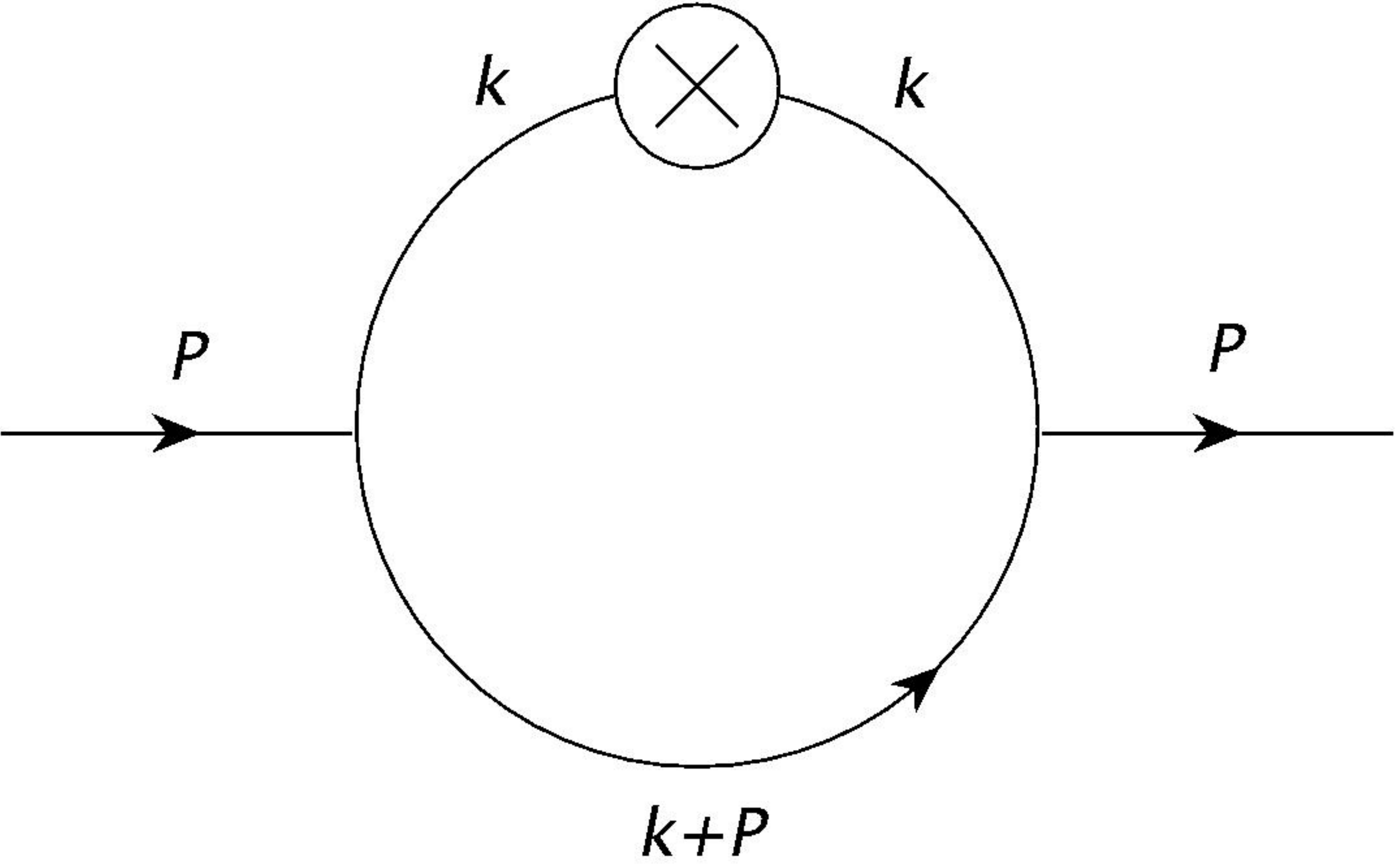}
\par\end{centering}
\caption{{\small One-loop contribution to the two-point function with an insertion
of the operator in $g\phi^{3}$}.}
\label{fig:gf3}
\end{figure}
In this appendix, the generalization to non-zero external momentum is
presented,  where the one-loop correction to the two-point function
with an  insertion of the
smeared operator is considered in scalar $g\phi^{3}$ theory, see Fig.~\ref{fig:gf3}. 
The loop integral to be evaluated is
\begin{equation}
J_{LM}
\ =\ 
\frac{3}{4\pi
  N^{3}}\sum_{\bf n}^{|{\bf n}|\le N}
\int_{-\frac{\pi}{a}}^{\frac{\pi}{a}}\frac{d^{4}k}{\left(2\pi\right)^{4}}
\frac{e^{i\mathbf{k}\cdot\mathbf{n}a}}{\left(\hat{k}^{2}+m^{2}\right)^{2}
\left(\left(\widehat{k+P}\right)^{2}+m^{2}\right)}
\ Y_{LM}\left(\Omega_{\mathbf{n}}\right)
,
\label{eq:66}
\end{equation}
where 
\begin{eqnarray}
\hat{k}^{2}& = & 
{4\over a^2}\ \sum_{\mu}
\sin^{2}\left(k_{\mu}a\over 2\right)
,\ \ \ 
\left(\widehat{k+P}\right)^{2}
\ =\ 
{4\over a^2}\ \sum_{\mu}
\sin^{2}\left(\left(k_{\mu}+P_{\mu}\right)a\over 2\right).
\label{eq:momext}
\end{eqnarray}
Note that the operator is smeared over a physical region whose size
is small compared to the hadronic scale, and as a result the external momenta are small compared to
the the scale of the operator $\Lambda=1/Na$. 
Therefore one may perform a Taylor expansion of the loop integral in $P_{i}/\Lambda$ to obtain
\begin{eqnarray}
J_{LM} 
& = & 
\frac{3}{16\pi^4}i^{L}\frac{1}{\Lambda^{2}}
\ \int_{-\pi N}^{\pi N}\ dq_{4}\ d^{3}q
\ \left[\int_{0}^{1}dy\ y^{2}\ 
j_{L}\left(qy\right)\right]
\ Y_{LM}\left(\Omega_{\mathbf{q}}\right)
\nonumber\\
&&
\times
\left(4N^{2}\sum_{i=1}^{3}\sin^{2}\left(\frac{q_{i}}{2N}\right)+4N^{2}\sin^{2}
\left(\frac{q_{4}}{2N}\right)+\frac{m^{2}}{\Lambda^{2}}\right)^{-2}
\nonumber\\
&&
\times
\left(4N^{2}\sum_{i=1}^{3}\sin^{2}\left(\frac{q_{i}}{2N}\right)+4N^{2}\sin^{2}
\left(\frac{q_{4}}{2N}+\frac{P_{4}}{2N\Lambda}\right)+\frac{m^{2}}{\Lambda^{2}}\right)^{-1}
\nonumber\\
&&
\times
\sum_{k=0}^{\infty}\left[-\frac{4N^{2}\sum_{i=1}^{3}\frac{1}{2}\sin\left(\frac{q_{i}}{N}\right)
\sin\left(\frac{P_{i}}{N\Lambda}\right)+4N^{2}\sum_{i=1}^{3}\cos\left(\frac{q_{i}}{N}\right)
\sin^{2}\left(\frac{P_{i}}{2N\Lambda}\right)}{4N^{2}\sum_{i=1}^{3}\sin^{2}
\left(\frac{q_{i}}{2N}\right)+4N^{2}\sin^{2}\left(\frac{q_{4}}{2N}+\frac{P_{4}}{2N\Lambda}\right)
+\frac{m^{2}}{\Lambda^{2}}}\right]^{k},
\nonumber\\
\label{eq:67}
\end{eqnarray}
where $\mathbf{q}=\mathbf{k}/\Lambda$, $q_{4}=k_{4}/\Lambda$, and
only the leading term in the Poisson sum is retained.
As was shown before, the non-zero terms in the Poisson sum are suppressed
by at least $1/N^{2}$ compared to the continuum operator insertion
in the loop.

The first term in the above Taylor expansion corresponds to the zero
external momentum in the loop, therefore at LO, it contributes
to the $L=0$ operator, and the sub-leading rotational invariance
breaking operators can be easily shown to be suppressed by $1/N^{2}$
using the procedure described in section \ref{sec:Scalar}. Note that
the loop integrals one needs to deal with in $g\phi^{3}$ are
more convergent than comparable integrals in $\lambda\phi^{4}$ theory, 
which 
simplifies
the discussion of the scaling of the different contributions.

The next term in the Taylor expansion of the loop integral can be
expanded in large $N$ since the integral is convergent.
The numerator has an expansion of the form
\begin{eqnarray}
{\rm Num.} & \sim & 
4N^{2}\sum_{i=1}^{3}\frac{1}{2}\sin\left(\frac{q_{i}}{N}\right)\sin\left(\frac{P_{i}}{N\Lambda}\right)
+4N^{2}\sum_{i=1}^{3}\cos\left(\frac{q_{i}}{N}\right)\sin^{2}\left(\frac{P_{i}}{2N\Lambda}\right)
\nonumber\\
&&\ =\ 
\frac{2\mathbf{P}\cdot\mathbf{q}}{\Lambda}+\frac{\left|\mathbf{P}\right|^{2}}{\Lambda^{2}}
+\mathcal{O}\left(\frac{1}{N^{2}}\right),
\label{eq:68}
\end{eqnarray}
where the rotational invariance breaking terms are suppressed
by at least $1/N^{2}$, and the leading contribution to the above
integral modifies the $L=1$ matrix element, while the $L=0$ term is suppressed
by $1/\Lambda$ compared to the $L=1$ contribution. The next terms
in the Taylor expansion give rise to contributions to
the $L=2,3,...$ matrix elements at the LO 
in $1/\Lambda$, 
while the rotational invariance violating terms remain
suppressed by at least $1/N^{2}$ compared to the LO  contributions.

\section{Links on the Grid}
\label{app:linksongrid}

In this appendix, the method to evaluate the link at $\mathcal{O}\left(g\right)$
on a three-dimensional grid is outlined through an example, and the
result is generalized to other similar cases. The link is constructed 
to be the closest link to the continuum diagonal link in the continuum. This link is used in constructing the smeared QCD operators with smooth continuum limit properties in chapter \ref{chap:operators}.

Suppose that the link lies between points $x$ and $x+\mathbf{n}a$
on a cubic lattice where: $\mathbf{n}a=a_{0}\left(Q,1,0\right)$.
$Q$ is an arbitrary integer and $a_{0}$ is a finite number denoting
the original lattice spacing which is not necessarily small. Then
the paths which make minimal area with the diagonal path can be formed
easily. Among those, the paths which are symmetric under reflection
around the midpoint of the path are desired since they have somewhat simple 
forms. One such a path in shown in Fig.~\ref{fig:links}a for $Q=2$, where it is
straightforward to show that:
\begin{equation}
U_{\left(Q,1,0\right)}^{\left(1g\right)}\left(q\right)=iga_{0}e^{i\mathbf{q}\cdot\mathbf{n}a/2}
\left[A_{y}\left(q\right)+2A_{x}\left(q\right)\frac{\sin\left(Qq_{x}a_{0}/4\right)}{
\sin\left(q_{x}a_{0}/2\right)}\cos\left(\frac{Qq_{x}a_{0}}{4}+\frac{q_{y}a_{0}}{2}\right)\right].
\label{eq:69}
\end{equation}
If the lattice spacing is halved, the closest link to the diagonal
path can be obtained by adding up two paths each of the form above
with an appropriate phase factor and where $\mathbf{n}a$ is replaced
by $\mathbf{n}a/2$, Fig.~\ref{fig:links}b,
\begin{eqnarray}
U_{\left(Q,1,0\right)}^{\left(1g\right)}\left(q\right)
& = & 
ig\frac{a_{0}}{2}e^{i\mathbf{q}\cdot\mathbf{n}a/2}\frac{\sin\left(\frac{\mathbf{q}\cdot\mathbf{n}a}{2}\right)}{
\sin\left(\frac{\mathbf{q}\cdot\mathbf{n}a}{4}\right)}
\left[A_{y}\left(q\right)
\phantom{\frac{\sin\left(Qq_{x}a_{0}/8\right)}{\sin\left(q_{x}a_{0}/4\right)} }
\right.\nonumber\\
&& \left.\qquad\qquad
\ +\ 
2A_{x}\left(q\right)
\frac{\sin\left(Qq_{x}a_{0}/8\right)}{\sin\left(q_{x}a_{0}/4\right)}\cos\left(\frac{Qq_{x}a_{0}}{8}
+\frac{q_{y}a_{0}}{4}\right)\right].
\label{eq:70}
\end{eqnarray}
This process can be repeated  to build extended gauge links on finer grids. 
For the general
case, where the original lattice spacing is divided by $2^{K}$, it
is not hard to show that
\begin{eqnarray}
U_{2^{K}\left(Q,1,0\right)}^{\left(1g\right)}\left(q\right)
& = & 
ig\frac{a_{0}}{2^{K}}e^{i\mathbf{q}\cdot\mathbf{n}a/2}\frac{\sin\left(\frac{\mathbf{q}\cdot\mathbf{n}a}{2}\right)}{
\sin\left(\frac{\mathbf{q}.\mathbf{n}a}{2^{K+1}}\right)}
\left[A_{y}\left(q\right)
\phantom{\frac{\sin\left(Qq_{x}a_{0}/2^{K+2}\right)}{\sin\left(q_{x}a_{0}/2^{K+1}\right)}}
\right. \nonumber\\
&& \left. \qquad
+\ 2 A_{x}\left(q\right)\frac{\sin\left(Qq_{x}a_{0}/2^{K+2}\right)}{\sin\left(q_{x}a_{0}/2^{K+1}\right)}
\cos\left(\frac{Qq_{x}a_{0}}{2^{K+2}}+\frac{q_{y}a_{0}}{2^{K+1}}\right)\right].
\label{eq:71}
\end{eqnarray}
The continuum limit is obtained by taking $K\rightarrow\infty$,
which corresponds to $a=a_{0}/2^{K}\rightarrow0$, recovering Eq.~(\ref{eq:40}).
Note that after interchanging the gauge field indices properly, this
expression is applicable to a class of $\mathbf{n}$ vectors with
one zero component and $n_{i}/n_{j}=Q$ for the ratio of the remaining
components.

The above expression for the gauge link in Eq.~(\ref{eq:71}) can be
generalized easily to another class of $\mathbf{n}$ vectors with
one component being equal to $Q$ and the other two components each
being one. For example for $\mathbf{n}a=a_{0}\left(Q,1,1\right)$
one obtains
\begin{eqnarray}
U_{2^{K}\left(Q,1,1\right)}^{\left(1g\right)}\left(q\right)
& = & 
ig\frac{a_{0}}{2^{K}}e^{i\mathbf{q}\cdot\mathbf{n}a/2}\frac{\sin\left(\frac{\mathbf{q}\cdot\mathbf{n}a}{2}\right)}{
\sin\left(\frac{\mathbf{q}\cdot\mathbf{n}a}{2^{K+1}}\right)}
\left[A_{z}\left(q\right)e^{iq_{y}a_{0}/2^{K+1}}+A_{y}\left(q\right)e^{-iq_{z}a_{0}/2^{K+1}}
\right.
\nonumber\\
&& \left.
+2A_{x}\left(q\right)\frac{\sin\left(Qq_{x}a_{0}/2^{K+2}\right)}{\sin\left(q_{x}a_{0}/2^{K+1}\right)}
\cos\left(\frac{Qq_{x}a_{0}}{2^{K+2}}+\frac{q_{y}a_{0}}{2^{K+1}}+\frac{q_{z}a_{0}}{2^{K+1}}\right)\right] .
\label{eq:72}
\end{eqnarray}
However, since the vector $\mathbf{n}a$ is symmetric in its $y$
and $z$ components, the link has to respect this symmetry as well. 
In
fact, there exist an equivalent path which arises from
the first path by interchanging the steps in the $y$ direction and
the $z$ direction. 
Taking an average of  these two paths gives
a link which is symmetric in the $y$ and $z$ components,
\begin{eqnarray}
\bar{U}_{2^{K}\left(Q,1,1\right)}^{\left(1g\right)}\left(q\right)
& = & ig\frac{a_{0}}{2^{K}}e^{i\mathbf{q}\cdot\mathbf{n}a/2}\frac{
\sin\left(\frac{\mathbf{q}\cdot\mathbf{n}a}{2}\right)}{\sin\left(\frac{\mathbf{q}\cdot\mathbf{n}a}{2^{K+1}}\right)}
\left[A_{z}\left(q\right)\cos\left(\frac{q_{y}a_{0}}{2^{K+1}}\right)
+A_{y}\left(q\right)\cos\left(\frac{q_{z}a_{0}}{2^{K+1}}\right)
\right.
\nonumber\\
&& \left.
+2A_{x}\left(q\right)\frac{\sin\left(Qq_{x}a_{0}/2^{K+2}\right)}{
\sin\left(q_{x}a_{0}/2^{K+1}\right)}\cos\left(\frac{Qq_{x}a_{0}}{2^{K+2}}+\frac{q_{y}a_{0}}{2^{K+1}}
+\frac{q_{z}a_{0}}{2^{K+1}}\right)\right].
\label{eq:73}
\end{eqnarray}
Taking the $K\rightarrow\infty$ limit of the above link gives rise
to the rotational invariant link as well as non-continuum 
corrections which start at $\mathcal{O}\left(a^{2}\right)$.

Another class of $\mathbf{n}$ vectors are those where two components
are equal to $Q$ while the other one is equal one. For example for
$\mathbf{n}a=a_{0}\left(Q,Q,1\right)$ the link which is symmetric
with respect to $x$ and $y$ can be shown to have the form:
\begin{eqnarray}
&&\bar{U}_{2^{K}\left(Q\times Q\times1\right)}^{\left(1g\right)}
\ =\ 
ig\frac{a_{0}}{2^{K}}e^{i\mathbf{q}\cdot\mathbf{\Delta x}/2}
\frac{\sin\left(\frac{\mathbf{q}\cdot\mathbf{\Delta x}}{2}\right)}{
\sin\left(\frac{\mathbf{q}\cdot\mathbf{\Delta x}}{2^{K+1}}\right)}\times
\nonumber\\
&&
\left[2\left(A_{x}\left(q\right)
\cos\left(\frac{q_{y}a_{0}}{2^{K+1}}\right)
+A_{y}\left(q\right)
\cos\left(\frac{q_{x}a_{0}}{2^{K+1}}\right)\right)
\frac{\sin\left(
    Q(q_{x}a_{0}+q_{y}a_{0})/2^{K+2}\right)}{\sin\left((q_{x}a_{0}+q_{y}a_{0})/2^{K+1}\right)}
\right.\nonumber\\
&&\left.\qquad\qquad
\times
\cos\left(\frac{Qq_{x}a_{0}}{2^{K+2}}+\frac{Qq_{y}a_{0}}{2^{K+2}}+\frac{q_{z}a_{0}}{2^{K+1}}\right)+A_{z}\left(q\right)\right]
,
\label{eq:74}
\end{eqnarray}
where $\bar{U}$ the  average of two links which are identical
upon interchanging the $x$ and $y$ coordinate axes. This link recovers
the rotational invariant link up to corrections of ${\cal O}(a^{2})$.

For $\mathbf{n}$ vectors with equal components, $\mathbf{n}a=a_{0}\left(Q,Q,Q\right)$,
there are six equivalent links which are averaged over to obtain
\begin{align}
&
\bar{U}_{2^{K}\left(Q\times Q\times Q\right)}^{\left(1g\right)}
\ = \ 
ig\frac{a_{0}}{2^{K}}e^{i\mathbf{q}\cdot\mathbf{\Delta
    x}/2}\frac{\sin\left(\frac{\mathbf{q}\cdot\mathbf{\Delta x}}{2}\right)}
{\sin\left(\frac{\mathbf{q}\cdot\mathbf{\Delta x}}{2^{K+1}}\right)}\times
\nonumber\\
&
\left[2\left(A_{x}\left(q\right)\cos\left(\frac{q_{y}a_{0}}{2^{K+1}}+\frac{q_{z}a_{0}}{2^{K+1}}\right)+A_{y}\left(q\right)
\cos\left(\frac{q_{x}a_{0}}{2^{K+1}}+\frac{q_{z}a_{0}}{2^{K+1}}\right)
+A_{z}\left(q\right)\cos\left(\frac{q_{x}a_{0}}{2^{K+1}}+\frac{q_{y}a_{0}}{2^{K+1}}\right)\right)\right.
\nonumber\\
&
\qquad
\times\left.\frac{
\sin\left((Qq_{x}a_{0}+Qq_{y}a_{0}+Qq_{z}a_{0})/2^{K+2}\right)}
{\sin\left((q_{x}a_{0}+q_{y}a_{0}+q_{z}a_{0})/2^{K+1}\right)}
\cos\left(\frac{Qq_{x}a_{0}}{2^{K+2}}+\frac{Qq_{y}a_{0}}{2^{K+2}}+\frac{Qq_{z}a_{0}}{2^{K+2}}
\right)
\right],
\label{eq:75}
\end{align}
which results in $\mathcal{O}\left(a^{2}\right)$ corrections to the
rotational invariant continuum path.
It is the case that determining
the link for a general extended path  is  quite involved, but the
general trend that the  deviation  from the
rotationally invariant continuum path is  ${\cal O}(a^{2})$
is anticipated.


\chapter{Finite-volume Formalism for NN systems and the Deuteron}

\section{Quantization Conditions under $\mathbf{P}\rightarrow \mathbf{P}'$ Transformation when $\mathbf{P}$ and $\mathbf{P}'$ are Related by a Cubic Rotation\label{app:invariant}}
We aim to show that the master QC of chapter \ref{chap:NN}, Eq. (\ref{NNQC}), is invariant under a $\mathbf{P}\rightarrow \mathbf{P}'$ transformation where $\mathbf{P}$ and $\mathbf{P}'$ are two boost vectors that are related by a cubic rotation. Denoting such rotation by $R$, it is straightforward to show that
\begin{eqnarray}
c^{\mathbf{P}'}_{lm}=\sum_{m'=-l}^{l}\mathcal{D}^{(l)}_{mm'}(R)~c^{\mathbf{P}}_{lm'}.
\label{clm-trans-P}
\end{eqnarray}
Note that for $\mathbf{P}=0$ this relation reduces to Eq. (\ref{clm-trans}), while for a general non-zero boost vector, it only holds if the rotation $R$ corresponds to the symmetry operation of the cube. For example, such a transformation can take the $c^{\mathbf{P}}_{lm}$ function evaluated with $\textbf{d}=(0,0,1)$ to a $c^{\mathbf{P}'}_{lm}$ evaluated with $\textbf{d}=(1,0,0)$. To proceed let us rewrite the $\delta \mathcal{G}^{V}$ matrix elements as given in Eq. (\ref{deltaG}) in terms of the matrix elements of the $\mathcal{F}^{FV}$ that is defined in Eq. (\ref{F}) for the scalar sector,
\begin{align}
& \left[\delta\mathcal{G}^{V,\mathbf{P}}\right]_{JM_J,IM_I,LS;J'M_J',I'M_I',L'S'}=\frac{iM}{4\pi}\delta_{II'}\delta_{M_IM_I'}\delta_{SS'}\times
\nonumber\\
& \qquad ~ \times \left[k^*\delta_{JJ'}\delta_{M_JM_J'}\delta_{LL'} +i\sum_{M_L,M_L',M_S}\langle JM_J|LM_L,SM_S\rangle \langle L'M_L',SM_S|J'M_J'\rangle \mathcal{F}^{FV,\mathbf{P}}_{LM_L,L'M_L'} \right].
\nonumber\\
\label{deltaG-F}
\end{align}
Superscript $\mathbf{P}$ on $\delta\mathcal{G}^{V}$ and $\mathcal{F}^{FV}$ reflects the fact that they depend on both magnitude and direction of the boost vector. Now given the transformation of $c^{\mathbf{P}}_{lm}$ under a cubic rotation of the boost vector, Eq. (\ref{clm-trans-P}), one can write $\mathcal{F}^{FV,\mathbf{P}'}$ as following
\begin{align}
&\left[\mathcal{F}^{FV,\mathbf{P}'}\right]_{LM_L,L'M_L'}=\sum_{l,m}\sum_{m'=-l}^{l}\mathcal{D}^{(l)}_{mm'}(R)\frac{(4\pi)^{3/2}}{k^{*l}}c^{\mathbf{P}}_{lm'}(k^{*2})\int d\Omega~Y^*_{L,M_L}Y^*_{l,m}Y_{L',M_L'}
\nonumber\\
& \qquad \qquad \qquad \qquad = \sum_{\bar{M}_L=-L}^{L}\sum_{\bar{M}_L'=-L'}^{L'}\mathcal{D}^{(L)}_{\bar{M}_LM_L}(R^{-1})\left[\mathcal{F}^{FV,\mathbf{P}}\right]_{L\bar{M}_L,L'\bar{M}_L'}
\mathcal{D}^{(L')}_{M_L'\bar{M}_L'}(R),
\nonumber\\
\label{F-trans}
\end{align}
where in the last equality we have used the fact that under rotation
\begin{eqnarray}
\sum_{M'=-L}^{L}\mathcal{D}^{(L)}_{MM'}(R)~Y_{LM'}(\hat{\mathbf{r}})=Y_{LM}(R\hat{\mathbf{r}}).
\label{deltaG1}
\end{eqnarray}

Now one can obtain the relation between $\delta\mathcal{G}^{V,\mathbf{P}'}$ and $\delta\mathcal{G}^{V,\mathbf{P}}$ using Eqs. (\ref{deltaG-F}, \ref{F-trans}),
\begin{align}
&\left[\delta\mathcal{G}^{V,\mathbf{P}'}\right]_{JM_J,L;J'M_J',L'}=\frac{iM}{4\pi}\times \left[k^*\delta_{JJ'}\delta_{M_JM_J'}\delta_{LL'} +i \sum_{M_L,M_L',M_S}  \langle JM_J|LM_L,SM_S\rangle\right.
\nonumber\\
&  \left. \times \langle L'M_L',SM_S|J'M_J'\rangle \sum_{\bar{M}_L=-L}^{L}\sum_{\bar{M}_L'=-L'}^{L'}\mathcal{D}^{(L)}_{\bar{M}_LM_L}(R^{-1})\left[\mathcal{F}^{FV,\mathbf{P}}\right]_{L\bar{M}_L,L'\bar{M}_L'}
\mathcal{D}^{(L')}_{M_L'\bar{M}_L'}(R) \right],
\nonumber\\
\label{G-trans1}
\end{align}
where we have suppressed spin and isospin indices for the sake of compactness. Using the fact that
\begin{align}
&\langle JM_J|LM_L,SM_S\rangle=\sum_{\widetilde{M}_J,\widetilde{M}_L,\widetilde{M}_S}\mathcal{D}^{(J)}_{M_J\widetilde{M}_J}(R^{-1})\mathcal{D}^{(L)}_{M_L\widetilde{M}_L}(R)\mathcal{D}^{(S)}_{M_S\widetilde{M}_S}(R)
\langle J\widetilde{M}_J|L\widetilde{M}_L,S\widetilde{M}_S\rangle,
\label{CG-trans}
\end{align}
and given that Wigner $\mathcal{D}$-matrices are unitary, one easily arrives at
\begin{align}
&\left[\delta\mathcal{G}^{V,\mathbf{P}'}\right]_{JM_J,L;J'M_J',L'}=\sum_{\bar{M}_J=-J}^{J}\sum_{\bar{M}_J'=-J'}^{J'}\mathcal{D}^{(J)}_{\bar{M}_JM_J}(R^{-1})\left[\delta\mathcal{G}^{V,\mathbf{P}}\right]_{J\bar{M}_J,L;J'\bar{M}_J',L'}\mathcal{D}^{(J')}_{M_J'\bar{M}_J'}(R),
\label{G-trans2}
\end{align}
or in the matrix notation, $\delta\mathcal{G}^{V,\mathbf{P}'}=\mathcal{D}^*(R)\delta\mathcal{G}^{V,\mathbf{P}}\mathcal{D}^T(R)$. Given that the scattering amplitude is diagonal in the $|J,M_J\rangle$ basis, and that the quantization condition Eq. (\ref{NNQC}) is a determinant condition, one obtains
\begin{eqnarray}
\det\left[{(\mathcal{M}^{\infty})^{-1}+\delta\mathcal{G}^{V,\mathbf{P}'}}\right]&=&\det\left[\mathcal{D}^*(R)\left({(\mathcal{M}^{\infty})^{-1}+\delta\mathcal{G}^{V,\mathbf{P}}}\right)\mathcal{D}^T(R)\right]
\nonumber\\
&=&\det\left[{(\mathcal{M}^{\infty})^{-1}+\delta\mathcal{G}^{V,\mathbf{P}}}\right]=0.
\label{QC-trans}
\end{eqnarray}
As one would expect, although the FV functions are in general different for different boosts with the same magnitude within a given $A_1$ irrep of the cubic group, the spectrum does not depend on the choice of the direction of the boost vector. As discussed in Sec. \ref{sec: Reduction}, in order to extract the scattering parameters of NN systems from the QCs presented in chapter \ref{chap:NN} (and in Refs. \cite{Briceno:2013lba, BDLsupp, Briceno:2013rwa}), one needs to use the specific boost vectors that are studied in this there. However, the energy eigenvalues can be taken from the LQCD calculations that are performed with any other boost vector that is a cubic rotation of the specific boost vectors considered.

\section{Boosted Quantization Conditions for the Deuteron Channel\label{app: QC}}
\noindent
The NN FV QCs in the positive-parity isoscalar channels
that have an overlap with  $\siii$-$\diii$ coupled channels, and used in producing the FV deuteron spectrum in chapter \ref{chap:deuteron},
are listed in this appendix for a number of CM boosts.\footnote{We restate that for  systems composed of equal-mass NR particles,
these QCs can be also utilized for boosts of the form 
$(2n_1,2n_2,2n_3)$, $(2n_1,2n_2,2n_3+1)$, 
$(2n_1+1,2n_2+1,2n_3)$ and $(2n_1+1,2n_2+1,2n_3+1)$
where $n_1,n_2,n_3$ are integers, and all cubic rotations of these vectors~\cite{Briceno:2013lba}.}
 
With the notation introduced in Ref.~\cite{Briceno:2013lba},
the QC for the irrep $\Gamma_i$ can be written as
\begin{eqnarray}
\det\left({\mathbb{M}}^{(\Gamma_i)}+\frac{iMk^{*}}{4\pi
  }-\mathcal{F}^{(\Gamma_i),{\textbf{d}}}\right)=0
,
\label{QC-simplified}
\end{eqnarray} 
where
\begin{eqnarray}
\mathcal{F}^{(\Gamma_i),{\textbf{d}}}(k^{*2}; {\rm L} )
& = &
{M}\sum_{l,m}\frac{1}{k^{*l}}~{\mathbb{F}
}_{lm}^{(\Gamma_i)}~{c_{lm}^{\textbf{d}}(k^{*2};{\rm L})}
,
\nonumber\\
{\mathbb{M}}^{(\Gamma_i)}
& = & \left( \mathcal{M}^{-1}\right)_{\Gamma_i}
,
\label{def-F}
\end{eqnarray}
where 
the functions ${c_{lm}^{\textbf{d}}(k^{*2};{\rm L})}$ are defined in
Eq.~(\ref{clm}), $M$ is the nucleon mass and $k^*$ is the on-shell momentum of each nucleon in the CM frame.
It is straightforward to decompose $\mathcal{M}^{-1}$ into 
$\left( \mathcal{M}^{-1}\right)_{\Gamma_i}$
using the eigenvectors of the FV functions.
The matrices $\mathbb{F}$ and $\mathbb{M}$ are given in the following subsections.

For notational convenience,
$\mathcal{M}_{1,S(D)}$ denotes the scattering amplitude in the channel with total
angular momentum $J=1$ 
and orbital angular momentum $L=0~(L=2)$, 
$\mathcal{M}_{1,SD}$ is the amplitude between $S$ and $D$ partial
waves in the $J=1$ channel, 
and $\text{det} \mathcal{M}_1$ is the determinant of the $J=1$ sector of the scattering-amplitude matrix,
\begin{eqnarray}
\det\mathcal{M}_{1}=\det \left( \begin{array}{cc}
\mathcal{M}_{1,S}&\mathcal{M}_{1,SD}\\
\mathcal{M}_{1,DS}&\mathcal{M}_{1,D}\\
\end{array} \right)
.
\end{eqnarray}

\begin{small}
\subsubsection{$\mathbf{d}=(0,0,0)$}
\begin{align}
& \mathbb{T}_1: \hspace{.5cm}
\mathbb{F}_{00}^{(\mathbb{T}_1)}=\textbf{I}_{3},\hspace{.5cm}
\mathbb{F}_{40}^{(\mathbb{T}_1)}=
\left(
\begin{array}{ccc}
 0 & 0 & 0 \\
 0 & 0 & \frac{2 \sqrt{6}}{7} \\
 0 & \frac{2 \sqrt{6}}{7} & \frac{2}{7} \\
\end{array}
\right),\hspace{.5cm}
{\mathbb{M}}^{(\mathbb{T}_1)}=\left(
\begin{array}{ccc}
 \frac{\mathcal{M}_{1,D}}{{\det\mathcal{M}_{1}}} & -\frac{\mathcal{M}_{1,SD}}{\det\mathcal{M}_{1}} & 0 \\
 -\frac{\mathcal{M}_{1,SD}}{\det\mathcal{M}_{1}} & \frac{\mathcal{M}_{1,S}}{\det\mathcal{M}_{1}} & 0 \\
 0 & 0 &\mathcal{M}_{3,D}^{-1} \\
\end{array}
\right). \label{I000T1}
\end{align}
%

\subsubsection{$\mathbf{d}=(0,0,1)$}
\begin{align}
& \mathbb{A}_2:\hspace{.5cm}
\mathbb{F}_{00}^{(\mathbb{A}_2)}=\textbf{I}_{3}
\ \ ,\hspace{.5cm}
\mathbb{F}_{20}^{(\mathbb{A}_2)}=
\left(
\begin{array}{ccc}
 \frac{2}{\sqrt{5}} & 0 & -\frac{9}{7 \sqrt{5}} \\
 0 & -\frac{1}{\sqrt{5}} & \frac{6 }{7}\sqrt{\frac{2}{5}} \\
 -\frac{9}{7 \sqrt{5}} & \frac{6 }{7}\sqrt{\frac{2}{5}} & \frac{8}{7 \sqrt{5}} \\
\end{array}
\right),\hspace{.5cm}
\mathbb{F}_{40}^{(\mathbb{A}_2)}=\left(
\begin{array}{ccc}
 0 & 0 & -\frac{4}{7} \\
 0 & 0 & -\frac{2 \sqrt{2}}{7}  \\
 -\frac{4}{7} & -\frac{2 \sqrt{2}}{7} & \frac{2}{7} \\
\end{array}
\right),
\nonumber\\
&\hspace{1cm}
{\mathbb{M}}^{(\mathbb{A}_2)}= \left(
\begin{array}{ccc}
 \frac{2 \mathcal{M}_{1,S}+2 \sqrt{2}\mathcal{M}_{1,SD}+\mathcal{M}_{1,D}}{3
   \det\mathcal{M}_{1}} 
& \frac{\sqrt{2} \mathcal{M}_{1,S}-\mathcal{M}_{1,SD}-\sqrt{2}\mathcal{M}_{1,D}}{3 \det\mathcal{M}_{1}} & 0 \\
 \frac{\sqrt{2}
   \mathcal{M}_{1,S}-\mathcal{M}_{1,SD}-\sqrt{2}\mathcal{M}_{1,D}}{3
   \det\mathcal{M}_{1}} 
& \frac{\mathcal{M}_{1,S}-2 \sqrt{2}\mathcal{M}_{1,SD}+2\mathcal{M}_{1,D}}{3 \det\mathcal{M}_{1}} & 0 \\
 0 & 0 & \mathcal{M}_{3,D}^{-1}  \\
\end{array}
\right).
\label{I001A2}
\end{align}
\begin{align}
& E:\hspace{.5cm}\mathbb{F}_{00}^{(\mathbb{E})}=\textbf{I}_{5}
\ \ \ ,\hspace{.5cm}
\mathbb{F}_{20}^{(\mathbb{E})}=
\left(
\begin{array}{ccccc}
 \frac{1}{2 \sqrt{5}} & 0 & -\frac{\sqrt{3}}{2} & 0 & \frac{4 \sqrt{\frac{3}{5}}}{7} \\
 0 & -\frac{1}{\sqrt{5}} & 0 & 0 & -\frac{3}{7}   \sqrt{\frac{6}{5}} \\
 -\frac{\sqrt{3}}{2} & 0 & \frac{\sqrt{5}}{14} & 0 & -\frac{2}{7} \\
 0 & 0 & 0 & -\frac{2 }{7} \sqrt{5}  & 0 \\
 \frac{4 \sqrt{\frac{3}{5}}}{7} & -\frac{3}{7}   \sqrt{\frac{6}{5}} & -\frac{2}{7} & 0 & \frac{6}{7 \sqrt{5}} \\
\end{array}
\right),
\nonumber
\end{align}
\begin{align}
&\hspace{1cm}
\mathbb{F}_{40}^{(\mathbb{E})}=\left(
\begin{array}{ccccc}
 0 & 0 & 0 & 0 & \frac{\sqrt{3}}{7} \\
 0 & 0 & 0 & 0 & \frac{\sqrt{6}}{7} \\
 0 & 0 & \frac{8}{21} & 0 & -\frac{5 \sqrt{5}}{21} \\
 0 & 0 & 0 & \frac{1}{7} & 0 \\
 \frac{\sqrt{3}}{7} & \frac{\sqrt{6}}{7} & -\frac{5 \sqrt{5}}{21} & 0 & \frac{1}{21}
\end{array}
\right),
\hspace{.5cm}
\mathbb{F}_{44}^{(\mathbb{E})}=
\left(
\begin{array}{ccccc}
 0 & 0 & 0 & \sqrt{\frac{2}{7}} & 0 \\
 0 & 0 & 0 & \frac{2}{\sqrt{7}} & 0 \\
 0 & 0 & 0 & \sqrt{\frac{10}{21}} & 0 \\
 \sqrt{\frac{2}{7}} & \frac{2}{\sqrt{7}} & \sqrt{\frac{10}{21}} & 0 & \sqrt{\frac{2}{21}} \\
 0 & 0 & 0 & \sqrt{\frac{2}{21}} & 0
\end{array}
\right),
\nonumber\\
&\hspace{1cm}{\mathbb{M}}^{(\mathbb{E})}=
 \left(
\begin{array}{ccccc}
 \frac{\mathcal{M}_{1,S}-2 \sqrt{2}\mathcal{M}_{1,SD}+2\mathcal{M}_{1,D}}{3
   \det\mathcal{M}_{1}} 
& \frac{\sqrt{2} \mathcal{M}_{1,S}-\mathcal{M}_{1,SD}-\sqrt{2}\mathcal{M}_{1,D}}{3 \det\mathcal{M}_{1}} & 0 & 0 & 0 \\
 \frac{\sqrt{2}
   \mathcal{M}_{1,S}-\mathcal{M}_{1,SD}-\sqrt{2}\mathcal{M}_{1,D}}{3
   \det\mathcal{M}_{1}} 
& \frac{2 \mathcal{M}_{1,S}+2 \sqrt{2}\mathcal{M}_{1,SD}+\mathcal{M}_{1,D}}{3 \det\mathcal{M}_{1}} & 0 & 0 & 0 \\
 0 & 0 & \mathcal{M}_{2,D}^{-1} & 0 & 0 \\
 0 & 0 & 0 & \mathcal{M}_{3,D}^{-1} & 0 \\
 0 & 0 & 0 & 0 & \mathcal{M}_{3,D}^{-1} \\
\end{array}
\right).
\label{I001E}
\end{align}
%

\subsubsection{$\mathbf{d}=(1,1,0)$}
\begin{align}
& \mathbb{B}_1: \hspace{.5cm}
\mathbb{F}_{00}^{(\mathbb{B}_1)}=\textbf{I}_{5}\ \ \ ,\hspace{.5cm}
\mathbb{F}_{20}^{(\mathbb{B}_1)}=\left(
\begin{array}{ccccc}
 \frac{2}{\sqrt{5}} & 0 & 0 & -\frac{9}{7 \sqrt{5}} & 0 \\
 0 & -\frac{1}{\sqrt{5}} & 0 & \frac{6}{7}\sqrt{\frac{2}{5}} & 0 \\
 0 & 0 & -\frac{\sqrt{5}}{7} & 0 & -\frac{\sqrt{10}}{7} \\
 -\frac{9}{7 \sqrt{5}} & \frac{6}{7}\sqrt{\frac{2}{5}} & 0 & \frac{8}{7 \sqrt{5}} & 0 \\
 0 & 0 & -\frac{\sqrt{10}}{7} & 0 & 0 \\
\end{array}
\right),
\nonumber\\
&\hspace{1cm}\mathbb{F}_{40}^{(\mathbb{B}_1)}=\left(
\begin{array}{ccccc}
 0 & 0 & 0 & -\frac{4}{7} & 0 \\
 0 & 0 & 0 & -\frac{2 \sqrt{2}}{7}   & 0 \\
 0 & 0 & -\frac{2}{21} & 0 & \frac{5 \sqrt{2}}{21} \\
 -\frac{4}{7} & -\frac{2 \sqrt{2}}{7}   & 0 & \frac{2}{7} & 0 \\
 0 & 0 & \frac{5 \sqrt{2}}{21} & 0 & -\frac{1}{3} \\
\end{array}
\right), \hspace{.5cm} \mathbb{F}_{44}^{(\mathbb{B}_1)}=\left(
\begin{array}{ccccc}
 0 & 0 & 0 & 0 & 0 \\
 0 & 0 & 0 & 0 & 0 \\
 0 & 0 & -\frac{2}{3}\sqrt{\frac{10}{7}} & 0 & -\frac{2}{3}  \sqrt{\frac{5}{7}} \\
 0 & 0 & 0 & 0 & 0 \\
 0 & 0 & -\frac{2}{3}  \sqrt{\frac{5}{7}} & 0 & -\frac{\sqrt{\frac{10}{7}}}{3} \\
\end{array}
\right),
\nonumber\\
&\hspace{1cm}{\mathbb{M}}^{(\mathbb{B}_1)}=
\left(
\begin{array}{ccccc}
 \frac{2 \mathcal{M}_{1,S}+2
   \sqrt{2}\mathcal{M}_{1,SD}+\mathcal{M}_{1,D}}{3\det\mathcal{M}_{1}} 
& \frac{\sqrt{2} \mathcal{M}_{1,S}-\mathcal{M}_{1,SD}-\sqrt{2}\mathcal{M}_{1,D}}{3 \det\mathcal{M}_{1}} & 0 & 0 & 0 \\
 \frac{\sqrt{2}
   \mathcal{M}_{1,S}-\mathcal{M}_{1,SD}-\sqrt{2}\mathcal{M}_{1,D}}{3
   \det\mathcal{M}_{1}} 
& \frac{\mathcal{M}_{1,S}-2 \sqrt{2}\mathcal{M}_{1,SD}+2\mathcal{M}_{1,D}}{3 \det\mathcal{M}_{1}} & 0 & 0 & 0 \\
 0 & 0 & \mathcal{M}_{2,D}^{-1} & 0 & 0 \\
 0 & 0 & 0 & \mathcal{M}_{3,D}^{-1} & 0 \\
 0 & 0 & 0 & 0 & \mathcal{M}_{3,D}^{-1} \\
\end{array}
\right).
\label{I110B1}
\end{align}
\begin{align}
& \mathbb{B}_2:\hspace{.5cm}\mathbb{F}_{00}^{(\mathbb{B}_2)}=\textbf{I}_{5}\ \
\ ,\hspace{.5cm}
\mathbb{F}_{20}^{(\mathbb{B}_2)}=\left(
\begin{array}{ccccc}
 -\frac{1}{\sqrt{5}} & 0 & 0 & 0 & -\frac{3}{7}  \sqrt{\frac{6}{5}} \\
 0 & \frac{1}{2 \sqrt{5}} & -\frac{\sqrt{3}}{2} & 0 & \frac{4 \sqrt{\frac{3}{5}}}{7} \\
 0 & -\frac{\sqrt{3}}{2} & \frac{\sqrt{5}}{14} & 0 & -\frac{2}{7} \\
 0 & 0 & 0 & -\frac{2\sqrt{5}}{7}   & 0 \\
 -\frac{3}{7}  \sqrt{\frac{6}{5}} & \frac{4 \sqrt{\frac{3}{5}}}{7} & -\frac{2}{7} & 0 & \frac{6}{7 \sqrt{5}} \\
\end{array}
\right),
\nonumber\\
&\hspace{1cm}
\mathbb{F}_{40}^{(\mathbb{B}_2)}=\left(
\begin{array}{ccccc}
 0 & 0 & 0 & 0 & \frac{\sqrt{6}}{7} \\
 0 & 0 & 0 & 0 & \frac{\sqrt{3}}{7} \\
 0 & 0 & \frac{8}{21} & 0 & -\frac{5 \sqrt{5}}{21}  \\
 0 & 0 & 0 & \frac{1}{7} & 0 \\
 \frac{\sqrt{6}}{7} & \frac{\sqrt{3}}{7} & -\frac{5 \sqrt{5}}{21}  & 0 & \frac{1}{21} \\
\end{array}
\right), \hspace{0.25cm}
\mathbb{F}_{44}^{(\mathbb{B}_2)}=\left(
\begin{array}{ccccc}
 0 & 0 & 0 & \frac{2 i}{\sqrt{7}} & 0 \\
 0 & 0 & 0 & i \sqrt{\frac{2}{7}} & 0 \\
 0 & 0 & 0 & i \sqrt{\frac{10}{21}} & 0 \\
 -\frac{2 i}{\sqrt{7}} & -i \sqrt{\frac{2}{7}} & -i \sqrt{\frac{10}{21}} & 0 & -i \sqrt{\frac{2}{21}} \\
 0 & 0 & 0 & i \sqrt{\frac{2}{21}} & 0 \\
\end{array}
\right),
\nonumber\\
&\hspace{1cm}{\mathbb{M}}^{(\mathbb{B}_2)}=\left(
\begin{array}{ccccc}
 \frac{2 \mathcal{M}_{1,S}+2
   \sqrt{2}\mathcal{M}_{1,SD}+\mathcal{M}_{1,D}}{3\det\mathcal{M}_{1}} 
& \frac{\sqrt{2} \mathcal{M}_{1,S}-\mathcal{M}_{1,SD}-\sqrt{2}\mathcal{M}_{1,D}}{3 \det\mathcal{M}_{1}} & 0 & 0 & 0 \\
 \frac{\sqrt{2}
   \mathcal{M}_{1,S}-\mathcal{M}_{1,SD}-\sqrt{2}\mathcal{M}_{1,D}}{3
   \det\mathcal{M}_{1}} 
& \frac{\mathcal{M}_{1,S}-2 \sqrt{2}\mathcal{M}_{1,SD}+2\mathcal{M}_{1,D}}{3 \det\mathcal{M}_{1}} & 0 & 0 & 0 \\
 0 & 0 & \mathcal{M}_{2,D}^{-1} & 0 & 0 \\
 0 & 0 & 0 & \mathcal{M}_{3,D}^{-1} & 0 \\
 0 & 0 & 0 & 0 & \mathcal{M}_{3,D}^{-1} \\
\end{array}
\right).
\label{I110B2}
\end{align}
\begin{align}
&\mathbb{B}_3:\hspace{.5cm}
\mathbb{F}_{00}^{(\mathbb{B}_3)}=\textbf{I}_{5}\ \ \ ,\hspace{.5cm}
\mathbb{F}_{20}^{(\mathbb{B}_3)}=\left(
\begin{array}{ccccc}
 \frac{1}{2 \sqrt{5}} & 0 & -\frac{\sqrt{3}}{2} & \frac{4 \sqrt{\frac{3}{5}}}{7} & 0 \\
 0 & -\frac{1}{\sqrt{5}} & 0 & -\frac{3}{7} \sqrt{\frac{6}{5}} & 0 \\
 -\frac{\sqrt{3}}{2} & 0 & \frac{\sqrt{5}}{14} & -\frac{2}{7} & 0 \\
 \frac{4 \sqrt{\frac{3}{5}}}{7} & -\frac{3}{7} \sqrt{\frac{6}{5}} & -\frac{2}{7} & \frac{6}{7 \sqrt{5}} & 0 \\
 0 & 0 & 0 & 0 & -\frac{2 \sqrt{5}}{7}  \\
\end{array}
\right),
\nonumber\\
&\hspace{1cm}
\mathbb{F}_{40}^{(\mathbb{B}_3)}=\left(
\begin{array}{ccccc}
 0 & 0 & 0 & \frac{\sqrt{3}}{7} & 0 \\
 0 & 0 & 0 & \frac{\sqrt{6}}{7} & 0 \\
 0 & 0 & \frac{8}{21} & -\frac{5 \sqrt{5}}{21}  & 0 \\
 \frac{\sqrt{3}}{7} & \frac{\sqrt{6}}{7} & -\frac{5 \sqrt{5}}{21}  & \frac{1}{21} & 0 \\
 0 & 0 & 0 & 0 & \frac{1}{7} \\
\end{array}
\right), \hspace{0.25cm}
\mathbb{F}_{44}^{(\mathbb{B}_3)}=\left(
\begin{array}{ccccc}
 0 & 0 & 0 & 0 & -i \sqrt{\frac{2}{7}} \\
 0 & 0 & 0 & 0 & -\frac{2 i}{\sqrt{7}} \\
 0 & 0 & 0 & 0 & -i \sqrt{\frac{10}{21}} \\
 0 & 0 & 0 & 0 & -i \sqrt{\frac{2}{21}} \\
 i \sqrt{\frac{2}{7}} & \frac{2 i}{\sqrt{7}} & i \sqrt{\frac{10}{21}} & i \sqrt{\frac{2}{21}} & 0 \\
\end{array}
\right),
\nonumber
\end{align}
\begin{align}
&{\mathbb{M}}^{(\mathbb{B}_3)}=
\left(
\begin{array}{ccccc}
 \frac{\mathcal{M}_{1,S}-2 \sqrt{2}\mathcal{M}_{1,SD}+2\mathcal{M}_{1,D}}{3
   \det\mathcal{M}_{1}} 
& \frac{\sqrt{2} \mathcal{M}_{1,S}-\mathcal{M}_{1,SD}-\sqrt{2}\mathcal{M}_{1,D}}{3 \det\mathcal{M}_{1}} & 0 & 0 & 0 \\
 \frac{\sqrt{2}
   \mathcal{M}_{1,S}-\mathcal{M}_{1,SD}-\sqrt{2}\mathcal{M}_{1,D}}{3
   \det\mathcal{M}_{1}} 
& \frac{2 \mathcal{M}_{1,S}+2 \sqrt{2}\mathcal{M}_{1,SD}+\mathcal{M}_{1,D}}{3 \det\mathcal{M}_{1}} & 0 & 0 & 0 \\
 0 & 0 & \mathcal{M}_{2,D}^{-1} & 0 & 0 \\
 0 & 0 & 0 & \mathcal{M}_{3,D}^{-1} & 0 \\
 0 & 0 & 0 & 0 & \mathcal{M}_{3,D}^{-1} \\
\end{array}
\right).
\label{I110B3}
\end{align}
%

\subsubsection{$\mathbf{d}=(1,1,1)$}
\begin{align}
& \mathbb{A}_2:\hspace{.4cm}
\mathbb{F}_{00}^{(\mathbb{A}_2)}=\textbf{I}_{4}\ \ \ ,\hspace{.4cm}
\mathbb{F}_{40}^{(\mathbb{A}_2)}=
\left(
\begin{array}{cccc}
 0 & 0 & 0 & 0 \\
 0 & 0 & \frac{2 \sqrt{6}}{7} & 0 \\
 0 & \frac{2 \sqrt{6}}{7} & \frac{2}{7} & 0 \\
 0 & 0 & 0 & -\frac{4}{7}
\end{array}
\right), 
\nonumber\\
& \hspace{1.1cm}
{\mathbb{M}}^{(\mathbb{A}_2)}=\left(
\begin{array}{cccc}
 \frac{\mathcal{M}_{1,D}}{{\det\mathcal{M}_{1}}} & -\frac{\mathcal{M}_{1,SD}}{\det\mathcal{M}_{1}} & 0 & 0 \\
 -\frac{\mathcal{M}_{1,SD}}{\det\mathcal{M}_{1}} & \frac{\mathcal{M}_{1,S}}{\det\mathcal{M}_{1}} & 0 & 0 \\
 0 & 0 &\mathcal{M}_{3,D}^{-1} & 0 \\
 0 & 0 & 0 & \mathcal{M}_{3,D}^{-1} \\
\end{array}
\right).
\label{I111A2}
\\
& \mathbb{E}:\hspace{.5cm}\mathbb{F}_{00}^{(\mathbb{E})}=\textbf{I}_{6}\ \ \ ,\hspace{.5cm}
\mathbb{F}_{40}^{(\mathbb{E})}=
\left(
\begin{array}{cccccc}
 0 & 0 & 0 & 0 & 0 & 0 \\
 0 & 0 & \frac{2 \sqrt{6}}{7} & 0 & 0 & 0 \\
 0 & \frac{2 \sqrt{6}}{7} & \frac{2}{7} & 0 & 0 & 0 \\
 0 & 0 & 0 & \frac{8}{21} & -\frac{10 \sqrt{2}}{21} & 0 \\
 0 & 0 & 0 & -\frac{10 \sqrt{2}}{21} & -\frac{2}{21} & 0 \\
 0 & 0 & 0 & 0 & 0 & -\frac{4}{7}
\end{array}
\right),
\nonumber\\
&\hspace{1cm}{\mathbb{M}}^{(\mathbb{E})}=
 \left(
\begin{array}{cccccc}
  \frac{\mathcal{M}_{1,D}}{{\det\mathcal{M}_{1}}} & -\frac{\mathcal{M}_{1,SD}}{\det\mathcal{M}_{1}} & 0 & 0 & 0 & 0 \\
-\frac{\mathcal{M}_{1,SD}}{\det\mathcal{M}_{1}} & \frac{\mathcal{M}_{1,S}}{\det\mathcal{M}_{1}} & 0 & 0 & 0 & 0 \\
 0 & 0 & \mathcal{M}_{3,D}^{-1} & 0 & 0 & 0 \\
 0 & 0 & 0 & \mathcal{M}_{2,D}^{-1} & 0 & 0 \\
 0 & 0 & 0 & 0 & \mathcal{M}_{3,D}^{-1} & 0 \\
 0 & 0 & 0 & 0 & 0 & \mathcal{M}_{2,D}^{-1} \\
\end{array}
\right).
\label{I111E}
\end{align}
\end{small}

%
\section{The Finite-Volume $c^{\mathbf{d}}_{LM}$ Functions \label{app: clm}}
\begin{figure}[t!]
\begin{center}  
\subfigure[]{
\includegraphics[scale=0.235]{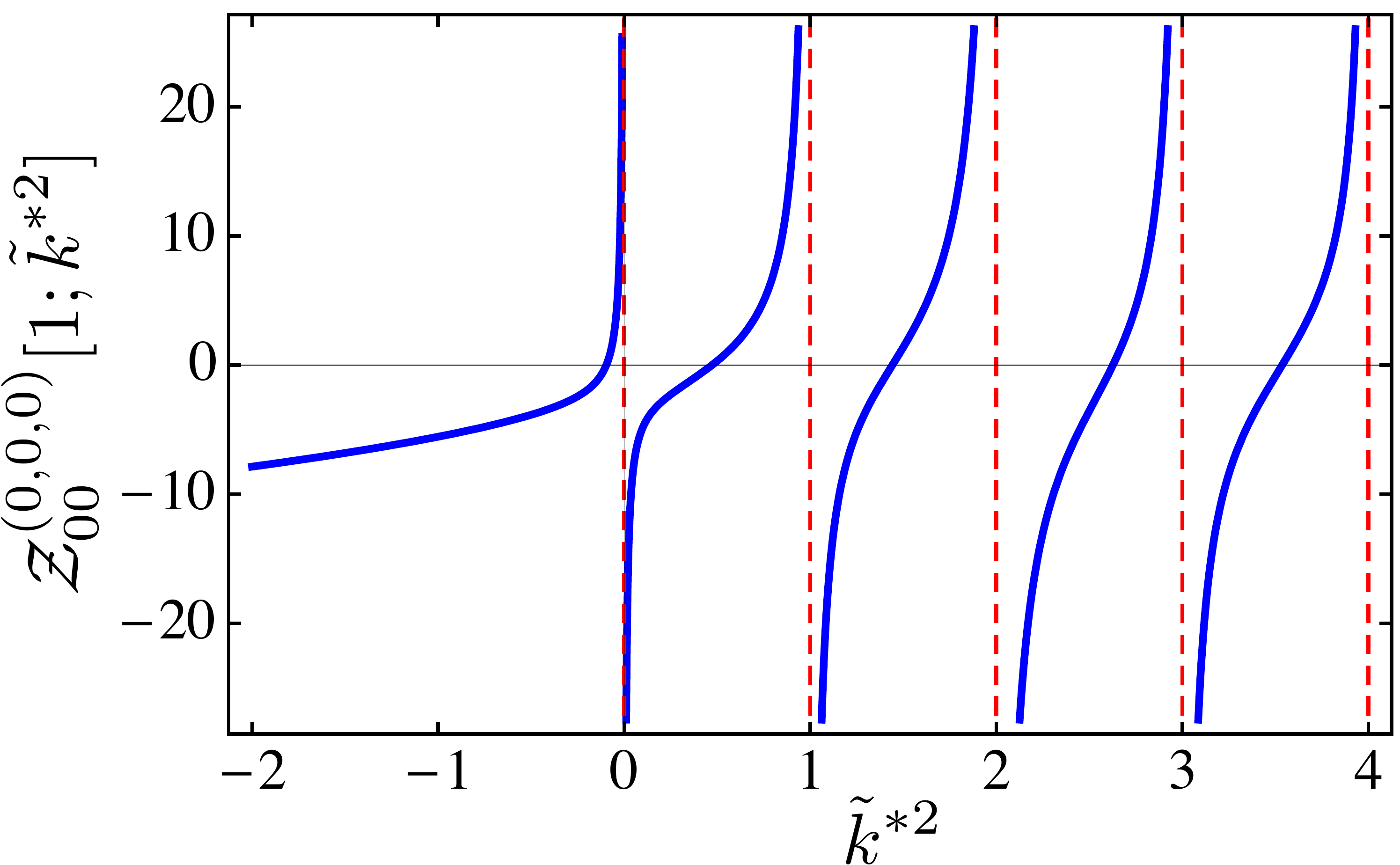}}
\subfigure[]{
\includegraphics[scale=0.235]{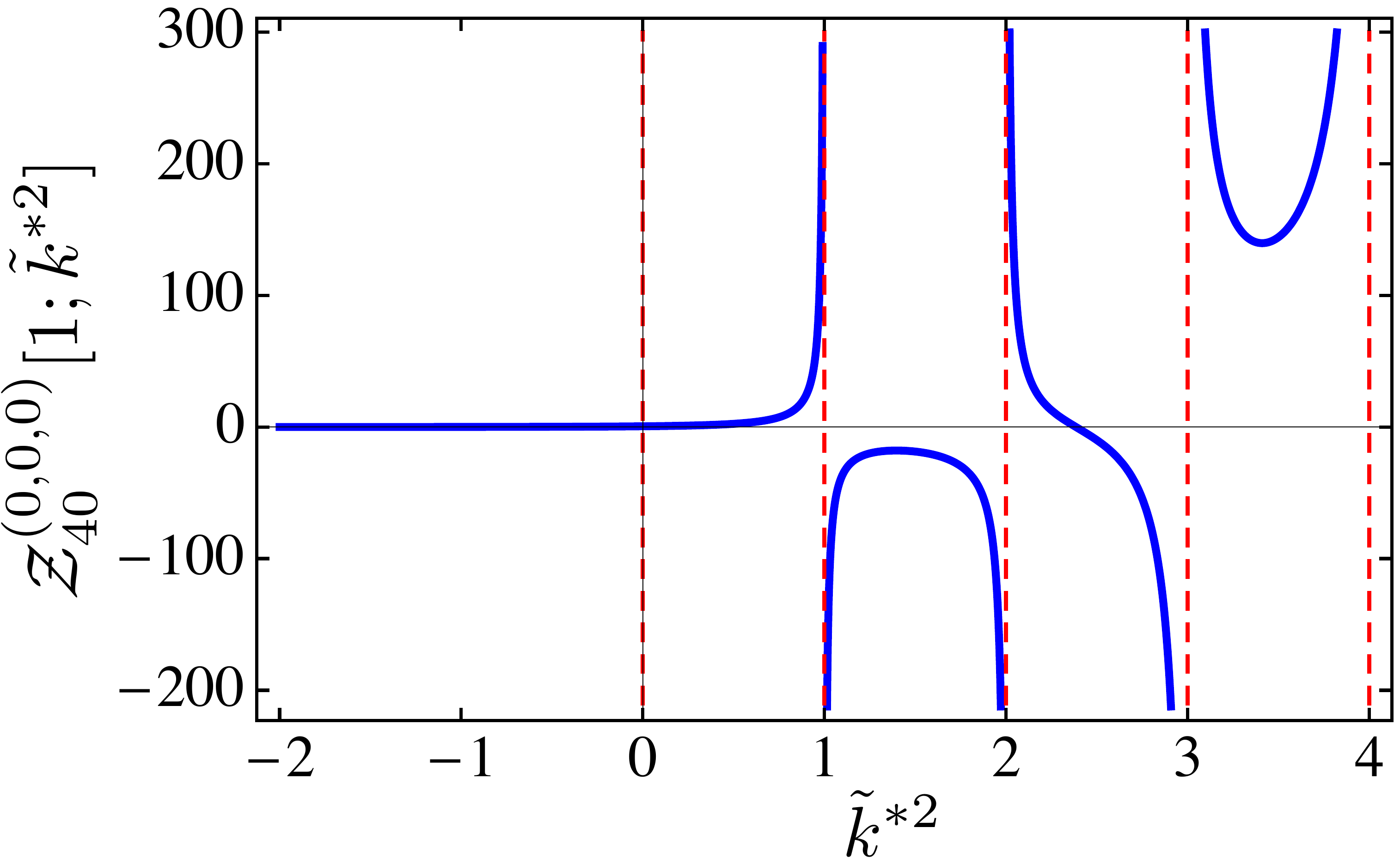}}
\caption{{\small a) $\mathcal{Z}_{00}^{\mathbf{d}}$ and b)
  $\mathcal{Z}_{40}^{\mathbf{d}}$ for $\mathbf{d}=(0,0,0)$ as a function of
  $\tilde{k}^{*2}=k^{*2} {\rm L}^2/4\pi^2$.}}
\label{fig:Z-func}
\end{center}
\end{figure}

\noindent
The FV NN energy spectra in chapter \ref{chap:deuteron} are determined by
the $c_{LM}^{\mathbf{d}}(k^{*2}; {\rm L})$ functions that are defined in
Eq.~(\ref{clm}). 
They are smooth analytic functions of $k^{*2}$ for negative
values of $k^{*2}$, but have poles at $k^{*2}=\frac{4\pi^2}{\rm L^2}(\mathbf{n}-\mathbf{d}/2)^2$,
where $\mathbf{n}$ is an integer triplet, corresponding to the
energy of two non-interacting 
nucleons in a cubic volume with the
PBCs. 
In obtaining  the spectra  in the
positive-parity isoscalar channels from the $\mathbb{T}_1$ irrep of the cubic
group that are shown in 
Fig.~\ref{T1specfull}, 
the $c_{00}^{(0,0,0)}(k^{*2}; {\rm L})
=
\mathcal{Z}_{00}^{(0,0,0)}[1;\tilde{k}^{*2}]/(2\pi^{3/2} {\rm L} )$ 
and $c_{40}^{(0,0,0)}(k^{*2}; {\rm
  L})=\mathcal{Z}_{40}^{(0,0,0)}[1;\tilde{k}^{*2}]/(8\pi^{5/2} {\rm L}^5)$ 
functions 
have been determined. 
The corresponding $\mathcal{Z}$ functions are shown 
in Fig.~\ref{fig:Z-func} as a function of $\tilde{k}^{*2}$,
see Ref.~\cite{Luu:2011ep}.  

When $k^{*2}=-\kappa^2\leq0$, the exponential volume dependence of the
$c_{LM}^{\mathbf{d}}$
can be made explicit 
by performing a Poisson resummation of Eq.~(\ref{clm}),
\begin{align}
& c^\mathbf{d}_{00}(-\kappa^2; {\rm L})=-\frac{\kappa}{4\pi}+\sum_{\textbf{n}\neq\textbf{0}}{e^{i \pi \textbf{n}\cdot\textbf{d}}}
\ \frac{e^{-n\kappa {\rm L}}}{4\pi n {\rm L}}
,
\label{c00-exp}
\\
& c^\mathbf{d}_{20}(-\kappa^2;{\rm L})=
-\kappa^2\sqrt{4\pi}\sum_{\textbf{n}\neq\textbf{0}}{e^{i \pi
    \textbf{n}\cdot\textbf{d}}}
\ Y_{20}(\hat{\mathbf{n}})\left(1+\frac{3}{n\kappa
    {\rm L}} +\frac{3}{n^2\kappa^{2}{\rm L}^2}\right)
\frac{e^{-n\kappa {\rm L}}}{4\pi n{\rm L}}
,
\label{c20-exp}
\\
& c^\mathbf{d}_{40(\pm4)}(-\kappa^2;{\rm L})=
\kappa^4\sqrt{4\pi}\sum_{\textbf{n}\neq\textbf{0}}{e^{i \pi
    \textbf{n}\cdot\textbf{d}}}
\ Y_{40(\pm4)}(\hat{\mathbf{n}})\left(1+\frac{10}{n\kappa {\rm L}} 
+\frac{45}{n^2\kappa^{2}{\rm L}^2}+\frac{105}{n^3\kappa^{3}{\rm L}^3}+\frac{105}{n^4\kappa^{4}{\rm L}^4}\right)
\frac{e^{-n\kappa {\rm L}}}{4\pi n{\rm L}}
,
\label{c40-exp}
\end{align}
where $\mathbf{n}$ is an integer triplet and $n=|{\bf n}|$. 
The expansions of 
$c_{20}^{\mathbf{d}}$ and $c_{40(\pm 4)}^{\mathbf{d}}$ start at
$\sim {1\over{\rm L}}e^{-\kappa {\rm L}}$, 
while $c_{00}^{\mathbf{d}}$ has a leading term that does not vanish in the
infinite-volume limit. 
It is also evident from these relations that $c_{20}^{\mathbf{d}}$ is
non-vanishing only for 
$\mathbf{d}=(0,0,1)$ and $(1,1,0)$, 
which gives rise to the $\mathcal{O}(\sin \epsilon_1)$ contributions to the
corresponding QCs given in Sec. \ref{sec:DeutFV}.

\begin{figure}[t!]
\begin{center}  
\subfigure[]{
\label{dZ00P000}
\includegraphics[scale=0.210]{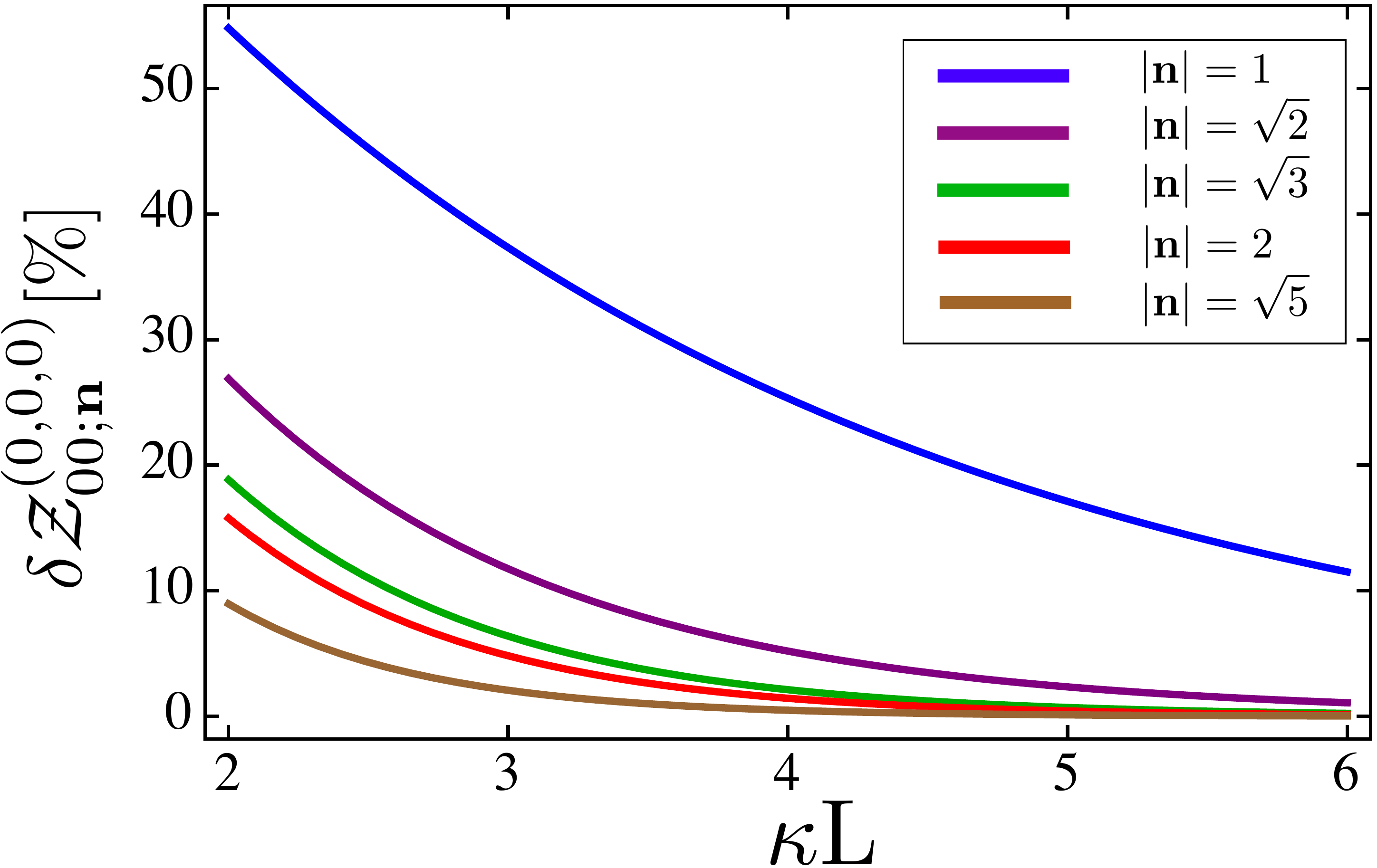}}
\subfigure[]{
\label{dZ00P001}
\includegraphics[scale=0.210]{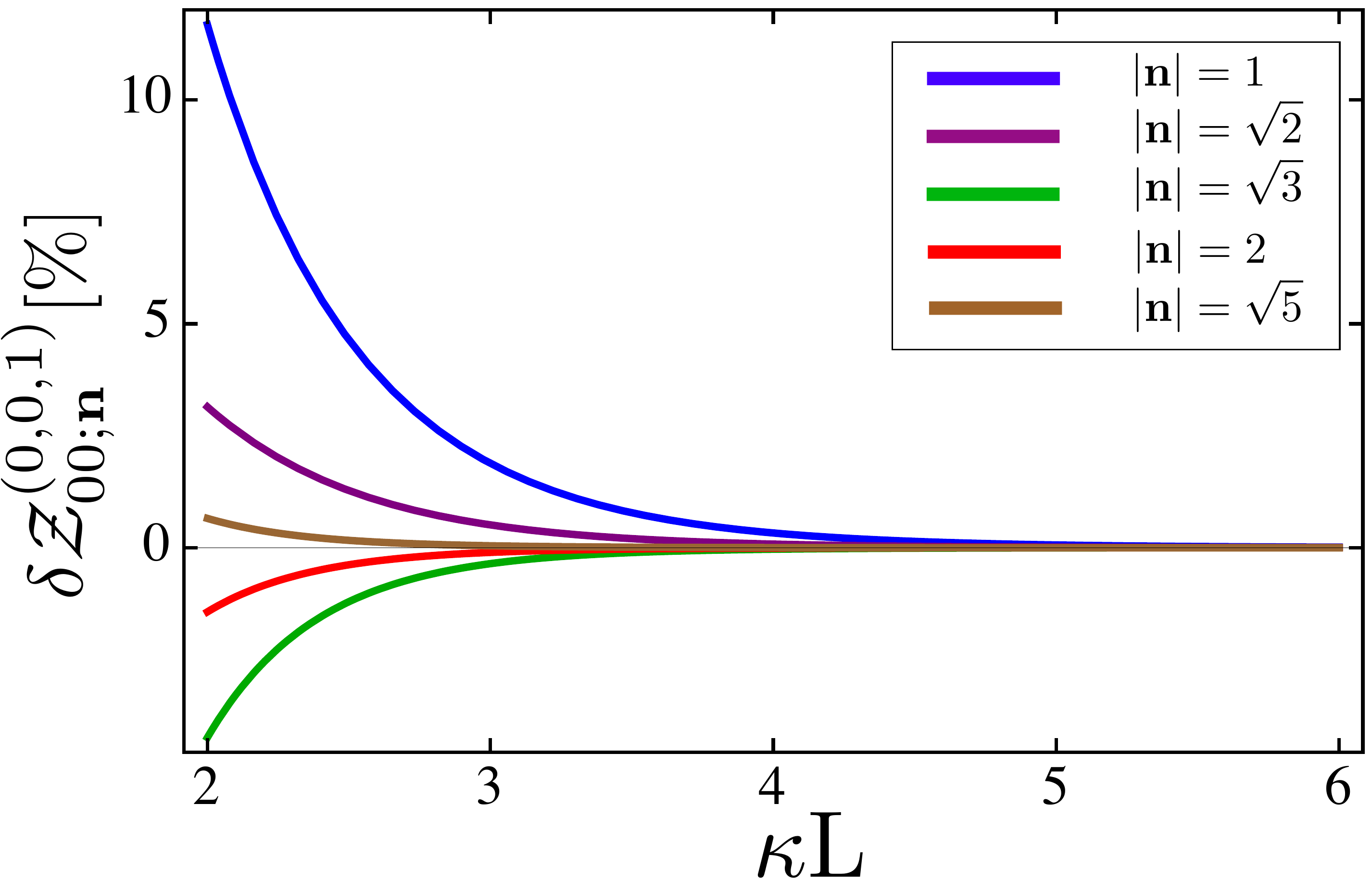}}
\subfigure[]{
\label{dZ00P110}
\includegraphics[scale=0.210]{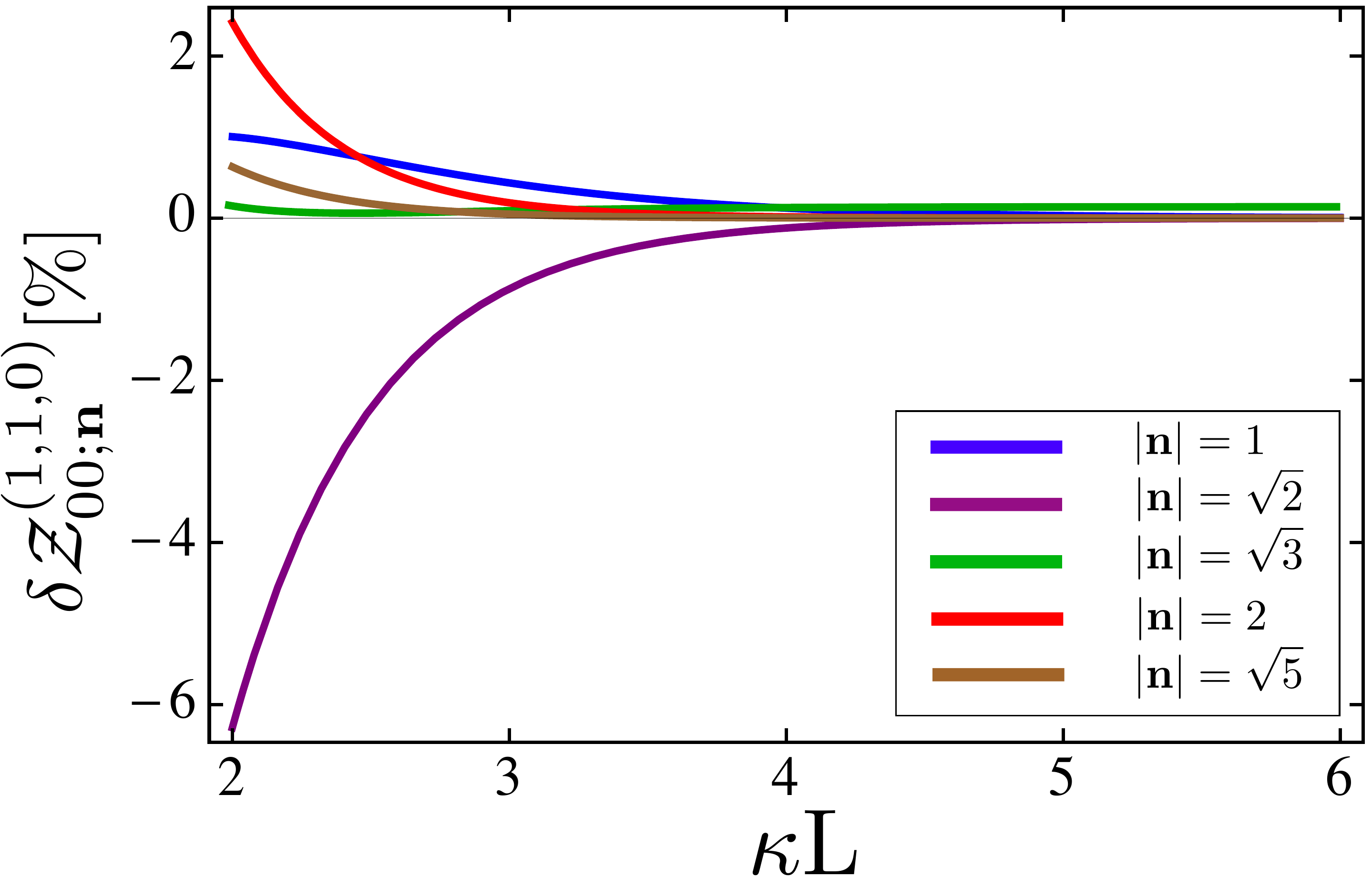}}
\subfigure[]{
\label{dZ00P111}
\includegraphics[scale=0.210]{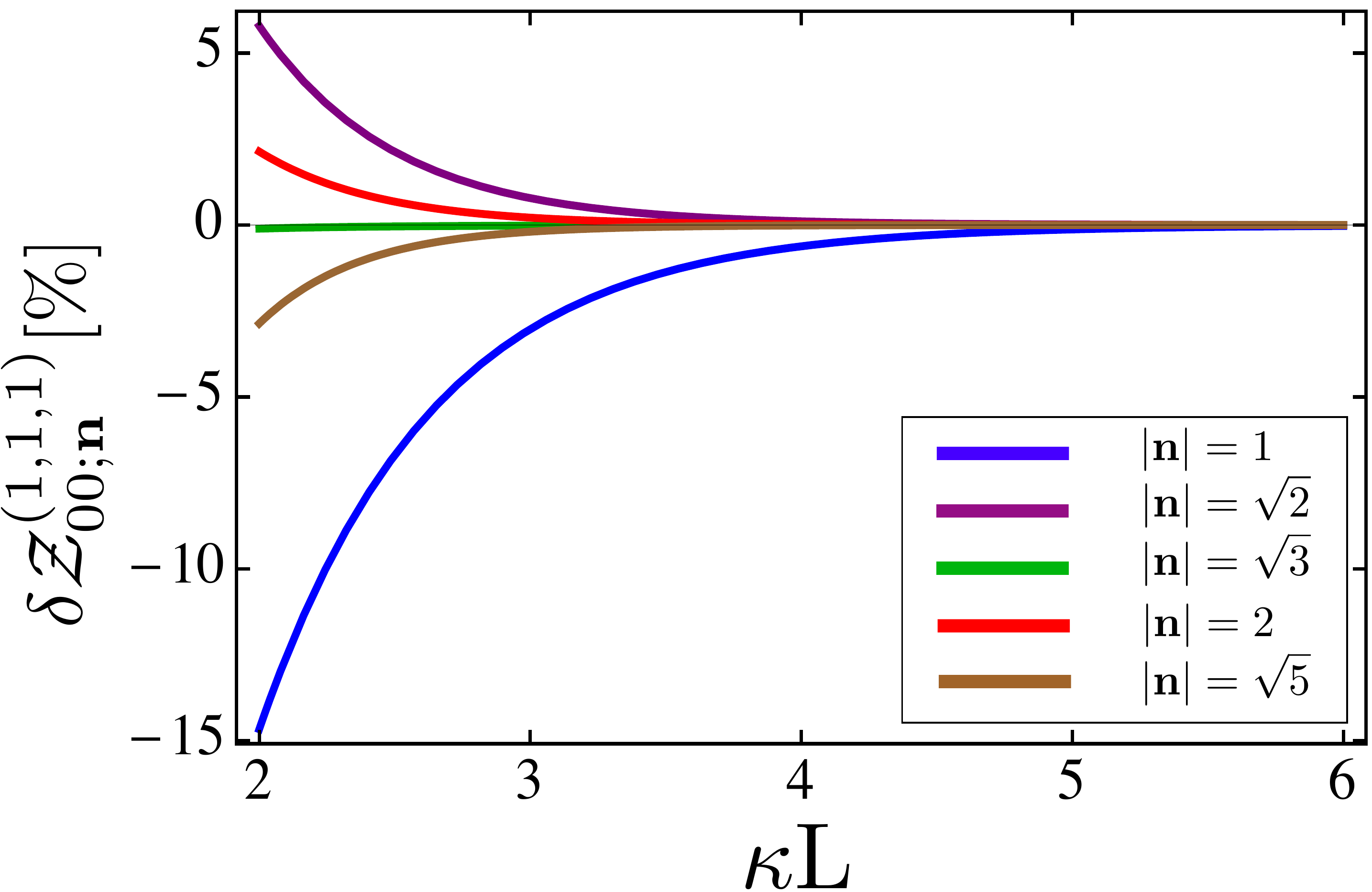}}
\caption{{\small The quantities
  $\delta\mathcal{Z}^{\mathbf{d}}_{00;\mathbf{n}}
\equiv\frac{1}{\mathcal{Z}_{00}^{\mathbf{d}}}(\mathcal{Z}_{00}^{\mathbf{d}}-\mathcal{Z}_{00;\mathbf{n}}^{\mathbf{d}})$
(in percent) as a function of $\kappa {\rm L}$ for different
boosts. $\mathcal{Z}_{00;\mathbf{n}}^{\mathbf{d}}$ 
denotes the value of the $\mathcal{Z}_{00}$-function when the sum in
Eq.~(\ref{c00-exp}) is truncated 
to a maximum shell $\mathbf{n}$.}}
\label{fig:dZ}
\end{center}
\end{figure}

Previous works~\cite{Beane:2006mx, Beane:2012vq,  Beane:2013br,
  Yamazaki:2012hi, Borasoy:2006qn, Lee:2008fa, Bour:2012hn, Davoudi:2011md},
have proposed extracting the infinite-volume deuteron binding energy from the
FV spectra using the S-wave QC expanded around
the infinite-volume deuteron pole, $\kappa_d^{\infty}$,
retaining only a 
finite number of terms in the expansion of the $c_{LM}^{\mathbf{d}}$. 
Fig. \ref{fig:dZ} shows the quantity
$\delta\mathcal{Z}^{\mathbf{d}}_{00;\mathbf{n}}
\equiv\frac{1}{\mathcal{Z}_{00}^{\mathbf{d}}}(\mathcal{Z}_{00}^{\mathbf{d}}-\mathcal{Z}_{00;\mathbf{n}}^{\mathbf{d}})$
as a function of $\kappa {\rm L}$ for different boosts. $\mathcal{Z}_{00;\mathbf{n}}^{\mathbf{d}}$ denotes the value of the
$\mathcal{Z}_{00}$-function when the sum in Eq.~(\ref{c00-exp}) is truncated to
a maximum shell $\mathbf{n}$. 
As can be seen, for modest volumes truncating the $\mathcal{Z}$-functions can lead to large deviations
from the exact values.

\section{Finite-Volume Deuteron Wavefunctions
\label{sec:Wavefunc}
}
\begin{figure}[b!]
\begin{center}  
\subfigure[]{
\label{WF-A2-L10}
\includegraphics[scale=0.215]{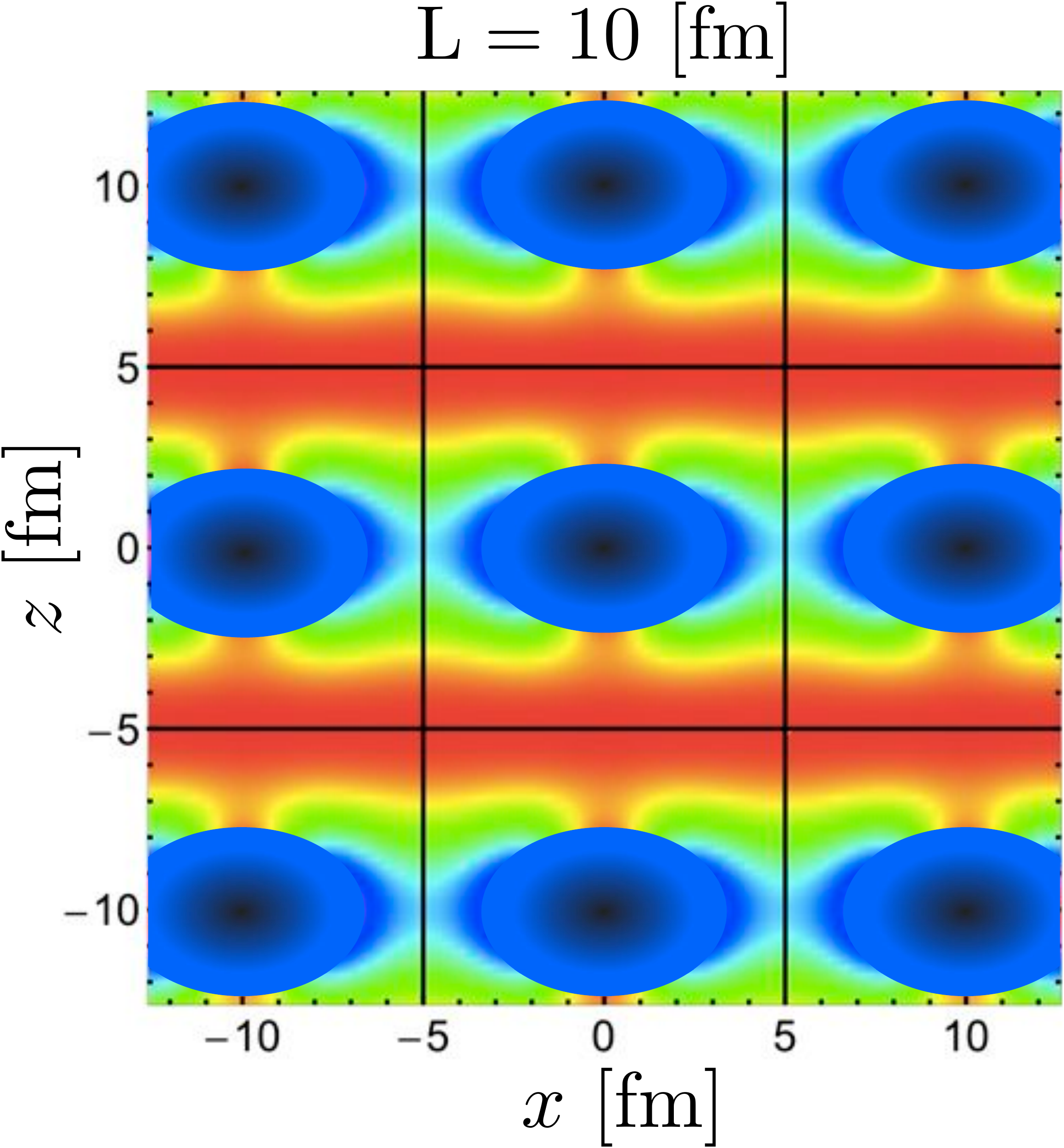}}
\subfigure[]{
\label{WF-A2-L15}
\includegraphics[scale=0.215]{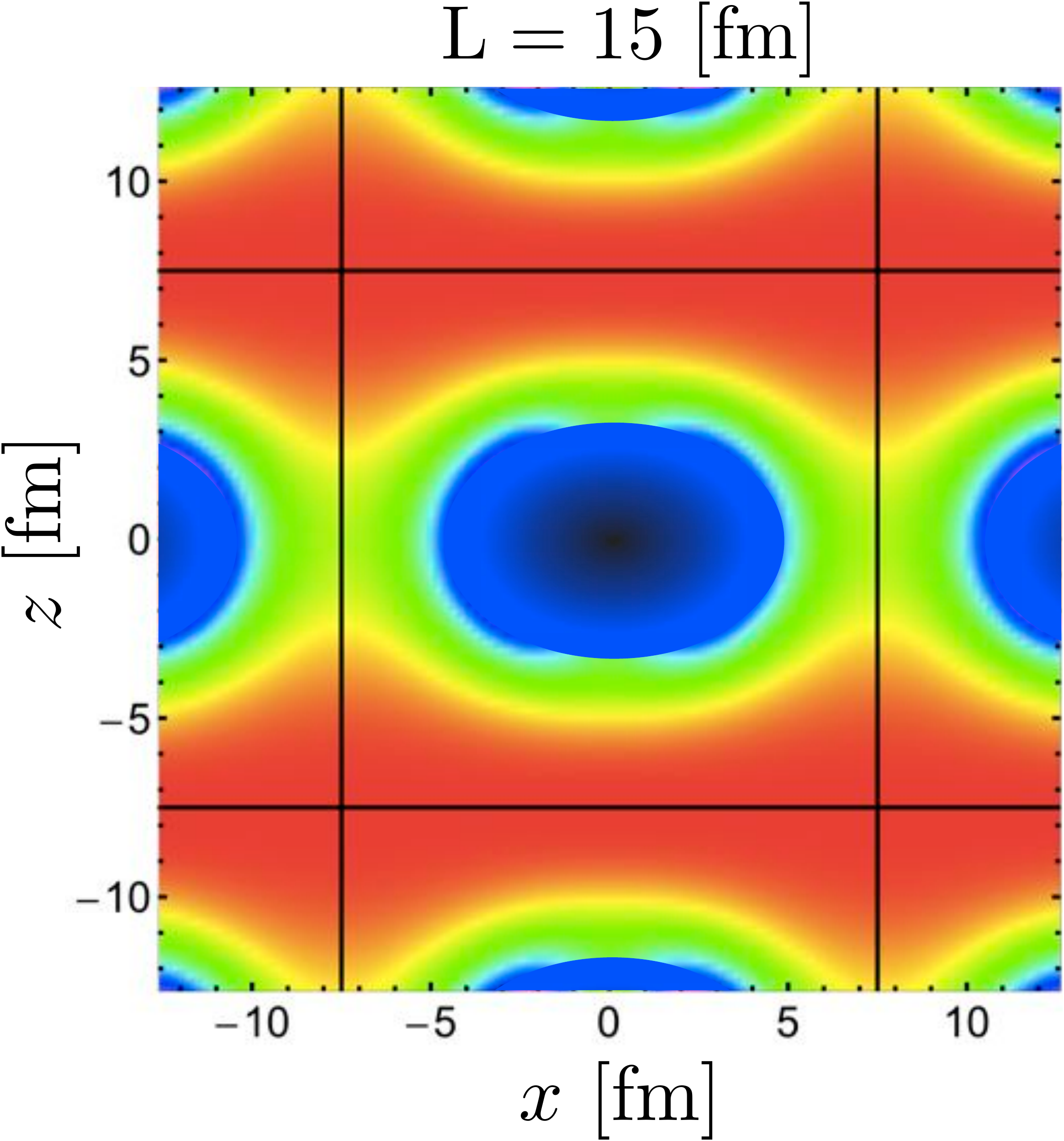}}
\subfigure[]{
\label{WF-A2-20}
\includegraphics[scale=0.215]{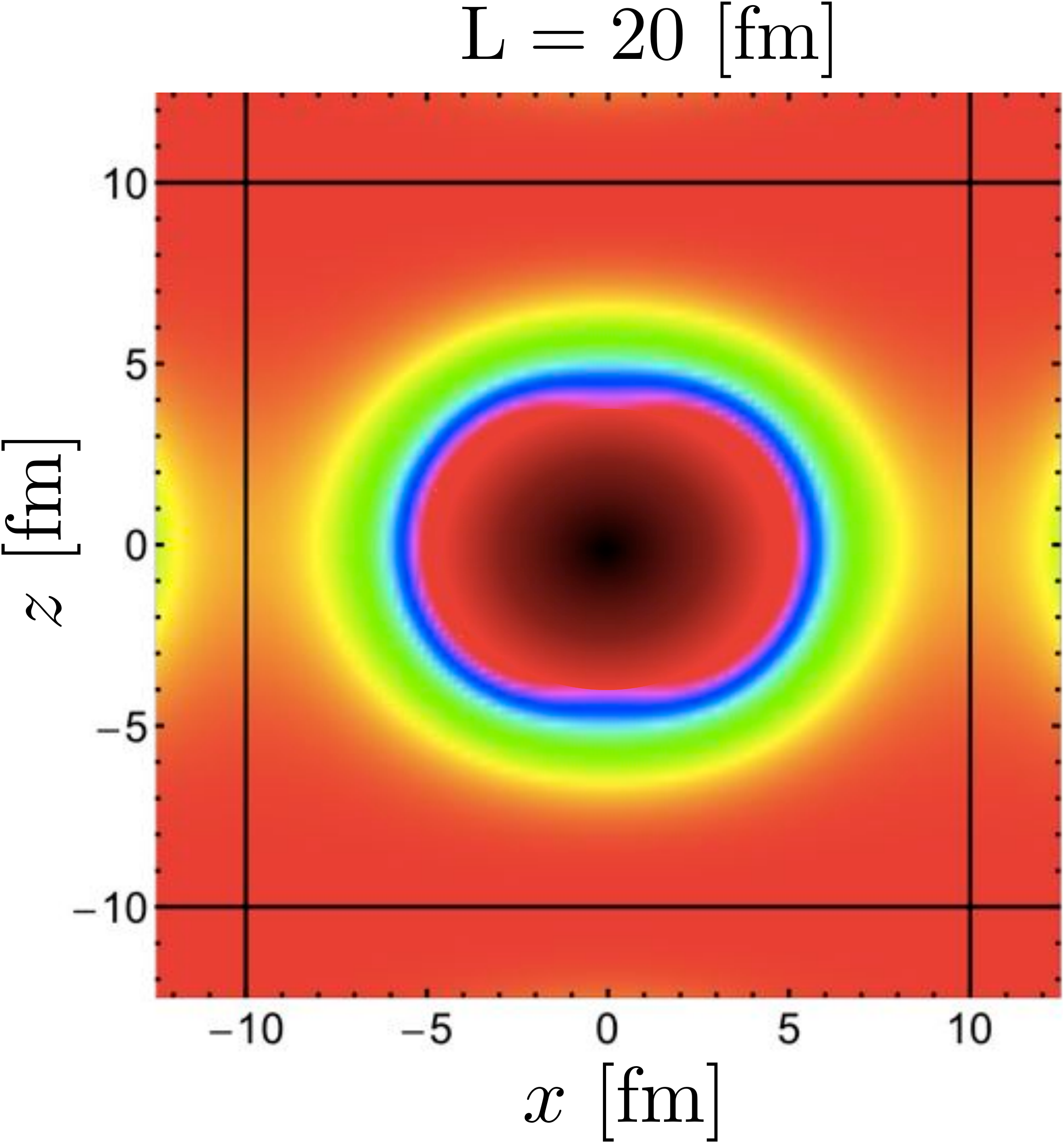}}
\subfigure[]{
\label{WF-A2-30}
\includegraphics[scale=0.215]{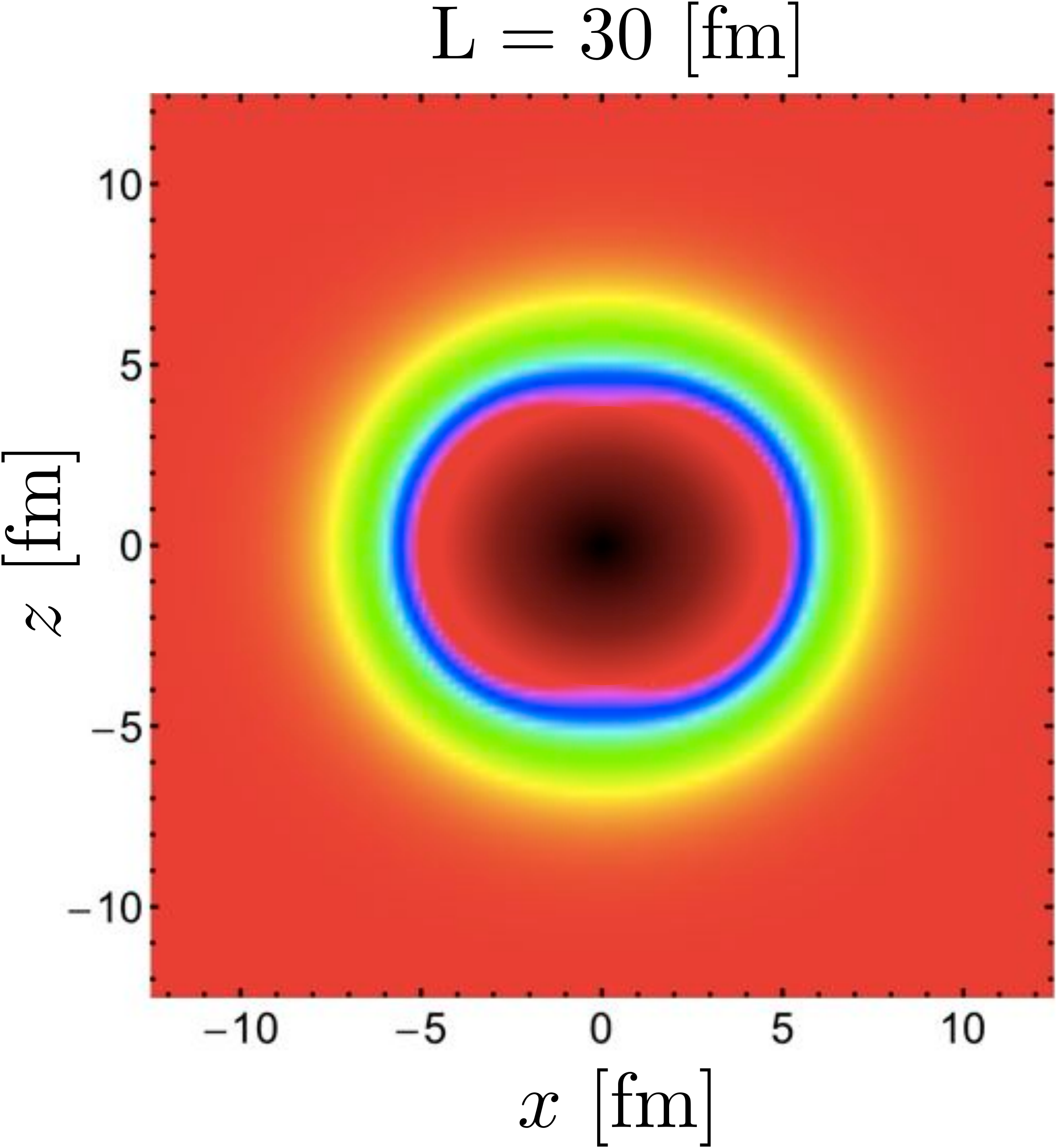}}
\caption{{\small 
The mass density  in the $xz$-plane from the $\mathbb{A}_2$ FV deuteron wavefunction with
  $\mathbf{d}=(0,0,1)$.
}}
\label{WF-A2}
\end{center}
\end{figure}
\noindent
It is useful to visualize how the deuteron is distorted within a finite volume,
and in this appendix, 
based on the asymptotic FV wavefunction of the deuteron given in Eq.~(\ref{psi-V}), 
we show the mass density  in the $xz$-plane from selected wavefunctions.  
As the interior region is not 
described by the asymptotic form of the wavefunction, 
it is ``masked'' by a shaded disk in the following figures. 
In each figure,
the black straight lines separate adjacent lattice volumes that contain the
periodic images of the wavefunction.

\begin{figure}[h!]
\begin{center}  
\subfigure[]{
\label{WF-E-L10}
\includegraphics[scale=0.215]{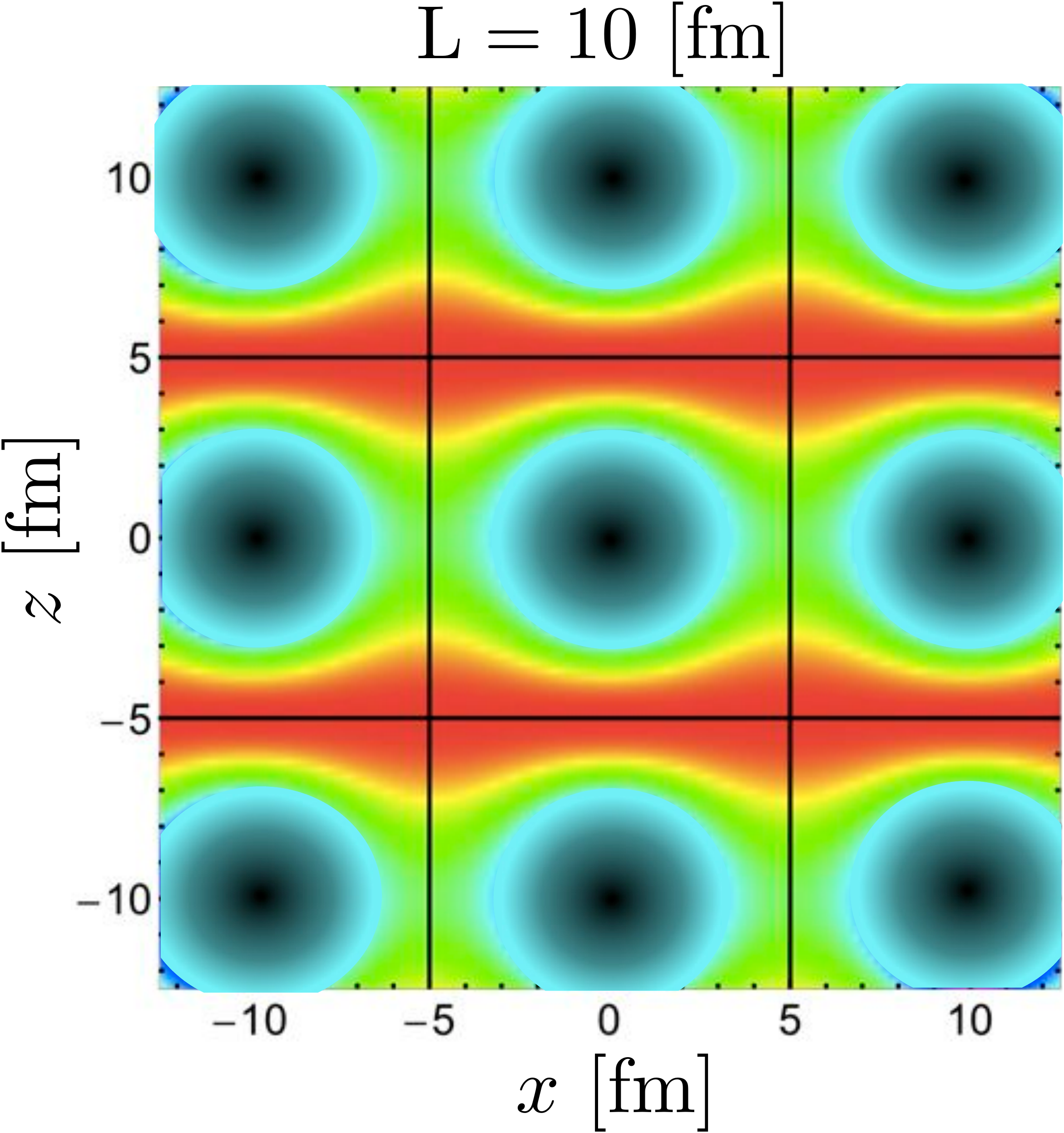}}
\subfigure[]{
\label{WF-E-L15}
\includegraphics[scale=0.215]{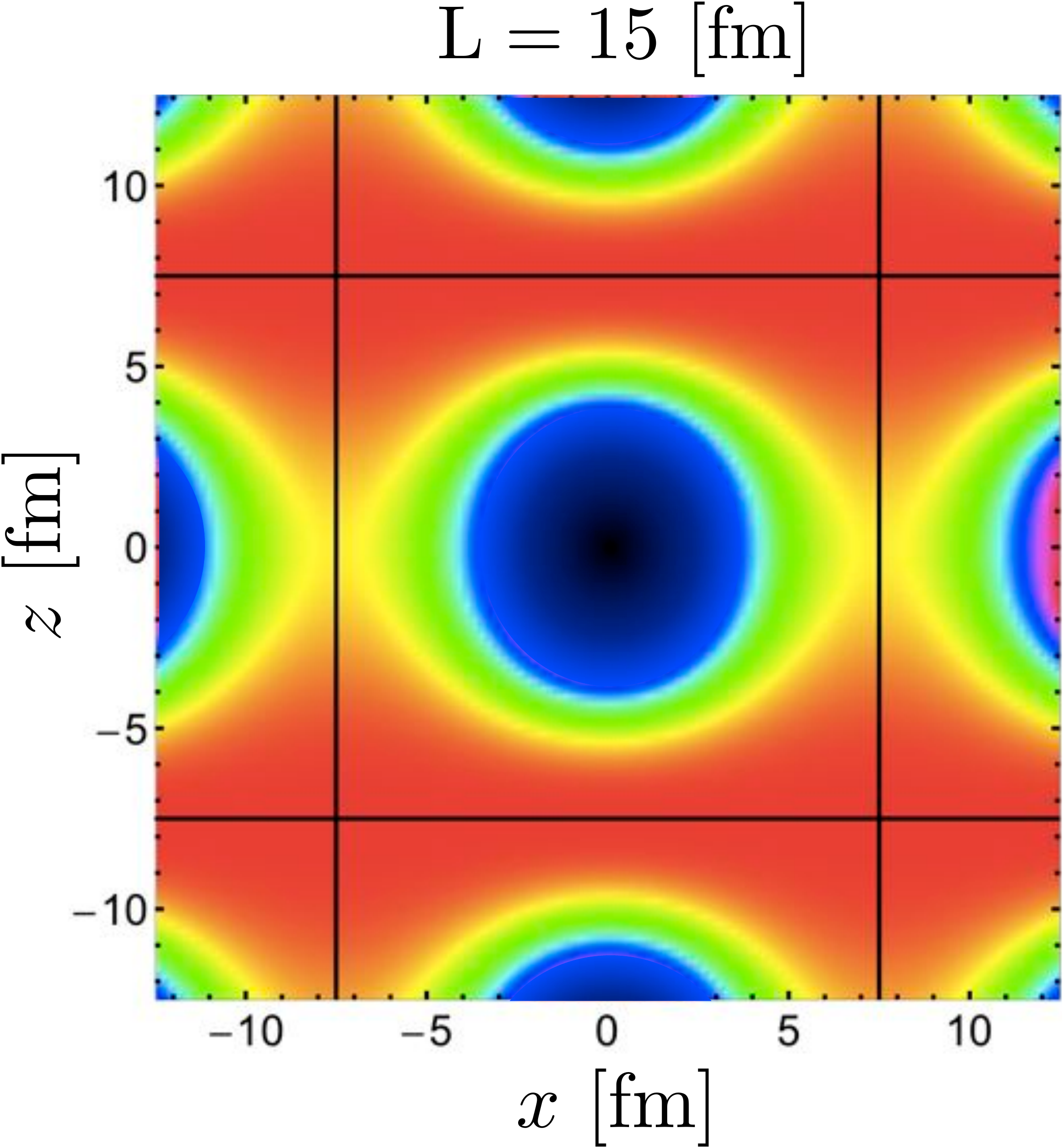}}
\subfigure[]{
\label{WF-E-20}
\includegraphics[scale=0.215]{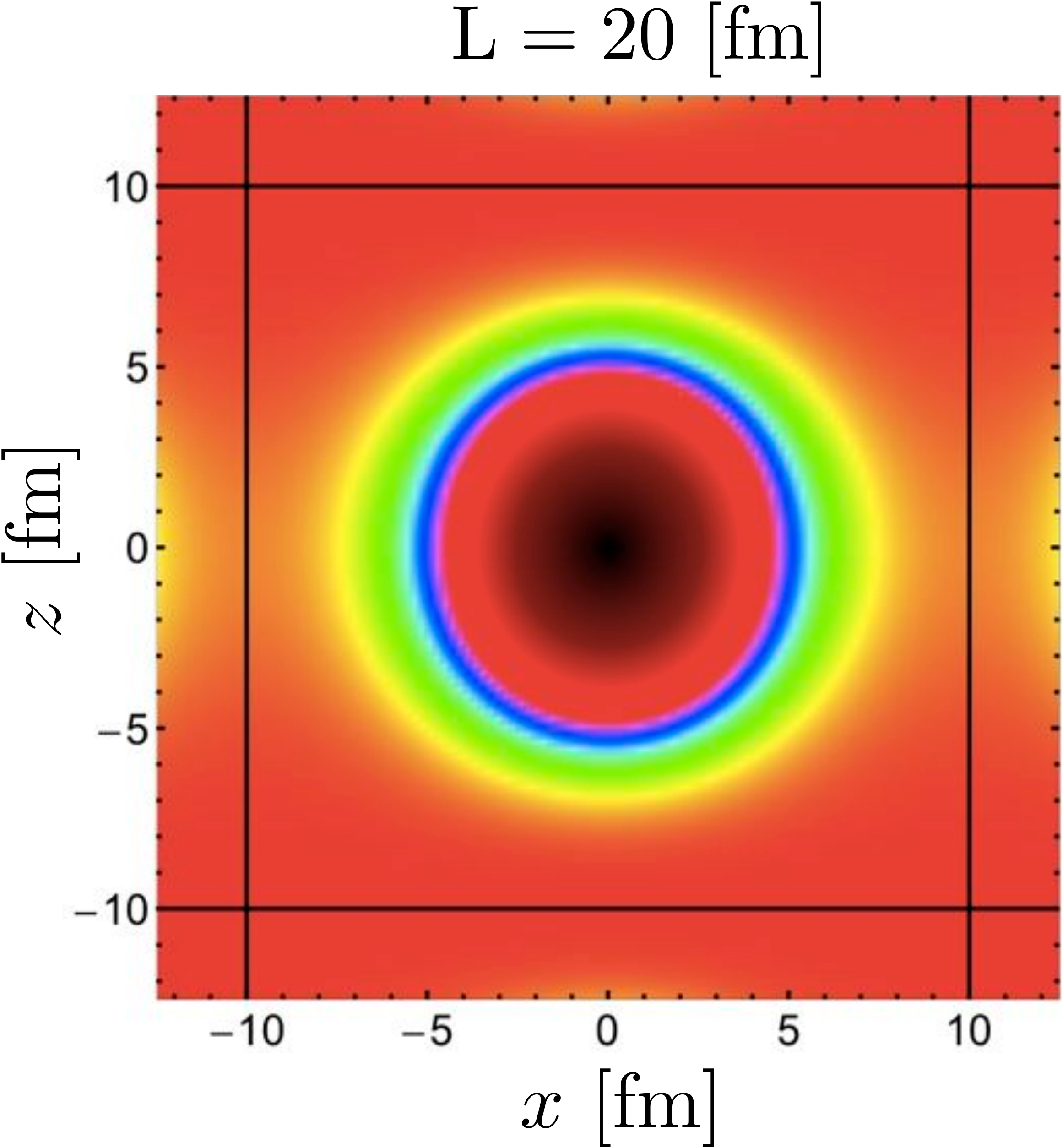}}
\subfigure[]{
\label{WF-E-30}
\includegraphics[scale=0.215]{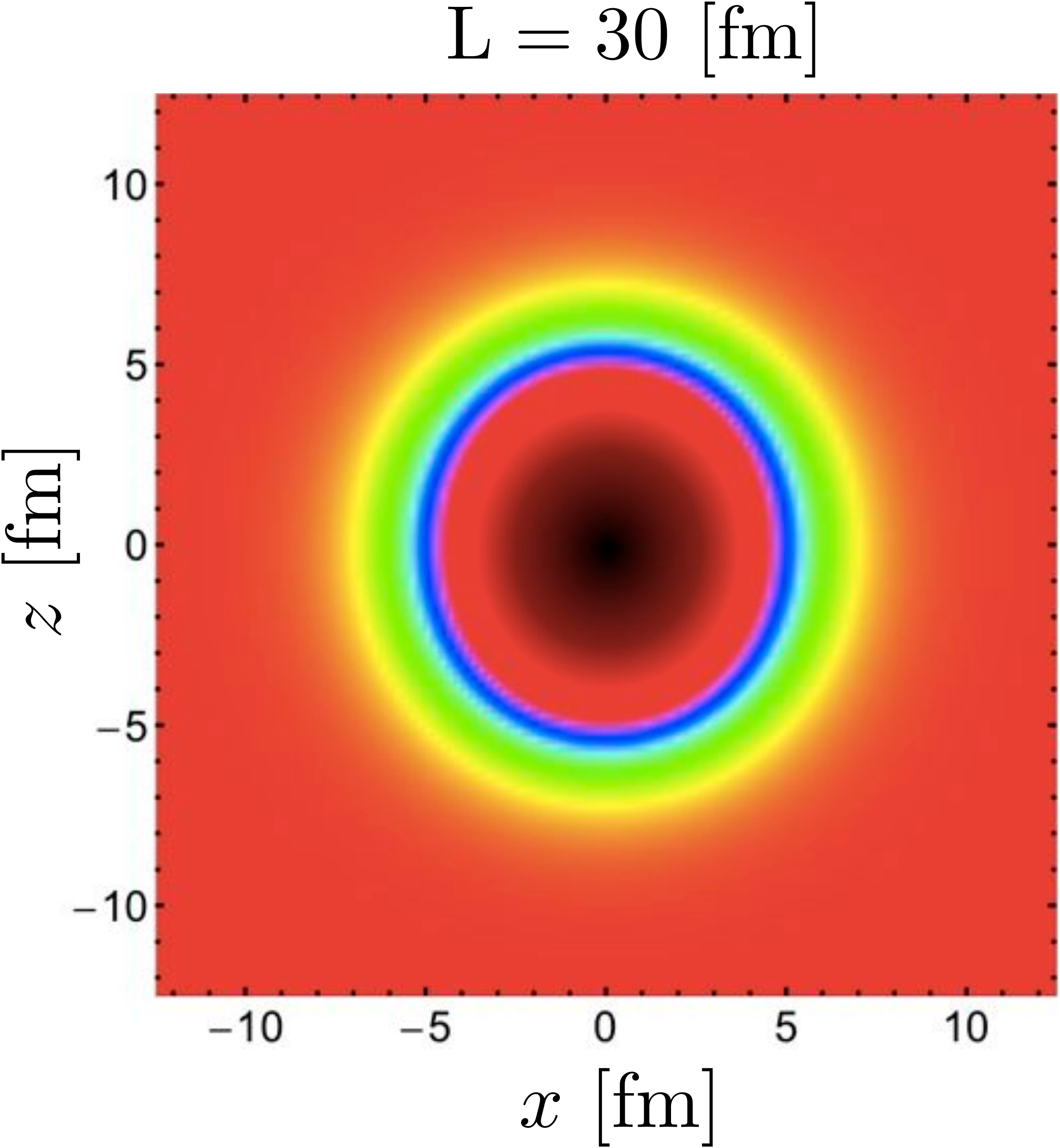}}
\caption{{\small 
The mass density  in the $xz$-plane from the $\mathbb{E}$ FV deuteron wavefunction with
  $\mathbf{d}=(0,0,1)$.
}}
\label{WF-E}
\end{center}
\end{figure}

\begin{figure}[h!]
\begin{center}  
\subfigure[]{
\label{WF-B1-L10}
\includegraphics[scale=0.215]{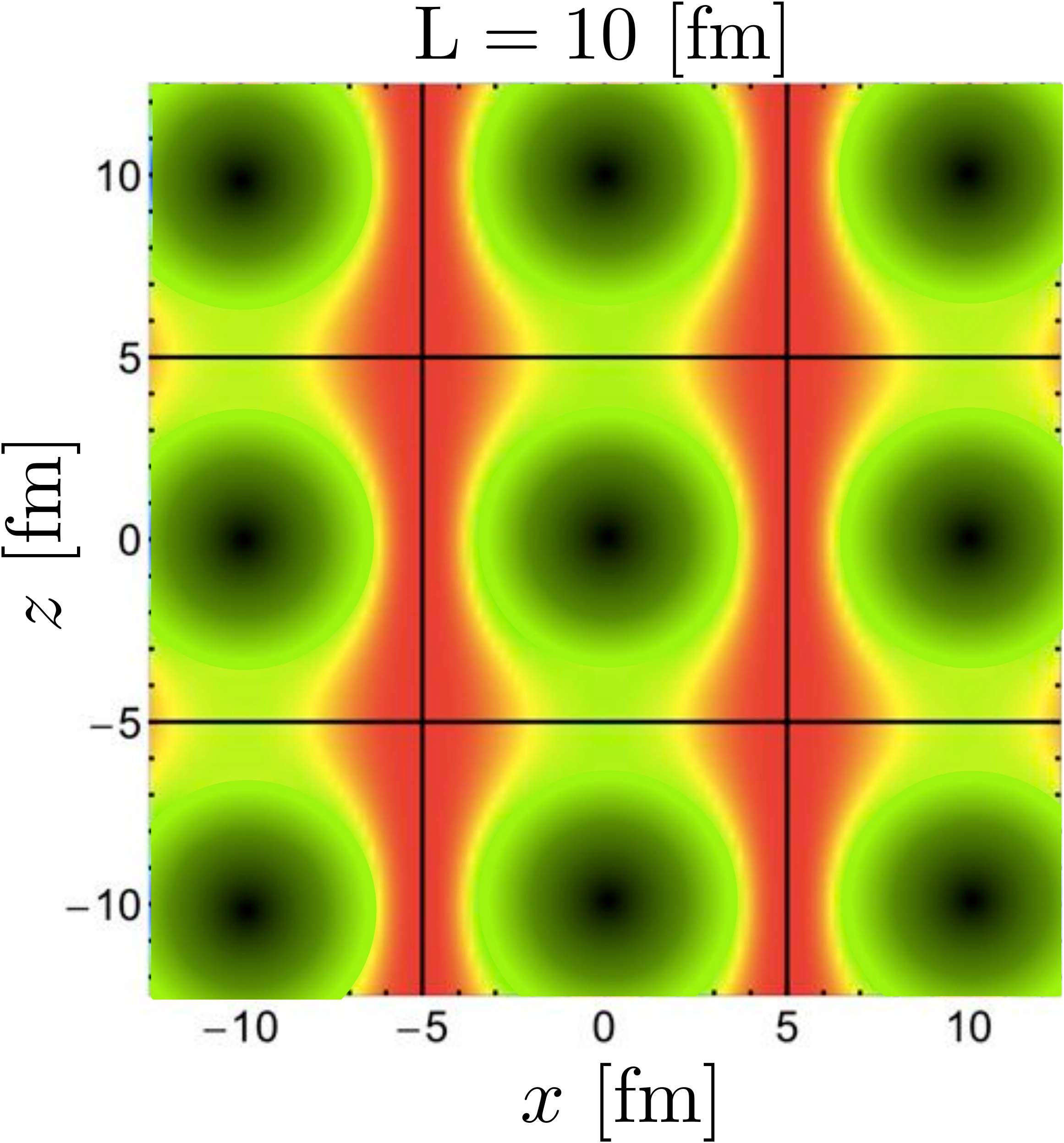}}
\subfigure[]{
\label{WF-B1-L15}
\includegraphics[scale=0.215]{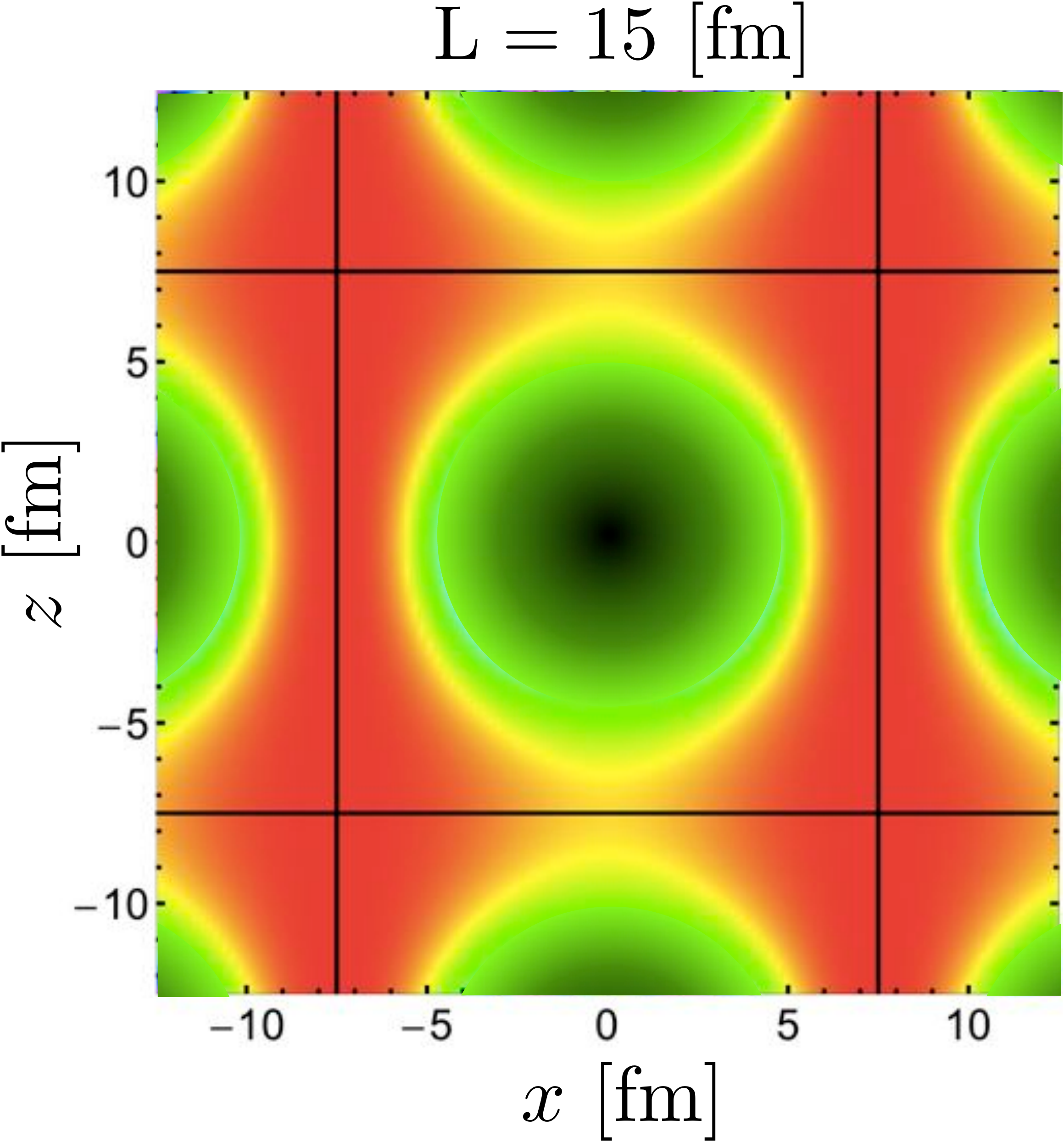}}
\subfigure[]{
\label{WF-B1-20}
\includegraphics[scale=0.215]{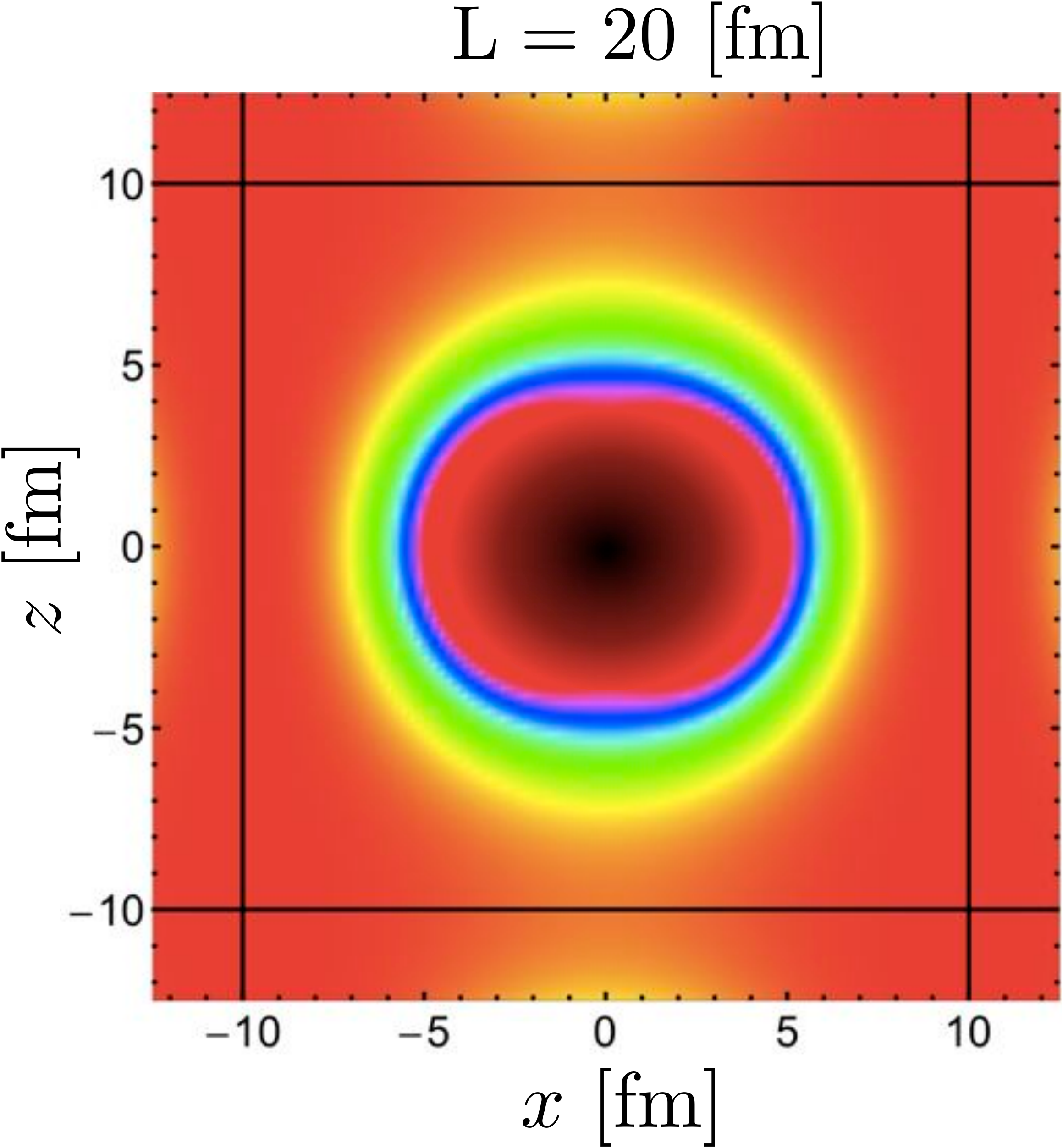}}
\subfigure[]{
\label{WF-B1-30}
\includegraphics[scale=0.215]{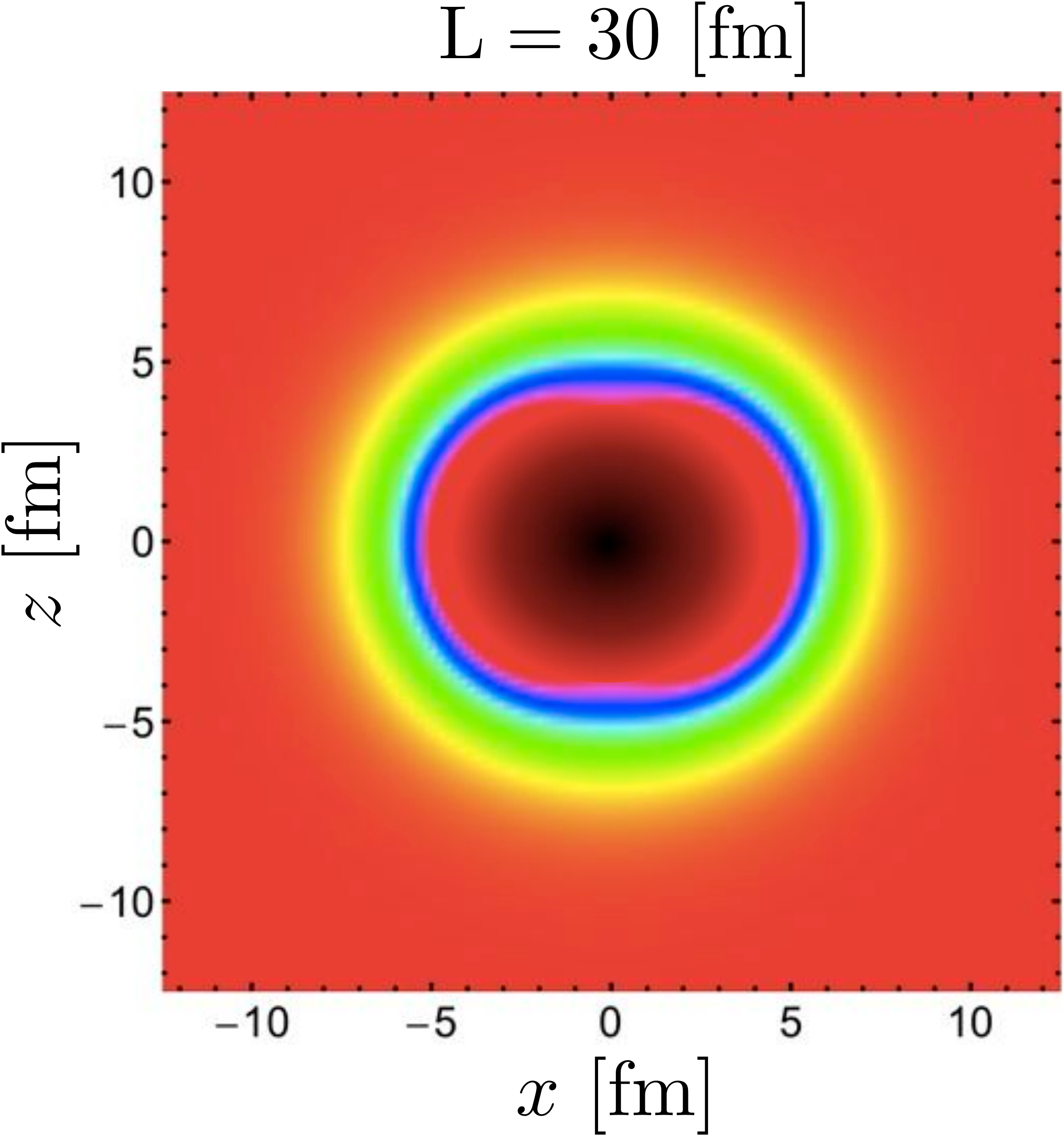}}
\caption{{\small 
The mass density  in the $xz$-plane from the $\mathbb{B}_1$ FV deuteron wavefunction with
  $\mathbf{d}=(1,1,0)$.
}}
\label{WF-B1}
\end{center}
\end{figure}

\begin{figure}[h!]
\begin{center}  
\subfigure[]{
\label{WF-B2B3-L10}
\includegraphics[scale=0.215]{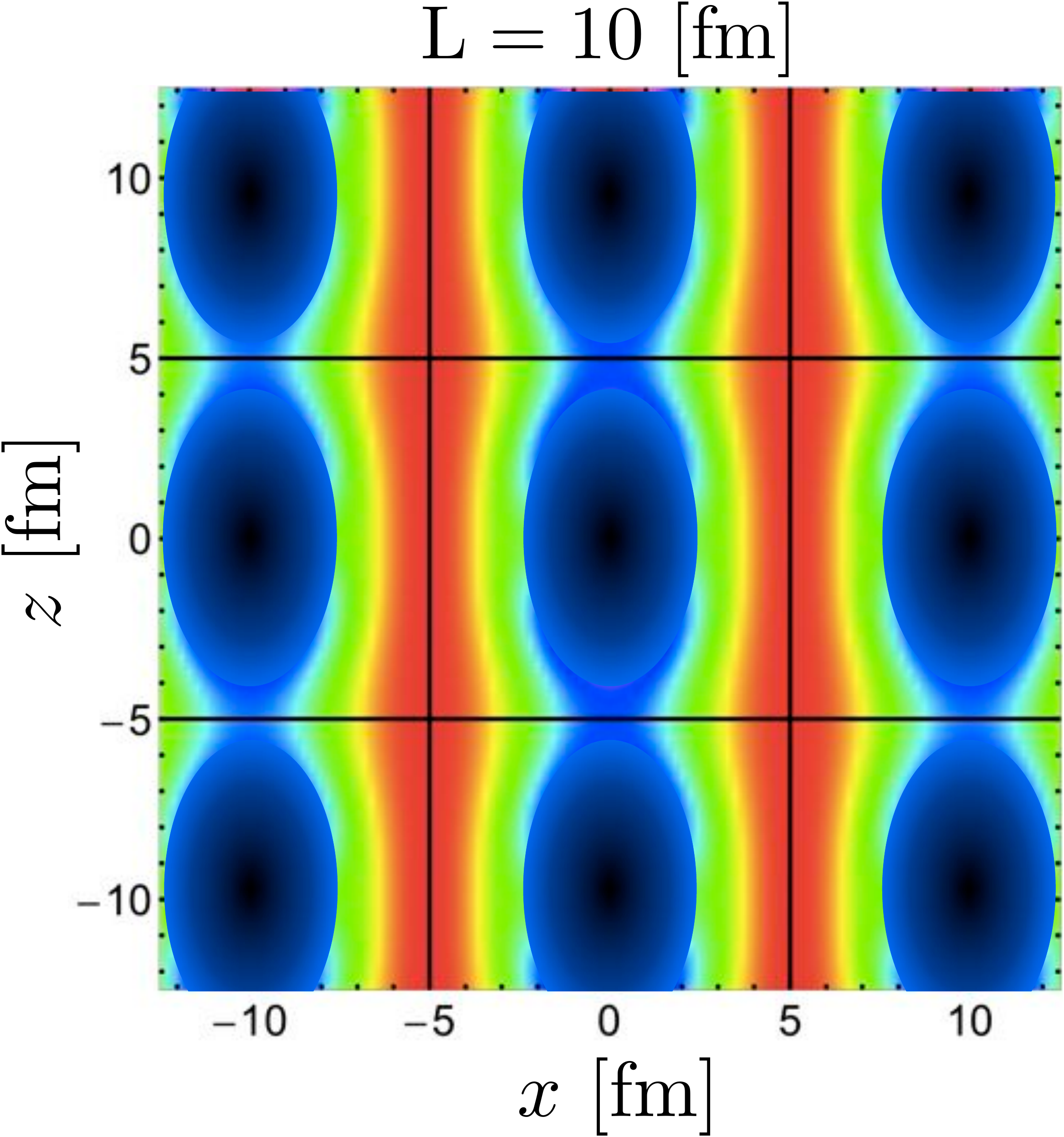}}
\subfigure[]{
\label{WF-B2B3-L15}
\includegraphics[scale=0.215]{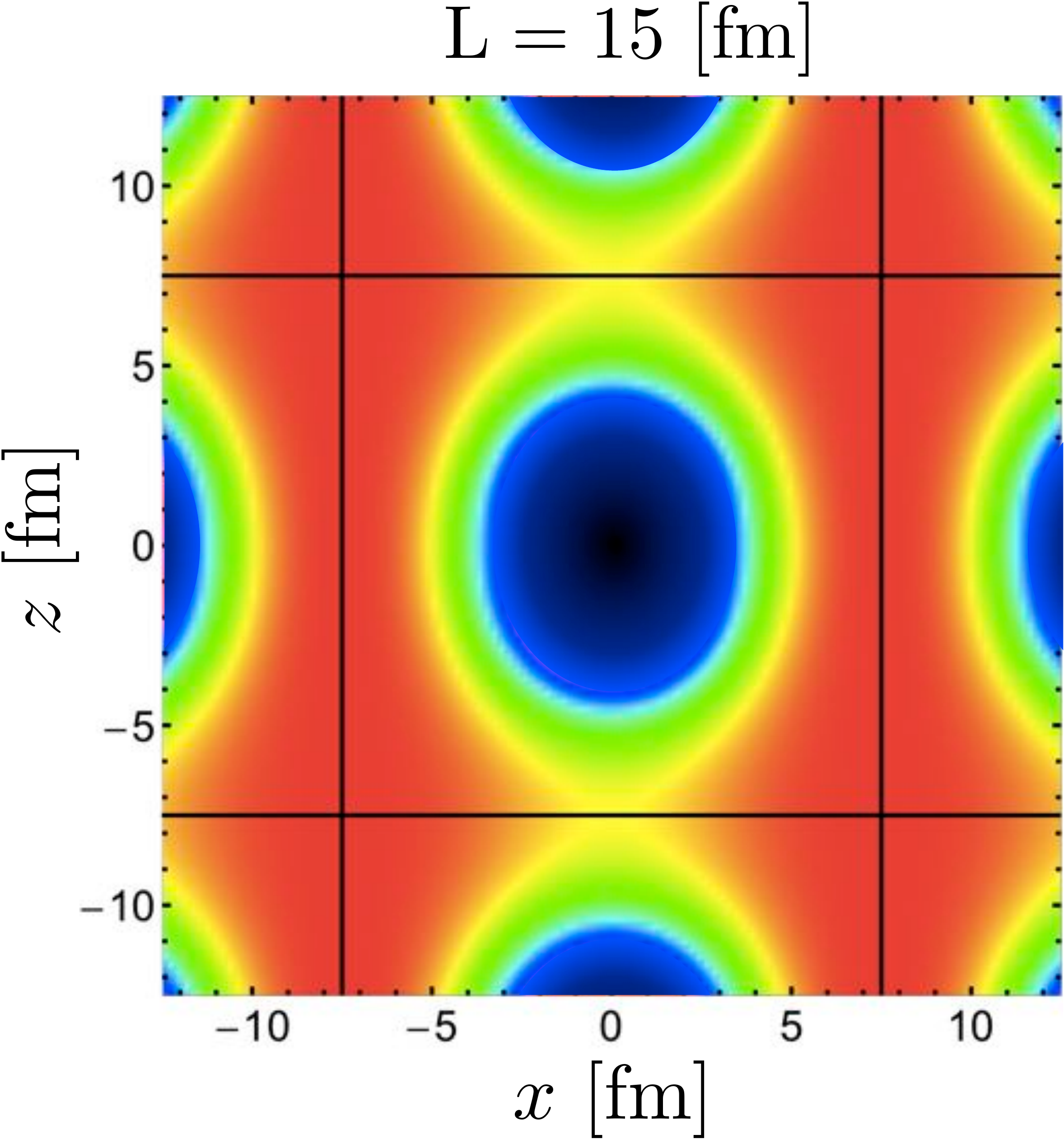}}
\subfigure[]{
\label{WF-B2B3-20}
\includegraphics[scale=0.215]{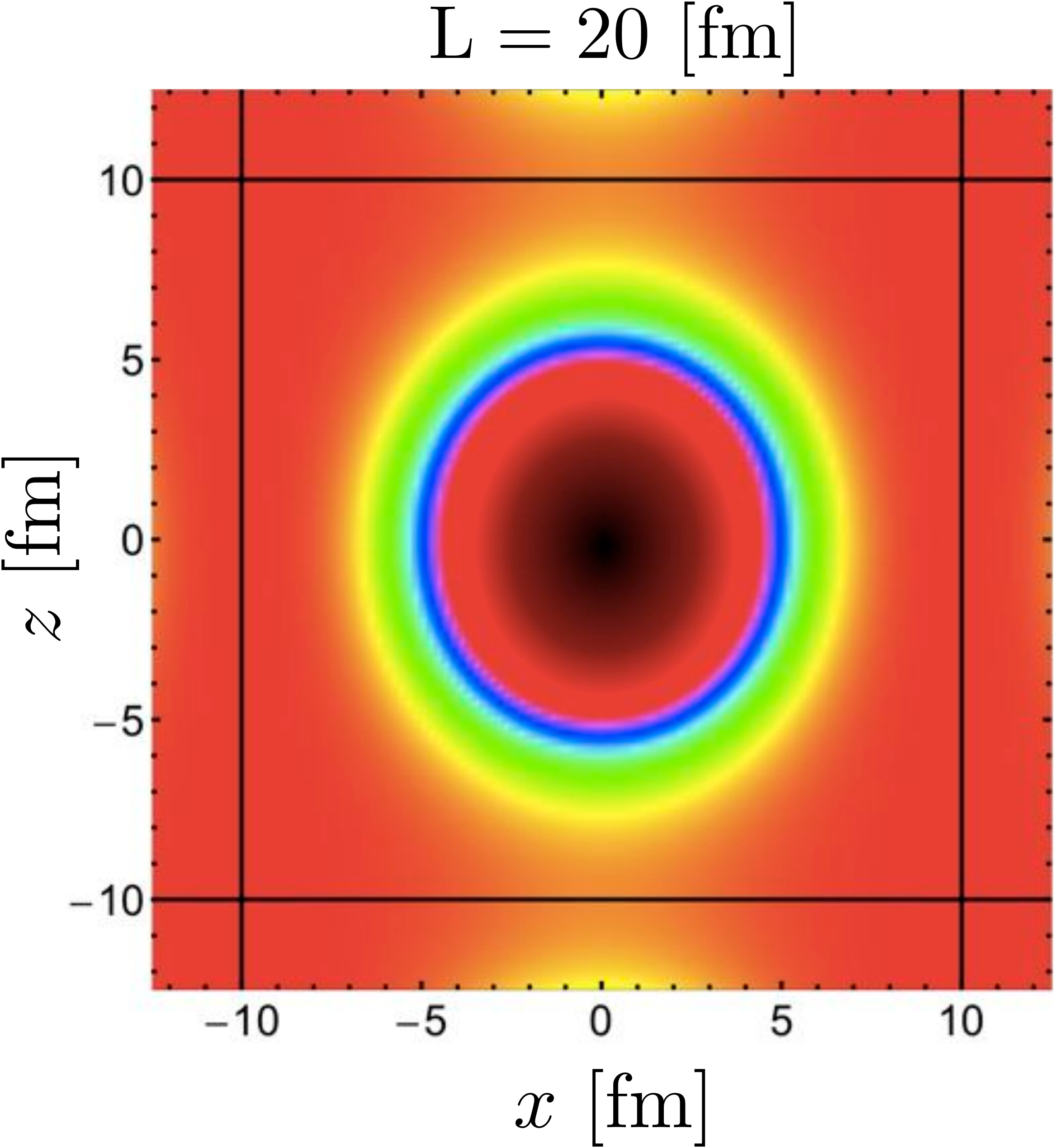}}
\subfigure[]{
\label{WF-B2B3-30}
\includegraphics[scale=0.215]{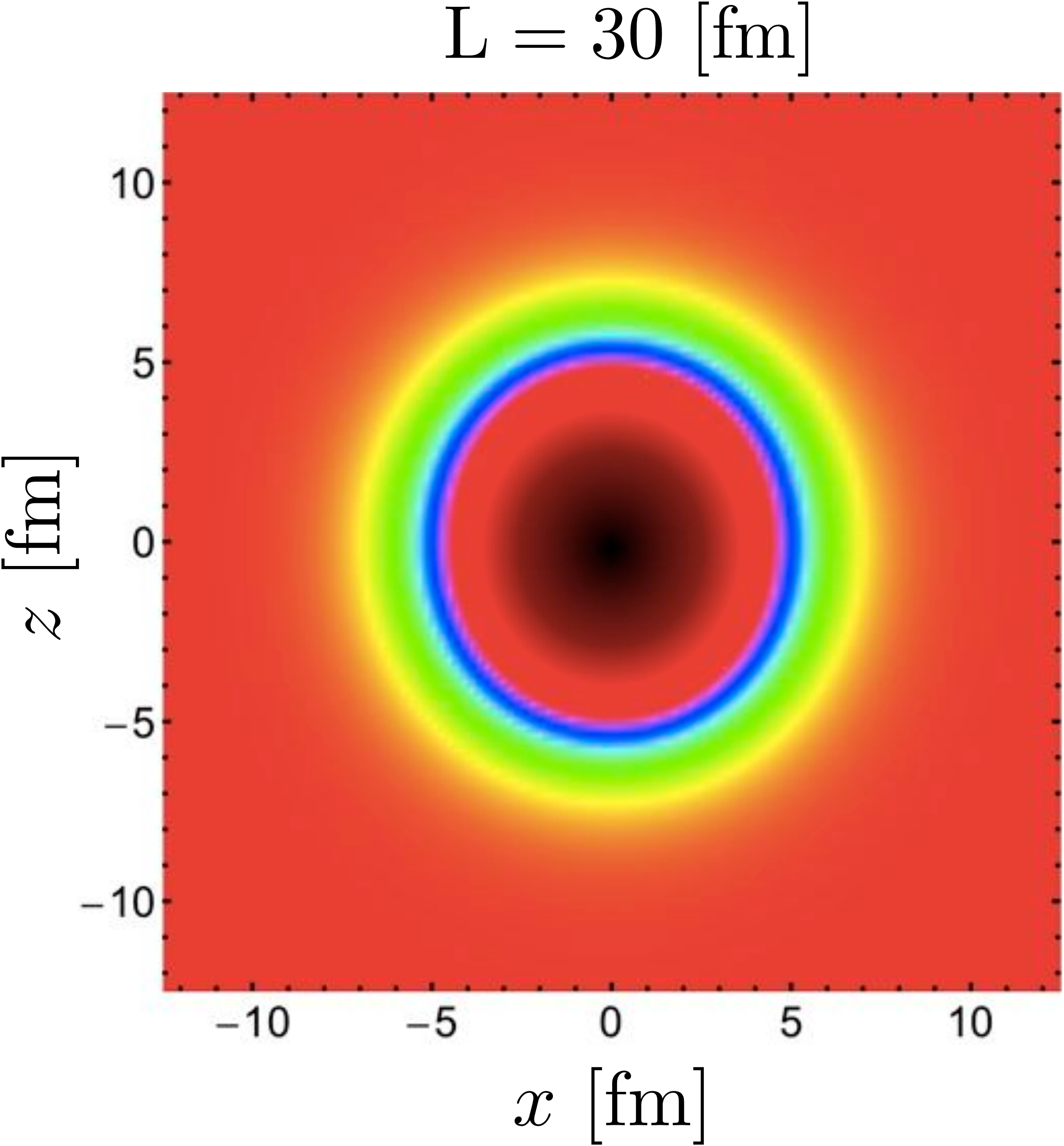}}
\caption{{\small 
The mass density  in the $xz$-plane from the $\mathbb{B}_2/\mathbb{B}_3$ FV deuteron wavefunction with
  $\mathbf{d}=(1,1,0)$.
}}
\label{WF-B2B3}
\end{center}
\end{figure}

\begin{figure}[h!]
\begin{center}  
\subfigure[]{
\label{WF-A2E-L10}
\includegraphics[scale=0.215]{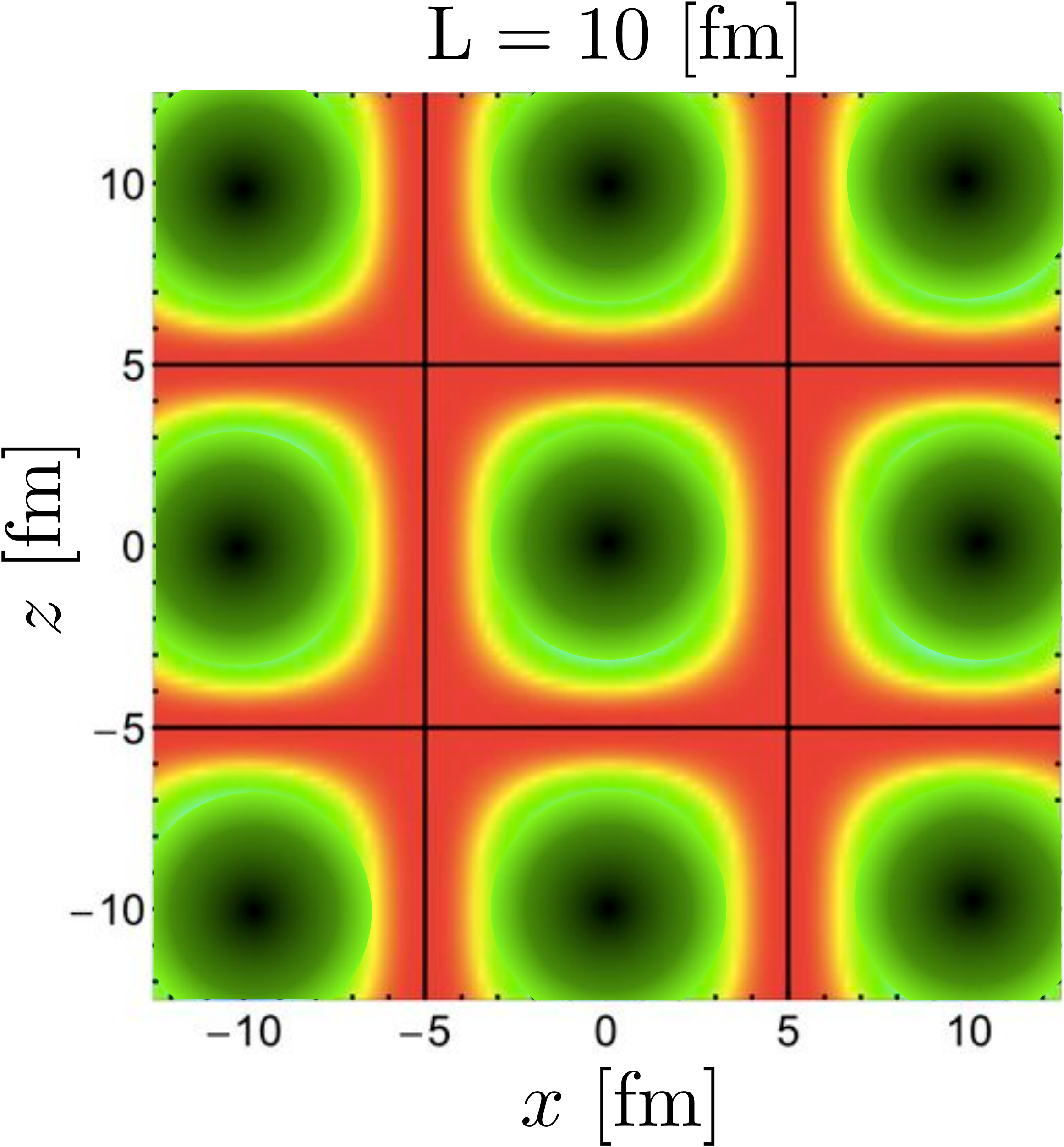}}
\subfigure[]{
\label{WF-A2E-L15}
\includegraphics[scale=0.215]{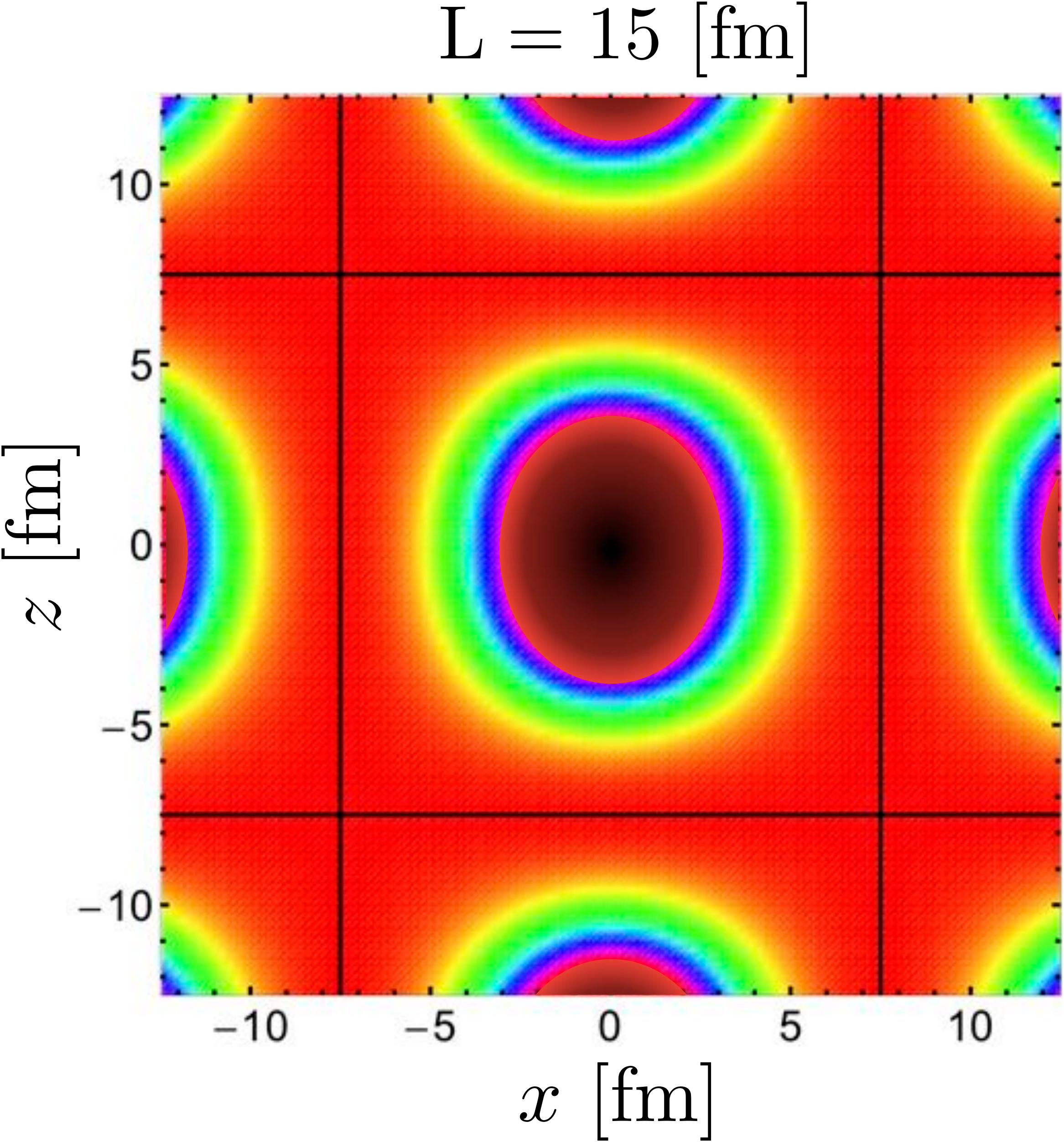}}
\subfigure[]{
\label{WF-A2E-20}
\includegraphics[scale=0.215]{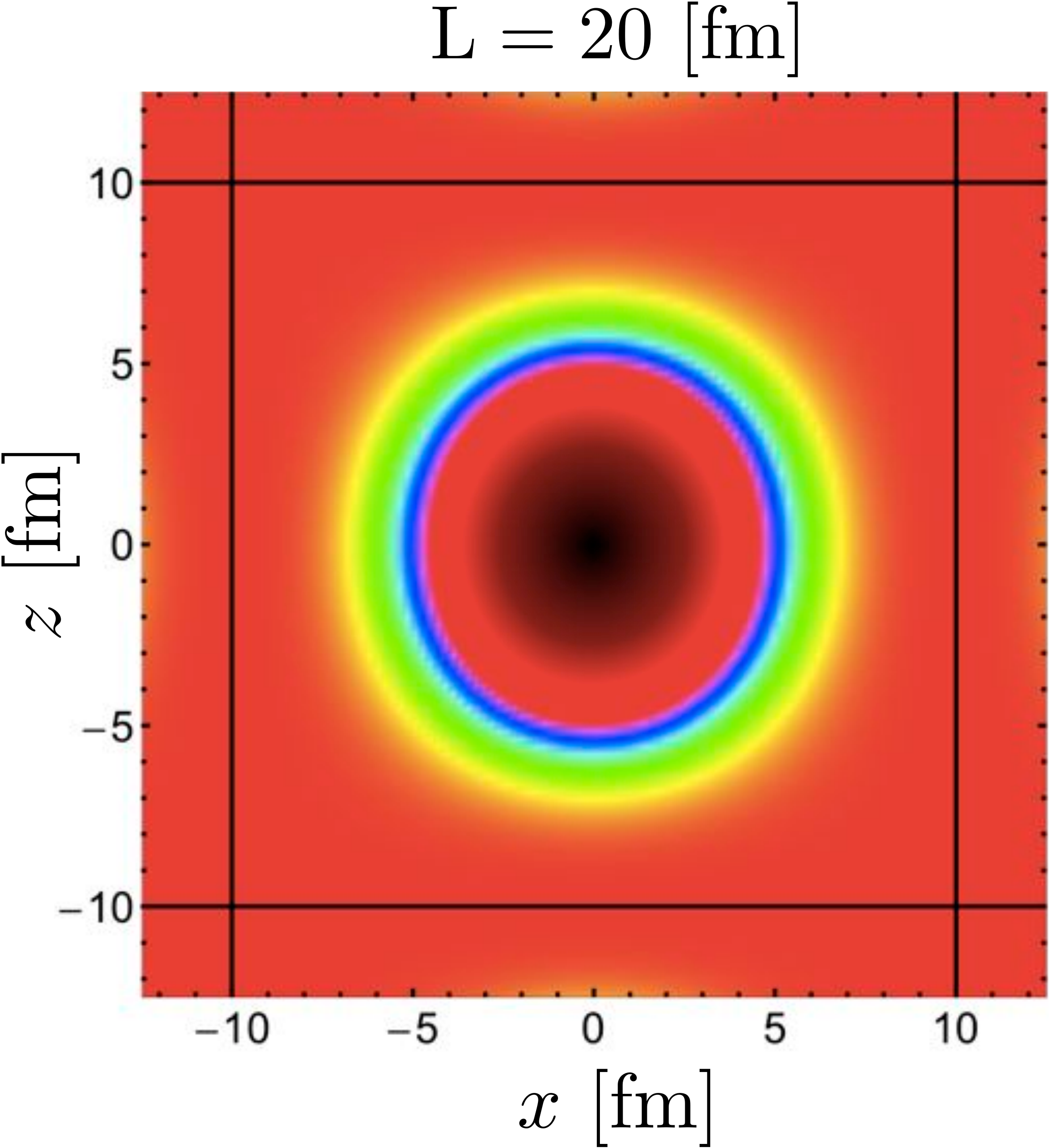}}
\subfigure[]{
\label{WF-A2E-30}
\includegraphics[scale=0.215]{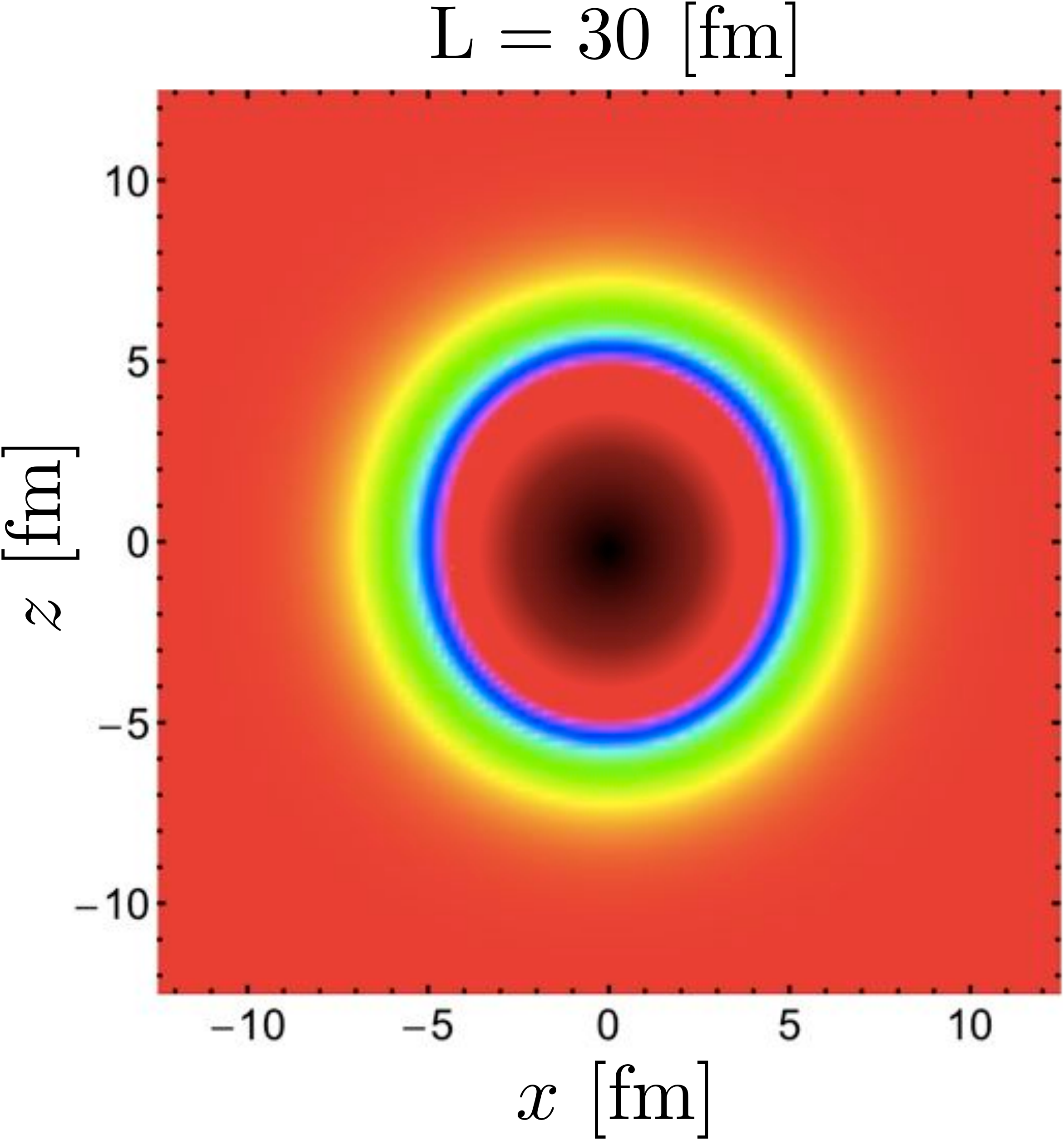}}
\caption{{\small 
The mass density  in the $xz$-plane from the $\mathbb{A}_2/\mathbb{E}$ FV deuteron wavefunction with
  $\mathbf{d}=(1,1,1)$.
}}
\label{WF-A2E}
\end{center}
\end{figure}


\chapter{Twisted boundary conditions}

\section{Twisted Images 
\label{app: TI}
}
\noindent
It is helpful for discussions in chapter \ref{chap:TBC} to make explicit the sums over the twist phases . 
Consider the sum  
\begin{eqnarray}
S({\bm\phi}) & = & 
\sum_{\mathbf{n}\neq \mathbf{0}} 
\frac{e^{-|\mathbf{n}| m_{\pi} {\rm L}}}{|\mathbf{n}|} 
\ e^{-i\mathbf{n}\cdot\bm{\phi}}
,
\label{eq:fullsum}
\end{eqnarray}
of which the first few terms are 
\begin{eqnarray}
S({\bm\phi}) & = & 
2\  e^{-m_{\pi} {\rm L}}\ 
\left( \cos\phi_x + \cos\phi_y + \cos\phi_z\right)
\nonumber\\
& + &  
2\sqrt{2}\  e^{-\sqrt{2} m_{\pi} {\rm L}}\ 
\left( \cos\phi_x  \cos\phi_y +  \cos\phi_x  \cos\phi_z +  \cos\phi_y  \cos\phi_z \right)
\nonumber\\
& + &  
{8\over\sqrt{3}}\ e^{-\sqrt{3} m_{\pi} {\rm L}}\ 
\cos\phi_x  \cos\phi_y  \cos\phi_z \ 
\nonumber\\
& + &  
e^{-2 m_{\pi} {\rm L}}\ 
 \left( \cos 2\phi_x + \cos 2\phi_y + \cos 2\phi_z\right)
\ +\ \cdots
.
\label{eq:partsum}
\end{eqnarray}
For PBCs, with ${\bm\phi}=(0,0,0)$, the first few terms in the 
sum in Eq.~(\ref{eq:fullsum}) and (\ref{eq:partsum}) are
\begin{eqnarray}
S({\bf 0}) & = & 
6 \ e^{-m_{\pi} {\rm L}}\ 
+
6\sqrt{2} \ e^{-\sqrt{2} m_{\pi} {\rm L}}\ 
+
{8\over\sqrt{3}} \ e^{-\sqrt{3} m_{\pi} {\rm L}}\ 
+
3 \ e^{-2 m_{\pi} {\rm L}}\ 
\ +\ \cdots
,
\label{eq:partsum}
\end{eqnarray}
while for APBs, with ${\bm\phi}=(\pi,\pi,\pi)$, the sum becomes
\begin{eqnarray}
S({\bm \pi}) & = & 
-6 \ e^{-m_{\pi} {\rm L}}\ 
+
6\sqrt{2} \ e^{-\sqrt{2} m_{\pi} {\rm L}}\ 
-
{8\over\sqrt{3}} \ e^{-\sqrt{3} m_{\pi} {\rm L}}\ 
+
3 \ e^{-2 m_{\pi} {\rm L}}\ 
\ -\ \cdots
.
\label{eq:partsum}
\end{eqnarray}
It is obvious that the leading terms vanish
in the average, with $\frac{1}{2}(S({\bf 0})+S({\bm \pi}))= 6\sqrt{2}\  e^{-\sqrt{2} m_{\pi} {\rm L}} +\dots~$.
A particularly interesting twist is ${\bm\phi}=({\pi\over 2},{\pi\over 2},{\pi\over 2})$, 
induced by i-PBCs, 
for which the first three terms in the sum
vanish, leaving 
\begin{eqnarray}
S({{\bm \pi}\over 2})
&  =   & 
-3\  e^{-2 m_{\pi} {\rm L}} 
\ +\ \cdots
.
\label{eq:magictwist}
\end{eqnarray}
Finally, twist averaging this function gives
\begin{eqnarray}
\langle S({\bm \phi}) \rangle_{\bm\phi}\ =\ 
\int\ {d^3{\bm\phi}\over (2\pi)^3}\ 
S({\bm \phi})
&  =   & 
0
.
\label{eq:twistAvS}
\end{eqnarray}
%


\section{Twisted Quantization Conditions for the Deuteron Channel\label{app: QC-TBC}}
\noindent
The NN FV QCs in the channels
that have an overlap with the $\siii$-$\diii$ coupled channels
are listed in this appendix for a selection of twist angles. 
With a similar notation as in Sec. \ref{app: QC}, paying attention to the differences due to the relativistic kinematics that we have chosen in this section, the QC for the irrep $\Gamma_i$ can be written as
\begin{eqnarray}
\det\left({\mathbb{M}}^{(\Gamma_i)}
\ +\  i \frac{k^*}{8\pi E^*}-\mathcal{F}^{(\Gamma_i),{\textbf{d},\bm{\phi}_1,\bm{\phi}_2}}\right)=0
,
\label{QC-simplified}
\end{eqnarray} 
where
\begin{eqnarray}
\mathcal{F}^{(\Gamma_i),{\textbf{d},\bm{\phi}_1,\bm{\phi}_2}}(k^{*2}; {\rm L} )
& = &
\frac{1}{2E^*}\sum_{l,m}\frac{1}{k^{*l}}~{\mathbb{F}
}_{lm}^{(\Gamma_i)}~{c_{lm}^{\textbf{d},\bm{\phi}_1,\bm{\phi}_2}(k^{*2};{\rm L})}
,
\nonumber\\
{\mathbb{M}}^{(\Gamma_i)}
& = & \left( \mathcal{M}^{-1}\right)_{\Gamma_i}
.
\label{def-F}
\end{eqnarray}
where 
${c_{lm}^{\textbf{d},\bm{\phi}_1,\bm{\phi}_2}(k^{*2};{\rm L})}$ functions are defined in
Eqs.~(\ref{clm}),(\ref{Zlm}) and (\ref{r-TBC}), 
$E^*$ is NN CM energy and $k^*$ is the on-shell momentum of each nucleon in the CM 
frame.~\footnote{
The relativistic normalization of states has been used such that for a single  S-wave channel with phase shift $\delta$, 
the scattering amplitude is
${\cal M} =  {8\pi E^*\over k^*} {\left( e^{2 i \delta}-1\right) \over 2i}$.
} 
In the summation over ``$m$'' in Eq.~(\ref{def-F}), only the 
${\mathbb{F}}_{lm}^{(\Gamma_i)}$ listed below are included as the other contributions have already 
been summed using the symmetries of the systems.
In the following we set $\bm{\phi}_1=-\bm{\phi}_2=\bm{\phi}$.
It is straightforward to decompose $\mathcal{M}^{-1}$ into 
$\left( \mathcal{M}^{-1}\right)_{\Gamma_i}$
using the eigenvectors of the FV functions~\cite{Thomas:2011rh,Dudek:2012gj}. The notation used for the scattering amplitude matrix has been introduced in Sec. \ref{app: QC}.

\begin{small}
\subsection{${ \phi}=(0,0,0)$}
\begin{footnotesize}
\begin{align}
& \mathbb{T}_1: \hspace{0.4cm}
\mathbb{F}_{00}^{(\mathbb{T}_1)}=\textbf{I}_{3},\hspace{0.15cm}
\mathbb{F}_{40}^{(\mathbb{T}_1)}=
\left(
\begin{array}{ccc}
 0 & 0 & 0 \\
 0 & 0 & \frac{2 \sqrt{6}}{7} \\
 0 & \frac{2 \sqrt{6}}{7} & \frac{2}{7} \\
\end{array}
\right),\hspace{0.15cm}
{\mathbb{M}}^{(\mathbb{T}_1)}=\left(
\begin{array}{ccc}
 \frac{\mathcal{M}_{1,D}}{{\det\mathcal{M}_{1}}} & -\frac{\mathcal{M}_{1,SD}}{\det\mathcal{M}_{1}} & 0 \\
 -\frac{\mathcal{M}_{1,SD}}{\det\mathcal{M}_{1}} & \frac{\mathcal{M}_{1,S}}{\det\mathcal{M}_{1}} & 0 \\
 0 & 0 &\mathcal{M}_{3,D}^{-1} \\
\end{array}
\right)
. 
\label{000T1}
\end{align}
\end{footnotesize}

\subsection{${ \phi}=(\frac{\pi}{2},\frac{\pi}{2},\frac{\pi}{2})$}
\begin{footnotesize}
\begin{align}
& \mathbb{A}_2:\hspace{0.4cm}
\mathbb{F}_{00}^{(\mathbb{A}_2)}=\textbf{I}_{6},\hspace{0.15cm}
\mathbb{F}_{10}^{(\mathbb{A}_2)}= \left(
\begin{array}{cccccc}
 0 & -1 & \sqrt{2} & 0 & 0 & 0 \\
 -1 & 0 & 0 & \sqrt{2} & 0 & 0 \\
 \sqrt{2} & 0 & 0 & -\frac{1}{5} & 0 & 0 \\
 0 & \sqrt{2} & -\frac{1}{5} & 0 & \frac{3 \sqrt{6}}{5} & 0 \\
 0 & 0 & 0 & \frac{3 \sqrt{6}}{5} & 0 & 0 \\
 0 & 0 & 0 & 0 & 0 & 0
\end{array}
\right)
\ ,
\nonumber\\
&\hspace{0.15cm}
\mathbb{F}_{22}^{(\mathbb{A}_2)}
= 
i  
\left(
\begin{array}{cccccc}
 0 & 0 & 0 & 2 \sqrt{\frac{3}{5}} & 0 & 0 \\
 0 & 0 & 2 \sqrt{\frac{3}{5}} & 0 & -3  \sqrt{\frac{2}{5}} & 0 \\
 0 & 2 \sqrt{\frac{3}{5}} & - \sqrt{\frac{6}{5}} & 0 & \frac{6 }{7 \sqrt{5}} & 0 \\
 2  \sqrt{\frac{3}{5}} & 0 & 0 & - \sqrt{\frac{6}{5}} & 0 & 0 \\
 0 & -3  \sqrt{\frac{2}{5}} & \frac{6 }{7 \sqrt{5}} & 0 & -\frac{8}{7}  \sqrt{\frac{6}{5}} & 0 \\
 0 & 0 & 0 & 0 & 0 & \frac{2  \sqrt{30}}{7}
\end{array}
\right)
\ ,
\
\mathbb{F}_{30}^{(\mathbb{A}_2)}=
\left(
\begin{array}{cccccc}
 0 & 0 & 0 & 0 & \frac{2}{\sqrt{7}} & \sqrt{\frac{5}{7}} \\
 0 & 0 & 0 & 0 & 0 & 0 \\
 0 & 0 & 0 & \frac{6 \sqrt{\frac{3}{7}}}{5} & 0 & 0 \\
 0 & 0 & \frac{6 \sqrt{\frac{3}{7}}}{5} & 0 & -\frac{4 \sqrt{\frac{2}{7}}}{5} & \sqrt{\frac{5}{14}} \\
 \frac{2}{\sqrt{7}} & 0 & 0 & -\frac{4 \sqrt{\frac{2}{7}}}{5} & 0 & 0 \\
 \sqrt{\frac{5}{7}} & 0 & 0 & \sqrt{\frac{5}{14}} & 0 & 0
\end{array}
\right)
\ ,
\nonumber\\
&\hspace{0.15cm}
\mathbb{F}_{32}^{(\mathbb{A}_2)}=
i  
\left(
\begin{array}{cccccc}
 0 & 0 & 0 & 0 &  \sqrt{\frac{10}{21}} & -2  \sqrt{\frac{2}{21}} \\
 0 & 0 & 0 & 0 & 0 & 0 \\
 0 & 0 & 0 & 3  \sqrt{\frac{2}{35}} & 0 & 0 \\
 0 & 0 & 3 \sqrt{\frac{2}{35}} & 0 & -\frac{4 }{\sqrt{105}} & -\frac{2 }{\sqrt{21}} \\
  \sqrt{\frac{10}{21}} & 0 & 0 & -\frac{4 }{\sqrt{105}} & 0 & 0 \\
 -2  \sqrt{\frac{2}{21}} & 0 & 0 & -\frac{2 }{\sqrt{21}} & 0 & 0
\end{array}
\right)
\ ,
\
\mathbb{F}_{40}^{(\mathbb{A}_2)}=
\left(
\begin{array}{cccccc}
 0 & 0 & 0 & 0 & 0 & 0 \\
 0 & 0 & 0 & 0 & 0 & 0 \\
 0 & 0 & 0 & 0 & \frac{4 \sqrt{\frac{2}{3}}}{7} & \frac{2 \sqrt{\frac{10}{3}}}{7} \\
 0 & 0 & 0 & 0 & 0 & 0 \\
 0 & 0 & \frac{4 \sqrt{\frac{2}{3}}}{7} & 0 & -\frac{4}{21} & \frac{4 \sqrt{5}}{21} \\
 0 & 0 & \frac{2 \sqrt{\frac{10}{3}}}{7} & 0 & \frac{4 \sqrt{5}}{21} & -\frac{2}{21}
\end{array}
\right)\ ,
\nonumber\\
& \mathbb{F}_{42}^{(\mathbb{A}_2)}=
i  
\left(
\begin{array}{cccccc}
 0 & 0 & 0 & 0 & 0 & 0 \\
 0 & 0 & 0 & 0 & 0 & 0 \\
 0 & 0 & 0 & 0 & \frac{8}{7}  \sqrt{\frac{5}{3}} & -\frac{1}{\sqrt{3}} \\
 0 & 0 & 0 & 0 & 0 & 0 \\
 0 & 0 & \frac{8}{7}  \sqrt{\frac{5}{3}} & 0 & -\frac{4  \sqrt{10}}{21} & -\frac{ \sqrt{2}}{3} \\
 0 & 0 & -\frac{1}{\sqrt{3}} & 0 & -\frac{ \sqrt{2}}{3} & -\frac{2  \sqrt{10}}{21}
\end{array}
\right)
\ ,
\ {\mathbb{M}}^{(\mathbb{A}_2)}=
\left(
\begin{array}{cccccc}
 \mathcal{M}_{0,P}^{-1} & 0 & 0 & 0 & 0 & 0 \\
 0 & \frac{\mathcal{M}_{1,D}}{{\det\mathcal{M}_{1}}} & -\frac{\mathcal{M}_{1,SD}}{{\det\mathcal{M}_{1}}} & 0 & 0 & 0 \\
 0 & -\frac{\mathcal{M}_{1,SD}}{{\det\mathcal{M}_{1}}} & \frac{\mathcal{M}_{1,S}}{{\det\mathcal{M}_{1}}} & 0 & 0 & 0 \\
 0 & 0 & 0 & \mathcal{M}_{2,P}^{-1} & 0 & 0 \\
 0 & 0 & 0 & 0 & \mathcal{M}_{3,D}^{-1} & 0 \\
 0 & 0 & 0 & 0 & 0 & \mathcal{M}_{3,D}^{-1}
\end{array}
\right)
\ .
\label{pi2pi2pi2A2}
\end{align}
\begin{align}
& \mathbb{E}:\hspace{0.4cm}\mathbb{F}_{00}^{(\mathbb{E})}=\textbf{I}_{9} ,\hspace{0.15cm}
\mathbb{F}_{10}^{(\mathbb{E})}=
\left(
\begin{array}{ccccccccc}
 0 & \sqrt{\frac{3}{2}} & 0 & 0 & \sqrt{\frac{3}{2}} & 0 & 0 & 0 & 0 \\
 \sqrt{\frac{3}{2}} & 0 & \frac{\sqrt{3}}{2} & 0 & 0 & 0 & \frac{3 \sqrt{\frac{3}{5}}}{2} & 0 & 0 \\
 0 & \frac{\sqrt{3}}{2} & 0 & 0 & -\frac{\sqrt{3}}{10} & 0 & 0 & 0 & 0 \\
 0 & 0 & 0 & 0 & 0 & \sqrt{\frac{3}{5}} & 0 & \sqrt{\frac{6}{5}} & 0 \\
 \sqrt{\frac{3}{2}} & 0 & -\frac{\sqrt{3}}{10} & 0 & 0 & 0 & \frac{\sqrt{\frac{3}{5}}}{2} & 0 & \frac{4 \sqrt{3}}{5} \\
 0 & 0 & 0 & \sqrt{\frac{3}{5}} & 0 & 0 & 0 & 0 & 0 \\
 0 & \frac{3 \sqrt{\frac{3}{5}}}{2} & 0 & 0 & \frac{\sqrt{\frac{3}{5}}}{2} & 0 & 0 & 0 & 0 \\
 0 & 0 & 0 & \sqrt{\frac{6}{5}} & 0 & 0 & 0 & 0 & 0 \\
 0 & 0 & 0 & 0 & \frac{4 \sqrt{3}}{5} & 0 & 0 & 0 & 0
\end{array}
\right)
\ ,
\nonumber\\
& \hspace{1.2cm}\mathbb{F}_{22}^{(\mathbb{E})}=
i  
\left(
\begin{array}{ccccccccc}
 0 & 0 & - \sqrt{\frac{3}{5}} & 0 & 0 & 0 & -\sqrt{3} & 0 & -2 \sqrt{\frac{3}{5}} \\
 0 & - \sqrt{\frac{3}{10}} & 0 & 0 & -3  \sqrt{\frac{3}{10}} & 0 & 0 & 0 & 0 \\
 - \sqrt{\frac{3}{5}} & 0 &  \sqrt{\frac{3}{10}} & 0 & 0 & 0 & - \sqrt{\frac{3}{2}} & 0 & \frac{2}{7}  \sqrt{\frac{6}{5}} \\
 0 & 0 & 0 &  \sqrt{\frac{6}{5}} & 0 & 0 & 0 & 0 & 0 \\
 0 & -3  \sqrt{\frac{3}{10}} & 0 & 0 & - \sqrt{\frac{3}{10}} & 0 & 0 & 0 & 0 \\
 0 & 0 & 0 & 0 & 0 & \frac{ \sqrt{30}}{7} & 0 & -\frac{2 \sqrt{15}}{7} & 0 \\
 -\sqrt{3} & 0 & - \sqrt{\frac{3}{2}} & 0 & 0 & 0 & -\frac{1}{7}  \sqrt{\frac{15}{2}} & 0 & -\frac{2  \sqrt{6}}{7} \\
 0 & 0 & 0 & 0 & 0 & -\frac{2 \sqrt{15}}{7} & 0 & 0 & 0 \\
 -2  \sqrt{\frac{3}{5}} & 0 & \frac{2}{7}  \sqrt{\frac{6}{5}} & 0 & 0 & 0 & -\frac{2  \sqrt{6}}{7} & 0 & -\frac{6}{7}  \sqrt{\frac{6}{5}}
\end{array}
\right)
\ ,
\nonumber\\
& \hspace{1.2cm}\mathbb{F}_{30}^{(\mathbb{E})}=
\left(
\begin{array}{ccccccccc}
 0 & 0 & 0 & 0 & 0 & 0 & 0 & 0 & 0 \\
 0 & 0 & 0 & 0 & 0 & -\sqrt{\frac{5}{14}} & -\frac{2}{\sqrt{35}} & \frac{\sqrt{\frac{5}{7}}}{2} & -\frac{2}{\sqrt{7}} \\
 0 & 0 & 0 & \frac{3}{\sqrt{14}} & -\frac{6}{5 \sqrt{7}} & 0 & 0 & 0 & 0 \\
 0 & 0 & \frac{3}{\sqrt{14}} & 0 & 0 & \frac{2}{\sqrt{35}} & -\sqrt{\frac{5}{14}} & 2 \sqrt{\frac{2}{35}} & \frac{1}{\sqrt{14}} \\
 0 & 0 & -\frac{6}{5 \sqrt{7}} & 0 & 0 & \sqrt{\frac{5}{14}} & -\frac{4}{\sqrt{35}} & -\frac{\sqrt{\frac{5}{7}}}{2} & -\frac{2}{5 \sqrt{7}} \\
 0 & -\sqrt{\frac{5}{14}} & 0 & \frac{2}{\sqrt{35}} & \sqrt{\frac{5}{14}} & 0 & 0 & 0 & 0 \\
 0 & -\frac{2}{\sqrt{35}} & 0 & -\sqrt{\frac{5}{14}} & -\frac{4}{\sqrt{35}} & 0 & 0 & 0 & 0 \\
 0 & \frac{\sqrt{\frac{5}{7}}}{2} & 0 & 2 \sqrt{\frac{2}{35}} & -\frac{\sqrt{\frac{5}{7}}}{2} & 0 & 0 & 0 & 0 \\
 0 & -\frac{2}{\sqrt{7}} & 0 & \frac{1}{\sqrt{14}} & -\frac{2}{5 \sqrt{7}} & 0 & 0 & 0 & 0
\end{array}
\right)
\ ,
\nonumber
\end{align}
\begin{align}
& \hspace{0.2cm}\mathbb{F}_{32}^{(\mathbb{E})}=
i  
\left(
\begin{array}{ccccccccc}
 0 & 0 & 0 & 0 & 0 & 0 & 0 & 0 & 0 \\
 0 & 0 & 0 & 0 & 0 & \frac{2 }{\sqrt{21}} & -\sqrt{\frac{2}{21}} & -\sqrt{\frac{2}{21}} & -\sqrt{\frac{10}{21}} \\
 0 & 0 & 0 & -2 \sqrt{\frac{3}{35}} & -\sqrt{\frac{6}{35}} & 0 & 0 & 0 & 0 \\
 0 & 0 & -2  \sqrt{\frac{3}{35}} & 0 & 0 &  \sqrt{\frac{2}{21}} & \frac{2 }{\sqrt{21}} & \frac{2 }{\sqrt{21}} & -\frac{2 }{\sqrt{105}} \\
 0 & 0 & -\sqrt{\frac{6}{35}} & 0 & 0 & -\frac{2 }{\sqrt{21}} & -2 \sqrt{\frac{2}{21}} &  \sqrt{\frac{2}{21}} & - \sqrt{\frac{2}{105}} \\
 0 & \frac{2 }{\sqrt{21}} & 0 &  \sqrt{\frac{2}{21}} & -\frac{2 }{\sqrt{21}} & 0 & 0 & 0 & 0 \\
 0 & - \sqrt{\frac{2}{21}} & 0 & \frac{2 }{\sqrt{21}} & -2  \sqrt{\frac{2}{21}} & 0 & 0 & 0 & 0 \\
 0 & - \sqrt{\frac{2}{21}} & 0 & \frac{2 }{\sqrt{21}} &  \sqrt{\frac{2}{21}} & 0 & 0 & 0 & 0 \\
 0 & - \sqrt{\frac{10}{21}} & 0 & -\frac{2 }{\sqrt{105}} & - \sqrt{\frac{2}{105}} & 0 & 0 & 0 & 0
\end{array}
\right)
\ ,
\nonumber\\
& \hspace{0.2cm}\mathbb{F}_{40}^{(\mathbb{E})}=
\left(
\begin{array}{ccccccccc}
 0 & 0 & 0 & 0 & 0 & 0 & 0 & 0 & 0 \\
 0 & 0 & 0 & 0 & 0 & 0 & 0 & 0 & 0 \\
 0 & 0 & 0 & 0 & 0 & 0 & 0 & \frac{2 \sqrt{5}}{7} & -\frac{2}{7} \\
 0 & 0 & 0 & 0 & 0 & 0 & 0 & 0 & 0 \\
 0 & 0 & 0 & 0 & 0 & 0 & 0 & 0 & 0 \\
 0 & 0 & 0 & 0 & 0 & \frac{4}{63} & -\frac{20 \sqrt{2}}{63} & \frac{10 \sqrt{2}}{63} & \frac{10 \sqrt{10}}{63} \\
 0 & 0 & 0 & 0 & 0 & -\frac{20 \sqrt{2}}{63} & -\frac{16}{63} & -\frac{10}{63} & -\frac{10 \sqrt{5}}{63} \\
 0 & 0 & \frac{2 \sqrt{5}}{7} & 0 & 0 & \frac{10 \sqrt{2}}{63} & -\frac{10}{63} & \frac{2}{9} & -\frac{4 \sqrt{5}}{63} \\
 0 & 0 & -\frac{2}{7} & 0 & 0 & \frac{10 \sqrt{10}}{63} & -\frac{10 \sqrt{5}}{63} & -\frac{4 \sqrt{5}}{63} & -\frac{2}{63}
\end{array}
\right)
\ ,
\nonumber\\
& \hspace{0.15cm}\mathbb{F}_{42}^{(\mathbb{E})}=
i  
\left(
\begin{array}{ccccccccc}
 0 & 0 & 0 & 0 & 0 & 0 & 0 & 0 & 0 \\
 0 & 0 & 0 & 0 & 0 & 0 & 0 & 0 & 0 \\
 0 & 0 & 0 & 0 & 0 & 0 & 0 & -\frac{1}{\sqrt{2}} & -\frac{2  \sqrt{10}}{7} \\
 0 & 0 & 0 & 0 & 0 & 0 & 0 & 0 & 0 \\
 0 & 0 & 0 & 0 & 0 & 0 & 0 & 0 & 0 \\
 0 & 0 & 0 & 0 & 0 & \frac{4  \sqrt{10}}{63} & \frac{2  \sqrt{5}}{9} & \frac{20  \sqrt{5}}{63} & -\frac{5 }{9} \\
 0 & 0 & 0 & 0 & 0 & \frac{2 \sqrt{5}}{9} & -\frac{16 \sqrt{10}}{63} & \frac{1}{9}  \sqrt{\frac{5}{2}} & -\frac{50 \sqrt{2}}{63} \\
 0 & 0 & -\frac{1}{\sqrt{2}} & 0 & 0 & \frac{20  \sqrt{5}}{63} & \frac{1}{9} \sqrt{\frac{5}{2}} & \frac{2  \sqrt{10}}{9} & \frac{\sqrt{2}}{9} \\
 0 & 0 & -\frac{2  \sqrt{10}}{7} & 0 & 0 & -\frac{5 }{9} & -\frac{50  \sqrt{2}}{63} & \frac{ \sqrt{2}}{9} & -\frac{2 \sqrt{10}}{63}
\end{array}
\right)
\ ,
\nonumber\\
&\hspace{0.2cm}{\mathbb{M}}^{(\mathbb{E})}=\left(
\begin{array}{ccccccccc}
 \frac{\mathcal{M}_{1,D}}{{\det\mathcal{M}_{1}}} & 0 & - \frac{\mathcal{M}_{1,SD}}{{\det\mathcal{M}_{1}}} & 0 & 0 & 0 & 0 & 0 & 0 \\
 0 & \mathcal{M}_{1,P}^{-1} & 0 & 0 & 0 & 0 & 0 & 0 & 0 \\
 - \frac{\mathcal{M}_{1,SD}}{{\det\mathcal{M}_{1}}} & 0 &  \frac{\mathcal{M}_{1,S}}{{\det\mathcal{M}_{1}}} & 0 & 0 & 0 & 0 & 0 & 0 \\
 0 & 0 & 0 & \mathcal{M}_{2,P}^{-1} & 0 & 0 & 0 & 0 & 0 \\
 0 & 0 & 0 & 0 & \mathcal{M}_{2,P}^{-1} & 0 & 0 & 0 & 0 \\
 0 & 0 & 0 & 0 & 0 & \mathcal{M}_{2,D}^{-1} & 0 & 0 & 0 \\
 0 & 0 & 0 & 0 & 0 & 0 & \mathcal{M}_{2,D}^{-1} & 0 & 0 \\
 0 & 0 & 0 & 0 & 0 & 0 & 0 & \mathcal{M}_{3,D}^{-1} & 0 \\
 0 & 0 & 0 & 0 & 0 & 0 & 0 & 0 & \mathcal{M}_{3,D}^{-1}
\end{array}
\right)
\ .
\label{pi2pi2pi2E}
\end{align}
\end{footnotesize}

\subsection{${ \phi}=(\pi,\pi,\pi)$}
\begin{footnotesize}
\begin{align}
& \mathbb{A}_2:\hspace{0.4cm}
\mathbb{F}_{00}^{(\mathbb{A}_2)}=\textbf{I}_{4} ,\hspace{0.15cm}
\mathbb{F}_{40}^{(\mathbb{A}_2)}=
\left(
\begin{array}{cccc}
 0 & 0 & 0 & 0 \\
 0 & 0 & \frac{2 \sqrt{6}}{7} & 0 \\
 0 & \frac{2 \sqrt{6}}{7} & \frac{2}{7} & 0 \\
 0 & 0 & 0 & -\frac{4}{7}
\end{array}
\right), \hspace{0.15cm}
{\mathbb{M}}^{(\mathbb{A}_2)}=\left(
\begin{array}{cccc}
 \frac{\mathcal{M}_{1,D}}{{\det\mathcal{M}_{1}}} & -\frac{\mathcal{M}_{1,SD}}{\det\mathcal{M}_{1}} & 0 & 0 \\
 -\frac{\mathcal{M}_{1,SD}}{\det\mathcal{M}_{1}} & \frac{\mathcal{M}_{1,S}}{\det\mathcal{M}_{1}} & 0 & 0 \\
 0 & 0 &\mathcal{M}_{3,D}^{-1} & 0 \\
 0 & 0 & 0 & \mathcal{M}_{3,D}^{-1} \\
\end{array}
\right)
\ .
\label{pipipiA2}
\\
& \mathbb{E}:\hspace{0.4cm}\mathbb{F}_{00}^{(\mathbb{E})}=\textbf{I}_{6},\hspace{0.15cm}
\mathbb{F}_{40}^{(\mathbb{E})}=
\left(
\begin{array}{cccccc}
 0 & 0 & 0 & 0 & 0 & 0 \\
 0 & 0 & \frac{2 \sqrt{6}}{7} & 0 & 0 & 0 \\
 0 & \frac{2 \sqrt{6}}{7} & \frac{2}{7} & 0 & 0 & 0 \\
 0 & 0 & 0 & \frac{8}{21} & -\frac{10 \sqrt{2}}{21} & 0 \\
 0 & 0 & 0 & -\frac{10 \sqrt{2}}{21} & -\frac{2}{21} & 0 \\
 0 & 0 & 0 & 0 & 0 & -\frac{4}{7}
\end{array}
\right)
\ ,
\nonumber\\
& \hspace{0.15cm}{\mathbb{M}}^{(\mathbb{E})}=
 \left(
\begin{array}{cccccc}
  \frac{\mathcal{M}_{1,D}}{{\det\mathcal{M}_{1}}} & -\frac{\mathcal{M}_{1,SD}}{\det\mathcal{M}_{1}} & 0 & 0 & 0 & 0 \\
-\frac{\mathcal{M}_{1,SD}}{\det\mathcal{M}_{1}} & \frac{\mathcal{M}_{1,S}}{\det\mathcal{M}_{1}} & 0 & 0 & 0 & 0 \\
 0 & 0 & \mathcal{M}_{3,D}^{-1} & 0 & 0 & 0 \\
 0 & 0 & 0 & \mathcal{M}_{2,D}^{-1} & 0 & 0 \\
 0 & 0 & 0 & 0 & \mathcal{M}_{3,D}^{-1} & 0 \\
 0 & 0 & 0 & 0 & 0 & \mathcal{M}_{2,D}^{-1} \\
\end{array}
\right)
\ .
\label{pipipiE}
\end{align}
\end{footnotesize}

\end{small}

\section{Twisted $c_{lm}^{ {\bf d},{\bf\phi}_1,{\bf\phi}_2}$  Functions for Systems at Rest
\label{app:TwistC}}
\noindent
To understand the relative contributions of phase shifts beyond the $\alpha$ wave to the 
deuteron binding energy in chapter \ref{chap:TBC}, it is helpful to 
consider the expansions of the $c_{lm}^{ {\bf d},{\bf\phi}_1,{\bf\phi}_2}$  functions.
As i-PBCs, with the twist angles $\bm{\phi}=(\frac{\pi}{2},\frac{\pi}{2},\frac{\pi}{2})$, lead to the most significant reduction 
in the FV corrections, we focus on these angles in the expansions, restricting ourselves to systems at rest.
The general form of the $c_{lm}^{ {\bf d},{\bm\phi}_1,{\bm\phi}_2}$  functions for $\mathbf{d}=\mathbf{0}$ and ${\bm\phi}_1=-{\bm\phi}_2=\bm\phi$ is
\begin{eqnarray}
{c_{lm}^{\textbf{0},\bm{\phi},-\bm{\phi}}(-\kappa^2;{\rm L})}
& = & 
\frac{i^l}{\pi^{3/2}}\sum_{\mathbf{n} \neq \mathbf{0}}
\ e^{-i \mathbf{n} \cdot \bm{\phi}}\ 
Y_{lm}(\hat{\mathbf{n}})
\int_{0}^{\infty}dk~\frac{k^{l+2}}{k^2+\kappa^2}~j_l(n k {\rm L})
,
\end{eqnarray}
where $n=|{\bf n}|$.
By direct evaluation of the integral,
it is straightforward to show that
\begin{align}
&
{c_{00}^{\textbf{0},\bm{\phi},-\bm{\phi}}}
\ =\ 
-{\kappa\over 4\pi}\ +\ 
 \sqrt{4\pi}~
\sum_{\mathbf{n} \neq \mathbf{0}}
\ e^{-i \mathbf{n} \cdot \bm{\phi}}\ 
Y_{00}(\hat{\mathbf{n}})~
\frac{e^{-n \kappa {\rm L}}}{4\pi n {\rm L}}
,
\nonumber\\
&
{c_{1m}^{\textbf{0},\bm{\phi},-\bm{\phi}}}=
(i\kappa)\sqrt{4\pi}~
\sum_{\mathbf{n} \neq \mathbf{0}}
\ e^{-i \mathbf{n} \cdot \bm{\phi}}\ 
Y_{1m}(\hat{\mathbf{n}})~\left(1+\frac{1}{n \kappa {\rm L}}\right)~
\frac{e^{-n \kappa {\rm L}}}{4\pi n {\rm L}}
,
\nonumber\\
&{c_{2m}^{\textbf{0},\bm{\phi},-\bm{\phi}}}=
(i\kappa)^2\sqrt{4\pi}~
\sum_{\mathbf{n} \neq \mathbf{0}}
\ e^{-i \mathbf{n} \cdot \bm{\phi}}\ 
Y_{2m}(\hat{\mathbf{n}})~
\left(1+\frac{3}{n \kappa {\rm L}}+\frac{3}{n^2 \kappa^2 {\rm L}^2}\right)~
\frac{e^{-n \kappa {\rm L}}}{4\pi n {\rm L}}
,
\nonumber\\
&{c_{3m}^{\textbf{0},\bm{\phi},-\bm{\phi}}}=
(i\kappa)^3\sqrt{4\pi}~
\sum_{\mathbf{n} \neq \mathbf{0}}
\ e^{-i \mathbf{n} \cdot \bm{\phi}}\ 
Y_{3m}(\hat{\mathbf{n}})~
\left(1+\frac{6}{n \kappa {\rm L}}+\frac{15}{n^2 \kappa^2 {\rm L}^2}+\frac{15}{n^3 \kappa^3 {\rm L}^3}\right)~
\frac{e^{-n \kappa {\rm L}}}{4\pi n {\rm L}}
,
\nonumber\\
&{c_{4m}^{\textbf{0},\bm{\phi},-\bm{\phi}}}=
(i\kappa)^4\sqrt{4\pi}~
\sum_{\mathbf{n} \neq \mathbf{0}}
\ e^{-i \mathbf{n} \cdot \bm{\phi}}\ 
Y_{4m}(\hat{\mathbf{n}})~
\left(1+\frac{10}{n\kappa {\rm L}} 
+\frac{45}{n^2\kappa^{2}{\rm L}^2}+\frac{105}{n^3\kappa^{3}{\rm L}^3}+\frac{105}{n^4\kappa^{4}{\rm L}^4}\right)~
\nonumber\\
&\qquad \qquad \qquad \qquad \qquad \qquad \qquad \qquad \qquad \qquad \qquad \qquad \qquad \qquad \qquad
\times \frac{e^{-n\kappa {\rm L}}}{4\pi n{\rm L}}.
\end{align}
\begin{table}[b!]
\begin{centering}
\renewcommand{\arraystretch}{0.0}
\scalebox{1.0}{
\begin{tabular}{|c||c|c|c|c|}
\hline 
$(l,m)$ & $\alpha^{(1)}_{lm}$ & $\alpha^{(\sqrt{2})}_{lm}$ & $\alpha^{(\sqrt{3})}_{lm}$ & $\alpha^{(2)}_{lm}$\tabularnewline
\hline 
\hline 
$(0,0)$ & $0$ & $0$ & $0$ & $-\frac{3}{\sqrt{\pi}}$\tabularnewline
\hline
$(1,0)$ & $-i\sqrt{\frac{3}{\pi}}$ & $0$ & $0$ & $0$\tabularnewline
\hline 
$(2,2)$ & $0$ & $-i\sqrt{\frac{15}{2\pi}}$ & $0$ & $0$\tabularnewline
\hline 
$(3,0)$ & $-i\sqrt{\frac{7}{\pi}}$ & $0$ & $0$ & $0$\tabularnewline
\hline 
$(3,2)$ & $0$ & $0$ & $-\frac{2}{3}\sqrt{\frac{70}{\pi}}$ & $0$\tabularnewline
\hline 
$(4,0)$ & $0$ & $0$ & $0$ & $-\frac{21}{4\sqrt{\pi}}$\tabularnewline
\hline 
$(4,2)$ & $0$ & $i\frac{3}{2}\sqrt{\frac{5}{2\pi}}$ & $0$ & $0$\tabularnewline
\hline 
\end{tabular}
}
\caption{Coefficients of independent, non-vanishing terms in the expansion of $F_{lm}$ given in Eq.~(\protect\ref{eq:Flm})}
\label{tab:alphas}
\par\end{centering}
\end{table}
These functions are of the form
\begin{eqnarray}
F_{lm} & = & 
\sum_{\mathbf{n} \neq \mathbf{0}} e^{-i \mathbf{n} \cdot \bm{\phi}}
Y_{1m}(\hat{\mathbf{n}})f(n)
= \alpha^{(1)}_{lm} f(1)+\alpha^{(\sqrt{2})}_{lm}  f(\sqrt{2})+\alpha^{(\sqrt{3})}_{lm} f(\sqrt{3}) + \alpha^{(2)}_{lm} f(2)
+ \cdots
.
\nonumber\\
\label{eq:Flm}
\end{eqnarray}
The independent and non-vanishing coefficients $\alpha^{(n)}$ are presented in Table~\ref{tab:alphas} for the twist angles 
$\bm{\phi}=(\frac{\pi}{2},\frac{\pi}{2},\frac{\pi}{2})$. The remaining coefficients are dictated by  symmetry,
\begin{eqnarray}
&&F_{1\pm 1}\ = \ \mp e^{\pm i \pi/4}\ F_{10}
,
\nonumber\\
&& F_{2+2} \ = \  -F_{2-2} \ =\ -{1\over\sqrt{2}}\ e^{\pm i \pi/4}\ F_{2\pm 1}
,
\nonumber\\
&& F_{30} \ = \ \mp {4\over\sqrt{10}} e^{\pm i \pi/4} \ F_{3\pm 3}
\ =\  
\pm {4\over\sqrt{6}} e^{\mp i \pi/4} \ F_{3\pm 1}
,\ \ 
F_{3-2}\ =\ -F_{3+2}
,
\nonumber\\
&& F_{4 +2} \ = \ -F_{4 -2} \ =\ 
-{2\over\sqrt{7}}  e^{\mp i \pi/4} \ F_{4\pm 3} 
\ =\  2  e^{\pm i \pi/4} \ F_{4\pm 1} 
,\ \ 
F_{40}\ =\ {\sqrt{14\over 5}}\ F_{4\pm4}
.
\nonumber\\
\label{eq:alpharels}
\end{eqnarray}
The coefficients presented in Table~\ref{tab:alphas} and Eq.~(\ref{eq:alpharels}) show that the leading volume dependences of the 
$c_{lm}^{ {\bf 0},{\bm\phi},-{\bm\phi}}$  functions for i-PBCs are
$c_{00} =-\frac{\kappa}{4\pi}+\mathcal{O}({e^{-2\kappa {\rm L}}}/{{\rm L}})$,
$c_{10} =\mathcal{O}({e^{-\kappa {\rm L}}}/{{\rm L}})$,
$c_{22} =\mathcal{O}({e^{-\sqrt{2}\kappa {\rm L}}}/{{\rm L}})$,
$c_{30} =\mathcal{O}({e^{-\kappa {\rm L}}}/{{\rm L}})$,
$c_{32} =\mathcal{O}({e^{-\sqrt{3}\kappa {\rm L}}}/{{\rm L}})$,
$c_{40} =\mathcal{O}({e^{-2\kappa {\rm L}}}/{{\rm L}})$
and
$c_{42} =\mathcal{O}({e^{-\sqrt{2}\kappa {\rm L}}}/{{\rm L}})$.
As the P-wave contribution to the FV spectra is due to non-zero $c_{1m}$ and $c_{3m}$ functions,
they provide the dominant corrections to the approximate QC in Eq.~(\ref{eq:QC3by3}).

A numerical comparison between these expansions and an exact evaluation of the $c_{lm}^{ {\bf d},{\bm\phi}_1,{\bm\phi}_2}$ functions
reveals that the expansions are only slowly convergent~\cite{Briceno:2013bda}.
Precisions extractions of the energy eigenvalues require the use of the exact evaluations, even in modest volumes.


\vita{Zohreh was born in summer of 1985 in a small mountain-side town in the beautiful Province of Lorestan, western Iran. She was raised by two wonderful  parents who happened to be two successful teachers at work and whose passionate for teaching and intellectually nourishing small humans would not go away when come to home to their children. As a result Zohreh's childhood home was no different than a fun school, with lots of playing times and joyful moments with her five siblings, as well as lots of pondering and studying times. 

While most little girls grow up playing with dolls and reading fairy tales, Zohreh never got the chance to experience that kind of childhood. Instead of dolls and Barbies she ended up playing with, and taking care of, real babies (her siblings) whom she and her older sister were put in charge of while their parents were at work. Instead of being told fairy tales, she ended up reading through his father's textbook and lecture notes as he had gone back to collage to get a degree in theoretical physics and would fill the whole house with mysterious books on quantum mechanics and Einstein's relativity. Surely the little girl could not stand to stay away from such interesting materials, sadly though, the only parts of the books she ended up, more or less, comprehending was the beginning of the chapters -- before complicated equations start filling the pages! Gladly, that turned out to be sufficient to inspire her with the beauties that physics had to offer. These unique experiences all together started to give her a different perspective of life.

Zohreh chose, not to surprise of her parents, theoretical physics as her major in college. This was despite the overwhelming encouragements from her teachers and friends to become an electrical/computer engineer given how high prestige and high demand those majors were among college students in Iran at the time. It was not a matter of doubt to her though that being a physicist is what makes her happy and self-satisfied throughout her life, and that Sharif University of Technology in the capital city of Tehran is the best place in the country to start this long journey from. Coming from a small town, with limited resources and opportunities, to Sharif, and being surrounded by lots of young smart people that were as excited and as driven about Physics as she was turned out to be one of the best experiences of her life.

After getting a baccalaureate (2007) and a masters (2009) degree from Sharif, the desire of coming to the scientists' dreamland, the US, and being able to learn from and work with world-class physicists, and to experience the frontline of research and discovery, left Zohreh no choice to leave behind her beloved family, her incredible professors at Sharif and her sweet memories of a country full of joy and peace,\footnote{Despite all political complications.} and move to the US. She arrived to the US in summer of 2009 along with her husband to start her PhD. After a year in the ultra-cold state of Minnesota, she moved to the University of Washington and met some of the most influential people in her life. These were physicists, and above all of whom her advisor, from whom she learned how to think properly about the most interesting problems in nature, how to work and collaborate effectively with others and how to develop appreciation for all the good work that is done in every area of physics. In her long-started journey from town to town, city to city and country to country, Zohreh is, once again, leaving a lot of good people and good memories in Seattle and is moving to Boston to start her post-doc position in the Center for Theoretical Physics at MIT.}

\end{document}